\documentclass[12pt,twoside]{book}
\pdfoutput=1

\usepackage{graphicx}
\usepackage{amssymb,amsmath}
\usepackage{multirow}
\usepackage{cite,color,url}
\usepackage{framed,fancybox,xcolor}
\usepackage[titletoc]{appendix}
\usepackage{xfrac} 
\usepackage{arydshln}
\usepackage{tikz-cd}
\usepackage{float}
\usepackage{enumitem}
\usepackage{bbold} 

\usepackage[nottoc,notlot,notlof]{tocbibind}

\usepackage[colorlinks=true
,urlcolor=blue
,anchorcolor=blue
,citecolor=blue
,filecolor=blue
,linkcolor=blue
,menucolor=blue 
,linktocpage=true
,pdfproducer=medialab
,pdfa=true
]{hyperref}

\usepackage[square,numbers,compress,sort]{natbib}
\allowdisplaybreaks

\usepackage{fancyhdr}

\usepackage{soul}
\usepackage{slashed}
\usepackage[makeroom]{cancel}
\usepackage[tableposition=top]{caption}
\usepackage{booktabs}
\usepackage{titletoc}

\def\be{\begin{equation}}
\def\ee{\end{equation}}
\def\ba{\begin{align}}
\def\ea{\end{align}}

\newcommand{\non}{\nonumber}
\renewcommand\eqref[1]{Eq.~(\ref{#1})}
\newcommand\eqrefs[1]{Eqs.~(\ref{#1})}
\newcommand\figref[1]{Fig.~\ref{#1}}
\newcommand\figrefs[1]{Figs.~\ref{#1}}
\newcommand\tabref[1]{Table~\ref{#1}}

\newcommand\appref[1]{App.~\ref{#1}}
\newcommand\chref[1]{Chapter~\ref{#1}}
\newcommand\secref[1]{Section~\ref{#1}}

\graphicspath{{./figs/}}
\begin{document}



\pagestyle{headings}
\pagestyle{empty}

\vspace*{-2.5cm}
\newcommand{\HRule}{\rule{\linewidth}{1mm}}
\setlength{\parindent}{1cm}
\setlength{\parskip}{1mm}
\noindent
\HRule
\begin{center}
\Huge{\textbf{Lepton flavor violation from low scale seesaw neutrinos with masses reachable at the LHC}}
 \\ [5mm]
\end{center}
\HRule

\vspace{.3cm}

\begin{center}

	   Memoria de tesis doctoral realizada por \\\textbf{Xabier Marcano Imaz} \\
	   presentada ante el Departamento de F\'isica Te\'orica\\
	   de la Universidad Aut\'onoma de Madrid\\
	   para optar al T\'itulo de Doctor en F\'isica
	   
\vspace{0.5cm}

Trabajo dirigido por la\\
\textbf{Dra. Mar\'ia Jos\'e Herrero Solans} \\
Profesora Catedr\'atica del Departamento de F\'isica Te\'orica \\
Universidad Aut\'onoma de Madrid\\

\vspace{0.3cm}

\end{center}

\begin{figure}[ht]
\centering
\begin{tabular}{lr}

\includegraphics[scale=0.12]{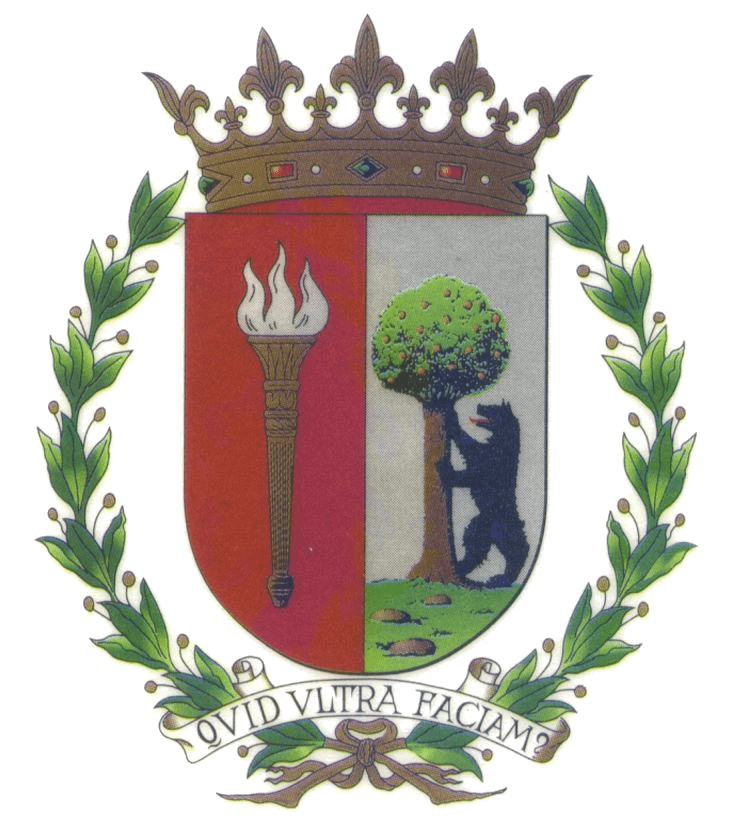}\qquad\qquad\qquad
&
\qquad\qquad\qquad\includegraphics[scale=0.54]{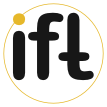}

\end{tabular}
\end{figure}


\begin{center}
{\Large {Departamento de F\'isica Te\'orica\\ Universidad Aut\'onoma de Madrid }}\\
\vspace{0.1cm}
{\Large {Instituto de F\'isica Te\'orica\\ \vspace{0.1cm} UAM-CSIC}}\\

\end{center}

\begin{center}
Madrid, septiembre de 2017
\end{center}

\newpage
\thispagestyle{empty}
\phantom{asdfaf}


\newpage
\thispagestyle{empty}
\vspace*{5cm}
\begin{flushright}
{\it
Nire aitxitxari,\\
honaino bere erruz heldu naizelako.
}
\end{flushright}
%

%
%
%



\newpage
\thispagestyle{empty}
\tableofcontents

\newpage


\pagestyle{fancy}
\fancyhead{}
\fancyfoot{}

\fancyfoot[C]{\thepage}
\renewcommand{\headrulewidth}{0.3pt}
\renewcommand{\footrulewidth}{0pt}

\pagenumbering{arabic}
\setcounter{page}{1}


\chapter*{Introducci\'on}
\label{Introduccion}
\addcontentsline{toc}{chapter}{Introducci\'on}
\fancyhead[RO] {\scshape Introducci\'on }

El 4 de julio de 2012, los experimentos ATLAS y CMS, situados en el gran colisionador de hadrones (LHC) del CERN, anunciaron el descubrimiento~\cite{Aad:2012tfa,Chatrchyan:2012xdj} del eslab\'on perdido en el Modelo Est\'andar (SM)~\cite{Glashow:1961tr,GellMann:1964nj,Weinberg:1967tq,Salam:1968rm} de las interacciones fundamentales, el bos\'on de Higgs. 
Este descubrimiento complet\'o una de las teor\'ias m\'as exitosas en los anales de la F\'isica, un modelo predictivo capaz de describir con extraordinaria precisi\'on la mayor\'ia de los fen\'omenos conocidos en la F\'isica de Part\'iculas.
Sin embargo, existen algunas evidencias experimentales, como las masas de los neutrinos, la materia oscura o la asimetr\'ia bari\'onica del universo, y ciertos problemas te\'oricos, tales como el problema de las jerarqu\'ias, el problema de $CP$ fuerte o el puzzle de sabor, que el SM no es capaz de explicar, invit\'andonos a un viaje hacia  nueva f\'isica m\'as all\'a del SM.

El SM es una teor\'ia cu\'antica de campos, basada en el grupo de simetr\'ia gauge  $SU(3)_C \times SU(2)_L\times U(1)_Y$, que describe tres de las cuatro interacciones fundamentales que conocemos entre las part\'iculas elementales, las denominadas interaci\'on fuerte, d\'ebil y electromagn\'etica. 
En su espectro de part\'iculas podemos encontrar a los fermiones, constituyentes de la materia, a los bosones de esp\'in uno, mensajeros de las fuerzas, y al bos\'on escalar de Higgs, el remanente del proceso que genera la masa de las part\'iculas elementales, el mecanismo de Brout-Englert-Higgs (BEH)~\cite{Higgs:1964ia,Englert:1964et,Guralnik:1964eu,Higgs:1966ev}. 
Este mecanismo explica c\'omo la ruptura espont\'anea de la simetr\'ia electrod\'ebil (EWSB) genera las masas que observamos para los bosones gauge $W$ y $Z$, as\'i como para todos los fermiones salvo los neutrinos, que son tratados como part\'iculas sin masa por el SM. 

Hist\'oricamente, los experimentos dedicados a la b\'usqueda de procesos con violaci\'on de sabor han sido una pieza fundamental en los avances te\'oricos en la F\'isica de Part\'iculas, tales como el mecanismo de Glashow-Iliopoulos-Maiani (GIM)~\cite{Glashow:1970gm}, para explicar por qu\'e no se ve\'ian las corrientes neutras con cambio de sabor, o la matriz de Cabibbo-Kobayashi-Maskawa (CKM)~\cite{Cabibbo:1963yz,Kobayashi:1973fv} de mezcla de los quarks para las corrientes cargadas.
Asimismo, la evidencia experimental m\'as clara de la existencia de nueva f\'isica proviene de la f\'isica del sabor, en concreto de la observaci\'on de violaci\'on de sabor lept\'onico (LFV) en el sector de los neutrinos. 
Tal y como acabamos de decir, el SM se construy\'o asumiendo que los neutrinos no ten\'ian masa.
Sin embargo, las evidencias experimentales de LFV en las oscilaciones de neutrinos, observadas por primera vez por las colaboraciones de Super-Kamiokande~\cite{Fukuda:1998mi} y SNO~\cite{Ahmad:2001an,Ahmad:2002jz}, han demostrado que los neutrinos s\'i tienen masa, estableciendo as\'i la necesidad de modificar el SM para explicar tal hecho.
Del mismo modo, si alguno de los experimentos actuales o futuros detectase alguna se\~nal de los procesos con LFV en el sector cargado, todav\'ia no observados en la naturaleza, se abrir\'ia una nueva ventana a la f\'isica m\'as all\'a del SM,  m\'as all\'a incluso de la f\'isica de la masa de los neutrinos. 

Estudiando el espectro de part\'iculas del SM, puede comprobarse que la ausencia de neutrinos dextr\'ogiros (RH) en la teor\'ia proh\'ibe que los neutrinos interact\'uen con el campo de Higgs y, por lo tanto, que adquieran masa despu\'es del EWSB. 
As\'i, la manera m\'as simple de introducir masas para los neutrinos en el SM ser\'ia a\~nadir los neutrinos RH, $\nu_R$, que le faltan.
De esta forma, los neutrinos podr\'ian interactuar con el campo de Higgs a trav\'es de un acoplamiento de tipo Yukawa, $Y_\nu$, y obtener una masa de Dirac $m_D=vY_\nu$  tras el EWSB, como el resto de fermiones, siendo \'esta proporcional al valor esperado del vac\'io (vev) del Higgs, cuya normalizaci\'on tomamos como $v=174$~GeV. 
Sin embargo, esta extensi\'on m\'inima del SM plantea nuevas preguntas: ?`Por qu\'e son los neutrinos tan diferentes a los dem\'as fermiones?
?`Por qu\'e son sus masas tan peque\~nas comparadas con el resto de las masas de los otros fermiones?
Adem\'as, dado que los neutrinos no tienen carga de color ni carga el\'ectrica, podr\'ian ser fermiones de Majorana, i.e., podr\'ian ser sus propias antipart\'iculas.
De ser \'este el caso, supondr\'ia una gran novedad para la F\'isica de Part\'iculas. 

Escudri\~nando m\'as profundamente los nuevos campos $\nu_R$, podemos darnos cuenta de que son singletes bajo todo el grupo de simetr\'ia gauge del SM y, por tanto, no existe nada que proh\'iba escribir  un t\'ermino de masa Majorana, $m_M$, para ellos. 
Estas dos masas distintas, $m_D$ y $m_M$, son los ingredientes b\'asicos del bien conocido modelo del {\it seesaw} de tipo-I~\cite{Minkowski:1977sc,GellMann:1980vs,Yanagida:1979as,Mohapatra:1979ia,Schechter:1980gr}, que normalmente asume una escala para la masa de Majorana $m_M$ mucho m\'as pesada que $m_D$ y que la escala electrod\'ebil $v$. 
En tal caso, el espectro de neutrinos f\'isicos consistir\'ia en un neutrino de Majorana pesado y otro ligero por cada generaci\'on, con masas del orden de $m_M$ y $m_D^2/m_M$, respectivamente. 
As\'i, el {\it seesaw} de tipo-I explica de forma elegante por qu\'e la masa de los neutrinos que observamos es tan peque\~na, ya que \'esta surgir\'ia del cociente entre dos escalas muy distantes, $m_D$ y $m_M$.
Analizando la escala de la masa ligera $m_\nu\sim m_D^2/m_M$, puede comprobarse que para tener $m_\nu\sim\mathcal O$(eV), como sugieren los experimentos, con acoplamientos Yukawa grandes $Y_\nu\sim\mathcal O(1)$, se necesitan masas pesadas del orden de $m_M\sim\mathcal O(10^{14}~{\rm GeV})$. 
Por otro lado, masas de neutrinos dextr\'ogiros en el rango de los TeV requerir\'ian acoplamientos muy peque\~nos $Y_\nu\sim\mathcal O(10^{-5})$.
De una manera o de otra, la mayor parte de la fenomenolog\'ia en el {\it seesaw} tipo-I est\'a muy suprimida.
Por ejemplo, correcciones peque\~nas en este tipo de modelos a la masa del Higgs han sido encontradas en un trabajo complementario no incluido en esta Tesis~\cite{Heinemeyer:2014hka}.
Por tanto, la simplicidad de este argumento es lo que hace este modelo atractivo y, a su vez, dif\'icil de testar en otros observables de baja energ\'ia m\'as all\'a de la propia masa de los neutrinos ligeros.

Los modelos {\it seesaw} de baja escala son variaciones interesantes del {\it seesaw} tipo-I que ofrecen una fenomenolog\'ia mucho m\'as rica.
En este tipo de modelos, se recurre a una nueva simetr\'ia con el objetivo de evitar que los neutrinos ligeros adquieran una masa demasiado grande, permitiendo por tanto que los neutrinos pesados puedan tener masas m\'as ligeras y acoplamientos Yukawa grandes al mismo tiempo.
Una realizaci\'on particular de este tipo de modelos, en el que se centra esta Tesis, es el {\it seesaw} inverso (ISS)~\cite{Mohapatra:1986aw,Mohapatra:1986bd,Bernabeu:1987gr}, que asume una simetr\'ia aproximada de conservaci\'on de n\'umero lept\'onico (LN).
En el l\'imite en el que el LN est\'a conservado de forma exacta, los neutrinos ligeros no tienen masa.
Sin embargo, si esta simetr\'ia est\'a rota por cierto par\'ametro de masa, los neutrinos adquieren una peque\~na masa de Majorana proporcional a este par\'ametro de violaci\'on del LN.
Por otro lado, podemos esperar que este par\'ametro sea peque\~no de forma natural, en el sentido de 't Hooft~\cite{tHooft:1979rat}, ya que si fuese cero la simetr\'ia del modelo ser\'ia mayor.
Por lo tanto, en el modelo ISS la ligereza de los neutrinos est\'a relacionada con una peque\~na violaci\'on del LN.

En el modelo ISS, la condici\'on de que la violaci\'on del LN sea peque\~na se cumple introduciendo nuevos singletes fermi\'onicos en pares $(\nu_R, X)$ con LN opuesto, y asumiendo que la conservaci\'on del LN se viola s\'olo por un peque\~na masa de Majorana $\mu_X$ para los campos $X$. 
De este modo, el espectro f\'isico consiste en un neutrino de Majorana ligero con masa suprimida por el peque\~no valor de $\mu_X$, y dos neutrinos de Majorana pesados y casi degenerados, por generaci\'on. \'Estos \'ultimos forman pares de fermiones pseudo-Dirac y de hecho se comportan pr\'acticamente como fermiones de Dirac, al contrario que los neutrinos pesados del {\it seesaw} de tipo-I.
De forma interesante, como la escala $\mu_X$ asegura la ligereza de los neutrinos observados, los neutrinos pesados pueden tener grandes acoplamientos de Yukawa a los neutrinos del SM y al mismo tiempo masas del orden de varios TeV o menores, siendo as\'i accesibles en el LHC.
Esto hace del ISS un modelo atractivo e interesante con una fenomenolog\'ia muy rica que puede ser estudiada en experimentos presentes y del futuro cercano.
Esto incluye, entre otros, estudios de violaci\'on de universalidad de sabor lept\'onico en desintegraciones lept\'onicas y semilept\'onicas de mesones~\cite{Abada:2012mc, Abada:2013aba}, momentos dipolares el\'ectricos de los leptones cargados~\cite{Abada:2015trh,Abada:2016awd}, momentos magn\'eticos de los leptones~\cite{Abada:2014nwa}, producci\'on de neutrinos pesados en colisionadores \cite{Chen:2011hc,BhupalDev:2012zg,Das:2012ze,Das:2014jxa, Bambhaniya:2014kga,Bambhaniya:2014hla,Das:2015toa, Das:2016hof}, 
materia oscura~\cite{Abada:2014zra}, leptog\'enesis~\cite{Garayoa:2006xs,Abada:2015rta} y  procesos con LFV en el sector cargado~\cite{Deppisch:2004fa,Deppisch:2005zm,Dev:2009aw,Hirsch:2009ra,Abada:2011hm,Abada:2012cq,Hirsch:2012ax,Abada:2014cca,Abada:2014kba,Abada:2015oba,Abada:2015zea,Abada:2016vzu}.

Desde el punto de vista te\'orico, el SM tambi\'en posee algunas propiedades indeseadas, como el llamado problema de las jerarqu\'ias.
Este problema se refiere a la inestabilidad del sector de Higgs frente a correcciones radiativas en presencia de nueva f\'isica a una escala muy grande.
Para ilustrar esta idea, podemos considerar la masa del bos\'on de Higgs, $m_H$, y calcular sus correcciones radiativas bajo la hip\'otesis de que no hay nueva f\'isica hasta la masa de Planck $M_{\rm P}\sim10^{19}$~GeV, donde los efectos gravitacionales empiezan a desempe\~nar un papel importante.
Haciendo esto, encontramos que las correcciones cu\'anticas $\Delta m_H^2$ crecen como el cuadrado de la escala de la nueva f\'isica, lo que proporciona un valor muy alejado del medido experimentalmente $m_H=125.09\pm0.21({\rm stat.})\pm0.11({\rm syst.})$~GeV~\cite{Aad:2015zhl}.
As\'i, para obtener una predicci\'on compatible con el experimento, se requiere un ajuste muy fino en la cancelaci\'on entre la masa desnuda y las correcciones cu\'anticas, un hecho que no resulta muy natural sin una nueva explicaci\'on o simetr\'ia.

Una de las soluciones m\'as populares y elegantes a este problema la proporciona la supersimetr\'ia (SUSY)~\cite{Golfand:1971iw,Volkov:1973ix,Wess:1974tw}, una nueva simetr\'ia que relaciona fermiones y bosones.
En la extensi\'on m\'as simple y m\'inima del SM, conocida como {\it Minimal Supersymmetric Standard Model} (MSSM)~\cite{Haber:1984rc,Gunion:1984yn,Gunion:1986nh}, cada fermi\'on del SM tiene un compa\~nero de esp\'in cero, llamado sfermi\'on, con la misma masa y los mismos n\'umeros cu\'anticos que el fermi\'on original; de la misma manera, todos los bosones del SM tienen su compa\~nero de esp\'in un medio. 
El hecho de que existan fermiones y bosones con los mismos acoplamientos y masas cancela completamente las peligrosas correcciones cu\'anticas a la masa del bos\'on de Higgs, proporcionando as\'i una soluci\'on elegante al problema de las jerarqu\'ias. 
Obviamente, duplicar el espectro de part\'iculas del SM es una predicci\'on fenomenol\'ogica muy importante, pues supone la existencia de nuevas part\'iculas que deber\'ian observarse en los experimentos.
Desafortunadamente, \'estos no han descubierto ning\'un miembro supersim\'etrico del MSSM todav\'ia. 
Este hecho implica que, si la SUSY existe en la naturaleza, \'esta no puede ser una simetr\'ia exacta, debe estar rota, de tal forma que las part\'iculas SUSY sean m\'as pesadas que sus compa\~neras del SM.
Esta ruptura, sin embargo, no puede arruinar completamente la soluci\'on al problema de las jerarqu\'ias, por lo que la SUSY debe estar rota de manera suave~\cite{Girardello:1981wz}, i.e., las correcciones a $\Delta m_H^2$ que dependen cuadr\'aticamente con la escala de la nueva f\'isica han de seguir cancel\'andose, aunque podr\'ia quedar todav\'ia una dependencia logar\'itmica. 

Una vez a\~nadidos estos t\'erminos que rompen la SUSY de manera suave, el MSSM es una teor\'ia viable con consecuencias fenomenol\'ogicas muy interesantes. 
Sin embargo, al estar constru\'ida a partir del SM, todav\'ia requiere de un mecanismo que explique la generaci\'on de la masa de los neutrinos.
Retomando la discusi\'on anterior, podemos considerar de nuevo el modelo ISS y adaptarlo al contexto SUSY, a\~nadiendo al espectro del MSSM nuevos neutrinos y sneutrinos, cuyas masas puedan estar en el entorno del TeV y, al mismo tiempo, tener acoplamientos grandes.  
As\'i, este modelo SUSY-ISS combina las cualidades m\'as interesantes de ambos contextos, tanto de la SUSY como del ISS. 

La forma \'optima de demostrar experimentalmente la validez de cualquier modelo m\'as all\'a del SM ser\'ia detectando las nuevas part\'iculas que predice. 
No obstante, esta tarea puede ser muy tediosa en muchos casos, sobre todo si estas nuevas part\'iculas son demasiado pesadas como para producirlas directamente en los experimentos actuales, por lo que un primer indicio sobre su existencia prodr\'ia venir de sus implicaciones indirectas sobre observables de baja energ\'ia.
En el caso particular de los modelos de generaci\'on de masas para los neutrinos, los procesos con LFV podr\'ian ser de nuevo los observables id\'oneos para ver tales implicaciones indirectas, concretamente en el sector de leptones cargados, pues podr\'ian ser inducidos cu\'anticamente por {\it loops} de neutrinos pesados. 

Los procesos con LFV en el sector cargado (cLFV) est\'an proh\'idos en el SM si los neutrinos no tienen masa, y extremamente suprimidos en el caso de a\~nadir {\it ad-hoc} las peque\~nas masas de los neutrinos necesarias para explicar las oscilaciones de neutrinos. 
Por tanto, una se\~nal positiva en cualquiera de los experimentos en busca de procesos con cLFV implicar\'ia autom\'aticamente la existencia de nueva f\'isica m\'as all\'a del SM incluso con la masa de los neutrinos a\~nadida de manera m\'inima. 
A pesar de que ning\'un proceso de este tipo ha sido observado todav\'ia, se trata de un campo muy activo, explorado continuamente por un gran n\'umero de experimentos que han sido capaces de poner cotas superiores a la probabilidad de que este tipo de procesos con cLFV puedan ocurrir. 
A d\'ia de hoy, las cotas m\'as estrictas han sido encontradas para las transiciones del tipo $\mu$-$e$, como la desintegraci\'on radiativa $\mu\to e\gamma$ o la conversi\'on $\mu$-$e$ en n\'ucleos pesados, cuyas probabilidades de desintegraci\'on han sido acotadas por las colaboraciones de MEG~\cite{TheMEG:2016wtm}  y SINDRUM II~\cite{Bertl:2006up}, respectivamente, a ocurrir menos del $4.2\times10^{-13}$ y $7.0\times10^{-13}$  de las veces.
Adem\'as, se espera que la siguiente generaci\'on de experimentos sean capaces de mejorar la sensibilidad a estas transiciones $\mu$-$e$, alcanzando el impresionante rango de $10^{-18}$ para la conversi\'on $\mu$-$e$ en n\'ucleos en el experimento PRISM en el J-PARC~\cite{Alekou:2013eta}. 
Por otro lado, las cotas actuales a las transiciones en los sectores $\tau$-$\mu$ y $\tau$-$e$ son mucho m\'as suaves, siendo, por ejemplo, del orden de $10^{-8}$ para las desintegraciones con LFV del tau seg\'un los experimentos de BABAR~\cite{Aubert:2009ag} y BELLE~\cite{Hayasaka:2010np,Hayasaka:2010et}.
Esto quiere decir que existe algo m\'as de espacio para transiciones con LFV en estos sectores que en el de $\mu$-$e$, que podr\'an ser exploradas en un futuro cercano por el experimento BELLE-II~\cite{Aushev:2010bq}.

As\'imismo, en la actualidad el LHC est\'a en funcionamiento y tambi\'en tiene mucho que decir sobre procesos con cLFV. 
En primer lugar, el hecho de haber descubierto una nueva part\'icula, el bos\'on de Higgs, abre una nueva ventana a posibles desviaciones del SM que definitivamente debe ser explorada.
En particular, este descubrimiento a\~nade  al mercado tres nuevos canales con cLFV, las desintegraciones LFV del bos\'on de Higgs a dos leptones de distinto sabor, $H\to\ell_k\bar\ell_m$, que de hecho ya han sido buscadas por los experimentos de CMS~\cite{Khachatryan:2015kon,Khachatryan:2016rke,CMS:2017onh} y ATLAS~\cite{Aad:2016blu}. 
A pesar de que CMS observ\'o un peque\~no pero interesante exceso en el canal $H\to\tau\mu$ tras analizar los datos de la etapa-I~\cite{Khachatryan:2015kon}, \'este no ha sido confirmado con los nuevos datos de la etapa-II y, actualmente, han extra\'ido una cota superior a este proceso de  $2.5\times10^{-3}$~\cite{CMS:2017onh}.
Estas b\'usquedas de las desintegraciones con LFV del Higgs, al igual que otras exploraciones del sector de Higgs, seguir\'an en el LHC y  que ser\'an mejoradas con m\'as datos. 

Otros observables interesantes que el LHC tambi\'en est\'a busc\'ando son las desintegraciones con LFV del bos\'on $Z$ en dos leptones de diferente sabor,  $Z\to\ell_k\bar\ell_m$~\cite{Aad:2014bca,Aad:2016blu}.
Es interesante destacar que, tras la etapa-I, ATLAS ha conseguido alcanzar ya las sensibilidades anteriores del experimento LEP~\cite{Akers:1995gz,Abreu:1996mj}, e incluso de mejorar las cotas para el canal $Z\to\mu e$.
Al igual que las b\'usquedas del Higgs, las desintegraciones con LFV del $Z$ tambi\'en continuar\'an durante las nuevas etapas, por lo que podemos esperar nuevos resultados interesantes por parte de ATLAS y CMS.
A\'un as\'i, la mejor sensibilidad a las desintegraciones con LFV del $Z$ se espera que la provea la siguiente generaci\'on de colisionadores de leptones, dado que pueden operar como factor\'ias de producci\'on de bosones $Z$ en un entorno muy limpio. 
En particular, en los futuros colisionadores lineales, que esperan alcanzar sensibilidades de hasta $10^{-9}$~\cite{Wilson:I, Wilson:II}, o en los futuros colisionadores circulares $e^+ e^-$ (como el FCC-ee (TLEP)~\cite{Blondel:2014bra}), donde se estima que se podr\'ian producir un total de $10^{13}$ bosones  $Z$ y que las sensibilidades a las desintegraciones LFV del $Z$ podr\'ian ser mejoradas hasta  $10^{-13}$.

La motivaci\'on principal de esta Tesis es, por tanto, la de explorar la conexi\'on entre la existencia de nuevos neutrinos dextr\'ogiros con masas en el entorno del TeV, accesibles en el LHC, y la existencia de LFV en el sector de los leptones cargados. 
Estamos particularmente interesados en los dos modelos descritos anteriormente, el ISS y el SUSY-ISS, ya que contienen precisamente neutrinos dextr\'ogiros con masas en el rango del TeV y, al mismo tiempo, con acoplamientos grandes.  
La fenomenolog\'ia asociada al cLFV en modelos con neutrinos masivos ha sido estudiada anteriormente
\cite{Mann:1983dv,Dittmar:1989yg,Pilaftsis:1992st,Korner:1992zk,Korner:1992an,Ilakovac:1994kj,Ilakovac:1999md,Illana:1999ww,Illana:2000ic,Casas:2001sr,Illana:2003pj,Arganda:2004bz,Arganda:2005ji,Antusch:2006vw,Abada:2007ux,Arganda:2007jw,Arganda:2008jj,Abada:2008ea,Gavela:2009cd,Ilakovac:2009jf,Herrero:2009tm,Casas:2010wm,Dinh:2012bp,Alonso:2012ji,Ilakovac:2012sh,Herrero-Garcia:2016uab},
 tambi\'en en el ISS
\cite{Deppisch:2004fa,Deppisch:2005zm,Dev:2009aw,Hirsch:2009ra,Abada:2011hm,Abada:2012cq,Hirsch:2012ax,Abada:2014cca,Abada:2014kba,Abada:2015oba,Abada:2015zea,Abada:2016vzu}.
En esta Tesis, nos concentramos mayormente en el estudio de las predicciones a las desintegraciones con LFV de los bosones de Higgs y $Z$, procesos que, como dec\'iamos, son extremadamente oportunos de estudiar a la luz de los nuevos datos del LHC.
Adem\'as, tambi\'en realizamos nuevas predicciones para otros procesos  con cLFV, tales como $\ell_m\to\ell_k\gamma$ o $\ell_m\to\ell_k\ell_k\ell_k$, junto con otros procesos que preservan el sabor lept\'onico y que son relevantes en el contexto de los modelos que consideramos. 

Estudiamos, de manera completa y por primera vez, las desintegraciones con LFV del bos\'on de Higgs en presencia de neutrinos dextr\'ogiros en el modelo ISS, as\'i como en presencia de sneutrinos en el modelo SUSY-ISS.
Realizamos este estudio siguiendo dos estrategias diferentes. 
Primero presentamos, bas\'andonos en los resultados de la Ref.~\cite{Arganda:2004bz} para el modelo {\it seesaw} de tipo-I, los resultados del c\'alculo completo a nivel de un {\it loop} realizado en la base de masa de los neutrinos f\'isicos.
En segundo lugar, usamos la t\'ecnica de la aproximaci\'on de inserci\'on de masa (MIA), que se basa en trabajar en la base electrod\'ebil y permite obtener f\'ormulas \'utiles y simples para las desintegaciones con LFV del Higgs. 
En la evaluaci\'on num\'erica, demostramos que es posible obtener tasas de desintegraci\'on grandes para los procesos con LFV en direcciones particulares del espacio de par\'ametros donde las transiciones $\mu$-$e$, las m\'as acotadas experimentalmente, est\'an muy suprimidas.
En esta l\'inea, proponemos una nueva forma para construir este tipo de escenarios fenomenol\'ogicamente interesantes que suprimen las transiciones en el sector $\mu$-$e$. 
Estos escenarios est\'an basados en la parametrizaci\'on $\mu_X$, que tambi\'en es una nueva y genu\'ina contribuci\'on de esta Tesis. 
Por otro lado, las desintegraciones con LFV del bos\'on $Z$ en el modelo ISS han sido estudiadas previamente en la Ref.~\cite{Abada:2014cca}.  
As\'i pues, en este caso nos centramos directamente en el estudio de los escenarios particulares con las transiciones $\mu$-$e$ suprimidas, ya que sabemos que las predicciones para las tasas de desintegraci\'on son mayores. 
En este sentido, realizamos un estudio complementario al hecho con anterioridad en la literatura. 

Tal y como hemos dicho antes, nuestro inter\'es en los modelos con neutrinos dextr\'ogiros con masas del orden del TeV radica en el hecho de que podr\'ian ser producidos en el LHC. 
En el modelo ISS, debido al car\'acter pseudo-Dirac de los neutrinos pesados, las b\'usquedas est\'andares de neutrinos de Majorana~\cite{Pilaftsis:1991ug,Datta:1993nm,delAguila:2007qnc,Atre:2009rg}, basadas en procesos con violaci\'on de n\'umero lept\'onico en el estado final de dos leptones con la misma carga, no son eficientes. 
Por tanto, aqu\'i proponemos el uso alternativo de estados finales con violaci\'on de sabor lept\'onico para discriminar los eventos de producci\'on y desintegraci\'on de neutrinos de modelos de {\it seesaw} de baja escala. 
En particular, nos centramos en eventos ex\'oticos con estados finales de $\tau^\pm\mu^\mp jj$ o $\tau^\pm e^\mp jj$ y sin energ\'ia transversa perdida, que podr\'ian ser producidos en escenarios en los que los $\nu_R$ tienen masas en y por debajo de la escala del TeV y donde la LFV est\'a favorecida en el sector $\tau$-$\mu$ o $\tau$-$e$, respectivamente.

Esta Tesis est\'a organizada de la siguiente manera. 
En el Cap\'itulo~\ref{Models} repasamos la f\'isica de las oscilaciones de neutrinos y su conexi\'on con la necesidad de introducir masas para los mismos. 
Discutimos algunos de los modelos de generaci\'on de masas de neutrinos m\'as populares, centr\'andonos especialmente en los modelos con neutrinos dextr\'ogiros con masas en el entorno del TeV, como es el caso del modelo del ISS  y de su versi\'on SUSY. 
Estos son los modelos que consideraremos para explorar en detalle su fenomenolog\'ia con LFV en los Cap\'itulos posteriores.
Adem\'as, durante el estudio del sector de los neutrinos en estos modelos, presentamos una nueva parametrizaci\'on, a la que nos referimos como la parametrizaci\'on $\mu_X$, que resultar\'a extremadamente \'util a la hora de explorar el espacio de par\'ametros del modelo asegurando siempre el acuerdo con los datos de las oscilaciones de neutrinos. 

En el Cap\'itulo~\ref{PhenoLFV} abordamos la importancia de la b\'usqueda de nueva f\'isica en los procesos con LFV en el sector cargado y resumimos el estado experimental actual. 
Posteriormente, repasamos las desintegraciones con LFV de los leptones, en concreto las radiativas $\ell_m\to\ell_k\gamma$ y las de tres cuerpos $\ell_m\to\ell_k\ell_k\ell_k$, con $k\neq m$.
El estudio de estos procesos nos permitir\'a aprender sobre el comportamiento de la LFV con los par\'ametros del ISS, as\'i como resaltar las ventajas a la hora de usar nuestra parametrizaci\'on $\mu_X$.
Como resultado de este estudio, encontraremos algunos escenarios fenomenol\'ogicamente interesantes, motivados por las cotas experimentales actuales, en los que se favorecen las transiciones con LFV en el sector $\tau$-$\mu$ o $\tau$-$e$ a la vez que se suprimen las del sector $\mu$-$e$, y que nos resultar\'an muy \'utiles a la hora de explorar las desintegraciones con LFV del $H$ y el $Z$.
Adem\'as, en este Cap\'itulo tambi\'en discutimos las implicaciones de los neutrinos dextr\'ogiros con masas del orden del TeV en otros observables de baja energ\'ia, que acotar\'an nuestras b\'usquedas en los siguientes Cap\'itulos de las tasas de desintegraci\'on m\'aximas permitidas por los datos experimentales actuales. 

El Cap\'itulo~\ref{LFVHD} est\'a dedicado al estudio de las desintegraciones con LFV del Higgs (LFVHD) en los modelos del ISS y SUSY-ISS.
En \'el presentamos los resultados del c\'alculo completo a nivel de un {\it loop} de las tasas de LFVHD en el modelo ISS y estudiamos sistem\'aticamente su dependencia con los par\'ametros de este modelo.
Con el fin de entender mejor estos resultados, realizamos un c\'alculo completo e independiente de estas tasas de desintegraci\'on usando la aproximaci\'on de inserci\'on de masa, lo que nos permitir\'a obtener una f\'ormula muy simple para el v\'ertice efectivo $H\ell_k\ell_m$ con LFV, extremadamente \'util para quien desee realizar una estimaci\'on r\'apida de las LFVHD en este tipo de modelos. 
Igualmente, exploramos estos procesos en el modelo SUSY-ISS, demostrando que los nuevos {\it loops} SUSY que incluyen sleptones y sneutrinos pueden aumentar notablemente los valores m\'aximos permitidos, llegando a valores cercanos a las sensibilidades experimentales actuales.

En el Cap\'itulo~\ref{LFVZD} estudiamos las desintegraciones con LFV del bos\'on $Z$ (LFVZD) en el modelo ISS.  
Centramos nuestro an\'alisis en los escenarios fenomenol\'ogicos introducidos previamente, donde pueden obtenerse tasas de desintegraci\'on con  LFV grandes en los sectores $\tau$-$\mu$ y $\tau$-$e$. 
Comparamos nuestros resultados con los trabajos previos en la literatura y demostramos que, en estas interesantes direcciones del espacio de par\'ametros del ISS, se pueden conseguir tasas de desintegraci\'on altas, perfectamente alcanzables por la siguiente generaci\'on de experimentos en busca de las LFVZD.

En el Cap\'itulo~\ref{LHC} nos centramos en la producci\'on en el LHC de los neutrinos de los modelos {\it seesaw} de baja escala.
Estudiamos la posibilidad de detectar la producci\'on y desintegraci\'on de los neutrinos pesados del ISS buscando eventos ex\'oticos con LFV, $\ell_k^\pm\ell_m^\mp jj$ con $k\neq m$, y sin energ\'ia perdida transversa. 
De forma alternativa a las b\'usquedas est\'andares de neutrinos de Majorana en los colisionadores, que no son eficientes en el modelo ISS, nuestra propuesta trata de aprovechar el hecho de que los nuevos neutrinos pesados pueden tener una estructura de sabor no trivial, produciendo as\'i este tipo de se\~nales interesantes con LFV.
Concretamente, aplicaremos esta idea a la producci\'on de eventos ex\'oticos $\tau\mu jj$ en los escenarios introducidos anteriormente, encontrando resultados prometedores para las  futuras etapas del LHC.

Finalmente, resumiremos las conclusiones m\'as relevantes de este trabajo en la parte final de este documento.

Los contenidos presentados en esta Tesis, contemplados a lo largo de los Cap\'itulos 1-5, las Conclusiones y los Ap\'endices, son trabajos originales que han sido publicados en los art\'iculos de revista de las Refs.~\cite{Arganda:2014dta,Arganda:2015naa,Arganda:2015ija,DeRomeri:2016gum,Arganda:2017vdb} y en el los art\'iculos de actas de conferencias de las Refs.~\cite{Arganda:2014via,Arganda:2014oga,Arganda:2016snv,DeRomeri:2017ucv}.


\chapter*{Introduction}
\label{Introduction}
\addcontentsline{toc}{chapter}{Introduction}
\fancyhead[RO] {\scshape Introduction }

The 4th of July of 2012, the ATLAS and CMS collaborations  at the CERN Large Hadron Collider (LHC) announced the discovery~\cite{Aad:2012tfa,Chatrchyan:2012xdj} of the last missing piece of the Standard Model (SM)~\cite{Glashow:1961tr,GellMann:1964nj,Weinberg:1967tq,Salam:1968rm} of fundamental interactions, the Higgs boson. 
This discovery completed one of the most successful theories in the annals of Physics, a predictive model able to describe with an extraordinary precision most of the known phenomena in Particle Physics.
Nevertheless, there are at present some experimental evidences, as neutrino masses, dark matter or the baryon asymmetry of the universe, and theoretical issues, like the hierarchy problem, the strong-CP problem or the flavor puzzle, which the SM fails to explain, inviting us to a journey towards new physics beyond the SM (BSM).

The SM is a quantum field theory based on the $SU(3)_C \times SU(2)_L\times U(1)_Y$ gauge symmetry that describes three of the four known fundamental interactions among elemental particles, i.e, the strong, weak and electromagnetic interactions. 
To its particle spectrum belong the fermions, constituents of matter, the spin one bosons, force carriers, and the Higgs scalar boson, the remnant of the mass generation procedure via the Brout-Englert-Higgs (BEH) mechanism~\cite{Higgs:1964ia,Englert:1964et,Guralnik:1964eu,Higgs:1966ev}. 
This mechanism explains how the spontaneous electroweak symmetry breaking (EWSB) generates the observed masses for the gauge $W$ and $Z$ bosons, as well as for all the fermions but the neutrinos, which remain massless in the SM. 

Historically, experimental searches for flavor violating processes have been essential for the theoretical developments in Particle Physics, as the Glashow-Iliopoulos-Maiani (GIM) mechanism~\cite{Glashow:1970gm} for explaining the lack of signal from flavor changing neutral currents or the Cabibbo-Kobayashi-Maskawa (CKM) quark mixing matrix~\cite{Cabibbo:1963yz,Kobayashi:1973fv} for the flavor changing charged currents.
Likewise,  the most clear experimental evidence for new physics at present comes from lepton flavor violation (LFV) in the neutrino sector. 
As we just said, neutrinos are massless by construction in the SM. 
However, experimental evidences of LFV in neutrino oscillations, first observed by the Super-Kamiokande~\cite{Fukuda:1998mi} and SNO~\cite{Ahmad:2001an,Ahmad:2002jz} collaborations, have showed that neutrinos do have masses, implying that the SM needs to be modified.
Moreover, if ongoing or future experiments could detect a positive signal from the yet not observed LFV processes in the charged sector, a new window to physics beyond the SM and beyond neutrino masses would be opened. 

Looking at the SM particle spectrum, we see that the absence of  right-handed (RH) neutrino fields in the theory forbids the neutrinos from interacting with the Higgs field and, thus, from acquiring a mass after the EWSB. 
Therefore, the simplest way of incorporating neutrino masses to the SM would be adding the missing RH neutrino fields, $\nu_R$.
This way, neutrinos could interact with the Higgs field via a Yukawa coupling $Y_\nu$ and obtain, after the EWSB, a Dirac mass $m_D=vY_\nu$  proportional to the Higgs vacuum expectation value (vev), which we normalize as $v=174$~GeV.
Nonetheless, this minimal extension of the SM sets out further questions: why are neutrinos so different than the rest of the fermions? Why are their masses much smaller with respect to other fermion masses?
Moreover, since neutrinos have no color nor electric charge, they could be Majorana fermions, i.e., they could be their own antiparticles. If true, this would be certainly a novelty in Particle Physics. 

Having a closer look to the new added $\nu_{R}$ fields, we realize that they are singlets under the full SM gauge group and, therefore, there is nothing that forbids them from having a Majorana mass $m_M$. 
These two different masses, $m_D$ and $m_M$, are the basic ingredients of the well known type-I seesaw model~\cite{Minkowski:1977sc,GellMann:1980vs,Yanagida:1979as,Mohapatra:1979ia,Schechter:1980gr}, which usually assumes that the Majorana mass scale $m_M$ is much heavier than $m_D$ and than the electroweak scale $v$.
In such case, the physical neutrino spectrum consists on one heavy and one light Majorana neutrino per generation, with masses of the order of $m_M$ and $m_D^2/m_M$, respectively.
Therefore, the type-I seesaw elegantly explains the smallness of the observed light neutrino masses as the ratio of two very distinct mass scales $m_D$ and $m_M$.
Inspecting the light neutrino mass scale $m_\nu\sim m_D^2/m_M$, we also see that in order to have the experimentally suggested $m_\nu\sim\mathcal O$(eV) with large Yukawa couplings $Y_\nu\sim\mathcal O(1)$, we need very heavy type-I seesaw masses of $m_M\sim\mathcal O(10^{14}~{\rm GeV})$. 
On the other hand, lighter right-handed neutrino masses at the TeV range would demand small couplings $Y_\nu\sim\mathcal O(10^{-5})$.
One way or the other, most of the phenomenology is suppressed in this type-I seesaw model. 
For instance, small corrections to the mass of the Higgs in this kind of models have been found in a complementary work that has not been included in this Thesis~\cite{Heinemeyer:2014hka}.
Therefore, the simplicity of this argument is what makes this model appealing and, at the same time, what makes it difficult to be tested in other low energy observables beyond the light neutrino masses themselves.

Interesting variations of this simple type-I seesaw model that have a much richer phenomenology are the low scale seesaw models. 
In this kind of models, some new symmetry is invoked with the aim of protecting the light neutrinos of having large masses and, therefore, allowing the new heavy neutrinos to have lower masses and large Yukawa couplings at the same time. 
A particular realization of these low scale seesaw models, on which we will focus this Thesis, is the inverse seesaw (ISS) model~\cite{Mohapatra:1986aw,Mohapatra:1986bd,Bernabeu:1987gr}, which assumes an approximately conserved total lepton number (LN) symmetry. 
In the limit of exact LN conservation, the light neutrinos will be massless.
 However, if this symmetry is broken by some mass parameter, the light neutrinos have a small Majorana mass proportional to this LN breaking parameter. 
On the other hand, we can expect this parameter to be naturally small,  in the sense of 't Hooft~\cite{tHooft:1979rat}, since setting it to zero increases the symmetry of the model. 
Therefore, in the ISS model the lightness of the neutrino masses is related to the smallness of a LN symmetry breaking mass parameter.

In the ISS model, the above demanded small LN breaking is obtained by introducing new fermionic singlets in pairs ($\nu_R, X)$ of opposite LN, and assuming that the LN conservation is only violated by a small Majorana mass  $\mu_X$ for the $X$ fields. 
Then, the physical spectrum consists of light Majorana neutrinos with masses suppressed by the smallness of $\mu_X$, and two heavy nearly degenerate Majorana neutrinos per generation, which form pseudo-Dirac pairs and indeed behave almost as Dirac fermions, contrary to the heavy neutrinos of the standard type-I seesaw.
Interestingly, since the $\mu_X$ scale ensures the smallness of light neutrino masses, the heavy neutrino states can have, at the same time, both large Yukawa couplings to the SM neutrinos and masses of the order of a few TeV or below, being therefore reachable at the LHC. 
This makes the ISS model an appealing model, with a rich phenomenology that can be tested at present or near-future experiments.
These include, among others, studies of lepton flavor universality violation in meson leptonic and semileptonic decays~\cite{Abada:2012mc, Abada:2013aba}, lepton electric dipole moments~\cite{Abada:2015trh,Abada:2016awd}, lepton magnetic moments~\cite{Abada:2014nwa}, heavy neutrino production at colliders \cite{Chen:2011hc,BhupalDev:2012zg,Das:2012ze,Das:2014jxa, Bambhaniya:2014kga,Bambhaniya:2014hla,Das:2015toa, Das:2016hof}, 
dark matter~\cite{Abada:2014zra}, leptogenesis~\cite{Garayoa:2006xs,Abada:2015rta} and  charged LFV processes~\cite{Deppisch:2004fa,Deppisch:2005zm,Dev:2009aw,Hirsch:2009ra,Abada:2011hm,Abada:2012cq,Hirsch:2012ax,Abada:2014cca,Abada:2014kba,Abada:2015oba,Abada:2015zea,Abada:2016vzu}.

On the theoretical side, the SM also suffers from some undesired properties, as the so-called hierarchy problem.
This problem refers to the instability of the Higgs sector under radiative corrections if some new physics at a large scale is introduced. 
In order to illustrate this idea we can consider the Higgs boson mass, $m_H$, and compute its radiative corrections under the assumption that there is no new physics until the Planck mass $M_{\rm P}\sim10^{19}$~GeV, where the gravitational effects start playing a role. 
By doing this, we find that the quantum corrections $\Delta m_H^2$ grow as the square of the new physics scale, which gives a value very far from the experimentally measured value $m_H=125.09\pm0.21({\rm stat.})\pm0.11({\rm syst.})$~GeV~\cite{Aad:2015zhl}.
Thus, in order to obtain a prediction that is compatible with this experimental value, a very fine tuned cancellation among the bare mass and the quantum corrections is needed, which is not very natural without any further explanation nor extra symmetry.

One of the most popular and elegant solutions to this problem is provided by supersymmetry (SUSY)~\cite{Golfand:1971iw,Volkov:1973ix,Wess:1974tw}, a new symmetry that relates fermions and bosons. 
In its simplest and minimal extension of the SM, known as the Minimal Supersymmetric Standard Model (MSSM)~\cite{Haber:1984rc,Gunion:1984yn,Gunion:1986nh}, each fermion of the SM has a spin-zero partner, called sfermion, with the same mass and quantum numbers as the original fermion; equivalently, all the SM bosons have spin one-half partners. 
The fact that there are fermions and bosons with the same couplings and masses gives the needed cancellation of the dangerous quantum corrections to the Higgs boson mass, providing an elegant solution to the hierarchy problem. 
Of course, doubling the SM spectrum is a strong prediction that experiments have tested and, unfortunately, the SUSY part of the MSSM spectrum has not been found yet.
This means that, if SUSY exists in Nature, it cannot be an exact symmetry and it must be broken, such that the SUSY particles must be heavier than the SM ones. 
This breaking, however, cannot spoil completely the nice solution to the hierarchy problem, so SUSY needs to be softly broken~\cite{Girardello:1981wz}, i.e., the dominant quadratic dependence on the new physics scale of $\Delta m_H^2$ still cancels out, although a logarithmic dependence remains.

Once these soft SUSY breaking terms are included, the MSSM is still a viable model with a very appealing phenomenology. 
Nevertheless, since it is constructed from the SM, it also demands a mechanism for neutrino mass generation. 
Following the previous discussion, we can consider again the ISS model and embed it in a supersymmetric context, adding to the MSSM spectrum new neutrinos and sneutrinos which can have both masses at a few TeV scale or below and with large couplings. 
This way, this SUSY-ISS model combines the appealing features of both frameworks, the SUSY and the ISS ones.

The best way of experimentally proving that any model beyond the SM is correct would be directly detecting the new particles that it predicts.
Nevertheless, this task can in many cases be very difficult if these new particles are too heavy as to be directly produced in present experiments, so a first indication of their existence could come from their indirect implications to some other low energy observables.
In the particular case of neutrino mass models, one of the optimal places for this purpose is again looking for LFV processes, concretely in the charged lepton sector, which can be quantumly induced   via heavy neutrino loop effects. 

Processes involving charged lepton flavor violation (cLFV) are forbidden in the SM if neutrinos are massless, and extremely suppressed if the small neutrino masses from oscillation data are {\it ad-hoc} added to the SM. 
Consequently, a positive signal in any of the experimental searches for cLFV processes would automatically imply the existence of new physics, and it must be indeed beyond the SM with minimally added neutrino masses. 
Although no such processes have been observed yet, this is a very active field that is being explored by many experiments which have set upper bounds to this kind of cLFV processes. 
At present, the strongest bounds have been found in the $\mu$-$e$ transitions, as the radiative  $\mu\to e\gamma$ decay or $\mu$-$e$ conversion in heavy nuclei, whose branching ratios have been bounded to be below $4.2\times10^{-13}$ and $7.0\times10^{-13}$ by the MEG~\cite{TheMEG:2016wtm} and SINDRUM II~\cite{Bertl:2006up} collaborations, respectively. 
Moreover, next generation of experiments are expected to improve in several orders of magnitude the sensitivities for LFV $\mu$-$e$ transitions, reaching the impressive range of $10^{-18}$ for $\mu$-$e$ conversion in nuclei by the PRISM experiment in J-PARC~\cite{Alekou:2013eta}. 
On the other hand, present bounds on transitions in the $\tau$-$\mu$ and $\tau$-$e$ sectors are less constraining, with, for example, upper bounds of about $10^{-8}$ for LFV  tau decays from BABAR~\cite{Aubert:2009ag} and BELLE~\cite{Hayasaka:2010np,Hayasaka:2010et}. 
Therefore, there is some more room in these sectors than in the $\mu$-$e$ one for having large LFV signals that new experiments as BELLE-II~\cite{Aushev:2010bq} would be able to test in the near future. 

Additionally, the currently running LHC has also many things to say about cLFV. 
First of all, the fact that a new particle, the Higgs boson, has been discovered opens a new window for possible deviations from the SM that definitely needs to be explored. 
In particular, three new cLFV channels are introduced in the cLFV market, the  LFV Higgs boson decays into two leptons of different flavor, $H\to\ell_k\bar\ell_m$, $k\neq m$, which have already been searched by the CMS~\cite{Khachatryan:2015kon,Khachatryan:2016rke,CMS:2017onh} and ATLAS~\cite{Aad:2016blu} experiments. 
Even though CMS saw a small but intriguing excess in the $H\to\tau\mu$ channel after run-I~\cite{Khachatryan:2015kon}, it has not been confirmed yet with run-II data and, at present, they have been able to set an upper bound of $2.5\times10^{-3}$~\cite{CMS:2017onh}.
These searches of LFV Higgs decays, as well as other explorations of the Higgs sector, will continue and surely  be improved with more data after new LHC runs.

Other interesting observables that the LHC is also looking for are the LFV $Z$ boson decays into two leptons of different flavor $Z\to\ell_k\bar\ell_m$~\cite{Aad:2014bca,Aad:2016blu}.
Interestingly, after the run-I, ATLAS has already reached the previous sensitivities from the LEP experiment~\cite{Akers:1995gz,Abreu:1996mj}, even improving the bound for the $Z\to\mu e$ channel.
As for the Higgs searches, LFV $Z$ decays will certainly continue during the new runs, so hopefully new interesting data will come from ATLAS and CMS.
Nevertheless, the best sensitivities for LFV $Z$ decays are expected from next generation of lepton colliders, as long as they can work as $Z$ factories with a very clean environment.  
In particular, at future linear colliders, with an expected sensitivity of $10^{-9}$~\cite{Wilson:I, Wilson:II}, or at a Future Circular  $e^+ e^-$ Collider (such as FCC-ee (TLEP)~\cite{Blondel:2014bra}), where it is estimated that up to $10^{13}$ $Z$ bosons would be produced and the sensitivities to LFV $Z$ decay rates could be improved up to $10^{-13}$.

The main motivation of this Thesis, therefore, is to explore the connection between the existence of new right-handed neutrino particles with masses of a few TeV or below, reachable at the LHC, and  the existence of LFV in the charged lepton sector. 
We are particularly interested in the two models above described, the ISS and the SUSY-ISS model, which are very appealing models since they can provide right-handed neutrino states with masses at the TeV range and, at the same time, with large couplings. 
Charged LFV phenomenology within massive neutrino models has been studied before
\cite{Mann:1983dv,Dittmar:1989yg,Pilaftsis:1992st,Korner:1992zk,Korner:1992an,Ilakovac:1994kj,Ilakovac:1999md,Illana:1999ww,Illana:2000ic,Casas:2001sr,Illana:2003pj,Arganda:2004bz,Arganda:2005ji,Antusch:2006vw,Abada:2007ux,Arganda:2007jw,Arganda:2008jj,Abada:2008ea,Gavela:2009cd,Ilakovac:2009jf,Herrero:2009tm,Casas:2010wm,Dinh:2012bp,Alonso:2012ji,Ilakovac:2012sh,Herrero-Garcia:2016uab},
 also in the ISS
\cite{Deppisch:2004fa,Deppisch:2005zm,Dev:2009aw,Hirsch:2009ra,Abada:2011hm,Abada:2012cq,Hirsch:2012ax,Abada:2014cca,Abada:2014kba,Abada:2015oba,Abada:2015zea,Abada:2016vzu}.
In this Thesis, we concentrate mainly in studying the predictions for the LFV Higgs and $Z$ boson decays, which as we said are extremely timely to explore in the light of the recent discovery of the Higgs boson and the new LHC data on the LFV $Z$ decays. 
In addition, we also make new predictions for other cLFV processes like $\ell_m\to\ell_k\gamma$ and $\ell_m\to\ell_k\ell_k\ell_k$, as well as for other lepton flavor preserving observables that will be also relevant in the context of the models we consider here. 

We fully study for the first time the LFV $H$ decays in presence of right-handed neutrinos in the ISS model, as well as in presence of  sneutrinos in the SUSY-ISS model. 
We perform this study following two different approaches. 
First we present, based on the results for the type-I seesaw model in Ref.~\cite{Arganda:2004bz}, the results for the full one-loop computation done in the physical neutrino mass basis. 
Second, we use the mass insertion approximation technique, which works in the electroweak basis and allows us to obtain useful and simple formulas for the LFV H decay rates. 
For the numerical evaluation, we show that large LFV rates can be obtained focusing on particular directions of the parameter space where $\mu$-$e$ transitions, the experimentally most constrained ones, are highly suppressed. 
Along this same line of research, we provide a new proposal for the building of these phenomenologically  interesting scenarios with suppressed $\mu$-$e$ transitions.
These scenarios are based on the $\mu_X$ parametrization, which is a new parametrization proposed in this Thesis. 
On the other hand, LFV $Z$ decay rates in the ISS model have been first explored in Ref.~\cite{Abada:2014cca}.  
Therefore, in the case of these observables, we directly present a more specific study in those particular directions of the parameters space with suppressed $\mu$-$e$ transitions, that give large allowed LFVZD rates. 
In this sense, we perform a complementary study of that previously done. 

As we said, we are interested in models with right-handed neutrinos at the TeV range, since they belong to the scale of energies that the LHC is now probing. 
In the ISS model, due to the pseudo-Dirac character of the heavy neutrinos, standard collider searches looking for lepton number violating final states with two same-sign leptons, the `smoking gun' signature of Majorana fermions~\cite{Pilaftsis:1991ug,Datta:1993nm,delAguila:2007qnc,Atre:2009rg}, are not efficient. 
Therefore, here we propose to use instead lepton flavor violating final states in order to discriminate events from the production and decay of the low scale seesaw heavy neutrinos. 
In particular, we focus on exotic $\tau^\pm\mu^\mp jj$ or $\tau^\pm e^\mp jj$ final states with no missing transverse energy, which could be produced in scenarios with $\nu_R$ masses at and below the TeV range and where LFV is favored in the $\tau$-$\mu$ or $\tau$-$e$ sector, respectively.

This Thesis is organized as follows. 
In \chref{Models} we review neutrino oscillation physics and its connection with the need of introducing neutrino masses. 
We discuss some popular neutrino mass models, paying special attention to models with right-handed neutrinos with TeV range masses, as the inverse seesaw model and  its SUSY version.
These are the two models that we will consider for exploring in detail the LFV phenomenology in the following Chapters.
Moreover, when studying the neutrino sector of these models, we present a new parametrization, which we refer to as the $\mu_X$ parametrization, that will turn out to be very useful for exploring the parameter space while being always in agreement with neutrino oscillation data. 

In \chref{PhenoLFV} we address the importance of charged LFV processes in the search of new physics and summarize the experimental status. 
Then, we revisit the LFV lepton decays in the ISS model, meaning the radiative $\ell_m\to\ell_k\gamma$ and three-body $\ell_m\to\ell_k\ell_k\ell_k$ decays with $k\neq m$.
This new study of these processes will allow us to learn about the behavior of the LFV as a function of the ISS parameters, as well as to emphasize the advantages of using our $\mu_X$ parametrization.
As a result of this study, we will find some interesting phenomenological scenarios, well motivated by present experimental bounds, where LFV $\tau$-$\mu$ or $\tau$-$e$ transitions are favored while keeping the $\mu$-$e$ transitions highly suppressed.
These will be useful for exploring LFV $H$ and $Z$ decays.
Furthermore, we also discuss in this Chapter the implications of right-handed neutrinos with TeV masses to other relevant low energy observables, which we will consider as constraints when looking for maximum allowed rates in the next Chapters.

\chref{LFVHD} is devoted to the study of the LFV Higgs decay (LFVHD) rates in the ISS and SUSY-ISS models. 
We present the results of the full one-loop calculation of the LFVHD rates in the ISS model and systematically study their dependence with the different parameters of this model. 
In order to better understand the results, we perform a complete and independent calculation of these rates using the mass insertion approximation, which will allow us to derive a simple expression for an effective LFV $H\ell_k\ell_m$ vertex, very useful for any author that wishes to make a fast estimation of the LFVHD rates in this kind of models. 
This complete analysis will serve us to conclude on the maximum LFVHD rates allowed by present experimental constraints. 
Moreover, we explore these rates in the SUSY-ISS model, showing that the new SUSY loops including sleptons and sneutrinos may considerably enhance the maximum allowed rates, reaching values close to the present experimental sensitivities.

In \chref{LFVZD} we study the LFV $Z$ boson decays (LFVZD) in the ISS model.
We focus our analysis on the previously introduced phenomenological scenarios, where large allowed LFV rates in the $\tau$-$\mu$ and $\tau$-$e$ sector can be achieved. 
We compare our finding to previous works in the literature and show that, in these interesting directions of the ISS parameter space, large allowed rates can be obtained, well within the reach of next generation of experiments searching for LFVZD.

In \chref{LHC} we focus on low scale seesaw neutrino production at the LHC. 
We study the possibility of detecting the production and decay of the ISS heavy neutrinos searching for exotic LFV $\ell_k^\pm\ell_m^\mp jj$ events, with $k\neq m$, and with no missing transverse energy. 
Alternatively to standard Majorana neutrino searches at colliders that are not relevant for the ISS model, our proposal explores the fact that the new heavy neutrino states can have non-trivial flavor structure, leading to this kind of interesting LFV signals.
Concretely, we will apply this idea to the production of exotic $\tau\mu jj$ events within the previously introduced scenarios, finding promising results for the future LHC runs. 

Finally, we summarize the main conclusions at the end of this document. 

The contents presented in this Thesis, summarized along the Chapters 1-5, the Conclusions and the Appendices, are original works that have been published in Refs.~\cite{Arganda:2014dta,Arganda:2015naa,Arganda:2015ija,DeRomeri:2016gum,Arganda:2017vdb} and in the conference proceedings~\cite{Arganda:2014via,Arganda:2014oga,Arganda:2016snv,DeRomeri:2017ucv}.

%

\chapter{Seesaw models with heavy neutrinos at the TeV energy range}
\fancyhead[LE] { \scshape \chaptername\ \thechapter}
\fancyhead[RO] {\scshape Seesaw models with heavy neutrinos at the TeV energy range}
\label{Models}

In this Chapter we motivate the need of going beyond the Standard Model for explaining lepton flavor changing neutrino oscillation data and review some of the most popular models  for this task, the seesaw models. 
We will concentrate specially in the so-called low scale seesaw models, one of which will be of special relevance for this Thesis: the inverse seesaw model. 
Finally, we will introduce a Supersymmetric version of the latter, the SUSY-ISS, an interesting model that combines the appealing features of the Minimal Supersymmetric Standard Model and the inverse seesaw model. 
Along this Chapter, we derive the $\mu_X$ parametrization in \secref{sec:ISSmodel}, useful for accommodating neutrino oscillation data, which is a genuine contribution of this Thesis and was first published in Ref.~\cite{Arganda:2014dta}.
The implementation of the ISS model in the SUSY framework, as given in \eqrefs{Msnu}-(\ref{thistwo}), and the derivation of the interaction Lagrangian in the physical SUSY-ISS basis in  \eqref{SUSYintLagrangian} and in \appref{App:LFVHD_SUSY} are original works of this Thesis that have been published in Ref.~\cite{Arganda:2015naa}.

\section{Neutrino oscillations}

In the Standard Model (SM) the neutrinos, and antineutrinos, come in three different flavors.
When they are produced by the standard charged current, they are always produced together with a charged lepton, which is the one that labels them: 
if the neutrino is produced with an $e^+$ or $e^-$, we name it as electron-neutrino ($\nu_e$) or electron-antineutrino ($\bar\nu_e$), respectively; if it is produced with a $\mu^+$ or $\mu^-$, we have a $\nu_\mu$ or $\bar\nu_\mu$; and if it is produced with $\tau^+$ or $\tau^-$, it is a $\nu_\tau$ or $\bar\nu_\tau$.
This one-to-one identification with the charged lepton sector is, at the same time, what allows us to detect and identify these elusive particles. 
These three neutrino flavor states $\nu_\ell\equiv\big(\nu_e, \nu_\mu, \nu_\tau\big)$ form a basis that we will refer to as the interaction basis.

Neutrinos only suffer from weak interactions and, consequently, they can travel long distances without interacting with anything. 
Their evolution is given by the  Schr\"odinger equation, whose solutions are plane waves with energies defined by the eigenvalues of the $3\times3$ neutrino mass matrix. 
These stationary solutions define a new basis, the so-called mass or physical basis\footnote{Assuming the simplest scenario where only three physical neutrinos exist.}  $\nu_\alpha\equiv\big(\nu_1, \nu_2, \nu_3\big)$, which in general does not coincide with the above introduced interaction basis.
This misalignment is the origin of the neutrino oscillation phenomena. 

The relation between the two bases can be written as: 
\begin{equation}\label{NeuRotation}
\nu_\ell = \sum_{\alpha=1}^{3} \big(U_{\rm PMNS}\big)_{\ell \alpha} \nu_\alpha\,,
\end{equation}
where the $U_{\rm PMNS}$ is a unitary $3\times3$ rotation matrix, analogous to the CKM matrix in the quarks sector, whose name comes from Pontecorvo, who proposed neutrino oscillations~\cite{Pontecorvo:1957cp}, and from Maki-Nakagawa-Sakata, who introduced the mixing matrix~\cite{Maki:1962mu}.

When a neutrino is produced, it is in a specific flavor state, which can be expressed as a superposition of the mass eigenstates. 
If neutrinos were massless or degenerate in mass, all the mass eigenstates would have the same time evolution and, consequently, the initial flavor state would remain unchanged. 
In such a situation, we could say that the individual lepton flavor numbers, i.e., $L_e$, $L_\mu$ and $L_\tau$, were preserved. 
On the contrary, if physical neutrinos had non-degenerate masses, each of the mass eigenstates would evolve differently in time, modifying the initial superposition and therefore the flavor of the initial neutrino state. 
This process, which is a direct consequence of non-degenerate neutrino masses, is known as neutrino oscillation and implies that individual lepton flavor numbers are not conserved.
In the ultrarelativistic limit, the oscillation probability in vacuum  from a flavor $\nu_\ell$ to a flavor $\nu_{\ell'}$ is given by~\cite{GiuntiKim}:
\begin{equation}\label{NeuOscillation}
\mathcal P_{\nu_{\ell} \to \nu_{\ell'}} \big(L,E\big) = \sum_{\alpha,\beta=1}^3 U_{\ell \alpha}^* U_{\ell'\alpha}^{} U_{\ell\beta}^{} U_{\ell'\beta}^* \exp\left( -i \frac{\Delta m_{\alpha\beta}^2 L}{2E}\right)\,,
\end{equation}
where  $U\equiv U_{\rm PMNS}$  to shorten the notation, $E\sim|\boldsymbol p|$ is the neutrino energy, $L$ is the distance between the source and the detector, and $\Delta m^2_{\alpha\beta}\equiv m_\alpha^2-m_\beta^2$ are the squared mass differences.

Several experiments involving solar, atmospheric, reactor and accelerator neutrinos have established the evidences for neutrino oscillations and, therefore, for neutrino masses (see Ref.~\cite{Olive:2016xmw} for a review).  
Nevertheless, and in spite of this experimental effort, there are still open issues related to neutrino oscillations and masses. 

First of all, we do not know the absolute neutrino mass scale, although we know that it is at the eV scale or below from the upper limits on the effective electron neutrino mass in $\beta$ decays, given by the Mainz~\cite{Kraus:2004zw} and Troitsk~\cite{Aseev:2011dq} experiments:
\begin{equation}
 m_\beta < 2.05\;\mathrm{eV}\quad\mathrm{at}\;95\%\;\mathrm{C.L.}
\end{equation}
Additional information on the absolute neutrino mass scale, can be obtained from cosmological observations, which are sensitive to the sum of the light neutrino masses.
At present, there are only upper bounds on this quantity, being the most constraining ones provided by the Planck collaboration~\cite{Ade:2013zuv}:
\begin{equation}\label{numassMAX}
\sum m_\nu < 0.23~{\rm eV}\,.
\end{equation}

\begin{figure}[t!]
\begin{center}
\includegraphics[width=.95\textwidth]{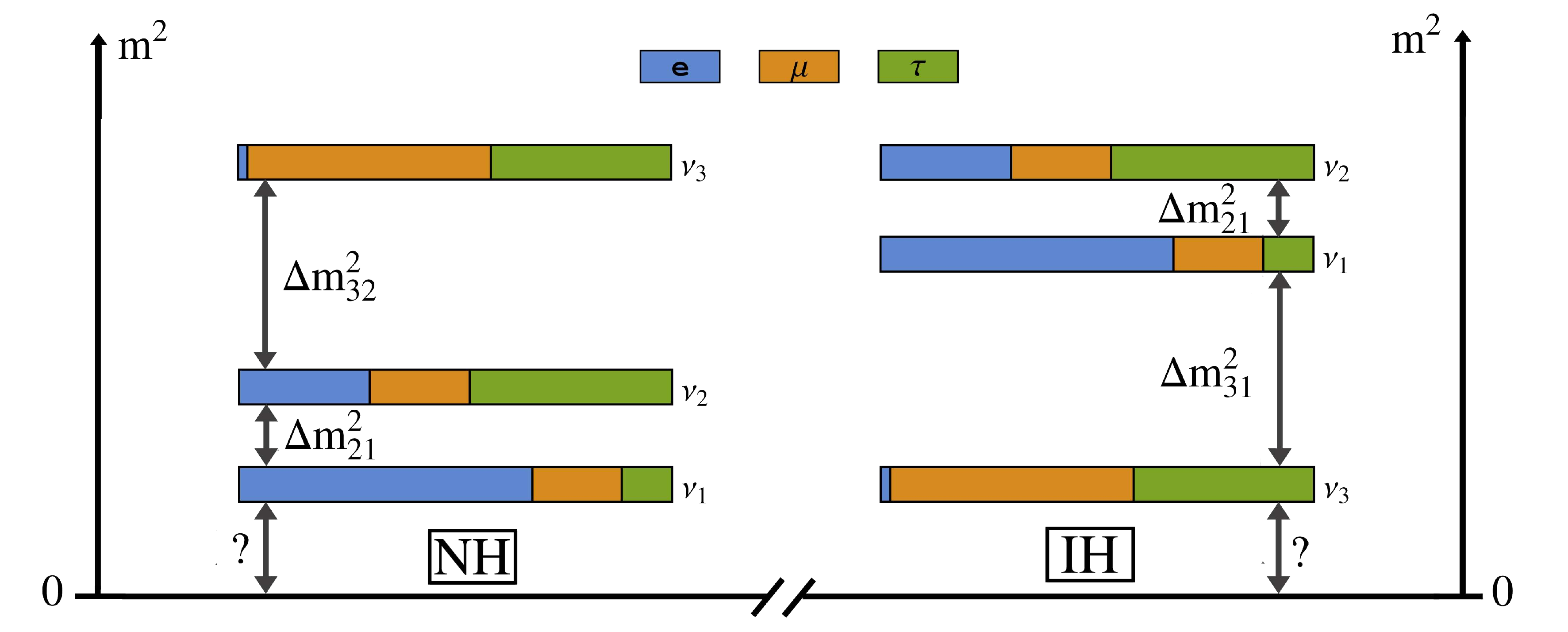}
\caption{The two possible neutrino mass orderings, known as normal (left) and inverted (right) hierarchies.
The colors represent the flavor composition of each of the physical neutrinos: blue for $\nu_e$, orange for $\nu_\mu$ and green for $\nu_\tau$. }\label{Hierarchies}
\end{center}
\end{figure}

Measuring neutrino oscillations in vacuum allows us to know the mass differences $|\Delta m^2_{21}|$ and $|\Delta m^2_{31}|$, but it does not tell us anything about  neither the absolute neutrino mass scale nor the neutrino mass hierarchy.
Additional measurements of matter effects or the Mikheev-Smirnov-Wolfenstein (MSW) effects~\cite{Wolfenstein:1977ue,Wolfenstein:1979ni,Mikheev:1986gs,Mikheev:1986wj,Mikheev:1986if} in neutrino oscillations can help solving the sign degeneracies.
Nowadays, matter effects in the sun have made possible to know that $\Delta m^2_{21}>0$, but the sign of $\Delta m^2_{31}$ is a mystery yet. 
Therefore, two orderings are still possible, as shown if \figref{Hierarchies}: a Normal Hierarchy (NH) where $m_{\nu_1}<m_{\nu_2}<m_{\nu_3}$ and an Inverted Hierarchy (IH) where $m_{\nu_3}<m_{\nu_1}<m_{\nu_2}$.
Solving this degeneracy is one of the most important open issues in neutrino physics.

Second, being neutrinos the only electrically neutral fermions in the SM,  they could be Majorana particles, i.e., they could be their own antiparticles.
This would be in contrast to the rest of the SM model Dirac fermions, for which their antiparticle is a different state. 
This hypothetical Majorana character of the neutrinos, although very common in theoretical models as we will see later, does not have any impact on neutrino oscillations and, therefore, new observables to discern between Majorana and Dirac fermions need to be considered. 
The fact that a lepton can be its own antiparticle is directly related to the conservation of the total Lepton Number (LN) violation, since a Majorana mass terms  breaks LN in two units. 
Consequently, LN violating processes are usually considered as the smoking gun signatures for Majorana neutrinos, like like neutrinoless double beta decay (for a review on $0\nu\beta\beta$, see for instance Refs.~\cite{Vergados:2012xy,DellOro:2016tmg}).  
Unfortunately, no experimental evidence has been found yet for any LN violating processes, so knowing if neutrinos are Majorana or Dirac fermions is still an open issue. 

Third, massive neutrinos add new $CP$-violating phases to the SM parameters. 
For the case of three massive neutrinos, the PMNS matrix in \eqref{NeuRotation} introduces one $CP$-violating phase if neutrinos are Dirac fermions, known as the Dirac $CP$ phase, and two extra Majorana $CP$-violating phases if neutrinos are Majorana fermions. 
However, neutrino oscillations are only sensitive to the Dirac phase and this dependence appears via a particular combination of several oscillation parameters, known as the Jarlskog invariant~\cite{Jarlskog:1985cw} $J_{CP}=  {\rm Im} \big( U_{\mu3} U_{e3}^*U_{e2} U_{\mu2}^*\big)$.
This fact makes difficult to measure the Dirac phase in oscillation experiments, but  experiments as T2K, NO$\nu$A, or future Hyper-K or Dune, may be able to do it in the next years. 
On the other hand, the extra Majorana phases do not play any role in neutrino oscillations.
Nevertheless, they can be important for trying to explain the matter-antimatter asymmetry of the universe though the mechanism known as Baryogenesis via Leptogenesis (for a review see, for instance, Ref.~\cite{Davidson:2008bu}).

Finally, we may wonder how many neutrinos do exist in Nature. 
Despite the fact that there are some anomalies~\cite{Acero:2007su,Giunti:2010zu,Mueller:2011nm,Huber:2011wv,Mention:2011rk,Aguilar:2001ty,AguilarArevalo:2007it,AguilarArevalo:2010wv,Aguilar-Arevalo:2013pmq} pointing towards the existence of eV-KeV scale sterile neutrinos, we will assume in this Thesis that there are only three light neutrinos, which is the minimal requirement to fit neutrino oscillation observations. 
In this situation, the unitary PMNS matrix in \eqref{NeuRotation} can be parametrized in its standard form as: 
\begin{equation}\label{PMNS}
U_{\rm PMNS} = \left(\begin{array}{ccc}
c_{12}c_{13} &s_{12}c_{13}&s_{13}e^{-i\delta}\\
-s_{12}c_{23}-c_{12}s_{23}s_{13}e^{i\delta} & c_{12}c_{23}-s_{12}s_{23}s_{13}e^{i\delta}&s_{23}c_{13}\\
s_{12}s_{23}-c_{12}c_{23}s_{13}e^{i\delta} &-c_{12}s_{23}-s_{12}c_{23}s_{13}e^{i\delta}&c_{23}c_{13}
\end{array}\right)\cdot P\,,
\end{equation}
where $c_{ij}=\cos\theta_{ij}$ and $s_{ij}=\sin\theta_{ij}$ with $\theta_{ij}$ the neutrino mixing angles, $\delta$ is the  CP-violating Dirac phase and the diagonal matrix $P={\rm diag}(1,e^{i\phi_1},e^{i\phi_2})$ accounts for the two extra CP-phases that do not play any role in neutrino oscillations, as we said before.
In order to consider all the experimental neutrino oscillation data in a consistent way, we will take the results from the global fit analysis done by the NuFIT group~\cite{Esteban:2016qun}.
Assuming a Normal Hierarchy, they obtain at the $1\sigma$ level:
\begin{align}\label{NuFit}
\sin^2\theta_{12}&=0.306^{+0.012}_{-0.012}\,, &
\Delta m^2_{21}&=7.50^{+0.19}_{-0.17}\times10^{-5}{\rm eV}^2\,, \nonumber\\
\sin^2\theta_{23}&=0.441^{+0.027}_{-0.021}\,, &
\Delta m^2_{31}&=2.524^{+0.039}_{-0.040}\times10^{-3}{\rm eV}^2\,, \nonumber\\
\sin^2\theta_{13}&=0.02166^{+0.00075}_{-0.00075} \,, &
\delta &= 261^{+51}_{-59}\,,
\end{align}
while for the Inverted Hierarchy they give
\begin{align}\label{NuFitIH}
\sin^2\theta_{12}&=0.306^{+0.012}_{-0.012}\,, &
\Delta m^2_{21}&=7.50^{+0.19}_{-0.17}\times10^{-5}{\rm eV}^2\,, \nonumber\\
\sin^2\theta_{23}&=0.587^{+0.020}_{-0.024}\,, &
\Delta m^2_{32}&=-2.514^{+0.038}_{-0.041}\times10^{-3}{\rm eV}^2\,, \nonumber\\
\sin^2\theta_{13}&=0.02179^{+0.00076}_{-0.00076} \,, &
\delta &= 277^{+40}_{-46}\,.
\end{align}

\section{Seesaw models for neutrino masses}

At the time that the SM was built, there were no evidences for neutrino masses.
Moreover, experiments showed that neutrinos produced in charged weak interactions were left-handed (LH) fields, while antineutrinos were right-handed (RH).  
These facts were minimally satisfied in the SM by a chiral LH flavor field $\nu_{\ell_L}\equiv (\nu_\ell)_L$, which together with the LH charged lepton field forms a $SU(2)_L$ doublet, as required by the SM gauge symmetry.
As a result, right-handed neutrino fields were left out and neutrinos were treated as massless fields by the SM.  

Nowadays the status has changed. 
The strong experimental evidences of neutrino oscillations, as mentioned in the previous section, have established that neutrinos do have masses, claiming for new physics beyond the SM to accommodate this new situation.
In a very simple and minimalistic choice, one could reconsider the addition of RH neutrino fields to the SM, in such a way that neutrinos could obtain their masses via their Yukawa interaction with the Higgs field, mimicking the mass generation for the rest of the SM fermions,  
\begin{equation}\label{YukLag}
\mathcal L_{\rm Yuk} = -Y_{\nu}^{ij} \overline{L_i}\widetilde\Phi \nu_{R_j} + h.c.
\end{equation}
where $Y_\nu$ is the neutrino Yukawa coupling matrix, $L=(\nu_L ~\ell_L)^T$ is the $SU(2)_L$ lepton doublet and $\widetilde\Phi= i\sigma_2\Phi^*$ with $\Phi$ the Higgs $SU(2)_L$ doublet:
\begin{equation}\label{HiggsDoublet}
\Phi =\left(\begin{array}{c} G^+ \\ v+\dfrac1{\sqrt2}(H+ i G^0 )\end{array}\right)
\,,\quad
\widetilde\Phi= i\sigma_2\Phi^*=\left(\begin{array}{c} v+\dfrac1{\sqrt2}(H- i G^0)\\-G^-\end{array}\right).
\end{equation}
After the Electroweak Symmetry Breaking (EWSB), this Lagrangian term leads to a Dirac mass term  for neutrinos $m_\nu=m_D=v Y_\nu$, with $v=174$GeV. 
In order for this mass to be at the eV scale, as suggested by neutrino oscillations, the Yukawa coupling needs to be very small, of the order of $10^{-11}$.
This value would extremely suppress any kind of phenomenology beyond neutrino oscillations.
Moreover, such a tiny neutrino Yukawa coupling is five orders of magnitude smaller than the electron Yukawa coupling and eleven orders with respect to the top Yukawa coupling, so it would make even worse the flavor puzzle problem of understanding the hierarchy of the fermion masses.  

On the other hand, having a closer look to the new added $\nu_{R}$ fields, we realize that they are singlets under the full SM gauge group and, therefore, there is nothing that protects them from having a Majorana mass term.
In that situation, physical neutrinos will be Majorana particles.

In order to work with Majorana fermions, it is very useful to introduce the particle-antiparticle conjugation operator $\hat C$, which is defined as~\cite{Akhmedov:2014kxa,GiuntiKim}:
\begin{equation}\label{Cconjugation}
\hat C: \psi \to \psi^C = \mathcal C \bar\psi^T\,.
\end{equation}
This matrix $\mathcal C$ fulfills: 
\begin{equation}
\mathcal C^{-1} \gamma_\mu \mathcal C = -\gamma_\mu^T \,,
\quad
\mathcal C^{-1} \gamma_5 \mathcal C = \gamma_5^T\,,
\quad
\mathcal C^{\dagger}=\mathcal C^{-1}=-\mathcal C^{*}\,,
\end{equation}
which, in the Weyl representation, can be satisfied by chosing $\mathcal C =i \gamma_2\gamma_0$. 
Consequently, this operator $\hat C$ flips the chirality of chiral fields:
\begin{equation}\label{ChiralityFlipC}
\hat C: \psi_L \to (\psi_L)^C=(\psi^C)_R\,,\quad \psi_R\to(\psi_R)^C=(\psi^C)_L\,,
\end{equation}
meaning that the antiparticle of a LH field is a RH field. 
Moreover, the fact that Majorana fermions coincide with their antiparticle can be expressed in terms of this operator as: 
\begin{equation}
\psi^C = \psi\,.
\end{equation}
These relations will be very useful when considering models with Majorana neutrinos, as the standard seesaw models. 

On a more model independent ground, we could make use of the effective Lagrangians formalism in order to try a bottom up approach to the neutrino mass problem. 
In this approach, we assume that the SM is an effective theory of a more complete but unknown theory, which in general will contain new symmetries and fields at a heavy scale $\Lambda$.
If we knew the complete theory at high energies, we could integrate out all the heavy fields with masses above the electroweak scale and obtain a low energy Lagrangian in terms only of the SM fields. 
The modifications with respect to the SM Lagrangian would then be a series of new non-renormalizable operators suppressed by the heavy scale $\Lambda$, which encode all the new physics effects at low energy. 
Unfortunately, we do not know the complete theory to follow this top down way. 
Therefore, with the aim of covering any possible high energy theory for neutrino physics, we can alternatively write the most general non-renormalizable $SU(2)_L\times U(1)_Y$ invariant Lagrangian that involves neutrino and other SM fields, the low energy fields. 
This will lead us to an effective Lagrangian that  can be generically written as the SM Lagrangian extended with a series of higher order non-renormalizable operators,
\begin{equation}\label{Leff}
\mathcal L_{\rm eff} = \mathcal L_{\rm SM} +\delta\mathcal L_{d=5} +\delta\mathcal L_{d=6} + \dots
\end{equation}
where $d$ stands for the dimension of the operators in $\delta\mathcal L_d$, which will be suppressed by $d-4$ inverse powers of the heavy scale $\Lambda$.

It is illustrative to look at the $d=5$ Lagrangian. 
Since neutrinos are members of a $SU(2)_L$ doublet, there is only one possible operator, first written by Weinberg~\cite{Weinberg:1979sa} and named after him, contributing to  $\delta\mathcal L_5$, given by:
\begin{equation}\label{WeinbergOperator}
\delta\mathcal L_{d=5} = \frac12\frac{c_{ij}}{\Lambda}\, \big(\overline{L^C_i} \widetilde\Phi^* \big) \big( \widetilde \Phi^\dagger L_j\big)\,,
\end{equation}
where $c_{ij}$ are dimensionless complex coefficients and $i,j=1,2,3$. 
After the EWSB, this operator gives a Majorana mass term for the neutrinos:
\begin{equation}\label{MajoranaEffmass}
\delta\mathcal L_{d=5} \to  \frac{v^2 c_{ij}}{2\Lambda} \Big(\,  \overline{\nu_i^C}\nu_j + h.c.\Big)\,.
\end{equation}
Interestingly, this Majorana mass term is naturally small, suppressed by the new physics scale. 
The higher the scale $\Lambda$, the lower the neutrino mass, as if they were playing with a seesaw. 
This is the idea behind the so-called seesaw models, the simplest renormalizable models leading to this relation after integrating out the new heavy particles responsible of generating neutrino masses at the tree level. 

\begin{figure}[t!]
\begin{center}
\includegraphics[width=\textwidth]{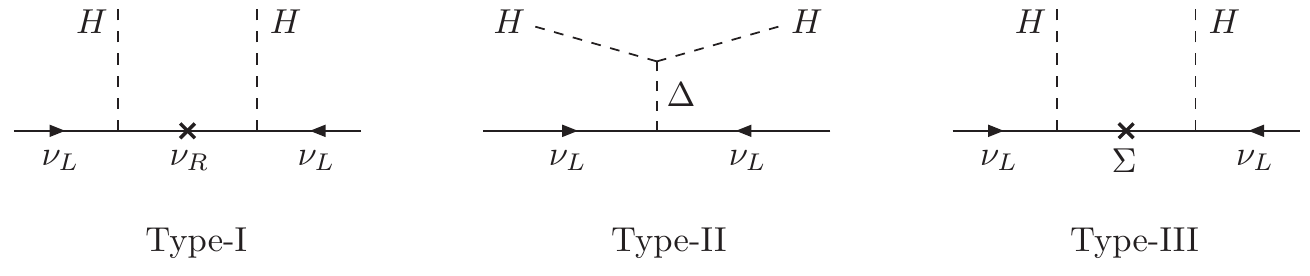}
\caption{Diagrams for the tree level light Majorana mass generation in the type-I (left), type-II (middle) and type-III (right) seesaw models. }\label{SeesawMasses}
\end{center}
\end{figure}

Looking at the Weinberg operator, we can already learn some properties about the new particles of the seesaw models.
In order to be gauge invariant, these new particles can be either singlets or triplets of $SU(2)_L$, since they need to couple to two $SU(2)$ doublets in a gauge invariant way.
On the other hand, they can be either fermions or scalars, so we can define three\footnote{The choice of a scalar singlet is not possible due to the structure of the Weinberg operator.} possible seesaw models according to what new type of fields they add to the SM: fermionic singlets (type-I), scalar triplets (type-II) or fermionic triplets (type-III), as shown in \figref{SeesawMasses}.

Before going to the details of these seesaw models, we want to emphasize that  any high energy theory that introduces a Majorana mass term for the neutrinos leads to the Weinberg operator in \eqref{WeinbergOperator} when integrating out the heavy fields, as it is the only one that can be written at lowest order.
This implies that, in order to distinguish between the different neutrino theories, we need to consider their implications beyond the neutrino mass generations. 
In terms of the effective Lagrangian in \eqref{Leff}, this means looking for the effect of the $d=6$ operators~\cite{Abada:2007ux} or higher. 
As we will see later, lepton flavor violating phenomenology is one of the optimal places for studying this task.

\subsection*{Type-I seesaw model}

The type-I seesaw model~\cite{Minkowski:1977sc,GellMann:1980vs,Yanagida:1979as,Mohapatra:1979ia,Schechter:1980gr} extends the SM with right-handed neutrinos $\nu_R$, which are fermionic singlets under the full SM gauge group. 
As mentioned above, these new fields allow  the neutrinos to have a Yukawa interaction with the Higgs field, as well as a Majorana mass term for themselves.
The Lagrangian of the type-I seesaw can be thus written as:
\begin{equation}\label{typeI}
\mathcal L_{\rm type-I} = -Y_{\nu}^{ij} \overline{L_i}\widetilde\Phi \nu_{Rj} - \frac12 m_M^{ij}\overline{\nu_{Ri}^{C}}\nu_{Rj}+ h.c.
\end{equation}
where the first term is the Yukawa Lagrangian as in \eqref{YukLag} and in the second term $m_M$ is a symmetric Majorana mass matrix.
If we assign to the $\nu_R$ fields the same lepton number than to the $L$ fields, we realize that the Yukawa interaction conserves LN, while the Majorana mass term breaks it in two units.
Therefore, this model introduces a new scale that explicitly breaks LN.
After the EWSB, \eqref{typeI} leads to a neutrino mass Lagrangian that, in the electroweak interaction basis reads as:
\begin{equation}\label{typeI-MassLag}
\mathcal L_{\rm type-I}^{\rm mass} = - m_D^{ij} \overline{\nu_{Li}^C} \nu_{Rj} -\frac12 m_M^{ij} \overline{\nu_{Ri}^C}\nu_{Rj} +h.c. = 
-\frac12 \overline{N_L} \left(\begin{array}{cc} 0 & m_D \\ m_D^T & m_M \end{array}\right) N_L^C + h.c. 
\end{equation}
where we have defined the LH fields as a column vector  $N_L=(\nu_{Li},\nu_{Ri}^{\, C})^T$. 
From this mass Lagrangian, we identify the neutrino Majorana mass matrix in the EW basis:
\begin{equation}\label{typeImatrix}
M_{\rm type-I}^\nu=\left(\begin{array}{cc} 0 & m_D \\ m_D^T & m_M\end{array}\right).
\end{equation}

It is illustrative to consider first the case where there is only one generation of neutrinos. 
In that context, all the parameters in \eqref{typeImatrix} are just numbers and $M_{\rm type-I}^\nu$ is a $2\times2$ mass matrix which, in the seesaw limit, defined by choosing the two involved scales very distant, $m_D\ll m_M$, has the following two eigenvalues:  
\begin{equation}\label{mnuTypeI-1G}
m_\nu\simeq - \dfrac{m_D^2}{m_M}=- \dfrac{v^2 Y_\nu^2}{m_M}\,,\quad m_N\simeq m_M\,.
\end{equation}
One of the physical neutrinos is then a heavy state $N$ with a mass close to the LN breaking scale $m_M$, while the other is a light state $\nu$ with a small mass $m_\nu$ suppressed by  $m_M^{-1}$. 
This desired suppression is a particular realization of \eqref{MajoranaEffmass}, and can be understood as the tree level processes in the left plot of \figref{SeesawMasses}.

For a more realistic model, we can add three\footnote{Although the addition of two neutrinos is enough for explaining neutrino oscillation data, we prefer to add one RH per generation.} RH fields to the SM spectrum. 
In that case, $M_{\rm type-I}^\nu$ is a $6\times6$ matrix which, again in the seesaw limit, can be block-diagonalized by the following $6\times6$ approximate unitary matrix:
\begin{equation}
 U_\xi^\nu=\left(\begin{array}{cc}(\mathbb{1}-\frac12\xi^*\xi^T) &\xi^*(\mathbb{1}-\frac12\xi^T\xi^*)\\
-\xi^T(\mathbb{1}-\frac12\xi^*\xi^T)&(\mathbb{1}-\frac12\xi^T\xi^*)\end{array}\right)+\mathcal{O}(\xi^4)\ ,
\end{equation}
where we have introduced the small seesaw matrix parameter $\xi=m_Dm_M^{-1}$.
This rotation leads to two separated $3\times3$ mass matrices, given by:
\begin{equation}
U_\xi^{\nu T} M^\nu_{\rm type-I}  U_\xi^\nu\simeq
\left(\begin{array}{cc} -m_D m_M^{-1} m_D^T & 0 \\ 0 &m_M\end{array}\right)\equiv \left(\begin{array}{cc}m_\nu&\\&m_N\end{array}\right)\ .
\end{equation}
We see again that there are two sectors, one light sector whose mass matrix $m_\nu$ is suppressed by $m_M^{-1}$ and a heavy sector with masses close to $m_M$.

Without lost of generality, we can decide to work in the basis where $m_M$ is already diagonal and, therefore, we only need to diagonalize the light mass matrix.
In order to be in agreement with neutrino oscillation data, we can impose this light matrix to be diagonalized by the proper $U_{\rm PMNS}$ matrix and to have the right eigenvalues.
This can be done by requiring:
\begin{equation}\label{mnu}
m_\nu\simeq -m_D m_M^{-1} m_D^T \equiv U_{\rm PMNS}^* m_\nu^{\rm diag} U_{\rm PMNS}^\dagger.
\end{equation}
In this equation, $m_\nu^{\rm diag}$ contains the masses of the physical light neutrino states and the $U_{\rm PMNS}$ matrix the mixing angles, as explained in the previous section. 
\eqref{mnu} can be solved for $m_D$, leading to the Casas-Ibarra parametrization~\cite{Casas:2001sr} of the Dirac mass matrix:
\begin{equation}\label{CasasIbarra}
m_D^T=i \sqrt{m_N^{\rm diag}} R \sqrt{m_\nu^{\rm diag}} U_{\rm PMNS}^\dagger\,,
\end{equation}
where we have used the relation $m_N^{\rm diag}\simeq m_M$ so as to express everything in terms of the physical masses.
This parametrization allow us to use as input parameters the physical masses, the experimentally measured mixing angles and an unknown $3\times3$ complex orthogonal matrix $R$.

As we said, the type-I seesaw model is a simple extension of the SM that explains the smallness of the neutrino masses as a ratio of two very distant scales, the low Dirac mass and the high Majorana mass scales.
In order to accommodate light neutrino masses in the eV scale, this ratio needs to be very small. 
Following \eqref{mnuTypeI-1G}, we see that large $Y_\nu\sim1$ couplings imply GUT scale Majorana masses of $m_M\sim10^{14}$~GeV.
On the other hand, TeV scale heavy neutrinos require very small Yukawa couplings, of the order of $10^{-5}$.
As a consequence, one way or the other, most of the new phenomenology related with these new heavy neutrinos, as well as their direct production at colliders, will be then very suppressed.

\subsection*{Type-II seesaw model}

The type-II seesaw model~\cite{Magg:1980ut,Schechter:1980gr,Wetterich:1981bx,Lazarides:1980nt,Mohapatra:1980yp} addes  a new scalar $SU(2)_L$ triplet with hypercharge 2 to the SM.
In contrast with the fermionic singlets of the type-I seesaw, this new triplet does not interact only with the neutrino fields, but also with the rest of the SM fields, so the full Lagrangian is more involved. 
Nevertheless, for the purpose of our discussion of neutrino mass generation, it is enough to consider  the following relevant terms: 
\begin{equation}\label{LagTypeII}
\mathcal L_{\rm type-II}=-\frac12Y_\Delta^{ij} \overline{L_i^C}\, \widetilde\Delta L_j 
-\mu\Phi^T i\sigma_2\Delta^\dagger  \Phi 
- \frac12 M_\Delta^2 {\rm Tr} \big(\Delta^\dagger\Delta\big) +h.c.
\end{equation}
Here, $\Delta$ stands for the new scalar $SU(2)$ triplet, defined as:
\begin{equation}
\Delta=\left(\begin{array}{cc}
\Delta^+/\sqrt{2} & \Delta^{++} \\
\Delta^0 & -\Delta^+/\sqrt2
\end{array}\right)\,.
\end{equation}
The first term in \eqref{LagTypeII} is a Yukawa like interaction between the SM $SU(2)_L$ leptonic doublet $L$ and the scalar triplet, with coupling $Y_\Delta$ and $\widetilde\Delta=i\sigma_2\Delta^*$.
The other two terms are part of the new scalar potential in this model, with $M_\Delta$ the mass of the triplet and $\mu$ the coupling of the triplet scalar with two Higgs doublets.
The full scalar potential, given in Ref.~\cite{Arhrib:2011uy} for instance, sets the vacuum expectation values for the neutral components of both the doublet and triplet scalars. 
In the limit  where the triplet is heavy, $M_\Delta\gg v$, its vev is given by,
\begin{equation}
v_\Delta\simeq \mu \frac{ v^2}{M_\Delta^2}\,,
\end{equation}
which then gives a Majorana mass for the neutrinos of
\begin{equation}
m_\nu \simeq v_\Delta Y_\Delta\,. 
\end{equation}
Diagrammatically, this mass term can be understood as the tree level process in the middle of \figref{SeesawMasses}.

We can now understand how the type-II seesaw can explain the smallness of the neutrino masses.
Looking at these equations, we see that, on one hand, small Yukawa couplings or heavy triplet masses is a possibility, as in the type-I seesaw.
On the other hand, in these equations there is an extra scale $\mu$, which in the case of being small, could explain the smallness of $m_\nu$ even for large $Y_\Delta$ and low $M_\Delta$.
Furthermore, assigning to the triplet a leptonic number $L=2$, we see that the only LN violating term is precisely the one proportional to this $\mu$ parameter and, therefore, it is natural~\cite{tHooft:1979rat} to consider $\mu$ to be small, as setting it to zero would increase the symmetry of the model. 
As a result, the smallness of neutrino masses is related somehow to a small breaking of a symmetry. 

However, the addition of a new scalar to the SM spectrum will in general modify the Higgs sector or contribute to electroweak precision observables, which are constrained by experiments. 
For instance, precision measurements of the electroweak $\rho$ parameter set an upper bound on the new scalar triplet vev of $v_\Delta\lesssim 3$~GeV.
Fortunately, such a small $v_\Delta$ can be again explained by a small LN violating mass scale $\mu$.
Additionally, the (double) charged components of the scalar triplet could also induce potentially large tree level flavor changing processes, not observed yet by any experiment.

\subsection*{Type-III seesaw model}

The type-III seesaw model~\cite{Foot:1988aq} explains the neutrino masses by adding a new fermionic $SU(2)_L$ triplet to the SM spectrum, which is defined as: 
\begin{equation}
\Sigma=\left(\begin{array}{cc}\Sigma^0/\sqrt2 & \Sigma^+ \\ \Sigma^- & \Sigma^0/\sqrt2\end{array}\right)\,.
\end{equation}
As the $\nu_R$ of the type-I seesaw model,  this $\Sigma$ couples to the LH neutrinos and to the Higgs doublet, with a  Lagrangian  given by
\begin{equation}
\mathcal L_{\rm type-III} = - Y_\Sigma^{ij} \overline{L_i} \widetilde\Phi \Sigma  - \frac12 M_\Sigma^{ij} {\rm Tr}\big(\overline{\Sigma_i^C}\Sigma_j\big) +h.c.
\end{equation}
This Lagrangian is very similar to \eqref{typeI}, with the Yukawa coupling $Y_\Sigma$ and the Majorana mass $M_\Sigma$.
Consequently, in the seesaw limit $M_\Sigma\gg v Y_\Sigma$, the obtained light neutrino mass matrix is equivalent to the one in \eqref{mnu}:
\begin{equation}
m_\nu\simeq -v Y_\Sigma M_\Sigma^{-1} v Y_\Sigma^T\,.
\end{equation}
The tree level diagram generating this  neutrino mass is shown in the right panel of \figref{SeesawMasses}, which is the same as in the type-I seesaw with $\Sigma$ instead of $\nu_R$.
Actually, in this model the neutral component of $\Sigma$ behaves like the right-handed neutrino of the type-I seesaw. 
Nevertheless, being the $\Sigma$ a $SU(2)_L$ triplet, it also has gauge couplings to the SM.
This fact can lead to new phenomenology, such as tree-level flavor changing currents mediated by the charged components  $\Sigma^{\pm}$, which so far have not been observed experimentally.

These three types of seesaw mechanisms are  some examples of models for generating light neutrino masses. 
There are many other proposals in the literature, which try different approaches to explain the lightness of neutrino masses. 
For instance, in the models known as radiative seesaw models, the tree level neutrino masses are forbidden, so they need to be generated at the loop level and, therefore, they are naturally suppressed with respect to the rest of fermion masses (see Ref.~\cite{Cai:2017jrq} for a recent review). 
This is the case in the Zee-Babu model~\cite{Zee:1985id,Babu:1988ki}. 
Another option could be to assume that there is a symmetry protecting the neutrinos from having a tree level mass term, which is spontaneously broken with a small vev that generates small neutrino masses, as  some $R$-parity violating supersymmetric models~\cite{Aulakh:1982yn,Hall:1983id,Barbier:2004ez}. 
For a review of neutrino mass models, see for instance Refs.~\cite{Mohapatra:2005wg,Ma:2016gsc}.

In this Thesis, we are interested in the  phenomenology of right-handed neutrinos with masses at the TeV scale, such that they can lead to not very suppressed low energy effects.
We are also interested in the possibility of producing them at colliders. 
As we said, the type-I seesaw model introduces $\nu_R$ fields that can indeed be at the TeV, although in this case their Yukawa coupling is so small than it suppresses most of the phenomenological implications.
An interesting way out of this situation is to invoke a symmetry that protects the light neutrino masses even in the case of low right-handed masses and large Yukawa couplings. 
This is the main idea behind low scale seesaw models, as the inverse seesaw model, that we describe in full detail in the following.

\section{The inverse seesaw model and its parametrizations}
\label{sec:ISSmodel}

As we discussed above, the original type-I seesaw model cannot have right-handed neutrinos with masses at the TeV scale and large Yukawa couplings and, at the same time, accommodate light neutrino masses at the eV range. 
In low scale seesaw models this interesting situation is accomplished by making use of extra symmetries, for instance a common assumption is global lepton number conservation. 
In order to better understand this idea, we can first consider the simplified situation of one generation, where there is only one left-handed neutrino as in the SM, and add two extra fermionic singlets, $\nu_R$ and $X$. 
The $\nu_R$ field is like in the type-I seesaw model, a right-handed partner of the $\nu_L$, singlet under the full SM group and with lepton number $L=-1$, while the new singlet $X$ has the opposite lepton number $L=1$.
In the limit where LN is conserved, the only terms that we can add to the SM Lagrangian are:
\begin{equation}
\mathcal L = - Y_\nu \overline{L} \widetilde{\Phi} \nu_{R} - M_R \overline{\nu_{R}^C} X\,,
\end{equation}
which, after the EWSB, leads to the following neutrino Majorana mass matrix in the EW $(\nu_L, \nu_R^C,X^C)$ basis: 
\begin{equation}
M^\nu=\left(\begin{array}{ccc} 0 & m_D & 0 \\ m_D & 0 & M_R\\ 0 & M_R & 0\end{array}\right)\,.
\end{equation}
Diagonalizing this matrix we obtain two degenerate Majorana neutrinos, which form one single Dirac neutrino, and one massless neutrino. 
This means that imposing exact LN conservation gives rise to massless neutrinos and, therefore, we need to include a LN breaking in order to generate neutrino masses.  
If this breaking is small, the neutrino masses will be small. 
As a result, in these low scale seesaw models the smallness of neutrino masses is related to a small breaking of a symmetry, which is natural in the sense of 't Hooft~\cite{tHooft:1979rat}.  

Different models can be defined following this idea, depending on where we introduce the  small LN violating scale that generates the masses for the light neutrinos. 
Including  Majorana mass terms for the new fermionic singlets, we end up with the inverse seesaw (ISS) model~\cite{Mohapatra:1986aw,Mohapatra:1986bd,Bernabeu:1987gr}, whereas a LN violating interaction between the $\nu_L$ and $X$ fields defines the linear seesaw (LSS) model~\cite{Akhmedov:1995ip,Akhmedov:1995vm}. 
In this Thesis, we will work in the framework of the inverse seesaw model, although most of our results will also apply  to other  models as the linear seesaw, since we will see that they can be easily related. 

We consider a realization of the inverse seesaw model adapted to three generations where three\footnote{Although the minimal realization of the ISS model that accounts for oscillation data only needs two fermionic pairs~\cite{Abada:2014vea}, usually referred to as the (2,2)-ISS, we prefer to add one pair for each SM family, usually denoted by (3,3)-ISS.} pairs of fermionic singlets $(\nu_R,X)$ are added to the SM. 
The ISS Lagrangian in this case is given by
\begin{equation}
 \label{LagrangianISS}
 \mathcal{L}_\mathrm{ISS} = - Y^{ij}_\nu \overline{L_{i}} \widetilde{\Phi} \nu_{Rj} - M_R^{ij} \overline{\nu_{Ri}^C} X_j - \frac{1}{2} \mu_{R}^{ij} \overline{\nu_{Ri}^C} \nu_{Rj} - \frac{1}{2} \mu_{X}^{ij} \overline{X_{i}^C} X_{j}  + h.c.\,,
\end{equation}
with the favor indices $i,j$ running from 1 to 3. 
Here, $Y_\nu$ is the $3\times3$ neutrino Yukawa coupling matrix, as in \eqref{typeI}, $M_R$ is a lepton number conserving complex $3\times3$ mass matrix, and $\mu_R$ and $\mu_X$ are Majorana complex $3\times3$ symmetric mass matrices that violate LN conservation by two units.
As we said, these two LN violating scales are naturally small in this model,  as setting them to zero would restore the conservation of LN, thus enlarging  the symmetry of the model.
After the EWSB, we obtain the complete $9\times 9$ neutrino mass matrix, which again in the EW $(\nu_{L_i}^{}, \nu_{R_i}^C,X_i^C)$ basis, reads,
\begin{equation}
\label{ISSmatrix}
 M^\nu_{\rm{ISS}}=\left(\begin{array}{c c c} 0 & m_D & 0 \\ m_D^T & \mu_R & M_R \\ 0 & M_R^T & \mu_X \end{array}\right)\,.
\end{equation}
Since this complex mass matrix is symmetric, it can be diagonalized using a $9\times 9$ unitary matrix $U_\nu$ according to
\begin{equation}
U^T_\nu M^\nu_{\rm{ISS}} U_\nu = \text{diag}(m_{n_1},\dots,m_{n_9})\,, 
\end{equation}
where $n_i$ are the nine physical neutrino Majorana states, with masses $m_{n_i}$, respectively, and related to the electroweak eigenstates through the rotation $U_\nu$ as:
\begin{equation}
  \left(\begin{array}{c} \nu_L^C \\ \nu_R \\ X \end{array} \right) = U_\nu P_R \left(\begin{array}{c} n_1 \\ \vdots \\ n_9 \end{array} \right)\,,\quad
    \left(\begin{array}{c} \nu_L \\ \nu_R^C \\ X^C \end{array} \right) = U_\nu^* P_L \left(\begin{array}{c} n_1 \\ \vdots \\ n_9 \end{array} \right)\,.
\end{equation}

As we did for the type-I seesaw, it is interesting to consider first the simplified scenario where there is only one generation of ($\nu_L, \nu_R, X$). 
In that case, all the parameters in \eqref{ISSmatrix} are just numbers and $M_{\rm ISS}^\nu$ is a $3\times3$ mass matrix which, in the $\mu_X\ll m_D,M_R$ limit of approximate LN conservation, has the following three eigenvalues:  
\begin{align}
 m_\nu & \simeq \frac{m_{D}^2}{m_{D}^2+M_{R}^2} \mu_X\,\label{mnuISS},\\
 m_{N_1,N_2} &\simeq \pm \sqrt{M_{R}^2+m_{D}^2} + \frac{M_{R}^2 \mu_X}{2 (m_{D}^2+M_{R}^2)}+\frac{\mu_R}2\,.\label{mN}
\end{align}
\begin{figure}[t!]
\begin{center}
\includegraphics[width=.5\textwidth]{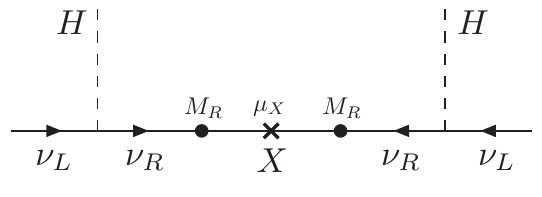}
\caption{Diagram for the tree level light Majorana mass generation in the ISS model. }\label{SeesawMassISS}
\end{center}
\end{figure}
We see that the mass $m_\nu$ of one of the states is small, since it is proportional to the small parameter $\mu_X$, and, therefore, it can be associated to the light neutrino states observed in neutrino oscillations. 
Notice that, contrary to the type-I seesaw model where the lightness of $m_\nu$ is related to a suppression of a large LN breaking scale, here it is proportional to a small LN breaking scale, $\mu_X$, motivating the model name of inverse seesaw.
In the seesaw limit $m_D\ll M_R$, it can be further simplified to $m_\nu\sim \mu_X m_D^2/M_R^2$, which can be understood as the result of the diagram in \figref{SeesawMassISS}. 
The other two mass eigenstates have almost degenerate heavy masses, $m_{N_{1},N_{2}}$, which combine to form a pseudo-Dirac pair. 
Notice that the $\mu_R$ Majorana mass term for the $\nu_R$ fields does not enter in \eqref{mnuISS}, meaning that $\mu_R$ does not generate light neutrino masses at the tree level. 
The effects of this new scale appear only at one-loop level in the light neutrino masses and are, consequently, more suppressed. 
Therefore, we will set $\mu_R$ to zero for the rest of this Thesis and consider a small $\mu_X$ as the only lepton number violating parameter leading to the light neutrino masses. 

\begin{figure}[t!]
\begin{center}
\includegraphics[width=\textwidth]{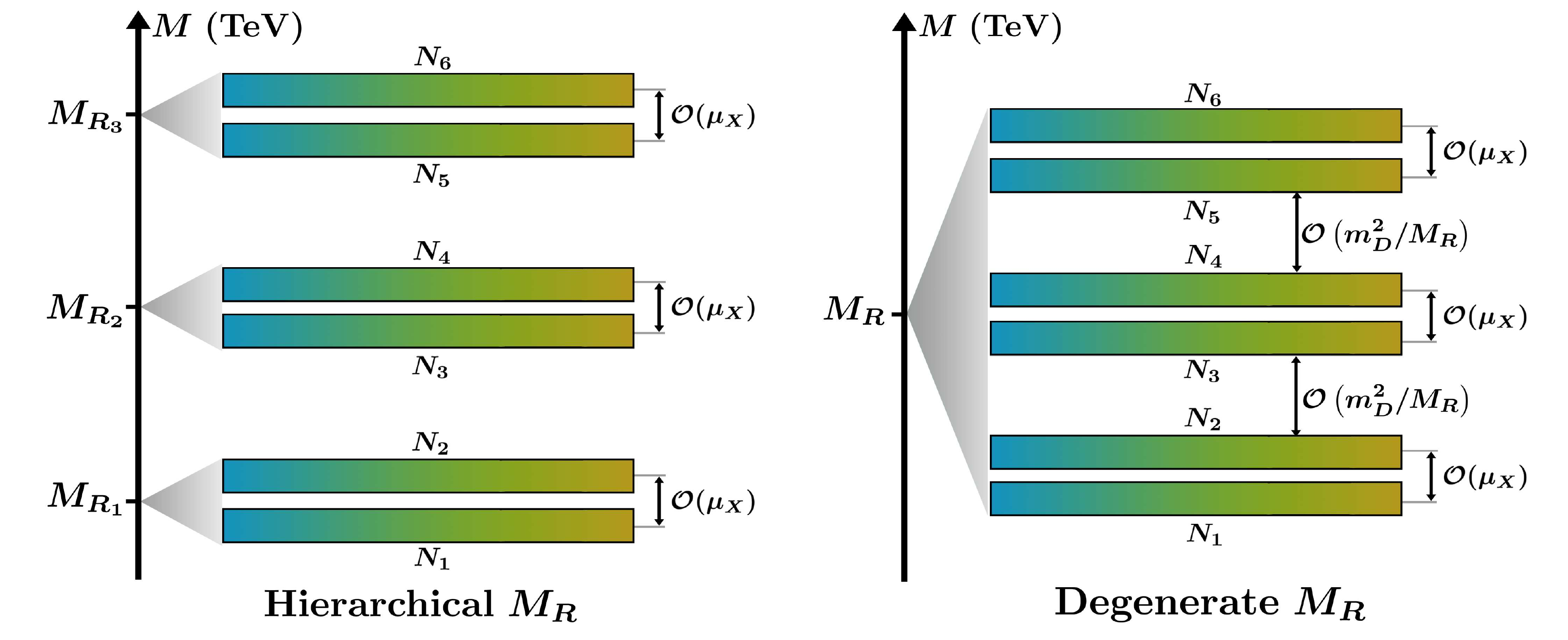}
\caption{Schematic distribution for the mass spectrum of the six ISS heavy neutrinos, $N_{i=1,\dots,6}$ with $m_{N_1}<\dots<m_{N_6}$. 
For hierarchical $M_R$, with $M_{R_1}<M_{R_2}<M_{R_3}$ here, there are three quasidegenerate states with masses $m_{N_1}\simeq m_{N_2}\simeq M_{R_1}$, $m_{N_3}\simeq m_{N_4}\simeq M_{R_2}$ and $m_{N_5}\simeq m_{N_6}\simeq M_{R_3}$.
For degenerate $M_R$, meaning $M_{R_{1,2,3}}\equiv M_R$, all the heavy neutrino masses are close to $M_R$, with small separations between the pairs of $\mathcal O(m_D^2/M_R)$.
In both cases,  there are small $\mathcal O(\mu_X)$ splittings  between the quasidegenerate neutrinos. 
We demand that the light neutrino sector is as in \figref{Hierarchies} by using the parametrizations described in this section.
The color code represents the fact that heavy neutrinos can have a non-trivial flavor structure, which we will further study in \chref{LHC}.
}\label{HeavyMasses}
\end{center}
\end{figure}

A similar pattern of neutrino masses occurs in our three generations case, with one light and two nearly degenerate heavy neutrinos per generation.
In the mass range of our interest with $\mu_X \ll m_D \ll M_R$, the mass matrix $M^\nu_{\mathrm{ISS}}$ can be diagonalized by blocks~\cite{GonzalezGarcia:1988rw}, leading to six heavy neutrinos that form quasidegenerate pairs with masses approximately given by the eigenvalues of $M_R$ and with splittings of order $\mathcal O(\mu_X)$. 
We display schematically the heavy neutrino mass spectrum in \figref{HeavyMasses}.
Regarding the light neutrino sector, it contains three light Majorana neutrinos, whose $3 \times 3$  mass matrix is as follows:
\begin{equation} \label{Mlight}
M_{\rm{light}} \simeq m_D {M_R^T}^{-1} \mu_X M_R^{-1} m_D^T\,.
\end{equation}
In the same manner as we did in \eqref{mnu}, we can ensure the agreement with neutrino oscillation data by demanding:
\begin{equation} \label{mnulight}
 M_{\mathrm{light}}  \equiv U_{\rm PMNS}^* m_\nu^{\rm diag} U_{\rm PMNS}^\dagger\,.
\end{equation}
Then, we can solve this equation for $m_D$, as we did before to obtain the Casas-Ibarra parametrization. 
In fact, this can be easily done by analogy with the type-I seesaw.
Defining a new $3 \times 3$ mass matrix as
\begin{equation}
M=M_R \mu_X^{-1} M_R^T\,,
\label{heavymasses} 
\end{equation}
the light neutrino mass matrix can be written similarly to \eqref{mnu}:
\begin{equation}
 M_{\mathrm{light}}\simeq m_D M^{-1} m_D^T\,.
\end{equation}
Therefore, we can  modify the Casas-Ibarra parametrization in \eqref{CasasIbarra} to define a new version for the ISS given by:
\begin{equation}
 m_D^T= V^\dagger \sqrt{M_{\phantom{a}}^{\rm diag}} R \sqrt{m_\nu^{\rm diag}} U_{\rm PMNS}^\dagger\,,
\label{CasasIbarraISS}
\end{equation}
where $V$ is a unitary matrix that diagonalizes $M$ according to $ M=V^\dagger M^{\rm diag} V^*$ and, as before, $R$ is an unknown complex orthogonal matrix that can be written as
\begin{equation}
\label{R_Casas}
R = \left( \begin{array}{ccc} c_{2} c_{3} 
& -c_{1} s_{3}-s_1 s_2 c_3& s_{1} s_3- c_1 s_2 c_3\\ c_{2} s_{3} & c_{1} c_{3}-s_{1}s_{2}s_{3} & -s_{1}c_{3}-c_1 s_2 s_3 \\ s_{2}  & s_{1} c_{2} & c_{1}c_{2}\end{array} \right)\,,
\end{equation}
with $c_i\equiv \cos \theta_i$, $s_i\equiv \sin\theta_i$ and
$\theta_1$, $\theta_2$, and $\theta_3$ are arbitrary complex angles. 

The Casas-Ibarra parametrization has been extensively used in the literature for accommodating light neutrino data in this type of models. 
We propose here an alternative parametrization that we find very useful to ensure the agreement with oscillation data in the ISS model\footnote{Nevertheless, this idea could be generically applied  to any model with a new scale responsible of explaining the smallness of neutrino masses.}. 
The main idea is to use precisely the new scale $\mu_X$  to codify light neutrino masses and mixings. 
This can be done by solving \eqref{mnulight} for $\mu_X$ instead of $m_D$, which defines our $\mu_X$ parametrization:
\begin{equation}\label{MUXparam}
\mu_X=M_R^T ~m_D^{-1}~ U_{\rm PMNS}^* m_\nu U_{\rm PMNS}^\dagger~ {m_D^T}^{-1} M_R\,.
\end{equation}
Notice that we have assumed that the Dirac mass matrix, or equivalently the Yukawa coupling matrix, is not singular so it can be inverted. 

Of course, physics does not depend on the parametrization one chooses, however the efficiency in exploring the model parameter space does. 
When using this new parametrization, we see two important advantages. 
First, \eqref{mnulight} is quadratic in $m_D$, so solving it leads to a redundancy reflected in the orthogonal matrix $R$ that introduces six new unknown parameters to scan over. 
In contrast, \eqref{mnulight}  is linear in $\mu_X$, and therefore the $\mu_X$ parametrization does not introduce any new parameter and the parameter space can be easily explored.
Second, the $\mu_X$ parametrization considers $Y_\nu$ and $M_R$ as input parameters, while the Casas-Ibarra parametrization works with $\mu_X$ as input instead of $Y_\nu$.
As we will see, the relevant parameters for the LFV and related phenomenology in these models are the  Yukawa coupling and main source of the LFV, $Y_\nu$, and the heavy mass scale $M_R$.
Therefore, being able to treat them as independent input parameters is important for studying the LFV phenomenology in these models.   
Furthermore, this latter point will be also important in order to understand the decoupling behavior of the different observables in the limit of very heavy right-handed neutrinos. 
All these issues  will become clear in the following Chapters.

In order to complete the theoretical set up of the model, we specify next all the relevant neutrino interactions in their mass basis.
These include the neutrino Yukawa couplings, the gauge couplings of the charged and neutral gauge bosons, $W^\pm$ and $Z$, and the couplings of the corresponding Goldstone bosons, $G^{\pm}$ and $G^0$:
\begin{align}
\mathcal L_W&=-\dfrac g{\sqrt{2}} \sum_{i=1}^{3}\sum_{j=1}^{9} W^-_\mu \bar\ell_i B_{\ell_i n_j} \gamma^\mu P_L n_j + h.c., \\
\mathcal L_Z &= -\dfrac g{4c_W} \sum_{i,j=1}^{9}Z_\mu\, \bar n_i \gamma^\mu \Big[C_{n_i n_j} P_L - C_{n_in_j}^* P_R\Big]n_j ,\\
\mathcal L_H &=-\dfrac g{2m_W}  \sum_{i,j=1}^{9} H\,\bar n_i C_{n_in_j}\Big[m_{n_i}P_L+m_{n_j} P_R\Big]n_j,\\
\mathcal{L}_{G^{\pm}} &= -\frac{g}{\sqrt{2} m_W}\sum_{i=1}^{3}\sum_{j=1}^{9} G^{-}\bar{\ell_i} B_{\ell_i n_j} \Big[m_{\ell_i} P_L - m_{n_j} P_R \Big]n_j  + h.c\,,\\ 
\mathcal{L}_{G^{0}} & =-\dfrac {ig}{2 m_W} \sum_{i,j=1}^{9}G^0\, \bar n_i  C_{n_in_j} \Big[m_{n_i}  P_L - m_{n_j} P_R\Big]n_j ,  
\end{align}
where $P_{R,L}=(1\pm \gamma^5)/2$  are the usual chirality projectors and, 
\begin{align}
\label{eq:BCmatrices}
B_{\ell_in_j}&=U_{ij}^{\nu*}, \\
C_{n_in_j}&=\sum_{k=1}^{3} U^\nu_{ki} U^{\nu*}_{kj}\,.
\end{align}
These coupling matrices follow the subsequent interesting identities~\cite{Ilakovac:1994kj}:
\begin{align}
&\sum_{k=1}^{N} B_{\ell_1 n_k}^{} B_{\ell_2 n_k}^* = \delta_{\ell_1 \ell_2}\,,&
&\sum_{k=1}^{N} B_{\ell n_k}^{} C_{n_k n_i}^{} = B_{\ell n_i}^{}\,, &
&\sum_{k=1}^{N} C_{n_i n_k}^{} C^*_{n_j n_k} = \sum_{k=1}^{3} B_{\ell_k n_i}^* B_{\ell_k n_j} ^{}= C_{n_i n_j}^{} \,,\nonumber\\
&\sum_{k=1}^{N} m_{n_k}^{} C_{n_in_k}^{}C_{n_jn_k}^{}=0\,,&
&\sum_{k=1}^{N} m_{n_k}^{} B_{\ell n_k}^{} C^*_{n_kn_i}=0\,,&
&\sum_{k=1}^{N} m_{n_k}^{} B_{\ell_1 n_k}^{} B_{\ell_2 n_k}^{}=0\,,
\end{align}
where $N$ stands for the total number of neutrinos, i.e., $N=9$ for the ISS realization we are considering.

\subsection{The linear seesaw model}

Finally, we want to briefly comment on the other low scale seesaw model above mentioned, the linear seesaw model, and its relation with the ISS model.
As we said, in the LSS model the LN violating mass scale is introduced via a Yukawa coupling, $\widetilde Y_\nu$, between the $\nu_L$ and the $X$ singlets, in contrast to the ISS model where the LN violating scale is introduced via the Majorana mass terms. 
For simplicity, we focus this discussion on the one generation case, although it can be easily generalized to more generations.
In this case, the neutrino mass matrix in the linear seesaw,  in the same EW basis $(\nu_L, \nu_R^C,X^C)$,  reads as:
\begin{equation}\label{MnuLSS}
M_{\rm LSS}^\nu = \left(\begin{array}{ccc}
0& m_D& \widetilde m_D \\ m_D & 0 & M_R \\ \widetilde m_D&M_R&0\end{array}\right)\,,
\end{equation}
where $\widetilde m_D=v\widetilde Y_\nu$ and the other parameters are as before. 
Assuming an approximate LN symmetry, this new Yukawa interaction will be small, leading to a light neutrino mass. 
This is given, for $\widetilde m_D\ll m_D\ll M_R$, by,
\begin{equation}
m_\nu\simeq -2 \widetilde m_D\frac{m_D}{M_R}\,,
\end{equation}
which is linear in $m_D$, motivating the name of the model. 

In order to show the relation between the linear and inverse seesaw models, first we may notice that they are identical in the limit of massless neutrinos, i.e., in the limit of exact LN conservation.
When LN is violated in the linear seesaw model via a non-zero $\widetilde m_D$, we can redefine the fermionic singlets  in such a way that $\widetilde m_D$ is rotated away.
This can be done by  performing a rotation~\cite{Ma:2009du} of $\tan\theta=\widetilde m_D/m_D$ between the ($\nu_R, X)$ fields. 
Then
\begin{equation}
\hspace{-.25cm}
\left(\begin{array}{ccc}1&0&0 \\ 0 &c&s\\0&-s&c\end{array}\right)
\left(\begin{array}{ccc} 0& m_D&\widetilde m_D\\ m_D&0&M_R\\ \widetilde m_D& M_R&0 \end{array}\right)
\left(\begin{array}{ccc} 1&0&0 \\ 0 &c&-s\\0&s&c \end{array}\right)=
\left(\begin{array}{ccc} 0 & m_D/c & 0 \\ m_D/c & M_R s_2 & M_R c_2 \\ 0 & M_R c_2 & -M_R s_2\end{array}\right),
\end{equation}
where $c=\cos\theta$, $s=\sin\theta$, $c_2=\cos2\theta$ and $s_2=\sin2\theta$.
Since $\widetilde m_D\ll m_D$, $\tan\theta$ is small and, therefore, the mass matrix in this rotated basis has the same pattern as the ISS mass matrix in \eqref{ISSmatrix}.
Given this relation, we emphasize again that the results and general conclusions of this Thesis, computed and studied for the particular case of the ISS model, could be applicable to other low scale seesaw models, like the linear seesaw model. 

It is important to notice that in this kind of low scale seesaw models, in contrast to the standard type-I seesaw, there are three different scales which play different roles.
In the particular case of the ISS model, $\mu_X$ controls the smallness of the light neutrino masses, $M_R$ the masses of the new heavy neutrinos and $m_D=vY_\nu$ the interaction between the new added neutrino sector and the SM $\nu_L$ fields.
Since they are independent, we can have at the same time large Yukawa couplings, $Y_\nu^2/4\pi\sim\mathcal O(1)$, and right-handed neutrinos in the TeV range, i.e., reachable at present experiments like the LHC. 
These two properties make the ISS model in particular, and low scale seesaw models in general, interesting models with a rich phenomenology that we wish to explore further in this Thesis.

\subsection{The SUSY Inverse Seesaw model}

The Higgs mechanism of the SM introduces a fundamental scalar particle, the recently found Higgs boson, whose mass has been  experimentally set to $m_H=125.09\pm0.21\, ({\rm stat.})\pm0.11\, ({\rm syst.})$~GeV~\cite{Aad:2015zhl}.
Being a fundamental scalar, it will receive huge contributions via quantum corrections if a new heavy scale associated to some new physics is introduced, for instance, if gravitational effects are introduced at the Planck mass scale $M_P\sim10^{19}$ GeV.
This instability of the Higgs sector under radiative corrections is known as the hierarchy problem and it is one of the main theoretical problems of the SM.
One of the most popular and elegant solutions to this problem is provided by supersymmetry (SUSY)~\cite{Golfand:1971iw,Volkov:1973ix,Wess:1974tw}.
This is a new symmetry that relates fermions with bosons, such that their contributions to the radiative corrections to the Higgs boson mass exactly cancel. 
Therefore, it introduces a new symmetry that protects the scalar sector from unnaturally large radiative corrections.

The minimal SUSY extension of the SM is the Minimal Supersymmetric Standard Model (MSSM)~\cite{Haber:1984rc,Gunion:1984yn,Gunion:1986nh}.
We can understand the way this model extends the SM in two steps. 
First, for consistency reasons, it adds a second scalar $SU(2)_L$ doublet to the SM, in a way that one of the doublets is responsible of giving masses to the up-type fermions and the other one to the down-type fermions.
Then, it doubles all the particles in the spectrum introducing a SUSY partner for each field. 
For a full review of SUSY and the MSSM, we refer the reader to Refs.~\cite{Haber:1993wf,Martin:1997ns}.
However, in this minimal extension of the SM, neutrinos remain massless as in the SM and, therefore, it needs to be further extended in order to explain neutrino oscillation data. 
Here, we shortly summarize the most relevant aspects of the simplest SUSY version of the inverse seesaw model, introducing the new neutrino and sneutrino sectors needed for our forthcoming phenomenological studies, in particular for the LFV Higgs decay computations in \secref{sec:LFVHDSUSY}.

In the simplest supersymmetric realization of the ISS model, the SUSY-ISS model in short, the MSSM superfield content is supplemented by three pairs of gauge singlet chiral superfields $\widehat N_i$ and $\widehat X_i$ with opposite lepton
numbers ($i=1,2,3$). 
As in the original ISS model, this allows to write a Yukawa interaction term for the neutrinos, with coupling $Y_\nu$, a heavy mass term $\widetilde M_R$ between the extra singlets, and a Majorana mass term $\widetilde \mu_X$ for the $X$ fields. 
The SUSY-ISS model is then defined by the following superpotential:
\begin{equation}
 W=W_\mathrm{MSSM} + \varepsilon_{ab} \widehat N Y_\nu \widehat H^b_2 \widehat L^a + \widehat N \widetilde M_R \widehat X + \frac{1}{2} \widehat X \widetilde \mu_X \widehat X\,,
\end{equation}
with $\varepsilon_{12}=1$ and
\begin{equation}
 W_\mathrm{MSSM}=\varepsilon_{ab} \left[ \widehat E Y_e \widehat H^a_1 \widehat L^b + \widehat D Y_D \widehat H^a_1 \widehat Q^b + \widehat U Y_U \widehat H^b_2 \widehat Q^a- \mu \widehat H^a_1 \widehat H^b_2 \right]\,.
\end{equation}
Here, all chiral superfields are taken to be left-handed.
This means that $\widehat Q$ and $\widehat L$ are respectively the chiral superfields involving the left-handed $SU(2)_L$ doublets  $Q_L$ and $L_L$, as well as their SUSY partners. 
For instance, the the spin $0$ and spin $\frac {1}{2}$ components of $\widehat L$ are $(\widetilde\nu_L, \widetilde e_L)$ and $(\nu_L,e_L)$.
On the other hand, the chiral superfields  $\widehat D\,, \widehat U\,, \widehat E$ contain the $d_R^c$, $u_R^c$ and $e_R^c$ fields and their partners, for example, we have $\widehat E = \big[(\widetilde{e_R})^*\,, (e_R)^c\big]$.
The down- and up-type Higgs bosons, $\widehat H_1$ and $\widehat H_2$ respectively, are defined as
\begin{equation}
 \widehat H_1 = \binom{\hat h^0_1}{\hat h^-_1}\,,	\quad \widehat H_2 = \binom{\hat h^+_2}{\hat h^0_2}\,,
\end{equation}
and $\mu$ is the Higgs superfield mass parameter. 
The couplings $Y_{e,D,U}$ are the Yukawa coupling matrices.
The generation indices have been suppressed for simplicity and it should be understood in a tensor notation as $\widehat N Y_\nu \widehat H^b_2 \widehat L^a=\widehat N_i (Y_\nu)_{ij} \widehat H^b_2 \widehat L^a_j$.

Exact SUSY invariance requires any particles to have the same mass of its SUSY partner, meaning that the SM spectrum should have a SUSY copy with the same corresponding masses.
Nevertheless, the lack of any signal from such new particles in the experiments has excluded this situation and, therefore, if SUSY exists in Nature, it must be broken. 
This breaking, however, cannot spoil the above commented solution to the hierarchy problem, so it needs to be softly broken.

Even if we do not know the SUSY breaking mechanism, we can parametrize it  by introducing a set of well established soft SUSY breaking terms~\cite{Girardello:1981wz}.
In the SUSY-ISS model, this soft SUSY breaking Lagrangian is given by
\begin{align}
 -\mathcal{L}_\mathrm{soft}=&-\mathcal{L}_\mathrm{soft}^\mathrm{MSSM} 
 + \widetilde \nu_R^T\, m_{\tilde \nu_R}^2\, \widetilde \nu_R^* 
 + \bigg[\widetilde \nu_R^\dagger (A_{\nu} Y_\nu) \widetilde \nu_L h_2^0 - \widetilde \nu_R^\dagger (A_{\nu} Y_\nu) \widetilde e_L h_2^+ + h.c.\bigg]\nonumber\\ 
& + \widetilde X^T m_{\tilde X}^2 \widetilde X^* 
   + \bigg[\widetilde X^\dagger (B_{X} \widetilde \mu_X) \widetilde X^* + \widetilde \nu_R^\dagger (B_{R} \widetilde M_R) \widetilde X^* + h.c.\bigg]\,,
\end{align}
with
\begin{align}
 -\mathcal{L}_\mathrm{soft}^\mathrm{MSSM}&= \widetilde e_R^T\, m_{\tilde e}^2\, \widetilde e_R^* +  \widetilde d_R^T\, m_{\tilde d}^2\, \widetilde d_R^* + \widetilde u_R^T\, m_{\tilde u}^2\, \widetilde u_R^* 
		      + m_{H_1}^2 |H_1|^2 + m_{H_2}^2 |H_2|^2 \nonumber \\
		      & + \delta_{ab} (\widetilde Q^a)^\dagger m_{\tilde Q}^2 \widetilde Q^b + \delta_{ab} (\widetilde L^a)^\dagger m_{\tilde L}^2 \widetilde L^b 
		       + \frac{1}{2} \bigg[ M_1 \bar \lambda_b \lambda_b + M_2 \bar \lambda^\alpha_W \lambda^\alpha_W 
		       +  M_3 \bar \lambda^\alpha_g \lambda^\alpha_g + h.c.\bigg] \nonumber \\
		      & + \varepsilon_{ab} \bigg[ (\widetilde u_R^\dagger (A_u Y_u) \widetilde Q^a H^b_2
		       + \widetilde d_R^\dagger (A_d Y_d) \widetilde Q^b H^a_1
		       + \widetilde e_R^\dagger (A_e Y_e) \widetilde L^b H^a_1 + B \mu H_2^a H_1^b + h.c. \bigg]. 
\end{align}
These SUSY breaking Lagrangians introduce a set of new unknown parameters to the SUSY-ISS model.
From the MSSM Lagrangian, we have the squark and slepton soft masses, $m^2_{\tilde e,\tilde d, \tilde u, \tilde Q, \tilde L}$, the Higgs sector soft masses $m^2_{H_{1,2}}$ and $B \mu$, the gaugino soft masses $M_{1,2,3}$ and the trilinear couplings for squarks and sleptons, $A_{u,d,e}$.
In the SUSY-ISS model, there are some extra soft parameters related to the new added fields, in particular, the soft masses  $m^2_{\tilde\nu_R,\tilde X}, B_X\widetilde \mu_X$ and $B_R \widetilde M_R$, and the trilinear coupling $A_\nu$.

All these new parameters could, in principle, have a non-trivial flavor structure.
Nevertheless, when studying the SUSY-ISS model, and for the sake of simplicity, we will take all soft SUSY breaking masses, as well as the lepton number conserving mass term $\widetilde{M_R}$, to be flavor diagonal.
This way, the only sources of flavor violation are the neutrino Yukawa coupling $Y_\nu$, and the lepton number violating mass term $\widetilde \mu_X$. 
The only exception will be $m_{\tilde L}^2$, which receives radiative corrections via the renormalization group equations (RGE), from a heavy scale $M$ with universal soft SUSY breaking parameters down to the heavy neutrino scale $M_R$.
These corrections are also governed by $Y_\nu$ and, for phenomenological purposes, can be described as~\cite{Hisano:1995cp}:
\begin{equation} \label{SleptonMixing}
 \big(\Delta m_{\tilde L}^2\big)_{ij} = -\frac{1}{8\pi^2} (3 M_0^2 + A_0^2) \Big(Y_\nu^\dagger \log  \frac{M}{M_R} Y_\nu^{}\Big)_{ij}\,.
\end{equation}

After the EWSB, the neutrino sector is as in the previous ISS model, therefore the analysis of the mass matrix diagonalization and the discussion about using the Casas-Ibarra or the $\mu_X$ parametrization for accommodating oscillation data is the same as before. 
Hence, it is enough to describe the new SUSY sector, and the sneutrino mass matrix $M^2_{\tilde \nu}$, which is defined by
\begin{equation}\label{Msnu}
 -\mathcal{L}^{\tilde \nu}_{\mathrm{mass}}=\frac{1}{2}\left(\tilde \nu_L^\dagger\,, \tilde \nu_L^T\,, \tilde \nu_R^T\,, \tilde \nu_R^\dagger\,, \tilde X^T\,, \tilde X^\dagger\right) 
 M^2_{\tilde \nu} \left( \begin{array}{c} \tilde \nu_L \\ \tilde \nu_L^*\\ \tilde \nu_R^*\\ \tilde \nu_R \\ \tilde X^* \\ \tilde X \end{array} \right)\,,
\end{equation}
where $\tilde \nu_L$, $\tilde\nu_R$ and $\tilde X$ are vectors made of 3 weak eigenstates each and they are defined in a similar fashion, e.g. $\tilde \nu_L= (\tilde \nu_L^{(e)}\,,\tilde \nu_L^{(\mu)}\,, \tilde \nu_L^{(\tau)})^T$. 
The complete  $18\times18$ sneutrino mass matrix is then expressed in terms of $3\times3$ submatrices, giving:
\begin{equation}
 M^2_{\tilde \nu}=\left( \begin{array}{cccccc}
                          M^2_{LL} & 0 & 0 & M^2_{LR} & m_D M_R^* & 0 \\
                          0 & (M^2_{LL})^T & (M^2_{LR})^* & 0 & 0 & m_D^* M_R \\
                          0 & (M^2_{LR})^T & M^2_{RR} & 0 & M_R \mu_X^* & (B_R M_R^*)^* \\
                          (M^2_{LR})^\dagger & 0 & 0 & (M^2_{RR})^T & B_R M_R^* & M_R^* \mu_X \\
                          M_R ^T m_D^\dagger & 0 & \mu_X M_R^\dagger & (B_R M_R^*)^\dagger & M^2_{XX} & 2 (B_X \mu_X^*)^\dagger \\
                          0 & M_R^\dagger m_D^T & (B_R M_R^*)^T & \mu_X^* M_R^T & 2 (B_X \mu_X^*) & (M^2_{XX})^T
                         \end{array} \right)\,,
\end{equation}
with
\begin{align}
 M^2_{LL} &= m_D m_D^\dagger + m^2_{\tilde L} + \mathbb{1} \frac{m_Z}{2} \cos 2\beta\,, \\
 M^2_{LR} &= -\frac{\mu}{\tan \beta} m_D + m_D A_\nu^\dagger\,, \\
 M^2_{RR} &= m_D^T m_D^* + M_R M_R^\dagger + m^2_{\tilde \nu_R}\,, \\
 M^2_{XX} &= M_R^T M_R^* + \mu_X \mu_X^* + m^2_{\tilde X}\,.
\end{align}
Then, the sneutrino mass matrix is diagonalized by using:
\begin{equation}
 \tilde U^\dagger M^2_{\tilde \nu} \tilde U = M^2_{\tilde n} = \mathrm{diag}(m^2_{\tilde n_1}\,, ...\,, m^2_{\tilde n_{18}} ) \,,
\end{equation}
which corresponds to the following rotation between the interaction and mass basis
\begin{equation}
 \left( \begin{array}{c} \tilde \nu_L \\ \tilde \nu_L^*\\ \tilde \nu_R^*\\ \tilde \nu_R \\ \tilde X^* \\ \tilde X \end{array} \right) =
 \tilde U \left( \begin{array}{c} \tilde n_1 \\ \vdots \\ \vdots  \\ \vdots \\ \tilde n_{18} \end{array} \right)\,. \label{snuRot}
\end{equation}
Notice that the basis used in \eqref{Msnu} uses the sneutrino electroweak eigenstates, and their complex conjugate states and they fulfill:
\begin{align}
 \tilde \nu_i &= \tilde U_{i,j} \tilde n_j\,, \\
 \tilde \nu_i^* &= \tilde U_{3+i,j} \tilde n_j\,, \label{thisone}
\end{align}
but at the same time:
\begin{equation}
 (\tilde \nu_i)^* = \tilde U_{i,j}^* \tilde n_j\,, \label{thistwo}
\end{equation}
since the physical sneutrinos are real scalar fields. While both \eqrefs{thisone} and (\ref{thistwo}) are equally valid, we choose \eqref{thisone}. 

The mass matrices of the other SUSY particles, namely the charginos, neutralinos, and charged sleptons, are the same as in the SUSY type-I seesaw studied in Ref.~\cite{Arganda:2004bz}, so we will use their definitions of the corresponding rotation matrices, which in turn were based on the conventions of Ref.~\cite{Gunion:1984yn} for the charginos and neutralinos. 
Specifically, $U$ and $V$ will be the matrices that rotate the chargino states and $N$ the one that rotates the neutralino states.
In addition, combinations of  rotation matrices for the neutralinos are defined as
\begin{align}\label{Na1a2}
 N'_{a1}&=\phantom{-}N_{a1}\cos\theta_W+N_{a2}\sin\theta_W\,, \nonumber \\
 N'_{a2}&=-N_{a1}\sin\theta_W+N_{a2}\cos\theta_W\,. 
\end{align}
As for the charged sleptons, they are diagonalized by
\begin{equation}\label{sleptonRot}
 \tilde \ell^{'} = R^{(\ell)} \tilde \ell\,,
\end{equation}
where $\tilde \ell^{'}=(\tilde e_L,\tilde e_R,\tilde\mu_L,\tilde\mu_R,\tilde\tau_L\,, \tilde \tau_R)^T$ are the weak eigenstates
and $\tilde \ell=(\tilde \ell_1\,, ...\,, \tilde \ell_6)^T$ are the mass eigenstates.

Finally, we introduce the relevant interaction terms from the Lagrangian that will be needed later in the study of the LFV Higgs decays.
Following again the notation in Ref.~\cite{Arganda:2004bz}, these terms are given in the mass basis by 
\begin{align}
\mathcal{L}_{\tilde \chi_j^- \ell \tilde \nu_{\alpha} } &= 
-g\, \bar{\ell} \left[ A_{L \alpha j}^{(\ell)} P_L + 
A_{R \alpha j}^{(\ell)} P_R \right] \tilde \chi_j^- \tilde \nu_{\alpha} + h.c.\,, \nonumber \\
\mathcal{L}_{\tilde \chi_a^0 \ell \tilde \ell_{\alpha} } &= 
-g\bar{\ell} \left[ B_{L \alpha a}^{(\ell)} P_L + 
B_{R \alpha a}^{(\ell)} P_R \right] \tilde \chi_a^0 \tilde \ell_{\alpha} + h.c.\,, \nonumber \\ 
\mathcal{L}_{H_x \tilde s_{\alpha} \tilde s_{\beta}}&=
-i H_x \left[ g_{H_x\tilde \nu_{\alpha} \tilde \nu_{\beta}} \tilde \nu_{\alpha}^* 
\tilde \nu_{\beta}
+ g_{H_x\tilde \ell_{\alpha} \tilde \ell_{\beta}} \tilde \ell_{\alpha}^* \tilde \ell_{\beta}
\right]\,,\nonumber\\
\mathcal{L}_{H_x \tilde \chi_i^- \tilde \chi_j^-}&=
- g H_x\bar{\tilde{\chi}}_i^- 
\left[ W_{Lij}^{(x)}P_L+ W_{Rij}^{(x)}P_R \right] \tilde{\chi}_j^-\,,\nonumber\\
\mathcal{L}_{H_x \tilde \chi_a^0 \tilde \chi_b^0}&=
- \frac{g}{2} H_x\bar{\tilde{ \chi}}_a^0 
\left[ D_{Lab}^{(x)}P_L+ D_{Rab}^{(x)}P_R \right] \tilde {\chi}_b^0\,, \nonumber \\
\mathcal{L}_{H_x\ell\ell} &= 
-g H_x\bar{\ell} \left[ S_{L,\ell}^{(x)} P_L + S_{R,\ell}^{(x)} P_R \right]  \ell \,. \label{SUSYintLagrangian}
\end{align}
The coupling factors here are given in terms of the SUSY-ISS model parameters in \appref{App:LFVHD_SUSY}.

\vspace{.5cm}
Summarizing, in this Chapter we have seen that the experimental evidences for neutrino masses have established the need of new physics in order to add neutrino masses to the SM. 
We have reviewed some popular neutrino mass generation models, paying special attention to two low scale seesaw models, the ISS and the SUSY-ISS models, which add heavy neutrinos with masses at the reach of the LHC. 
We have discussed in detail the neutrino sector of these models and introduce a new parametrization, the $\mu_X$ parametrization, that allows to accommodate neutrino oscillation data while choosing as input parameters the Yukawa coupling matrix and the heavy neutrino mass matrix $M_R$, i.e., the most relevant parameters for our forthcoming study of the charged LFV observables.

\chapter{Phenomenological implications of low scale seesaw neutrinos on LFV}
\fancyhead[RO] {\scshape Phenomenological implications of low scale seesaw neutrinos on LFV  }
\label{PhenoLFV}

In this Chapter we revisit some of the most relevant phenomenological implications of right-handed neutrinos with TeV scale masses, paying special attention to their lepton flavor violating consequences.
After reviewing the experimental status of charged LFV searches, we discuss in detail the LFV radiative and three-body lepton decays in presence of right-handed neutrinos at the TeV scale.
Furthermore, we  also study other observables that are modified by the new neutrino sector, such as electroweak precision observables, or processes with lepton number violation or lepton flavor universality violation.
This Chapter will be very useful and illustrative to learn the general ideas about LFV from TeV right-handed neutrinos and to describe the main points of our analysis, as well as to introduce the set of observables that we will consider as potential constraints when studying maximum allowed predictions for LFV Higgs and Z decays in the forthcoming Chapters. 
The content of this Chapter, except \secref{sec:LFVexp}, is original work of this Thesis.
It includes the proposal of new scenarios with suppressed $\mu$-$e$ transitions, the geometrical interpretation of the associated neutrino Yukawa coupling matrix, as well as the phenomenological consequences of these mentioned scenarios. 
All these new contributions have been published in Refs.~\cite{Arganda:2014dta,DeRomeri:2016gum}. 

\section{Experimental status of charged LFV and constraints}
\label{sec:LFVexp}

Lepton flavor violating processes are forbidden in the SM due to the assumption of massless neutrinos, therefore any observation of lepton flavor violation would automatically imply the presence of new physics beyond the SM.
This was the case when LFV was observed in neutrino oscillations, what led to the need of extending the SM to include neutrino masses, as we discussed in \chref{Models}.
Interestingly, if neutrino masses and mixings are minimally added to the SM, they radiatively induce LFV in the charged lepton sector (cLFV), although with extremely small ratios, suppressed by the smallness of the  neutrino masses. 
For instance, using neutrino oscillation data in \eqref{NuFit}, the predictions for the $\mu\to e\gamma$ ratio are of the order of $10^{-50}$.
Consequently, a positive experimental signal for cLFV would open a new window for BSM physics and could also help throwing light on the question of what is the mechanism that generates  the neutrino masses.

\begin{figure}[t!]
\begin{center}
\includegraphics[width=.5\textwidth]{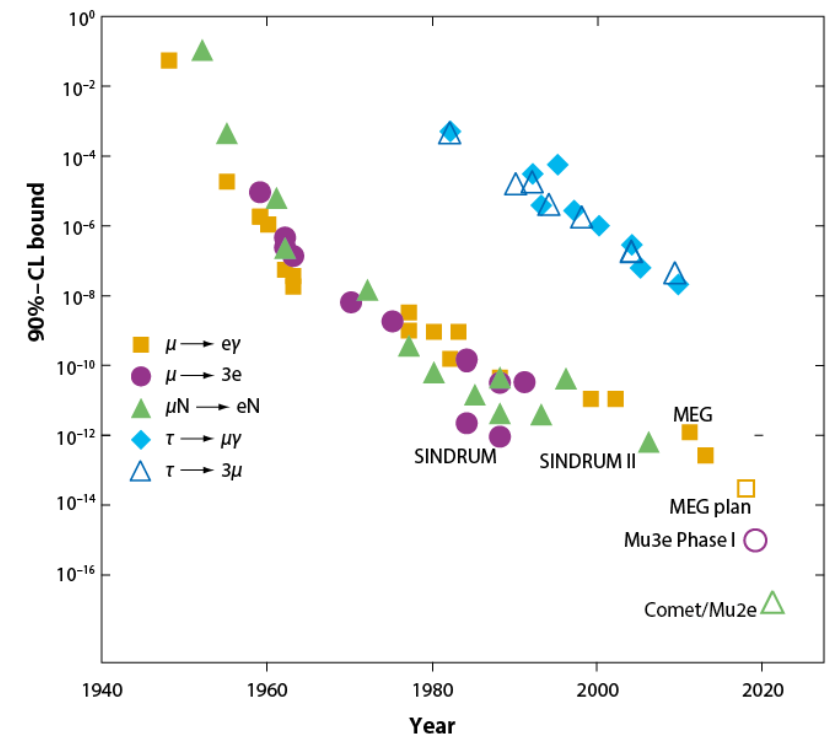}
\caption{Evolution of cLFV transition upper limits  from several experiments, including expected sensitivities for some next generation experiments. Figure borrowed from Ref.~\cite{Bravar:2016nxt}.}\label{cLFVevolution}
\end{center}
\end{figure}

Unfortunately, cLFV has not been observed yet in Nature, although there is an extensive experimental program developing different strategies to look for new physics signals in the charged lepton sector and, indeed, there are already  very competitive upper bounds on several cLFV processes.
One of the standard searches focuses on the radiative decay of a muon into a electron and a photon, concretely on the $\mu^+\to e^+\gamma$ channel, which has impressively evolved from the first bound of less than a $10\%$ by Hincks and Pontecorvo~\cite{Hincks:1948vr} in 1947, to the latest upper bound by the MEG collaboration~\cite{TheMEG:2016wtm} of $4.2\times10^{-13}$ in 2016.
We schematically show this evolution in \figref{cLFVevolution}, borrowed from Ref.~\cite{Bravar:2016nxt}. For a complete historical review see, for instance, Refs.~\cite{Kuno:1999jp,Bernstein:2013hba}.
In the next few years, the upgrade from MEG to MEG-II~\cite{Baldini:2013ke} is expected to improve the sensitivity to this cLFV channel in one order of magnitud. 

Another possible LFV decay channel of the muon is $\mu\to 3e$, complementary to the $\mu\to e\gamma$ channel and very interesting for many BSM models. 
The best current upper bound is provided by SINDRUM~\cite{Bellgardt:1987du} which sets BR$(\mu^+\to e^+e^+e^-)<1.0\times10^{-12}$, although a huge improvement is expected to be obtained by the Mu3e experiment~\cite{Blondel:2013ia}.
This experiment has been proposed at PSI, with the aim of reaching decay rates up to $\mathcal O(10^{-15})$  with the current muon beamline and $\mathcal O(10^{-16})$  if an upgrade to a High Intensity Muon Beam (HiMB) is achieved at PSI.

\begin{table}[t!]
\begin{center}
\caption{Present upper bounds and future expected sensitivities for cLFV transitions.}\label{LFVsearch}
{\small
\begin{tabular}{lllrl}
\toprule
\toprule
LFV Obs. & \multicolumn{2}{l}{Present Bound  $(90\%CL)$} & \multicolumn{2}{l}{Future Sensitivity} \\
\midrule
BR$(\mu\to e\gamma)$ &  $4.2\times10^{-13}$ &MEG (2016)\cite{TheMEG:2016wtm} & $6\times 10^{-14} $& MEG-II~\cite{Baldini:2013ke}\\
BR$(\tau\to e\gamma)$ & $3.3\times10^{-8}$ &BABAR (2010)~\cite{Aubert:2009ag} & $ 10^{-9}$ &BELLE-II~\cite{Aushev:2010bq}\\
BR$(\tau\to \mu\gamma)$ & $4.4\times10^{-8}$ &BABAR (2010)~\cite{Aubert:2009ag} & $ 10^{-9}$ &BELLE-II~\cite{Aushev:2010bq}\\
BR$(\mu\to eee)$ &   $1.0\times10^{-12}$ & SINDRUM (1988)~\cite{Bellgardt:1987du} & $10^{-16}$ &Mu3E (PSI)~\cite{Blondel:2013ia} \\
BR$(\tau\to eee)$ & $2.7\times10^{-8}$ &BELLE (2010)~\cite{Hayasaka:2010np} & $10^{-9,-10}$ &BELLE-II~\cite{Aushev:2010bq}\\
BR$(\tau\to \mu\mu\mu)$ & $2.1\times10^{-8}$ &BELLE (2010)~\cite{Hayasaka:2010np} & $10^{-9,-10}$ &BELLE-II~\cite{Aushev:2010bq}\\
BR$(\tau\to \mu\eta)$ & $2.3\times10^{-8}$ &BELLE (2010)~\cite{Hayasaka:2010et} & $10^{-9,-10}$ &BELLE-II~\cite{Aushev:2010bq}\\
CR$(\mu-e,{\rm Ti})$ &  $4.3\times10^{-12}$ &SINDRUM II (2004)\cite{Dohmen:1993mp}&$10^{-18}$ &PRISM (J-PARC)~\cite{Alekou:2013eta}\\
CR$(\mu-e,{\rm Au})$ & $7.0\times10^{-13}$ &SINDRUM II (2006)\cite{Bertl:2006up}& \\
CR$(\mu-e,{\rm Al})$ &&&$3.1\times10^{-15}$ &COMET-I (J-PARC)\cite{Kuno:2013mha}\\
&&&$2.6\times10^{-17}$ &COMET-II (J-PARC)\cite{Kuno:2013mha} \\
&&&$2.5\times10^{-17}$ &Mu2E (Fermilab)~\cite{Carey:2008zz} \\
\bottomrule
\bottomrule
\end{tabular}
}
\end{center}
\end{table}

Alternatively to muon decays, LFV between muons and electrons has been extensively searched for in $\mu\to e$ conversion experiments. 
Here, muons are stopped in a thin layer and form muonic atoms, in which a muon can be converted into an electron,
\begin{equation}
\mu^- + N \to e^- + N\,.
\end{equation}
Such a conversion in the field of the nucleus has as a clear signal the emission of a monochromatic electron of $E\simeq 100$~MeV, where the precise value of its energy depends on the nucleus~\cite{Measday:2001yr}. 
Current upper bound comes from the SINDRUM II collaboration, which is set to $7.0\times10^{-13}$ using gold atoms~\cite{Bertl:2006up}. 
Nevertheless, a strong experimental effort is planned in this direction, implying extraordinary expected sensitivities of $\mathcal O(10^{-18})$ for the next generation of $\mu$-$e$ conversion experiments, as  PRISM~\cite{Alekou:2013eta} at J-PARC or Mu2E at Fermilab~\cite{Carey:2008zz}.
Interestingly, these $\mu$-$e$ conversion experiments can also be used to look for a muon conversion into a positron 
\begin{equation}
\mu^- + N(A,Z) \to e^+ + N(A,Z-2)\,,
\end{equation}
in a process that violates, besides lepton flavor, total lepton number in two units.
These searches are complementary to other experiments looking for LN violation, like neutrinoless double beta decay ($0\nu\beta\beta$), and therefore, they are very interesting for testing the Majorana character of the neutrinos.
Although the experimental signal is not as clear as in the $\mu^-$- $e^-$ conversions, the SINDRUM II collaboration was able to set a very compelling upper bound on the $\mu^-+{\rm Ti} \to e^+ + {\rm Ca(g.s.)}$ transition of $4.3\times10^{-12}$ at the $90\%$ CL~\cite{Dohmen:1993mp}.
Furthermore, additional cLFV evidences are being looked for, like $\mu^+ e^-\to \mu^- e^+$ transitions of muonium atoms.
We refer to Ref.~\cite{Kuno:1999jp} for a complete review of cLFV in muon transitions.

\begin{figure}[t!]
\begin{center}
\includegraphics[width=.9\textwidth]{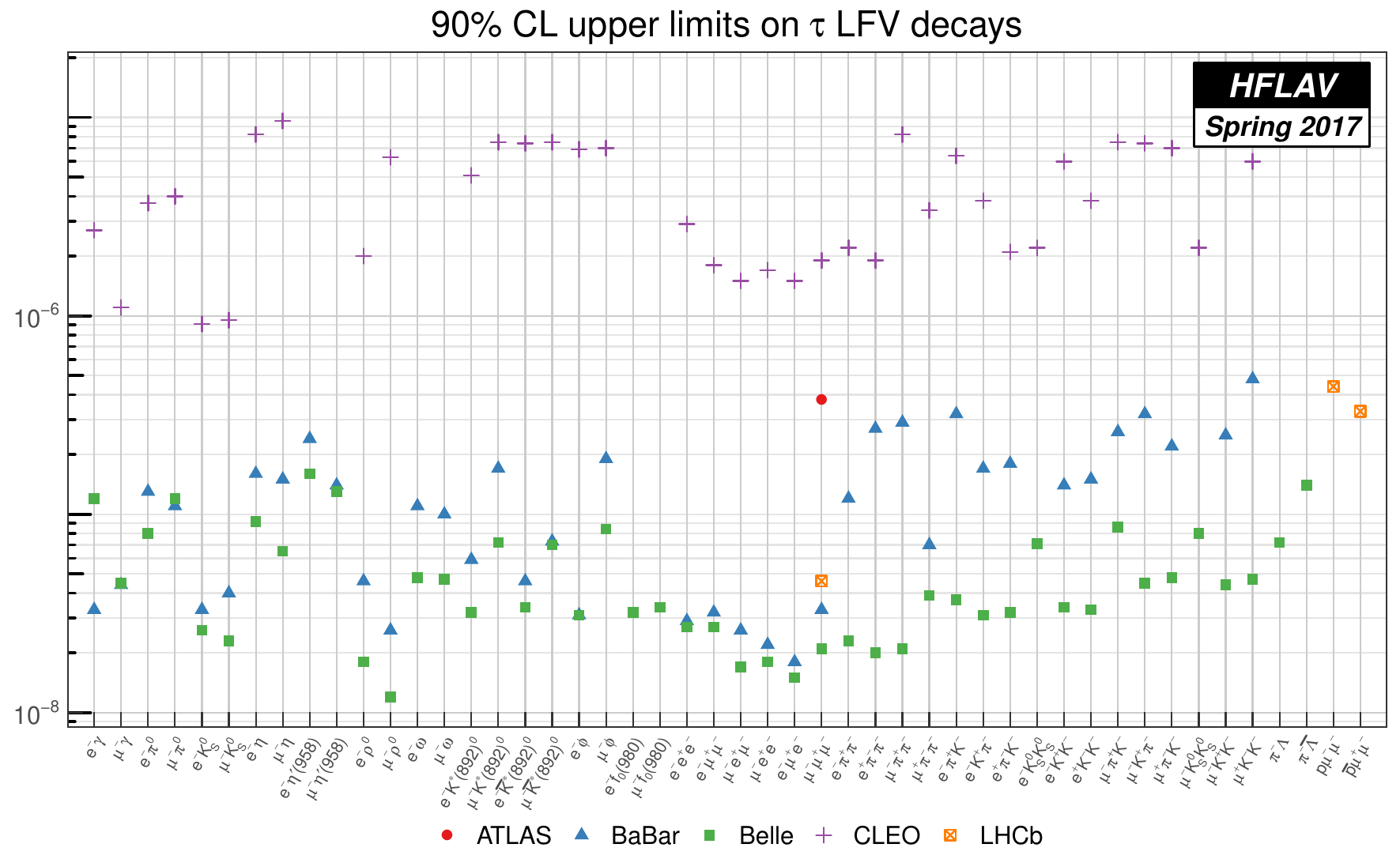}
\caption{Summary on present  LFV $\tau$ decay upper limits by the HFLAV group~\cite{Amhis:2016xyh}.}\label{HFLAV}
\end{center}
\end{figure}
In order to study LFV processes in the tau sector, we need to look at B factories. 
The BABAR collaboration put the most constraining upper bounds of LFV radiative tau decays, setting  BR$(\tau^\pm\to\mu^\pm\gamma)<4.4\times10^{-8}$ and BR$(\tau^\pm\to e^\pm\gamma)<3.3\times10^{-8}$~\cite{Aubert:2009ag}.
On the other hand, BELLE was able to constraint LFV three body decays~\cite{Hayasaka:2010np}, setting BR$(\tau^-\to\mu^-\mu^+\mu^-)<2.1\times10^{-8}$, BR$(\tau^-\to e^-e^+e^-)<2.7\times10^{-8}$, and similar bounds for mixed combinations of muons and electrons in the final state. 
The LHCb collaboration has also performed searches for $\tau^-\to\mu^-\mu^+\mu^-$ decays, finding an upper bound of $4.6\times10^{-8}$~\cite{Aaij:2014azz}. 
This analysis, done with $1 {\rm fb}^{-1}$ of proton-proton collision at $\sqrt{s}=7$~TeV and $2 {\rm fb}^{-1}$ at 8~TeV, already has a sensitivity close to the best current bounds, so we could expect that LHCb can tell us something new about this decay channel in a near future. 
Moreover, in 2018, the BELLE-II experiment~\cite{Aushev:2010bq} will start its operation with the aim of improving the sensitivities to both types of LFV $\tau$ decays in up to two orders of magnitude. 
We summarize the current upper bounds, as well as future expected sensitivities, on cLFV transitions in \tabref{LFVsearch}. 

An interesting feature about $\tau$ leptons is that their mass is large enough to decay into hadrons. 
This fact opens a new experimental window looking for semileptonic LFV tau decays. 
We summarize current upper bounds on the LFV $\tau$ decays in \figref{HFLAV}, taken from the Heavy Flavor Average (HFLAV) group~\cite{Amhis:2016xyh}, which are  around $10^{-8}$-$10^{-7}$.
Nevertheless, we can expect that BELLE-II will improve these sensitivities in about two orders of magnitude. 

\begin{table}[t!]
\begin{center}
\caption{Present upper bounds for some LFV/LNV meson decays.}\label{LFVsearchMesons}
{\small
\begin{tabular}{lll}
\toprule
\toprule
LFV/LNV Obs. & \multicolumn{2}{l}{Present Bound  $(90\%CL)$}  \\
\midrule
BR$(\pi^0\to\mu e)$ & $3.59\times10^{-10}$ & KTeV (2008)~\cite{Abouzaid:2007aa}\\
BR$(K_L\to\mu e)$ & $4.7\times10^{-12}$ & BNL E871 (1998)~\cite{Ambrose:1998us}\\
BR$(K_L\to\pi^0\mu e)$ & $7.56\times10^{-11}$ & KTeV (2008)~\cite{Abouzaid:2007aa}\\
BR$(K_L\to\pi^0\pi^0\mu e)$ & $1.64\times10^{-10}$ & KTeV (2008)~\cite{Abouzaid:2007aa}\\
BR$(K^+\to \pi^+\mu^+e^-)$ & $1.3\times10^{-11}$ & BNL E777/865 (2005)~\cite{Sher:2005sp}\\
BR$(K^+\to \pi^+\mu^-e^+)$ & $5.2\times10^{-10}$ & BNL E865 (2000)~\cite{Appel:2000tc}\\
BR$(K^+\to \pi^-\mu^+e^+)$ & $5.0\times10^{-10}$ & BNL E865 (2000)~\cite{Appel:2000tc}\\
BR$(K^+\to \pi^-e^+e^+)$ & $6.4\times10^{-10}$ & BNL E865 (2000)~\cite{Appel:2000tc}\\
BR$(K^+\to \pi^-\mu^+\mu^+)$ & $1.1\times10^{-9}$ & NA48/2 (2010)~\cite{Batley:2011zz}\\
BR$(B^+\to K^-\mu^+\mu^+)$ & $5.4\times10^{-8}$ & LHCb (2012)~\cite{Aaij:2011ex}\\
BR$(B^+\to \pi^-\mu^+\mu^+)$ & $5.8\times10^{-8}$ & LHCb (2012)~\cite{Aaij:2011ex}\\
\bottomrule
\bottomrule
\end{tabular}
}
\end{center}
\end{table}
Experimental searches of meson decays are also extremely interesting for looking for both LFV and LNV processes. 
LFV neutral kaon decay, $K_L\to\mu e$, has been extensively searched at experiments (see Ref.~\cite{Bernstein:2013hba} for a review) and currently the most stringent bound by the BNL collaboration sets BR$(K_L\to\mu^\pm e^\mp)<4.7\times10^{-12}$~\cite{Ambrose:1998us}.
Other LFV or LNV  $K_L$ or $\pi^0$ decays, as well as charged kaon $K^+\to\pi^\pm\ell_1^+\ell_2^\mp$  decays, have been searched at several experiments, although no positive signal has been found yet. 
We summarize in \tabref{LFVsearchMesons} some of these upper bounds.

We conclude this overview about cLFV experimental searches by commenting on possible $Z$ and Higgs boson LFV decays. 
LEP, as a $Z$ factory, searched for LFV $Z\to\ell\ell'$ decays with no luck~\cite{Akers:1995gz,Abreu:1996mj}. 
Thus, it established upper limits to these processes, as we summarize in \tabref{LFVsearchII}, which are still the most constraining ones when a tau lepton is involved. 
At present, these processes are being searched at the LHC and ATLAS is already at the level of LEP results for the LFV $Z$ decay rates, and even better for $Z\to\mu e$ channel~\cite{Aad:2014bca}.
Thus, we can expect that new LHC runs would help testing these channels. 
Moreover, the sensitivities to LFV $Z$ decay rates are expected to highly improve at the next generation of colliders.
In particular, following the discussion in Ref.~\cite{Abada:2014cca},  the future linear colliders are expected to reach sensitivities of $10^{-9}$ \cite{Wilson:I, Wilson:II}, and at a Future Circular  $e^+ e^-$ Collider (such as FCC-ee (TLEP)\cite{Blondel:2014bra}), where it is estimated that up to $10^{13}$ $Z$ bosons would be produced, the sensitivities to LFVZD rates could be improved up to $10^{-13}$.
\begin{table}[t!]
\begin{center}
\caption{Present upper bounds at $95\%$ CL on LFV decays of $Z$ and $H$ bosons.}\label{LFVsearchII}
{\small
\begin{tabular}{lllll}
\toprule
\toprule
LFV Obs. & \multicolumn{4}{c}{Present Bounds  $(95\%CL)$} \\
\midrule
BR$(Z\to\mu e)$ & $1.7\times10^{-6}$ &LEP  (1995)~\cite{Akers:1995gz}\hspace{.7cm} & $7.50\times10^{-7}$ &ATLAS (2014)~\cite{Aad:2014bca}\\
BR$(Z\to\tau e)$ & $9.8\times10^{-6}$ &LEP (1995)~\cite{ Akers:1995gz} \\
BR$(Z\to\tau\mu )$ &$1.2\times10^{-5}$ &LEP (1995)~\cite{Abreu:1996mj}& $1.69\times10^{-5}$ &ATLAS (2014)~\cite{Aad:2016blu}\\
BR$(H\to\mu e)$ & $3.5\times10^{-4}$ &CMS (2016)~\cite{Khachatryan:2016rke}\\
BR$(H\to\tau e)$ &  $6.1\times10^{-3}$  &CMS (2017)~\cite{CMS:2017onh}& $1.04\times10^{-2}$ &ATLAS (2016)~\cite{Aad:2016blu} \\
BR$(H\to\tau\mu )$ & $2.5\times10^{-3}$ &CMS (2017)~\cite{CMS:2017onh} & $1.43\times10^{-2}$ &ATLAS (2016)~\cite{Aad:2016blu}\\
\bottomrule
\bottomrule
\end{tabular}
}
\end{center}
\end{table}
More recently, the discovery of the Higgs boson has opened new channels for looking for LFV in the charged sector with the LFV H decays $H\to\ell\ell'$. 
The first search of this kind was done by CMS for $H\to\tau\mu$ at $\sqrt{s}=7$~TeV~\cite{Khachatryan:2015kon} and, interestingly, a $2.4\sigma$ excess was found with a best fit value of
\begin{equation}
{\rm BR}(H\to\tau\mu)= 8.4_{-3.7}^{+3.9}\times10^{-3}\,,
\end{equation}
which coincided with a smaller excess of around $1\sigma$ at ATLAS. 
Unfortunately, this excess has not been confirmed with more data at $\sqrt{s}=8$~TeV neither with the LHC run II.
Therefore, both ATLAS and CMS have constraint these ratios at the $\mathcal O(10^{-3,-4})$, as summarized in \tabref{LFVsearchII}.
All these bounds have been obtained using the run-I data at $\sqrt{s}=7$ and 8 TeV, with the exception of the very recent result by CMS~\cite{CMS:2017onh}, where the upper bounds  BR($H\to\tau e)<0.61\%$ and BR($H\to\tau \mu)<0.25\%$ have been set after analyzing $35.9 ~{\rm fb}^{-1}$ of data at $\sqrt{s}=13$~TeV.
These upper bounds improve previous constraints from indirect measurements at LHC \cite{Harnik:2012pb} by roughly one order of magnitude (see also \cite{Blankenburg:2012ex}), and it is close to the previous estimates in \cite{Davidson:2012ds} that predicted sensitivities  of $4.5 \times 10^{-3}$ (see also, \cite{Bressler:2014jta}).
The future perspectives for LFVHD searches are encouraging due to the expected high statistics of Higgs events at future hadronic and leptonic colliders. 
Although, to our knowledge, there is no realistic study, including background estimates, of the expected future experimental sensitivities for these kinds of rare LFVHD events, a naive extrapolation from the present situation can be done.  
For instance, the future LHC runs with $\sqrt s=14$~TeV and total integrated luminosity of first $300~\rm{fb}^{-1}$ and later $3000~\rm{fb}^{-1}$ expect the production of about 25 and 250  millions of Higgs events, respectively, to be compared with 1 million Higgs events that the LHC produced after the first runs~\cite{AtlasProjection,CMS:2013xfa,DeRoeck}. 
These large numbers suggest an improvement in the  long-term sensitivities to BR$(H\to\ell_k\bar\ell_m)$ of at least two orders of magnitude with respect to the present sensitivity.
Similarly, at the planned lepton colliders, like the international linear collider (ILC) with\footnote{We thank J. Fuster for private communication with the updated ILC perspectives.} $\sqrt{s}=1$ TeV and $2.5~\rm{ab}^{-1}$~\cite{Baer:2013cma}, and the future electron-positron circular collider (FCC-ee) as the TLEP with $\sqrt{s}=350$ GeV and $10~\rm{ab}^{-1}$~\cite{Gomez-Ceballos:2013zzn},
the expectations are of about 1 and 2 million Higgs events, respectively,  with much lower backgrounds due to the cleaner environment, which will also allow for a large improvement in LFV Higgs decay searches with respect to the current sensitivities.

Overall, we see that an incredible experimental effort is being made in searching for charged lepton flavor violating processes. 
As we said, any positive signal will automatically imply the existence of new physics even beyond the SM model with a minimal {\it ad-hoc} addition of neutrino masses. 
Nowadays, the lack of such a signal has allowed to several experiments to establish upper bounds on this kind of processes, specially in $\mu$-$e$ transitions, where the bounds are in general several orders of magnitude stronger than the equivalent ones for $\tau$-$e$ or $\tau$-$\mu$ sectors. 
Nevertheless, we hope that the expected improved sensitivities for next generation of experiments will find evidences for new physics in the form of charged lepton flavor violation.

\section[Study of $\ell_m\to\ell_k\gamma$ and $\ell_m\to\ell_k\ell_k\ell_k$ in the ISS]{Study of $\boldsymbol{\ell_m\to\ell_k\gamma}$ and $\boldsymbol{\ell_m\to\ell_k\ell_k\ell_k}$ in the ISS}

In order to better understand the implications on cLFV phenomenology of the TeV scale right-handed neutrinos, we first explore in this section the LFV lepton decays that, as we can see in \tabref{LFVsearch}, are one of the most constrained cLFV observables. 
Concretely, we will consider the ISS model as a specific realization of low scale seesaw models and study in this section the LFV radiative decays $\ell_m\to\ell_k \gamma$ and the LFV three body decays $\ell_m\to\ell_k\ell_k\ell_k$ with $k\neq m$.
The numerical estimations will be done using the full one-loop formulas given in Refs.~\cite{Ilakovac:1994kj,Alonso:2012ji}, which we collect in \appref{App:LFVdecays} for completeness. 

In all the forthcoming study, we will always impose agreement with light neutrino data\footnote{We will show our results for the case of a Normal Hierarchy, although similar results have been obtained for an Inverted Hierarchy.} in \eqref{NuFit}.
For that purpose, we will make use of one of the two parametrizations described in \secref{sec:ISSmodel} at a time, i.e., the Casas-Ibarra parametrization given in \eqref{CasasIbarraISS} or the $\mu_X$ parametrization in \eqref{MUXparam}.
By comparing the results when using these two parametrizations we will learn about the advantages and disadvantages of using one parametrization or the other for exploring the parameter space of the model. 

\begin{figure}[t!]
\begin{center}
\includegraphics[width=0.49\textwidth]{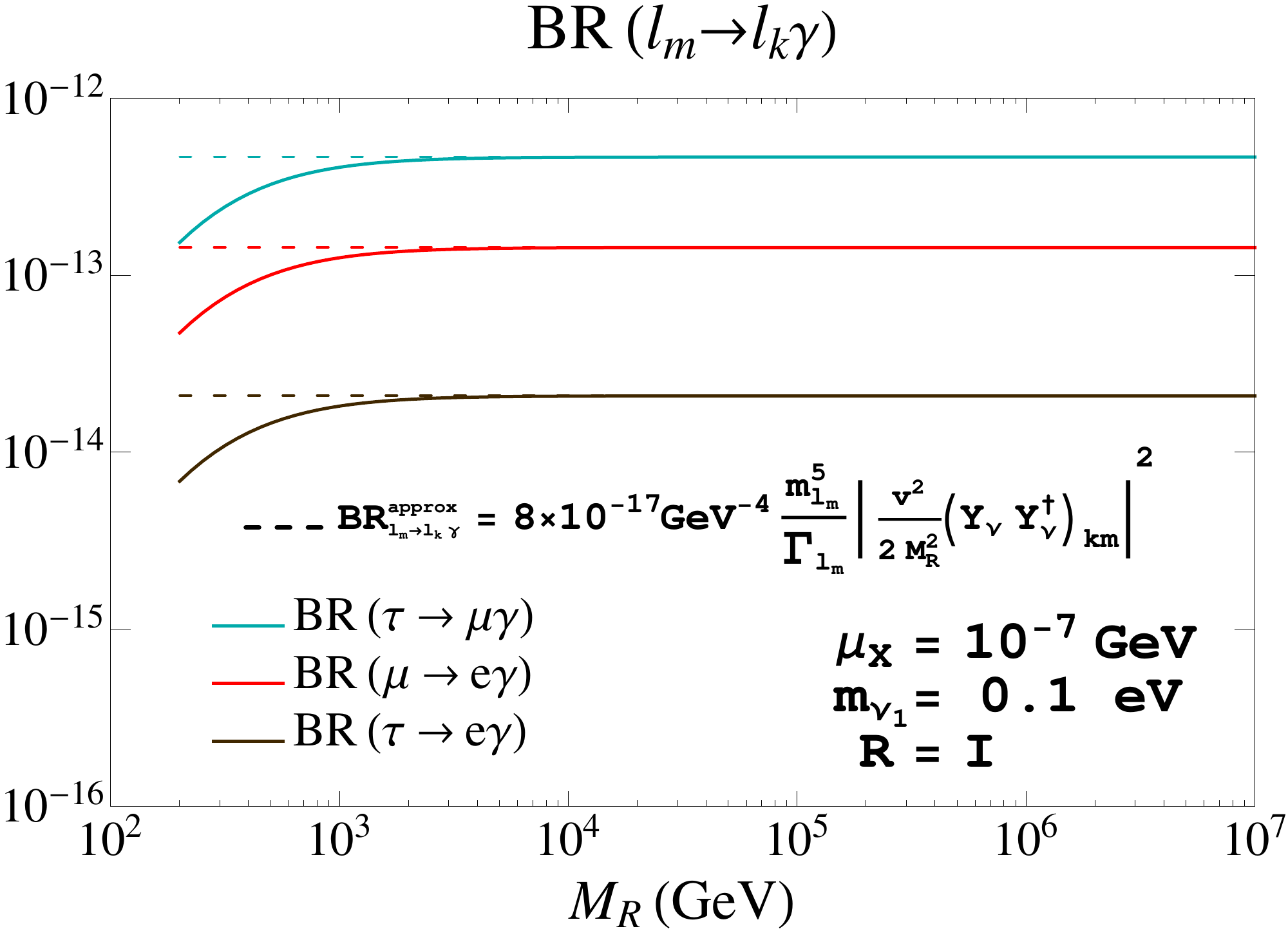} 
\includegraphics[width=0.49\textwidth]{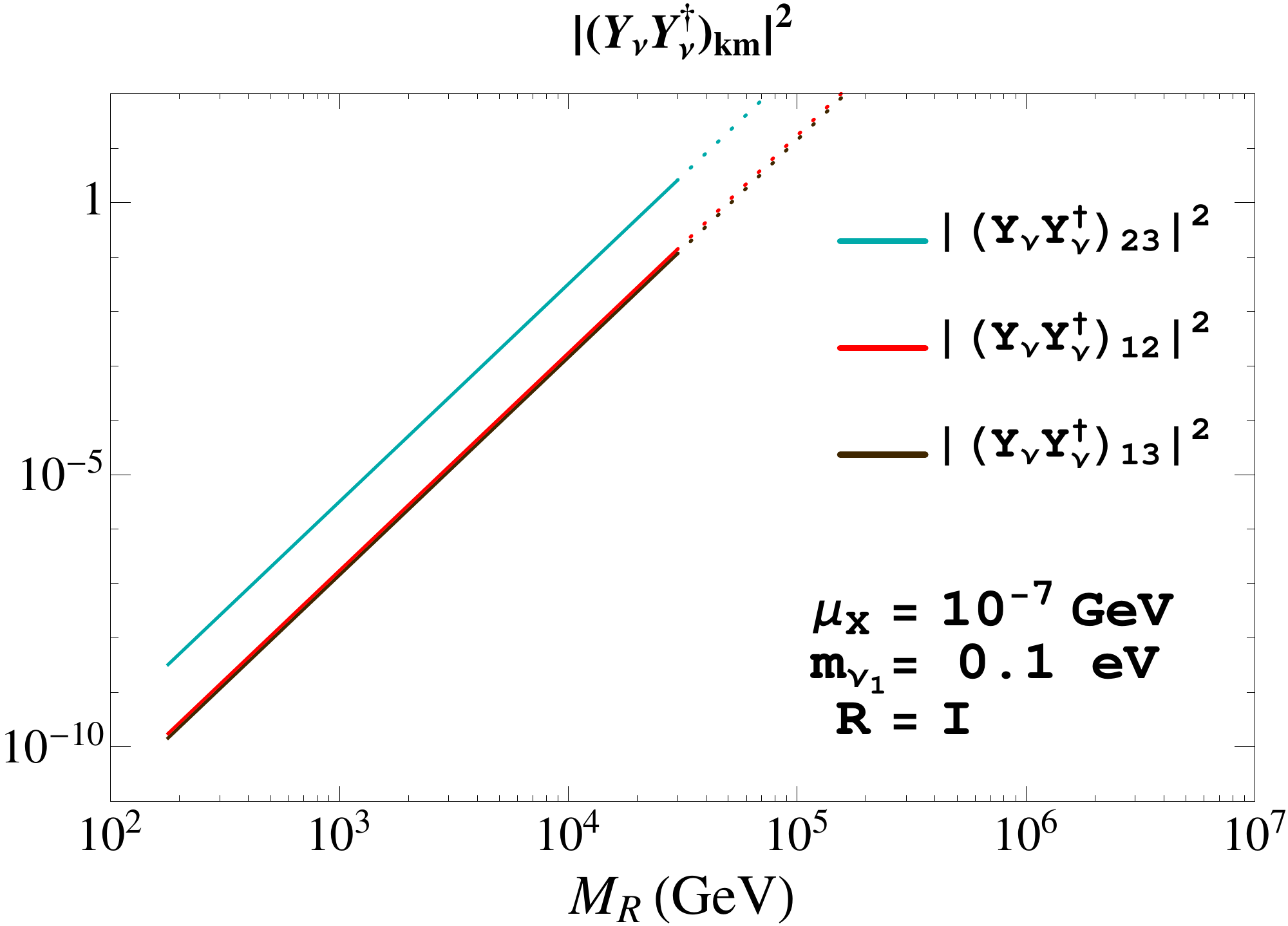}
\caption{Comparison of the full one-loop (solid lines) and approximate (dashed lines) rates for the radiative decays $\ell_m \to \ell_k \gamma$ as functions of $M_R$ and their relation with the $(Y_\nu^{} Y_\nu^\dagger)_{km}$ non-diagonal matrix elements in the Casas-Ibarra parametrization. 
Dotted lines in the right panel indicate non-perturbative Yukawa coupling according to $|Y_\nu^{ij}|^2/4\pi>1.5$.
The other input parameters are set to $\mu_X= 10^{-7} \, {\rm GeV}$, 
$m_{\nu_1}=0.1 \, {\rm eV}$, and $R=\mathbb1$.}\label{LFVradCasas}
\end{center}
\end{figure}

We focus first on the LFV radiative decays, since their analytical expressions are simpler and therefore very useful to gain intuition about cLFV processes in this kind of models. 
We stress that all the numerical estimates and plots are made using the full formulas in \appref{App:LFVdecays}. 
Nevertheless, for the purpose of the discussion, we have derived the following simple but useful approximated expression:
\begin{equation}\label{RadApprox}
{\rm BR}(\ell_m\to \ell_k\gamma)\approx \frac{\alpha_W^3 s_W^2}{1024\pi^2 m_W^4}\frac{m_{\ell_m}^5}{\Gamma_{\ell_m}} \frac{v^4}{M_R^4} \Big|\big(Y_\nu^{} Y_\nu^\dagger\big)_{km} \Big|^2\,,
\end{equation}
which works quite well for a single heavy mass scale, $M_R=M_R\mathbb1$, and in the seesaw limit $v Y_\nu \ll M_R$,  as we will see. 

We display in \figref{LFVradCasas} the numerical results for the LFV $\ell_m\to\ell_k\gamma$ decay rates when using the Casas-Ibarra parametrization to accommodate neutrino oscillation data. 
As explained before, this parametrization builds the Yukawa coupling matrix taking $M_R$, $\mu_X$, $m_\nu^{\rm diag}$ and the orthogonal matrix $R$ as input parameters in \eqref{CasasIbarraISS}.
In this first plot, we consider a simplified scenario where both $M_R$ and $\mu_X$ matrices are diagonal and degenerate, i.e., $M_R\equiv M_R \mathbb 1$ and $\mu_X\equiv\mu_X\mathbb1$. 
Concretely, we set $\mu_X=10^{-7}$~GeV, $m_{\nu_1}=0.1$~eV and $R=\mathbb1$, while we vary $M_R$ from 200 to $10^7$~GeV.
The left panel of \figref{LFVradCasas} illustrates the numerical predictions of BR($\ell_m\to\ell_k\gamma$) rates in this scenario, using both the full analytical expression in \eqref{BRradiative} (solid lines) and the approximated expression in \eqref{RadApprox} (dashed lines).
This plot already shows that the approximated expression works very well for large enough values of $M_R$, as we anticipated. 
Therefore, we can make use of \eqref{RadApprox} in order to understand the analytical dependence of these ratios with de heavy neutrino mass $M_R$.

In this left panel of \figref{LFVradCasas} we clearly see that these rates saturate to a constant value for increasing $M_R$, leading to an apparent non-decoupling behavior with the mass of the heavy neutrinos. 
Nevertheless, this non-decoupling effect is an artifact of the Casas-Ibarra parametrization, since it induces a $M_R$ dependence in the Yukawa coupling matrix, although they are both in principle independent parameters. 
More precisely, looking at \eqref{CasasIbarraISS} we can see that the elements of the relevant  combination in \eqref{RadApprox}, $(Y_\nu^{} Y_\nu^\dagger)$, grow with $M_R$ approximately as $M_R^2$, as can be seen in the right panel of \figref{LFVradCasas}.
Therefore, the final prediction for BR($\ell_m\to\ell_k\gamma$) is constant with $M_R$. 

We will deal with this kind of apparent non-decoupling effects every time we use the Casas-Ibarra parametrization. The growing of the $Y_\nu$ coupling with $M_R$ will compensate the suppression coming from the heavy mass scale running in the loops, leading to a fake violation of the decoupling theorem~\cite{Appelquist:1974tg}.
Nevertheless, we have checked that for constant value of the Yukawa coupling the predictions for the different one-loop processes considered in this Thesis decrease with $M_R$, as can be seen for the LFV radiative decay rates in \eqref{RadApprox}. 
This expected decoupling behavior will become manifest when using the $\mu_X$ parametrization, since it works with $Y_\nu$ and $M_R$ as independent input parameters. 

Finally, we can compare the results for the LFV radiative decay rates in \figref{LFVradCasas} with the present experimental upper bounds in \tabref{LFVsearch}.
We see that, for this choice of parameters, $\mu\to e\gamma$ is the closest one to its upper bound of $4.2\times10^{-13}$~\cite{TheMEG:2016wtm}. 
Being this bound much stronger that the ones on LFV radiative $\tau$ decays, we found that $\mu\to e\gamma$ is the most constraining radiative decay for most of the parameter space of the model.
Moreover, looking at \tabref{LFVsearch}, we see that in general strongest cLFV bounds will come from processes involving a $\mu$-$e$ transition, not only from the mentioned $\mu\to e\gamma$, but also from $\mu\to eee$ or $\mu$-$e$ transitions in nuclei. 
Consequently, we can try to find areas in the parameter space where these strong bounds are avoided by a suppressed prediction of $\mu$-$e$ transitions. 
We can then expect that the largest allowed $\tau$-$\mu$ and $\tau$-$e$ transitions will lie precisely in these areas.

\subsection[Proposal of scenarios with suppressed $\mu$-$e$ transitions]{Proposal of scenarios with suppressed $\boldsymbol{\mu}$-$\boldsymbol{e}$ transitions} 
\label{sec:scenarios}

Motivated by the fact that experimental searches in \tabref{LFVsearch} show much more constrained cLFV processes in the $\mu$-$e$ sector than in the other $\tau$-$\mu$ and $\tau$-$e$ sectors, we look for phenomenological scenarios where $\mu$-$e$ transitions are suppressed.
In order to do this, we find more useful to consider the $\mu_X$ parametrization instead of the Casas-Ibarra one, since it allow us to consider the Yukawa coupling directly as an input parameter.   
Looking again at \eqref{RadApprox}, we learn that, for diagonal and degenerate $M_R$ matrix, the relevant Yukawa combination for cLFV processes is $Y_\nu^{} Y_\nu^\dagger$, which is simplified to $Y_\nu^{} Y_\nu^T$ in the case of real matrices\footnote{In the following derivation of the $\mu$-$e$ suppressed scenarios, we will assume the situation of having real matrices in order to avoid potential constraints from lepton electric dipole moments.}.
Then, it will be very useful and instructive to consider a geometrical interpretation of the Yukawa matrix where its entries are interpreted as the components of three generic ($\boldsymbol{n}_e,\boldsymbol{n}_\mu,\boldsymbol{n}_\tau$) neutrino vectors in flavor space,
\begin{equation}\label{Ynuvectors}
Y_\nu=\left(\begin{array}{ccc}
 Y_\nu^{11} & Y_\nu^{12} & Y_\nu^{13}\\
 Y_\nu^{21} & Y_\nu^{22} & Y_\nu^{23}\\
 Y_\nu^{31} & Y_\nu^{32} & Y_\nu^{33}\end{array}\right)\equiv f 
 \left(\begin{array}{c} \boldsymbol{n}_e \\ \boldsymbol{n}_\mu \\ \boldsymbol{n}_\tau \end{array}\right),
\end{equation}
which for the  relevant combination in cLFV  processes give:
 \begin{equation}
 \label{Yneusq}
Y_\nu^{} Y_\nu^T= f^2 \left(\begin{array}{ccc}
 |\boldsymbol{n}_e|^2 & \boldsymbol{n}_e\cdot\boldsymbol{n}_\mu   & \boldsymbol{n}_e\cdot\boldsymbol{n}_\tau \\
 \boldsymbol{n}_e\cdot\boldsymbol{n}_\mu & |\boldsymbol{n}_\mu|^2 &  \boldsymbol{n}_\mu\cdot\boldsymbol{n}_\tau\\
 \boldsymbol{n}_e\cdot\boldsymbol{n}_\tau &   \boldsymbol{n}_\mu\cdot\boldsymbol{n}_\tau  & |\boldsymbol{n}_\tau|^2\end{array}\right).
 \end{equation}
This means that the 9 input parameters determining the $Y_\nu$ matrix can be seen as  the 3 modulus of these three vectors ($|\boldsymbol{n}_e|,|\boldsymbol{n}_\mu|,|\boldsymbol{n}_\tau|$), the 3 relative {\it flavor} angles between them ($\theta_{\mu e},\theta_{\tau e},\theta_{\tau \mu}$), with $\theta_{ij}\equiv \widehat{\,\boldsymbol{n}_i \boldsymbol{n}_j}$, and 3 extra angles ($\theta_1,\theta_2,\theta_3$) that parametrize a global rotation $\mathcal O$ of these 3 vectors that does not change their relative angles. In addition, we have introduced an extra parameter $f$ that characterizes the global Yukawa coupling strength and that will be useful in forthcoming analysis. 
Since the combination $Y_\nu^{} Y_\nu^T/f^2$ is symmetric, it only depends on 6 parameters that we take to be the 3 modulus ($|\boldsymbol{n}_e|,|\boldsymbol{n}_\mu|,|\boldsymbol{n}_\tau|$) and the cosine of the three flavor angles ($c_{\mu e},c_{\tau e},c_{\tau \mu}$), with $c_{ij}\equiv\cos\theta_{ij}$.
The names of the angles are motivated by the fact that the cosine of the angle $\theta_{ij}$ controls the LFV transitions in the $\ell_i$-$\ell_j$ sector, which we write in short as ${\rm LFV}_{\ell_i \ell_j}$. It is interesting to notice that 
the global rotation $\mathcal O$ does not enter in the $Y_\nu^{} Y_\nu^T$ combination and, therefore, it will not affect any of the cLFV  processes studied in this work. 

\begin{figure}[t!]
\begin{center}
\includegraphics[width=0.49\textwidth]{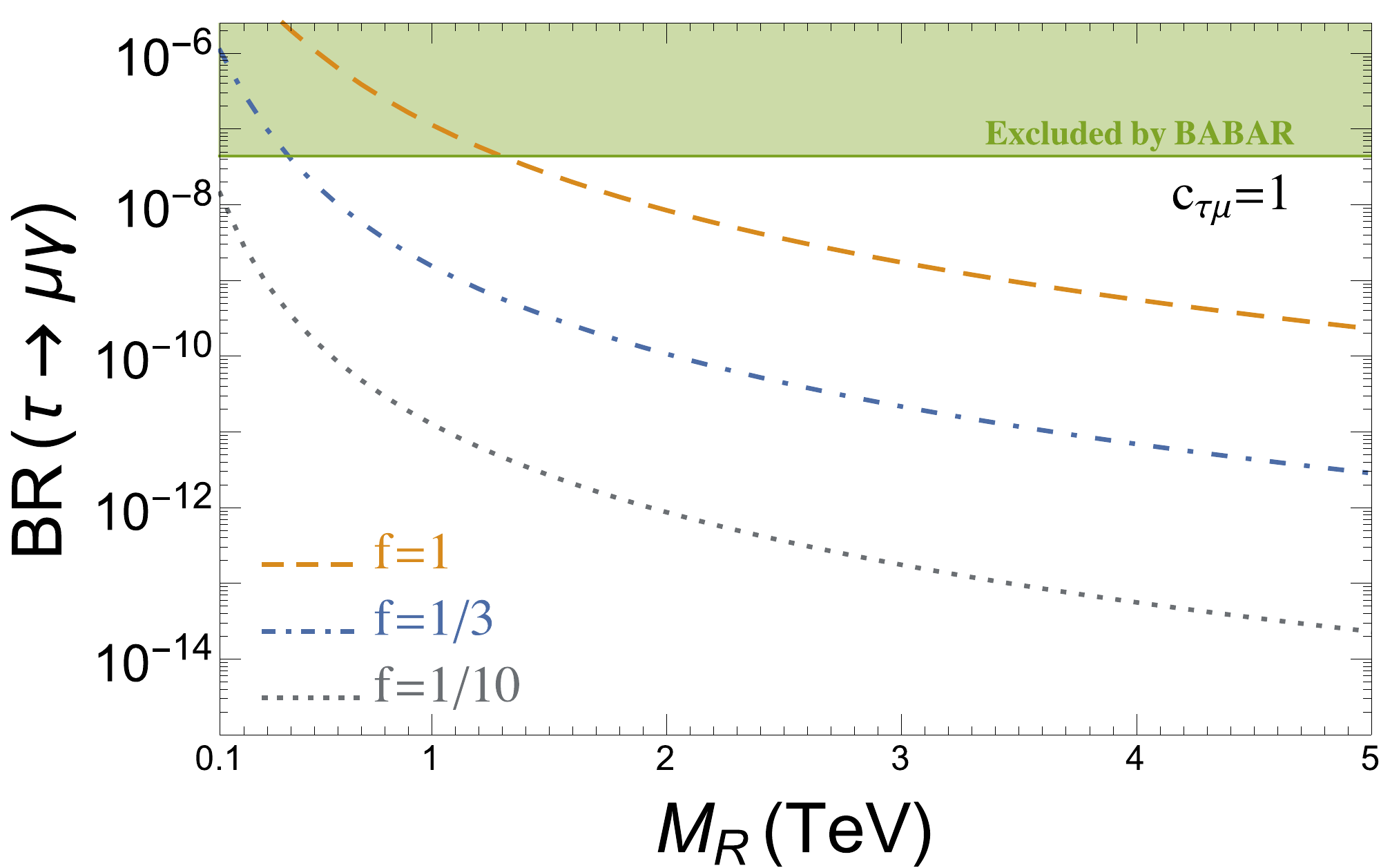} 
\includegraphics[width=0.49\textwidth]{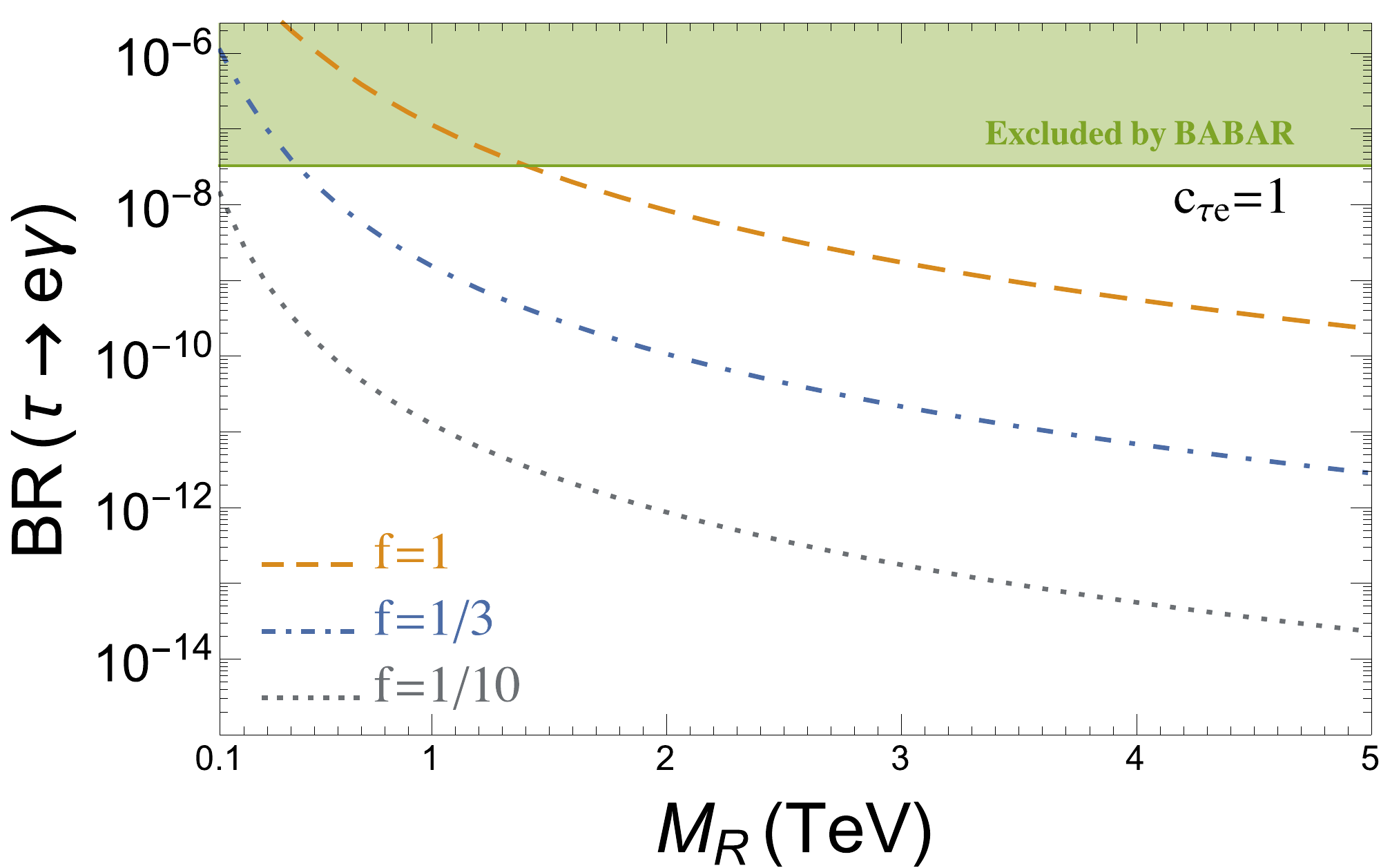}
\caption{Predictions for the BR($\tau\to\mu\gamma$) and BR($\tau\to e\gamma$) as functions of $M_R$ in the $\mu_X$ parametrization. 
The input $Y_\nu$ matrix is written as explained in the text, with different values of the strength parameter $f$. 
In the left (right) panel $\tau$-$\mu$ ($\tau$-$e$) transitions are maximized by setting $c_{\tau\mu} (c_{\tau e})=1$.
The other cosines are set to zero, the modulus $|\boldsymbol n_{e,\mu,\tau}|=1$ and $\mathcal O=\mathbb1$.}\label{LFVradMUX}
\end{center}
\end{figure}

Before going to the scenarios with suppressed $\mu$-$e$ transitions, we study in \figref{LFVradMUX} the predictions for the LFV radiative decay rates in terms of the most relevant parameters in this parametrization. 
In the left panel, we show BR($\tau\to\mu\gamma$) in a scenario where $\tau$-$\mu$ transitions are maximized by setting\footnote{We actually set $c_{\tau\mu}=0.99$ for this plot, since the $\mu_X$ parametrization in \eqref{MUXparam} requests a non-singular $Y_\nu$. Nevertheless, the results for cLFV processes are basically the same as setting $c_{\tau\mu}=1$.} $c_{\tau\mu}=1$, while they are suppressed in the $\tau$-$e$ and $\mu$-$e$ sectors,  $c_{\tau e}=c_{\mu e}=0$. 
Equivalently, the right panel displays BR($\tau\to e\gamma$) in a scenario where $\tau$-$e$ transitions are favored.  
Both plots show the dependence of these observables on the heavy neutrino mass scale $M_R$ for several values of the Yukawa coupling strength factor $f$. 

First of all, we see that both observables show the expected decoupling behavior with $M_R$. 
This is in contrast with the apparent non-decoupling effects we saw in \figref{LFVradCasas} when using the Casas-Ibarra parametrization. 
As we explained before, the difference is that now the Yukawa coupling is treated as an independent parameter and, therefore, the dominant dependence on the $M_R$ comes from the mass in the propagators of the right-handed neutrinos running in the loops and whose effects decrease as $M_R$ becomes heavier. 
On the other hand, the rates are bigger the larger the Yukawa coupling strength $f$ is, as expected. 
Moreover, although not shown here, we have checked that the rates for $\tau\to\mu\gamma$ ($\tau\to e\gamma$) grow with $c_{\tau\mu}$ ($c_{\tau e}$),  $|\boldsymbol{n}_\tau|$ and $|\boldsymbol{n}_\mu|$  ($|\boldsymbol{n}_e|$), while being independent of the other parameters, in particular the rotation matrix $\mathcal O$.
In summary, the full radiative decays rates follow the behavior of the approximated formula in \eqref{RadApprox}.

 Second, we learn that the predictions for $\tau\to\mu\gamma$ in the left panel are the same than those for $\tau\to e\gamma$ in the right panel. 
 The reason for this similarity is that they are related by the interchange $\boldsymbol{n}_\mu\leftrightarrow\boldsymbol{n}_e$ in \eqref{Ynuvectors} and, therefore, we expect to have basically the same results in both scenarios. 
 Based on this relation, we will show most of our results only for the $\tau$-$\mu$ sector, knowing that the conclusions for the $\tau$-$e$ sector can be obtained by just exchanging $\boldsymbol{n}_\mu$ with $\boldsymbol{n}_e$.
 
 Third, we observe that the predictions can reach the present experimental upper bounds from BABAR (see \tabref{LFVsearch}) and, therefore, they will constrain our parameter space when exploring other observables. 
 In particular, for these scenarios with favored $\tau$-$\mu$ or $\tau$-$e$ transitions,  maximum  values of $f\sim {\cal O}(0.5$-$1)$ for $M_R=$1 TeV, or minimum values for $M_R$  of $\sim  {\cal O}$(1-2) TeV for $f=1$, are allowed. 
 In the future, searches at BELLE-II are expected to improve the sensitivity up to $10^{-9}$ for these channels, so they will be able to probe values of $f\gtrsim0.3$ for 1~TeV neutrinos.

As we have seen, using this geometrical interpretation of the Yukawa matrix the $\mu$-$e$ suppression can be easily realized by just assuming that $\boldsymbol{n}_e$ and $\boldsymbol{n}_\mu$ are orthogonal vectors, i.e., $c_{\mu e}=0$.  
Such  condition defines a family of ISS scenarios that can be parametrized using the following Yukawa matrix:
\begin{equation}\label{YukawaAmatrix}
Y_\nu=A\cdot \mathcal O \quad{\rm with}\quad 
A\equiv f \left(\begin{array}{ccc} |\boldsymbol{n}_e| & 0 & 0 \\ 0 & |\boldsymbol{n}_\mu| & 0 \\ |\boldsymbol{n}_\tau| c_{\tau e} & |\boldsymbol{n}_\tau| c_{\tau \mu} & |\boldsymbol{n}_\tau|\sqrt{1-c_{\tau e}^2-c_{\tau \mu}^2}\end{array}\right),
\end{equation}
where $\mathcal O$ is the above commented orthogonal rotation matrix, which does not enter in the product $Y_\nu^{} Y_\nu^T$, and  $f$ is again the parameter controlling the global strength of the Yukawa coupling matrix. 
The fact that we are  assuming real and non-singular $Y_\nu$ imposes the condition $c_{\tau e}^2+c_{\tau\mu}^2<1$.
Notice that the $Y_\nu$ matrix in \eqref{YukawaAmatrix} is the most general one that satisfies the condition $\big(Y_\nu^{} Y_\nu^T \big)_{\mu e}=0$. 

\begin{figure}[t!]
\begin{center}
\includegraphics[width=.49\textwidth]{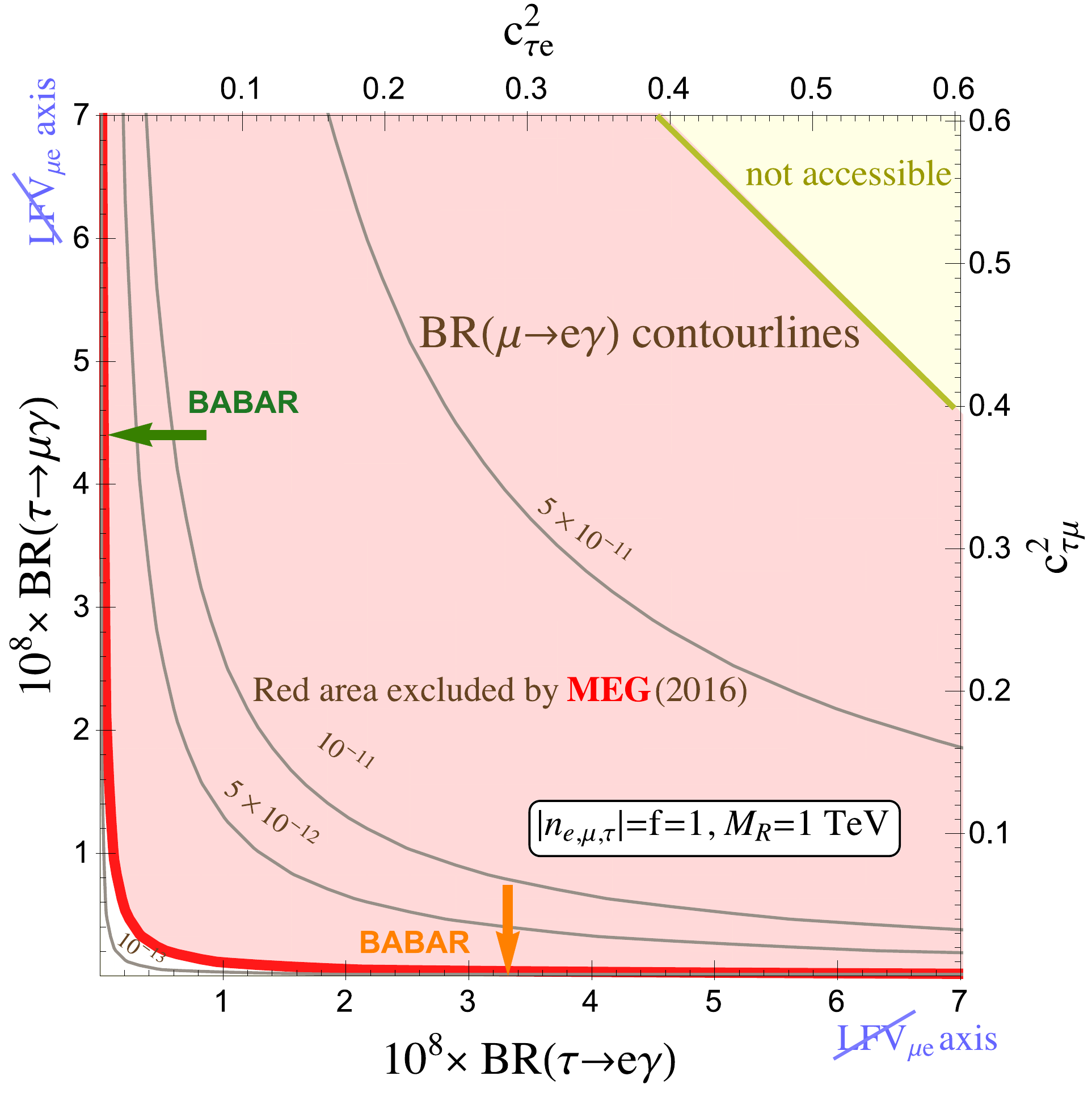}
\includegraphics[width=.5\textwidth]{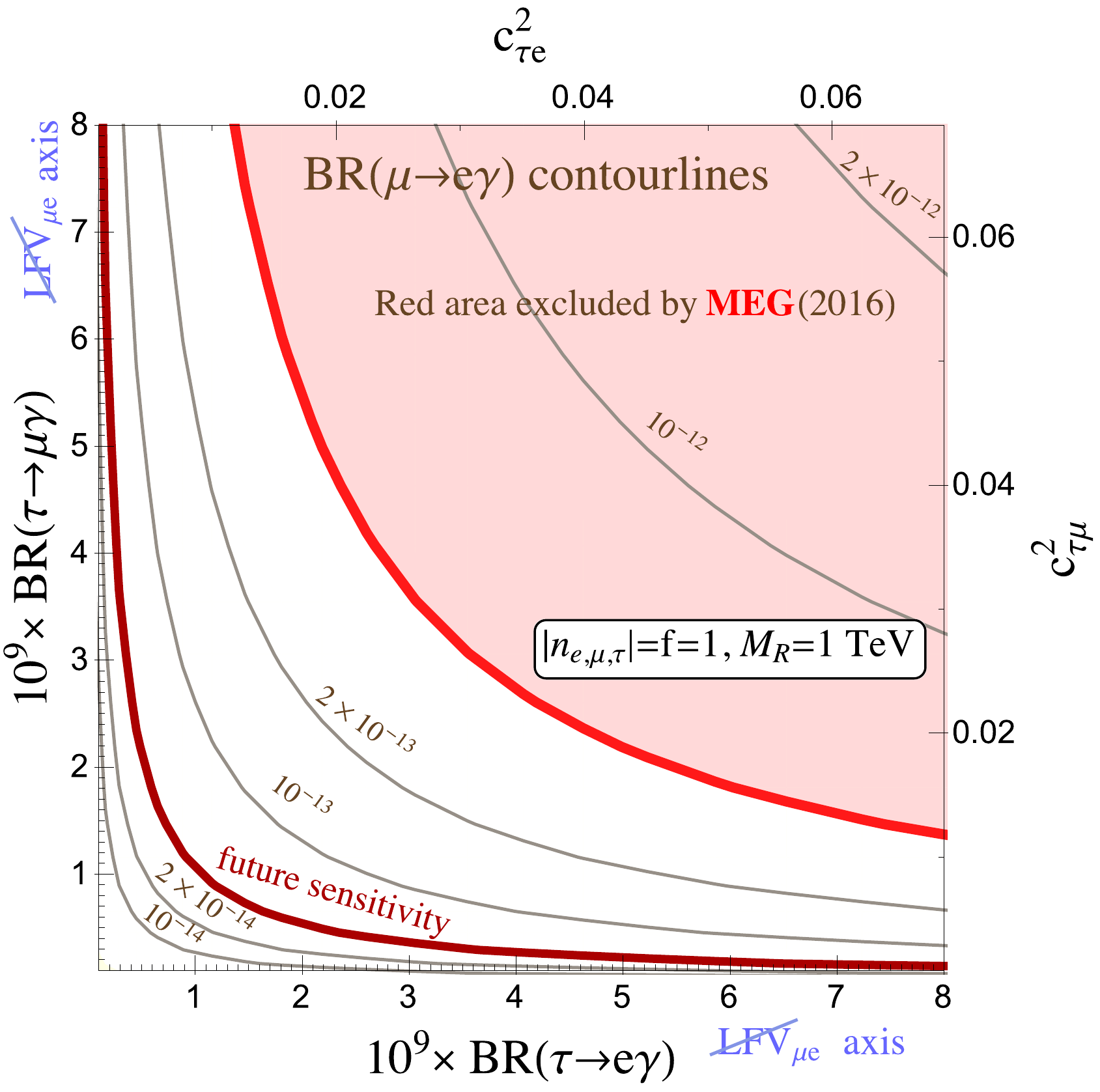}
\caption{
Left panel: Contour lines for BR($\mu\to e\gamma$) as a function of BR($\tau\to e\gamma$) and BR($\tau\to \mu\gamma$) rates, for fixed $M_R=1$~TeV, $|\boldsymbol{n}_{e,\mu,\tau}|=f=1$ values and varying $c_{\tau e}^2$ and $c_{\tau\mu}^2$ from 0 to $0.6$, as shown in the right and top axes. The yellow area represents the region that cannot be accessed with real Yukawa matrices. The red area is excluded by the upper bound on $\mu\to e\gamma$ from MEG, while the orange (green) arrow marks the present upper bound on  $\tau\to e\gamma$ ($\tau\to\mu\gamma$) from BABAR, see \tabref{LFVsearch}.
 Right panel: Zoom on the lower left corner of the plot in the left panel which allows for a better reading of the region allowed by present experimental data. The extra darker red line represents the future expected sensitivity of $4\times10^{-14}$ by MEG-II~\cite{Baldini:2013ke}.
} \label{raddecayscosines}
\end{center}
\end{figure}

We can now explore the predictions for BR($\ell_m\to\ell_k\gamma$) when this kind of scenarios are considered. 
Looking at \eqref{RadApprox}, we can easily see that the LFV radiative decays of the $\tau$ lepton depend on the most relevant parameters, $f$, $M_R$ and $c_{\tau\ell}$ as follows:
\begin{equation}\label{RadiativeApproxTAU}
{\rm BR}(\tau\to \ell\gamma)\sim \frac{v^4f^4}{M_R^4}~c_{\tau\ell}^2\quad{\rm with }~\ell=e,\mu.
\end{equation}
The case of $\mu\to e\gamma$ is different, since the assumption $c_{\mu e}=0$ cancels the leading order contribution given by the approximate formula in \eqref{RadApprox}, and therefore the first relevant contribution in this observable is of higher order in the expansion series in powers of the Yukawa coupling over $M_R$. 
Specifically, it is of the type $v^4 (Y_\nu^{} Y_\nu^T Y_\nu^{} Y_\nu^T)/M_R^4$. 
Consequently, it is suppressed with respect to \eqref{RadiativeApproxTAU} and the predicted rates for this observable turn out to depend on the product of both $c_{\tau e}$ and $c_{\tau \mu}$:
\begin{equation}\label{RadiativeApproxMUEG}
{\rm BR}(\mu\to e\gamma)\sim \frac{v^8f^8}{M_R^8}~c_{\tau e}^2 c_{\tau \mu}^2\,.
\end{equation}
Therefore, in order to define a scenario where all $\mu$-$e$ transitions are completely suppressed, i.e., $\big(Y_\nu^{} Y_\nu^T\big)_{\mu e}=\big(Y_\nu^{} Y_\nu^T Y_\nu^{} Y_\nu^T\big)_{\mu e}=\dots=0$, we see that the condition $c_{\mu e}=0$ is not enough and that we also need $c_{\tau e}=0$ or $c_{\tau\mu}=0$.

These behaviors of the BR($\ell_m\to\ell_k\gamma$) rates are numerically illustrated in \figref{raddecayscosines}, where the full one-loop formulas in \appref{App:LFVdecays} have been used.
These plots show contourlines for BR($\mu\to e\gamma$) in terms of the other radiative decay rates. 
It also displays the above commented correlations between the BR($\tau \to \mu \gamma$) and BR($\tau \to e \gamma$)  rates and the parameters $c_{\tau \mu}$ and $c_{\tau e}$, respectively. 
The contour lines for BR($\mu\to e\gamma$) are obtained by varying $c_{\tau \mu}^2$ and $c_{\tau e}^2$ within the interval $(0,0.6)$, which in turn provide predictions for BR($\tau \to \mu \gamma$) and BR($\tau \to e \gamma$) that are represented in the vertical and horizontal axes respectively. 
This is for the simple case with $|\boldsymbol{n}_{e,\mu,\tau}|=f=1$, $M_R=1$~TeV and $\mathcal O=\mathbb 1$ (although we checked again that the rates do not depend on $\mathcal O$), but similar qualitative conclusions  can be obtained for other choices of these parameters. 
Notice that the above mentioned condition of $c_{\tau e}^2+c_{\tau \mu}^2<1$ from the Yukawa matrix in \eqref{YukawaAmatrix} makes the yellow area, where $c_{\tau e}^2+c_{\tau \mu}^2\geq1$, not accesible to our analysis. 
We also find that the rates for $\tau\to \mu\gamma$ ($\tau\to e\gamma$) can in general be large and, for the values of the parameters selected in this plot, they are of the order of the present upper bounds from BABAR~\cite{Aubert:2009ag}, marked here with a green (orange) arrow. 
Moreover, we see that they depend just on $c_{\tau \mu}^2$ ($c_{\tau e}^2$), in agreement with the approximate expression in \eqref{RadiativeApproxTAU}.

We also learn that the predictions for BR($\mu\to e\gamma$) are between 3 and 4 orders of magnitude smaller than the $\tau$ radiative decay rates, as expected from \eqref{RadiativeApproxMUEG}. 
Nevertheless, they are still above the upper bound from the MEG experiment for most of the parameter space. 
In fact, the MEG bound excludes everything but the area close to the axes, since the BR($\mu\to e\gamma$) goes asymptotically to zero when approaching the axes, as can be seen in the zoom over the lower left corner, shown in the right panel of \figref{raddecayscosines}.
When lying just on top of these axes, the predictions for BR($\mu\to e\gamma$) completely vanish, as seen in \eqref{RadiativeApproxMUEG}, implying that BR($\tau\to e \gamma$) must be small in order to allow for large BR($\tau\to\mu\gamma$),  and viceversa.

\begin{table}[t!]
\begin{center}
\caption{TM scenarios  for numerical estimates of large $\tau$-$\mu$ transitions. 
Notation '$\simeq$' means $c_{\tau \mu}=0.99$ instead of 1 in order to have non-singular $Y_\nu$ matrices, see \eqref{MUXparam}.
The notation $Y_{\tau\mu}^{(1-3)}$ corresponds to the original one introduced in Ref.~\cite{Arganda:2014dta}.
Equivalent TE scenarios are easily obtained by exchanging $\mu$ and $e$ in these TM ones. }
{\fontsize{11}{5}
\begin{tabular}{lccccl}
\toprule
\toprule
 Scenario Name& $c_{\tau \mu}$ & $|\boldsymbol{n}_e|$ & $|\boldsymbol{n}_\mu|$ & $|\boldsymbol{n}_\tau|$ & 
  Example 	\\
\midrule
TM-1 & $1/\sqrt2$ & 1 & 1 & 1 & 
$Y_\nu=f \left(\begin{array}{ccc}1 & 0 & 0\\ 0&1&0\\0&\sfrac{1}{\sqrt2} &\sfrac{1}{\sqrt2}\end{array}\right)$\\ [3ex]
TM-2 & $1$ & 1 & 1 & 1 & 
$Y_\nu\simeq f\left(\begin{array}{ccc}1 & 0 & 0\\ 0&1&0\\0&1 &0\end{array}\right)$\\ [3ex]
TM-3 & $1/\sqrt2$ & 0.1 & 1 & 1 & 
$Y_\nu=f\left(\begin{array}{ccc}0.1 & 0 & 0\\ 0&1&0\\0&\sfrac{1}{\sqrt2} &\sfrac{1}{\sqrt2}\end{array}\right)$\\ [3ex]
TM-4 & $1$ & 0.1 & 1 & 1 & 
$Y_\nu\simeq f\left(\begin{array}{ccc}0.1 & 0 & 0\\ 0&1&0\\0&1 &0\end{array}\right)$\\ [3ex]
TM-5 & $1$ & $\sqrt2 $& $1.7$ &$ \sqrt3 $&
 $Y_\nu\equiv Y_{\tau\mu}^{(1)}=f \left(\begin{array}{ccc} 0&1&-1 \\ 0.9&1&1 \\ 1& 1& 1 \end{array}\right)$\\ [3ex]
TM-6 & $1/3$ & $\sqrt2 $& $\sqrt3$ & $\sqrt3 $& 
$Y_\nu \equiv Y_{\tau\mu}^{(2)}= f \left(\begin{array}{ccc}0&1&1 \\ 1&1&-1 \\ -1& 1& -1  \end{array}\right)$\\ [3ex]
TM-7 & $0.1$ & $\sqrt2 $& $\sqrt3 $& $1.1 $& 
$Y_\nu\equiv Y_{\tau\mu}^{(3)}=f\left(\begin{array}{ccc}0&-1&1 \\ -1&1&1 \\ 0.8& 0.5& 0.5  \end{array}\right)$\\ [3ex]
TM-8 & $1$ & $1/2$ & $1/3$ & $1/4$ & 
$Y_\nu\simeq f\left(\begin{array}{ccc}1 & 0 & 0\\ 0&0.5&0\\0&0.08 &0.32\end{array}\right)$\\ [3ex]
TM-9 & $0.77$ & $0.1$ & $0.46$ & $\sqrt2$ & 
$Y_{\nu}=f\left(\begin{array}{ccc}0.1&0&0\\0&0.46&0.04\\0&1&1\end{array}\right)$\\ [3ex]
TM-10 & $0.64$ & $0.1$ & $0.94$ & $\sqrt2$ & 
$Y_\nu= f\left(\begin{array}{ccc}0&0.1&0\\ 0.94&0&0.08\\1&0&-1\end{array}\right)$\\ [3ex]
\bottomrule
\bottomrule
\end{tabular}
}
\label{TMscenarios}
\end{center}
\end{table}

Therefore, we can identify our phenomenological scenarios with suppressed $\mu$-$e$ transitions, which we will refer to as ISS-$\cancel{\rm LFV}\hskip-.1cm_{\mu e}$, with the two axes in \figref{raddecayscosines}.
We then consider two classes of scenarios: the TM scenarios along the ${\rm LFV}_{\tau\mu}$ axis ($c_{\tau e}=0$) that may give sizable rates for $\tau$-$\mu$ transitions, but always give negligible contributions to ${\rm LFV}_{\mu e}$ and ${\rm LFV}_{\tau e}$; and the TE scenarios along  the ${\rm LFV}_{\tau e}$ axis ($c_{\tau \mu}=0$) that may lead to large rates only for the $\tau$-$e$ transitions.
In table \ref{TMscenarios} we list some specific examples that we will use along this Thesis for the numerical estimates of our selected TM scenarios.
Equivalent examples for the TE scenarios are obtained by exchanging $\mu$ and $e$ everywhere in these  TM scenarios. 
Notice that we introduced the notation $Y_{\tau\mu}^{(1)}$, $Y_{\tau\mu}^{(2)}$ and $Y_{\tau\mu}^{(3)}$ for the Yukawa matrices in the scenarios TM-5, TM-6 and TM-7, respectively, which corresponds to the original notation in Ref.~\cite{Arganda:2014dta}. We will indistinguishably use both notations along this Thesis.

\begin{figure}[t!]
\begin{center}
\includegraphics[width=0.49\textwidth]{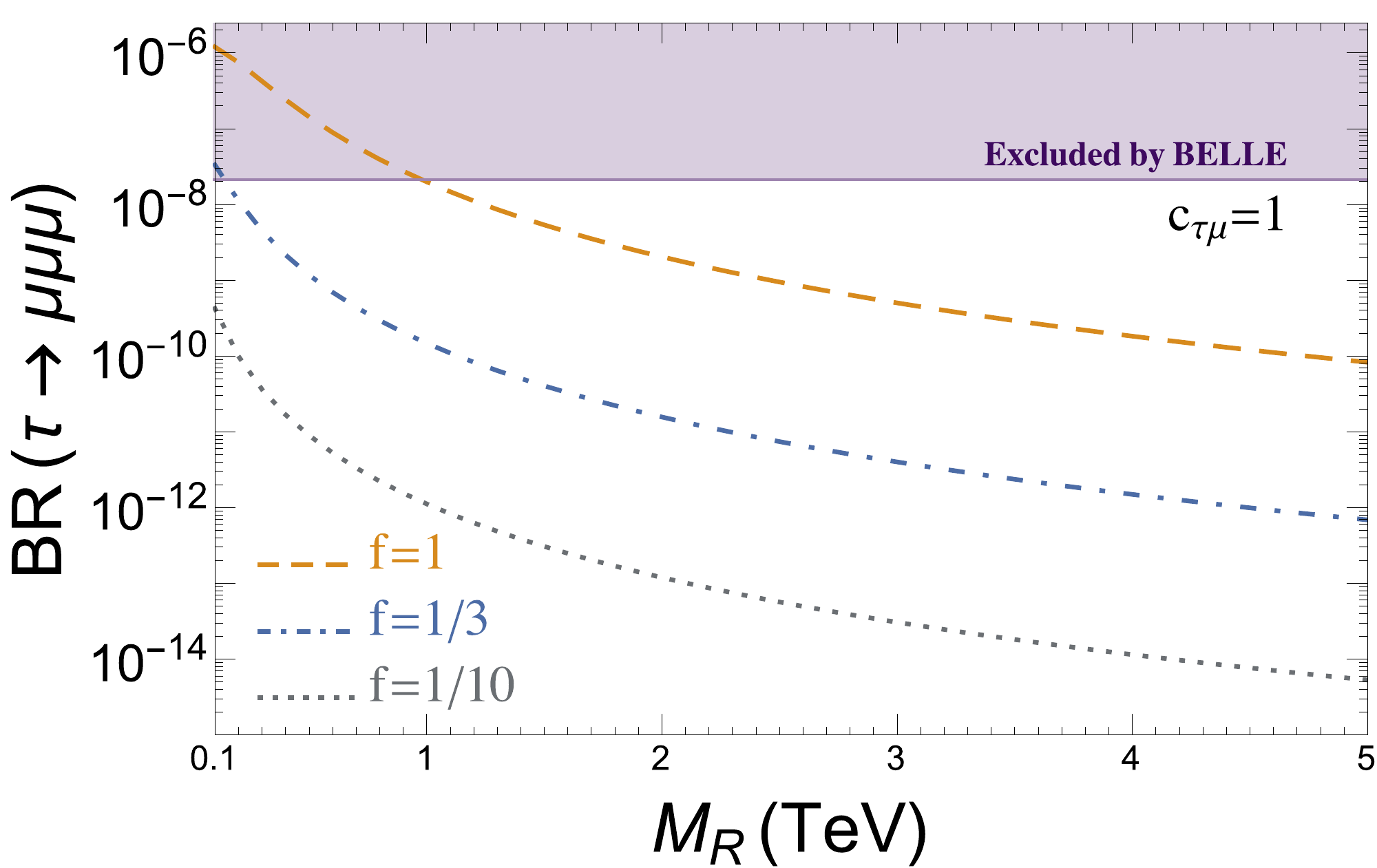} 
\includegraphics[width=0.49\textwidth]{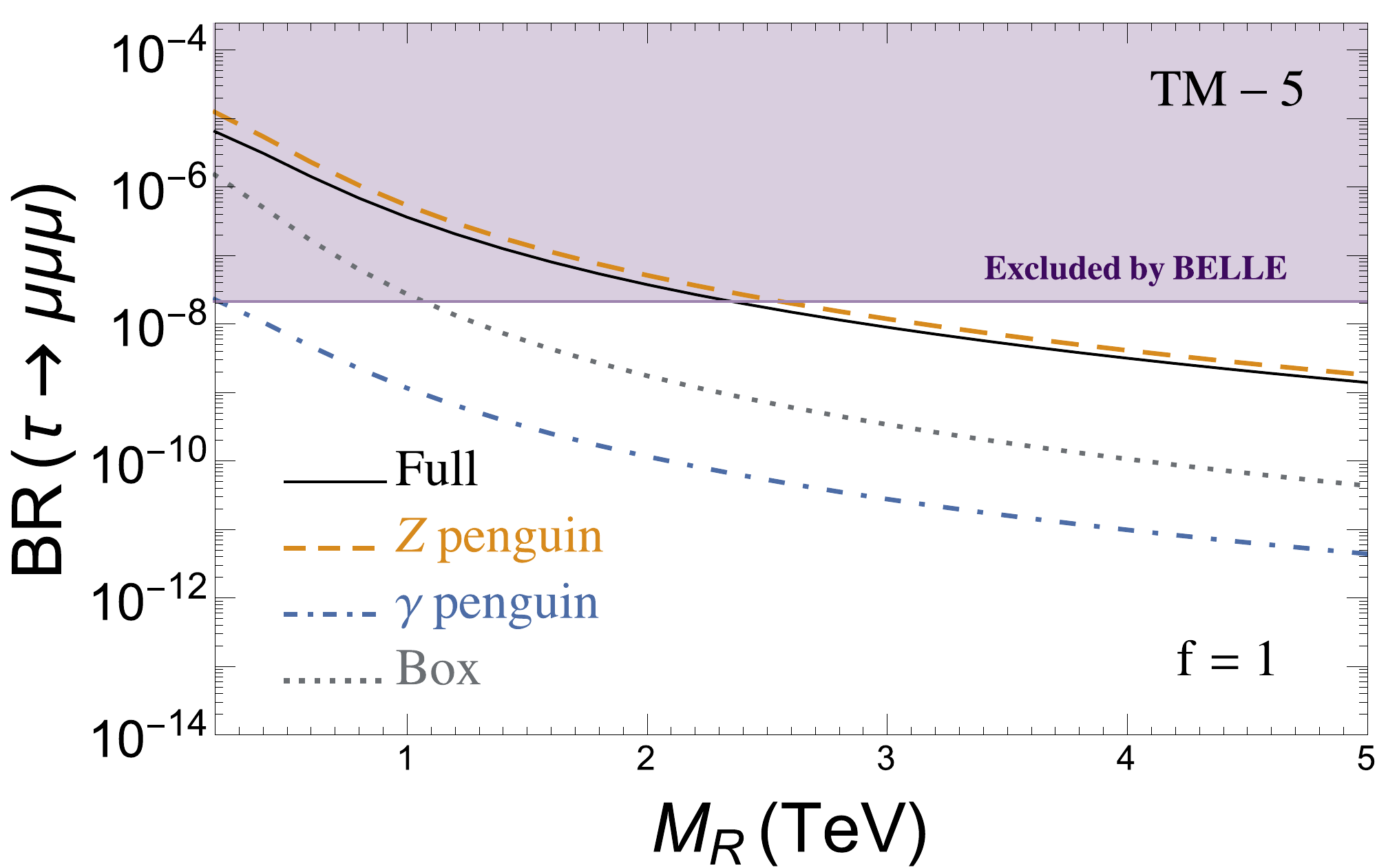}
\caption{BR($\tau\to\mu\mu\mu$) as a function of $M_R$ in two TM-like scenarios with $c_{\tau\mu}=1$ and $c_{\tau e}=0$. Left panel: Full predictions for $|\boldsymbol n_{e, \mu,\tau}|=1$ and three values of $f$.
Right panel: Full prediction (solid line) in the TM-5 scenario with $f=1$, decomposed in its contributions from $\gamma$ penguin (blue dot-dashed), boxes (gray dotted) and $Z$ penguin (yellow dashed), the dominant one. 
Purple shadowed area is excluded by BELLE.}\label{LFVtau3muMUX}
\end{center}
\end{figure}

We can next study the predictions for the LFV three body decays $\ell_m\to\ell_k\ell_k\ell_k$ in this kind of scenarios. 
\figref{LFVtau3muMUX}  shows the BR($\tau\to\mu\mu\mu$) rates in a TM-like scenario of maximized $\tau$-$\mu$ transitions, i.e., $c_{\tau\mu}=1$ and $c_{\tau e}=c_{\mu e}=0$, although the general conclusions are the same for BR($\tau\to eee$) in a TE scenario.
In the left panel, the dependence of the full decay rate is displayed as a function of $M_R$ for different values of $f$. 
We clearly see that the rates are again large, within present and future experimental sensitivities, specially for large $f$ values and low $M_R$. 
For example, in this particular case of $c_{\tau\mu}=1$, the present bound from BELLE of BR$(\tau\to\mu\mu\mu)<2.1\times10^{-8}$ excludes large couplings of $f\gtrsim1$ for heavy neutrinos below 1~TeV. 
Future expected sensitivities at BELLE-II may be able to probe values of $M_R$ up to 2-3 TeV for $f=1$.

As in the case of the radiative decay $\tau\to\mu\gamma$, we see again that the rates decrease with increasing $M_R$, manifesting the decoupling behavior expected when using the $\mu_X$ parametrization. 
Although we do not show all the plots here, we have explored how the rates for $\tau\to\mu\mu\mu$ depend on the most relevant parameters finding that they grow with $f$, $c_{\tau\mu}$, $|\boldsymbol{n}_\tau|$ and $|\boldsymbol{n}_\mu|$, whereas they are independent of $c_{\tau e}$, $c_{\mu e}$, $|\boldsymbol{n}_ e|$ and the rotation $\mathcal O$.

Nevertheless, the dependence of $\tau\to\mu\mu\mu$ on these parameters is not exactly equal to that of $\tau\to\mu\gamma$, since the former receives contributions from different types of diagrams, namely, the $\gamma$-penguin, $Z$-penguin and box diagrams. 
In order to better understand this, we display separately the contributions from each type of diagram to the total decay rate in the right panel of \figref{LFVtau3muMUX}.
We choose the TM-5 scenario from \tabref{TMscenarios}, although similar qualitative results are found for other TM scenarios. 
The dependences on $f$, $M_R$ and $c_{\tau\mu}$ are slightly different for each of the contributions, leading to a more complicated dependence for the total decay rate. 
Moreover, we see that, for this value of $f$, the dominant contribution is mostly coming from the $Z$-penguin. 
This fact will be important when studying the LFV Z decay rates $Z\to\ell_k\bar\ell_m$ in \chref{LFVZD}.

In the rest of this Thesis we will consider the two parametrizations described in \chref{Models}.
We will consider more generic searches using the Casas-Ibarra parametrization for scanning the ISS parameter space, although in that case it will be difficult to access to these particular directions and, therefore, to conclude on maximum allowed rates.
Therefore, we will focus on the scenarios in \tabref{TMscenarios} for studying maximum allowed LFV rates involving $\tau$ leptons making sure that we are not generating potentially constrained $\mu$-$e$ transitions.

\section{Other implications from low scale seesaw neutrinos}
\label{sec:otherconstraints}

Generically, the addition of heavy Majorana neutrinos to the particle content of the SM has a phenomenological impact on several low energy observables via their mixing with the active neutrinos. 
These observables can be related to lepton flavor violation, as the ones above studied, lepton number violation, lepton universality or others. 
Therefore, we  want to ensure that our forthcoming analysis in Chapters~\ref{LFVHD}, \ref{LFVZD} and \ref{LHC} comply  with the relevant theoretical and experimental constraints in all the regimes of the considered right handed neutrino masses and couplings. 
We briefly discuss in the following the constraints that we have found to be the most relevant ones for the present Thesis and which we consequently include in our analysis. 
For this study we have used our own \textit{Mathematica} code which includes all the relevant formulas for the constraining observables that are taken from the literature and that we include in \appref{App:constraints} for completeness. 

\begin{figure}[t!]
\begin{center}
\includegraphics[width=0.49\textwidth]{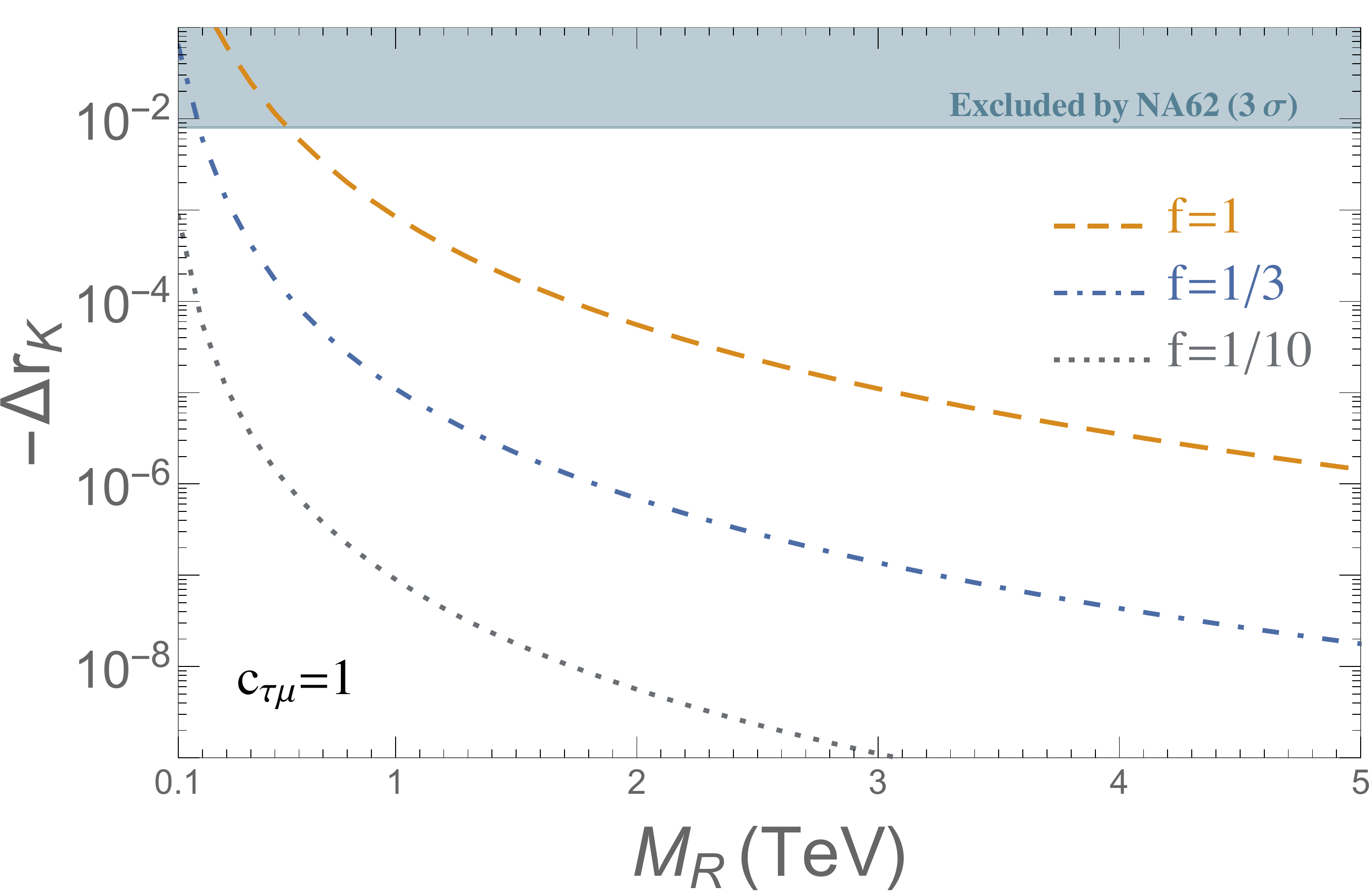} 
\includegraphics[width=0.49\textwidth]{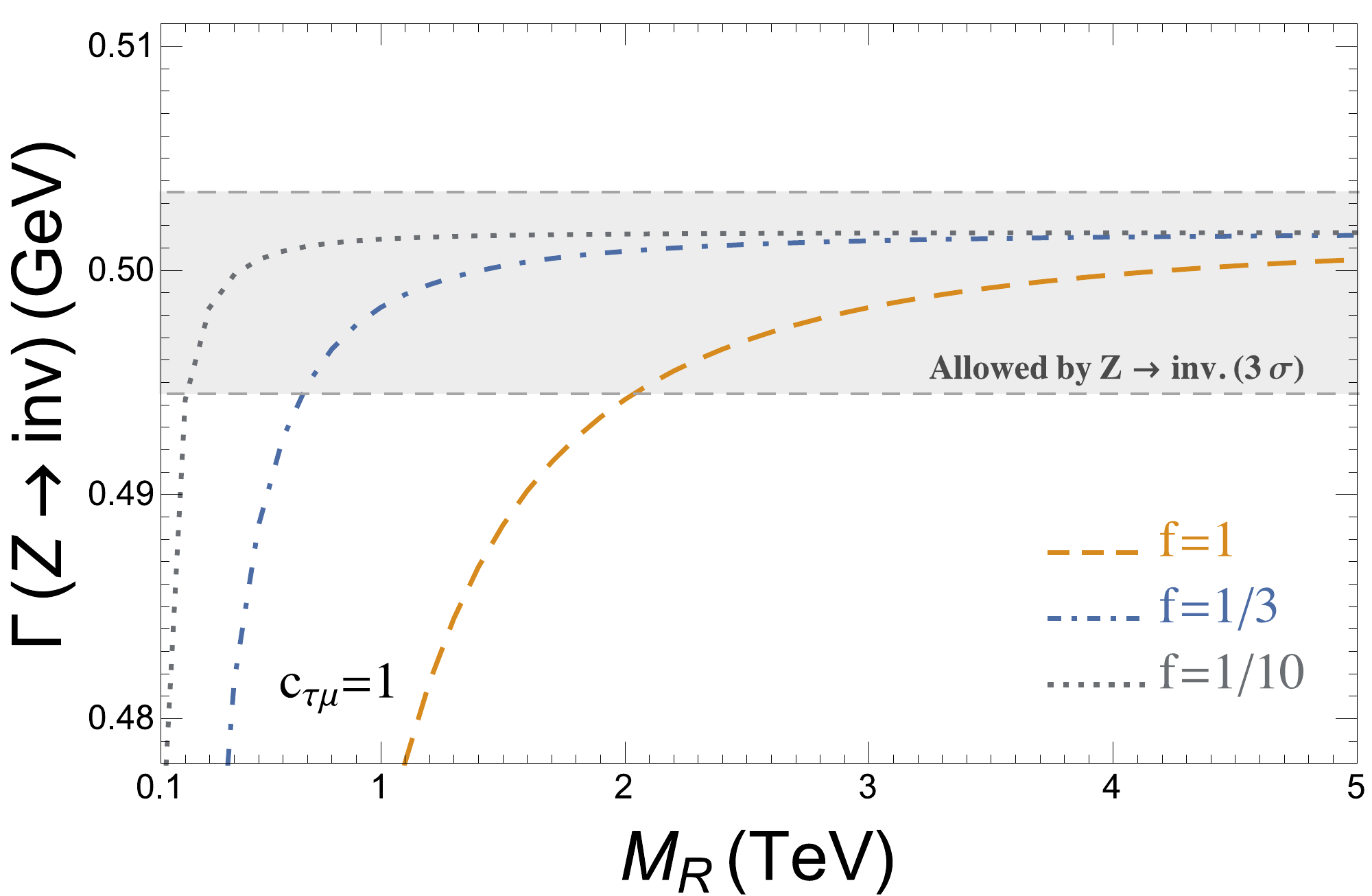}
\caption{Predictions for $\Delta r_k$ and $\Gamma(Z\to{\rm inv.})$ as functions of $M_R$. In both plots we set  $|\boldsymbol n_{e,\mu,\tau}|=1$, $c_{\tau\mu}=1$, $c_{\mu e}=c_{\tau e}=0$, $\mathcal O=\mathbb 1$ and consider three values of $f$. The shadowed band in the left (right) plot is the present excluded (allowed) region at $3\sigma$.}\label{LFV_deltark_Zinv}
\end{center}
\end{figure}

\subsection{Lepton flavor universality}

Leptonic and semileptonic decays of pseudoscalar mesons ($\pi$, $K, D$, $ D_s$, $B$) could  put important constraints on the mixing between the active and the sterile neutrinos in the ISS model, as it has been shown  in Refs.~\cite{Abada:2012mc, Abada:2013aba}.
In particular, the most severe bounds arise from the violation of lepton universality in leptonic kaon decays\footnote{We do not consider other lepton universality tests in view of the fact that they give similar bounds, as in the case of $\Delta r_\pi$, or they are less constraining, like the ones involving $\tau$ leptons \cite{Abada:2013aba}.}.
Following these references, we consider the contributions of the sterile neutrinos to the $\Delta r_k$ parameter, defined as:
\begin{equation}
\Delta r_k = \dfrac{R_K}{R_K^{\rm SM}}-1\quad{\rm with}\quad R_K=\dfrac{\Gamma(K^+\to e^+\nu)}{\Gamma(K^+\to \mu^+\nu)}.
\label{deltarkandRK}
\end{equation}
The comparison of the theoretical calculation in the SM~\cite{Cirigliano:2007xi,Finkemeier:1995gi}  with the recent measurements from the NA62 collaboration~\cite{Goudzovski:2011tc,Lazzeroni:2012cx} shows that the experimental measurements agree with the SM prediction within $1\sigma$:
\begin{equation}
\Delta r_k=(4\pm4)\times 10^{-3}.
\label{deltarkbound}
\end{equation}
We compute the new physics contributions to $\Delta r_k$  using the formulas listed in \appref{App:constraints} that we take from Ref.~\cite{Abada:2012mc} and compare the results with the bound in \eqref{deltarkbound} at the $3\sigma$ level. 

We display in \figref{LFV_deltark_Zinv} our numerical findings using the $\mu_X$ parametrization.
In particular, we choose for this plot a maximized $\tau$-$\mu$ scenario with $c_{\tau\mu}=1$, $c_{\tau e}=c_{\mu e}=0$, $|\boldsymbol n_{e,\mu,\tau}|=1$, $\mathcal O=\mathbb1$ and three different values of $f$. 
We see that $\Delta r_k$ is always negative, meaning that $R_K<R_K^{SM}$.
Nevertheless, $R_K$ tends to $R_K^{SM}$, and hence $\Delta r_k\to0$, for large values of $M_R$, as the new physics effects decouple with the heavy scale. 
We also see that the deviations from the SM values are larger for larger values of $f$, implying that the bound from NA62 can exclude the parameter space region of low $M_R$ and large $f$. 
Furthermore, we have found that this observable is also independent of the rotation matrix $\mathcal O$ and very sensitive to the modulus $|\boldsymbol n_e|$ and $|\boldsymbol n_\mu|$, as expected. 
Actually, the bound from this observable becomes an important constraint at low values of $M_R$ when the ratio between $|\boldsymbol n_e|$ and $|\boldsymbol n_\mu|$ is different from one.

\subsection{The invisible decay width of the Z boson}

The invisible decay width of the $Z$ boson puts very strong constraints on how many neutrinos with masses below $m_Z$ are present. 
The $Z$ invisible decay width was measured in LEP to be~\cite{Agashe:2014kda}: 
\begin{equation}
\Gamma(Z\to {\rm inv.})_{\rm Exp}= 499\pm1.5~{\rm MeV},
\label{Zinvbound}
\end{equation}
which is about 2$\sigma$ below the SM prediction:
\begin{equation}
\Gamma(Z\to {\rm inv.})_{\rm SM}= \sum_{\nu}\Gamma(Z\to \nu\bar\nu)_{\rm SM}= 501.69\pm0.06~{\rm MeV}.
\end{equation}
Although we are not considering the possibility of having $m_N<m_Z$ here, the presence of sterile neutrinos affects the tree level predictions of the $Z$ invisible width even if they are above the kinematical threshold, since they modify the couplings of the active neutrinos to the $Z$ boson. 
We compute the tree level predictions using the formulas provided in Ref.~\cite{Abada:2013aba}, which we collect in \appref{App:constraints} for completeness, and we further include the $\rho$ parameter that accounts for the part of the radiative corrections coming from SM loops, i.e.,
\begin{equation}\label{ZinvISS}
\Gamma(Z\to{\rm inv.})_{\rm ISS} = \sum_{\substack{i,j=1\\ i\leq j}}^{3} \Gamma(Z\to n_i n_j)_{\rm ISS} = \rho \Gamma(Z\to {\rm inv.})_{\rm ISS}^{\rm tree}\, ,
\end{equation}
where $n_i$ runs over all kinematically allowed neutrinos and $\rho$ is evaluated as:
\begin{equation}
 \rho =\frac{\Gamma(Z\to {\rm inv.} )_{\rm SM}}{\Gamma(Z\to {\rm inv.})_{\rm SM}^{\rm tree}}.
\end{equation}
We have also estimated the size of the extra loop corrections induced by the new heavy neutrino states using the formulas of Ref.~\cite{Fernandez-Martinez:2015hxa} and found out that they are numerically very small compared with the SM loop corrections, in agreement with Ref.~\cite{Fernandez-Martinez:2015hxa}, and therefore we will neglect them in the following.

We show our numerical results as a function of $M_R$ in \figref{LFV_deltark_Zinv}, for $c_{\tau\mu}=1$ and three values of $f$. 
We see again that the deviations from the SM value decrease with $M_R$ while they grow with $f$. 
Moreover, we found that the $Z$ invisible width only depends on $M_R$, $f$ and the modulus $|\boldsymbol{n}_{e,\mu,\tau}|$, while it is not dependent on $\mathcal O$ and on the flavor angles ($c_{\tau\mu}, c_{\tau e}$), as it was expected, since when adding all the possible neutrino final states in \eqref{ZinvISS} the dependence on $\mathcal O$ and on the flavor angles appearing in each channel disappears in the sum.
When comparing with data we require our predictions to be within the 3$\sigma$ experimental band (\eqref{Zinvbound}). 
As we can see in \figref{LFV_deltark_Zinv}, the $Z$ invisible width provides in general quite strong constraints, indeed comparable or even tighter  in some cases than the  previous constraints from the LFV lepton decays.  
For instance, for this scenario with $c_{\tau \mu}=1$ and $f=1$, this observable also excludes $M_R$ values lower than around 1-2 TeV, similar to the constraints from $\tau \to \mu \gamma$ and $\tau\to\mu\mu\mu$.

\subsection{Neutrinoless double beta decay}
\begin{figure}[t!]
\begin{center}
\includegraphics[width=0.49\textwidth]{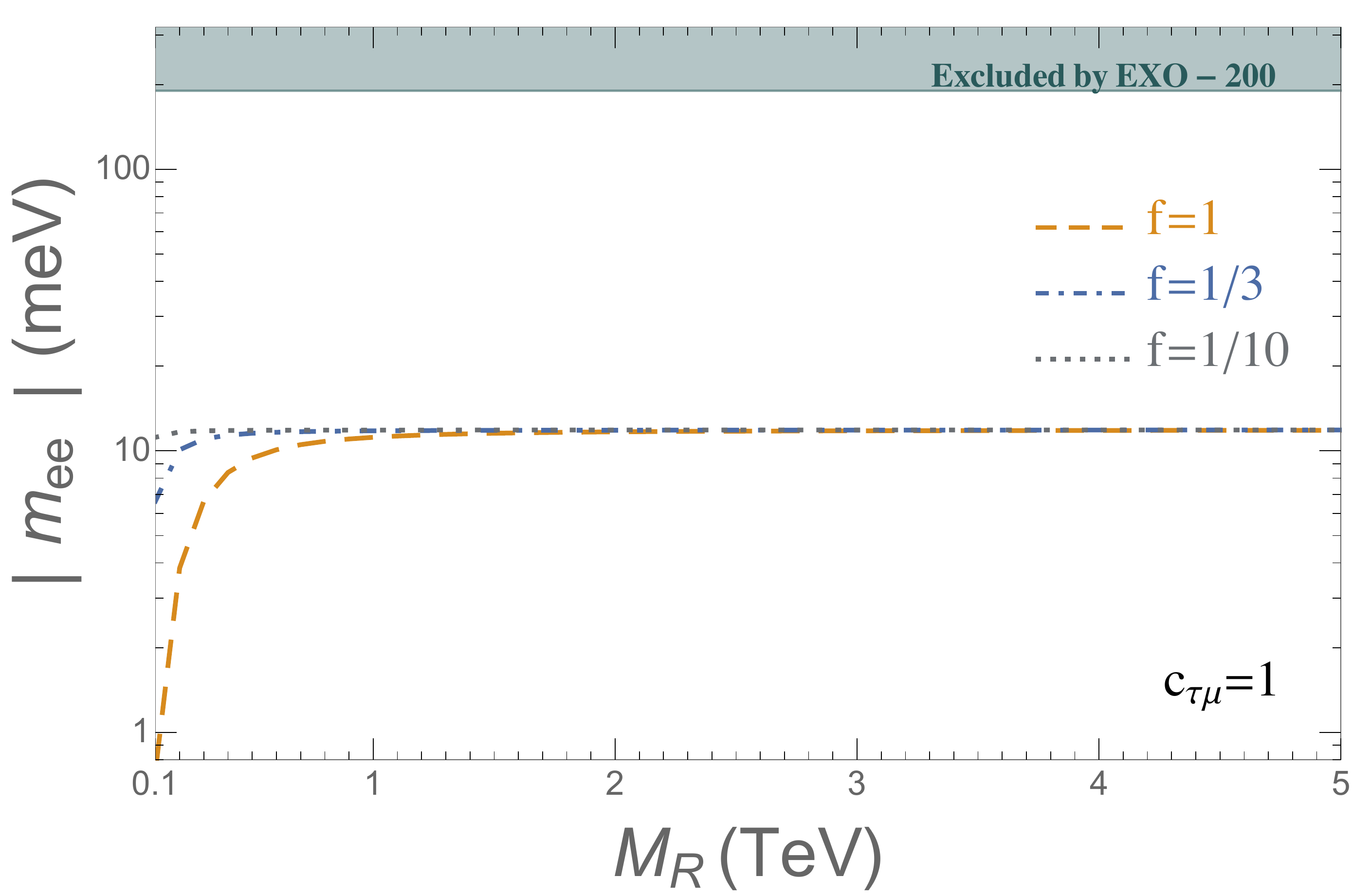} 
\includegraphics[width=0.49\textwidth,height=5.35cm]{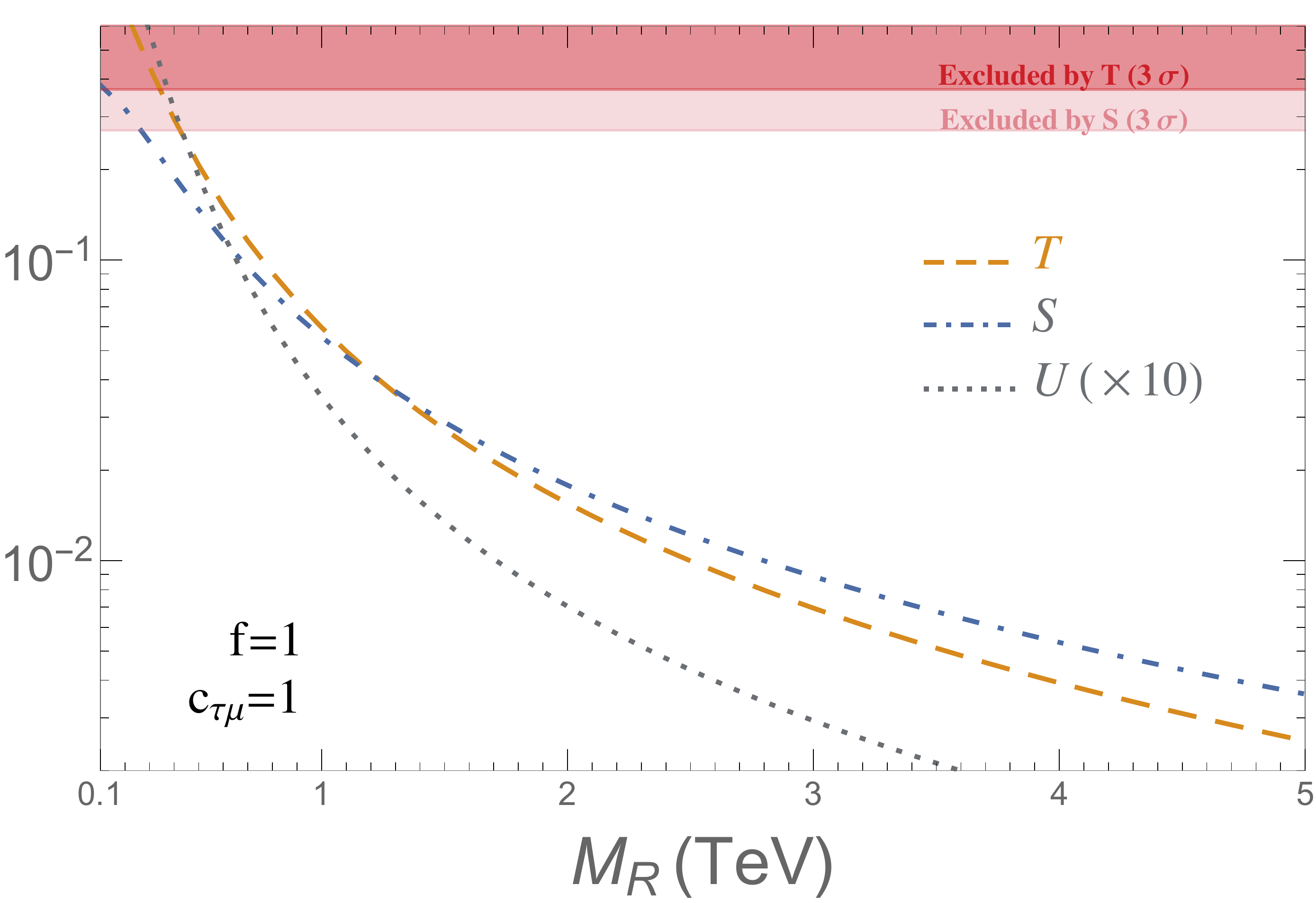}
\caption{Predictions for $|m_{ee}|$ and the Electroweak Precision Parameters $S$, $T$, $U$ (the latter enhanced by a factor of 10 to see it more clearly) as functions of $M_R$. 
In both plots we set $|\boldsymbol n_{e,\mu,\tau}|=1$, $c_{\tau\mu}=1$, $c_{\mu e}=c_{\tau e}=0$, $\mathcal O=\mathbb 1$ and for three different values of $f$. The shadowed bands are the present excluded regions at $3\sigma$.}\label{LFV_neutrinoless_STU}
\end{center}
\end{figure}

Models that introduce  neutrinos with Majorana mass terms allow for lepton number violating processes, such as neutrinoless double beta decay~\cite{Benes:2005hn}. 
Within the ISS framework with $6$ sterile fermions added to the SM particle content, the effective neutrino
mass $m_{ee}$ is given  by~\cite{Abada:2014nwa,Blennow:2010th,Abada:2014vea} 
\begin{equation} \label{22bbdecay}
 m_{ee}\,\simeq \,\sum_{i=1}^{9}  (B_{e n_i})^2 \,p^2
\frac{m_{n_i}}{p^2-m_{n_i}^2} \simeq 
\left(\sum_{i=1}^3 (B_{e n_i})^2 \, m_{n_i}\right)\, 
+ p^2 \, \left(\sum_{i=4}^{9} 
(B_{e n_i})^2  \,\frac{m_{n_i}}{p^2-m_{n_i}^2}\right)\,,
\end{equation}
where $p^2 \simeq - (125 \mbox{ MeV})^2$ is an average estimate over different values from different decaying nucleus of the virtual momentum of the neutrino exchanged in the process.

Although current experiments are searching for neutrinoless double beta decay, it has not been observed yet.
This lack of signal has allowed to the experiments with highest sensitivity such as GERDA~\cite{Agostini:2013mzu}, 
EXO-200~\cite{Auger:2012ar,Albert:2014awa} and KamLAND-ZEN~\cite{Gando:2012zm}  to set strong bounds on the neutrino effective mass. 
These bounds on the effective neutrino Majorana mass in \eqref{22bbdecay} lie in the range 
\begin{equation}
| m_{ee}| \lesssim 140\text { meV} - 700\text { meV}\,.
\end{equation}
In our analysis, we will apply the most recent constraint of $|m_{ee}| \lesssim 190$ meV from  Ref.~\cite{Albert:2014awa}.

\figref{LFV_neutrinoless_STU} displays the behavior of $m_{ee}$ with $M_R$ for different values of the Yukawa strength $f$.
As can be seen in this plot, a  maximum value of $|m_{ee}| \sim 10$ meV is reached at large $M_R \gtrsim$~1~TeV and for all studied values of $f$. 
We have checked that this asymptotic value depends linearly on the mass of the light active neutrinos, i.e.,
\begin{equation}
m_{\nu_1} \sim 0.01 (0.1) \rm eV \rightarrow |m_{ee}| \sim 0.01 (0.1) eV\,.
\end{equation}
As a conclusion, we learn that the prediction for this observable will be below the current experimental bound and, therefore, it will not impose an important constraint to our parameter space. 

\subsection{Electroweak precision observables}

The addition of new physics in the neutrino sector will, in general, modify the prediction of Electroweak Precision Observables (EWPO), which are well determined by experiments. 
We take into account the constraints to the ISS model from EWPO by computing the $S$, $T$ and $U$ parameters~\cite{Peskin:1991sw} and comparing our predictions to the experimental results~\cite{Agashe:2014kda}:
\begin{equation}
S=-0.03\pm0.10\,, \qquad T=0.01\pm0.12\,, \qquad U=0.05\pm0.10\,.
\end{equation}
We use the formulas from Ref.~\cite{Akhmedov:2013hec} (which we report in  \appref{App:constraints}) and compare them with the $3\sigma$ experimental bands. 

We show in  \figref{LFV_neutrinoless_STU} the prediction for $S$, $T$ and $U$ versus $M_R$, choosing again a scenario with maximized $\tau$-$\mu$ transitions. 
As can be seen, the predictions rapidly decrease  with $M_R$ and, in consequence, the constraints from these observables are in general weaker than from the LFV lepton decays and from the $Z$ invisible width. 
In this an most of the studied scenarios, we have found that the most constraining EWPO is the $T$ parameter and next, although quite close, the $S$ parameter. 
For instance, for $f=1$ and $c_{\tau \mu}=1$ we find that $M_R$ below around $300$ GeV are excluded by $T$. 

\subsection{Heavy neutrino decay widths}
\label{sec:HeavyWidhts}
\begin{figure}[t!]
\begin{center}
\includegraphics[width=0.49\textwidth]{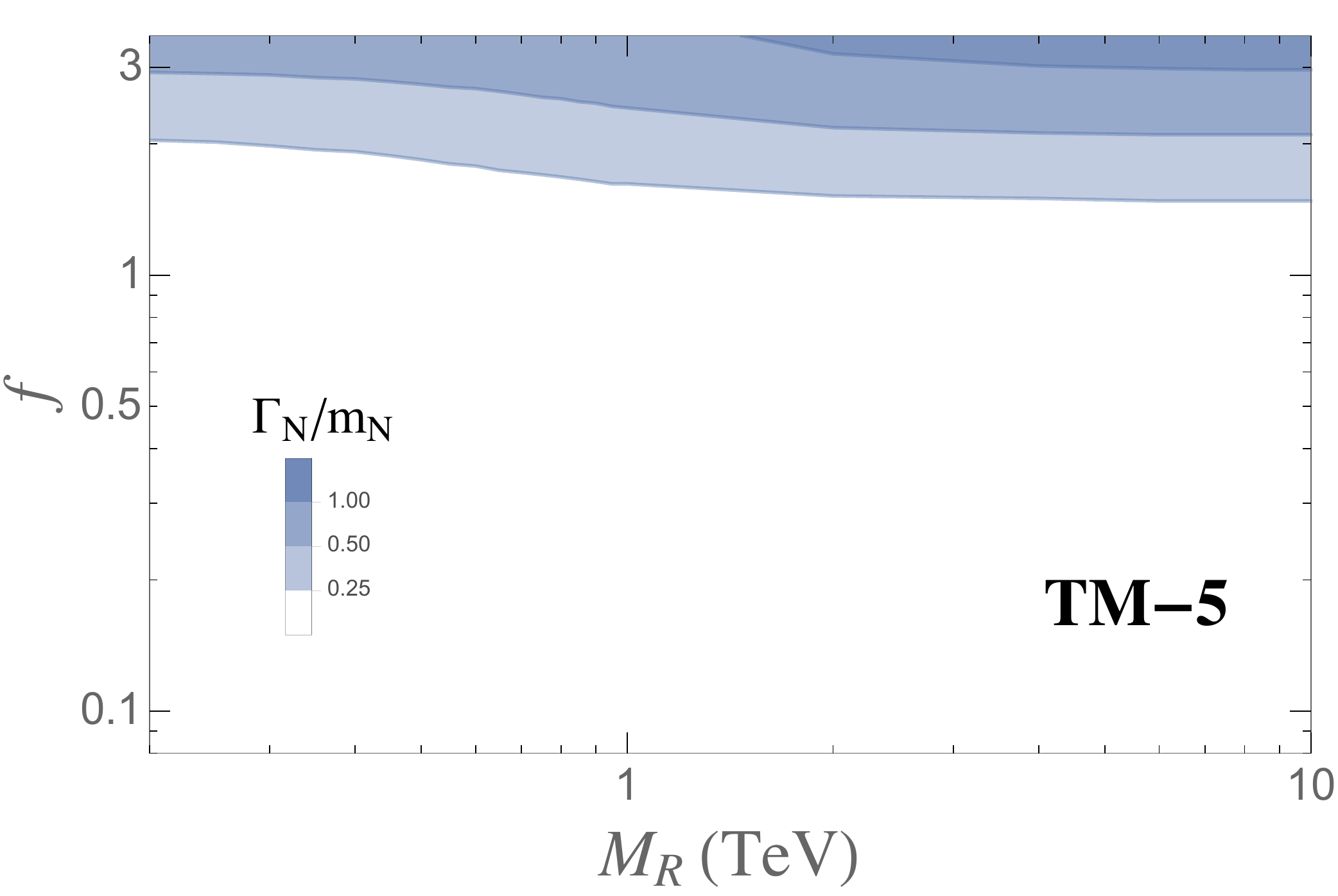} 
\includegraphics[width=0.49\textwidth,height=5.35cm]{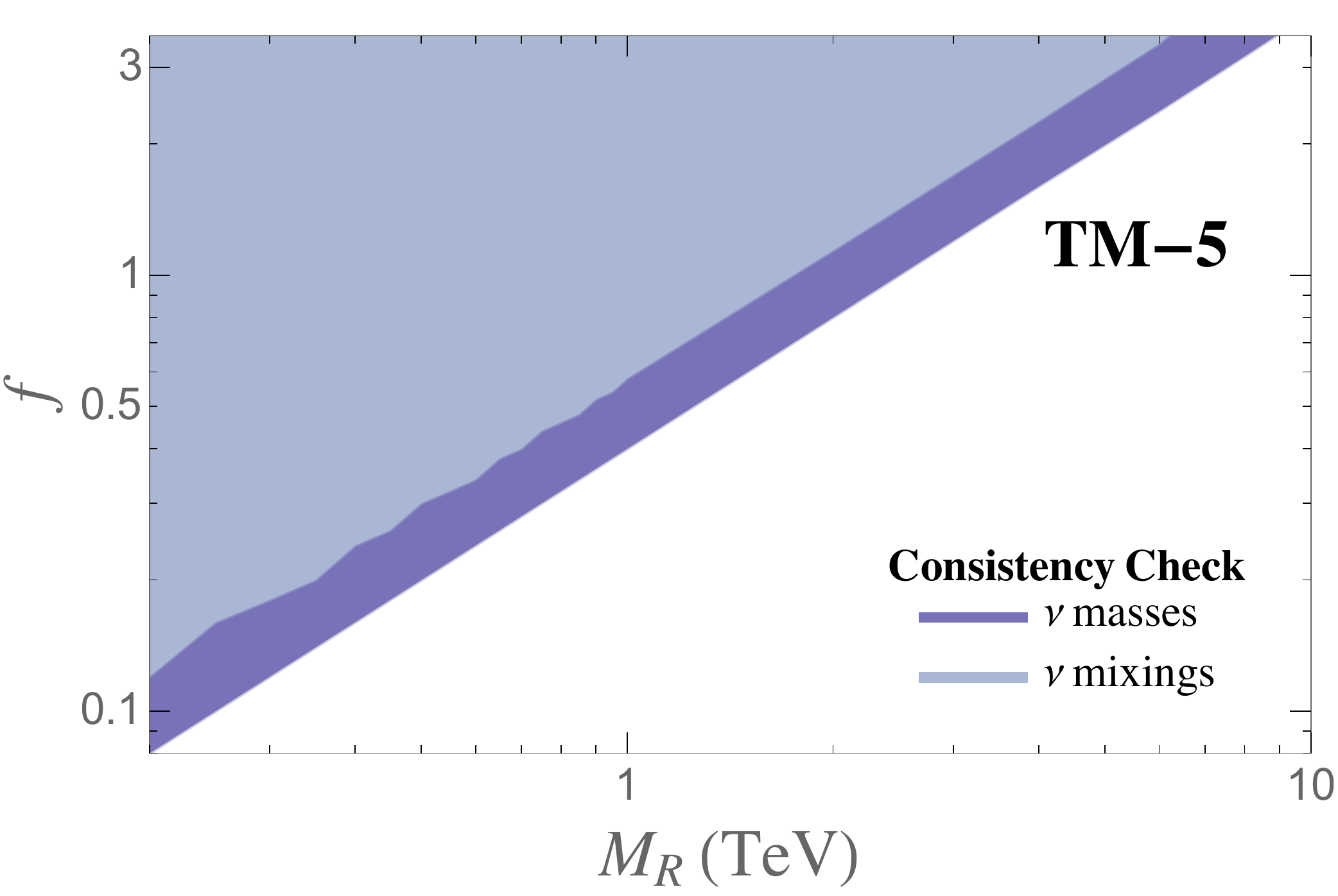}
\caption{Theoretical constraints from the requirement of perturbativity (left panel) and from the consistency of the $\mu_X$ parametrization (right panel), for the scenario TM-5. The regions excluded by the constraints are the shadowed areas.}\label{LFV_Nwidth_muX}
\end{center}
\end{figure}

In this Thesis, we are considering sizable neutrino Yukawa couplings, so we should check that they are still within the perturbative regime. 
In order to impose perturbativity we will either choose a direct constraint on the maximum allowed size of the Yukawa matrix entries, for instance $|Y_\nu^{ij}|^2 /(4 \pi)<1$ or,  alternatively, we will apply a constraint on an observable that grows with this Yukawa coupling, like it is the case of the total width of the heavy neutrinos.  
When choosing  the second method, we will require that the total decay width of each  heavy neutrino is always somehow smaller than the corresponding heavy neutrino mass. 

The computation of the total decay width, in the limit $M_R\gg m_D$ that we work with, is reduced to a few possible decay channels.
In this limit, the masses of all the heavy neutrinos are almost degenerate, close to $M_R$ with small differences of $\mathcal O(m_D^2M_R^{-1})$ between the different pseudo-Dirac pairs, see \figref{HeavyMasses}, and therefore, their potential decays into other heavy neutrinos are suppressed. 
In consequence, the dominant decay channels are simply $N_j\to Z\nu_i, H\nu_i$ and  $W^\pm \ell_i^\mp$, and the total neutrino width can be then easily computed by adding the corresponding partial widths of these four decays.
The partial decay width of the decay channel $N_j \to W \ell_i$ is given by:
\begin{equation}\label{Nwidth}
\Gamma_{N_j \to W \ell_i}=\frac{\sqrt{\big(m_{N_j}^2-m_{\ell_i}^2-m_W^2\big)^2-4 m_{\ell_i}^2 m_W^2}}{16\pi  m_{N_j}^3}~\big| \overline {F_W} \big|^2\,.
\end{equation}
The other channels have similar expressions and the corresponding form factors are defined as,
\begin{align}
\big| \overline {F_H} \big|^2&= \frac{g^2 m_{N_j}^4}{4m_W^2}\Big\{  \big(1-\sqrt{x_i}\big)^2\Big[ \big(1-\sqrt{x_i}\big)^2 -x_H\Big]  \big|C_{n_in_j}\big|^2 + 4\sqrt{x_i} \Big(2+2x_i -x_H\Big) \Big({\rm Re} C_{n_in_j}\Big)^2 \bigg\}\,,\non\\
\big| \overline {F_Z} \big|^2 &= \frac{g^2m_{N_j}^4}{4m_W^2} \bigg\{\bigg[ \Big(1-x_i\Big)^2 + x_Z \Big(1+x_i -6\sqrt{x_i}\Big) - 2 x_Z^2\bigg] \big|C_{n_in_j}\big|^2 
+12 x_Z \sqrt{x_i}  \Big({\rm Re} C_{n_in_j}\Big)^2  \bigg\}\,,\non\\
\big| \overline {F_{W}} \big|^2 &=\frac{g^2 m_{N_j}^4}{4m_W^2} \big|B_{\ell_iN_j}\big|^2 
 \Big\{ \big( 1-x_i\big)^2+x_W (1+x_i)-2x_W^2\Big\}\,,
 \label{NwidthFF}
\end{align}
where $x_H\equiv m_H^2/m_{N_j}^2$, $x_Z\equiv m_Z^2/m_{N_j}^2$, $x_W\equiv m_W^2/m_{N_j}^2$ and $x_i\equiv m_{i}^2/m_{N_j}^2$ with $m_i$ the mass of the corresponding lepton.
The total width is then computed as
\begin{equation}
\Gamma_{N_j}= \sum_{i=1}^3 \Big(\Gamma_{N_j\to h \nu_i} + \Gamma_{N_j\to Z \nu_i}+ 2~ \Gamma_{N_j\to W^+ \ell_i^-} \Big)\,.
\end{equation}
When summing over all flavors $i=1,2,3$ in the final state, the four ratios turn out to be approximately equal~\cite{Atre:2009rg}: 
\begin{equation}
{\rm BR}(N_j \to H \nu)={\rm BR}(N_j \to Z \nu) =
{\rm BR}(N_j \to W^+ \ell^-) ={\rm BR}(N_j \to W^- \ell^+)=25\%\,.
\end{equation} 
We have explored three different assumptions to comply with the perturbative unitary condition. In particular we have taken:
\begin{equation}
\frac{\Gamma_{N_i}}{m_{N_i}}<1,\frac12, \frac{1}{4}\quad {\rm for}~ i=1,\dots,6\,.
\label{widthconstraint}
\end{equation}
The results for the TM-5 scenario from \tabref{TMscenarios} are displayed in the left panel of  \figref{LFV_Nwidth_muX}, although similar qualitative results are found for other scenarios. 
Here, we show the areas in the $(M_R, f)$ plane that are excluded by the different assumptions in \eqref{widthconstraint}.  
We find that this perturbativity requirement is not much sensitive to $M_R$, giving an excluded area in the $(M_R, f)$ plane that is a  nearly horizontal band located at the top, which constrains basically just the size of the global Yukawa coupling $f$, in the most restricted scenarios, to be below order 2-3.  

For the rest of this Thesis, we will take the  second choice, $1/2$, in \eqref{widthconstraint} when we decide to use heavy neutrino widths as perturbativity  criteria. 

\subsection[Validity range of the $\mu_X$ parametrization]{Validity range of the $\boldsymbol{\mu_X}$ parametrization}
\label{sec:MUXcheck}

As explained before, one of the novelties of this Thesis is the introduction and use of the $\mu_X$ parametrization as a tool for exploring the model parameter space being always in agreement with oscillation data. 
In order to check the validity range of this parametrization, we require that both the predicted light neutrino mass squared differences and the neutrino mixing angles  that we obtain from the diagonalization of the full neutrino mass matrix in \eqref{ISSmatrix}, lie within the 3$\sigma$ experimental  bands~\cite{Esteban:2016qun,Tortola:2012te,Fogli:2012ua,Forero:2014bxa,Adhikari:2016bei}.  

More specifically, we demand that the corresponding entries of the $U_\nu$ matrix that refer to the light neutrino  sub-block agree with the $3\sigma$ range given in Ref.~\cite{Esteban:2016qun}:
\begin{equation}
|U_{\rm PMNS}^{\,3\sigma}|=\left(\begin{array}{ccc}
0.801\to0.845 & 0.514\to0.580 & 0.137\to0.158 \\
0.225\to0.517 & 0.441\to0.699 & 0.614\to0.793 \\
0.246\to0.529 & 0.464\to0.713 & 0.590\to0.776
\end{array}
\right)\,.
\end{equation}

We show in the right panel of \figref{LFV_Nwidth_muX} the predictions for the constraints found  in the ($M_R,f$) plane for the  TM-5 scenario.
As can be seen in this figure,  the bounds obtained from the constraints on the active neutrinos squared mass differences are in this scenario stronger than the ones from the light neutrino mixing matrix entries.   
For other scenarios, like TM-8, we have checked that this can  be reversed,  i.e., the constraints from the neutrino mixings can be stronger than from the neutrino masses.  
Additionally, we have also compared the range of validity of this parametrization for two values of the input lightest neutrino mass, $0.1$eV and $0.01$eV (the chosen value for  \figref{LFV_Nwidth_muX}), and we have concluded that the $\mu_X$ parametrization works better for the case with a smaller value of the light neutrino mass.

In general, we found that the area in the $(M_R,f)$ parameter space that is allowed by all the experimental bounds above studied is also allowed by the consistency checks of the $\mu_X$ parametrization, meaning that the parametrization works well for the  parameter space allowed by data.
Nevertheless, the validity of this parametrization can be improved by considering next order contributions to $M_{\rm light}$ in \eqref{Mlight}, as it was done in Ref.~\cite{Baglio:2016bop}.

\vspace{.5cm}

Summarizing, in this Chapter we have learnt that the presence of right-handed neutrinos at the TeV scale can have a large impact in many low energy observables if their couplings are sizable. 
Since the aim of this Thesis is to study LFV consequences of these $\nu_R$ fields, we focused mostly on the LFV radiative and three-body decays of the leptons, i.e., $\ell_m\to\ell_k\gamma$ and $\ell_m\to\ell_k\ell_k\ell_k$ with $k\neq m$.
This study allowed us to acquire some general ideas about LFV processes in this kind of models, to discuss the advantages and disadvantages of using the two parametrizations described in \chref{Models}, as well as to introduce the phenomenological TM and TE scenarios in \tabref{TMscenarios}, where the experimentally most constrained $\mu$-$e$ transitions are {\it ad-hoc} suppressed. 
Additionally, we also reviewed the effects of the TeV neutrinos in other observables, including processes with lepton number violation, lepton flavor universality violation, precision physics or theoretical implications, as perturbativity of the new Yukawa coupling. 
All these observables will be important in the following Chapters, as they will constraint the allowed parameter space for our study of maximum LFV H and Z decays in presence of TeV right-handed neutrinos.

\chapter{LFV Higgs decays from low scale seesaw neutrinos}
\fancyhead[RO] {\scshape LFV Higgs decays from low scale seesaw neutrinos  }
\label{LFVHD}

The recent discovery of the Higgs boson has opened a new experimental area to search for new physics beyond the SM, in particular with new LFV Higgs decay (LFVHD) channels.
As we discussed before, LFV transitions are forbidden in the SM, therefore any observation of a LFV Higgs decay would automatically imply the existence of new physics. 

The ATLAS and CMS collaborations are actively searching for these LFV Higgs boson decay processes. 
Interestingly, the CMS collaboration saw an excess on the $H\to\tau\mu$ channel after the run-I, with a significance of $2.4\sigma$ and a value of BR$(H\to\tau\mu)=(0.84^{+0.39}_{-0.37})\%$.
Unfortunately, neither this excess, nor other positive LFVHD signal, have been observed at the present run-II, so ATLAS and CMS have set bounds on these processes, as summarized in \tabref{LFVsearchII}. 
At present, ATLAS has released their results after analyzing $20.3~{\rm fb}^{-1}$ of data at a center of mass energy of $\sqrt{s}=8$~TeV, reaching sensitivities of the order of $10^{-2}$ for the $H\to\tau\mu$ and $H\to\tau e$ channels~\cite{Aad:2016blu}.
On the other hand,  CMS has also searched for the $H\to\mu e$ channel after the run-I~\cite{Khachatryan:2016rke} and has further improved the sensitivities of the $H\to\tau\mu$ and $H\to\tau e$ channels with new run-II data~\cite{CMS:2017onh} of $\sqrt{s}=13$~TeV, setting the most stringent upper bounds for the LFV Higgs decays, which at the $95\%$ CL are given by:
\begin{align}
{\rm BR}(H\to\mu e)&<3.5\times10^{-4}\,,\\
{\rm BR}(H\to\tau e)&<6.1\times10^{-3}\,,\\
{\rm BR}(H\to\tau\mu)&<2.5\times10^{-3}\,.
\end{align}
Additional indirect constraints on LFVHD rates have been also  derived using other LFV transitions~\cite{Harnik:2012pb,Blankenburg:2012ex}.
For instance, the upper bounds on $\mu\to e\gamma$  can be translated into a very strong upper bound of BR$(H\to\mu e)<\mathcal O(10^{-8})$.
On the contrary, much weaker indirect upper bounds, of order $\mathcal O(10\%)$, are derived for the other two channels.
This fact further motivates the kind of phenomenological scenarios that we introduced in \secref{sec:scenarios}, where $\mu$-$e$ transitions are {\it ad-hoc} suppressed.

From the theoretical point of view, there are many extensions of the SM that naturally predict large ratios for these LFV Higgs decays. 
For instance, they have been studied in the context of 
supersymmetric models~\cite{Han:2000jz,Curiel:2002pf,DiazCruz:2002er,Curiel:2003uk,Brignole:2003iv, Brignole:2004ah,Parry:2005fp,DiazCruz:2008ry,Crivellin:2010er,Giang:2012vs,Arhrib:2012mg,Arhrib:2012ax,Arana-Catania:2013xma,Arganda:2015uca,Aloni:2015wvn,Vicente:2015cka,Alvarado:2016par,Hammad:2016bng}, 
in composite Higgs models~\cite{Agashe:2009di},
two Higgs doublet models~\cite{Davidson:2010xv,Sierra:2014nqa,Omura:2015nja,Bizot:2015qqo,Botella:2015hoa},
the Zee model~\cite{Herrero-Garcia:2017xdu},
minimal flavor violation~\cite{Dery:2013aba,Dery:2014kxa,He:2015rqa,Baek:2016pef},
Randall-Sundrum models~\cite{Perez:2008ee,Azatov:2009na}, 
using effective Lagrangians~\cite{DiazCruz:1999xe,deLima:2015pqa,Buschmann:2016uzg,Herrero-Garcia:2016uab} 
and  many others~\cite{Falkowski:2013jya,Dery:2013rta,Heeck:2014qea,Campos:2014zaa,Dorsner:2015mja,Crivellin:2015mga,Hernandez:2015dga,Altmannshofer:2016oaq,Huitu:2016pwk,Baek:2016kud,DiIura:2016wbx}.
Likewise, the addition of new right-handed neutrinos to the SM can induce large LFVHD rates, specially if they are allowed to have large Yukawa interactions, which is in fact the interaction to the Higgs boson. 
As we saw in \chref{Models}, this is precisely the case in the low scale seesaw models, as the ISS or the SUSY-ISS models, where neutrinos can have large Yukawa couplings and moderately heavy neutrino masses. 
Consequently, we consider extremely timely to study the LFV Higgs boson decays in these low scale seesaw models. 

The LFV Higgs decays were analyzed in the context of the SM enlarged with heavy Majorana neutrinos for the first time in Refs.~\cite{Pilaftsis:1992st,Korner:1992zk}. 
Later, they were computed in the context of the type-I seesaw model in Ref.~\cite{Arganda:2004bz}, and they were found to lead to extremely small rates due to the strong suppression from the very heavy right-handed neutrino masses, at $10^{14-15}$ GeV, in that case. 
This motivates our study of the LFV Higgs decays in the case where the right-handed neutrino masses lie in contrast at the ${\cal O}({\rm TeV})$ energy scale and, at the same time, can have large Yukawa couplings. 
As we saw in \chref{Models}, the ISS model contemplates this possibility and, therefore, the rates are expected to be larger than in the type-I seesaw model case. 

In this Chapter we perform a detailed study of lepton flavor violating Higgs boson decays $H\to\ell_k\bar\ell_m$.
We consider the inverse seesaw model as an explicit realization of a low scale seesaw model and analyze, both analytically and numerically, the one-loop induced LFV H decays.
Furthermore, we make use of the mass insertion approximation to compute a simple effective LFV $H\ell_k\ell_m$ vertex that allows to rapidly estimate these rates in models with right-handed neutrinos.
Additionally, we also explore a supersymmetric realization of the ISS model and study the new one-loop contributions to the LFV H decays coming from sneutrinos with TeV masses. 
The results presented in this Chapter have been published in Refs.~\cite{Arganda:2014dta,Arganda:2015naa,Arganda:2017vdb}.

\section{LFV H decays in the ISS model} 
\label{sec:LFVHDanal}

LFV Higgs decay rates within the SM with new heavy Majorana neutrinos were first studied in Refs.~\cite{Pilaftsis:1992st,Korner:1992zk,Arganda:2004bz}.
In this Section, we analyze these rates in the context of the ISS model with three pairs of fermionic singlets added to the SM, and fully study their one-loop contributions to the LFVHD rates.
As we did for the LFV radiative decays in the previous Chapter, we use the two parametrizations introduced in \secref{sec:ISSmodel}, the Casas-Ibarra and the $\mu_X$ parametrization, to explore these LFVHD rates and discuss the main differences of using one parametrization versus the other.

\begin{figure}[t!]
\begin{center}
\includegraphics[width=.8\textwidth]{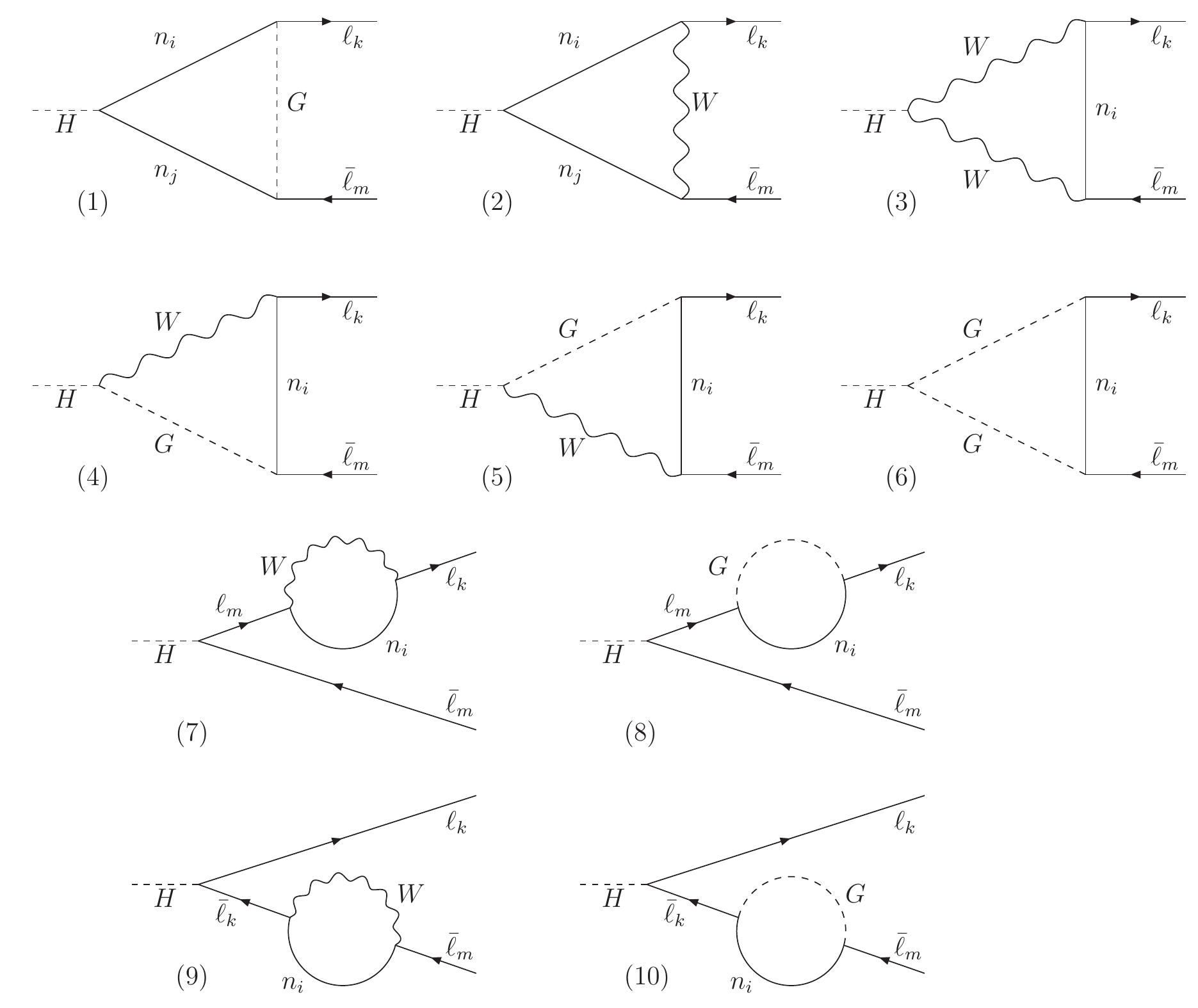}
\caption{One-loop diagrams  contributing to the full computation of $H\to\ell_k {\bar \ell_m}$ decays in the physical neutrino mass eigenstate basis and in the Feynman-'t Hooft gauge.}\label{diagsLFVHDphysbasis}
\end{center}
\end{figure}

The decay amplitude of the process $H(p_1) \to \ell_k(-p_2) \bar \ell_m (p_3)$  can be generically decomposed in terms of two form factors $F_{L,R}$ by:
\begin{equation}
i {\cal M} = -i g \bar{u}_{\ell_k} (-p_2) (F_L P_L + F_R P_R) v_{\ell_m}(p_3) \, , 
\label{LFVHDamp}
\end{equation}
and the partial decay width is then written as follows:
\begin{align}
\Gamma (H \to {\ell_k} \bar{\ell}_m)& = \frac{g^2}{16 \pi m_{H}} 
\sqrt{\left(1-\left(\frac{m_{\ell_k}+m_{\ell_m}}{m_{H}}\right)^2\right)
\left(1-\left(\frac{m_{\ell_k}-m_{\ell_m}}{m_{H}}\right)^2\right)} \nonumber\\
& \times \Big((m_{H}^2-m_{\ell_k}^2-m_{\ell_m}^2)\big(|F_L|^2+|F_R|^2\big)- 4 m_{\ell_k} m_{\ell_m} {\rm Re}(F_L F_R^{*})\Big) \, . 
\label{LFVHDwidth}
\end{align}
Here, $p_1$, $-p_2$ and $p_3$ are the momenta of the ingoing Higgs boson, the outgoing lepton $\ell_k$ and the outgoing antilepton $\bar \ell_m$, respectively, and the conservation of momentum has been implemented as $p_1=p_3-p_2$. Moreover, $m_H$ stands for the Higgs mass and $m_\ell=v\, Y_\ell $ for lepton masses (with $v=174$ GeV). The widths of the CP-conjugate channels $H\to\ell_m \bar \ell_k$ are trivially related to the previous ones  and their numerical values will coincide for the case of real mass matrices, as will be the case for most of this Thesis.

In the calculation of the LFV Higgs decay rates, we will first work in the physical basis for the neutrinos and consider the full set of contributing one-loop diagrams in the Feynman-'t Hooft gauge, drawn in \figref{diagsLFVHDphysbasis}. The form factors can be written in this case as the sum of the different contributions:
\begin{equation}
F_L = \sum_{i=1}^{10} F_L^{(i)}\,,\,\qquad  F_R = \sum_{i=1}^{10} F_R^{(i)}\,.
\end{equation}
These form factors were computed in the context of the type-I seesaw in Ref.~\cite{Arganda:2004bz} and we have adapted them to our present ISS case. The complete results are given in the \appref{App:LFVHD}.

The process $H\to\ell_k\bar\ell_m$ with $k\neq m$ does not exist at the tree level, neither in the SM nor in the  context of the ISS model we study here, therefore the full one-loop process must be finite.  
We have explicitly checked that the diagrams (2)-(6) are finite and that the divergent terms from diagrams (7) and (9) are cancelled when adding up to all the neutrinos running in the loop. 
Hence, the only divergent contributions to the LFV Higgs decays arise from the diagrams (1), (8), and (10), and we have checked that  they cancel among each other, giving rise to a total finite result. 
This cancellation  is in agreement with the results for the type-I seesaw~\cite{Arganda:2004bz}.

The numerical estimates of these LFV Higgs form factors and the LFV Higgs partial decay widths have been done with our private {\it Mathematica} code. 
In order to get numerical predictions for the BR$(H \to {\ell_k} \bar{\ell}_m)$ rates we use $m_H=125\,{\rm GeV}$ and its corresponding SM total width is computed with {\it FeynHiggs}~\cite{Heinemeyer:1998yj,Heinemeyer:1998np,Degrassi:2002fi} including two-loop corrections.  

In order to be in agreement with present neutrino oscillation data in \eqref{NuFit}, we will make use of the two parametrizations presented in \chref{Models}.
We start this study by taking the matrices $M_R$ and $\mu_X$ as input parameters and reconstructing the Yukawa coupling by means of the Casas-Ibarra parametrization in \eqref{CasasIbarraISS}. 
Next, we will follow the idea behind the $\mu_X$ parametrization of choosing the $M_R$ and $Y_\nu$  matrices as input parameters and then building the proper $\mu_X$ matrix that leads to the right light neutrino masses and mixing angles. 

In order to compare the predictions of the LFVHD rates with other LFV observables, we also present here the predictions for the related radiative decay rates, BR$(\mu \to e \gamma)$, BR$(\tau \to e \gamma)$ and BR$(\tau \to \mu \gamma)$.  
We will consider a more complete set of constraining observables and the corresponding updated bounds when looking for the maximum allowed LFV H decay rates at the end of this Chapter.

\subsection{LFVHD with the Casas-Ibarra parametrization}

We present first our numerical results for the LFV Higgs decay rates, BR$(H \to \mu \bar \tau)$, BR$(H \to e \bar \tau)$ and BR$(H \to e \bar \mu)$, when using the Casas-Ibarra parametrization to accommodate light neutrino data. 
We start by considering the simplest scenario where both $M_R$ and $\mu_X$ matrices are diagonal at the same time, $M_R\equiv$ diag($M_{R_1},M_{R_2},M_{R_3}$) and $\mu_X\equiv$ diag($\mu_{X_1},\mu_{X_2},\mu_{X_3}$). Although this is not the most general case, it will be very illustrative to learn how these observables depend on the parameters of the model and to find an optimal strategy to study a most general scenario afterwards. 

We study the LFV rates as functions of the input ISS parameters in this case, namely, $M_{R_i}$, $\mu_{X_i}$, the lightest\footnote{We will show again our results for a Normal Hierarchy, varying the value of the lightest neutrino mass $m_{\nu_1}$ and setting the other two masses using the mass differences in \eqref{NuFit}. 
Although not shown here, we found similar results for the Inverted Hierarchy.} neutrino mass $m_{\nu_1}$ and the angles $\theta_i$ of the $R$ matrix in \eqref{R_Casas}, trying to localize the areas of the parameter space where the LFV Higgs decays can both be large and respect the constraints on the radiative decays.
For a given set of these input parameters, we will build the Yukawa coupling by using \eqref{CasasIbarraISS}. 
Nevertheless, since this procedure can generate arbitrarily large Yukawa couplings, we will enforce their perturbativity in this study by setting an upper limit on the entries of the neutrino Yukawa coupling matrix, given by
\be\label{Ymax15}
\frac{\big|Y_{ij}\big|^2}{4\pi}<1.5\,,\quad {\rm for}~i,j=1,2,3\,.
\ee
The results of this first case will be presented in two generically different scenarios for the heavy neutrinos: the case of (nearly) degenerate heavy neutrinos, and  the case of hierarchical heavy neutrinos.

\begin{figure}[t!]
\begin{center}
\includegraphics[width=.496\textwidth]{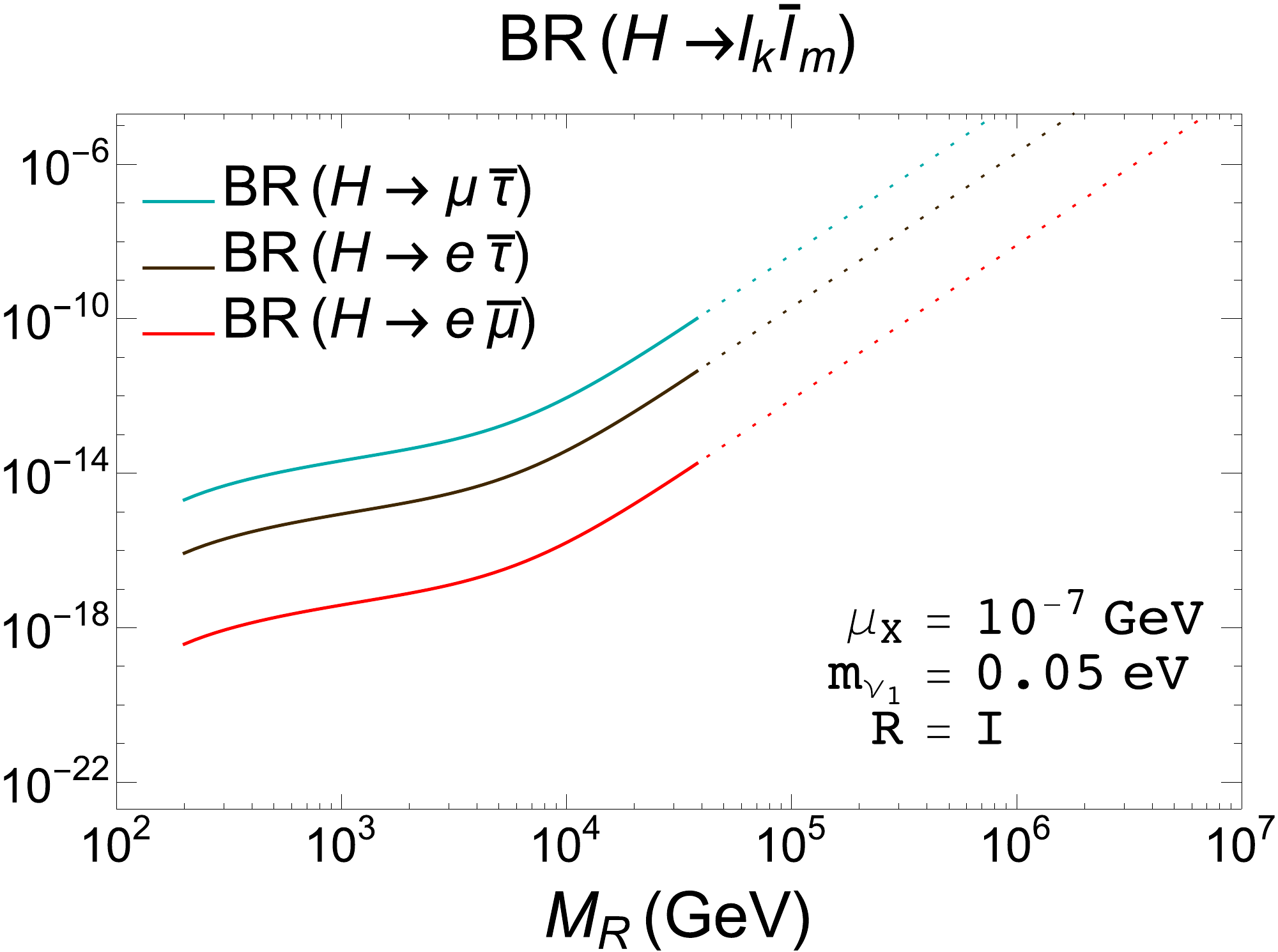}
\includegraphics[width=.496\textwidth]{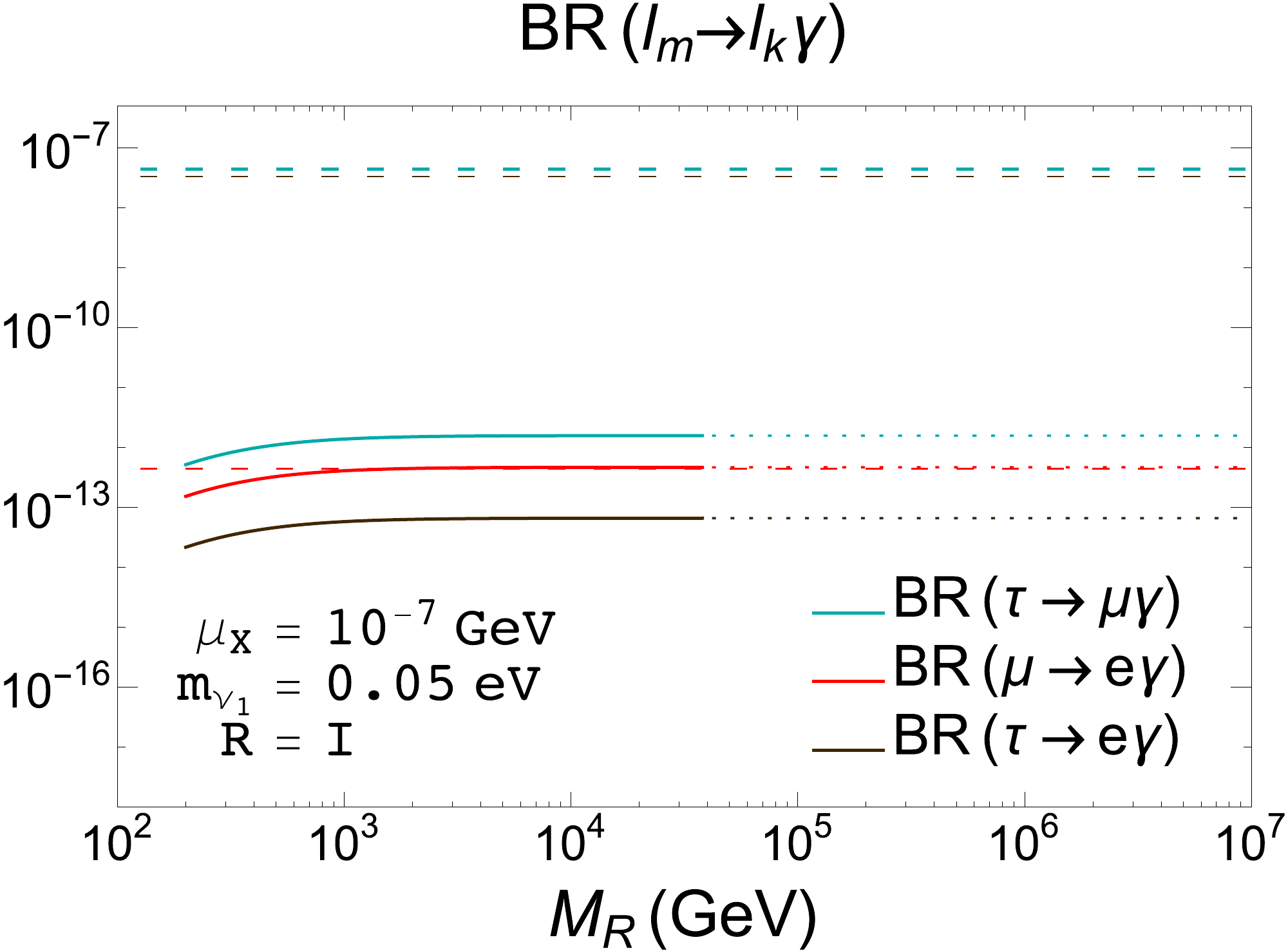}
\caption{Predictions for the LFV decay rates as functions of $M_R$ in the degenerate heavy neutrinos case with the Casas-Ibarra parametrization in \eqref{CasasIbarraISS}. 
Left panel: BR$(H \to \mu \bar \tau)$ (upper blue line), BR$(H \to e \bar \tau)$ (middle dark brown line), BR$(H \to e \bar \mu)$ (lower red line). 
Right panel: BR$(\tau \to \mu \gamma)$ (upper blue line), BR$(\mu \to e \gamma)$ (middle red line), BR$(\tau \to e \gamma)$ (lower dark brown line). 
The other input parameters are set to $\mu_X= 10^{-7} \, {\rm GeV}$, $m_{\nu_1}=0.05 \, {\rm eV}$, $R=\mathbb1$.  
Dotted lines  indicate non-perturbative $Y_\nu$, meaning that \eqref{Ymax15} is not fulfilled. 
Horizontal dashed lines in the right panel are the ($90\%$ C.L.) upper bounds: BR$(\tau \to \mu \gamma)<4.4 \times 10^{-8}$~\cite{Aubert:2009ag} (blue line), BR$(\tau \to e \gamma)<3.3 \times 10^{-8}$~\cite{Aubert:2009ag} (dark brown line), BR$(\mu \to e \gamma)< 4.2 \times 10^{-13}$~\cite{TheMEG:2016wtm} (red line).}\label{LFVdegenerateCasas}
\end{center}
\end{figure}

The case of (nearly) degenerate heavy neutrinos is implemented  by choosing  degenerate entries in $M_R$ and in $\mu_X$, i.e., by setting $M_{R_i} \equiv M_R$ and $\mu_{X_i} \equiv \mu_X$ for $i = 1, 2, 3$, see \figref{HeavyMasses}. 
First we show in \figref{LFVdegenerateCasas} the results for all the LFV rates as functions of the common right-handed neutrino mass parameter $M_R$.
The left panel shows the LFV Higgs decay channels, while the right panel displays de LFV radiative decays.  
Here we have fixed the other input parameters to $\mu_X=10^{-7}$ GeV, $m_{\nu_1}=0.05$ eV, and $R=\mathbb1$. 
We find that the largest LFV Higgs decay rates are for BR($H \to \mu\bar \tau$) and the largest radiative decay rates are for BR($\tau \to \mu \gamma$). 
We also see that, for this particular choice of input parameters, all the predictions for the LFV Higgs decays are allowed by the present experimental upper bounds on the three radiative decays (dashed horizontal lines in this and following plots for the radiative decays) for all explored values of $M_R$ in this interval of $(200, 10^7)\, {\rm GeV}$. 
Nevertheless, it shows clearly that the most constraining radiative decay at present is by far  $\mu \to e \gamma$. 

Regarding the $M_R$ dependence shown in \figref{LFVdegenerateCasas}, we manifestly  see that the LFVHD rates grow faster with $M_R$ than the radiative decays, which tend, as we already saw in \chref{PhenoLFV}, to a constant value for $M_R$ above $\sim 10^3$ GeV. 
In fact, the LFVHD rates can reach quite sizable values at the large $M_R$ region of these plots, yet  allowed by the constraints on the radiative decays. 
For example, we obtain BR$(H \to \mu\bar \tau) \sim 10^{-6}$ for $M_R \sim 10^6$ GeV. 
However, our requirement of perturbativity for the neutrino Yukawa coupling entries in \eqref{Ymax15} does not allow for such large $M_R$ values as they lead to too large $Y_\nu$ values in the framework of the Casas-Ibarra parametrization of \eqref{CasasIbarraISS}. 
This non-perturbative region is illustrated using dotted lines in these plots.

The qualitatively different functional behavior with $M_R$ of the LFVHD and the radiative rates shown in \figref{LFVdegenerateCasas} is an interesting feature that we wish to explore further.  
The results for the radiative decay rates can be understood with the approximated formula in \eqref{RadApprox}, valid for large values of $M_R$. 
As already explained in \chref{PhenoLFV}, the $(Y_\nu^{} Y_\nu^\dagger)_{km}$ elements grow with $M_R$ as $M_R^2$ when using the Casas-Ibarra parametrization in \eqref{CasasIbarraISS}, and therefore the radiative decay rates saturate to a constant value at large values of $M_R$, as can be seen in the plot on the right in \figref{LFVdegenerateCasas}. 
This simple behavior with $M_R$ is certainly not the case of the LFVHD rates, and we conclude that these do not follow this same behavior with $|(Y_\nu^{} Y_\nu^\dagger)_{km}|^2$.
This different functional behavior of BR$(H \to \ell_k \bar \ell_m)$ with $M_R$ will be further explored and clarified later by our study using the mass insertion approximation.
However, we want to emphasize again that the apparent non-decoupling behavior of the LFV rates with the heavy mass scale $M_R$ is an artifact of the Casas-Ibarra parametrization. 
As  we already said in \chref{PhenoLFV}, the expected decoupling behavior with $M_R$ will become manifest when using the $\mu_X$ parametrization, as we will see in \secref{sec:LFVHDmuX}.

\begin{figure}[t!] 
\begin{center}
\includegraphics[width=0.49\textwidth]{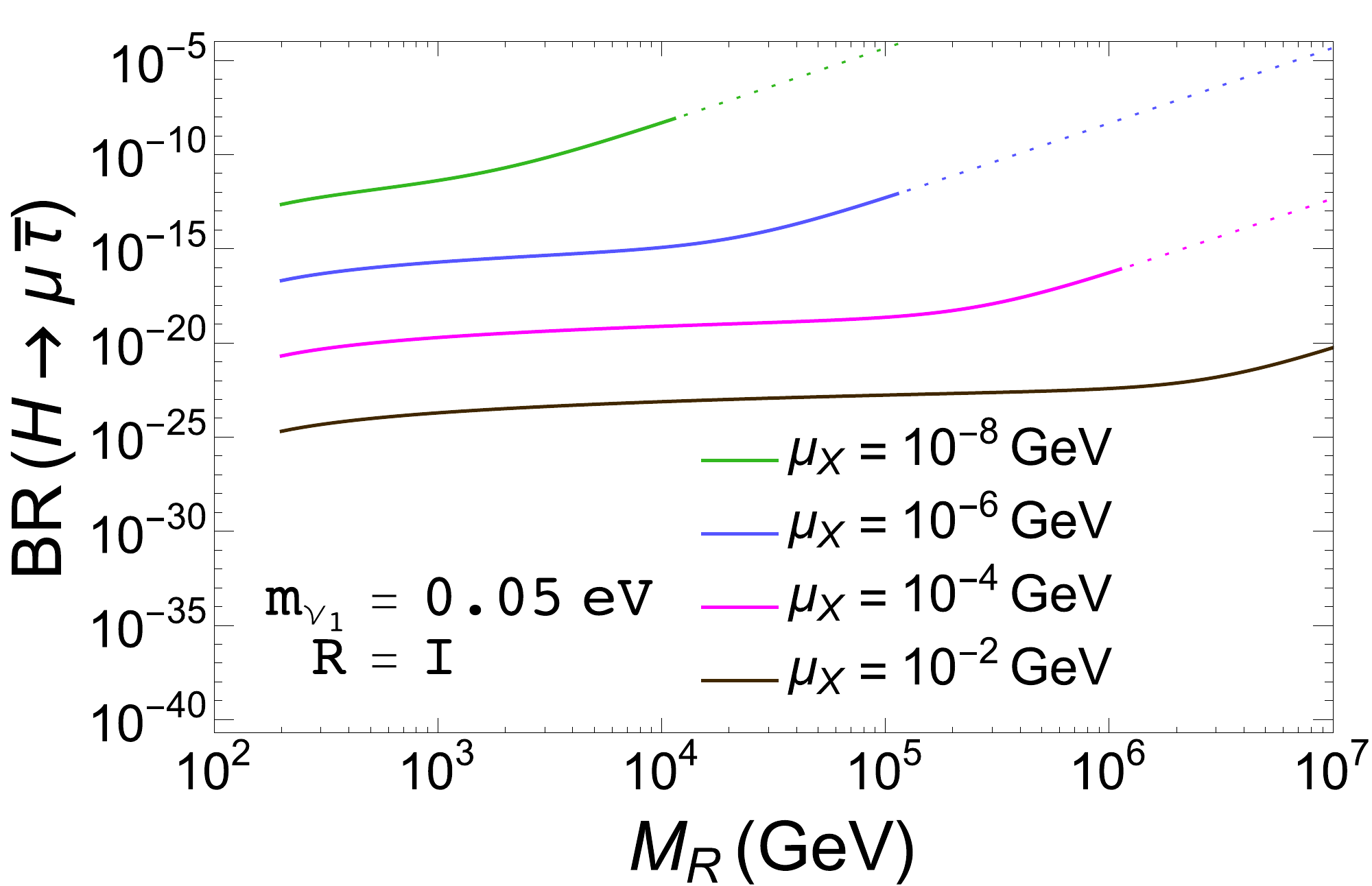} 
\includegraphics[width=0.49\textwidth]{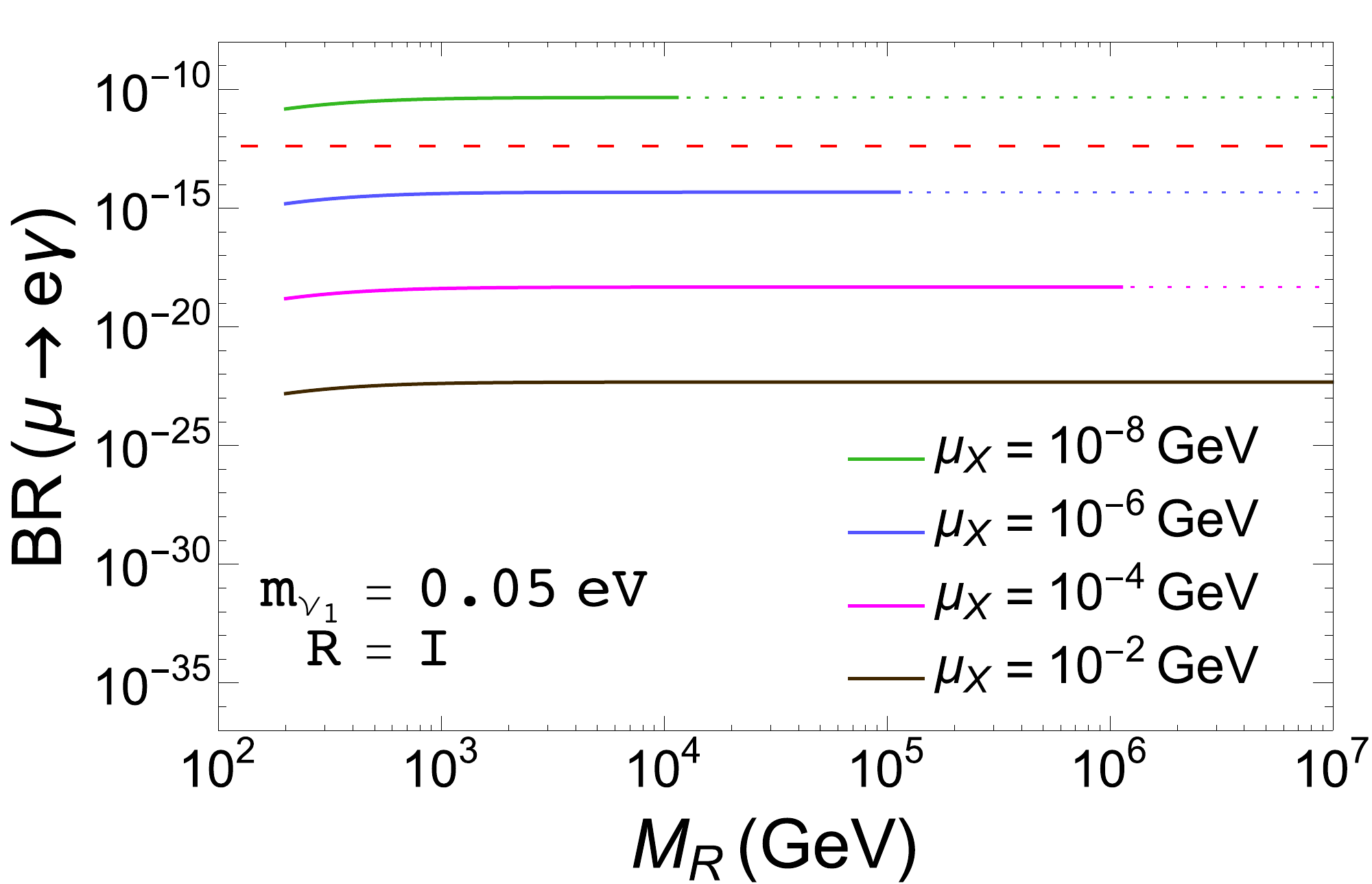}
\caption{BR($H \to \mu\overline\tau$) (left panel) and BR($\mu \to e \gamma$) (right panel) as functions of $M_R$ for different values of $\mu_X=(10^{-8},10^{-6},10^{-4},10^{-2})$ GeV from top to bottom. 
In both panels, $m_{\nu_1} = 0.05$ eV and $R = \mathbb1$. 
The horizontal red dashed line denotes the current experimental upper bound on $\mu \to e \gamma$ and dotted lines non-perturbative $Y_\nu$ as defined in \eqref{Ymax15}.}\label{Htaumu_MR_degenerate_Casas}
\end{center}
\end{figure}
Next we study the sensitivity of the LFV rates to other choices of $\mu_X$. For this study we focus on the largest LFVHD rates, BR($H \to \mu\bar\tau$), and on the most constraining BR$(\mu \to e \gamma)$ rates, although similar qualitative results are found for the other channels.  
In \figref{Htaumu_MR_degenerate_Casas} we show the predictions for the LFV rates for different values of $\mu_X=(10^{-8},10^{-6},10^{-4},10^{-2})$ GeV. The other input parameters have been fixed here to $m_{\nu_1} = 0.05$ eV and $R = \mathbb1$. 
On the left panel of \figref{Htaumu_MR_degenerate_Casas} we see again the increase of BR($H \to \mu\overline\tau$)  as $M_R$ grows, which is more pronounced in the region where $M_R$ is large and $\mu_X$ is low, i.e., where the Yukawa couplings are large (see \eqref{CasasIbarraISS}). 
We have checked that, in this region, the dominant diagrams are by far the divergent diagrams (1), (8) and (10), and that the BR($H \to \mu\overline\tau$) rates  grow as $M_R^4$.
Diagrams (2)-(6) have relevant contributions to BR($H \to \mu\overline\tau$)  only for low values of the Yukawa couplings, while diagrams (7) and (9) are subleading. 
Again, this will be further explored using the MIA in \secref{sec:LFVHDMIA}.
We also observe that the LFV Higgs rates grow as $\mu_X$ decreases from $10^{-2}$ GeV to $10^{-8}$ GeV. 
However, not all the values of $M_R$ and $\mu_X$ are allowed, because they may generate non-perturbative Yukawa entries, as we have already said, expressed again in this figure by dotted lines. 
Therefore, the largest LFV Higgs rates that are permitted by our perturbativity requirements in \eqref{Ymax15} are approximately of BR($H \to \mu\overline\tau$) $\sim 10^{-9}$, obtained for $\mu_X = 10^{-8}$ GeV and $M_R \simeq 10^4$ GeV.
Larger values of $M_R$, for this choice of $\mu_X$, would produce Yukawa couplings that are not perturbative.

On the other hand, we must also pay attention to the predictions of BR($\mu \to e \gamma$) for this choice of parameters, because the present experimental upper bound on this ratio is quite constraining.
We explore this observable in the right panel of  \figref{Htaumu_MR_degenerate_Casas}, where  the dependence of BR($\mu \to e \gamma$) on $M_R$ is depicted for the same parameter choice as in the left panel. 
The horizontal red dashed line denotes again its  present bound of BR($\mu \to e \gamma$) $< 4.2 \times 10^{-13}$~\cite{TheMEG:2016wtm}. 
In addition to what we have already learned about the constant behavior of BR($\mu\to e\gamma$) with $M_R$, we also learn from this figure about the generic behavior with $\mu_X$, which leads to increasing LFV rates for decreasing $\mu_X$ values.
This latter behavior  is actually true for both LFV observables.
In particular, we see that too small values of $\mu_X \leq \mathcal O(10^{-8}~\rm{GeV})$ lead to BR($\mu \to e \gamma$) rates that are excluded by the experimental upper bound.
Taking this into account, the largest value of BR($H \to \mu\overline\tau$), for this choice of parameters that is allowed by the BR($\mu \to e \gamma$) upper bound  is around $10^{-12}$, which is obtained for $M_R \sim 10^5$ GeV and $\mu_X \sim 10^{-6}$ GeV.

\begin{figure}[t!]
\begin{center}
\includegraphics[width=0.49\textwidth]{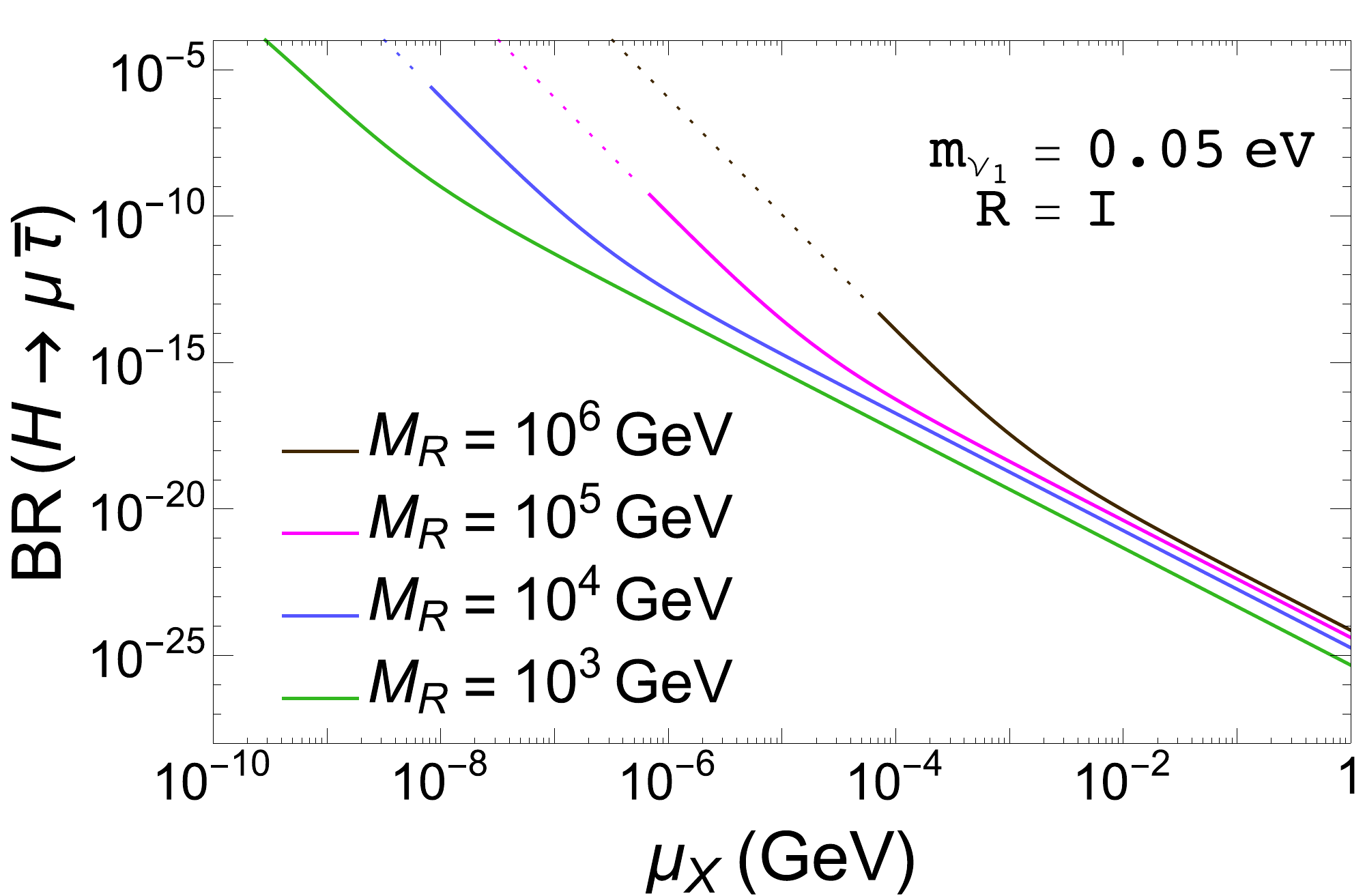}
\includegraphics[width=0.49\textwidth]{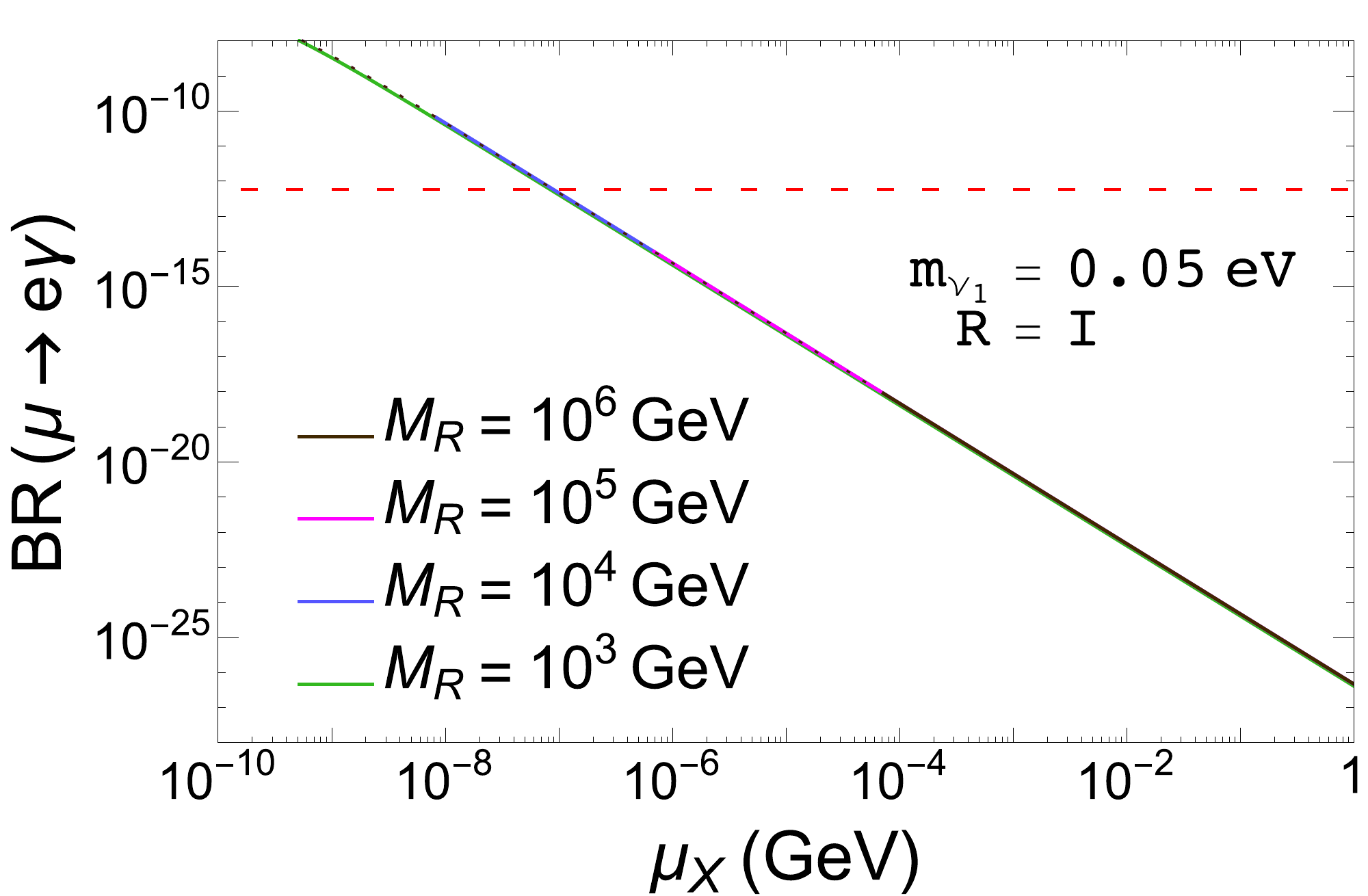}
\caption{Branching ratios of $H \to \mu\overline\tau$ (left) and $\mu \to e \gamma$ (right) as functions of $\mu_X$ for
different values of $M_R=(10^{6},10^{5},10^{4},10^{3})$ GeV from top to bottom. 
In both panels, $m_{\nu_1} = 0.05$ eV and $R = \mathbb1$. 
The horizontal red dashed line denotes the current experimental upper bound on $\mu \to e \gamma$ and dotted lines non-perturbative $Y_\nu$ as defined in \eqref{Ymax15}.}\label{Htaumu_muX_degenerate_Casas}
\end{center}
\end{figure}

The predictions of BR($H \to \mu\overline\tau$) and BR($\mu \to e \gamma$) as functions of $\mu_X$, for several values of $M_R$, $m_{\nu_1} = 0.05$ eV, and $R = \mathbb1$, are displayed in \figref{Htaumu_muX_degenerate_Casas}. As already seen in \figref{Htaumu_MR_degenerate_Casas}, both LFV rates decrease as $\mu_X$ grows; however, the functional dependence is not the same.
The LFV radiative decay rates decrease approximately as $\mu_X^{-2}$, in agreement with the approximate expression in \eqref{RadApprox}, while the LFVHD rates go as $\mu_X^{-4}$ in the regime of large Yukawa couplings.
For a fixed value of $\mu_X$, the larger $M_R$ is, the larger BR($H \to \mu\overline\tau$) can be, while the prediction for BR($\mu \to e \gamma$) is the same for all tested values of $M_R$. 
We have already learned this independence of the LFV radiative decays on $M_R$ from the previous figure, which can be easily confirmed on the right panel of \figref{Htaumu_muX_degenerate_Casas}, where all the lines for different values of $M_R$ are overplaced. 
We also see in this figure that the smallest value of $\mu_X$ that is allowed by the BR($\mu \to e \gamma$) upper bound is $\mu_X\sim 10^{-7}$ GeV, which is directly translated to a maximum allowed value of BR($H \to \mu\overline\tau$) $\sim 10^{-9}$, for $M_R \sim 10^4$ GeV.

\begin{figure}[t!]
\begin{center}
\includegraphics[width=0.49\textwidth]{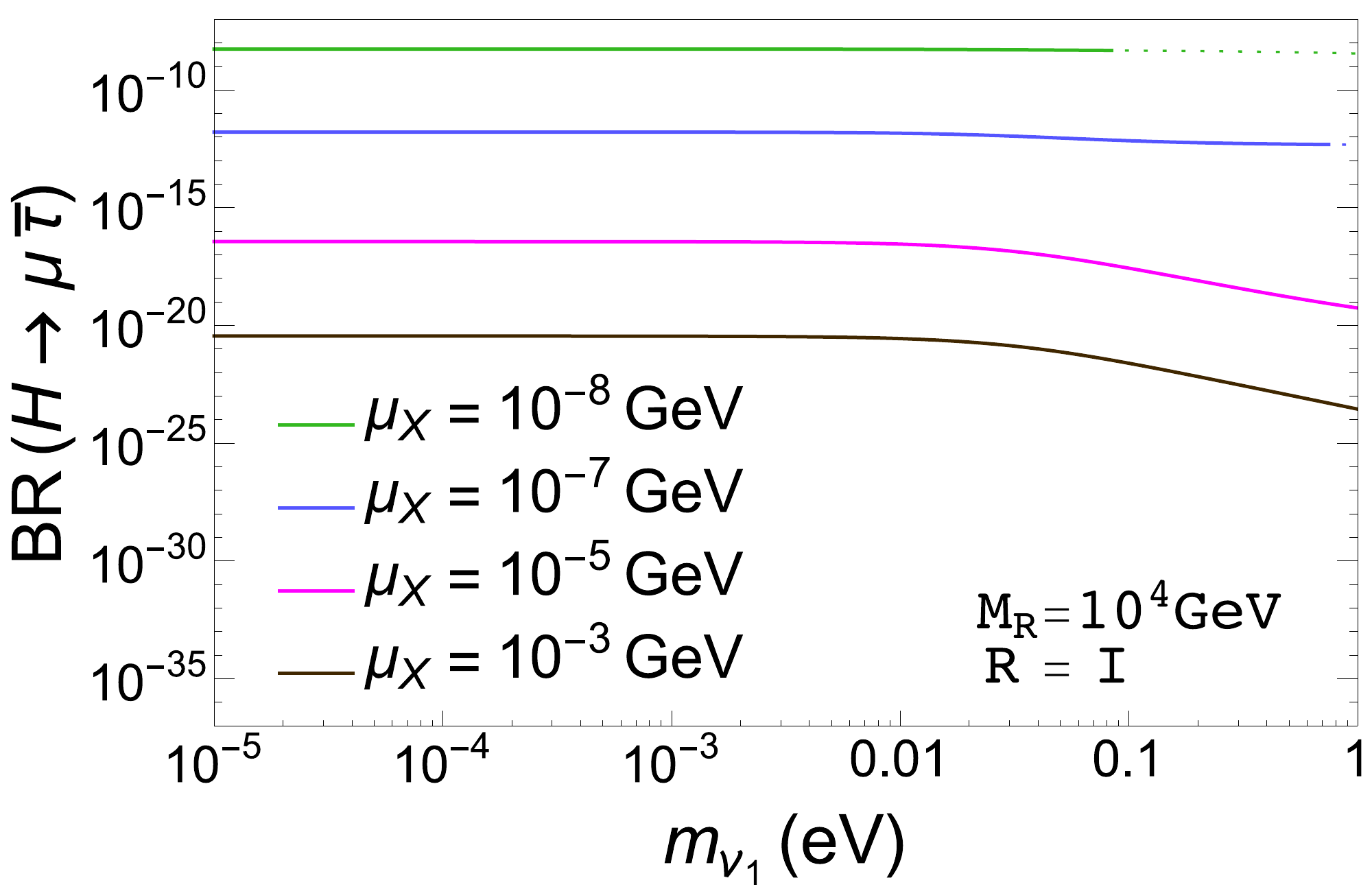} 
\includegraphics[width=0.49\textwidth]{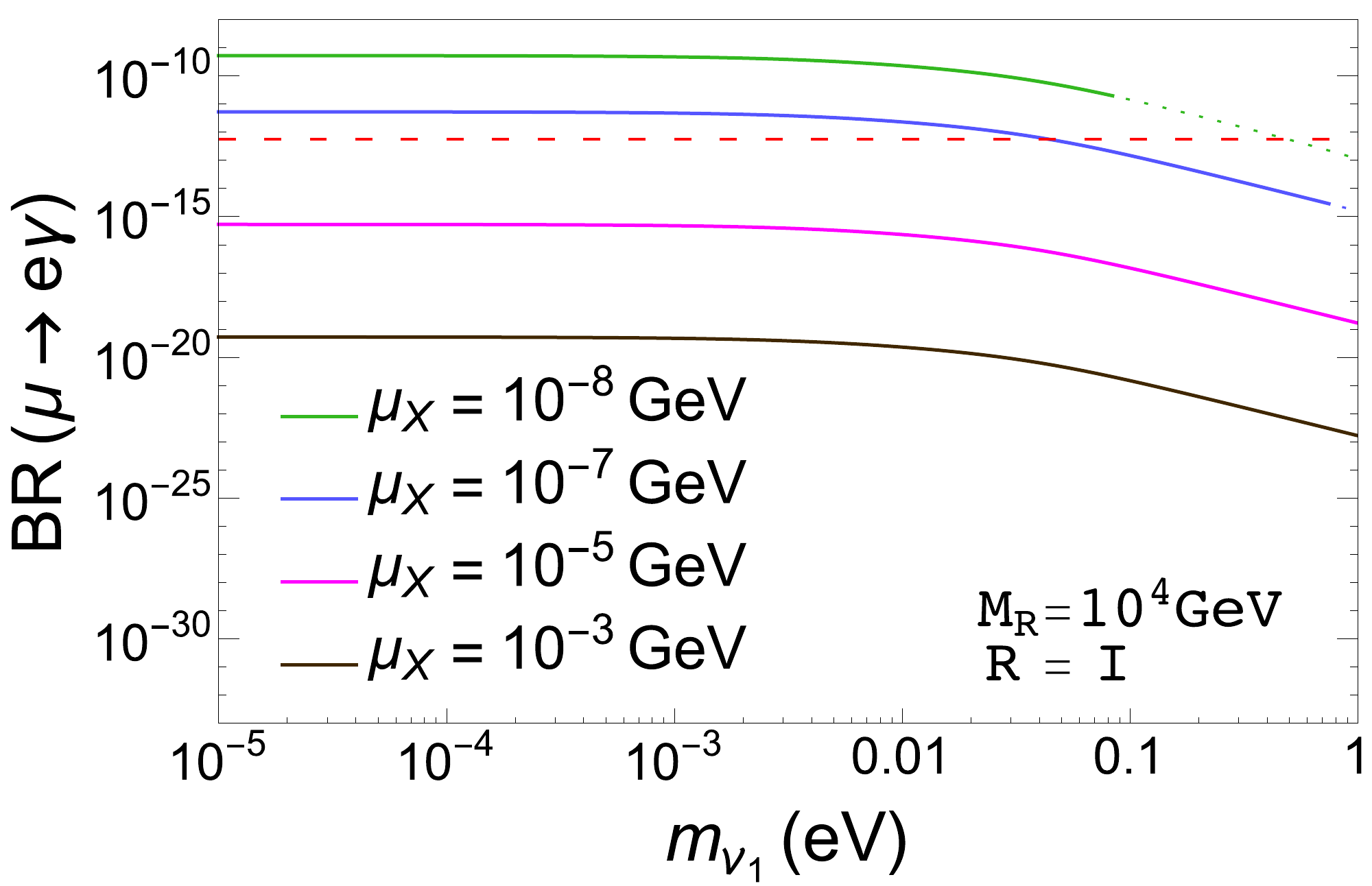}
\caption{Branching ratios of $H \to \mu\overline\tau$ (left panel) and $\mu \to e \gamma$ (right panel) as functions of $m_{\nu_1}$ for $M_R = 10^4$ GeV,  $R = \mathbb1$ and different values of $\mu_X=(10^{-8},10^{-7},10^{-5},10^{-3})$ GeV from top to bottom. 
The horizontal red dashed line denotes the current experimental upper bound on $\mu \to e \gamma$ and dotted lines non-perturbative $Y_\nu$ as defined in \eqref{Ymax15}.}\label{Htaumu_mnu1_degenerate_Casas}
\end{center}
\end{figure}

The dependence of BR($H \to \mu\overline\tau$) and BR($\mu \to e \gamma$) on the lightest neutrino mass $m_{\nu_1}$ is studied in \figref{Htaumu_mnu1_degenerate_Casas}, for different values of $\mu_X$ with $M_R = 10^4$ GeV and $R = \mathbb1$. 
For the chosen parameters, a similar dependence on $m_{\nu_1}$ is observed in both observables, in which there is a nearly flat behavior with $m_{\nu_1}$ until  $m_{\nu_1} \gtrsim$ 0.01 eV, where the LFV rates start to decrease.
The behavior of BR($\ell_m\to \ell_k\gamma$) with $m_{\nu_1}$ can be understood again from \eqref{RadApprox}. 
In this simplified case of real $R$ and $U_{\rm PMNS}$ matrices, and degenerate $M_R$ and $\mu_X$, we find the following simple expression for the non-diagonal $\big(Y_\nu^{} Y_\nu^\dagger\big)_{km}$ elements after using \eqref{CasasIbarraISS}:
\begin{equation}
\frac{v^2\Big(Y_\nu Y_\nu^\dag\Big)_{km}}{M_R^2}\approx
\left\{\begin{array}{ll}
\displaystyle{\frac{1}{\mu_X}\Big(U_{\rm PMNS} \sqrt{\Delta m^2}~ U_{\rm PMNS}^T\Big)_{km}} & ,\,{\rm for}~m_{\nu_1}^2\ll|\Delta m^2_{ij}| \,,\\\\
\displaystyle{\frac{1}{\mu_X}\frac{\Big(U_{\rm PMNS} \Delta m^2~ U_{\rm PMNS}^T\Big)_{km}}{2m_{\nu_1}} }& ,\,{\rm for}~ m_{\nu_1}^2\gg|\Delta m^2_{ij}|\,,
\end{array}\right.
\label{YnuYnudependence}
\end{equation}
where we have defined:
\begin{equation}\label{deltam}
\Delta m^2\equiv \rm{diag}(0, \Delta m_{21}^2,\Delta m_{31}^2)~,
\end{equation}
and we have expanded properly $m_{\nu_2}$ and $m_{\nu_3}$ in terms of $m_{\nu_1}$ and $\Delta m_{ij}^2$.
Using these equations, we conclude that the BR($\mu\to e\gamma$) rates have a flat behavior
with $m_{\nu_1}$ for low values of $m_{\nu_1}\lesssim 0.01~\rm{eV}$, but they decrease with $m_{\nu_1}$ for larger values, explaining the observed behavior in \figref{Htaumu_mnu1_degenerate_Casas}.
Again, the predictions for the BR($H\to\mu\bar\tau$) rates do not follow exactly the same pattern as for the  BR($\mu\to e\gamma$).

\begin{figure}[t!] 
\begin{center}
\includegraphics[width=.6\textwidth]{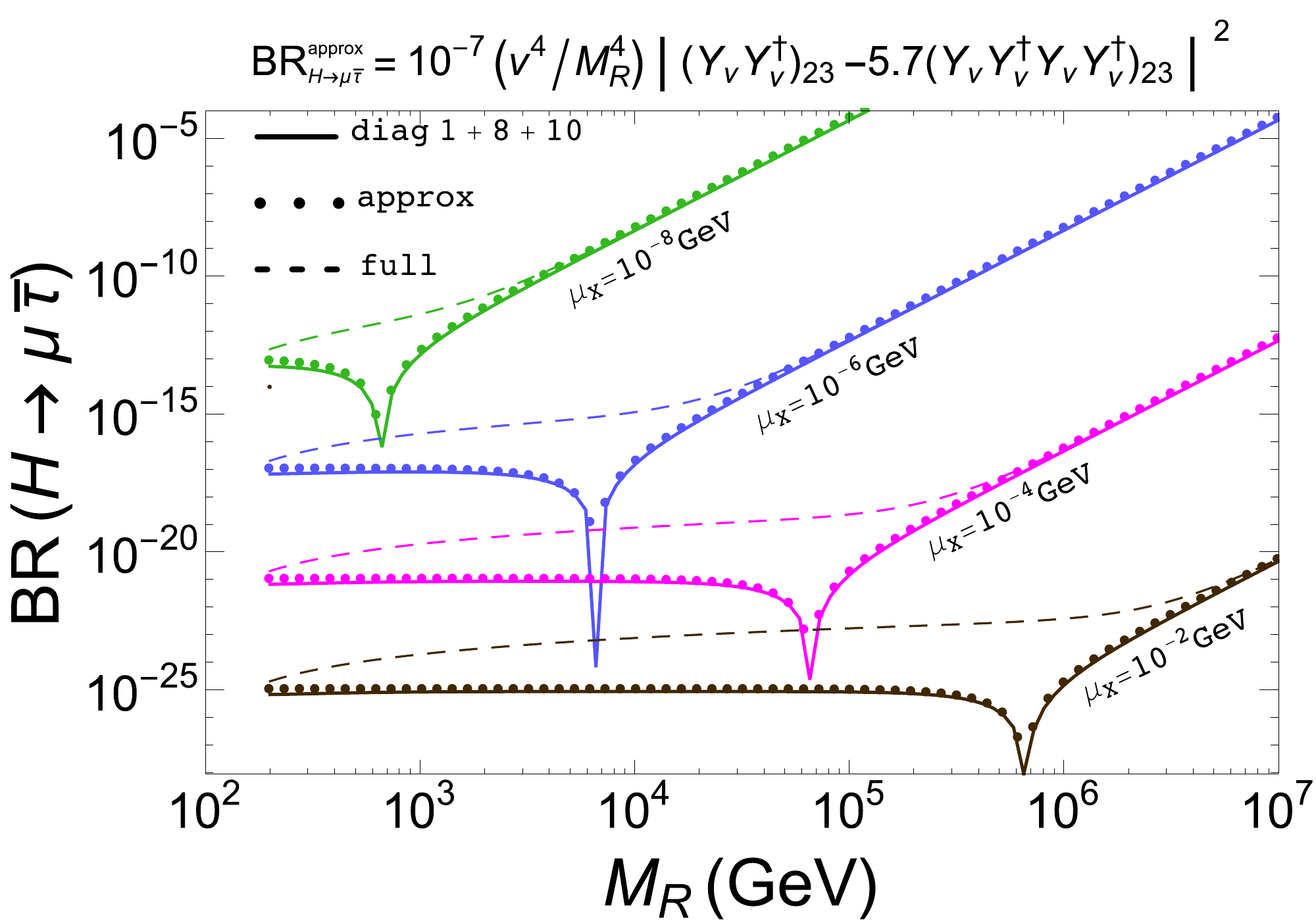} 
\caption{Comparison between the predicted rates for BR($H\to\mu\overline{\tau}$) computed with the full one-loop formulas (dashed lines), just the contributions from diagrams (1), (8), and (10) of \figref{diagsLFVHDphysbasis} (solid lines); and the approximate formula of \eqref{FIThtaumu} (dotted lines).
Here we set $m_{\nu_1} = 0.05$ eV and $R = \mathbb1$. }
\label{Htaumu_Fit_Casas}
\end{center}
\end{figure}

By taking into account all the behaviors learned above, we have tried to find an approximate simple formula that could explain the main features of the BR($H \to \mu\overline\tau$) rates. 
As we have already said, in contrast to what we have seen for the LFV radiative decays in \eqref{RadApprox}, a simple functional dependence being proportional to $|(Y_\nu^{} Y_\nu^\dagger)_{23}|^2$ is not enough to describe our results for the BR($H \to \mu\overline\tau$) rates.
Considering that, in the interesting region where the Yukawa couplings are large, the LFVHD rates are dominated by diagrams (1), (8) and (10), we have looked for a simple expression that could properly fit the contributions from these dominant diagrams. 
From this numerical fit we have found the following approximate formula:
\begin{equation}\label{FIThtaumu}
{\rm BR}^{\rm approx}_{H\to\mu\bar\tau}=10^{-7}\frac{v^4}{M_R^4}~\bigg|(Y_\nu^{} Y_\nu^\dagger)_{23}-5.7(Y_\nu^{} Y_\nu^\dagger Y_\nu^{} Y_\nu^\dagger)_{23}\bigg|^2,
\end{equation}
which turns out to work reasonably well. 
This particular analytical form in \eqref{FIThtaumu} will be justified later in \secref{sec:LFVHDMIA}.

In \figref{Htaumu_Fit_Casas} we show the predicted rates of BR($H\to\mu\overline{\tau}$) computed with the full one-loop formulas (dashed lines); taking just the contributions from diagrams (1), (8) and (10) of \figref{diagsLFVHDphysbasis} (solid lines); and using \eqref{FIThtaumu} (dotted lines). 
We see clearly that  \eqref{FIThtaumu} reproduces extremely well the contributions from  diagrams (1), (8), and (10) and approximates reasonably well the full rates, particularly in the fast growing as $M_R^4$ in the large $M_R$ region.

This approximate expression in \eqref{FIThtaumu} contains an extra contribution in the amplitude of $\mathcal O(Y_\nu^4)$ with respect to the one for radiative decays in \eqref{RadApprox}. By using again the parametrization in \eqref{CasasIbarraISS}, we obtain,
\begin{equation}\label{Ynu4dependence}
 \frac{v^2(Y_\nu^{} Y_\nu^\dagger Y_\nu^{} Y_\nu^\dagger)_{km}}{M_R^2}=\frac{M_R^2}{v^2\mu_X^2}\Big(U_{\rm PMNS} \Delta m^2~ U_{\rm PMNS}^T\Big)_{km}\,.
\end{equation}
Thus, we can clearly see from the above result that the second contribution in \eqref{FIThtaumu} is the one that dominates the LFVHD rates at large $M_R$ and low $\mu_X$, i.e., at large Yukawa couplings, and, indeed, it reproduces properly the behavior of BR$(H\to \mu\overline\tau)\propto M_R^4/\mu_X^{4}$ in this limit.  
It is also independent of $m_{\nu_1}$, explaining the flat behavior in \figref{Htaumu_mnu1_degenerate_Casas} for low values of $\mu_X$.
Moreover, since the two contributions in \eqref{FIThtaumu} have opposite signs, they  interfere destructively, leading to dips in the contribution from these diagrams to the full decay rates, as seen in \figref{Htaumu_Fit_Casas}.
As we said, the particular choice for the fitting functions will become clear from our posterior analysis in \secref{sec:LFVHDMIA}, where we will study in full analytical detail this observable by applying the MIA.

\begin{figure}[t!]
\begin{center}
\includegraphics[width=.49\textwidth]{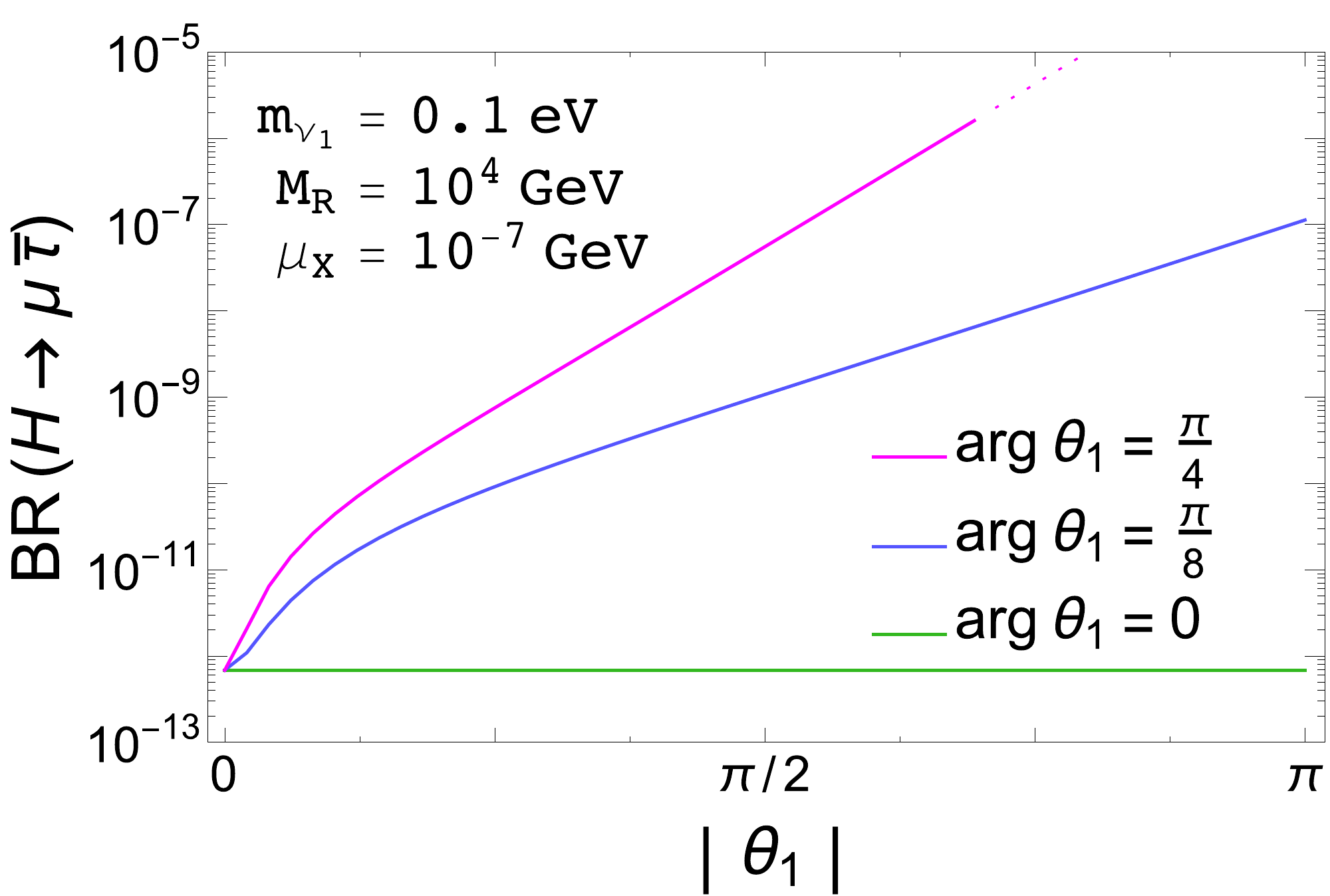} 
\includegraphics[width=.49\textwidth]{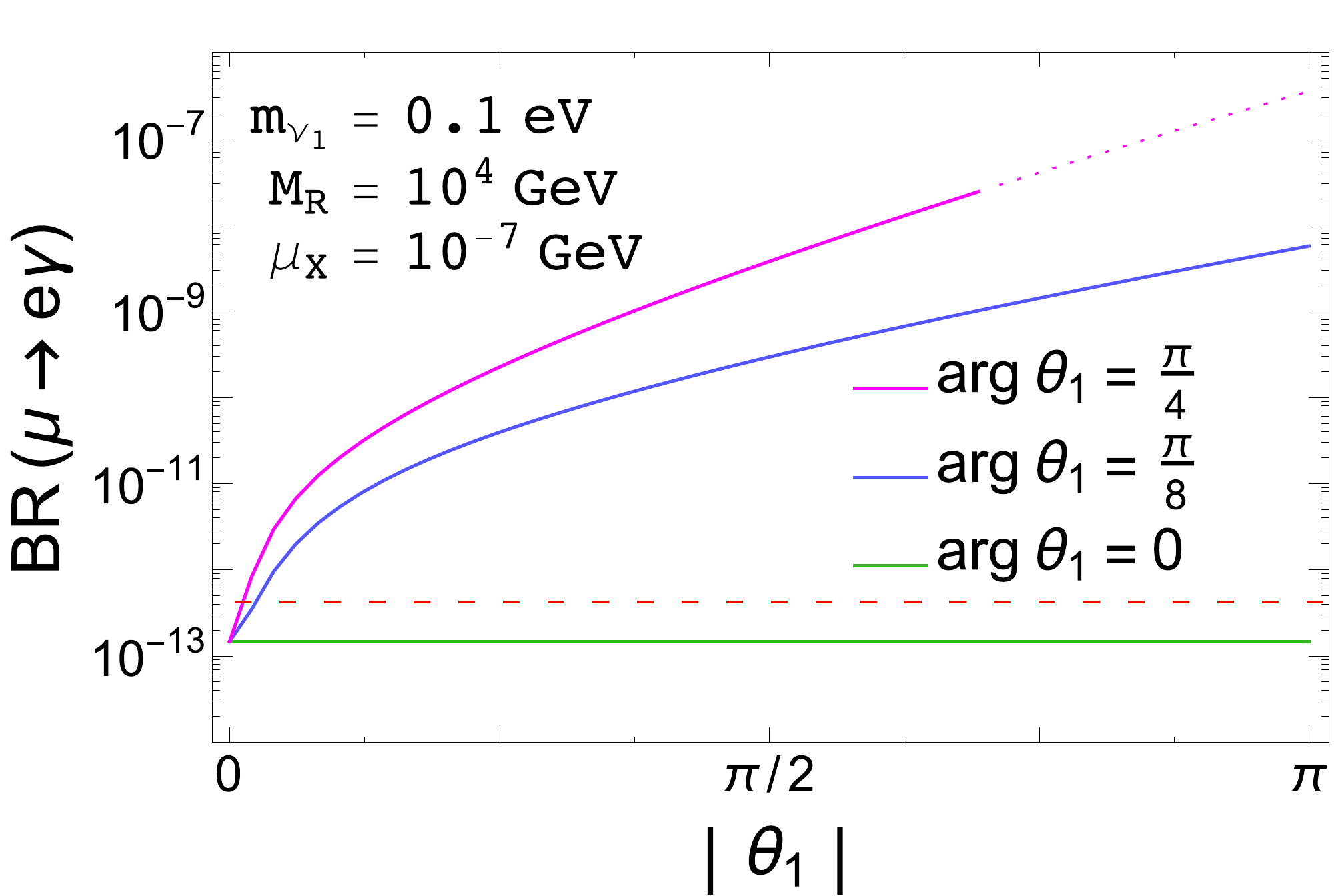}
\caption{Branching ratios of $H \to \mu\overline\tau $ (left panel) and $\mu \to e \gamma$ (right panel) as functions of $|\theta_1|$ for $M_R = 10^4$ GeV, $\mu_X = 10^{-7}$ GeV, $m_{\nu_1} =$ 0.1 eV and different values of arg$\theta_1$. 
The horizontal red dashed line denotes the current experimental upper bound on BR($\mu \to e \gamma$) and dotted lines represent non-perturbative $Y_\nu$ as defined in \eqref{Ymax15}.}\label{Htaumu_theta1_degenerate_Casas}
\end{center}
\end{figure}
Next, we study the effects of taking $R\neq\mathbb1$.
We display in \figref{Htaumu_theta1_degenerate_Casas} the dependence of the $H \to \mu\overline\tau$ and $\mu \to e \gamma$ decay rates on $|\theta_1|$ for different values of arg$\theta_1=0,\pi/8, \pi/4$, with $M_R = 10^4$ GeV, $\mu_X = 10^{-7}$ GeV, and $m_{\nu_1} =$ 0.1 eV.
First, we wish to highlight the flat behavior of both LFV rates with $|\theta_1|$ for real $R$ matrix (arg$\theta_1$ = 0).
This is a direct consequence of the degeneracy of $M_R$ and $\mu_X$, since the LFV rates for the degenerate heavy neutrinos case are independent of $R$ if it is real. 
Once we abandon the real case and consider values of arg$\theta_1$ different from zero, a strong dependence on $|\theta_1|$ appears. 
The larger $|\theta_1|$ and/or arg$\theta_1$ are, the larger the LFV rates. 
However, only values of $|\theta_1|$ lower than $\pi/32$ with arg$\theta_1 = \pi/8$ in this figure are allowed by the BR($\mu \to e \gamma$) constraint, which allows us to reach values of BR($H \to \mu \bar \tau$) $\sim 10^{-12}$ at the most. 
We have also explored the dependence on complex $\theta_2$ and $\theta_3$ and we have reached similar conclusions as for $\theta_1$. 
Therefore, we conclude that, in the case of degenerate $M_R$ and $\mu_X$ matrices, choosing complex $\theta_{1,2,3}$ does not increase the allowed LFVHD rates respect to te previous $R=\mathbb1$ case. 

\begin{figure}[t!] 
\begin{center}
\includegraphics[width=.49\textwidth]{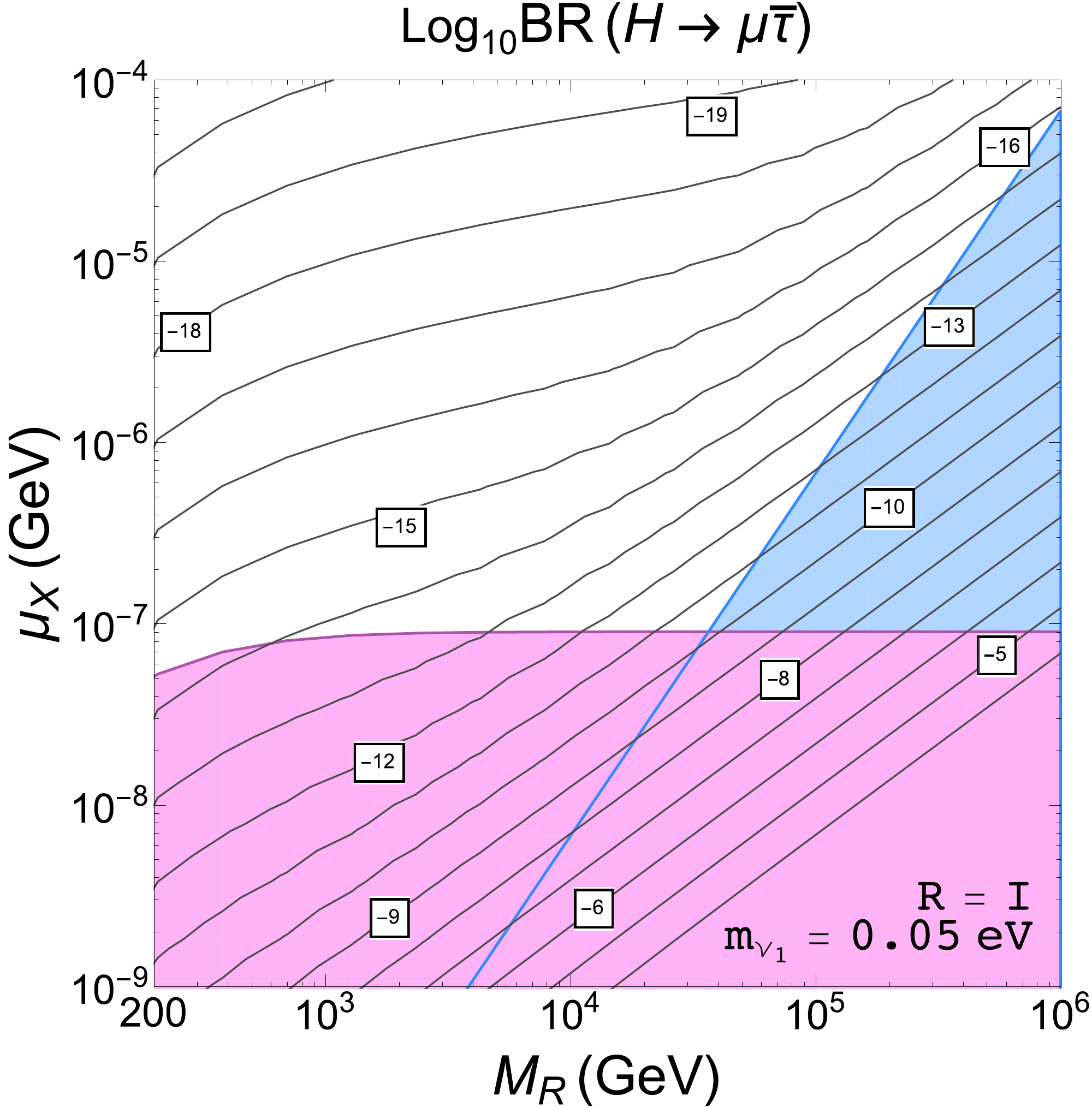} 
\includegraphics[width=.49\textwidth]{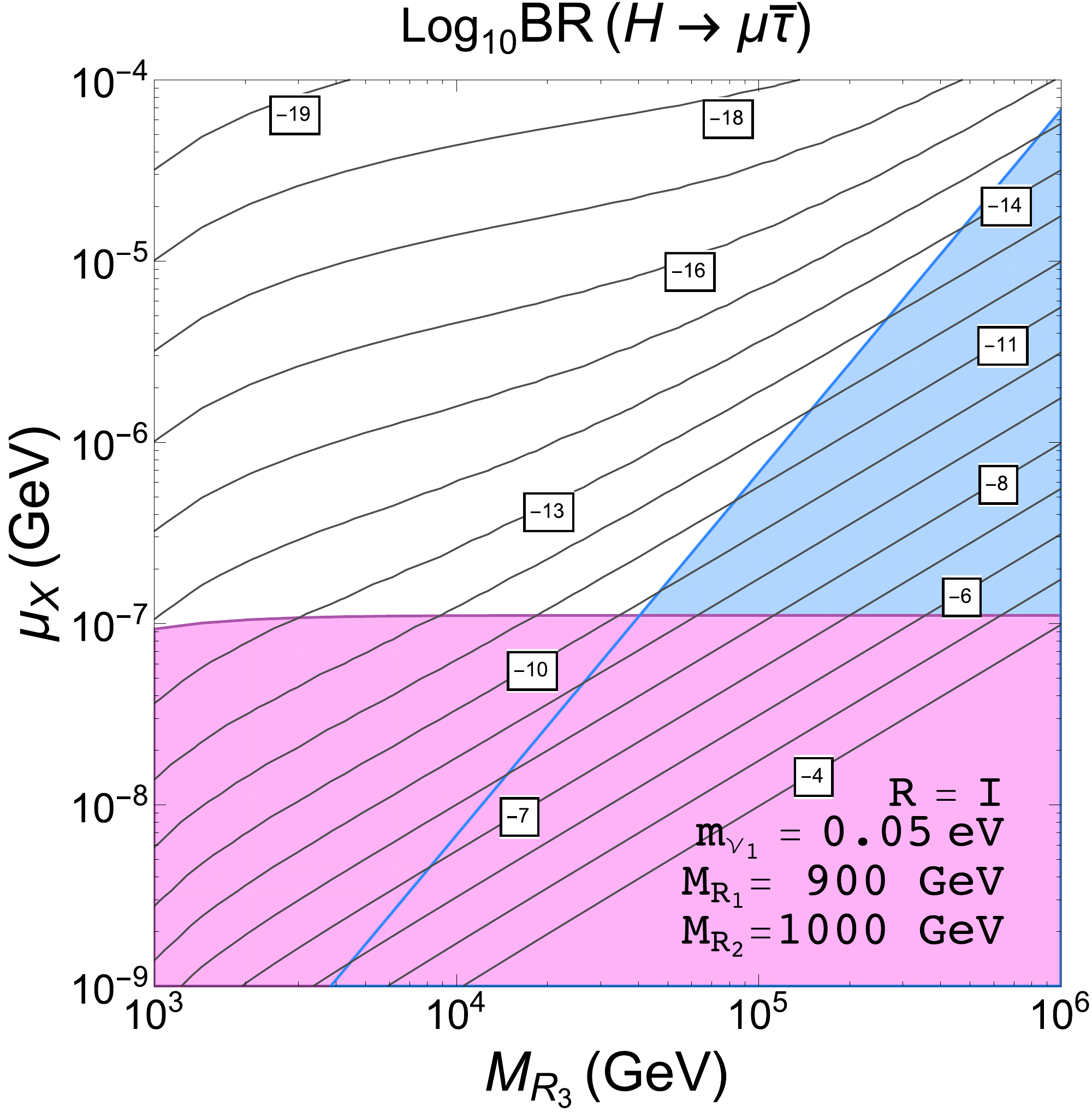} 
\caption{Left panel: Contour lines of BR($H \to \mu \bar \tau$) in the $(M_R,\mu_X)$ plane in the case of degenerate heavy neutrinos. 
Right panel: Contour lines of BR($H \to \mu \bar \tau$) in the $(M_{R_3},\mu_X)$ plane in the case of hierarchical heavy neutrinos with $M_{R_1}=900$ GeV and $M_{R_2}=1000$ GeV.
In both panels $R=\mathbb1$ and $m_{\nu_1}=0.05\,{\rm eV}$.  
Horizontal area in pink is excluded by the upper current bound on  $\mu \to e \gamma$ and the oblique area in blue is excluded by the perturbativity requirement for $Y_\nu$  in \eqref{Ymax15}.}
\label{Htaumu_ContourPlot_Casas}
\end{center}
\end{figure}

Once we have studied the behavior of all the LFV observables considered here with the most relevant parameters, we next present the concluding results for the maximum allowed LFV Higgs decay rates in the case of heavy degenerate neutrinos. 
The  left panel  in \figref{Htaumu_ContourPlot_Casas} shows the contour lines of BR($H \to \mu \bar \tau$) in the $(M_R, \mu_X)$ plane for $R=\mathbb1$ and $m_{\nu_1}=0.05\,{\rm eV}$. 
The horizontal area in pink is excluded by the upper bound on  BR$(\mu \to e \gamma)$.
The oblique area in blue is excluded by not respecting  the perturbativity requirement  of the neutrino Yukawa couplings, according to \eqref{Ymax15}. 
These contour lines summarize the previously learned behavior with $M_R$ and $\mu_X$, which lead to our findings for the largest values for the LFVHD rates that we localized at large $M_R$ and low $\mu_X$, i.e., in the bottom right-hand corner of the plot. 
As a conclusion  from this contour plot in the left panel of \figref{Htaumu_ContourPlot_Casas}, we learn that a maximum allowed LFVHD rates of approximately BR$(H \to \mu \bar \tau)\sim 10^{-10}$ are found for degenerate $M_R\sim 2\times 10^4\,{\rm GeV}$ and $\mu_X \sim   10^{-7} \, {\rm GeV}$. We have found similar conclusions for BR$(H \to e \bar \tau)$.

We can now move to the case of hierarchical heavy neutrinos. 
This case  refers here to the assumption of hierarchical masses among the heavy neutrino generations and it is implemented, still assuming diagonal $M_R$ and $\mu_X$ matrices,  but choosing instead hierarchical entries in the $M_R={\rm diag}(M_{R_1},M_{R_2},M_{R_3})$ matrix, see \figref{HeavyMasses}. 
As for the $\mu_X={\rm diag}(\mu_{X_1}, \mu_{X_2},\mu_{X_3})$ matrix that introduces the tiny splitting among the heavy masses in the same generation we choose them still to be degenerate, $\mu_{X_{1,2,3}}=\mu_X$.
We focus here on the normal hierarchy case $M_{R_1}<M_{R_2}<M_{R_3}$, since we have found similar conclusions for other hierarchies.

\begin{figure}[t!]
\begin{center}
\includegraphics[width=0.49\textwidth]{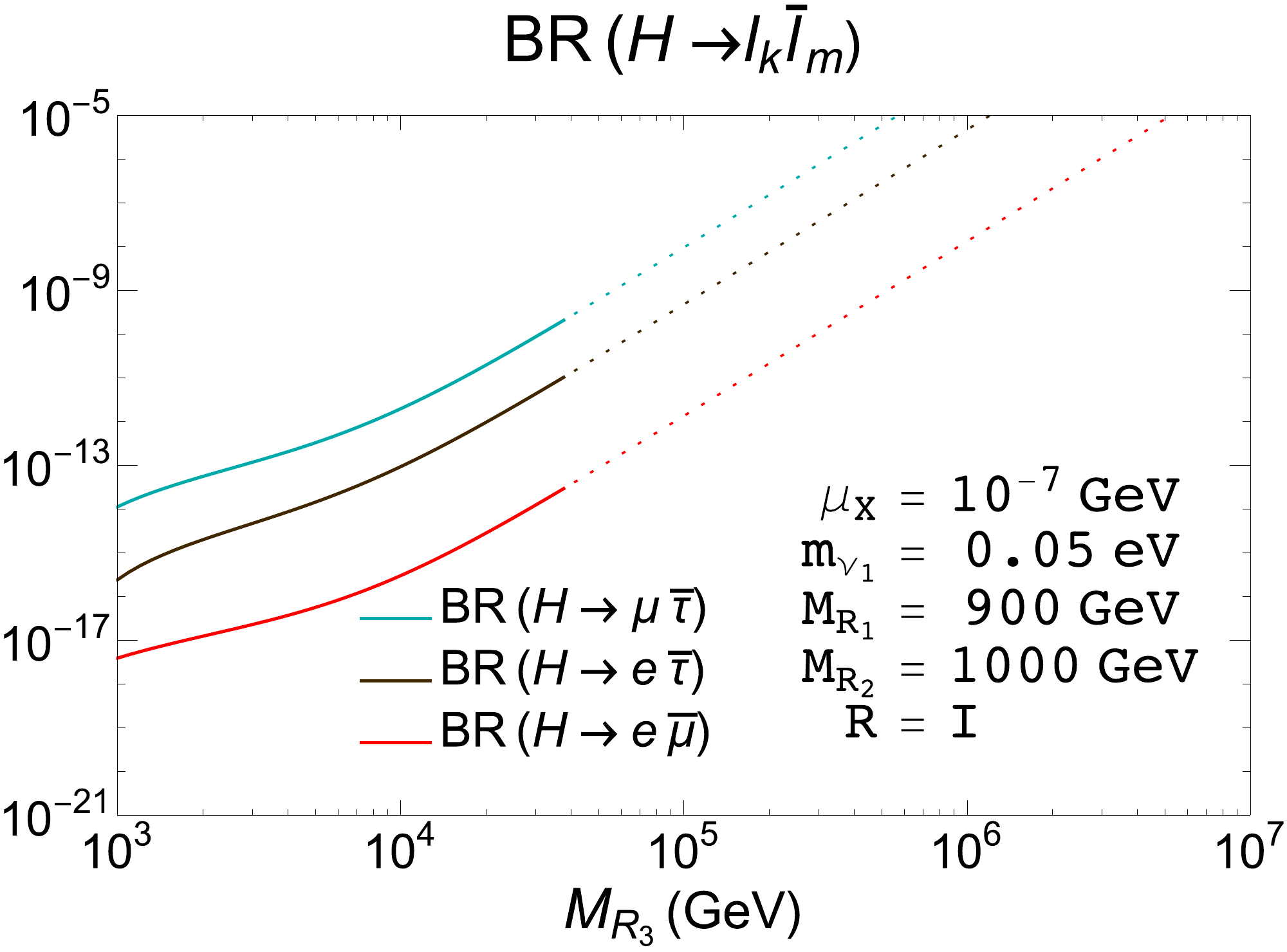} 
\includegraphics[width=0.49\textwidth]{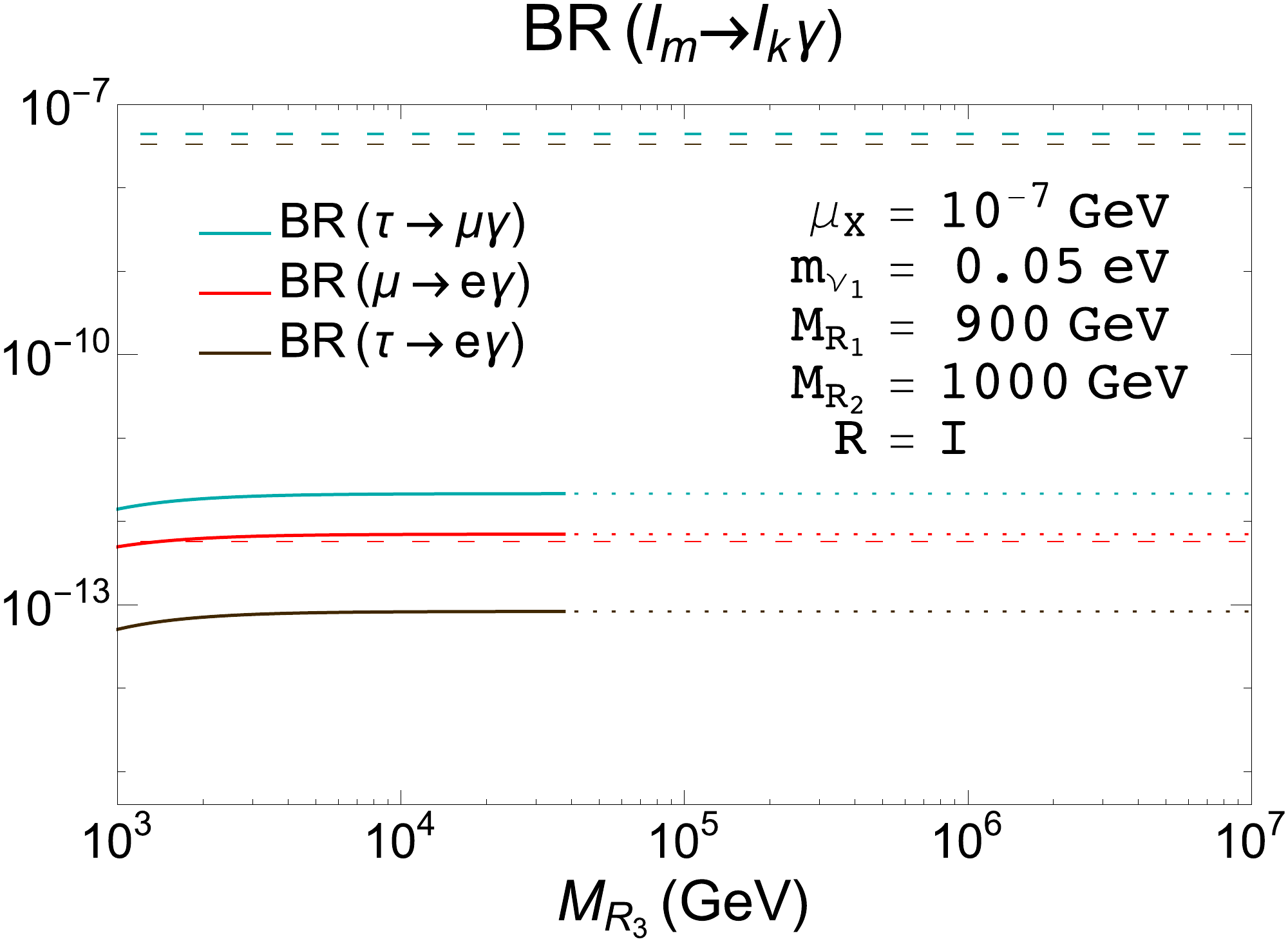}
\caption{
Predictions for the LFV Higgs (left panel) and radiative (right panel) decay rates as functions of $M_{R_3}$ in the hierarchical heavy neutrinos case with $M_{R_1}<M_{R_2}<M_{R_3}$. 
The other input parameters are set to $M_{R_1}=900$ GeV, $M_{R_2}=1000$ GeV, $\mu_X= 10^{-7} \, {\rm GeV}$, $m_{\nu_1}=0.05 \, {\rm eV}$ and $R=\mathbb1$.
The color code is as in \figref{LFVdegenerateCasas}.
}
\label{LFVhierarchicalCasas}
\end{center}
\end{figure}

The results of the LFV rates for both Higgs (left panel) and radiative (right panel) decays in this $M_{R_1}<M_{R_2}<M_{R_3}$ hierarchical case are shown in \figref{LFVhierarchicalCasas}.
This figure shows that the behavior of these rates in the hierarchical case with respect to the heaviest neutrino mass $M_{R_3}$ is similar to the previously found one for the degenerate case with respect to the common $M_{R}$. 
As before, the BR$(H \to \ell_k \bar \ell_m)$ rates grow fast with $M_{R_3}$ at large  $M_{R_3}> 3000\, {\rm GeV}$, whereas the BR$(\ell_m \to \ell_k \gamma)$ rates stay flat with $M_{R_3}$. 
We also see in this plot that, for the chosen parameters, the hierarchical scenario leads to larger BR($H\to\ell_k\bar\ell_m$) rates than the previous degenerate case. 
For instance, BR$(H \to \mu \bar \tau)$ reaches $10^{-9}$ at  $M_{R_3}=3\times 10^{4}\,{\rm GeV}$, to be compared with $10^{-11}$ at $M_{R}=3\times 10^{4}\,{\rm GeV}$ that we got in \figref{LFVdegenerateCasas} for the degenerate case.
We have found this same behavior of enhanced LFVHD rates by approximately one or two orders of magnitude  in the hierarchical case as compared to the degenerate case in most of the explored parameter space regions. 
This same enhancement can also be seen in the contour plot in the right panel of \figref{Htaumu_ContourPlot_Casas}, where the maximum allowed BR$(H \to \mu \bar \tau)$ rates reach larger values up to about  $10^{-9}$ for  $M_{R_1}=900\,{\rm GeV}$, $M_{R_2}=1000\,{\rm GeV}$, $M_{R_3}=3\times 10^{4}\,{\rm GeV}$, $\mu_X=10^{-7}\,{\rm GeV}$, $m_{\nu_1}=0.05$~eV and $R=\mathbb1$.\

\begin{figure}[t!]
\begin{center}
\begin{tabular}{cc}
\includegraphics[width=0.49\textwidth]{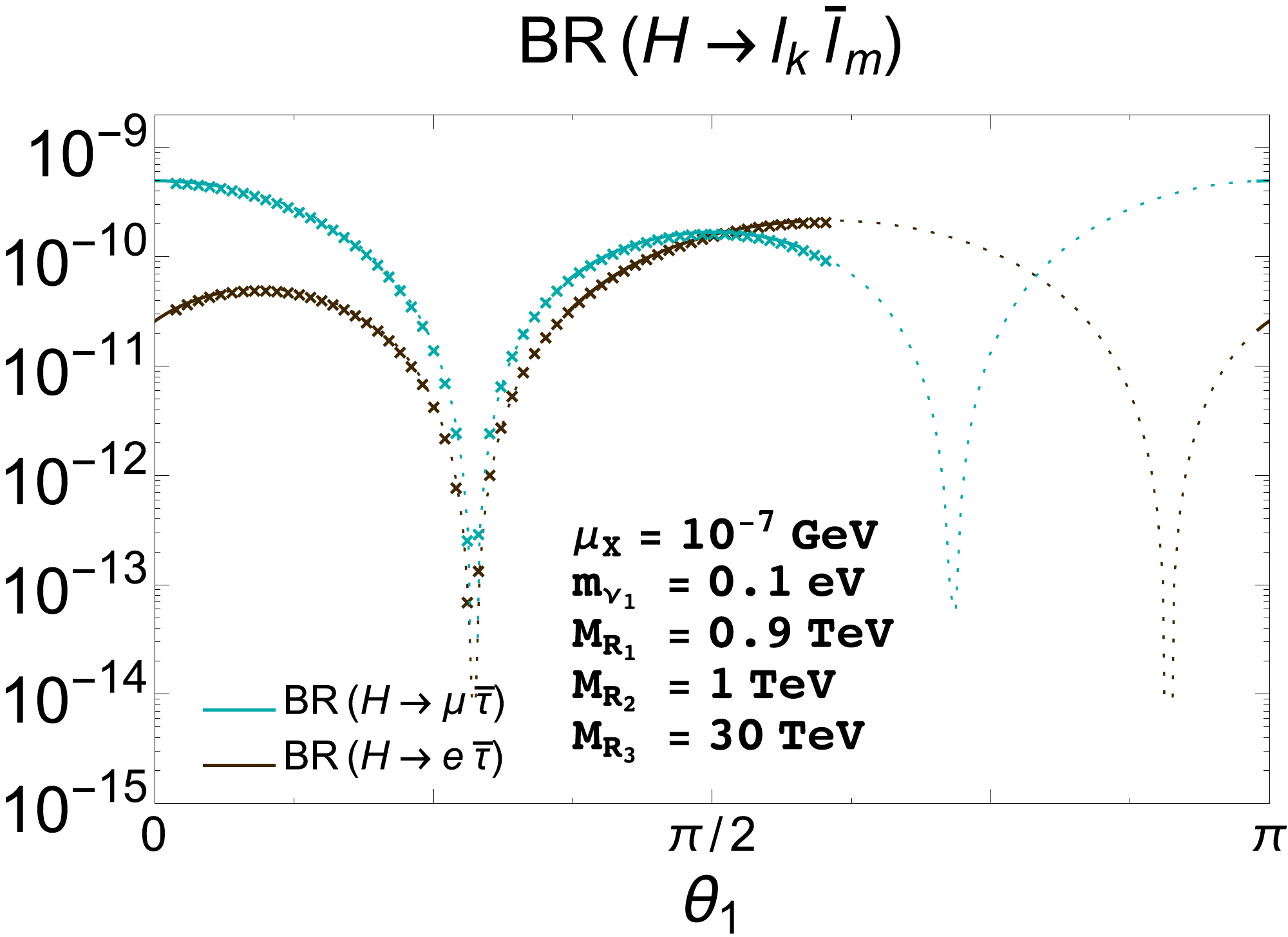} &
\includegraphics[width=0.49\textwidth]{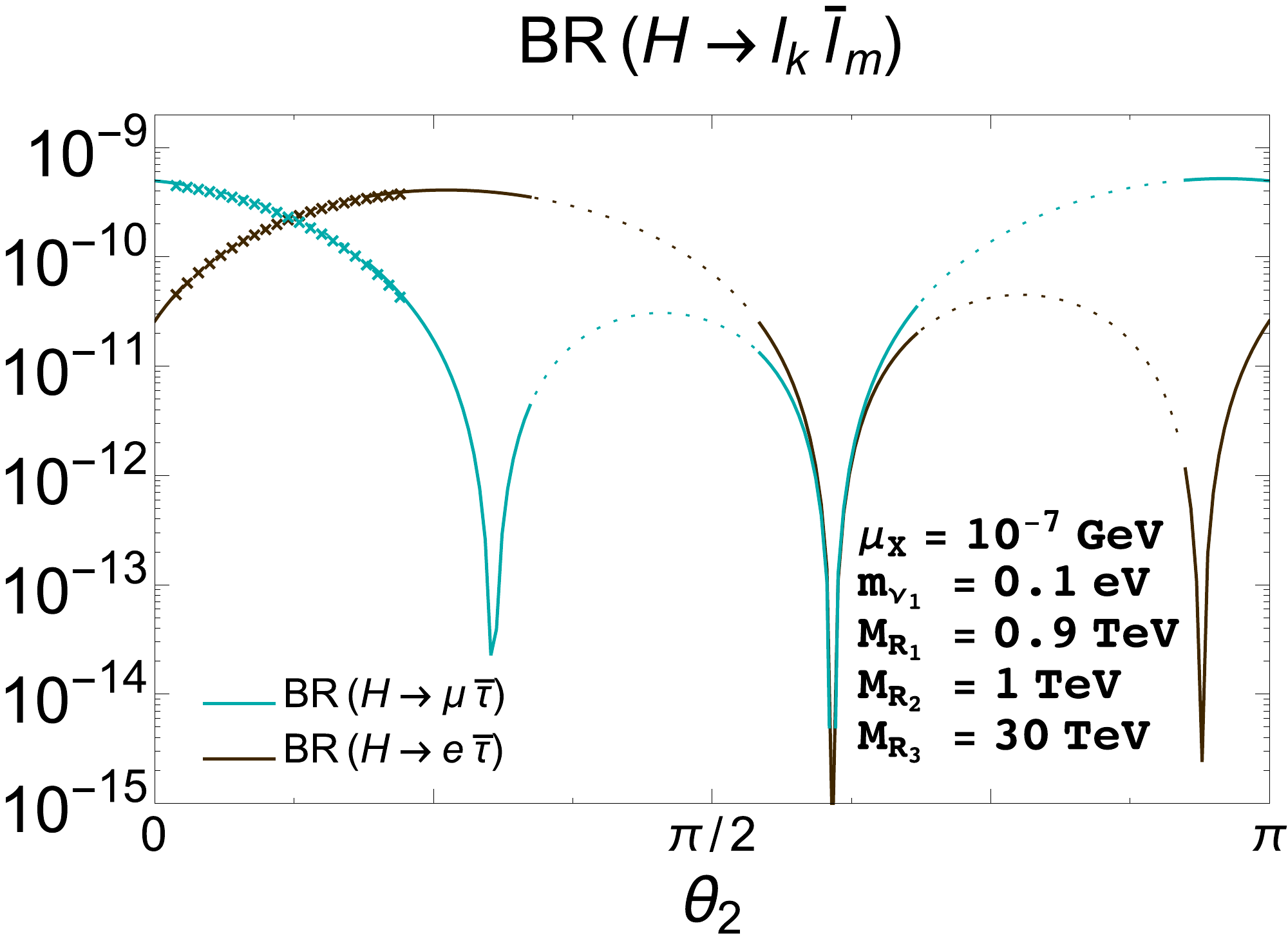}
\end{tabular}
\caption{Predictions for BR($H\to \mu\bar\tau$) (blue lines) and BR($H\to e\bar\tau$) (dark brown lines) rates as a function of real $\theta_1$ (left panel)
and $\theta_2$ (right panel). The other input parameters are set to $\mu_X=10^{-7}$ GeV, $m_{\nu_1}=0.1$ eV, $M_{R_1}=0.9$ TeV, $M_{R_2}=1$ TeV, $M_{R_3}=30$ TeV,
$\theta_2=\theta_3=0$ in the left panel and $\theta_1=\theta_3=0$ in the right panel. The dotted lines indicate non-perturbative neutrino Yukawa couplings and the crossed lines
are excluded by the present upper bound on BR($\mu\to e\gamma$). The solid lines are allowed by all the constraints.}
\label{LFVHD_Hierarchical_theta12_Casas}
\end{center}
\end{figure}
Finally, we study the effects of the $R$ matrix for this case of hierarchical heavy neutrinos. 
In contrast to the degenerate case,  there is a dependence on the $R$ matrix even if it is real.
Thus, we explore the behavior with the real $\theta_{1,2}$ angles.
We find that for this particular hierarchy of $M_{R_1}<M_{R_2}<M_{R_3}$, there is nearly an independence of $\theta_3$ but there is a clear dependence with $\theta_1$ and $\theta_2$, as it is illustrated in \figref{LFVHD_Hierarchical_theta12_Casas}. 
These plots also show that the BR$(H \to \ell_k \bar \ell_m)$ rates with $\theta_{1,2}\neq 0$ can indeed increase or decrease with respect to the reference $R=\mathbb1$ case. 
In particular, for  $0<\theta_{1}<\pi$ we find that BR($H \to \mu \bar \tau$) is always lower than for $R=\mathbb1$, whereas BR($H \to e \bar \tau$) can be one order of magnitude larger if  $\theta_{1}$ is near $\pi/2$. 
For the case of $0<\theta_{2}<\pi$, we find again that BR($H \to \mu \bar \tau$) is always lower than for $R=\mathbb1$, and BR($H \to e \bar \tau$) can be one order of magnitude larger than for $R=\mathbb1$ if  $\theta_{2}$ is near $\pi/4$. 
In this latter case, it is interesting to notice that in the region of $\theta_{2}$ close to $\pi/4$ BR($H \to e \bar \tau$) reaches the maximum value close to $10^{-9}$ and it is allowed by the constraints on the radiative decays and by the perturbativity condition. 
The results for the other decay channel BR($H \to e \bar \mu$) are not shown here because they again give much smaller rates, as in the degenerate case. 
We have also tried other choices for the hierarchies among the three heavy masses $M_{R_{1,2,3}}$, finding similar conclusions.

\subsection[LFVHD with the $\mu_X$ parametrization]{LFVHD with the $\boldsymbol{\mu_X}$ parametrization}\label{sec:LFVHDmuX}

Next, we explore the implications on LFV Higgs decays of going beyond the simplest previous hypothesis of diagonal $\mu_X$ and $M_R$ mass matrices in the ISS model.
Here, we will focus on the case of degenerate $M_R$ and will explore only the LFV Higgs decay
channels with the largest rates, namely, $H \to \mu \bar \tau$ and $H \to e \bar \tau$, looking for the maximum rates allowed by the radiative decays. 

In order to get an idea of how large the LFVHD rates could be, we first make a rough estimate of the expected maximal rates for the $H \to \mu \bar \tau$ channel by using our approximate formula of \eqref{FIThtaumu}, which is given just in terms of the neutrino Yukawa coupling matrix $Y_\nu$ and $M_R$.
 On the other hand, in order to keep the predictions for the radiative decays below their corresponding experimental upper bounds, we need to require a maximum value for the non-diagonal $(Y_\nu^{} Y_\nu^\dagger)_{ij}$ entries. 
 By using our approximate formula of \eqref{RadApprox} and the present bounds in \tabref{LFVsearch}, we obtain:
\begin{eqnarray}
v^2(Y_\nu^{} Y_\nu^\dagger)_{12}^{\rm max}/M_R^2 & \sim & 2.5\times10^{-5},\label{12max}\\
v^2(Y_\nu^{} Y_\nu^\dagger)_{13}^{\rm max}/M_R^2  & \sim  & 0.015,\label{13max}\\
v^2(Y_\nu^{} Y_\nu^\dagger)_{23}^{\rm max}/M_R^2 & \sim  & 0.017\label{23max}.
\end{eqnarray}
Then, in order to simplify our search, and given the above relative strong suppression of the 12 element, it seems reasonable to neglect it against the other off-diagonal elements. 
In that case,  by assuming $(Y_\nu^{} Y_\nu^\dagger)_{12}\simeq 0$ we have
\begin{equation}
(Y_\nu^{} Y_\nu^\dagger Y_\nu^{} Y_\nu^\dagger)_{23} \simeq  (Y_\nu^{} Y_\nu^\dagger)_{22} (Y_\nu^{} Y_\nu^\dagger)_{23}+ (Y_\nu^{} Y_\nu^\dagger)_{23} (Y_\nu^{} Y_\nu^\dagger)_{33},
\end{equation} 
and the approximate formula of \eqref{FIThtaumu} can then be rewritten as follows:
\begin{equation}
 {\rm BR}^{\rm approx}_{H\to\mu\bar\tau}=10^{-7}~\Big|\frac{v^2}{M_R^2}(Y_\nu^{} Y_\nu^\dagger)_{23}\Big|^2~\Big| 1-5.7 \Big( (Y_\nu^{} Y_\nu^\dagger)_{22}  +  (Y_\nu^{} Y_\nu^\dagger)_{33}\Big) \Big|^2.
 \end{equation} 
 
This equation clearly shows that the maximal BR($H\to\mu\bar\tau$) rates are obtained for the maximum allowed values of $(Y_\nu^{} Y_\nu^\dagger)_{23}$, $(Y_\nu^{} Y_\nu^\dagger)_{22}$, and $(Y_\nu^{} Y_\nu^\dagger)_{33}$. 
Thus, before going to any specific assumption for the $Y_\nu$ texture we can already conclude on these maximal rates, by setting the maximum allowed value for $v^2(Y_\nu^{} Y_\nu^\dagger)_{23}^{\rm max}/M_R^2$ to that given in \eqref{23max} and fixing the values of $(Y_\nu^{} Y_\nu^\dagger)_{22}$ and $(Y_\nu^{} Y_\nu^\dagger)_{33}$ to their maximum allowed values that are implied by our perturbativity condition in \eqref{Ymax15}.
This leads to our approximate prediction for the maximal rates:
\begin{equation}
 {\rm BR}^{\rm max}_{H\to\mu\bar\tau} \simeq 10^{-5}\,.
 \end{equation} 
We found similar conclusions for the $H\to e\bar\tau$ channel. 

For the purpose of reaching these large rates, we find more useful and effective to use the $\mu_X$ parametrization instead of the Casas-Ibarra one. 
As explained in \chref{PhenoLFV}, the main advantage when using this parametrization is that it allows us to consider $M_R$ and $Y_\nu$, the relevant parameters for the LFV processes, as the independent input parameters.
Furthermore, it makes very easy to focus our analysis directly on the TM and TE scenarios introduced in \tabref{TMscenarios}, which are designed to explore the parameter space directions where $\tau$-$\mu$ or $\tau$-$e$ transitions are maximized, respectively, whereas the $\mu$-$e$ transitions are extremely suppressed. 

\begin{figure}[t!]
\begin{center}
\includegraphics[width=.49\textwidth]{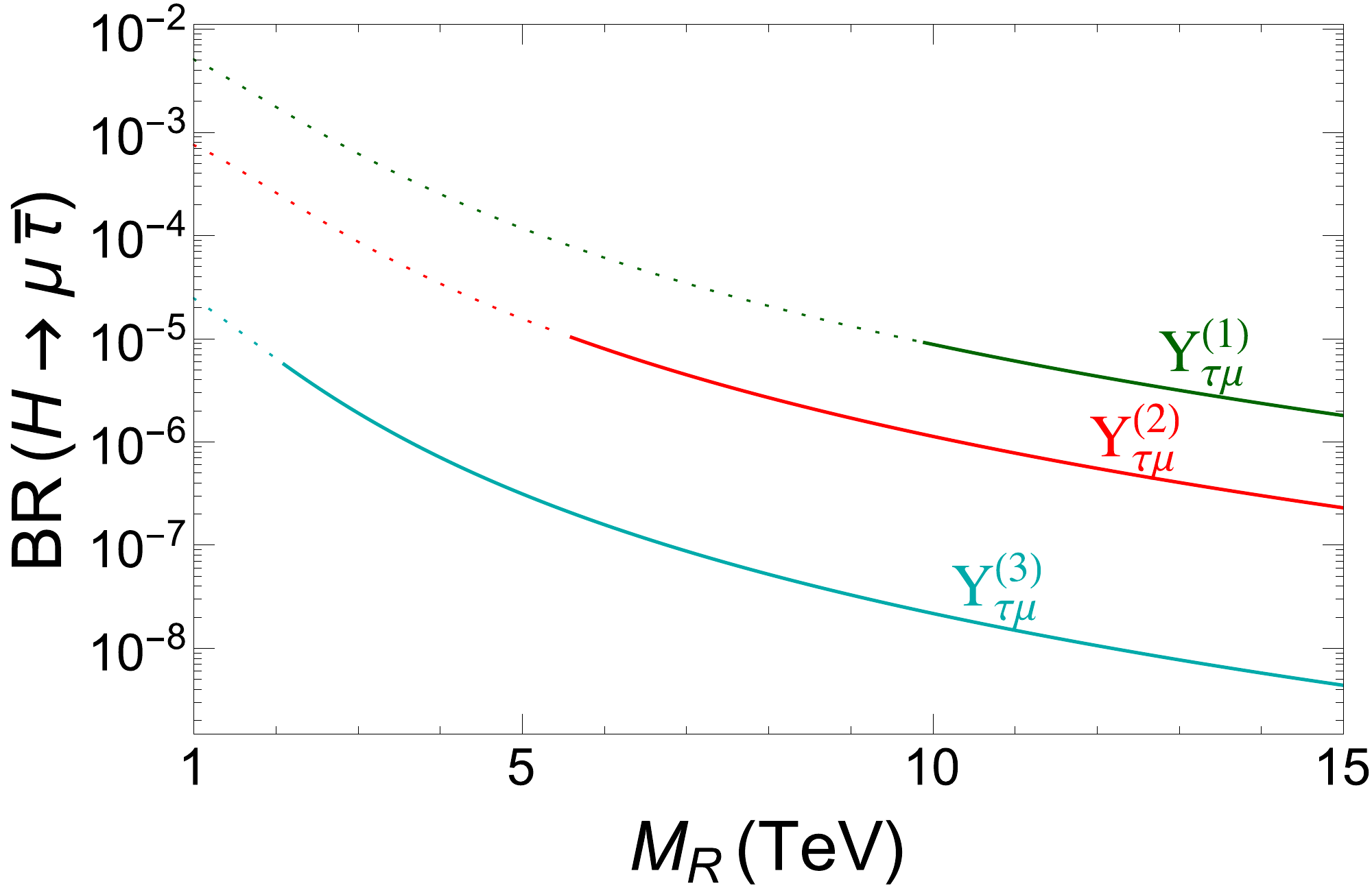}
\includegraphics[width=.49\textwidth]{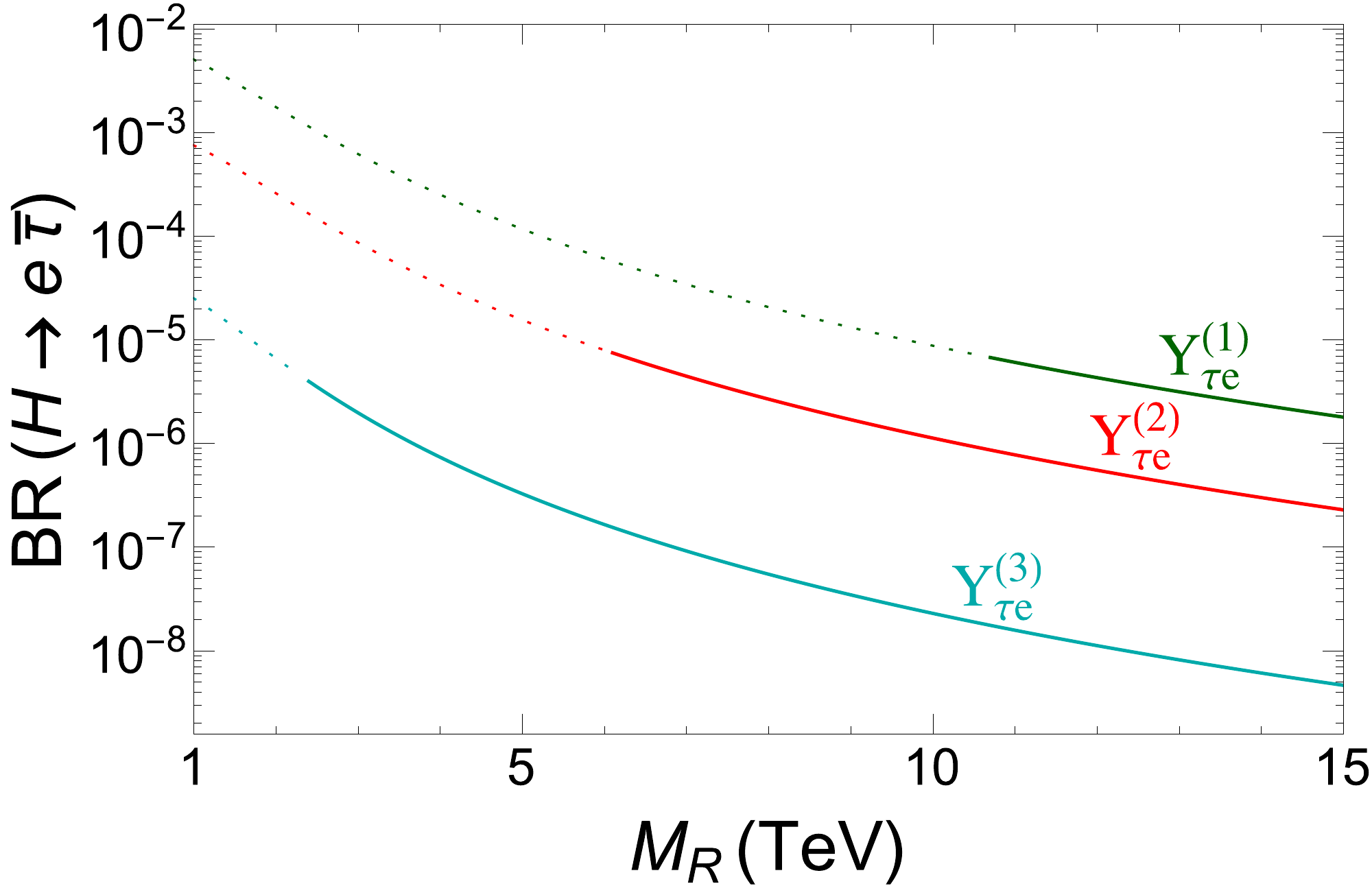}
\caption{Predictions for the LFVHD rates versus $M_R$ obtained when using the $\mu_X$ parametrization. 
Left panel:  BR$(H \to \mu \bar \tau)$  for $Y_{\tau \mu}^{(1)}$ (upper green line), $Y_{\tau \mu}^{(2)}$ (middle red line) and  $Y_{\tau \mu}^{(3)}$ (lower blue line) given in \tabref{TMscenarios} with $f=\sqrt{6\pi}$. 
Right panel:  BR$(H \to e \bar \tau)$ for the equivalent TE scenarios.
Solid (dotted) lines indicate input values allowed (disallowed) by upper bounds on radiative decays.}\label{LFVHD_degenerate_MUXparam}
\end{center}
\end{figure}

We present in \figref{LFVHD_degenerate_MUXparam} our predictions for the LFVHD as a function of the degenerate right-handed neutrino mass $M_R$ when using the $\mu_X$ parametrization for accommodating neutrino oscillation data.
Here, we show the results in the scenarios TM-5 $\big(Y_{\tau\mu}^{(1)}\big)$, TM-6 $\big(Y_{\tau\mu}^{(2)}\big)$ and TM-7 $\big(Y_{\tau\mu}^{(3)}\big)$ in \tabref{TMscenarios} for $f=\sqrt{6\pi}$, although similar results are found in other scenarios.
As before, we have used the full one-loop formulas, even though we have also checked that the approximate formula in \eqref{FIThtaumu}, and the equivalent one for $H\to e\bar\tau$ obtained by choosing the 13 entries instead of the 23 ones, gives a quite good estimate in the large $M_R$ region.
Notice again that the $\mu_X$ parametrization makes explicit the expected decoupling behavior with $M_R$, in contrast with the previous plots done with the Casas-Ibarra parametrization.

The main numerical conclusion from these plots is that in these scenarios one can indeed reach large LFVHD rates of the order of $10^{-5}$ and still be compatible with all the bounds from radiative decays, mainly $\tau\to\mu\gamma$ ($\tau\to e\gamma$) in the left (right) panel. 
The scenario with input $Y_{\tau \mu}$ ($Y_{\tau e}$) corresponding to lower $c_{\tau\mu}$ ($c_{\tau e}$) allows for lower $M_R$ values and vice versa. 
Thus, $Y_{\tau \mu}^{(1)}$ ($Y_{\tau e}^{(1)}$) leads to the maximum allowed BR$(H \to \mu \bar \tau)$ (BR$(H \to e \bar \tau)$) rates for $M_R$ around 10 TeV, $Y_{\tau \mu}^{(2)}$ ($Y_{\tau e}^{(2)}$) 6 TeV and $Y_{\tau \mu}^{(3)}$ ($Y_{\tau e}^{(3)}$) around 2 TeV.

 Summarizing, in this Section we have studied the LFV Higgs and radiative decays, by considering two different parametrizations to accommodate light neutrino data: the Casas-Ibarra and the $\mu_X$ parametrization.
 We have showed the advantages of using the latter for looking for large allowed LFVHD rates. 
 By doing this, we found larger rates for BR($H\to\mu\bar\tau$) and BR($H\to e\bar\tau$) of about $10^{-5}$ in the TM and TE scenarios, respectively. 
 Of course, in order to properly conclude on maximum allowed LFVHD rates, one must take into account a more complete set of constraining observables.
 We will do this at the end of this Chapter, in \secref{sec:LFVHDmax}.
 Furthermore, we have also learnt that LFVHD rates do not behave with the heavy $M_R$ mass as the LFV radiative decays, as can be seen from the plots presented in this Section and by comparing the respective approximated expressions in \eqrefs{FIThtaumu} and (\ref{RadApprox}).
We will study this particular behavior un more detail using the MIA in \secref{sec:LFVHDMIA}.

Nevertheless, before going to the full analysis of the right-handed neutrino contribution to the LFVHD rates, we want to explore these rates also in a different model, where the ISS is embedded in a supersymmetric context. 
In such a case, new important contributions could arise from diagrams with SUSY particles running in the loops and, as we will present next, they can considerably increase the LFVHD rates.

%
\section{LFV H decays in the SUSY-ISS model}
\label{sec:LFVHDSUSY}

We have seen in \chref{Models} that the ISS model can be easily embedded into a Supersymmetric context, leading to a new model that we refer to as the SUSY-ISS model.
Interestingly, previous studies have demonstrated that supersymmetric contributions usually enhance the LFV rates (see, for instance, Refs.~\cite{Deppisch:2004fa,Abada:2011hm,Abada:2012cq,Abada:2014kba,Arganda:2004bz}).
In particular, in the present SUSY-ISS model, we expect that a lower value of the heavy neutrino mass scale $M_R$ in the ISS, compared with the type-I seesaw, will enhance the slepton flavor mixing due to the RGE-induced radiative effects by the large neutrino Yukawa couplings, and this mixing will in turn generate via the slepton loops an enhancement in the LFVHD rates. 
On the other hand, new relevant couplings appear, like the neutrino trilinear coupling $A_{\nu}$, which for sneutrinos with $\mathcal{O}(1\,\mathrm{TeV})$ masses may lead to new loop contributions to LFVHD that could even dominate~\cite{Abada:2011hm}. 
This calls up for a new evaluation of the LFVHD rates in the SUSY-ISS model. 

Additionally, as we said in \secref{sec:LFVexp}, the CMS experiment saw an interesting excess in the $H\to\mu\tau$ channel after the LHC run-I~\cite{Khachatryan:2015kon} with a value of BR$(H\to\mu\tau)=8.4_{-3.7}^{+3.9}\times10^{-3}$ and a significance of $2.4\,\sigma$. 
Motivated by the fact that the  ISS model could not explain such a large ratio, we start by exploring the size of the new SUSY particle contribution to these LFVHD rates.
Of course, a more detailed analysis considering the full set of contributions in the SUSY-ISS model is needed and will come in a future work. 
Nonetheless, we expect that the dominant contributions will come from the new SUSY particles and, therefore, we  study first their impact on LFVHD rates.

\begin{figure}[t!]
\begin{center}
\includegraphics[width=\textwidth]{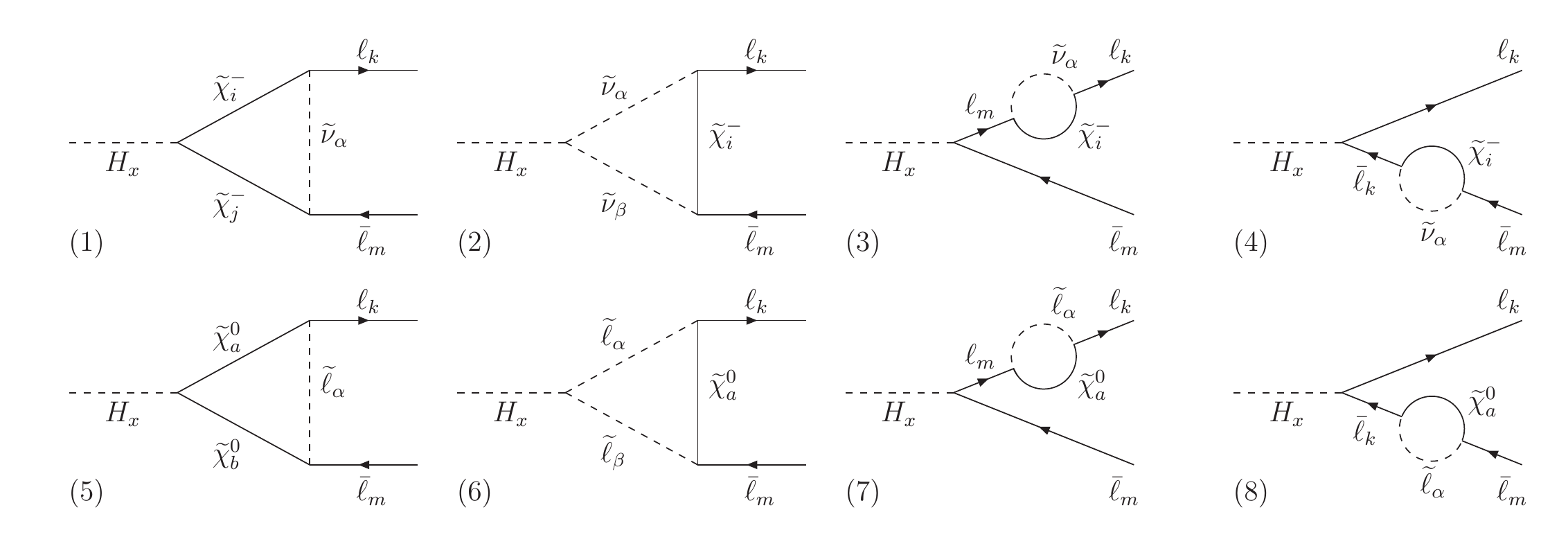}
\caption{One-loop supersymmetric diagrams contributing to the process $H_x \rightarrow \ell_k \bar \ell_m$.}\label{SUSYdiagrams}
\end{center}
\end{figure}

In this Section, then, we consider a full one-loop diagrammatic computation of all the supersymmetric loops within the SUSY-ISS model for $\mathrm{BR}(H_x\rightarrow \ell_k \bar \ell_m)$,
where $H_x$  refers in this Section to the three neutral MSSM Higgs bosons, $H_x=(h,H,A)$. 
This is in contrast to the previous estimate in Ref.~\cite{Abada:2011hm}, where an effective Lagrangian description of the Higgs mediated contributions to LFV processes was used, which was valid to capture just the relevant contributions at large $\tan\beta$, and where the mass insertion approach was used to incorporate, working in the electroweak basis, the flavor slepton mixing $(\Delta m_{\tilde L}^2)_{ij}$. 
However, an expansion up to the first order in the mass insertion approximation may not be enough for the type of scenarios studied here, due to the large value of the flavor-non-diagonal matrix entries. 
On the other hand, we are interested also in small and moderate $\tan\beta$ values and we also wish to explore more generic soft masses for the SUSY particles and scan over the relevant neutrino/sneutrino parameters, mainly $M_R$, $A_\nu$ and $m_{{\tilde \nu}_R}$, not focusing only on scenarios with universal or partially universal soft parameters nor fixing the relevant parameters to one particular value as in Ref.~\cite{Abada:2011hm}. 
Therefore, we perform the calculation instead in the more convenient mass basis for all the SUSY particles involved in the loops, i.e., the charged sleptons, sneutrinos, charginos, and neutralinos.

The decay amplitude for $H_x \rightarrow \ell_k \bar \ell_m$ can be written, similarly to \eqref{LFVHDamp}, as
\begin{equation}
i F_x = -i g \bar{u}_{\ell_k} (-p_2) (F_{L,x} P_L + F_{R,x} P_R) v_{\ell_m}(p_3)\,,\label{LFVHDampSUSY}
\end{equation}
where again $H_x=(h,H,A)$ and $p_1=p_3-p_2$ is the ingoing Higgs boson momentum.
The full LFVHD widths can be then obtained from \eqref{LFVHDwidth}, with the proper substitution of $m_H\to m_{H_x}$ for the Higgs masses and $F_{L/R}\to F_{L/R,x}$ in the form factors. 

Since we work in the mass basis, the set of diagrams contributing to the LFV Higgs decays is the same as in
the SUSY type-I seesaw model which was previously considered in Ref.~\cite{Arganda:2004bz}. 
We display this set of 8 diagrams, four diagrams with charginos and sneutrinos in the loops, and four more with neutralinos and charged sleptons, in \figref{SUSYdiagrams}.
The contributions of the SUSY diagrams are summed in $F_{L,x}$ and $F_{R,x}$ according to
\begin{equation}
F_{L,x} = \sum_{i=1}^{8} F_{L,x}^{(i)},\quad  F_{R,x} = \sum_{i=1}^{8} F_{R,x}^{(i)}\,.
\label{FormFactorsSUSY}
\end{equation}
We take the analytical expressions from Ref.~\cite{Arganda:2004bz} and properly adapt them to the SUSY-ISS model.
The resulting formulas are collected in \appref{App:LFVHD_SUSY}

As in the non-SUSY case, this observable does not exist at the tree level and, therefore, the full one loop contributions must give a finite result.
In the type-I seesaw SUSY model, we have checked that each diagram in \figref{SUSYdiagrams} gives a finite contribution to the $H_x\to\ell_k\bar\ell_m$ process when summing over all internal indexes, in agreement with Refs.~\cite{Arganda:2004bz,Arganda:2015uca}.
This is not the case in the SUSY-ISS model, where we have found that the new terms in the coupling factor $A_{R_{\alpha j}}^{(\ell)}$, see \eqref{SUSYISScouplings},  give rise to divergent terms in diagrams (3) and (4).
Nevertheless, these divergences cancel out when adding both diagrams, such that the full result is finite, as expected. 
Notice that the contributions from the sneutrino-chargino sector, adding diagrams (1)-(4), and the ones from the slepton-neutralino sector, adding diagrams (5)-(8), are finite separately and, therefore, it is legit to study both contributions separately, as we will do below. 

Next, we show the numerical results of the LFV decay rates of the lightest neutral Higgs boson, BR($h \to \tau \bar\mu$), as a function of the most relevant parameters for the full SUSY contribution to LFVHD, namely, $M_R$, $A_\nu$ and $m_{{\tilde \nu}_R}$. 
We will assume that the lightest CP-even Higgs boson, $h$, is the particle found by the LHC with a mass of 125 GeV, and explore, consequently, the process $h\to\tau\bar\mu$.
In order to ensure a Higgs boson mass in agreement with the experimental value, we will adjust the squark parameters, since they are irrelevant for the LFVHD studied here, and make sure that they lead to a supersymmetric spectrum allowed by ATLAS and CMS searches.
As in the non-SUSY case, we restrict ourselves to the case of heavy neutrinos and sneutrinos above the Higgs boson mass, with $M_R>m_h$, avoiding constraints from the invisible Higgs decay widths, and consider only real $U_{\rm PMNS}$ and mass matrices, making constraints from lepton electric dipole moments irrelevant.
Furthermore, this absence of CP violation makes BR($h\to\tau\bar\mu$)$=$BR($h\to\mu\bar\tau$) and, therefore, a factor of two should be added to our results when comparing with experimental data for BR($h\to\tau\mu$)$\equiv$ BR($h\to\tau\bar\mu$)$+$BR($h\to\mu\bar\tau$).

As in the previous section, we will also compute here, using the expressions in Ref.~\cite{Hisano:1995cp}, the LFV radiative decays as a good reference of the most relevant experimental constraints on LFV,  whose current upper limits at the $90\%$ C.L. are
\begin{align}
{\rm BR}(\mu\to e\gamma)&\leq 4.2\times 10^{-13}~\text{\cite{TheMEG:2016wtm}}\label{MUEGmax}\,,\\
{\rm BR}(\tau\to e\gamma)&\leq 3.3\times 10^{-8}~~\text{\cite{Aubert:2009ag}}\label{TAUEGmax}\,,\\
{\rm BR}(\tau\to \mu\gamma)&\leq 4.4\times 10^{-8}~~\text{\cite{Aubert:2009ag}}\label{TAUMUGmax}\,.
\end{align}
In the following plots, the points excluded by LFV radiative decays will be denoted by crosses, while triangles will represent the  allowed ones. 
We present here the predictions of BR($h \to \tau \bar \mu$) for three examples of the TM scenarios exposed in the \tabref{TMscenarios} that maximize $\tau$-$\mu$ transitions ensuring, at the same time, practically vanishing LFV in the $\mu$-$e$ sector, in particular, leading to BR$(\mu \to e \gamma)\sim 0$ and BR$(h \to e \bar \mu)\sim 0$.  
It should be noticed that in these TM scenarios LFV transitions in the $\tau$-$e$ sector are also substantially suppressed and, therefore, the most stringent radiative decay is that of the related LFV radiative decay $\tau \to \mu \gamma$. 
Although not shown here, we want to emphasize again that equivalent results are obtained for BR($h\to \tau\bar e$) in the TE scenarios.

\begin{figure}[t!]
\begin{center}
\includegraphics[width=0.475\textwidth]{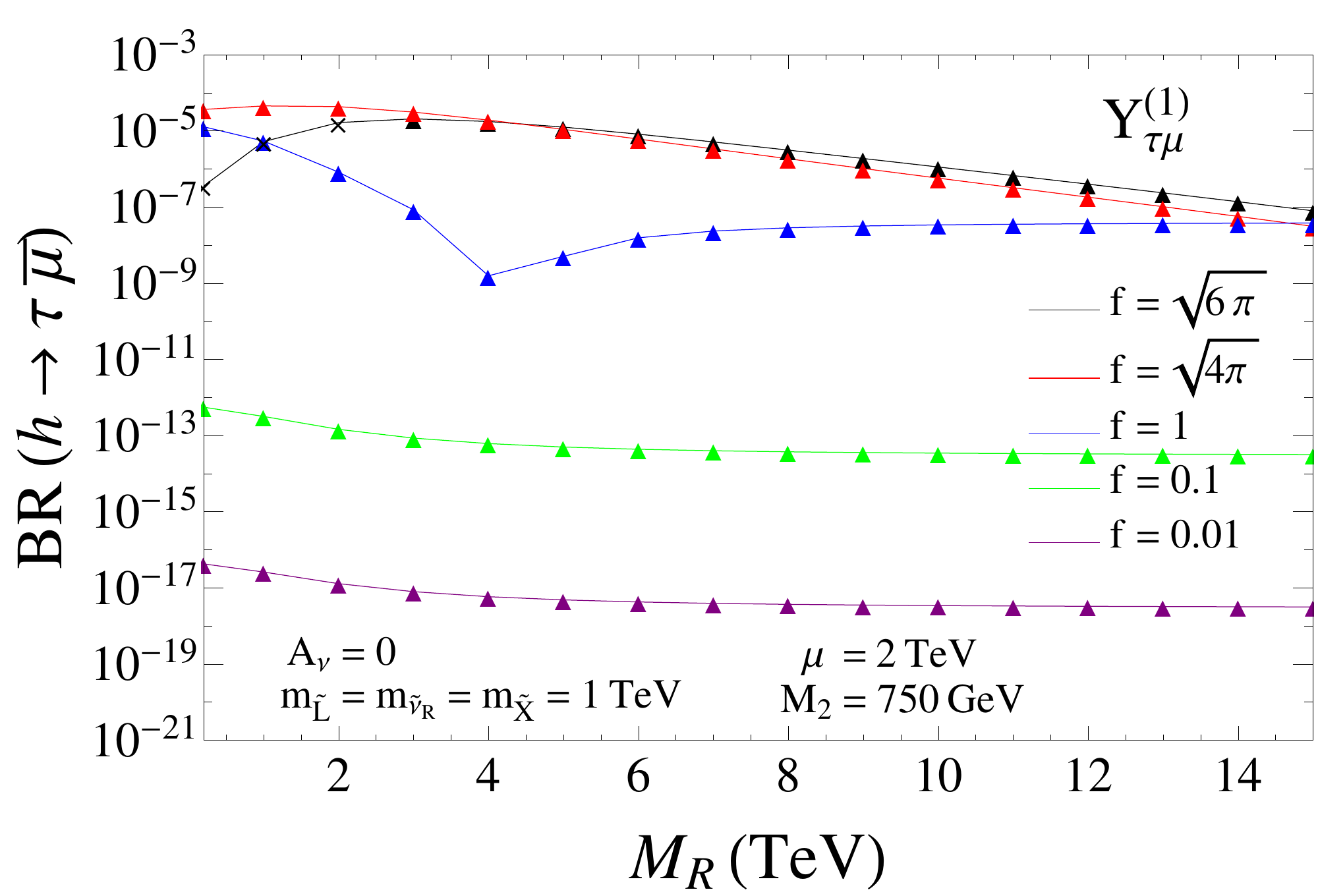}
\includegraphics[width=0.475\textwidth]{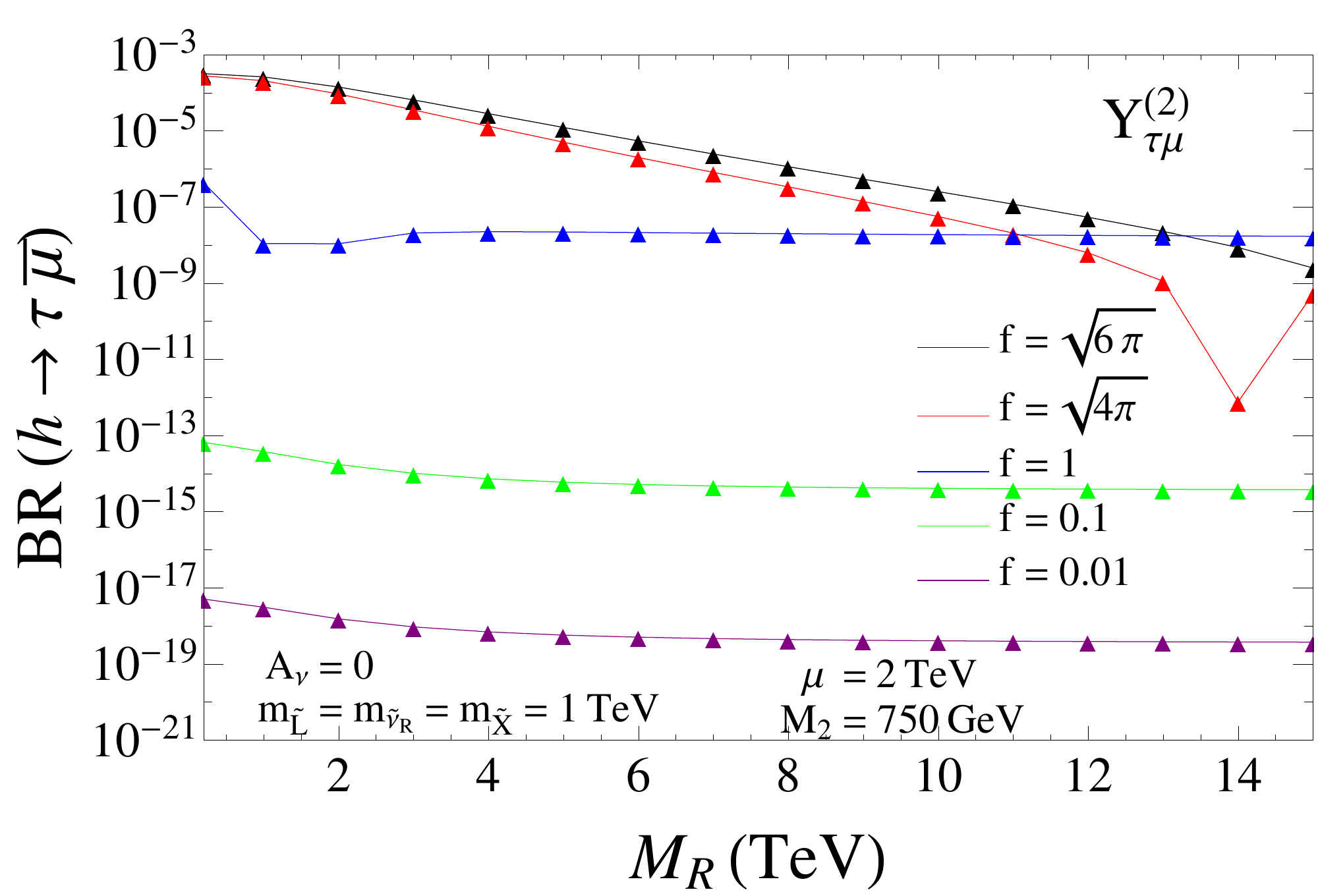}
\includegraphics[width=0.475\textwidth]{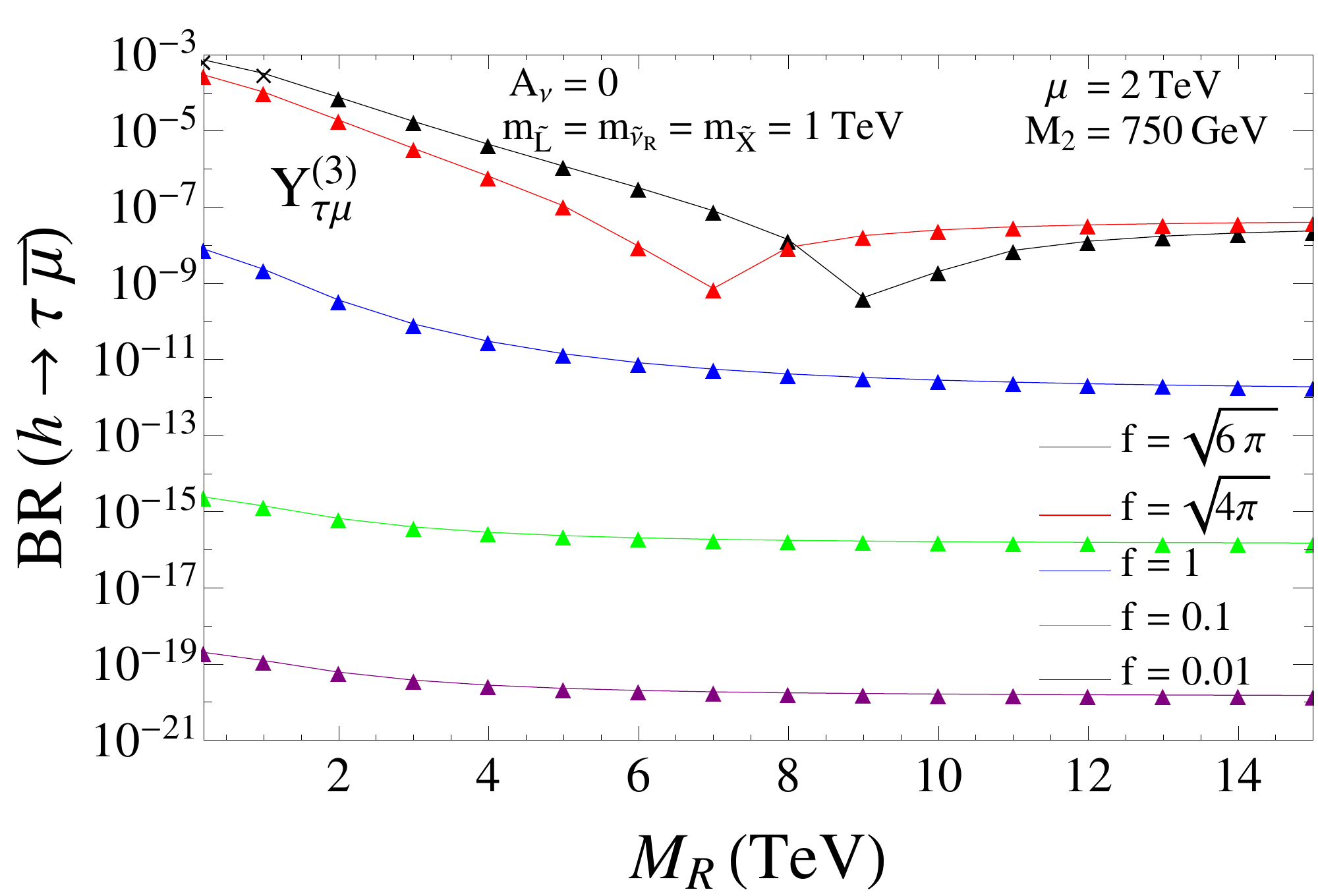}
\includegraphics[width=0.475\textwidth]{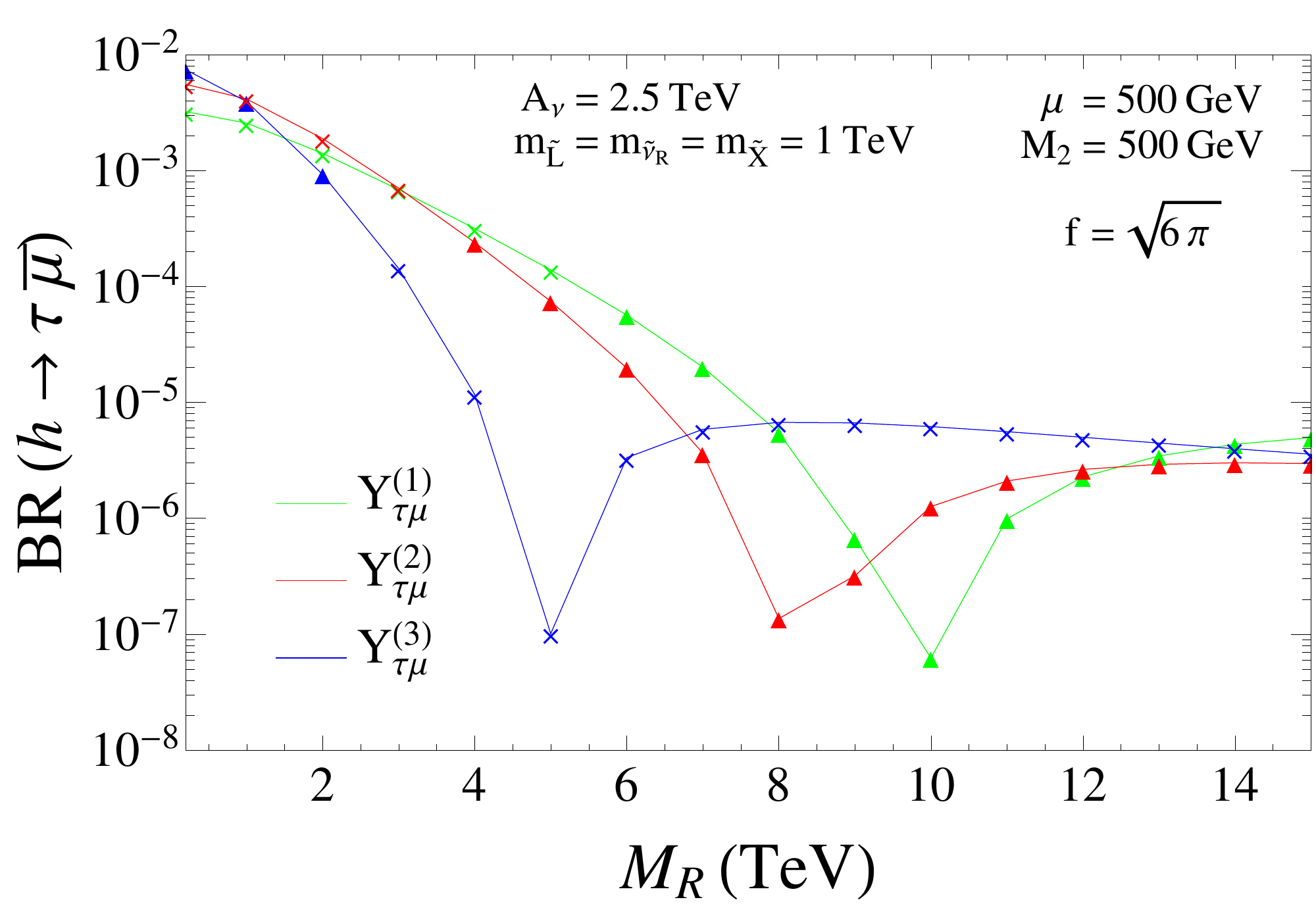}
\caption{BR($h \to \tau \bar \mu$) as a function of $M_R$ for $Y_{\tau\mu}^{(1)}$ (upper left panel), $Y_{\tau\mu}^{(2)}$ (upper right panel), and $Y_{\tau\mu}^{(3)}$ (lower left panel) with $M_2 =$ 750 GeV, $\mu =$ 2 TeV, $A_\nu =$ 0, $\tan\beta =$ 5 and different values of the scaling factor $f =$ 0.01, 0.1, 1, $\sqrt{4\pi}$, $\sqrt{6\pi}$. 
The lower right panel shows the results in these three scenarios with $M_2 = \mu =$ 500 GeV, $A_\nu =$ 2.5 TeV, $\tan\beta =$ 10 and $f = \sqrt{6\pi}$. 
In all  panels,  $m_{\tilde L} =$ $m_{\tilde e} =$ $m_{\tilde \nu_R} =$ $m_{\tilde X} =$ 1 TeV, $m_A =$ 800 GeV and $M_0=$ 1 TeV. 
We set, in these and all the figures of this Section, $A_0=A_e=B_X=B_R=0$, $M=10^{18}\,\mathrm{GeV}$ and the GUT inspired relation $M_1=5/3~ M_2\tan^2\theta_W$. Crosses (triangles) represent points in the SUSY-ISS parameter space excluded
(allowed) by the upper bound  BR($\tau \to \mu \gamma$) $< 4.4\times 10^{-8}$~\cite{Aubert:2009ag}.}
\label{LFVHD_SUSY_MR}
\end{center}
\end{figure}

In \figref{LFVHD_SUSY_MR}, we show the behavior of BR($h \to \tau \bar \mu$) as a function of $M_R$ in the above commented scenarios, $Y_{\tau\mu}^{(1)}$ from TM-5 (upper left panel), $Y_{\tau\mu}^{(2)}$ from TM-6 (upper right panel), and $Y_{\tau\mu}^{(3)}$ from TM-7 (lower left panel), for different values of the scaling factor $f$. 
First of all, we clearly see that, as expected, the larger the value of $f$ is, the larger the LFV rates are. 
We also observe qualitatively different behaviors of the LFV rates between small ($f < 1$) and large ($f > 1$) neutrino Yukawa couplings. 
This difference comes from the different behavior with the parameters of the two participating types of loops, the ones with charged sleptons, where the LFV is generated exclusively by the mixing $(\Delta m_{\tilde L}^2)_{ij}$, and the ones with sneutrinos where the LFV is generated by both $(\Delta m_{\tilde L}^2)_{ij}$ and $(Y_\nu)_{ij}$. 
For small values of $f$, the dominant contributions come from the slepton-neutralino loops, which only depend logarithmically on $M_R$, as can be seen from \eqref{SleptonMixing}, leading to the apparent flat behavior. 
Nevertheless, we checked that this flat behavior disappears when both $M_R$ and $M_0$, and consequently all slepton and sneutrino masses, increase simultaneously.
On the other hand, when the scale factor $f$ becomes larger, contributions from sneutrino-chargino loops become sizable and even dominate at low $M_R$. 
These contributions decrease with $M_R$, due to the increase in the singlet sneutrino masses, which explains the decrease in BR($h \to \tau \bar \mu$) observed in the plots in \figref{LFVHD_SUSY_MR}  for large $f>1$. 
In this latter situation, the appearance of dips due to negative interferences between the two types of loops marks the transition between the two regimes, with the main contribution coming from sneutrino-chargino loops at low $M_R$ and from slepton-neutralino loops at large $M_R$.

Regarding the numerical predictions, we find that, for these parameters, the largest BR($h \to \tau \bar \mu$) allowed by the $\tau \to \mu \gamma$ upper limit are obtained for $f =$ $\sqrt{4\pi}$ or $\sqrt{6\pi}$ and $M_R <$ 2 TeV, with a value of around $10^{-4}$ for the three shown scenarios, which could be probed in future runs of the LHC. 
Nonetheless, these predictions can have strong dependencies on the SUSY parameters, as we want to further explore next. 
In particular, we study the effects on these LFV observables of the trilinear coupling $A_\nu$, which had been set to zero up to now. 
On the lower right panel of \figref{LFVHD_SUSY_MR} we have chosen $A_\nu =$ 2.5 TeV and show the behavior of  BR($h \to \tau \bar \mu$) with $M_R$ for the three textures with a scaling factor $f = \sqrt{6\pi}$. 
This value of $A_\nu$ leads to an enhancement of the BR($h \to \tau \bar \mu$)  while simultaneously suppressing  the $\tau\rightarrow\mu\gamma$ rates. 
As a consequence, very large LFVHD branching ratios can be obtained for $Y_{\tau\mu}^{(3)}$ with low $M_R$ close to 1~TeV achieving values up to $7 \times 10^{-3}$ allowed by $\tau \to \mu \gamma$. 
These large rates are within the sensitivity of the present experiments.

\begin{figure}[t!]
\begin{center}
\begin{tabular}{cc}
\includegraphics[width=0.475\textwidth]{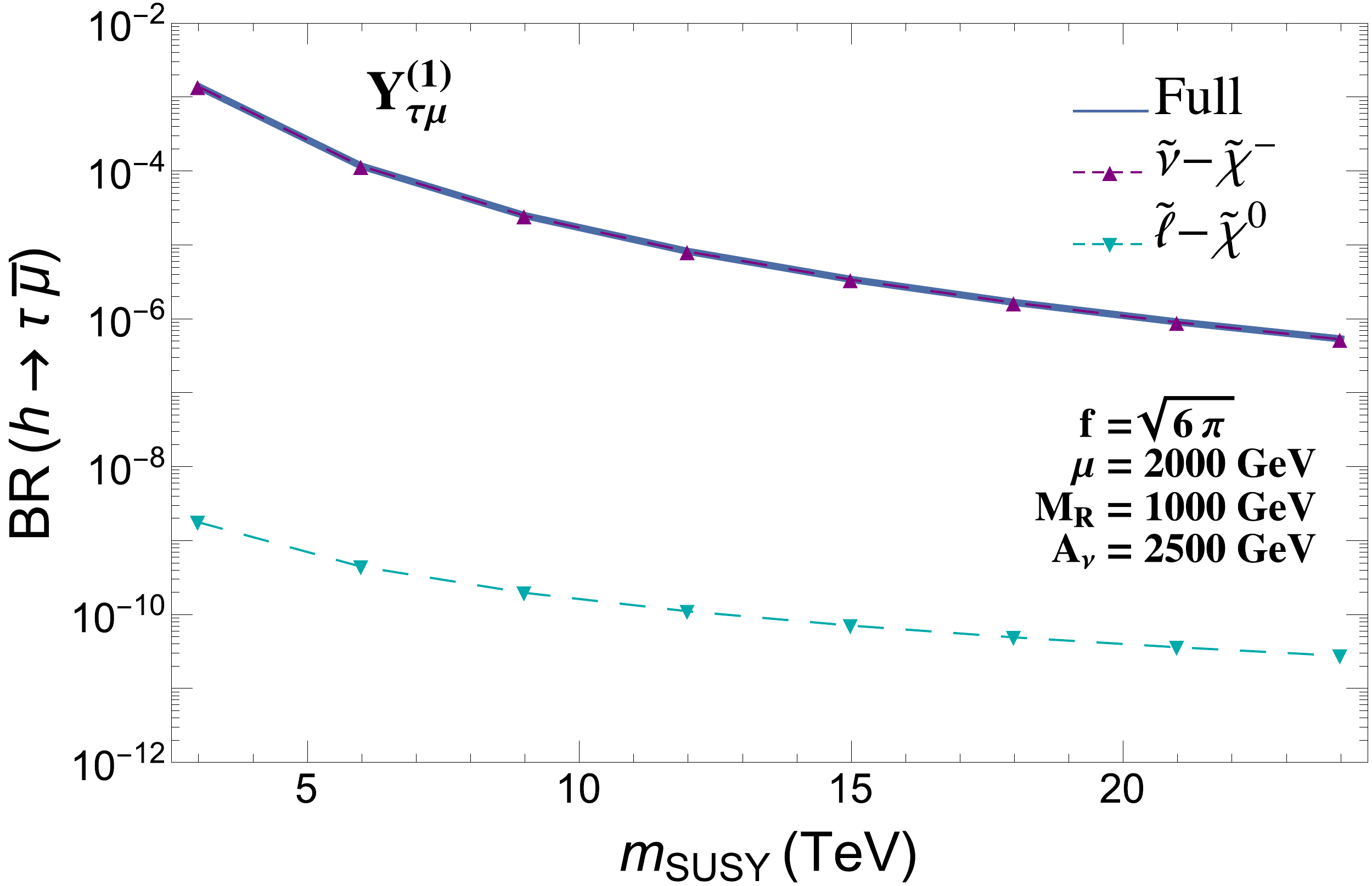}&
\includegraphics[width=0.475\textwidth]{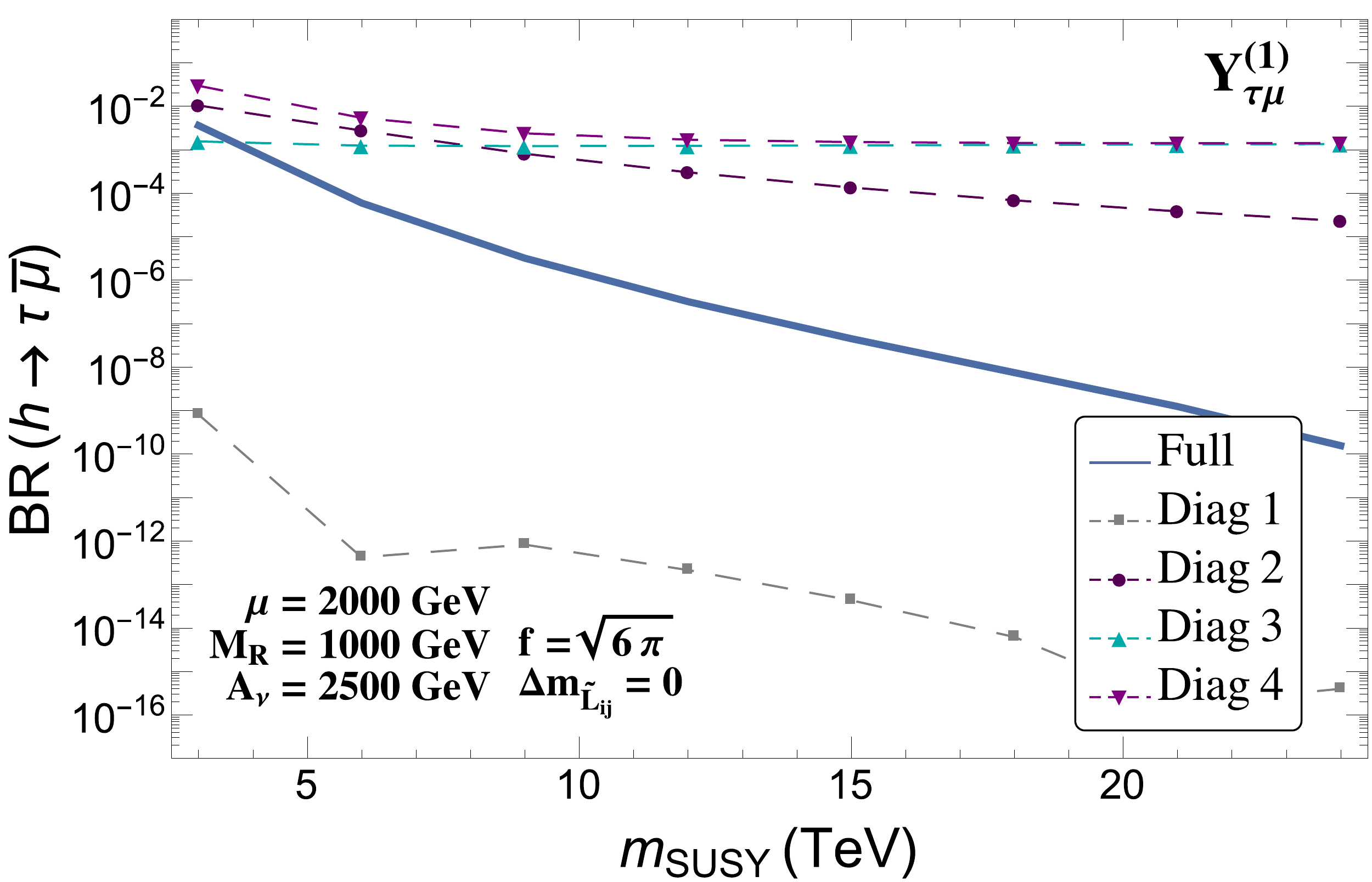}
\end{tabular}
\caption{BR($h\to\tau\bar\mu$) as a function of the common SUSY mass parameter $m_{\rm SUSY}$ in \eqref{msusy} for the TM-5 scenario $\big(Y_{\tau\mu}^{(1)}\big)$ with $M_R=1$~TeV, $f=\sqrt{6\pi}$,  $m_A=800$~GeV, $\mu=2$~TeV, $\tan\beta=10$ and $A_\nu=2.5$~TeV.
Left panel: Contributions from sneutrino-chargino loops, denoted by $\widetilde\nu$-$\widetilde\chi^-$, slepton-neutralino loops, denoted by $\widetilde\ell$-$\widetilde\chi^0$, and full results for BR($h\to\tau\bar\mu$). 
Right panel: Individual contributions from each $\widetilde\nu$-$\widetilde\chi^-$, diagrams (1)-(4) in \figref{SUSYdiagrams}, and full result in the case of $\Delta m_{\tilde L_{ij}}=0$, where the $\widetilde\ell$-$\widetilde\chi^0$ contributions vanish.}\label{HtaumuSUSYdecoulingplots}
\end{center}
\end{figure}

We next study the behavior of the $h\to\tau\bar\mu$ rates as a function of the SUSY mass scales in a simplified scenario where all the SUSY masses are equal to a common parameter $m_{\rm SUSY}$, namely,
\begin{equation}\label{msusy}
m_{\rm SUSY}=m_{\tilde L}=m_{\tilde e}=m_{\tilde \nu_R}=m_{\tilde X}=M_0=M_1=M_2\,.
\end{equation}
The left panel of \figref{HtaumuSUSYdecoulingplots}  shows the expected decoupling behavior with $m_{\rm SUSY}$,  where BR($h\to\tau\bar\mu$) decreases when increasing all the heavy sparticle masses. 
This plot is for the particular case of the TM-5 scenario $\big(Y_{\tau\mu}^{(1)}\big)$, but similar behaviors (not shown) are obtained for the other  scenarios.
In this figure we have included the full predictions for BR($h\to\tau\bar\mu$), as well as the separated contributions coming only from sneutrino-chargino loops, diagrams (1)-(4) in \figref{SUSYdiagrams}, and from slepton-neutralino loops, diagrams (5)-(8) in \figref{SUSYdiagrams}. 
We see that not only the full prediction but also the separated contributions from these two subsets decrease with $m_{\rm SUSY}$, showing that the decoupling occurs in both, the charginos-sneutrinos and the neutralinos-sleptons sectors, as expected from the decoupling theorem.

\begin{figure}[t!]
\begin{center}
\includegraphics[width=0.475\textwidth]{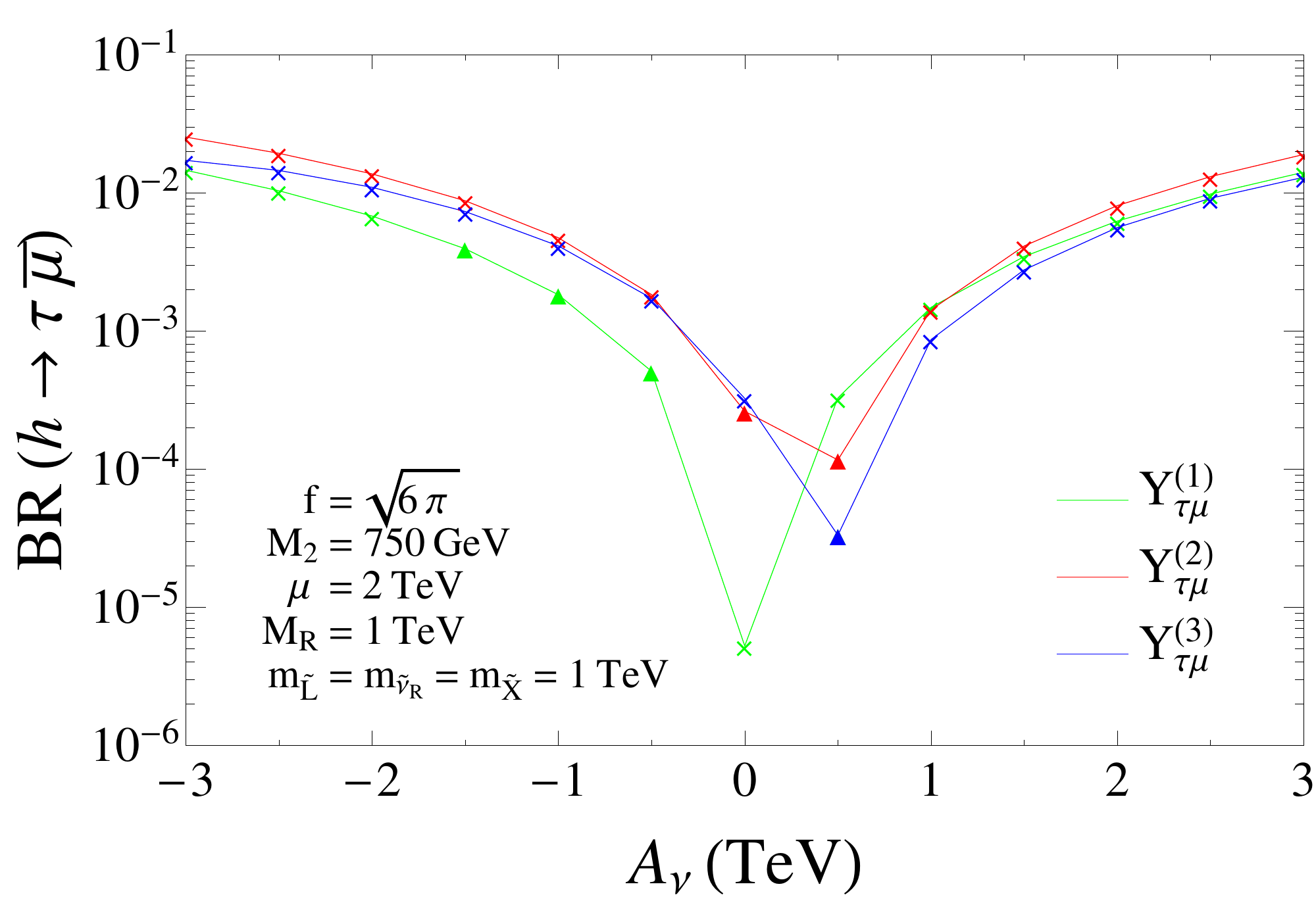}
\includegraphics[width=0.475\textwidth]{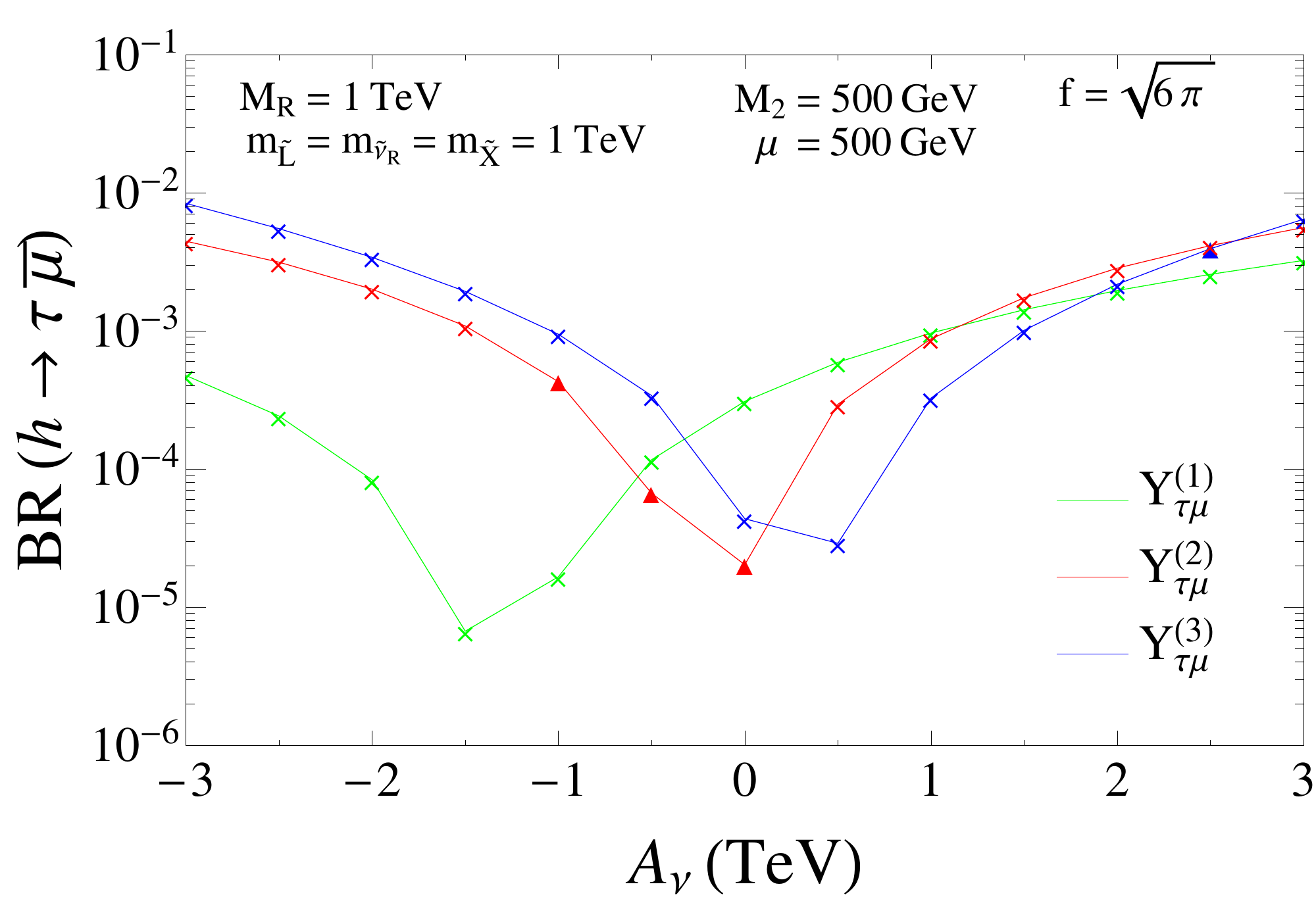}
\caption{Dependence of BR($h \to \tau \bar \mu$) on $A_\nu$ for scenarios TM-5 $\big(Y_{\tau\mu}^{(1)}\big)$, TM-6 $\big(Y_{\tau\mu}^{(2)}\big)$, and TM-7 $\big(Y_{\tau\mu}^{(3)}\big)$, with $M_2 =$ 750 GeV, $\tan\beta =$ 5 and $\mu =$ 2 TeV (left panel) or with $\tan\beta =$ 10 and $M_2 = \mu =$ 500 GeV (right panel).
On both panels, $m_A =$ 800 GeV, $M_0=$ 1 TeV, $M_R =$ $m_{\tilde L} =$ $m_{\tilde e} =$ $m_{\tilde \nu_R} =$ $m_{\tilde X} =$ 1 TeV, and the scaling factor $f = \sqrt{6\pi}$. Crosses (triangles) represent points
in the SUSY-ISS parameter space excluded (allowed) by the $\tau \to \mu \gamma$ upper limit, BR($\tau \to \mu \gamma$) $< 4.4\times 10^{-8}$~\cite{Aubert:2009ag}.}
\label{LFVHD_SUSY_Anu}
\end{center}
\end{figure}

In this heavy sparticle scenario, the full predictions are dominated by the contributions from the sneutrino-chargino sector, which is the one containing new sparticles with respect to the MSSM.
In order to better understand the contributions from this sector, we study the simple case of $\Delta m_{\tilde L_{ij}}=0$, where the contributions from the slepton-neutralino sector vanish.
We show in the right panel of \figref{HtaumuSUSYdecoulingplots} the full result in this situation, as well as the individual contributions from diagrams (1) to (4) in \figref{SUSYdiagrams}.
We see that the vertex correction, diagram (2), and the self-energies, diagrams (3) and (4), clearly compete in size and that their interference is destructive, manifesting a strong cancellation among them. 
The contributions from diagram (1), on the other hand, are subleading by several orders of magnitude.
Notice also that the finite contributions from the divergent diagrams (3) and (4) do not decouple individually with $m_{\rm SUSY}$, but their addition does, as expected.

As mentioned before, we have found that the LFVHD rates are indeed very sensitive to the particular value of the trilinear coupling $A_\nu$. 
Thus, we study in \figref{LFVHD_SUSY_Anu} the behavior of BR($h \to \tau \bar \mu$) with $A_\nu$ for the two SUSY scenarios considered in \figref{LFVHD_SUSY_MR}, with $M_R =$ 1 TeV and $f = \sqrt{6\pi}$. 
The strong dependence on $A_\nu$ is manifest in both panels, presenting deep dips in different positions that depend mainly on the values of $Y_\nu$, $\mu$, $m_A$ and $\tan\beta$. 
These parameters control, in particular, the $h$-$\tilde \nu_L$-$\tilde\nu_R$ coupling and the $\tilde\nu_L$-$\tilde\nu_R$ mixing, which would lead to the appearance of dips in the regime where contributions from sneutrino-chargino loops dominate, as it is the case of \figref{LFVHD_SUSY_Anu}.
It is interesting to note that, for this choice of parameters, practically all the parameter space is excluded by $\tau \to \mu \gamma$ except the points within the dips and surrounding them, where the LFV radiative decay $\tau \to \mu \gamma$ suffers also a strong reduction.
We find as a relevant feature that the location of the dips in BR($h \to \tau \bar \mu$) and BR($\tau \to \mu \gamma$) usually does not coincide, therefore allowing for large LFV Higgs decays rates, above $10^{-3}$ and within the reach of the LHC experiments, that are not excluded by $\tau \to \mu \gamma$.

\begin{figure}[t!]
\begin{center}
\includegraphics[width=0.49\textwidth]{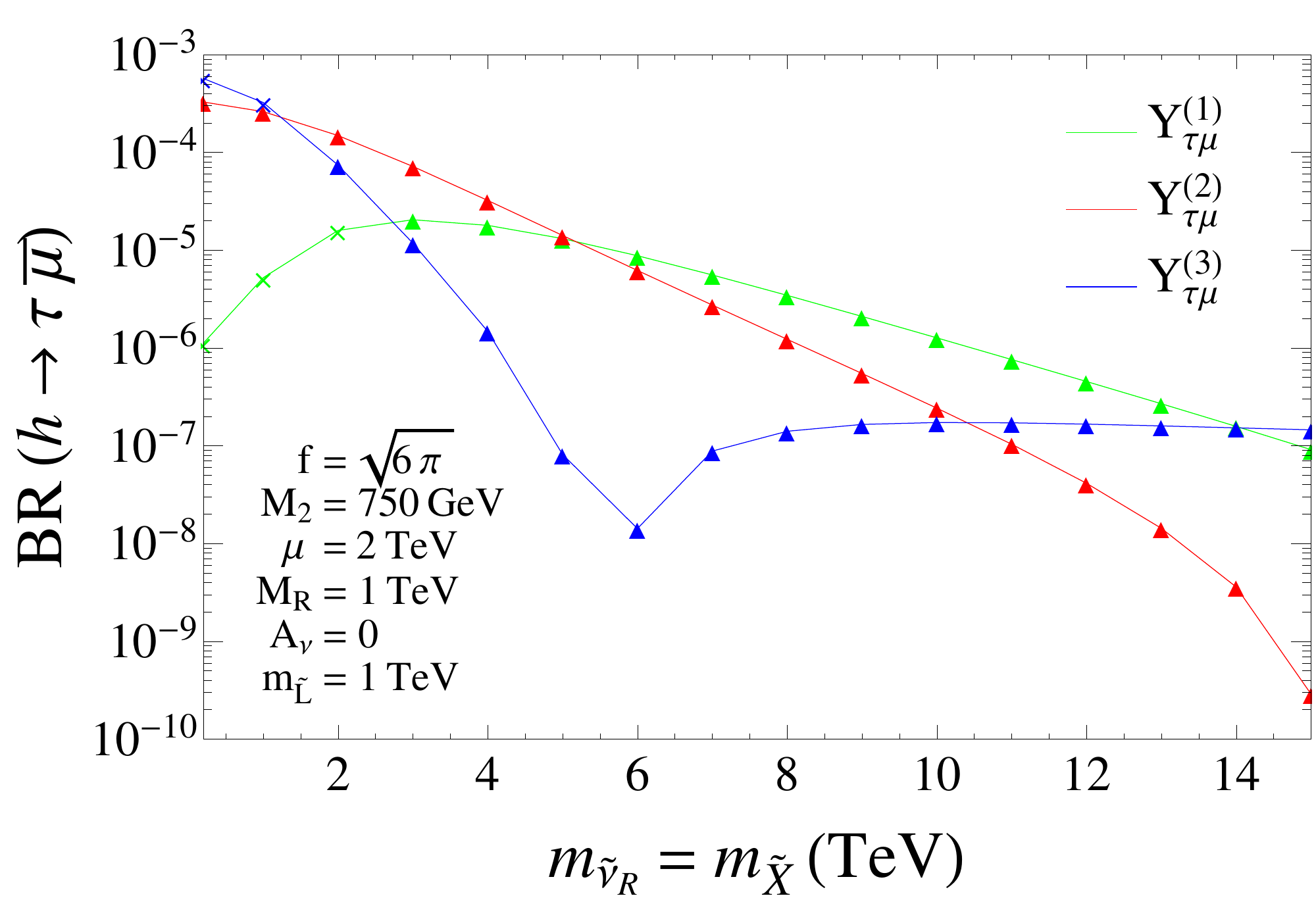}
\includegraphics[width=0.49\textwidth]{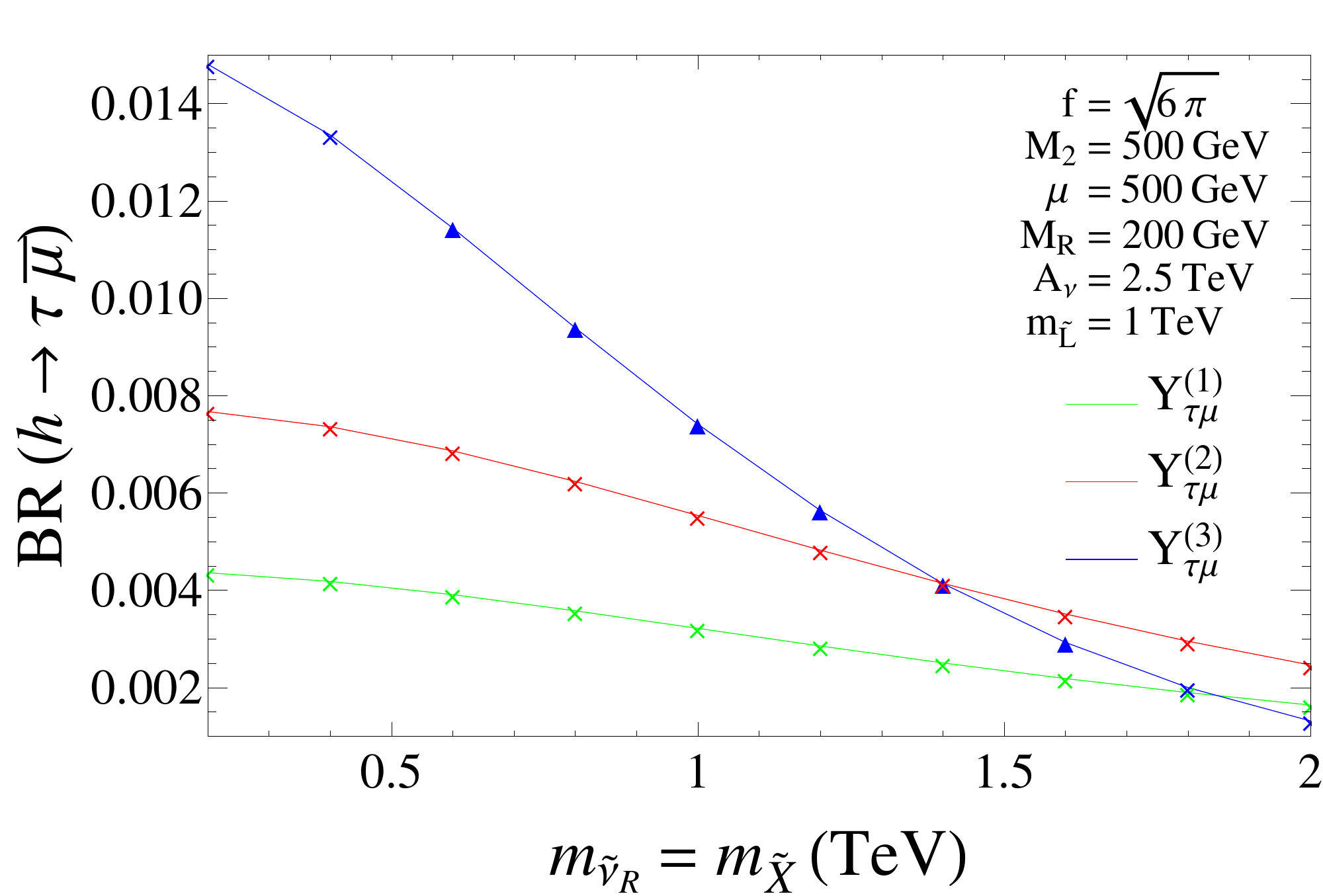}
\caption{Dependence of BR($h \to \tau \bar \mu$) on $m_{\tilde \nu_R}=m_{\tilde X}$ for scenarios TM-5 $\big(Y_{\tau\mu}^{(1)}\big)$, TM-6 $\big(Y_{\tau\mu}^{(2)}\big)$, and TM-7 $\big(Y_{\tau\mu}^{(3)}\big)$, with $M_R =$ $m_{\tilde L} =$ $m_{\tilde e} =$ 1 TeV, $M_2 =$ 750 GeV, $\mu =$ 2 TeV, $\tan\beta =$ 5 and $A_\nu =$ 0 (left panel) or
with $M_R =$ 200 GeV, $m_{\tilde L} =$ $m_{\tilde e} =$ 1 TeV, $M_2 = \mu =$ 500 GeV, $\tan\beta =$ 10 and $A_\nu =$ 2.5 TeV (right panel). On both
panels, $m_A =$ 800 GeV, $M_0=$ 1 TeV, and $f = \sqrt{6\pi}$. Crosses (triangles) represent points in the SUSY-ISS parameter space excluded (allowed)
by the $\tau \to \mu \gamma$ upper limit, BR($\tau \to \mu \gamma$) $< 4.4\times 10^{-8}$~\cite{Aubert:2009ag}.}
\label{LFVHD_SUSY_msoft}
\end{center}
\end{figure}

Finally, the dependence of the LFVHD rates on the new sneutrino soft SUSY breaking scalar masses, $m_{\tilde \nu_R}$ and $m_{\tilde X}$, is depicted in \figref{LFVHD_SUSY_msoft}, where we vary these parameters independently from the SUSY scale. 
As when modifying $M_R$, increasing $m_{\tilde \nu_R}$ and $m_{\tilde X}$ makes the singlet sneutrinos heavier and decreases the size of the chargino contribution. 
In the case of $Y_{\tau\mu}^{(1)}$ and $Y_{\tau\mu}^{(2)}$ which are dominated by this contribution, the BR($h \to \tau \bar \mu$) exhibits a strong decrease in the range explored in \figref{LFVHD_SUSY_msoft}, by more than five orders of magnitude in the case of $Y_{\tau\mu}^{(2)}$. 
For $Y_{\tau\mu}^{(3)}$ a dip can be observed, due again to cancellations between the chargino and neutralino contributions, with the latter dominating at large $m_{\tilde \nu_R}$. 
For the  benchmark point in the left panel, the largest $h \to \tau \bar \mu$ rates allowed by the $\tau \to \mu \gamma$ upper limit are obtained for $Y_{\tau\mu}^{(2)}$ with $m_{\tilde \nu_R} =$ 200 GeV, with a maximum value of $\sim 3 \times 10^{-4}$, just one order of magnitude below the present LHC sensitivity.
In the second benchmark point in the right panel of \figref{LFVHD_SUSY_msoft}, we found large LFVHD rates with $M_R =$ 200 GeV, $A_\nu =$ 2.5 TeV and low values of $m_{\tilde \nu_R}$. 
We observe a huge increase in BR($h \to \tau \bar \mu$) for the three Yukawa couplings $Y_{\tau\mu}^{(1)}$, $Y_{\tau\mu}^{(2)}$, and $Y_{\tau\mu}^{(3)}$, with maximum values of approximately  $4 \times 10^{-3}$, $ 8 \times 10^{-3}$ and $ 1.5 \times 10^{-2}$, respectively, due mainly to the low values of $m_{\tilde \nu_R}$ and $M_R$. 
Unfortunately,  the $\tau \to \mu \gamma$ upper limit excludes all the parameter space for the $Y_{\tau\mu}^{(1)}$ and $Y_{\tau\mu}^{(2)}$ cases. 
In contrast, most of the points for the $Y_{\tau\mu}^{(3)}$ texture are in agreement with this upper bound, since they are located in a region where the $\tau \to \mu \gamma$ rates suffer a strong suppression as a consequence of the value set for $A_\nu$, in this case, $A_\nu =$ 2.5 TeV. 
This fact allows us to obtain a maximum value of BR($h \to \tau \bar \mu$) $\sim 1.1\%$, completely within 
the reach of the current LHC experiments and large enough to explain the CMS  excess if confirmed by other experiments and/or future data.

Summarizing, in this Section we have studied the LFVHD rates in presence of the SUSY particles in the SUSY-ISS model.
We have seen that much larger contributions from the SUSY loops are obtained with respect to the predicted rates in the type-I seesaw due to large $Y_\nu^2/(4 \pi) \sim  {\cal O} (1) $, the presence of right-handed sneutrinos at the TeV scale and an increased RGE-induced slepton mixing from the GUT scale down to the $M_R$ scale.
We also demonstrated  that in the SUSY-ISS model new contributions coming from the SUSY particle loops can considerably enhance the LFVHD rates with respect to the non-SUSY ISS model. 
We have found particularly interesting the new contributions from the trilinear coupling $A_\nu$, since, in addition to enhance the LFVHD rates, it can lead to suppressions in the corresponding LFV radiative decay rates.
We find these results very promising and therefore this calls up for a more complete analysis.
We will therefore present in a future work a complete study including both the SUSY and non-SUSY contributions in the SUSY-ISS model, exploring also the heavy Higgs bosons LFV decays and considering a detailed analysis of experimental constraints beyond radiative LFV decays.

\section[Effective $H\ell_k\ell_m$ vertex from heavy $\nu_R$ within the Mass Insertion Approximation]{The effective LFV $\boldsymbol{H\ell_k\ell_m}$ vertex from heavy $\boldsymbol{\nu_R}$ within the mass insertion approximation}
\label{sec:LFVHDMIA}

As we have seen in \secref{sec:LFVHDanal}, LFV H decays in the ISS model do not behave as other LFV processes like the radiative decays.
By comparing the approximated expressions for the LFV radiative decays  in \eqref{RadApprox} and for the H decays in \eqref{FIThtaumu}, it is clear that the latter contains, in addition to the usual $\mathcal O(Y_\nu^2)$ contribution, an extra contribution of  $\mathcal O(Y_\nu^4)$ that is not present in the former.

In order to understand these results, we will perform a completely different and independent analysis of the LFVHD rates within the ISS model. 
Instead of using the physical neutrino basis, we will perform our computation of the LFVHD widths directly in the chiral electroweak interaction basis with left- and right-handed neutrinos being the fields propagating in the loops.
This will allow us to express the results explicitly in terms of the most relevant model parameters, namely,  the neutrino Yukawa coupling matrix $Y_\nu$ and the right-handed mass matrix $M_R$.  

We will do this new one-loop computation by using the mass insertion approximation (MIA), which turns out to be a very powerful tool  in presence of heavy right-handed neutrinos.  
In this context, the MIA provides the results in terms of a well defined expansion in powers of $Y_\nu$, which is the unique relevant coupling originating lepton flavor violation in this model, and therefore it is a very useful and convenient method for an easier and clearer interpretation of the related phenomenology.  

For the present study of the $H \to \ell_k \bar \ell_m$ decay amplitude  we will calculate this MIA expansion first to the leading order, ${\cal O}((Y_\nu^{} Y_\nu^\dagger)_{km})$, and second to the next to leading order ${\cal O}((Y_\nu^{} Y_\nu^\dagger Y_\nu^{} Y_\nu^\dagger)_{km})$. 
In addition, we will also use the MIA to compute the one-loop effective vertex ${H\ell_k\ell_m}$, that is the relevant one for these decays. 
For this purpose, we will explore the proper large $M_R$ mass expansion, which in the present case we must apply for the assumed mass hierarchy, 
\be\label{MIAmasshierarchy}
m_{\ell_{i,j}}\ll vY_\nu,m_W,m_H\ll M_R\,,
\ee
with $m_{\ell_{i,j}}$ the lepton masses, $v$ the Higgs vacuum expectation value and $m_W$ and $m_H$, the $W$ boson and Higgs particle masses, respectively.   
As we will see, the most appealing feature of our computation is that it provides very simple formulas, which turn out to work very well for both the one-loop effective ${H\ell_k\ell_m}$ LFV vertex  and the partial width  $\Gamma(H \to \ell_k \bar\ell_m)$  in terms of the most relevant parameters $Y_\nu$ and $M_R$.
These simple formulas could be easily used by other authors to rapidly estimate, without the need of a heavy numerical computation, the LFVHD rates with their own inputs for $Y_\nu$ and $M_R$.
Moreover, since these results are based only in the mass hierarchy in \eqref{MIAmasshierarchy} and the fact that $Y_\nu$ is the main source of LFV, they could presumably be used in alternative neutrino models that share these properties.
In order to make this statement clearer, we explain in more detail the hypothesis behind our calculation in the next Section.

\subsection{The proper basis and Feynman rules for a MIA computation}
\label{MIAsetup}

In order to use the MIA for the computation of the one-loop generated effective  ${H\ell_k\ell_m}$ vertex from right-handed neutrinos, it is important first to choose the proper EW interaction basis and to set up the necessary Feynman rules in terms of these fields. 
The main point of the MIA is precisely based on the use of the EW basis instead of the mass basis, which is the one usually used in the literature for the one-loop generated LFV observables in models with massive Majorana neutrinos. 
Nevertheless, the MIA computations can be even further simplified by choosing the proper EW basis for each model, as we  discuss in the following. 

In the case of the ISS model, the $9\times9$ mass matrix in \eqref{ISSmatrix}  provides all the relevant masses and mass insertions for the EW eigenstates that are needed for our computation. 
These mass insertions connect two different neutrino states, they are in general flavor non-diagonal, and can be expressed in terms of the three $3 \times 3$ matrices $m_D$, $\mu_X$ and $M_R$. 
Specifically, the mass insertion given by $m_D$ connects $\nu_L$ and $\nu_R$ fields, $M_R$ connects $\nu_R$ and $X$, and $\mu_X$ connects two  $X$. 
To simplify the computation, we will use again the freedom of redefining the new fields ($\nu_R, X$) in such a way that the $M_R$ matrix  is flavor diagonal. 
Thus, all the flavor violation is contained in the matrices $\mu_X$ and $m_D$. 
Nevertheless, since we are working with $\mu_X$ being extremely small as to accommodate the light neutrino masses, this mass matrix will be irrelevant for the LFV physics that we will study in this Thesis. 
Therefore, the only relevant flavor violating insertion will be provided by the $m_D$ matrix and, in consequence, by  the Yukawa coupling matrix $Y_\nu$.

On the other hand, it should be noticed that the flavor preserving
mass insertions given by $M_R$ can be very large if $M_R$ is taken to be heavy, as it will be our case with $M_R$ being at the TeV scale.   
Since we are finally interested in a perturbative MIA computation of the one-loop LFV Higgs form factors and effective vertices that are valid for heavy $M_R$ masses, we find convenient to use a different chiral basis where `the big insertions' given by  $M_R$ are resumed in such a way that the `large mass' $M_R$  appears effectively in the denominator of the propagators of the new states.
The key point in choosing this proper chiral basis is provided by the fact that for the quantities of our interest here, having $H,$ $\ell_k$ and $\ell_m$ as the external particles, the only neutrino states that interact with them are $\nu_L$ and $\nu_R$. The singlet  fields $X$ interact exclusively with the $\nu_R$ fields via the $M_R$ mass insertions and, therefore, they will only appear in the computation of the loop diagrams for LFV as internal intermediate states inside internal lines that start and end with $\nu_R$'s. This motivates clearly our choice of modified propagators for the $\nu_R$ fields which are built on purpose to include inside all the effects of the sequential insertions of the $X$ fields,  given each of these insertions by $M_R$. 
More concretely, we sum all the $M_R$ insertions and define two types of  modified propagators: one with the same initial and final particle, corresponding to an even number of $M_R$ mass insertions which we call {\it fat propagators}, and one with different initial and final particles, corresponding to an odd number of insertions. 
The {\it fat propagator},  which propagates a $\nu_R$ into a $\nu_R$ and contains the sum of all the infinite series of even number of $M_R$ insertions due to the interactions with $X$, is the one we need for the present computation. 
The details of the procedure to reach this proper chiral basis and the derivation of the modified propagators are explained in \appref{App:modifiedpropagators}.  
Similar results are obtained within the context of the Flavor Expansion Theorem\footnote{We warmly thank Michael Paraskevas for his kind comment about the similarities between our {\it fat propagators} and the results in the Flavor Expansion Theorem.}~\cite{Dedes:2015twa,Rosiek:2015jua}.

\begin{figure}[b!]
\begin{center}
\includegraphics[scale=1.04]{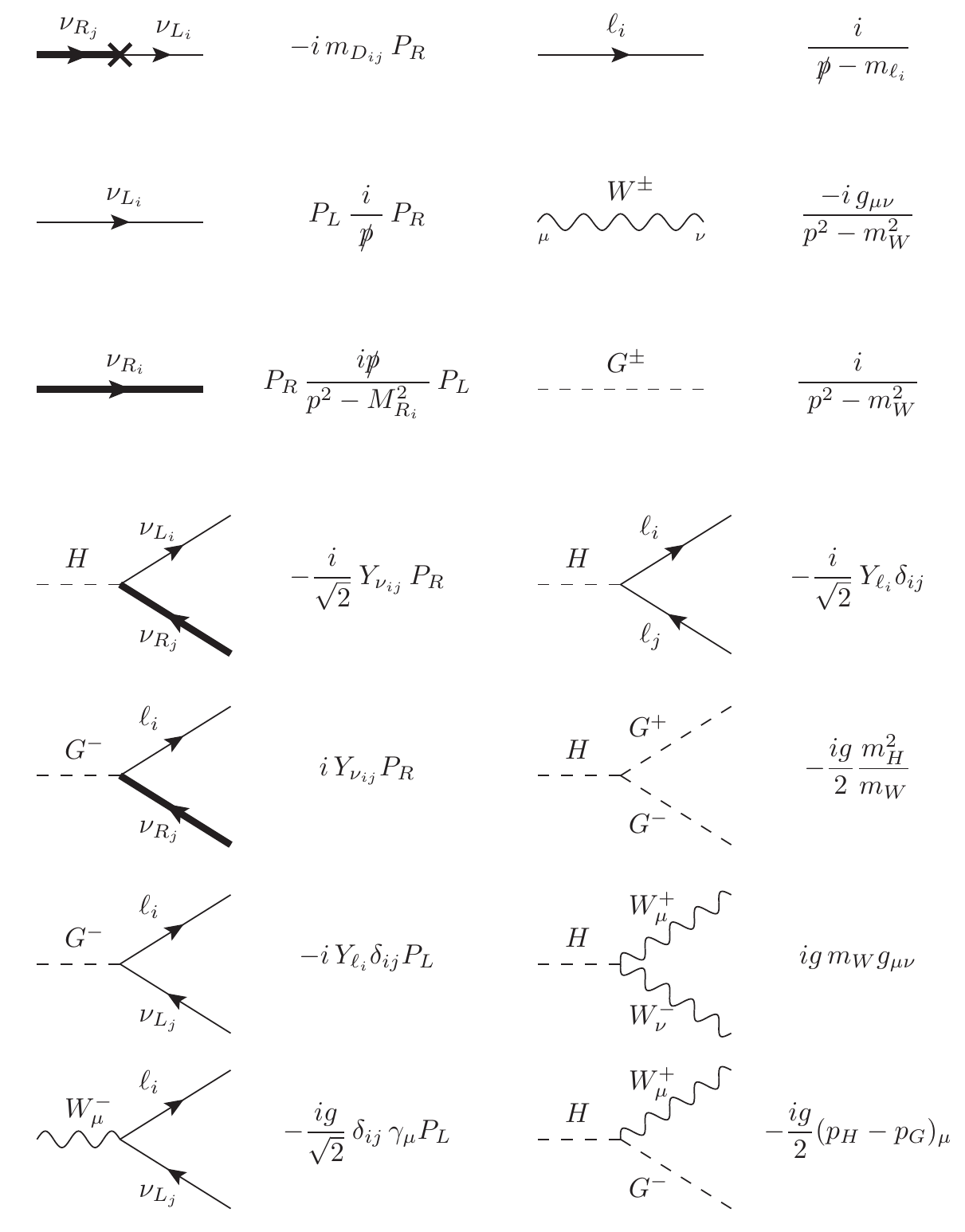}
\caption{Relevant Feynman rules for the present MIA computation of BR$(H\to\ell_k\bar\ell_m)$ in the Feynman-'t Hooft gauge. The rules involving neutrinos are written in terms of the proper EW chiral basis for $\nu_R$ and $\nu_L$, as defined in the text. For completeness, some additional SM Feynman rules that are needed are also included. The momenta $p_H$ and $p_G$ are incoming.
The thick solid line represents the {\it fat propagator} of $\nu_R$ introduced in the text.}\label{LFVHD_FR-MIA}
\end{center}
\end{figure}

In order to complete the set-up for our computations, we summarize the relevant Feynman rules in our previously chosen proper chiral basis in \figref{LFVHD_FR-MIA}. 
These include the relevant flavor changing mass insertions, given by $m_D$, the relevant propagators, both the usual SM EW propagators and the new {\it fat propagators} of the $\nu_R$'s, as well as the relevant interaction vertices needed for our computation, both the SM EW vertices and the new ones involving the $\nu_R$'s. 

Finally,  we want to stress that, although we have considered the ISS model to make our computations, our results could be applied in practice to any low scale seesaw model that leads to the same Feynman rules as in \figref{LFVHD_FR-MIA}. 
These are indeed quite generic Feynman rules in models with right-handed heavy neutrinos. 
The few specific requirements are that the only relevant LFV source is the Yukawa neutrino coupling matrix and that the heavy right-handed neutrino propagator is like our {\it fat propagator} introduced above. 
 
\subsection[$\Gamma(H\to\ell_k\bar\ell_m)$ to one-loop within the MIA]{$\boldsymbol{\Gamma(H\to\ell_k\bar\ell_m)}$ to one-loop within the MIA}
\label{computationwidthMIA}

Once we have introduced the set-up for our MIA calculation, we are ready to compute the LFVHD rates. 
We perform a diagrammatic MIA calculation of $\Gamma (H \to {\ell_k} \bar{\ell}_m)$  considering the following points:
\begin{enumerate}[noitemsep] 
\item We use the EW chiral neutrino basis.
\item We treat the external particles $H$, $\ell_k$ and $\bar{\ell}_m$ in their physical mass basis.
\item We use the {\it fat propagator} for the heavy right-handed neutrinos and the Feynman rules as described in \secref{MIAsetup}.
\item The LFVHD amplitude is evaluated at the one-loop order in the Feynman-'t Hooft gauge. In the  \appref{App:othergauge}  we show that the same result is obtained in the Unitary gauge. 
\item All the loops must contain at least one right-handed neutrino, since they are the only particles transmitting LFV through the flavor off-diagonal neutrino Yukawa matrix entries. 
\item According to the Feynman rules in \figref{LFVHD_FR-MIA},  these flavor changing Yukawa couplings,  appear just in two places, the mass insertions given by $m_D$ and the interactions of $H$ with $\nu_L$ and $\nu_R$, being proportional to $Y_\nu$.
\item All the one-loop diagrams will get an even number of powers of $Y_{\nu}$, since $Y_{\nu}$ appears twice for each $\nu_R$ in an internal line, and because of the absence of interactions containing two right-handed neutrinos. 
\item We further simplify our computation by considering that the diagonal matrix $M_R$ has degenerate entries, i.e., $M_{R_i}\equiv M_R$. The generalization to the non-degenerate case will be commented in \appref{App:MIA}. 
\end{enumerate}

In summary, taking into account all the points exposed above, the one-loop contributions to the LFV Higgs decay amplitude, as computed with the MIA, will then be given by an expansion in even powers of $Y_\nu$, concretely as $Y_\nu^{} Y_\nu^\dagger$. 
Therefore, the form factors defined in \eqref{LFVHDamp}, which we recall here for completeness,
\begin{equation}
i {\cal M} = -i g \bar{u}_{\ell_k} (-p_2) (F_L P_L + F_R P_R) v_{\ell_m}(p_3) \, , 
\label{LFVHDampBIS}
\end{equation}
can be written as follows:
\begin{equation}
F_{L,R}^{{\rm MIA}\;\; (Y^2+Y^4)} = \left(Y_{\nu}^{} Y_{\nu}^{\dagger} \right)^{km}f_{L,R}^{(Y^{2})} + \left(Y_{\nu}^{} Y_{\nu}^{\dagger} Y_{\nu}^{} Y_{\nu}^{\dagger}\right)^{km}f_{L,R}^{(Y^{4})} \, ,
\label{FFMIA}
\end{equation}
where ${\cal O} (Y_\nu Y_\nu^{\dagger})$ are the Leading Order (LO) terms and ${\cal O} (Y_\nu Y_\nu^{\dagger}Y_\nu Y_\nu^{\dagger})$ the Next to Leading Order (NLO) terms in our expansion. 
We expect that, in the perturbativity regime of the neutrino Yukawa couplings, the next terms in this expansion, i.e., those of ${\cal O}(Y_{\nu}^{6})$ and  higher, will be  very tiny and can be safely neglected.  
Furthermore, as will be explained in more detail below, considering this expansion in powers of $Y_{\nu}$ and working with the hypothesis of $M_R$ being the heaviest scale, also lead to an implicit ordering of the various contributions in powers of $v/M_R$. 
In fact, we will demonstrate, by an explicit analytical expansion of the form factors in the large  $M_R\gg v $ limit, that the dominant terms of the two contributions in \eqref{FFMIA}, the LO $f_{L,R}^{(Y^{2})}$ and the NLO $f_{L,R}^{(Y^{4})}$, indeed both scale  as $(v/M_R)^2$. 
In contrast, the next order contributions,  i.e., those of ${\cal O}(Y_{\nu}^{6})$, scale as $(v/M_R)^4$, and therefore they will be negligible for heavy right-handed neutrinos, even when the Yukawa couplings are sizable.
Thus, considering just these two first terms of the MIA expansion in \eqref{FFMIA}  will be sufficient to approach quite satisfactorily the full one-loop calculation of the neutrino mass basis in the case of $\mu_X \ll m_D \ll M_R$ that we are interested in.

In order to estimate the validity of the MIA results for the present study of the LFV Higgs decays we include a numerical comparison of these MIA results with those of the full one-loop computation in the physical neutrino basis presented in \secref{sec:LFVHDanal}.
For an easy comparison, we adopt in the MIA the same notation (i) (i=1,...,10) for the ten types of generic diagrams as in the full computation shown in \figref{diagsLFVHDphysbasis}.
They can be classified into diagrams with vertex corrections, i=1,..,6,  and diagrams with external leg corrections, i=7,..,10. 

\begin{figure}[b!]
\begin{center}
\includegraphics[scale=0.82]{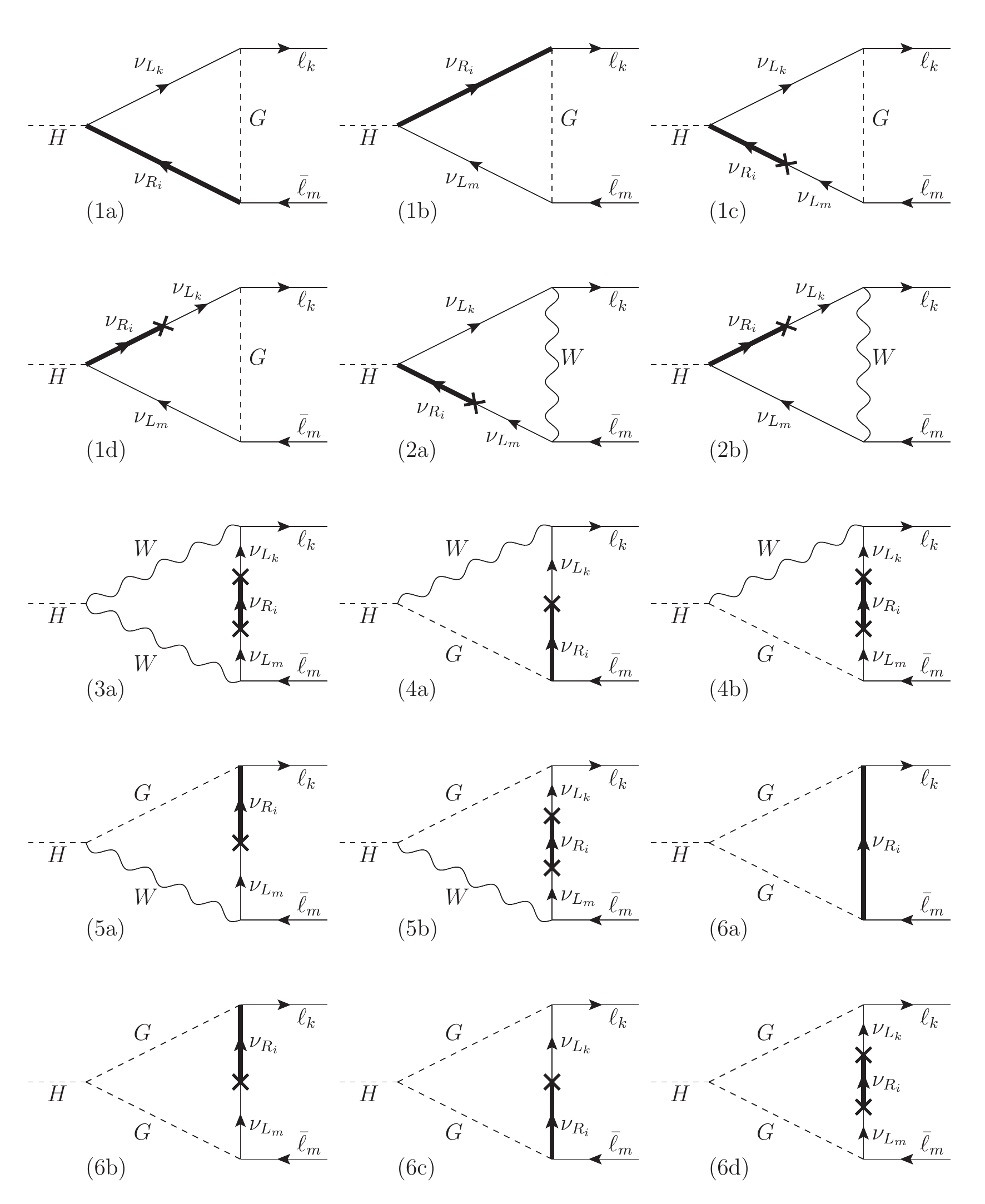}
\caption{Relevant vertex diagrams for the MIA form factors of LFVHD to ${\cal O}(Y_\nu^2)$.}
\label{LFVHDMIAdiagsvertexY2}
\end{center}
\end{figure}
\begin{figure}[t!]
\begin{center}
\includegraphics[scale=0.8]{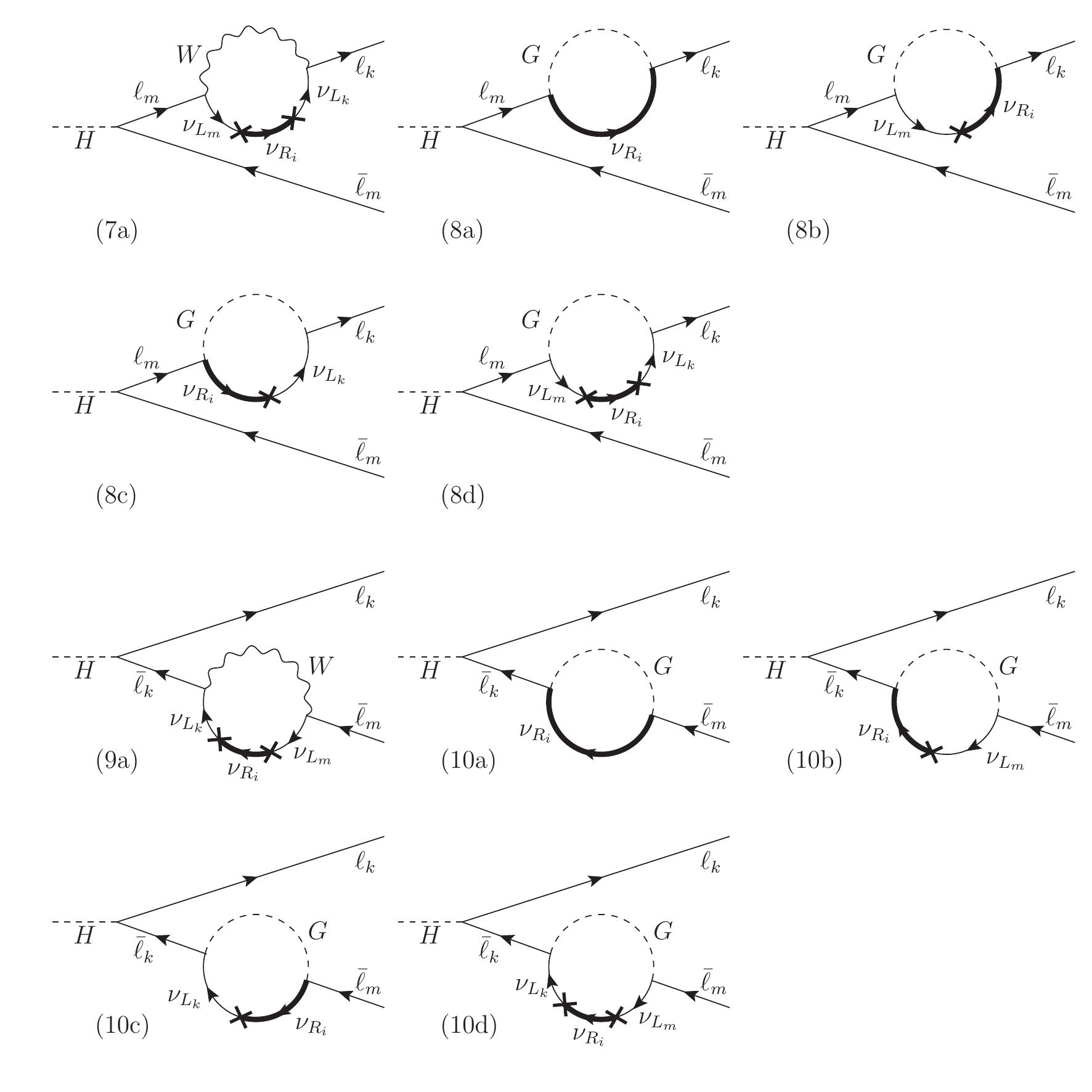}
\caption{Relevant external leg diagrams for the MIA form factors of LFVHD to ${\cal O}(Y_\nu^2)$.}
\label{LFVHDMIAdiagslegY2}
\end{center}
\end{figure}

\begin{figure}[t!]
\begin{center}
\includegraphics[scale=0.73]{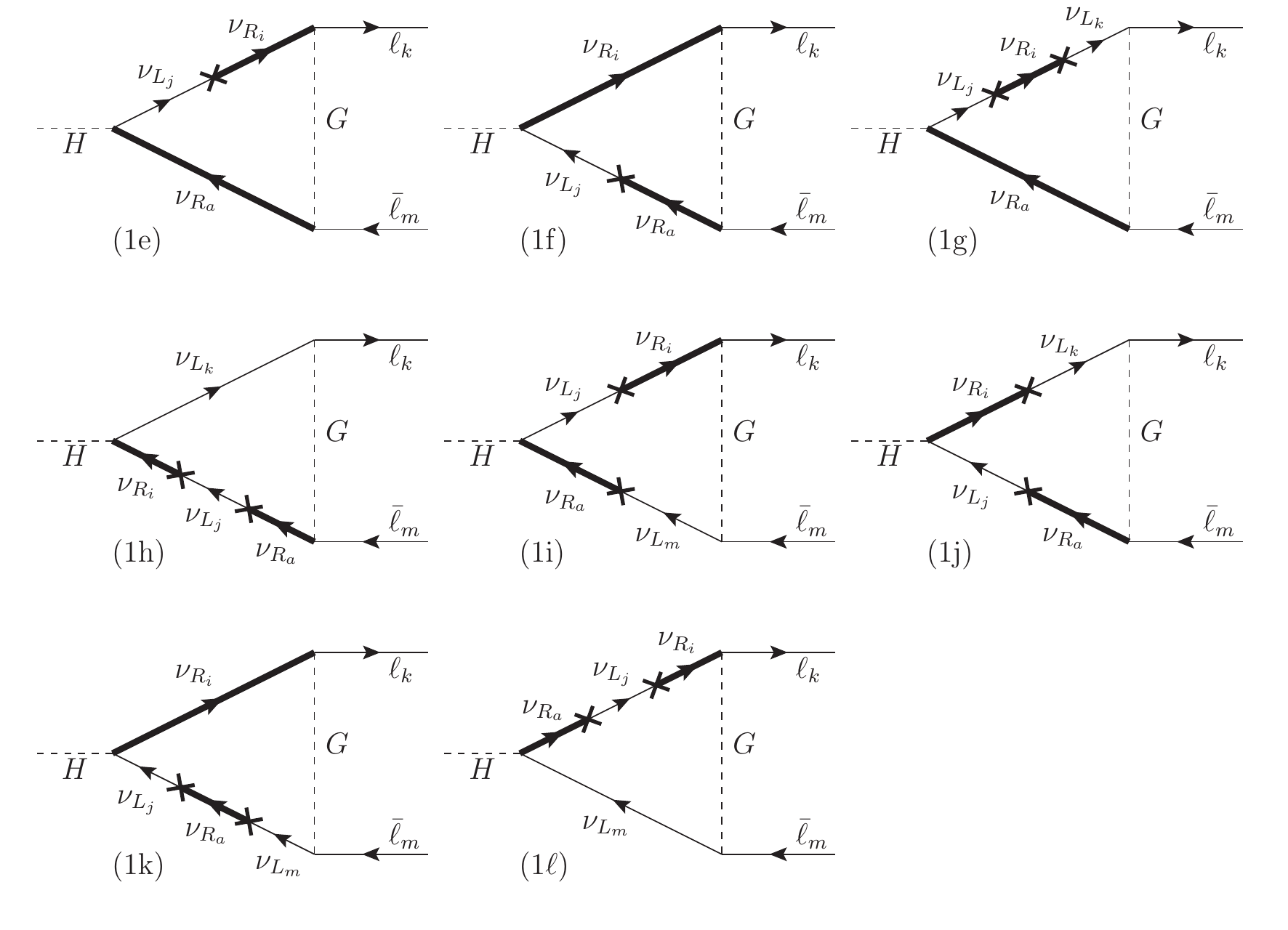}
\caption{Relevant vertex diagrams for the MIA form factors of LFVHD to ${\cal O}(Y_\nu^4)$.}
\label{LFVHDMIAdiagsvertexY4}
\end{center}
\end{figure}

\begin{figure}[t!]
\begin{center}
\includegraphics[scale=0.73]{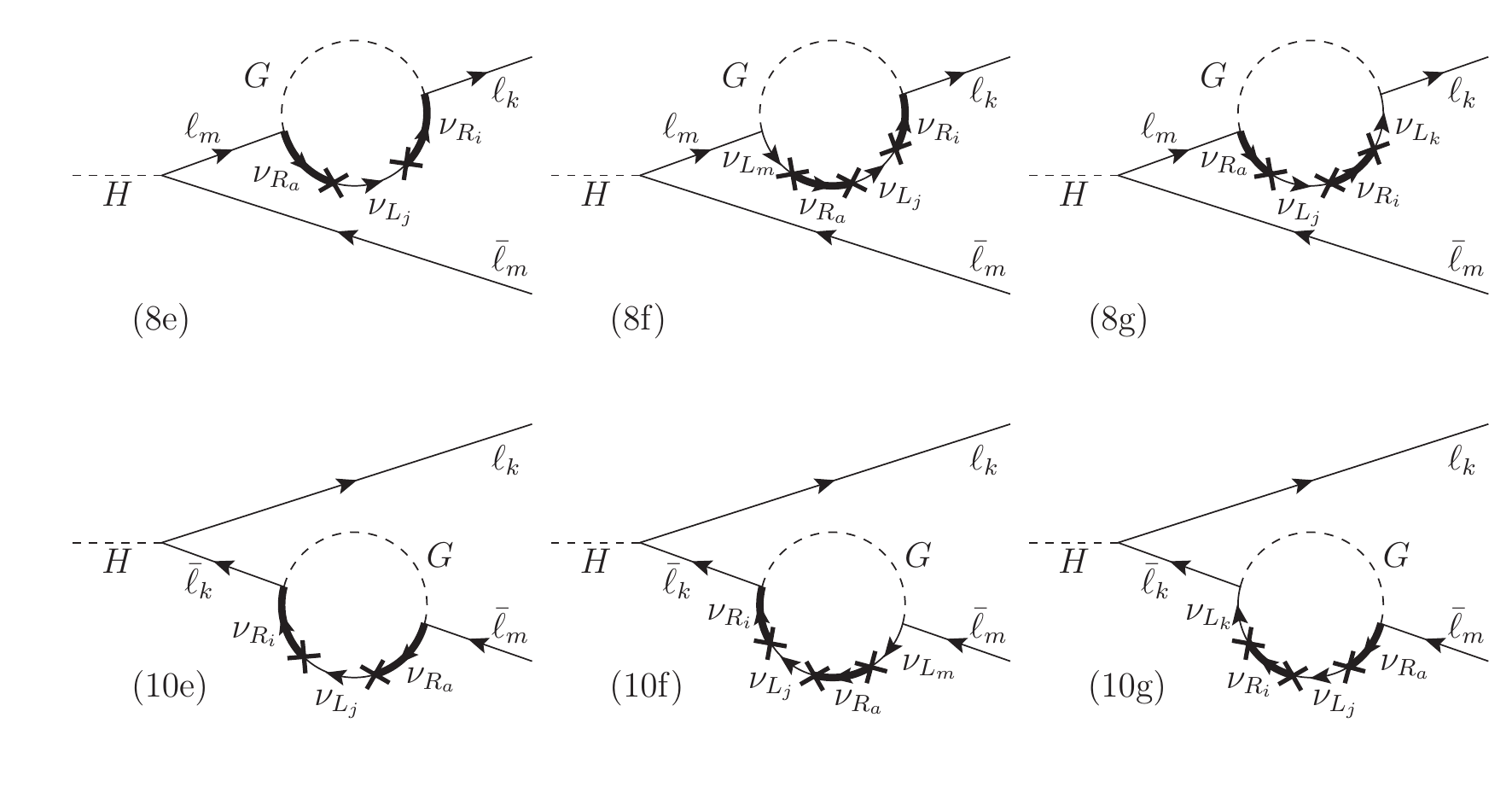}
\caption{Relevant external leg diagrams for the MIA form factors of LFVHD to ${\cal O}(Y_\nu^4)$.}
\label{LFVHDMIAdiagslegY4}
\end{center}
\end{figure}

For the one-loop computation in the MIA, we also follow a diagrammatic procedure that consists of the systematic insertion of right-handed  neutrino {\it fat-propagators} in all  possible places inside the loops, which are built with the interaction vertices and propagators of \figref{LFVHD_FR-MIA}. 
Generically, diagrams with one  right-handed neutrino propagator will contribute to the form factors of ${\cal O}(Y_{\nu}^{2})$, whereas diagrams with two right-handed neutrino propagators will contribute to the form factors of ${\cal O}(Y_{\nu}^{4})$.   
We show in \figrefs{LFVHDMIAdiagsvertexY2}, \ref{LFVHDMIAdiagslegY2}, \ref{LFVHDMIAdiagsvertexY4} and \ref{LFVHDMIAdiagslegY4} the relevant one-loop diagrams in the MIA corresponding to the dominant contributions of the LO and the NLO in \eqref{FFMIA}, respectively. 
These are also classified into those of vertex corrections  and those of leg corrections type. 
The MIA form factors are then obtained accordingly as the sum of all these contributions that can be summarized as follows:
\begin{equation}
F_{L,R}^{{\rm MIA}}=\sum_{{\rm i}=1}^{10} F_{L,R}^{\rm MIA (i)}\,.
\label{FFLRMIA}
\end{equation}

At LO, i.e., ${\cal O}(Y_{\nu}^{2})$,  each  $F_{L,R}^{\rm MIA (i)}$ receives contributions from all diagrams containing 1 right-handed neutrino propagator and one of these three combinations: i) 1 vertex with $\nu_R$ and 1 $m_D$ insertion,   ii)  0 vertices with $\nu_R$ and 2 $m_D$ insertions, iii)  2 vertices with $\nu_R$ and 0 $m_D$ insertions. 
This leads to the relevant  diagrams in \figrefs{LFVHDMIAdiagsvertexY2} and \ref{LFVHDMIAdiagslegY2} whose contributions are given, in an obvious  correlated notation, by:
\begin{align}
F_{L,R}^{{\rm MIA (1)\,\, (Y^2)}}&=F_{L,R}^{\rm  (1a)}+F_{L,R}^{\rm  (1b)}+F_{L,R}^{\rm  (1c)}+F_{L,R}^{\rm  (1d)}\,, \nonumber \\  
F_{L,R}^{{\rm MIA (2)\,\, (Y^2)}}&=F_{L,R}^{\rm  (2a)}+F_{L,R}^{\rm  (2b)} \,,
\nonumber \\  
F_{L,R}^{{\rm MIA (3)\,\, (Y^2)}}&=F_{L,R}^{\rm  (3a)} \,,
\nonumber \\ 
F_{L,R}^{{\rm MIA (4)\,\, (Y^2)}}&=F_{L,R}^{\rm  (4a)}+F_{L,R}^{\rm  (4b)} \,,
\nonumber \\ 
F_{L,R}^{{\rm MIA (5)\,\, (Y^2)}}&=F_{L,R}^{\rm  (5a)}+F_{L,R}^{\rm  (5b)} \,,
\nonumber \\ 
F_{L,R}^{{\rm MIA (6)\,\, (Y^2)}}&=F_{L,R}^{\rm  (6a)}+F_{L,R}^{\rm  (6b)}+F_{L,R}^{\rm  (6c)}+F_{L,R}^{\rm  (6d)} \,,
\nonumber \\ 
F_{L,R}^{{\rm MIA (7)\,\, (Y^2)}}&=F_{L,R}^{\rm  (7a)} \,,
\nonumber \\ 
F_{L,R}^{{\rm MIA (8)\,\, (Y^2)}}&=F_{L,R}^{\rm  (8a)}+F_{L,R}^{\rm  (8b)}+F_{L,R}^{\rm  (8c)}+F_{L,R}^{\rm  (8d)} \,,\nonumber \\  
F_{L,R}^{{\rm MIA (9)\,\, (Y^2)}}&=F_{L,R}^{\rm  (9a)} \,,
\nonumber \\ 
F_{L,R}^{{\rm MIA (10)\,\, (Y^2)}}&=F_{L,R}^{\rm  (10a)}+F_{L,R}^{\rm  (10b)}+F_{L,R}^{\rm  (10c)}+F_{L,R}^{\rm  (10d)} . 
\label{FFLRMIAY2}
\end{align}
The explicit analytical results for all these form factors are given in \eqrefs{FLtot_op2_Y2} and (\ref{FRtot_op2_Y2}) of the  \appref{App:FormFactorsMIA}.
These results are expressed in terms of the usual  one-loop Veltman-Passarino functions~\cite{Passarino:1978jh} of two points ($B_{0}$ and $B_{1}$), three points ($C_0$, $C_{11}$, $C_{12}$ and $\tilde{C}_{0}$) and four points ($D_{12}$, $D_{13}$ and $\tilde{D}_{0}$), whose definitions are given in \eqrefs{loopfunctionB}-(\ref{loopfunctionD}).

At NLO, i.e. ${\cal O}(Y_{\nu}^{4})$,  each  $F_{L,R}^{\rm MIA (i)}$ receives contributions from all diagrams  containing 2 right-handed neutrino propagators and one of these three combinations: i) 2 vertices with $\nu_R$ and 2 $m_D$ insertions,  ii)  3 vertices with $\nu_R$ and 1 $m_D$ insertion, iii)  1 vertex with $\nu_R$ and 3 $m_D$ insertions. 
Other possible combinations will provide subleading corrections in the heavy $M_R$ case of our interest,  since they will come with extra powers of $M_R$ in the denominator. 
Thus, we find that the most relevant diagrams are those of type (1), (8) and (10) summarized in  \figrefs{LFVHDMIAdiagsvertexY4} and \ref{LFVHDMIAdiagslegY4}, whose respective contributions are given by:
\begin{align}
F_{L,R}^{{\rm MIA (1)\,\, (Y^4)}}&=F_{L,R}^{\rm  (1e)}+F_{L,R}^{\rm  (1f)}+F_{L,R}^{\rm  (1g)}+F_{L,R}^{\rm  (1h)} 
+F_{L,R}^{\rm  (1i)}+F_{L,R}^{\rm  (1j)}+F_{L,R}^{\rm  (1k)}+F_{L,R}^{\rm  (1\ell)} \,,
\nonumber \\  
F_{L,R}^{{\rm MIA (8)\,\, (Y^4)}}&=F_{L,R}^{\rm  (8e)}+F_{L,R}^{\rm  (8f)}+F_{L,R}^{\rm  (8g)}  \,,
\nonumber \\  
F_{L,R}^{{\rm MIA (10)\,\, (Y^4)}}&=F_{L,R}^{\rm  (10e)}+F_{L,R}^{\rm  (10f)}+F_{L,R}^{\rm  (10g)}  .
\label{FFLRMIAY4}
\end{align}
Their explicit analytical results are collected in \eqrefs{FLtot_op2_Y4} and (\ref{FRtot_op2_Y4}) of the  \appref{App:FormFactorsMIA}.

\begin{figure}[b!]
\begin{center}
\includegraphics[width=.931\textwidth]{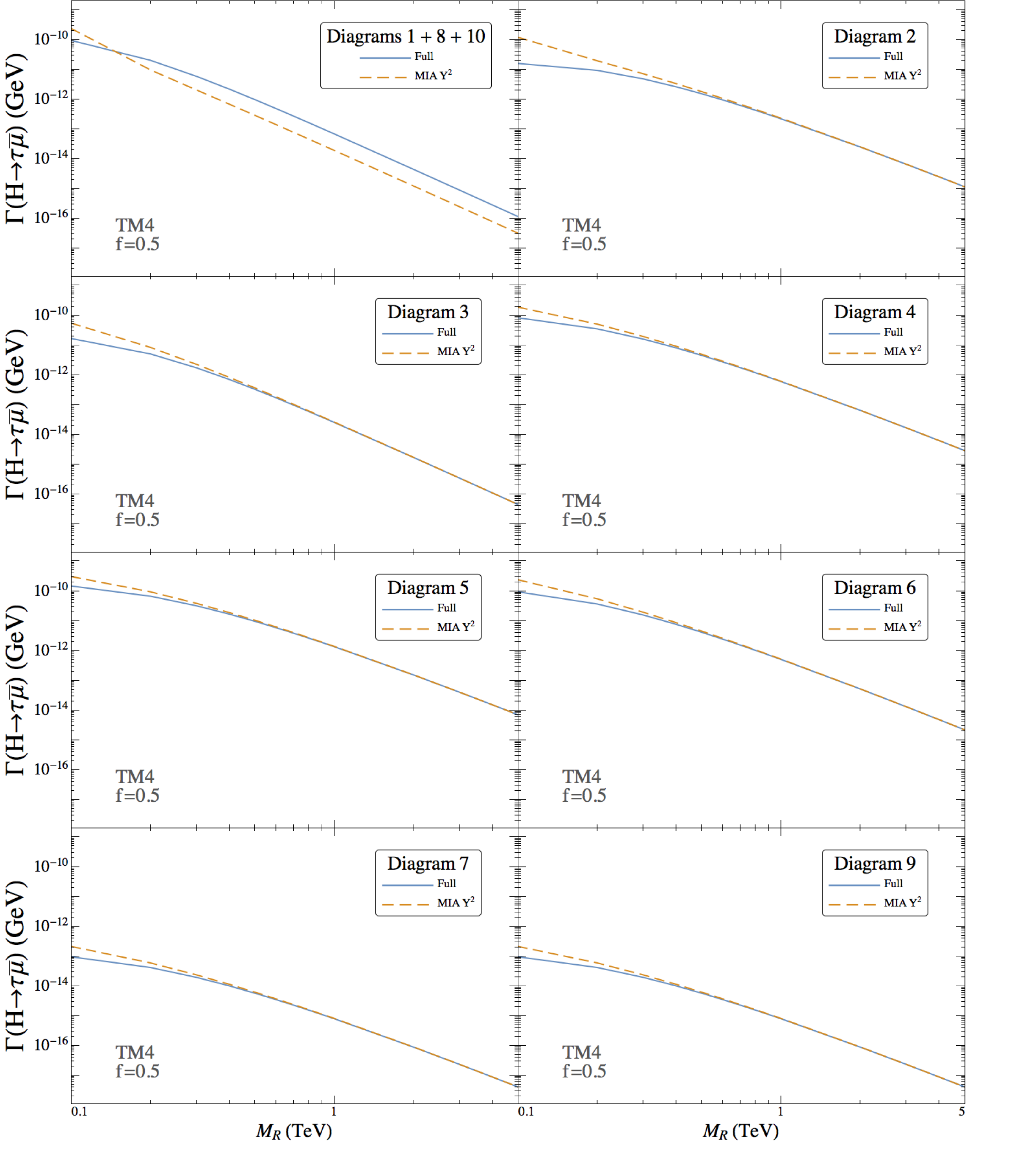}
\caption{Contributions from the various diagrams to $\Gamma (H \to \tau \bar \mu)$ as a function of $M_R$ in the TM-4 scenario from \tabref{TMscenarios} for $f=0.5$.   
Dashed lines are the predictions from the MIA to ${\cal O}(Y_\nu^2)$, while solid lines are from the full one loop computation in the mass basis.}\label{LFVHD_MIA_panelY2}
\end{center}
\end{figure}

 Some comments about the analytical properties of the previous MIA results are in order. 
 First, we analyze their  ultraviolet behavior.  
 From the results presented in \secref{sec:LFVHDanal}, we know that in the full one-loop computation of the mass basis, only the contributions to the amplitude from diagrams (1),  (8) and (10) of \figref{diagsLFVHDphysbasis} are ultraviolet divergent separately, and that the total sum from these diagrams (1)+(8)+(10) is finite, therefore providing a total one-loop amplitude that is ultraviolet finite as it must be. 
We have explored the divergences of the MIA diagrams and found the same results. 
Our calculation in the MIA also shows that diagrams of type (2), (3), (4), (5), (6), (7) and (9) are convergent separately, while each contribution of ${\cal O}(Y_{\nu}^{2})$ from diagrams (1), (8) and (10) is divergent, although their divergences cancel out again in their sum. 
For this reason, we will show (1)+(8)+(10) whenever we present results for each diagram in the next numerical analysis, which is convergent and therefore meaningful,  instead of the contributions from each of these diagrams separately. 
 
 Second, it is also worth to comment on the gauge invariance of our previous MIA results for the decay amplitude, computed  in the Feynman-'t Hooft gauge. 
 In order to prove the gauge invariance of our results, we have computed the amplitude also in other gauges and checked that we get the same result. 
 Specifically,  we have computed the form factors $F_{L,R}$  in the Unitary gauge and in an arbitrary $R_{\xi}$ gauge. 
 The details of the Unitary gauge computation are collected in \appref{App:othergauge}.

\begin{figure}[t!]
\begin{center}
\includegraphics[width=.49\textwidth]{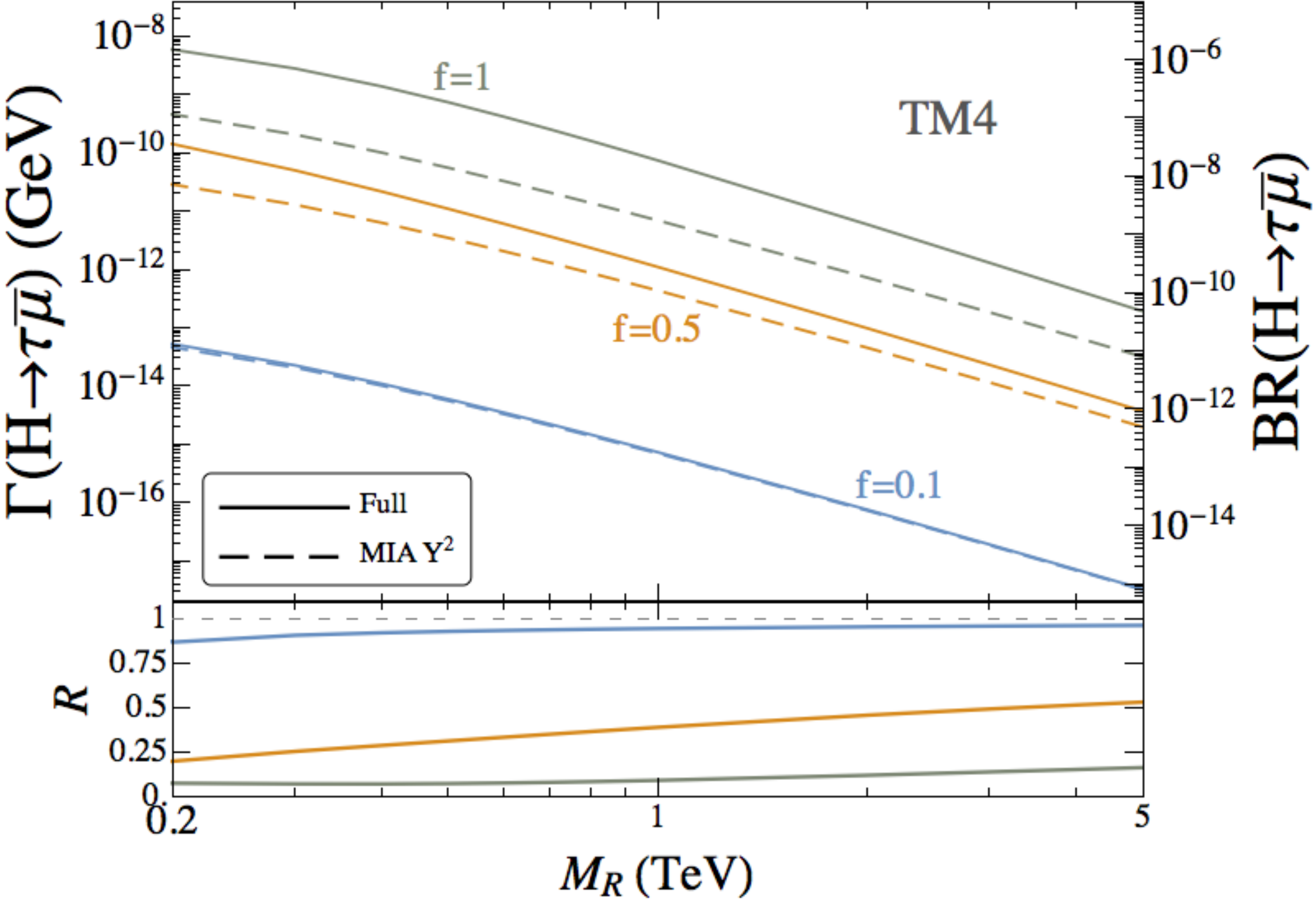}
\includegraphics[width=.49\textwidth]{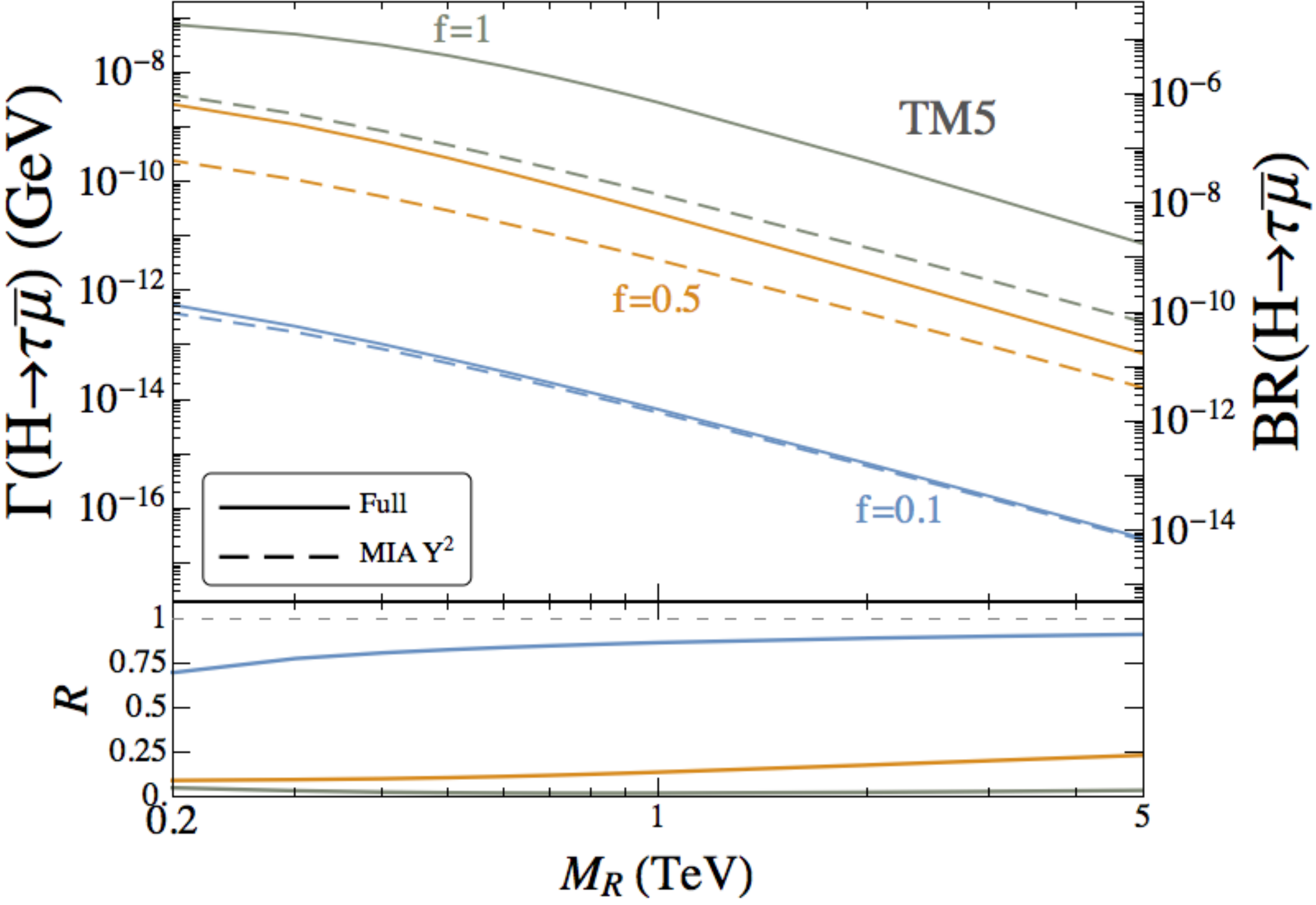}
\caption{Predictions for the partial width $\Gamma (H \to \tau \bar \mu)$ and branching ratio BR$(H \to \tau \bar \mu)$ as a function of $M_R$.  The dashed lines are the predictions from the MIA to ${\cal O}(Y_\nu^2)$. The solid lines are the predictions from the full one-loop computation in the mass basis. Here the scenarios TM4 (left panel) and TM5 (right panel) from \tabref{TMscenarios}   are chosen with f=0.1,0.5,1. In the bottom of these plots the ratio $R=\Gamma_{\rm MIA}/ \Gamma_{\rm full}$ is also shown. }\label{LFVHD_MIA_Y2}
\end{center}
\end{figure}

Next, we show our numerical results of these LO results in the MIA, together with the outcome from the full computation in the physical basis. 
We display our results only for scenarios TM-4 and TM-5 in \tabref{TMscenarios}, although we have found similar conclusions for other scenarios and input values of $Y_\nu$.

We first display in \figref{LFVHD_MIA_panelY2} the partial decay width of the full calculation together with our predictions from the MIA to ${\cal O}(Y_{\nu}^{2})$, separating the contributions from the various diagrams. 
We observe here that the contribution from each diagram (or group of diagrams in the case of (1)+(8)+(10))  to the form factor, and in consequence to the width, decreases with $M_R$. 
This behavior will be very well understood in \secref{sec:VeffMIA} with our simple formulas of the large $M_R$ expansions in \eqref{FLsimple_totdom}.
In particular, when adding the three contributions (1)+(8)+(10) in the MIA we will see explicitly the cancellation of the divergent contributions from $\Delta$ terms and the corresponding cancellation of the regularization $\mu$ scale dependent terms. 
The final behavior of the remaining finite terms in each form factor will go generically as $\sim (v^2/M_R^2)$, and in addition there will  also be logarithmic terms going as $\sim (v^2/M_R^2) ({\rm Log}(v^2/M_R^2))$. 

We see in \figref{LFVHD_MIA_panelY2} a consistent agreement between the MIA and full results for diagrams (2), (3), (4), (5), (6), (7) and (9). 
For the sum (1)+(8)+(10), the MIA reproduces the behavior of the full calculation very well but there is a mismatch in the partial decay width, which depends on the value of $f$. 
The larger $f$, the worse the discrepancy between them. 
This disagreement is then translated to the total partial width, as can be seen in \figref{LFVHD_MIA_Y2}.
In order to give a quantitative statement on this observation, we define the ratio $R=\Gamma_{\rm MIA}/ \Gamma_{\rm full}$. From the bottom of  \figref{LFVHD_MIA_Y2}, we have $R$ close to 1 for low values of $f$ ($f=0.1$) and large $M_R$ above 1 TeV. 
If we increase $f$ up to 1, poor values of $R$ far from 1 are obtained in the whole studied $M_R$ interval, so the MIA results to ${\cal O}(Y_{\nu}^{2})$ do not reproduce satisfactorily the results of the full calculation. 
We conclude that for large values of $f$, which are the interesting ones from a phenomenological point of view, the MIA only reproduces the functional behavior with $M_R$ but not the numerical values.
Thus, it is necessary to include in the MIA computation the next order contributions,  i.e., ${\cal O}(Y_{\nu}^{4})$.

\begin{figure}[t!]
\begin{center}
\includegraphics[width=.49\textwidth]{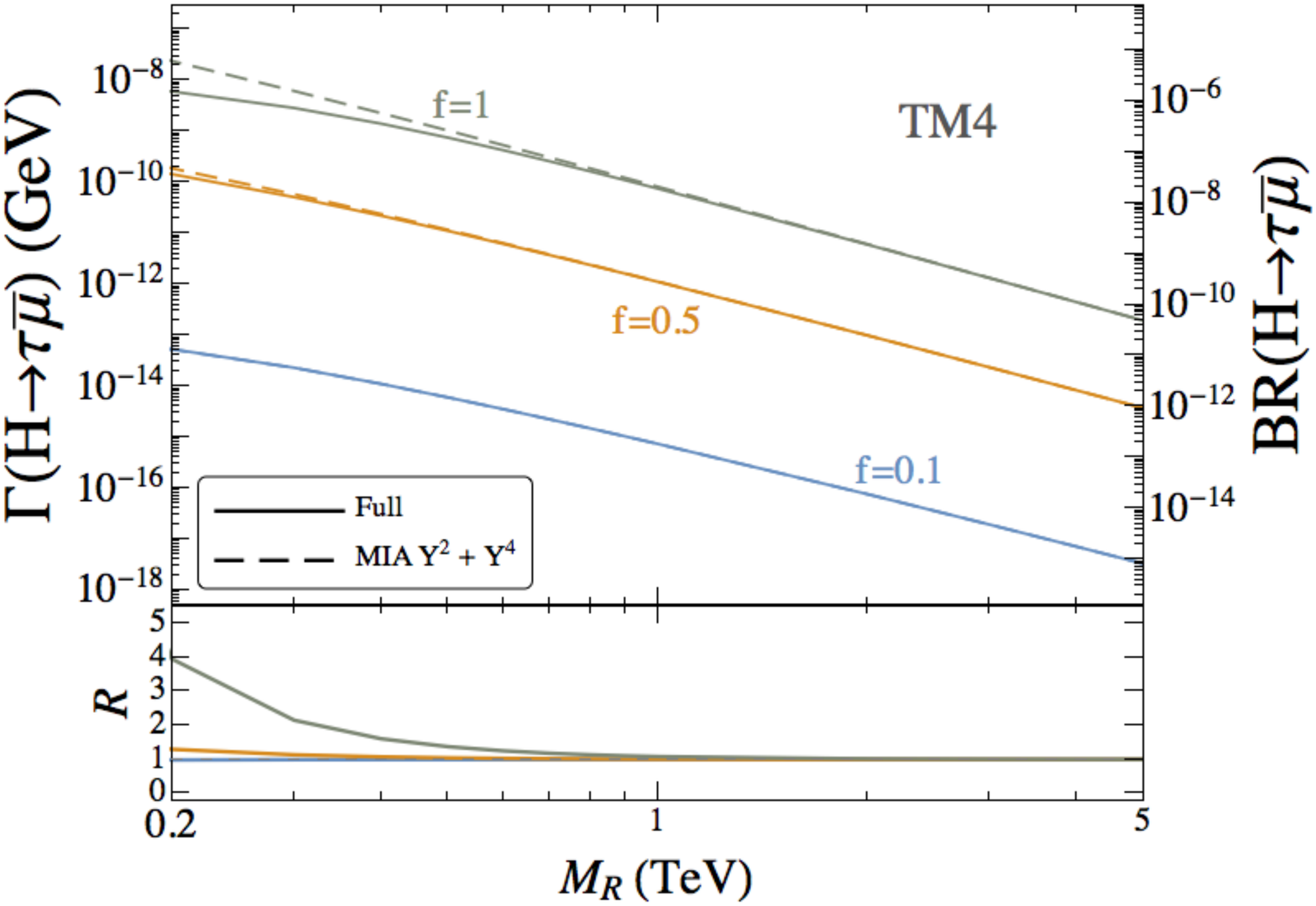}
\includegraphics[width=.49\textwidth]{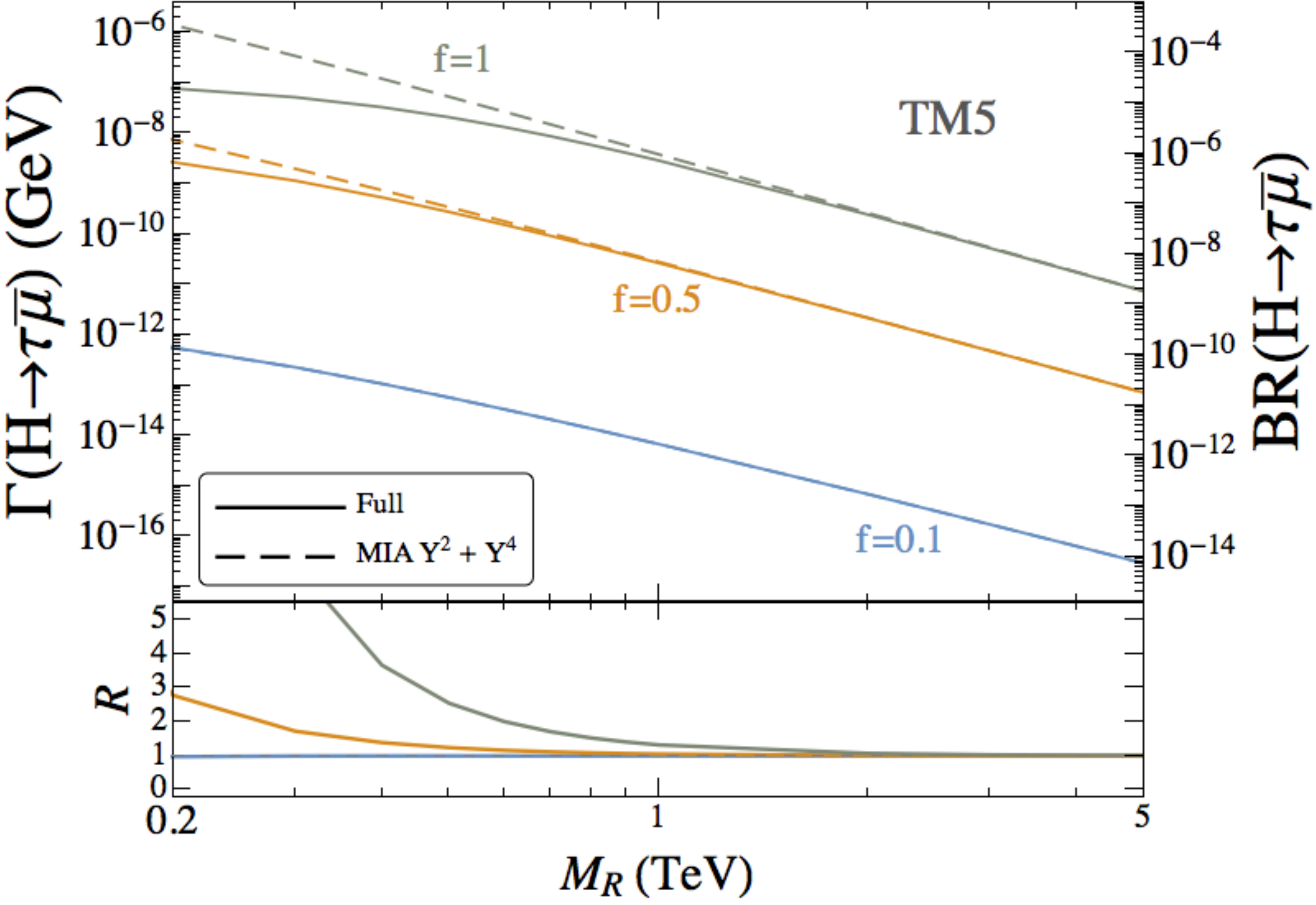}
\caption{Predictions for the partial width $\Gamma (H \to \tau \bar \mu)$ and branching ratio BR$(H \to \tau \bar \mu)$ as a function of $M_R$.  The dashed lines are the predictions from the MIA to ${\cal O}(Y_\nu^2+Y_\nu^4)$. The solid lines are the predictions from the full one-loop computation in the mass basis. Here the scenarios TM4 (left panel) and TM5 (right panel) from \tabref{TMscenarios}   are chosen with f=0.1,0.5,1. In the bottom of these plots the ratio $R=\Gamma_{\rm MIA}/ \Gamma_{\rm full}$ is also shown. }
\label{LFVHD_MIA_Y4}
\end{center}
\end{figure}

The results after including all the relevant  ${\cal O}(Y_{\nu}^{2}+Y_{\nu}^{4})$ contributions are shown in \figref{LFVHD_MIA_Y4}.
We can clearly see that the  sum of the MIA diagrams is in very good agreement with the full results for different values of $f$.
Therefore, we can conclude that our MIA calculation with the inclusion of the most relevant ${\cal O}(Y_{\nu}^{4})$ terms, corrects the ${\cal O}(Y_{\nu}^{2})$ contributions and achieves a better fit to the full numerical results for this process in the large $M_{R}\gg v Y_\nu$ mass range.  
In particular, we see this improvement with respect to ${\cal O}(Y_{\nu}^{2})$ contributions from the closeness of $R$ to 1 for different values of $f$. 
How large $M_{R}$ should be in order to get a good numerical prediction of the LFVHD rates  depends obviously on the size of the Yukawa coupling.  
For small Yukawa coupling, i.e., for small $f\lesssim 0.5$ the MIA works quite well for  $M_R$ above 400 GeV, whereas for larger couplings, say $f$ above 0.5,  the MIA also provides a good result but requires heavier $M_R$, above 1000 GeV.

\begin{figure}[t!]
\begin{center}
\includegraphics[width=.49\textwidth]{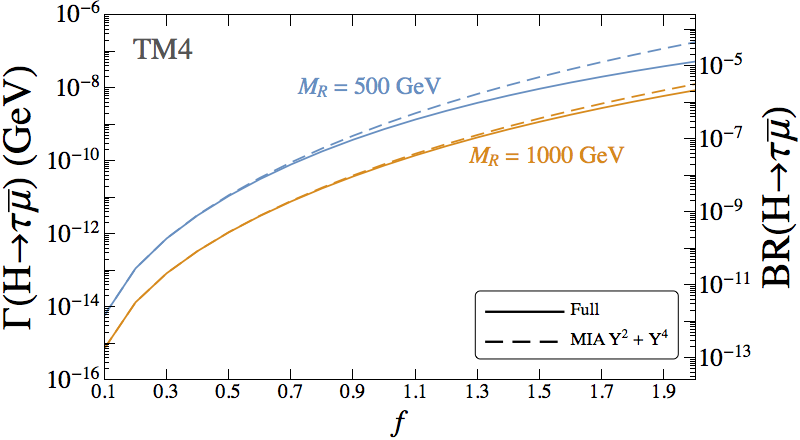}
\includegraphics[width=.49\textwidth]{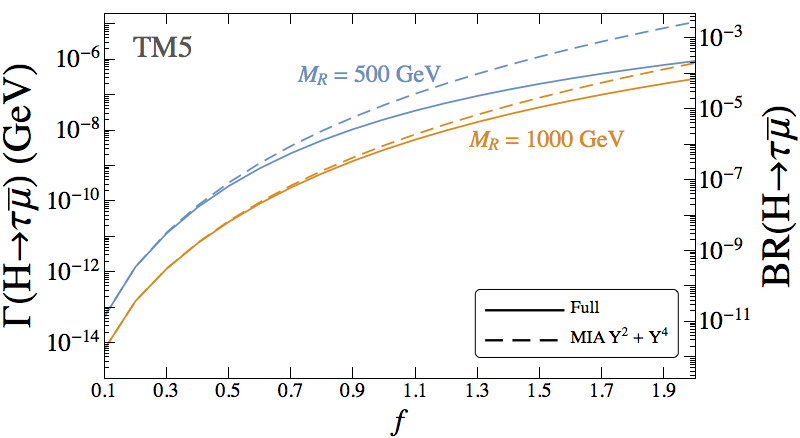}
\caption{Predictions for the partial width $\Gamma (H \to \tau \bar \mu)$ and branching ratio BR$(H \to \tau \bar \mu)$ as a function of the global Yukawa coupling strength $f$.  The dashed lines are the predictions from the MIA to ${\cal O}(Y_\nu^2+Y_\nu^4)$. The solid lines are the predictions from the full one-loop computation in the mass basis. Here the scenarios TM4 (left panel) and TM5 (right panel) from \tabref{TMscenarios} are chosen with $M_R=500,1000$ GeV.}
\label{LFVHD_MIA_f_Y4}
\end{center}
\end{figure}

 Before going to the derivation of the effective vertex in the large $M_R$ limit, we concentrate on the dependence of the branching ratios with $f$. 
In \figref{LFVHD_MIA_f_Y4}, we show the partial width and branching ratio as a function of $f$ for the scenarios TM4 and TM5 with two different values of $M_R$. 
In the perturbativity range of Yukawa couplings (implying approximately $f\lesssim 3.5$) we find a significant increase in the branching ratios up to ${\cal O}(10^{-4})$ for large $f\sim\mathcal O(2)$. 
However, for such large $f$ values the MIA provides an accurate prediction only for large $M_R$ values, say above 1000 GeV, as can be seen in  \figref{LFVHD_MIA_f_Y4}.
Overall, we can conclude that the results for the MIA form factors  to ${\cal O}(Y_\nu^2+Y_\nu^4)$ work reasonably well for  $M_R$ heavy enough, say above 1 TeV, and $f$ values not too large, such that $Y_\nu$ is within the perturbativity region, given by $Y_\nu^2/4\pi <1$. 

\subsection{Computation of the one-loop effective vertex for LFVHD}
\label{sec:VeffMIA}

In this Section we present our results in the large $M_R$ limit of the form factors involved in our computation of the LFVHD rates. 
In order to reach this simple expression for the effective vertex, valid in the large $M_R\gg v $ regime, we perform a systematic expansion  in powers of  $(v/M_R)$ of the one-loop  MIA  amplitude that we have computed in the previous Section.  
Generically,  the first order in this expansion is ${\cal O} (v^2/M_R^2)$, the next order is ${\cal O} (v^4/M_R^4)$, etc. 
There is also a logarithmic dependence with $M_R$, which is not expanded but left explicit in this calculation. 
In the final result for the effective vertex we will be interested just in the leading terms of ${\cal O} (v^2/M_R^2)$, which are by far the dominant ones for sufficiently heavy $M_R$.

We start with the formulas found in the previous Section and in \appref{App:FormFactorsMIA} for the one-loop LFVHD form factors in the MIA. 
Assuming the hierarchy $m_{\ell}\ll m_W, m_H\ll M_R$,  we may first ignore the tiny contributions that come from terms in the sum with factors of the lepton masses in our analytical results of \eqrefs{FLtot_op2_Y2}-(\ref{FRtot_op2_Y4}).
This leads to the following compact formula  for the total one-loop MIA form factors to ${\cal O}(Y_\nu^2+Y_\nu^4)$,  
\begin{align}
F_{L}^{{\rm MIA} }
&= \frac{1}{32 \pi^{2}} \frac{m_{\ell_k}}{m_{W}} \left(Y_{\nu}^{} Y_{\nu}^{\dagger}\right)^{km} \Big(  \tilde{C}_{0}(p_{2},p_{1},m_{W},0,M_{R}) - B_{0}(0,M_{R},m_{W})    \nonumber \\
&-  2m_{W}^{2} \big( (C_{0}+C_{11}-C_{12})(p_{2},p_{1},m_{W},0,M_{R}) + (C_{11}-C_{12})(p_{2},p_{1},m_{W},M_{R},0) \big)     \nonumber \\
&+  4m_{W}^{4} ( D_{12}-D_{13} )(0,p_{2},p_{1},0,M_{R},m_{W},m_{W})    \nonumber \\
&-  2m_{W}^{2}m_{H}^{2}D_{13}(0,p_{2},p_{1},0,M_{R},m_{W},m_{W}) 
 +2m_{W}^{2} \big(C_{0}+C_{11}-C_{12}\big)(p_{2},p_{1},M_{R},m_{W},m_{W})  \nonumber\\
&+  m_{H}^{2} \big(C_{0}+C_{11}-C_{12}\big)(p_{2},p_{1},M_{R},m_{W},m_{W})  \Big)  \nonumber\\
&+\frac{1}{32 \pi^{2}} \frac{m_{\ell_k}}{m_{W}} \left(Y_{\nu}^{} Y_{\nu}^{\dagger} Y_{\nu}^{} Y_{\nu}^{\dagger} \right)^{km} v^{2} \Big(-2(C_{11}-C_{12})(p_{2},p_{1},m_{W},M_{R},M_{R}) 
   \nonumber \\
&+   \tilde{D}_{0}(p_{2},0,p_{1},m_{W},0,M_{R},M_{R}) +\tilde{D}_{0}(p_{2},p_{1},0,m_{W},0,M_{R},M_{R}) -C_{0}(0,0,M_{R},M_{R},m_{W}) \Big). 
\label{FLdominant}
\end{align}
Here, we have ordered the various contributions as follows: the first line is from diagrams (1)+(8)+(10),  the second line from (2), the third line from (3), the fourth line from (4)+(5), the fifth line from (6) and the last two lines containing the ${\cal O}(Y_\nu^4)$ contribution are from (1)+(8)+(10). 
Notice that there are not final contributions from (7)+(9), since they cancel each other. 
Similarly, for the right-handed form factor we get:   
\begin{align}
F_{R}^{{\rm MIA}} &= \frac{1}{32 \pi^{2}} \frac{m_{\ell_m}}{m_{W}} \left(Y_{\nu}^{} Y_{\nu}^{\dagger}\right)^{km} \Big(  \tilde{C}_{0}(p_{2},p_{1},m_{W},M_{R},0) - B_{0}(0,M_{R},m_{W})    \nonumber \\
&  -2m_{W}^{2} ( C_{12}(p_{2},p_{1},m_{W},0,M_{R}) +(C_{0}+C_{12})(p_{2},p_{1},m_{W},M_{R},0) )    \nonumber \\
&  +4m_{W}^{4}  D_{13}(0,p_{2},p_{1},0,M_{R},m_{W},m_{W})    \nonumber \\
&  -2m_{W}^{2}m_{H}^{2}(D_{12}-D_{13})(0,p_{2},p_{1},0,M_{R},m_{W},m_{W}) +2m_{W}^{2} \big(C_{0}+C_{12}\big)(p_{2},p_{1},M_{R},m_{W},m_{W}) \nonumber\\
&  +m_{H}^{2} \big(C_{0}+C_{12}\big)(p_{2},p_{1},M_{R},m_{W},m_{W}) \Big)  \nonumber\\
&+\frac{1}{32 \pi^{2}} \frac{m_{\ell_m}}{m_{W}} \left(Y_{\nu}^{} Y_{\nu}^{\dagger} Y_{\nu}^{} Y_{\nu}^{\dagger} \right)^{km} v^{2} \Big(-2C_{12}(p_{2},p_{1},m_{W},M_{R},M_{R})
     \nonumber \\ 
&   +\tilde{D}_{0}(p_{2},p_{1},0,m_{W},M_{R},M_{R},0) +\tilde{D}_{0}(p_{2},0,p_{1},m_{W},M_{R},M_{R},0)-C_{0}(0,0,M_{R},M_{R},m_{W})
  \Big) \, ,
\label{FRdominant}
\end{align}
where the explanation for the various contributions in each line is as specified  for $F_L$. 
Note also that the right-handed form factor can be obtained from the left-handed one by exchanging $p_2$ and $m_{\ell_k}$ with $p_3$ and $m_{\ell_m}$, respectively.
From the previous compact formulas,  assuming the hierarchy $m_{\ell_{k}}\gg m_{\ell_{m}}$, it is also clear that the left-handed form factor is the dominant one for the decay mode $H \to \ell_{k} \bar{\ell}_{m}$. 
Conversely, the right-handed form factor will be the dominant one in the opposite case $m_{\ell_{m}}\gg m_{\ell_{k}}$. For the rest of this Section, we will assume $m_{\ell_{k}}\gg m_{\ell_{m}}$ and, therefore, we will focus on the dominant $F_L$.

The next step is to perform the large $M_R$ expansion of the loop integrals appearing in the MIA form factor. 
The details of how we perform these expansions are explained in \appref{App:LoopIntegrals}, where we also collect the results for both the loop integrals and the separate contributions to the form factors from all type of diagrams, i=1\dots10.
Finally, by plugging these large $M_R$ expansions into \eqref{FLdominant} we get the one-loop effective vertex, $F_L\simeq V_{H\ell_{k}\ell_{m}}^{\rm eff}$, which parametrizes the one-loop amplitude of  $H \to \ell_k {\bar \ell}_m$ as
\begin{equation}
i {\cal M} \simeq -i g \bar{u}_{\ell_k} V_{H\ell_{k}\ell_{m}}^{\rm eff} P_L  v_{\ell_m} \, , 
\label{amplitudeVeff}
\end{equation}
with the corresponding partial decay width: 
\be
\Gamma (H \to \ell_{k}\bar{\ell}_{m})\simeq \frac{g^{2}}{16\pi}m_{H}\big\vert V_{H\ell_{k}\ell_{m}}^{\rm eff}\big\vert^{2} \, .
\label{Gammaeff}
\ee
At the end, we find  the following simple result for the on-shell Higgs boson effective LFV vertex:
\be
\hspace{-.3cm}V_{H\ell_{k}\ell_{m}}^{\rm eff}=\frac{1}{64 \pi^{2}} \frac{m_{\ell_k}}{m_{W}}\left[  \frac{m_{H}^{2}}{M_{R}^{2}}
\left( r\Big(\frac{m_{W}^{2}}{m_{H}^{2}}\Big) +\log\left(\frac{m_{W}^{2}}{M_{R}^{2}}\right) \right) \left(Y_{\nu}^{} Y_{\nu}^{\dagger}\right)^{km} - \frac{3v^{2}} {M_{R}^{2}} \left(Y_{\nu}^{} Y_{\nu}^{\dagger} Y_{\nu}^{} Y_{\nu}^{\dagger} \right)^{km} \right],
\label{VeffMIA}  
\ee
where,
\begin{align}
r(\lambda)=&-\frac{1}{2} -\lambda -8\lambda^{2} +2(1-2\lambda +8\lambda^{2})\sqrt{4\lambda-1}\arctan\left(\frac{1}{\sqrt{4\lambda-1}}\right) \non\\
&+16\lambda^{2}(1-2\lambda)\arctan^2\left(\frac{1}{\sqrt{4\lambda-1}}\right). \label{lambdafunction}
\end{align}
 Notice that this solution is valid for $m_H<2m_W$ and that for the physical values of $m_H=125$~GeV and $m_W=80.4$~GeV we obtain numerically $r(m_W^2/m_H^2)\sim 0.31$.
 The partial width is then simplified correspondingly to:
\be
\Gamma (H \to \ell_{k}\bar{\ell}_{m})^{\rm MIA}= \frac{g^{2}m_{\ell_k}^{2}m_H}{2^{16} \pi^{5}m_{W}^{2}}  \bigg| \frac{m_{H}^{2}}{M_{R}^{2}}
\left( r\Big(\frac{m_{W}^{2}}{m_{H}^{2}}\Big) +\log\left(\frac{m_{W}^{2}}{M_{R}^{2}}\right) \right) \left(Y_{\nu}^{} Y_{\nu}^{\dagger}\right)^{km} - \frac{3v^{2}}{M_{R}^{2}} \left(Y_{\nu}^{} Y_{\nu}^{\dagger} Y_{\nu}^{} Y_{\nu}^{\dagger} \right)^{km} 
 \bigg|^{2} \, .
\label{widthsimple}
\ee
Some comments are in order. First we notice that the dominant behavior with $M_R$ of $V_{H\ell_{k}\ell_{m}}^{\rm eff}$  for large $M_R$ goes as $\log(M_{R}^{2})/M_{R}^{2}$ and the next dominant one goes as $1/M_{R}^{2}$. 
Second, the terms of ${\cal O}(Y_{\nu}^{2})$ depend on $m_H$, whereas the terms of  ${\cal O}(Y_{\nu}^{4})$ do not.  Notice also that the two contributions of  ${\cal O}(Y_{\nu}^{2})$ and ${\cal O}(Y_{\nu}^{4})$ get $M_R^2$ in the denominator and not $M_R^4$ as one could naively expect for the ${\cal O}(Y_{\nu}^{4})$ term. 
Third, we have also checked that we recover the simple phenomenological formula in \eqref{FIThtaumu}, which we obtained by a naive numerical fit of the dominant contributions at large $M_R$ from diagrams (1)+(8)+(10) in the physical mass basis.  Specifically, if we extract the contributions exclusively from diagrams (1), (8) and (10) in our MIA results in \eqref{widthsimple}, we get:
\begin{align}
{\rm BR} (H \to \ell_{k}\bar{\ell}_{m})^{\rm MIA}_{(1)+(8)+(10)}&= \frac{g^{2}m_{\ell_k}^{2}m_H}{2^{16} \pi^{5}m_{W}^{2}\Gamma_H}   \bigg\lvert \frac{m_{H}^{2}}{M_{R}^{2}}
  \left(Y_{\nu}^{} Y_{\nu}^{\dagger}\right)^{km} - \frac{3v^{2}}{M_{R}^{2}} \left(Y_{\nu}^{} Y_{\nu}^{\dagger} Y_{\nu}^{} Y_{\nu}^{\dagger} \right)^{km} 
 \bigg\rvert^{2} \, \nonumber \\
 &\simeq  10^{-7} \frac{v^4}{M_R^4}\Big\lvert \left(Y_{\nu}^{} Y_{\nu}^{\dagger}\right)^{km} -  5.7 \left(Y_{\nu}^{} Y_{\nu}^{\dagger} Y_{\nu}^{} Y_{\nu}^{\dagger} \right)^{km} 
\Big\rvert^{2},
\label{BR1810}
\end{align}
where in the last line we have used the numerical values of the physical parameters with $m_H=125$ GeV and $\Gamma_H$ given by the predicted value in the SM. 
 As announced,  we reach the previous  result in \eqref{FIThtaumu}.
 
It is also illustrative to compare our previous MIA result in \eqref{widthsimple} for the partial width in the large $M_R$ regime with the analytical approximate formula in Ref.~\cite{Pilaftsis:1992st} which was found after expanding the full one-loop computation in the physical neutrino mass basis in inverse powers of the physical heavy neutrino mass $m_N$.
Concretely  we compare our result in \eqref{widthsimple} with those in Eqs.~(26), (31)-(34) of Ref.~\cite{Pilaftsis:1992st}, which were obtained assuming $m_H^2/m_W^2\ll 4$ and $m_H^2/m_N^2\ll1$. 
By doing some algebra to express the physical neutrino couplings $B_{\ell_in_j}$ and $C_{n_in_j}$ appearing in those equations  in terms of the Yukawa couplings, and by extracting just the $m_H$ independent terms, we obtain complete agreement with our result in \eqref{widthsimple} in the $m_H\to0$ limit.
Specifically, by using
\begin{align}
\sum_{i\in{\rm Heavy}} B_{\ell_kn_i}B^*_{\ell_mn_i}&\simeq \frac{v^2}{m_N^2} \big(Y_\nu^{} Y_\nu^\dagger\big)^{km} \,,\\
\sum_{i,j\in{\rm Heavy}} B_{\ell_kn_i}C_{n_in_j}B^*_{\ell_mn_j}&\simeq \frac{v^4}{m_N^4} \big(Y_\nu^{} Y_\nu^\dagger Y_\nu^{} Y_\nu^\dagger\big)^{km} \,,
\end{align}
and neglecting ${\cal O}(1/m_N^4)$ and higher order terms, we get from Ref.~\cite{Pilaftsis:1992st}, 
\be
\Gamma (H \to \ell_{k}\bar{\ell}_{m})^{\rm full}= \frac{g^{2}m_{\ell_k}^{2}m_H}{2^{16} \pi^{5}m_{W}^{2}}  \bigg\lvert  -\frac{3m_W^2}{m_N^2}  \left(Y_{\nu}^{} Y_{\nu}^{\dagger}\right)^{km} - \frac{3v^{2}}{m_{N}^{2}} \left(Y_{\nu}^{} Y_{\nu}^{\dagger} Y_{\nu}^{} Y_{\nu}^{\dagger} \right)^{km} 
 \bigg\rvert^{2} \, ,
\label{widthmhzero}
\ee
which matches with our result in \eqref{widthsimple} after the substitution 
\be
(m_H^2/m_W^2) r(m_W^2/m_H^2) \to -3\,,
\ee
 corresponding to the limit $m_H \to 0$. 
In this sense, although a complete comparison is out of the scope of this work, we conclude that our MIA effective vertex and the effective vertex of the mass basis in Ref.~\cite{Pilaftsis:1992st} agree analytically in the  limit $m_H \to 0$.  
Nevertheless, we have checked by a numerical estimate of the LFVHD widths that the approximation of neglecting the Higgs boson mass in the effective vertex does not provide in general an accurate result and, therefore, in order to obtain a realistic estimate of these branching ratios, our final formula for the effective vertex in \eqref{VeffMIA} should be used, which is specific for on-shell Higgs decays and accounts properly for the Higgs boson mass effects. 

\begin{figure}[t!]
\begin{center}
\includegraphics[width=.496\textwidth]{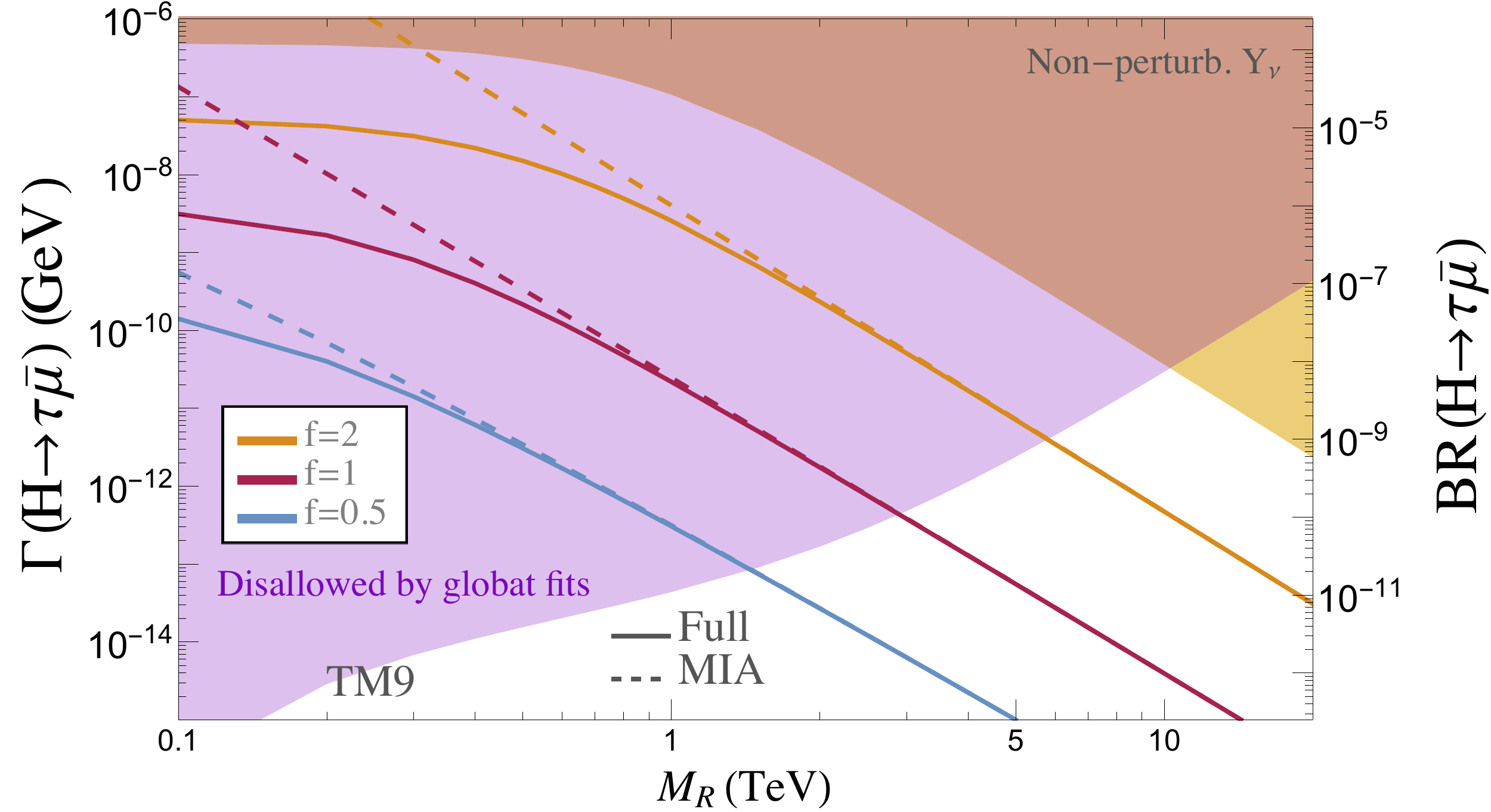}
\includegraphics[width=.496\textwidth]{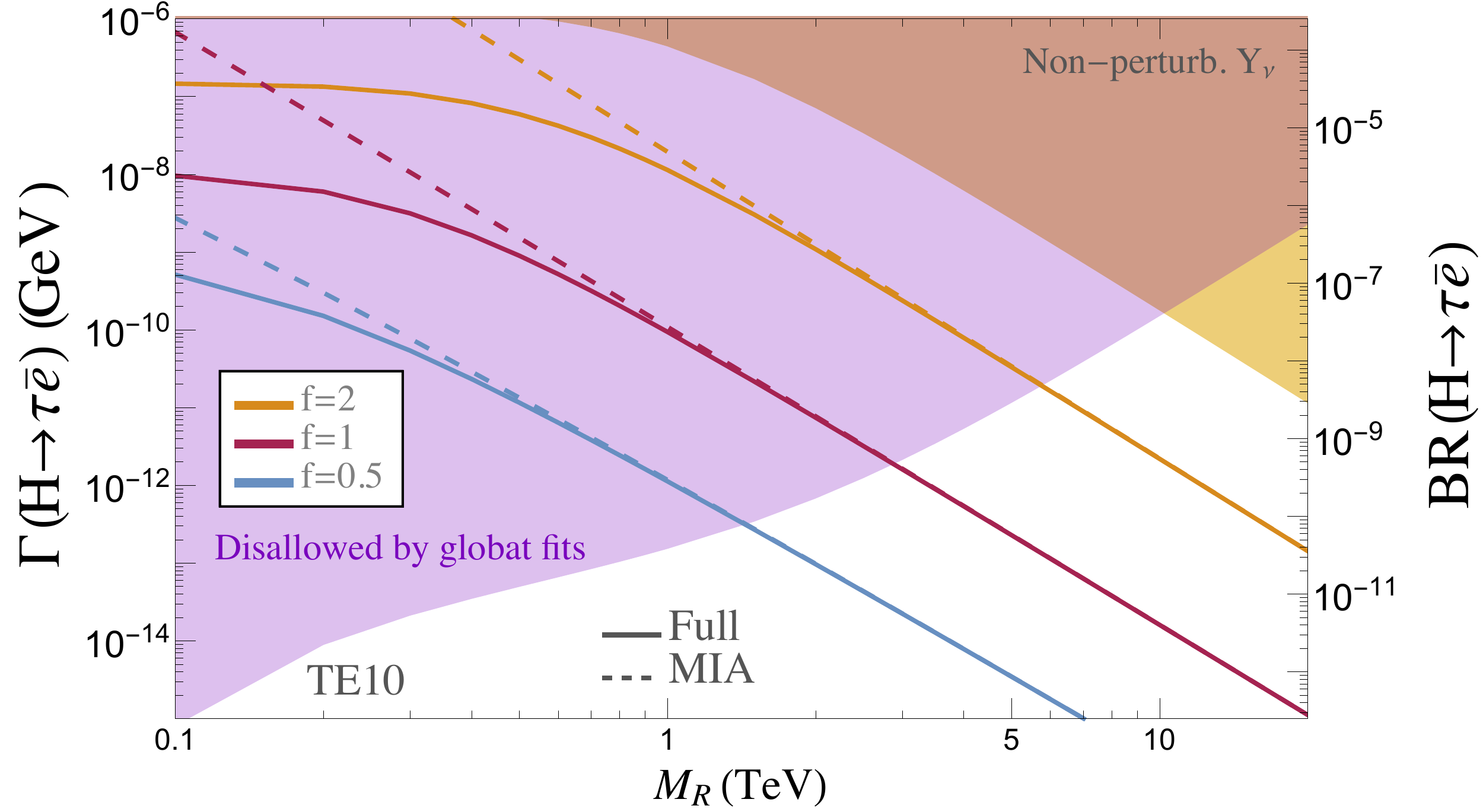}
\caption{Left panel: Predictions for $H \to \tau \bar \mu$ with the effective vertex computed with the MIA (dashed lines) for $Y_\nu^{\rm TM9}$. Right panel:
Predictions for $H \to \tau \bar e$ with the effective vertex computed with the MIA (dashed lines) for $Y_\nu^{\rm TE10}$. The chosen examples TM9 and TE10 are explained in the text. Solid lines are the corresponding predictions from the full one-loop computation in the mass basis. Shadowed areas to the left part of these plots (in purple) are disallowed by global fits. Shadowed areas to the right part of these plots (in yellow) give a non-perturbative Yukawa coupling. }
\label{LFVHDeffvTM9TE10}
\end{center}
\end{figure}

Finally, we wish to illustrate numerically the accuracy of our simple results of the effective vertex and the partial width in \eqrefs{VeffMIA} and (\ref{widthsimple}), respectively. 
For that purpose, we compare again our numerical predictions from these simple formulas  with the predictions from the full one-loop results of the mass basis in \figref{LFVHDeffvTM9TE10}.
Here, we display the results for the most interesting channels, $H \to \tau \bar\mu$ and $H \to \tau \bar e$, and for two scenarios, $Y_\nu^{\rm TM9}$ and $Y_\nu^{\rm TE10}$ in \tabref{TMscenarios}. 
Nonetheless, we have checked that the effective vertex in \eqrefs{VeffMIA} also works for other choices of scenarios. 
 
The plots in \figref{LFVHDeffvTM9TE10}  show the predictions of both the LFVHD partial widths and branching ratios, as functions of $M_R$ and three different values of $f=2,1,0.5$, with the colored areas being disallowed by global fit constraints (purple) or by non-perturbative Yukawa couplings (yellow).
Concretely, we have imposed the constraints on all the entries of the $\eta$ matrix (see \eqref{etamax3sigma}) that we have taken from the global analysis in Ref.~\cite{Fernandez-Martinez:2016lgt} at the three sigma level. 
For the perturbativity bound, we impose the condition $|Y_\nu^{ij}|^2/(4 \pi) <1 $ for every entry of the $Y_\nu$ coupling matrix. 
The areas in white are in consequence the regions that are allowed by the global fits and by perturbativity. 

From these plots we learn that the agreement between the full prediction and the MIA result obtained from the effective vertices computed in this Section is quite good for values of $M_R$ above 1 TeV and for all the explored Yukawa coupling examples.
In fact, the MIA works extremely well for the whole region of interest where both the global fits constraints and the perturbativity of the Yukawa coupling are respected. 
Consequently, we conclude that the simple expression for the effective vertex in \eqref{VeffMIA} is quite accurate and provides a good approximation for the partial width in \eqref{widthsimple}, in agreement with the full LFVHD rates. 
Therefore, it is a very useful formula for making fast numerical estimations of these rates in terms of the relevant model parameters $Y_\nu$ and $M_R$.


\section[Maximum allowed BR($H\to\ell_k\bar\ell_m$)]{Maximum allowed BR($\boldsymbol{H\to\ell_k\bar\ell_m}$)} 
\label{sec:LFVHDmax}

We conclude this Chapter by combining everything we have learnt about LFV H decays and by trying to conclude on the maximum rates allowed by a more complete set of present constraints, including both LFV and lepton flavor preserving observables. 
For this purpose, we consider the constraints obtained by the global fit analysis done in Ref.~\cite{Fernandez-Martinez:2016lgt}, where upper bounds on the $\eta$ matrix were derived. 
More concretely, these constraints define a maximum allowed by data $\eta$ matrix given by:
\begin{align}
 \eta_{3\sigma}^{\rm max}=\left(\begin{array}{ccc}
1.62\times 10^{-3}&1.51\times 10^{-5}&1.57\times 10^{-3}\\1.51\times 10^{-5}&3.92\times 10^{-4}&9.24\times 10^{-4}
\\1.57\times 10^{-4}&9.24\times 10^{-4}&3.67\times 10^{-3}
\end{array}\right)\,,
\label{etamax3sigma}
\end{align}

We can easily apply these bounds by means of the $\mu_X$ parametrization introduced in \eqref{MUXparam}.
As we said, this parametrization allows us to choose the $Y_\nu$ and $M_R$ matrices as input parameters of the model.
In our situation of degenerate $M_R$ matrix, the $\eta$ matrix is related to the Yukawa matrix approximately by,
\be\label{etaY2}
\eta=\frac{v^2}{2M_R^2}\, Y_\nu^{} Y_\nu^\dagger\,.
\ee
Therefore, we can combine \eqref{etamax3sigma} and (\ref{etaY2}) in order to find a Yukawa matrix that saturates the $\eta_{3\sigma}^{\rm max}$ bounds. 
Then, the \eqref{MUXparam} will ensure the agreement with oscillation data by building the proper $\mu_X$ matrix. 

A possible solution to this problem is given by our choice: 
\begin{align}
 Y_{\nu}^{\rm GF}=f\left(\begin{array}{ccc}
 0.33&0.83&0.6\\-0.5&0.13&0.1\\-0.87&1&1
 \end{array}\right)\,,
 \label{YnuGF}
\end{align}
which saturates the $\eta_{3\sigma}^{\rm max}$ bounds in a parameter space line given by the ratio $f/M_R= (3/10)\, {\rm TeV}^{-1}$, i.e., for  $(f,M_R)=(3,10\, {\rm TeV})$,  $(1,3.3\, {\rm TeV})$,  $(0.3,1\, {\rm TeV})$, \dots, etc. 
The Yukawa coupling matrix in \eqref{YnuGF} defines our last scenario, that we refer to as GF.

\begin{figure}[t!]
\begin{center}
\includegraphics[width=.496\textwidth]{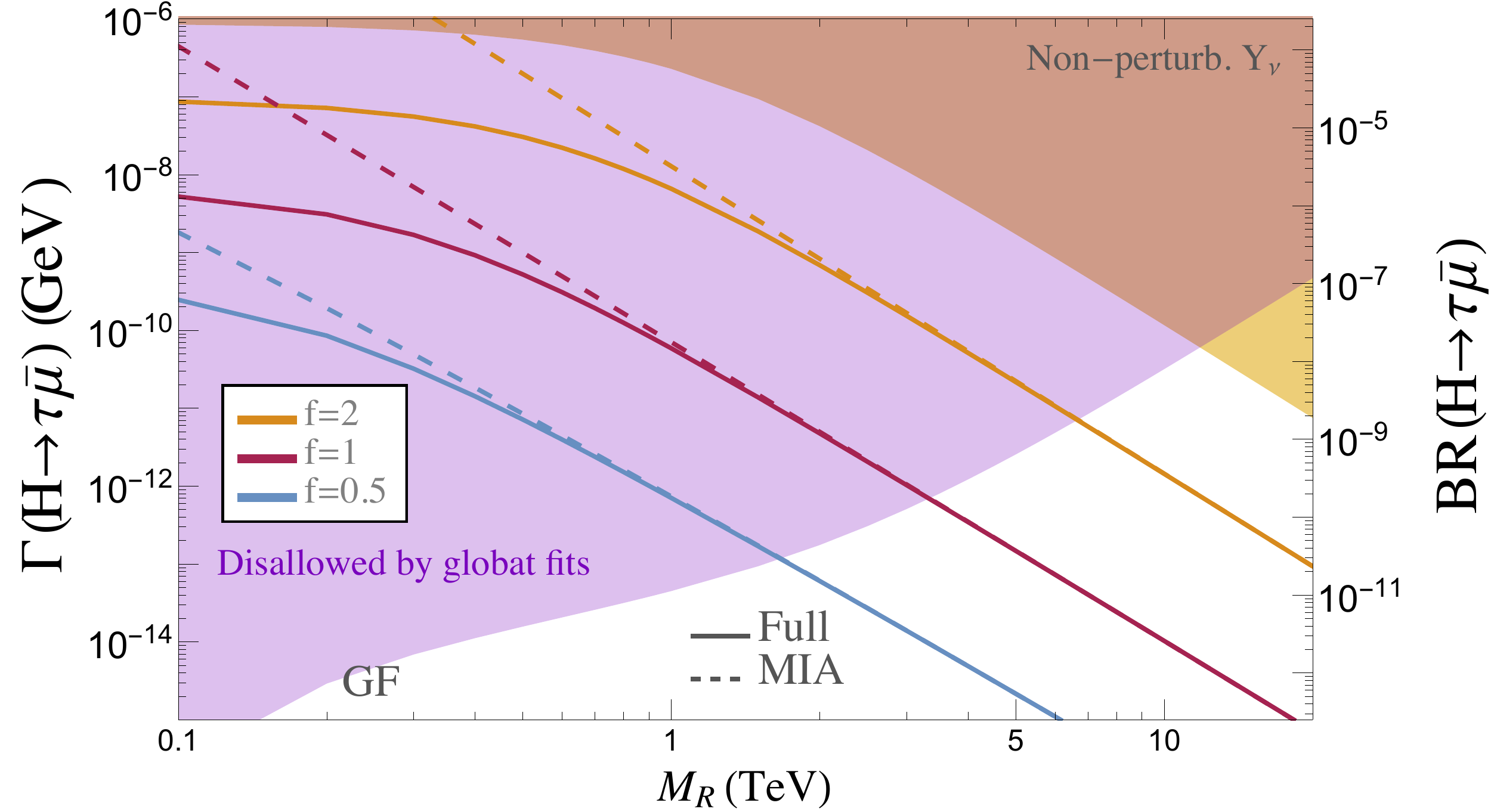}
\includegraphics[width=.496\textwidth]{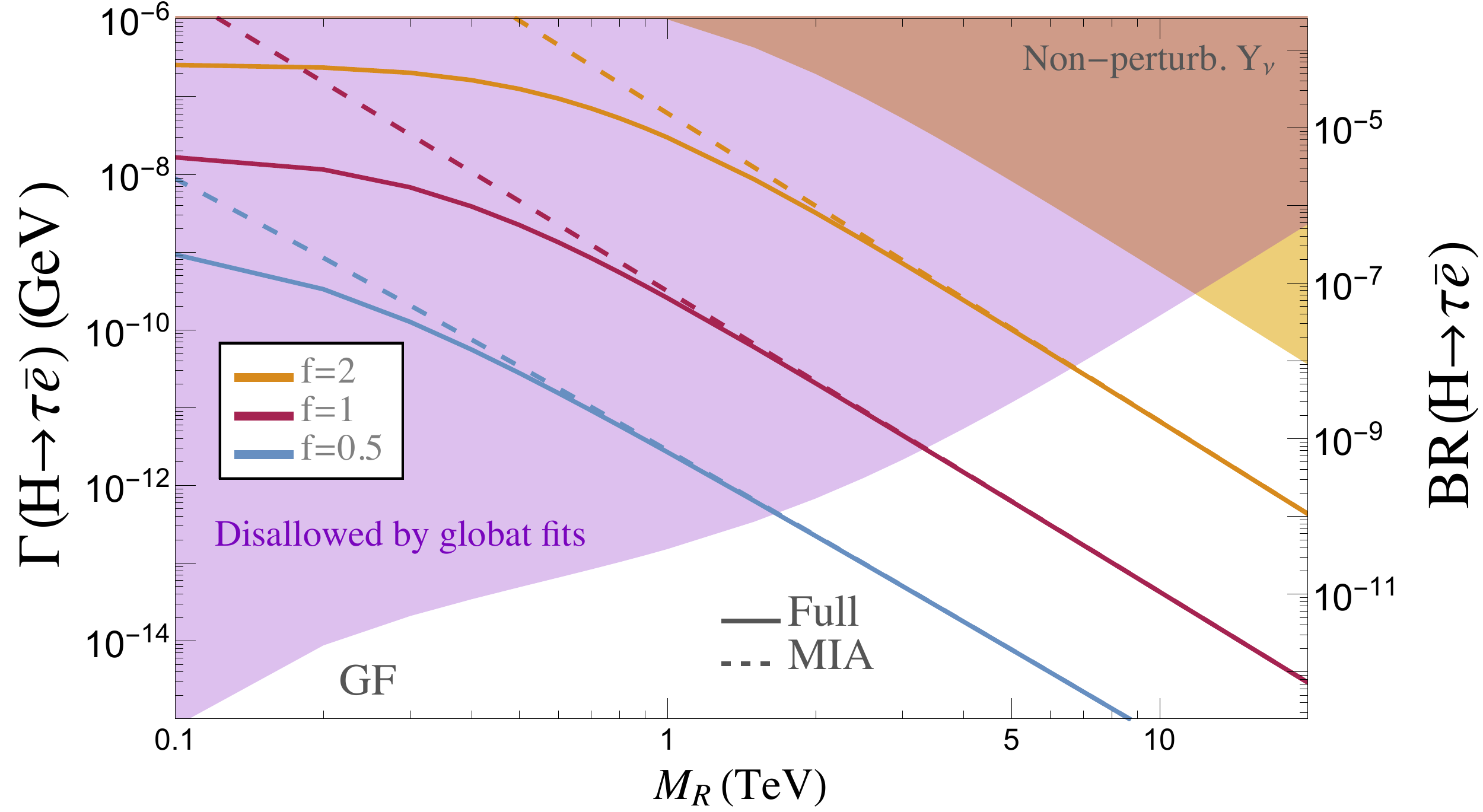}
\caption{Predictions for $H \to \tau \bar \mu$ (left panel) and 
$H \to \tau \bar e$ (right panel) with the effective vertex computed with the MIA (dashed lines) for $Y_\nu^{\rm GF}$. The chosen example GF is explained in the text. Solid lines are the corresponding predictions from the full one-loop computation in the mass basis. Shadowed areas to the left part of these plots (in purple) are disallowed by global fits. Shadowed areas to the right part of these plots (in yellow) give a non-perturbative Yukawa coupling.}
\label{LFVHD_GF}
\end{center}
\end{figure}

We show the numerical results in \figref{LFVHD_GF} for the most promising channels $H\to\tau\bar\mu$ and $H\to\tau\bar e$ in the GF scenario. We also computed the $H\to\mu\bar e$ channel, but we do not show the results for this case here, since the rates are extremely small due to strong bounds on $\eta_{\mu e}$.
As before, the purple area covers the parameter space region where the $\eta$ matrix is above the 3$\sigma$ bound in \eqref{etamax3sigma}, at least in one entry. The yellow area represents violation of the perturbativity bound
\be\label{Ymax1}
\frac{\big|Y_{ij}\big|^2}{4\pi}<1\,,\quad {\rm for}~i,j=1,2,3\,.
\ee
Our predictions are done with both full expressions in the mass basis, derived in \secref{sec:LFVHDanal}, and the effective vertex obtained in \secref{sec:LFVHDMIA}.
We see again that the agreement of the simple formula in \eqref{VeffMIA} is excellent in all the region allowed by both the global fit and the perturbativity constrains. 
Finally, we can conclude on the maximum LFVHD rates allowed by the complete set of present constraints as extracted from the approach of a global fit analysis. From \figref{LFVHD_GF} we learn that the maximum allowed branching ratios values are, respectively:
\begin{align}
{\rm BR}(H \to \tau \bar\mu)_{\rm max} &\sim 10^{-8}\,, \\  
{\rm BR}(H \to \tau \bar e)_{\rm max} &\sim 10^{-7}\,.
\end{align}

Finally, we shortly summarize  our findings in this Chapter, where we have explored in full detail the LFV H decays induced from right-handed neutrinos from the ISS model. 
We found that, having these neutrinos TeV scale masses and large Yukawa couplings at the same time, the LFVHD rates are manifestly enhanced with respect to the standard type-I seesaw, where rates of $\mathcal O(10^{-30})$ were obtained~\cite{Arganda:2004bz}.
After applying present bounds from a global fit analysis, we found maximum allowed rates of $10^{-7}$ and $10^{-8}$ for $H\to\tau\bar e$ and $H\to\tau\bar\mu$, respectively, which unfortunately are far from present LHC experimental sensitivities. 
However, in our aim of fully understanding the predictions for this observable, we have derived a one-loop effective vertex for the LFV interaction of our interest here, namely, the interaction of a Higgs boson with two leptons of different flavor $H\ell_k\ell_m$ with $k \neq m$.  
With such a simple expression for the involved effective vertex, one may perform a fast estimate of the LFV Higgs decay rates for many different input parameter values, mainly for  $Y_\nu$ and $M_R$, without the need of a diagonalization process to reach the physical neutrino basis, and thus avoiding the full computation of the one-loop diagrams in this basis, which is by far more computer time consuming. 
Moreover, the explicit dependence on the relevant parameters $Y_\nu$ and $M_R$ facilitates  the interpretation of the numerical results.
We find this simple formula useful also for other models that share the desired basic properties with the ISS model, since they can be easily used by other authors to obtain a fast estimate of the LFVHD rates, which are ready for an easy test with experimental data.

We have also explored these rates in the SUSY-ISS model, finding that new loops involving SUSY particles, sleptons and sneutrinos, could enhance the predictions for the allowed LFVHD rates close to the present or near future experimental sensitivities.


\chapter{LFV Z decays from low scale seesaw neutrinos}
\fancyhead[RO] {\scshape LFV Z decays from low scale seesaw neutrinos}
\label{LFVZD}

As we previously discussed, the LHC is providing new data on lepton flavor violating $Z$ boson decays,  $Z\to\ell_k\bar\ell_m$. 
After LHC run-I, the ATLAS experiment has improved previous bounds for $Z\to\mu e$ \cite{Aad:2014bca} and it is already at the level of the LEP results for the $Z\to\tau\mu $ channel.
These searches will continue in the next LHC runs with more luminosity and higher energies and, thus, the LHC will provide new data on these LFV observables. 
Furthermore, the sensitivity to LFVZD rates is expected to highly improve at future linear colliders, with an expected sensitivity of $10^{-9}$ \cite{Wilson:I, Wilson:II}, or at a Future Circular  $e^+ e^-$ Collider (such as FCC-ee (TLEP)\cite{Blondel:2014bra}), where it is estimated that up to $10^{13}$ $Z$ bosons would be produced and the sensitivities to LFVZD rates could be improved up to $10^{-13}$.
Therefore, we consider extremely timely to explore the predictions for these LFVZD rates in any new physics scenario that could be related to neutrino physics, as it has been previously done in beyond the Standard Model  frameworks like those with massive (Majorana and/or Dirac) neutrinos~\cite{Ilakovac:1994kj,Mann:1983dv,Bernabeu:1987gr,Dittmar:1989yg,Korner:1992an,Ilakovac:1999md,Illana:1999ww,Illana:2000ic,Abada:2014cca,Abada:2015zea}, or those using an effective field theory approach~\cite{Perez:2003ad,FloresTlalpa:2001sp,Delepine:2001di,Davidson:2012wn}.

In this Chapter, we consider again the inverse seesaw model with three pairs of right-handed neutrinos
as a specific realization of the low scale seesaw models, which, as we saw in previous Chapters, it is an appealing model with a very rich phenomenology, in particular for the charged LFV processes.
Nevertheless, as we discussed before, the present experimental upper bounds in  \tabref{LFVsearch} for cLFV processes involving $\mu$-$e$ transitions, here called ${\rm LFV}_{\mu e}$ in short, are much stronger that the ones in the other sectors, i.e., cLFV processes involving $\tau$-$\mu$ and $\tau$-$e$ transitions, named here in short ${\rm LFV}_{\tau \mu}$ and ${\rm LFV}_{\tau e}$, respectively. 
These very stringent constraints in the $\mu$-$e$ sector motivate the directions in the parameter space considered here, which incorporate automatically this suppression in their input.
Specifically, we will implement this $\mu$-$e$ suppression requirement within the context of the ISS, by working with the TM and TE scenarios we introduced in \tabref{TMscenarios}. 
These particular ISS settings with suppressed ${\rm LFV}_{\mu e}$ rates provide very interesting scenarios for exploring the relevant ISS parameter space directions that may lead to large cLFV rates in the other sectors, $\tau$-$\mu$ and/or $\tau$-$e$.

 Motivated by all the peculiarities exposed above, in this Chapter we perform a dedicated study of the LFVZD rates, in particular BR($Z\to\tau\mu$) and BR($Z\to\tau e$), in the context of these ISS scenarios with an {\it ad-hoc} suppression of ${\rm LFV}_{\mu e}$ rates, which will be called from now on ISS-$\cancel{\rm LFV}\hskip-.1cm_{\mu e}$ in short.
LFVZD processes in the presence of low scale heavy neutrinos have recently been studied considering the full one-loop contributions~\cite{Abada:2014cca} or computing the relevant Wilson coefficients~\cite{Abada:2015zea}.
In these works, maximum allowed LFVZD rates in the reach of future linear colliders were found when considering a minimal ``3+1'' toy model, with BR($Z\to\tau\mu$) up to $\mathcal O(10^{-8})$ for a neutrino mass in the few TeV range.
For more realistic models, like the (2,3) or (3,3) realizations of the ISS model, and after imposing all the relevant theoretical and experimental  bounds, smaller LFVZD rates  were achieved, BR$(Z\to\tau\mu)\lesssim\mathcal O(10^{-9})$, which would be below the reach of future linear colliders sensitivities and might be accessible only at future circular $e^+e^-$ colliders.
The main difference of our study with the ones previously done relies on the different settings of the ISS parameters, as we will focus on some specific directions that are more difficult to access with a random scan of the ISS parameter space.
We have learnt about this issue when studying the LFV Higgs boson decays in \chref{LFVHD}.
In the following, we will perform a complementary analysis to the one in Ref.~\cite{Abada:2014cca} and we will show that larger maximum allowed rates for BR($Z\to\tau\mu$) and BR($Z\to\tau e$) can be obtained by considering the particular TM and TE scenarios in \tabref{TMscenarios}, such that for some specific directions of the parameter space they could be reached at future linear colliders.
The results presented in this Chapter have been published in Ref.~\cite{DeRomeri:2016gum}.

\section{LFV Z decays in the ISS model}
\label{sec:LFVZDISS}

LFV $Z$ decays (LFVZD) in the context of right-handed neutrinos were first studied in Refs.~\cite{Ilakovac:1994kj,Illana:2000ic,Illana:1999ww}.
More recently, LFVZD processes in the presence of low-scale heavy neutrinos have been studied~\cite{Abada:2014cca,Abada:2015zea}, considering a simplified ``3+1'' model as well as different realizations of the ISS model.
In this Section, we revisit these decay rates in the ISS model with three pairs of fermionic singlets, focusing on the $\mu_X$ parametrization and the scenarios in \tabref{TMscenarios}, which have been proven to be useful for finding large rates in the case of LFV Higgs decays, as discussed in \chref{LFVHD}.

\begin{figure}[t!]
\begin{center}
\includegraphics[width=\textwidth]{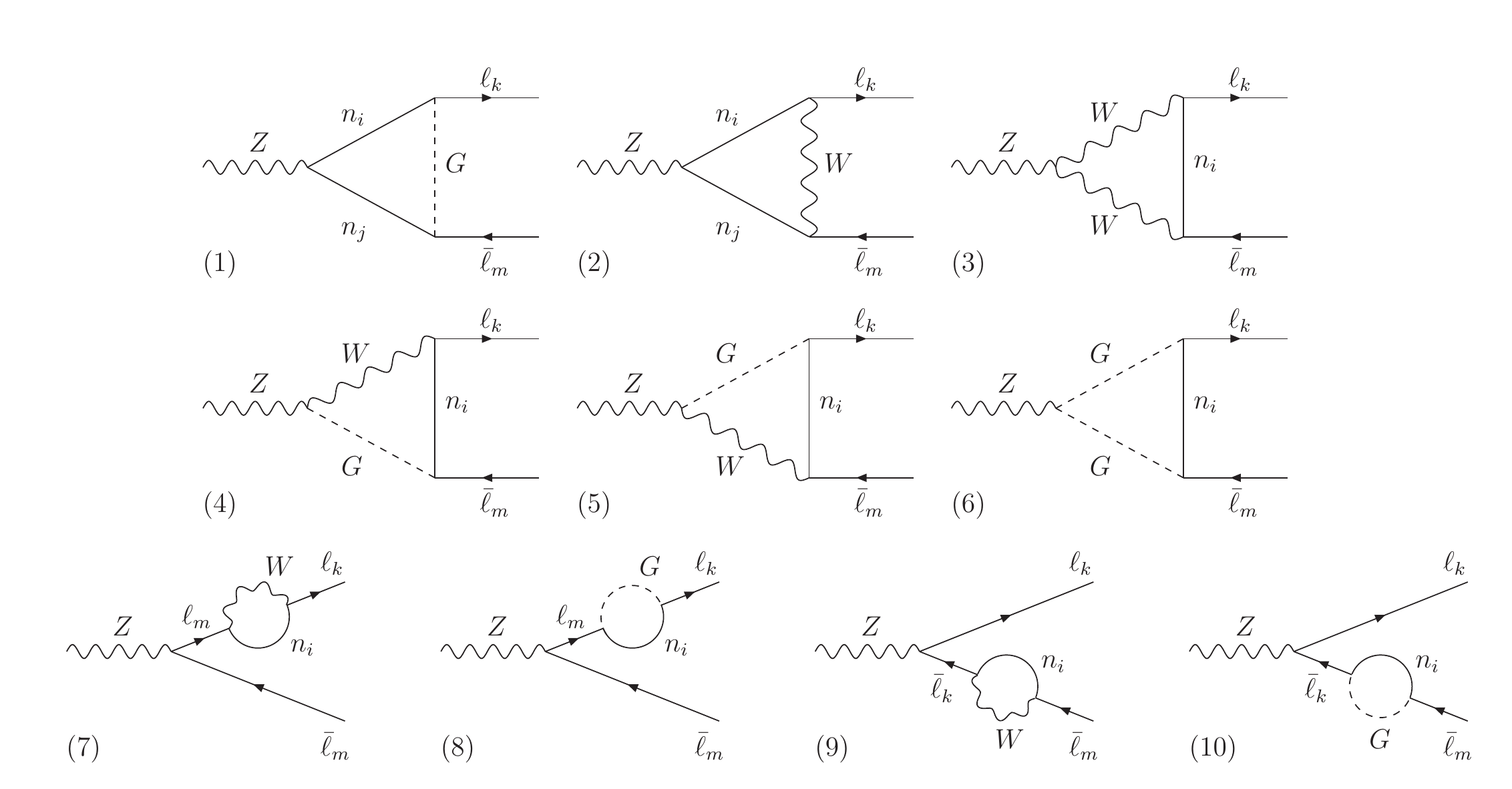}
\caption{One-loop diagrams in the Feynman-'t Hooft gauge for LFV Z decays with massive neutrinos.}\label{LFVZDDiagrams}
\end{center}
\end{figure}

Following Refs.~\cite{Illana:1999ww,Abada:2014cca,Illana:2000ic}, we can write the partial decay width for the LFV process $Z\to\ell_k\bar\ell_m$ process as,
\begin{equation}\label{LFVZDwidth}
\Gamma(Z\to \ell_k \bar\ell_m) = \frac{\alpha_W^3}{192\pi^2 c_W^2}\, m_Z\, \big| \mathcal F_Z \big|^2\,,
\end{equation}
after neglecting final state lepton masses. 
In the Feynman-'t Hooft gauge, the one-loop form factor $\mathcal F_Z$ receives contributions from the ten diagrams shown in \figref{LFVZDDiagrams}. 
Then, we can write it as, 
\begin{equation}\label{LFVZDFF}
\mathcal F_Z = \sum_{a=1}^{10} \mathcal F_{Z}^{(a)}\,.
\end{equation}
We have taken the full one-loop formulas from Ref.~\cite{Illana:1999ww} and we have adapted them to the ISS model we consider, rewriting them in terms of the proper physical neutrino masses and couplings introduced in \secref{sec:ISSmodel}. 
We include these formulas, for completeness, in \appref{App:LFVZD} where we have also adapted the loop functions to the usual notation in the literature.

For the numerical analysis of the BR($Z\to\ell_k\ell_m$) = BR($Z\to\ell_k\bar\ell_m$) + BR($Z\to\ell_m\bar\ell_k$) rates, we will evaluate these expressions with our code and with the help of the {\it LoopTools}~\cite{Hahn:1998yk} package for {\it Mathematica}.
As for the LFV H decays analysis, we will always demand a good agreement with experimental data coming from neutrino oscillations. 
This can be easily done by using any of the two parametrizations explained in \chref{Models}, i.e., the Casas-Ibarra parametrization in \eqref{CasasIbarraISS} or the $\mu_X$-parametrization in \eqref{MUXparam}.
As explained before, the choice of parametrization cannot change  physics, however  it can help to study the parameter space more efficiently or to design scenarios with  phenomenologically appealing features. 

As we said, the LFVZD in the context of the ISS model with three pairs of fermionic singlets were first  studied in Ref.~\cite{Abada:2014cca}.
In this work, the Authors used the Casas-Ibarra parametrization to accommodate the neutrino oscillation data and they scanned  over a large range of the parameter space. 
Concretely, the modulus of the entries of the input $M_R$ and $\mu_X$ matrices were randomly taken between $(0.1\, {\rm MeV}, 10^6\,{\rm GeV})$ and $(0.01\,{\rm eV}, 1\,{\rm MeV})$, respectively, with complex entries for $\mu_X$ and varying also the complex angles of the $R$ matrix in \eqref{R_Casas} between $0$ and $2\pi$.
After applying all the constraints, the Authors concluded from the scatter plots in Figs.~8-10 that maximum allowed rates of BR$(Z\to\tau\mu)\sim10^{-9}$ can be obtained in our same realization of the ISS model.

As we saw in the context of LFV Higgs decays in \chref{LFVHD}, random scans with the Casas-Ibarra parametrization allow one to explore a large region of the parameter space and to study the general features of the model, however, they are not always optimal to reach specific directions along the parameter space that are still allowed by experimental constraints and that can give indeed large predictions for some LFV observables. 
In the case of LFVHD, for instance, we found  maximum allowed rates for $H\to\tau\mu$ and $H\to\tau e$ when using the $\mu_X$ parametrization and the scenarios in \tabref{TMscenarios} that were two orders of magnitude larger than those found when using the Casas-Ibarra parametrization. 
In this sense, the study of this {\it ad-hoc} scenarios looking for maximum allowed rates is complementary to the general scan over the full parameter space. 
Therefore, in the following we focus on the LFV $Z$ decays along these particular directions in the parameters space, with the aim of complementing the study in Ref.~\cite{Abada:2014cca} covering points that a generic random scan could have missed. 

\begin{figure}[t]
\begin{center}
\includegraphics[width=.48\textwidth]{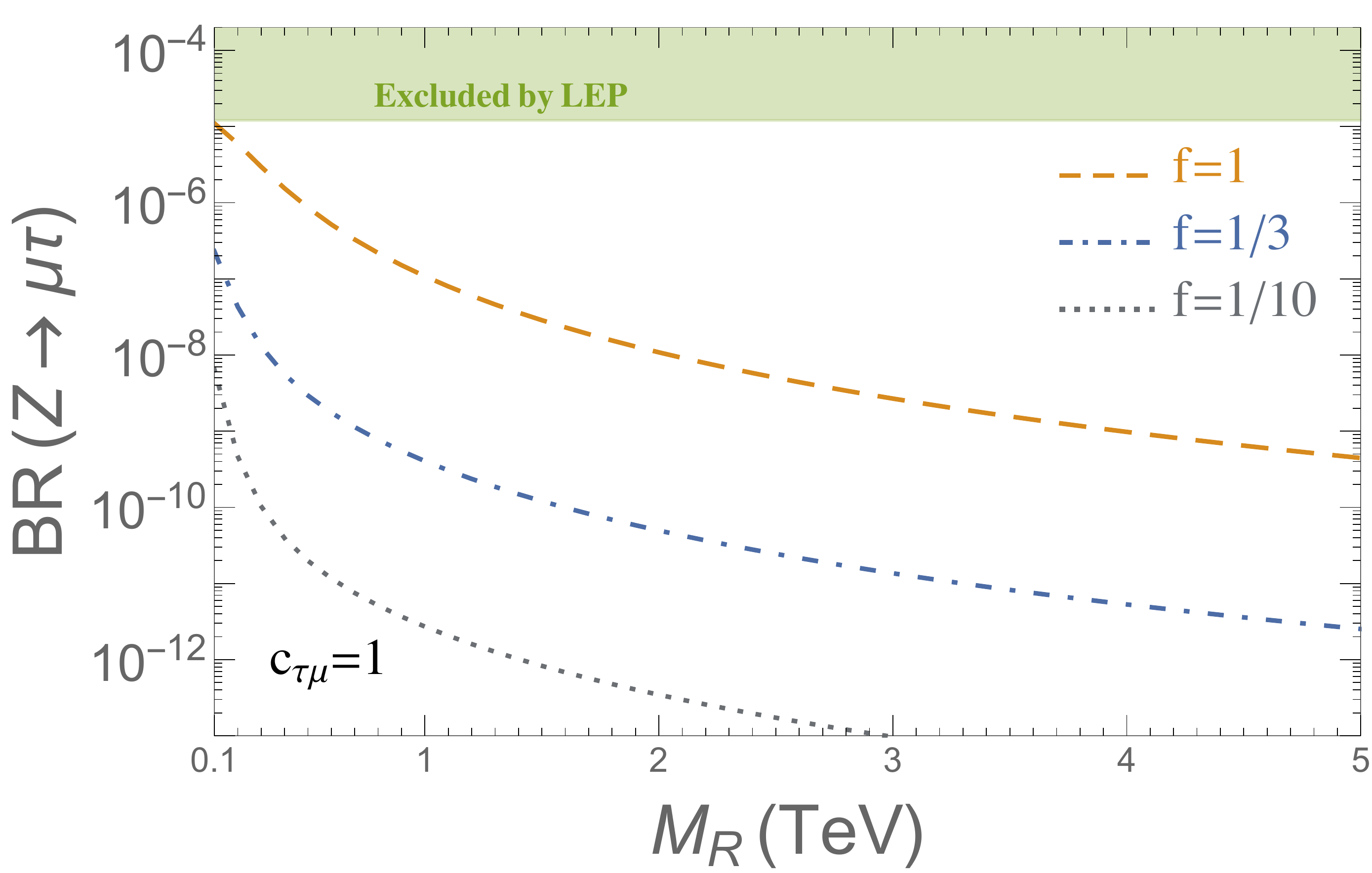}\, \,
\includegraphics[width=.48\textwidth]{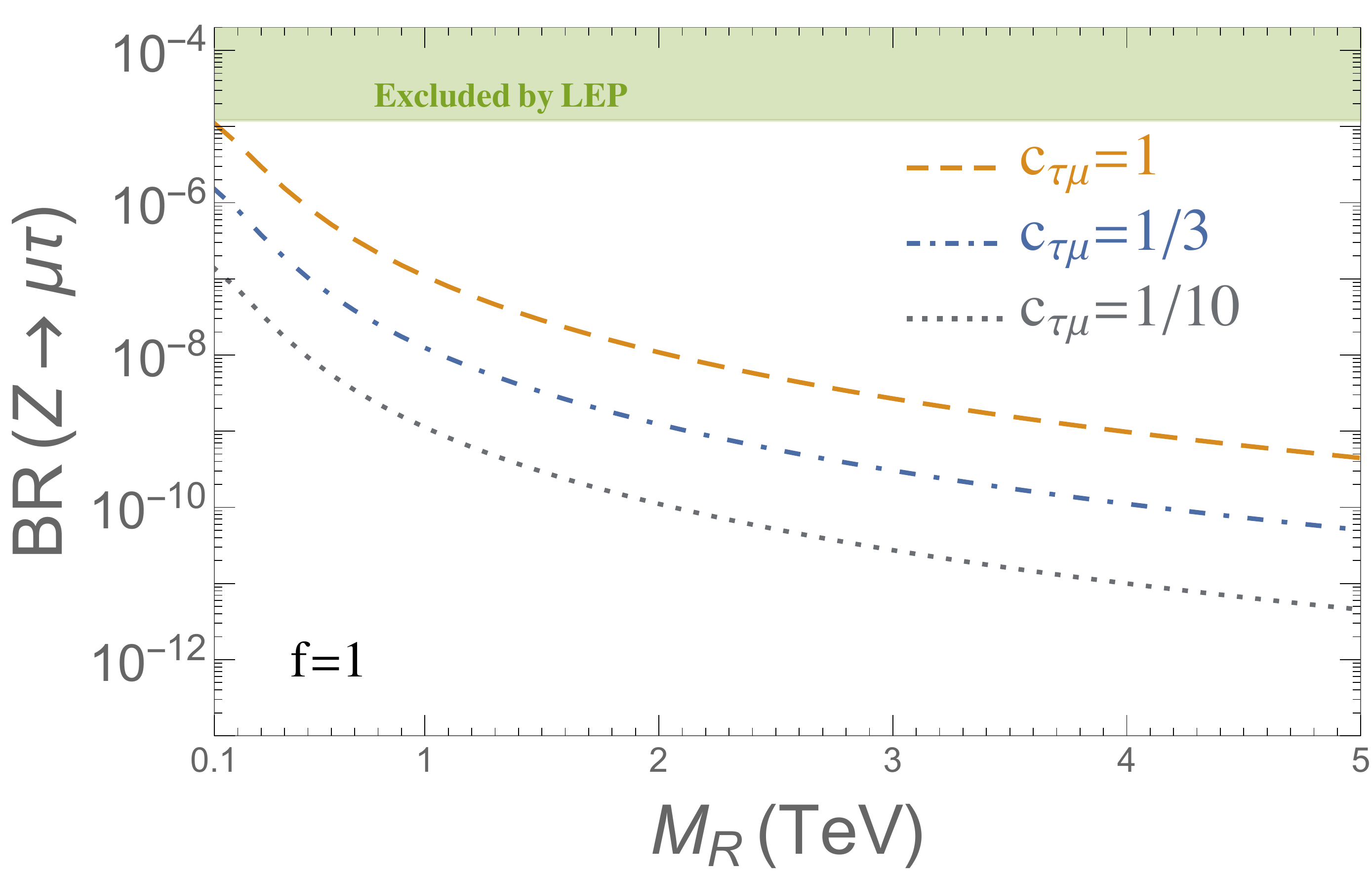}
\includegraphics[width=.48\textwidth]{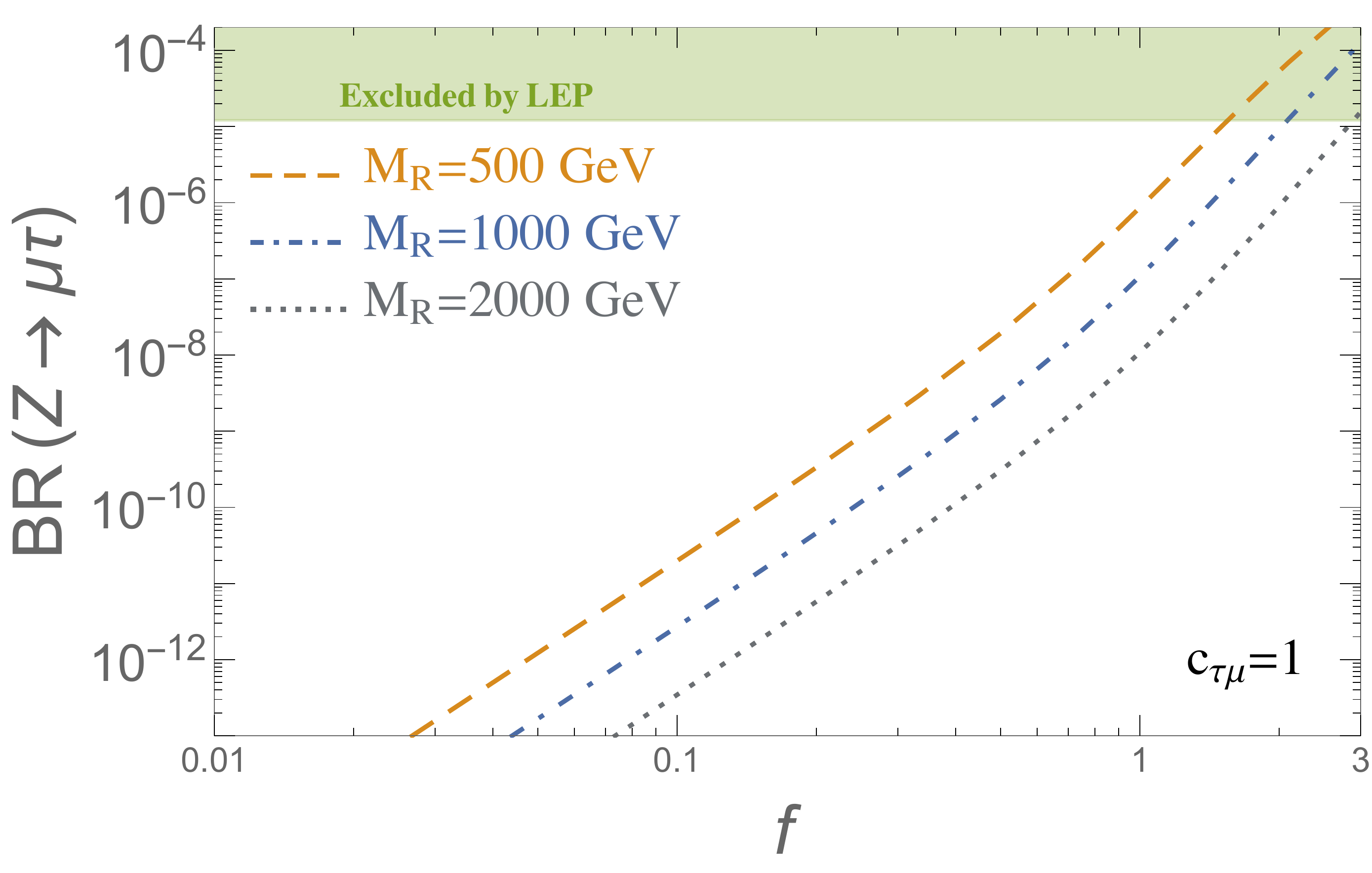}~~
\includegraphics[width=.48\textwidth]{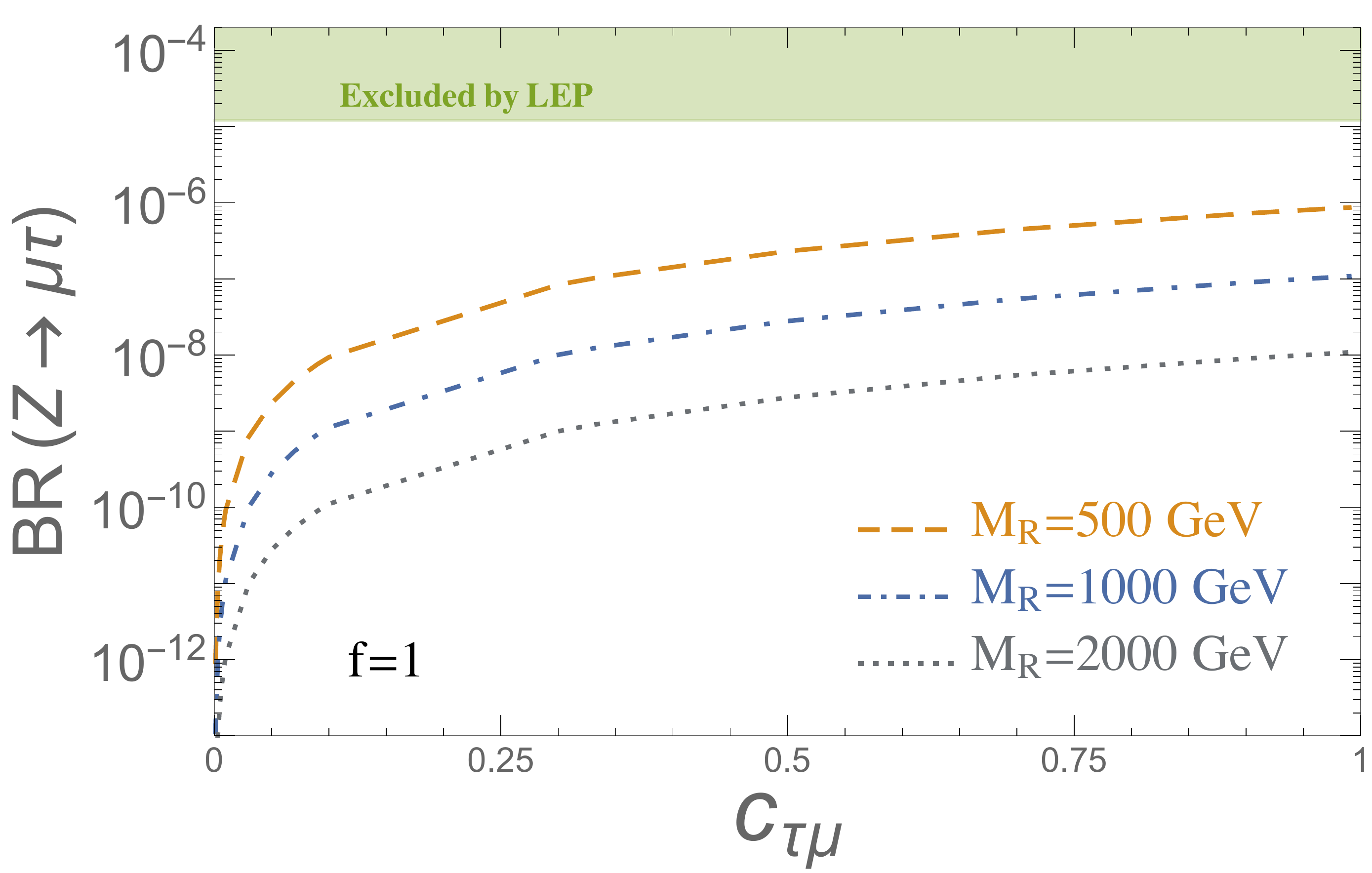}
\caption{Predictions for BR($Z\to\tau\mu$) within the ISS model as a function of the heavy neutrino mass parameter $M_R$ (two upper panels), the neutrino Yukawa coupling strength $f$ (lower left panel) and $c_{\tau \mu}$ (lower right panel) for various choices of the relevant parameters. In all plots we have fixed, $c_{\tau e}=0$ and $|\boldsymbol{n}_{e,\mu,\tau}|=1$.  
The upper shadowed areas (in green) are excluded by LEP~\cite{Abreu:1996mj}. 
We found similar results for BR($Z\to\tau e$) by exchanging $c_{\tau e}$ and $c_{\tau \mu}$.
}\label{ZtaumuParameters}
\end{center}
\end{figure}

Looking again at the present experimental upper bounds, summarized in \secref{sec:LFVexp}, we see that the constraints on cLFV processes involving $\mu$-$e$ transitions are much stronger that the ones in the other sectors, i.e., cLFV processes involving $\tau$-$\mu$ and $\tau$-$e$ transitions. 
These very stringent constraints in the $\mu$-$e$ sector motivated the class of scenarios introduced in \tabref{TMscenarios}, which incorporate automatically this suppression in their input.
We showed that these directions were not usually reached by the scans using the Casas-Ibarra parametrization, however they were easily implemented by using our $\mu_X$ parametrization and the geometrical interpretation of the Yukawa matrix introduced in \eqref{Ynuvectors}.
On the other hand, these particular ISS settings with suppressed ${\rm LFV}_{\mu e}$ rates provide very interesting scenarios for exploring the relevant ISS parameter space directions that may lead to large cLFV rates in the other sectors, $\tau$-$\mu$ and/or $\tau$-$e$, as we saw in the context of LFV Higgs decays in \chref{LFVHD}.
Thus, we will analyze the LFV Z decay rates $Z\to\tau\mu$ and $Z\to\tau e$ within these directions.

Before going to the analysis of maximum allowed LFV Z decay rates in these directions, we study how this observable depends on the most relevant parameters.
We show here our results for the particular case of BR($Z\to\tau\mu$) within the TM scenarios, which are defined by taking $c_{\tau e}=0$ in \eqref{YukawaAmatrix}, although similar results are found for BR($Z\to\tau e$) within the TE scenarios.
Along these directions, only the $Z\to\tau\mu$ channel gives relevant ratios and their predictions depend mainly on $M_R$, $f$, $|\boldsymbol{n}_\tau|$, $|\boldsymbol{n}_\mu| $ and $c_{\tau \mu}$. 

We display in \figref{ZtaumuParameters} the behavior of the BR($Z\to\tau\mu$) rates with the $M_R$, $f$ and $c_{\tau \mu}$ parameters for fixed values of $|\boldsymbol{n}_e|=|\boldsymbol{n}_\mu|=|\boldsymbol{n}_\tau|=1$, $c_{\tau e}=0$ and $\mathcal O=\mathbb 1$.  
As can be seen in this figure, these ISS directions give in general large rates for the LFV $Z\to\tau\mu$ decay, close, indeed, to the upper bound from LEP (and also close to the present LHC sensitivity) in the upper left corner of the two upper plots and in the upper right corner of the two lower plots. 
We also see that the rates decrease with the heavy scale $M_R$ and grow with the Yukawa coupling strength $f$, as expected. 
We found this growth to be approximately as $f^4$ in the low $f$ region and as $f^8$ in the high $f$ region of the studied interval of this parameter. 
This suggests that, in contrast to the radiative decays, two kinds of contributions $Y_\nu^{} Y_\nu^\dagger$ and $Y_\nu^{} Y_\nu^\dagger Y_\nu^{} Y_\nu^\dagger$ participate in this observable, similar to what we obtained for the LFV H decays (see \eqref{VeffMIA}).
These dependencies will be further studied by means of the mass insertion approximation in a forthcoming work.

In the lower right panel, we observe that the rates also grow with $c_{\tau \mu}$, albeit the dependence is milder, approximately as $c_{\tau \mu}^2$.
Although not shown here, we have also studied the dependence of the decay rates with the modulus of the vectors, $|{\bf n}_i|$, finding that the predictions for BR($Z\to\tau\mu$) grow with both $|\boldsymbol{n}_\mu|$ and $|\boldsymbol{n}_\tau|$, while they are constant with $|\boldsymbol{n}_e|$, as expected. 
Finally, we checked that the results do not depend on the global rotation $\mathcal O$, as argued when the parametrization for the $Y_\nu$ coupling matrix was motivated.

In order to conclude on the maximum allowed LFV $Z$ decay rates, we need to consider all the relevant constraints. 
Nevertheless, prior to the full study, we find interesting first to compare the predictions of these LFV $Z$ decays with the predictions of the three body LFV lepton decays in our particular ISS scenarios with suppressed $\mu$-$e$ transitions. 
Looking back to the right panel of \figref{LFVtau3muMUX}, we notice again that in these ISS directions, the $\tau\to\mu\mu\mu$ decay is mostly dominated by the $Z$ penguin.  
This fact implies a strong correlation between $\tau\to\mu\mu\mu$ and $Z\to\tau\mu$, as it was already found in Ref.~\cite{Abada:2014cca}. 
We have also checked in some examples of the ISS parameter space that our numerical predictions of these two observables are in agreement with that reference.  

\begin{figure}[t!]
\begin{center}
\includegraphics[width=.7\textwidth]{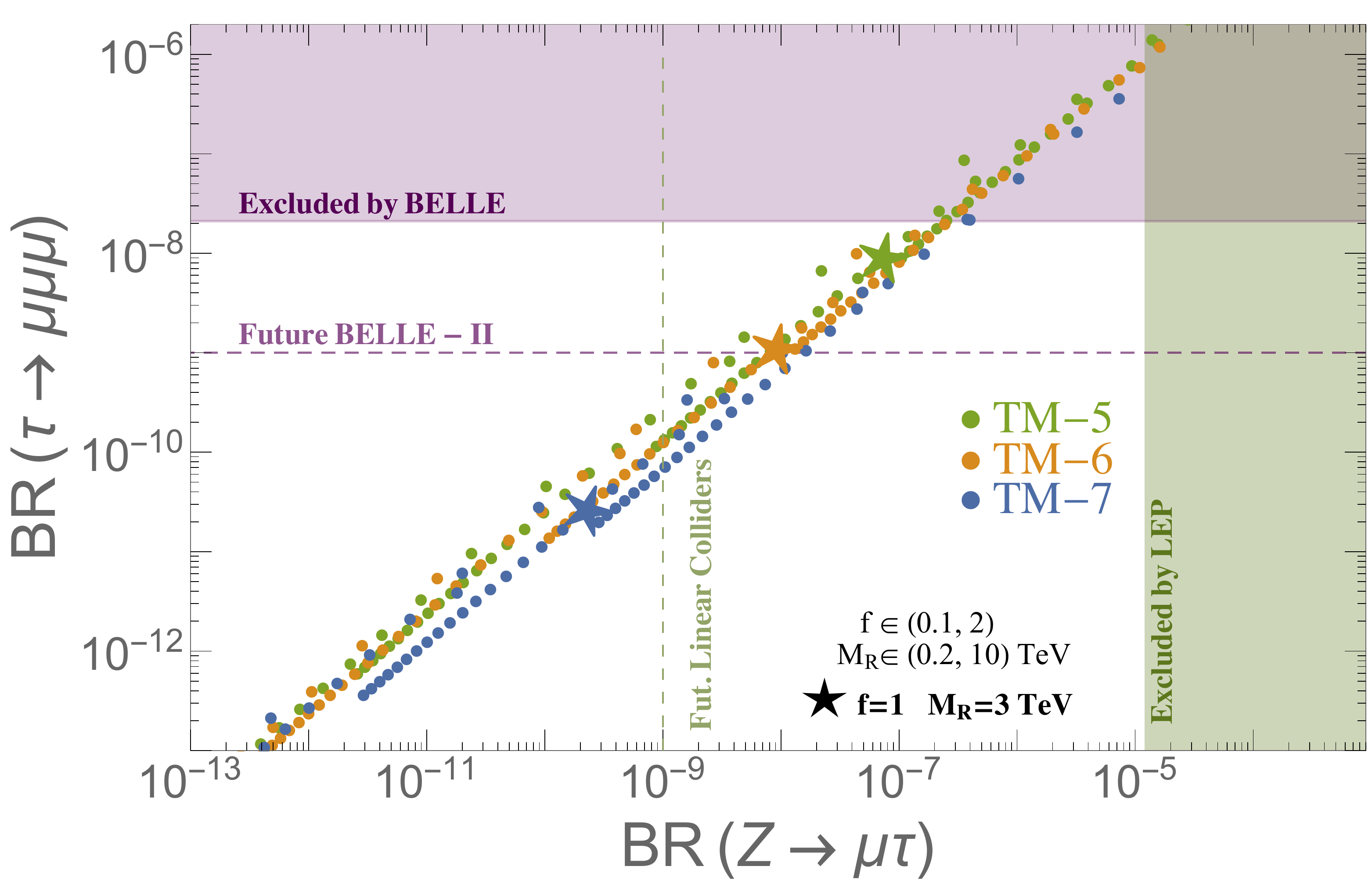}
\caption{Correlation plot for BR($Z\to\mu\tau$) and BR($\tau\to\mu\mu\mu$) for scenarios TM-5 (green), TM-6 (yellow) and TM-7 (blue) defined in \tabref{TMscenarios}. 
The dots are obtained by varying $f\in(0.1,2)$ and $M_R\in(0.2,10)$~TeV, while the stars are for the reference point $f=1$ and $M_R=3$~TeV.
Purple (green) shadowed area is excluded by BELLE~\cite{Hayasaka:2010np} (LEP~\cite{Abreu:1996mj}), while the dashed line denotes expected future sensitivity from BELLE-II (future linear colliders).
}\label{Ztaumutau3mu}
\end{center}
\end{figure}

We study this correlation in more detail in \figref{Ztaumutau3mu}, where we consider three of the scenarios given in \tabref{TMscenarios}, concretely TM-5, TM-6 and TM-7, and vary the values of the parameters within the ranges of $f\in(0.1,2)$ and $M_R\in(0.2,10)$~TeV. 
We see that both observables grow with $f$ and decrease with $M_R$ in approximately the same way, due to the already mentioned $Z$ penguin dominance in the three body decays. 
Although the predicted rates in each scenario are obviously different, see for instance the position of the reference points with $f=1$ and $M_R=3$~TeV, we clearly see that there is a strong correlation between the two observables in these ISS directions. 
We can also conclude from this plot that by considering just the constraints from the three body decays,  i.e., the present upper bound on $\tau\to\mu\mu\mu$ from BELLE, it already suggests a maximum allowed rate of BR($Z\to\tau\mu)\sim 2 \times 10^{-7}$, which is clearly within the reach of future linear colliders ($10^{-9}$ in the most conservative option).
 Interestingly, comparing the future expected sensitivities for both observables, we find some parameter space points where the LFVZD rates are in the reach of future linear colliders while the cLFV three body decay rates would not be accessible in other facilities, as BELLE-II.
This fact suggests that experiments looking for LFVZD would be able to provide additional information about the model that complements the results of other searches, like the ones in  \tabref{LFVsearch}.
We found a similar correlation between BR$(\tau\to eee)$ and BR($Z\to\tau e$) in the TE scenarios.

\section[Maximum allowed BR($Z\to\ell_k\bar\ell_m$)]{Maximum allowed BR($\boldsymbol{Z\to\ell_k\bar\ell_m}$)}
\label{sec:LFVZDmax}

In the following we present our full analysis of the LFVZD rates in the ISS scenarios with suppressed $\mu$-$e$ transitions introduced in \secref{sec:scenarios}, including all the most relevant constraints.
For this analysis we have explored the $(M_R, f)$ plane for the eight TM scenarios given in \tabref{TMscenarios} and provide numerical predictions for the BR($Z \to \ell_k \ell_m $) rates together with the predictions of the most constraining observables and their present bounds, which we reviewed in \chref{PhenoLFV}.  
Alternative checks of the allowed ISS parameter space can be made by using global fits results~\cite{Fernandez-Martinez:2015hxa,Antusch:2006vwa,FernandezMartinez:2007ms,delAguila:2008pw, Antusch:2008tz,Antusch:2014woa,Fernandez-Martinez:2016lgt}, but we prefer to make the explicit computations of the selected observables here and to compare them directly to their experimental bounds.

\begin{figure}[H]
\begin{center}
\includegraphics[height=.22\textheight]{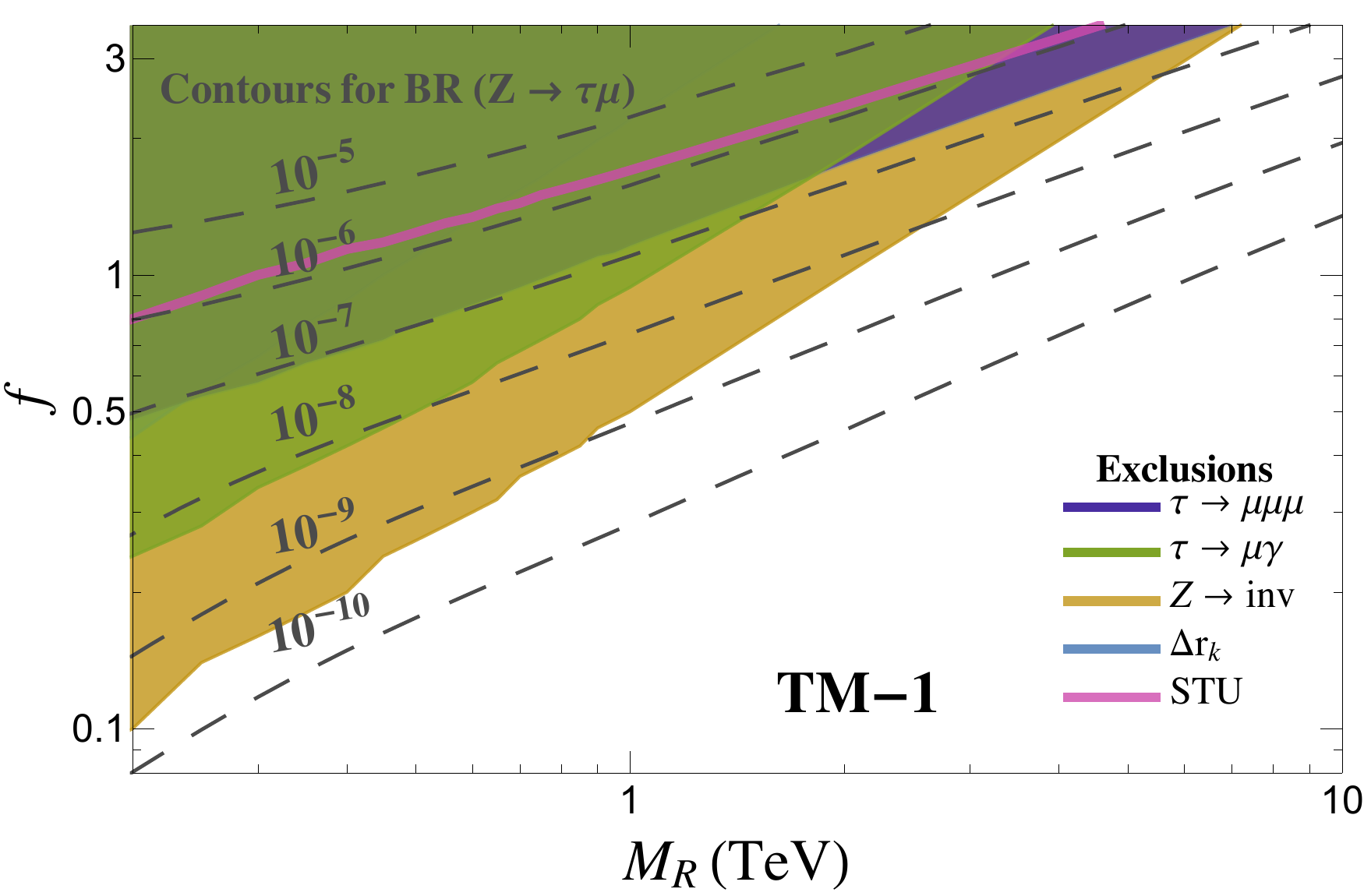}\qquad
\includegraphics[height=.22\textheight]{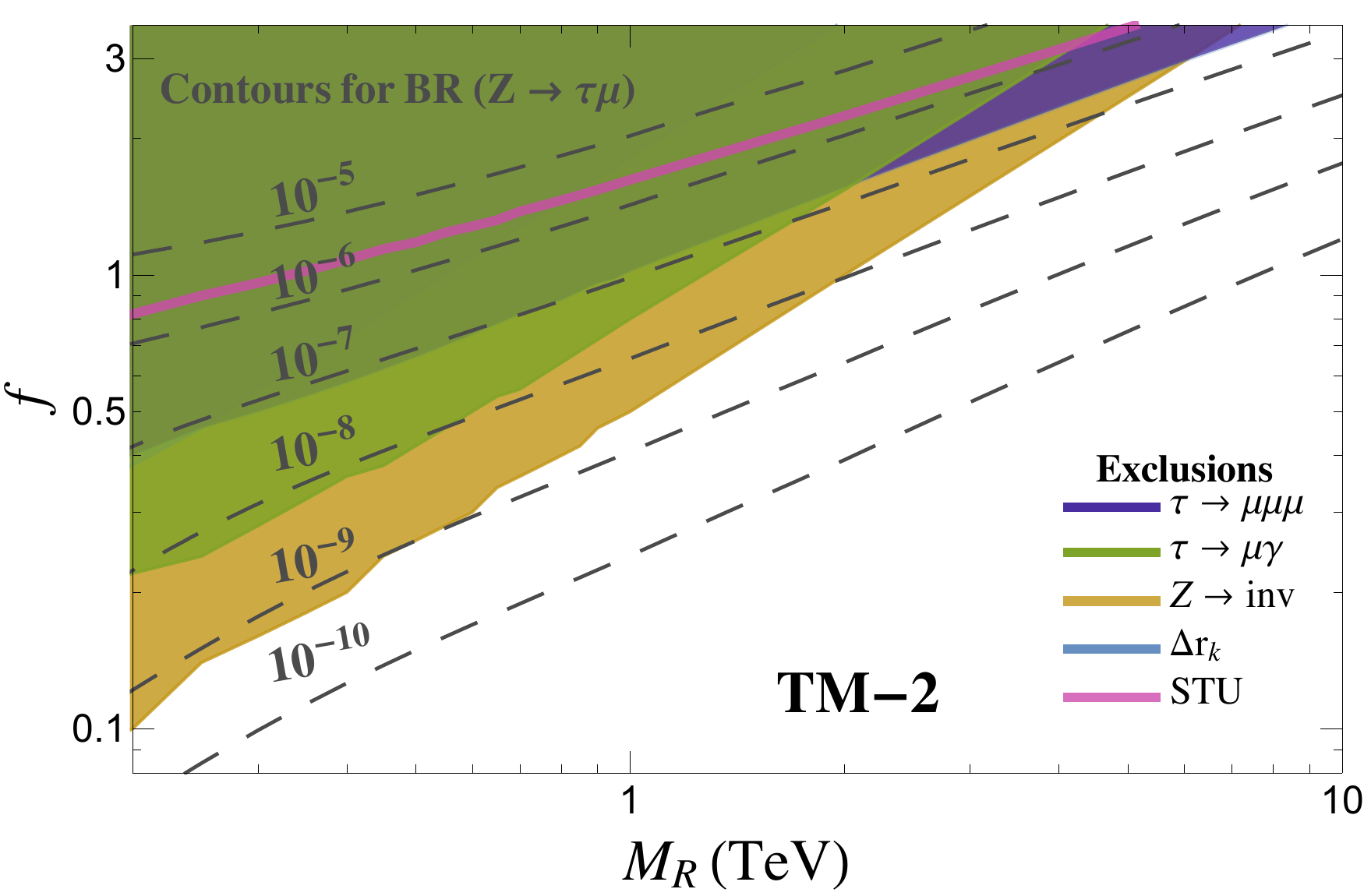}
\includegraphics[height=.22\textheight]{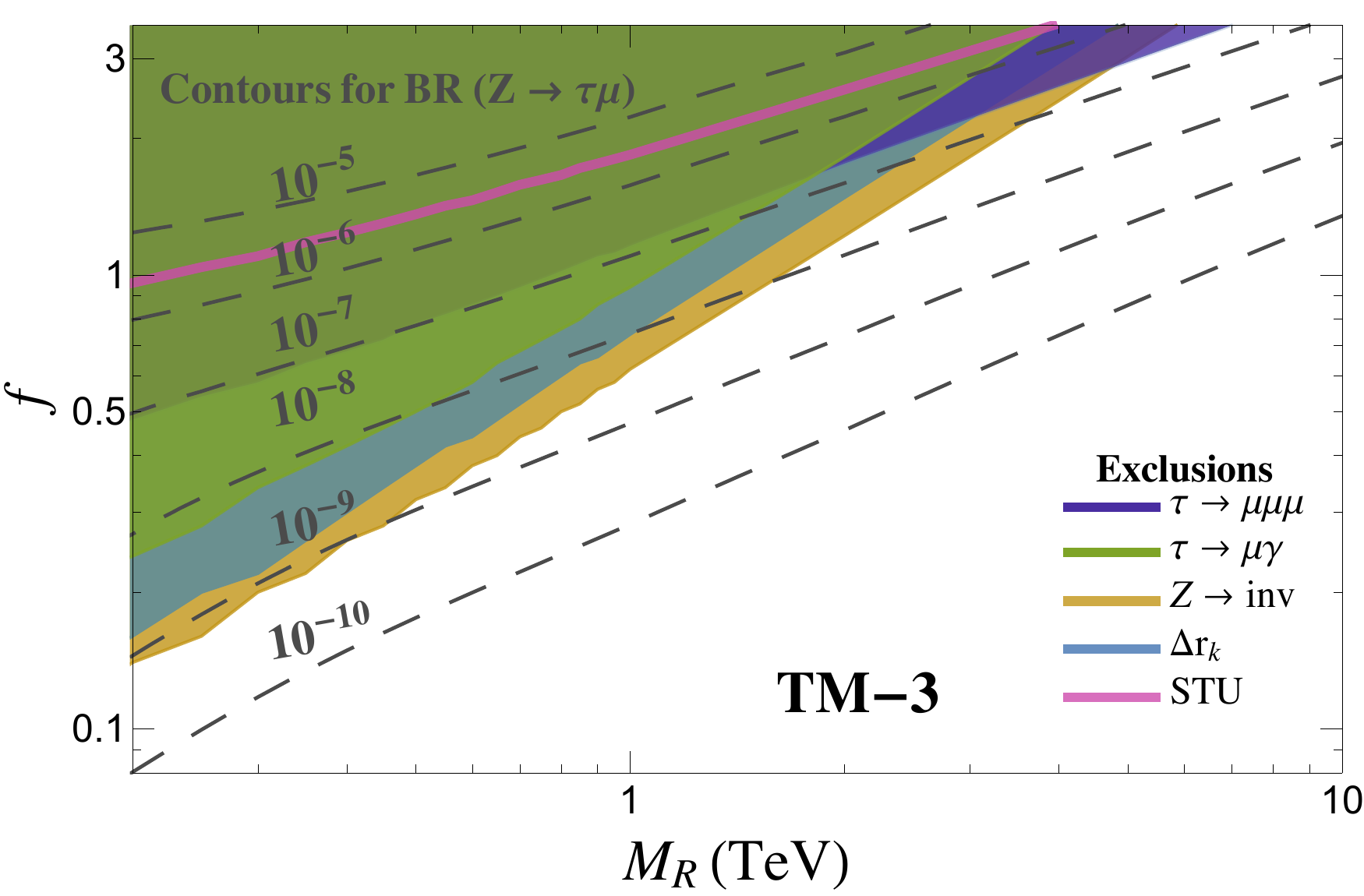}\qquad
\includegraphics[height=.22\textheight]{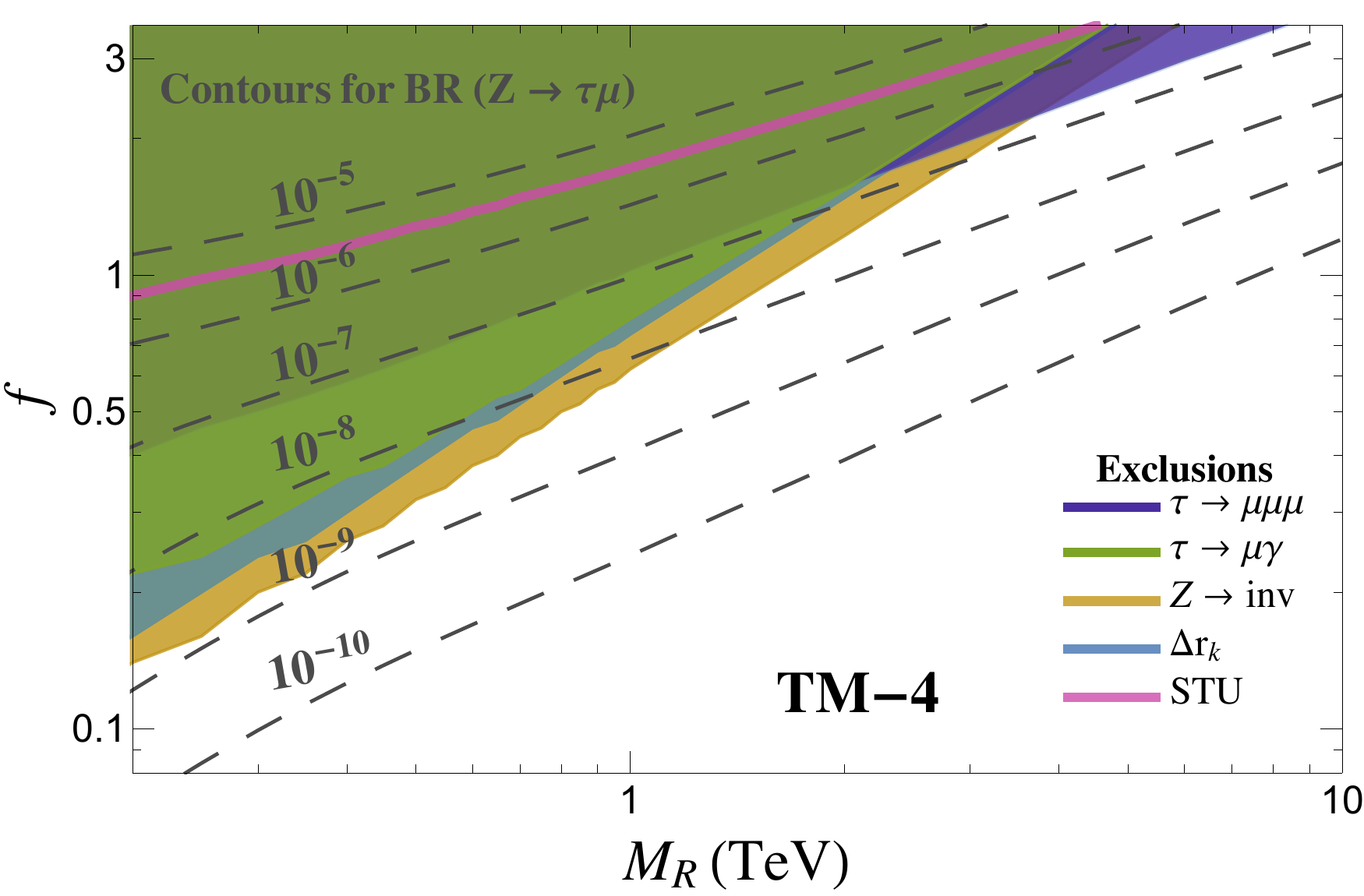}
\includegraphics[height=.22\textheight]{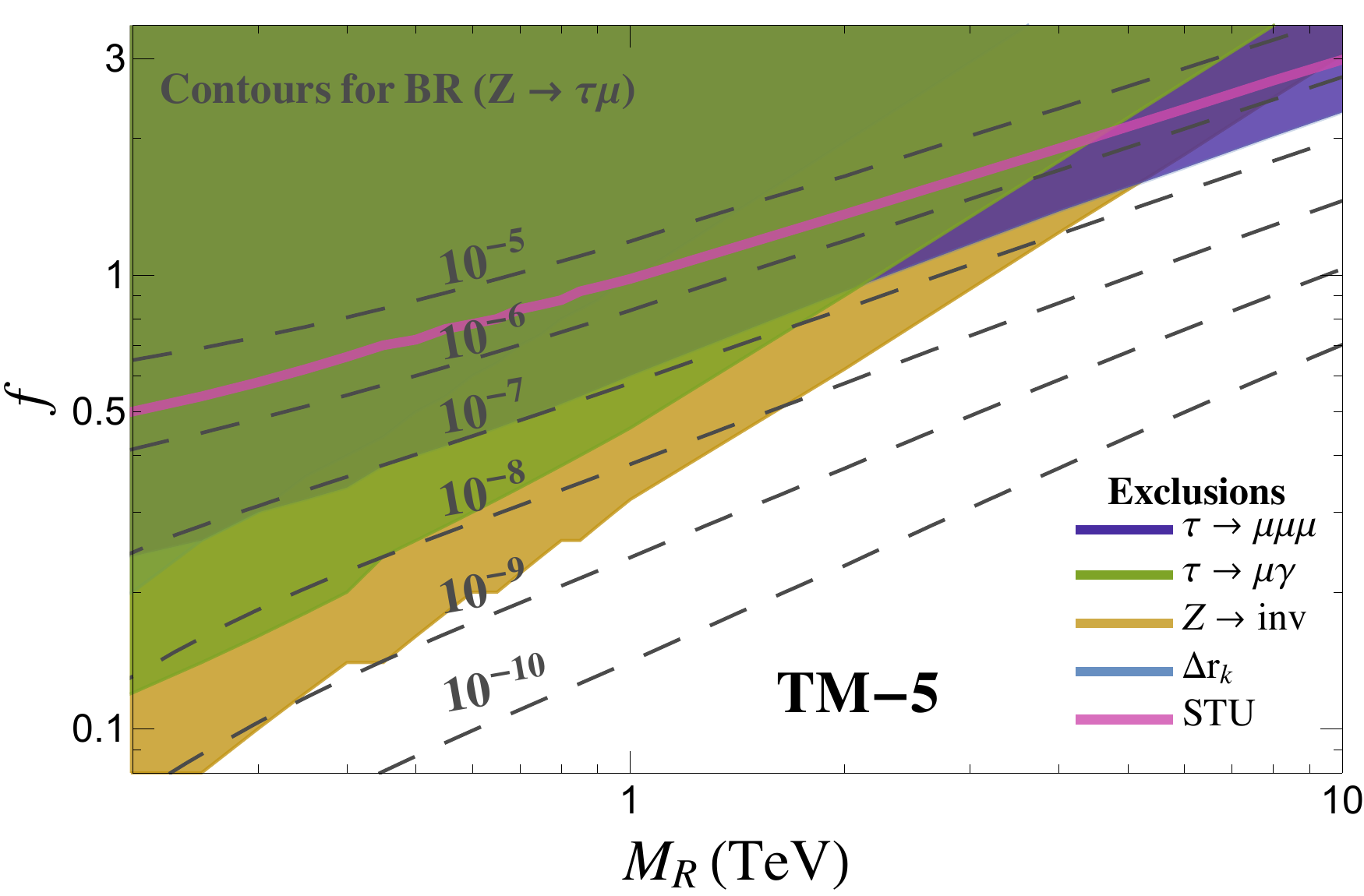}\qquad
\includegraphics[height=.22\textheight]{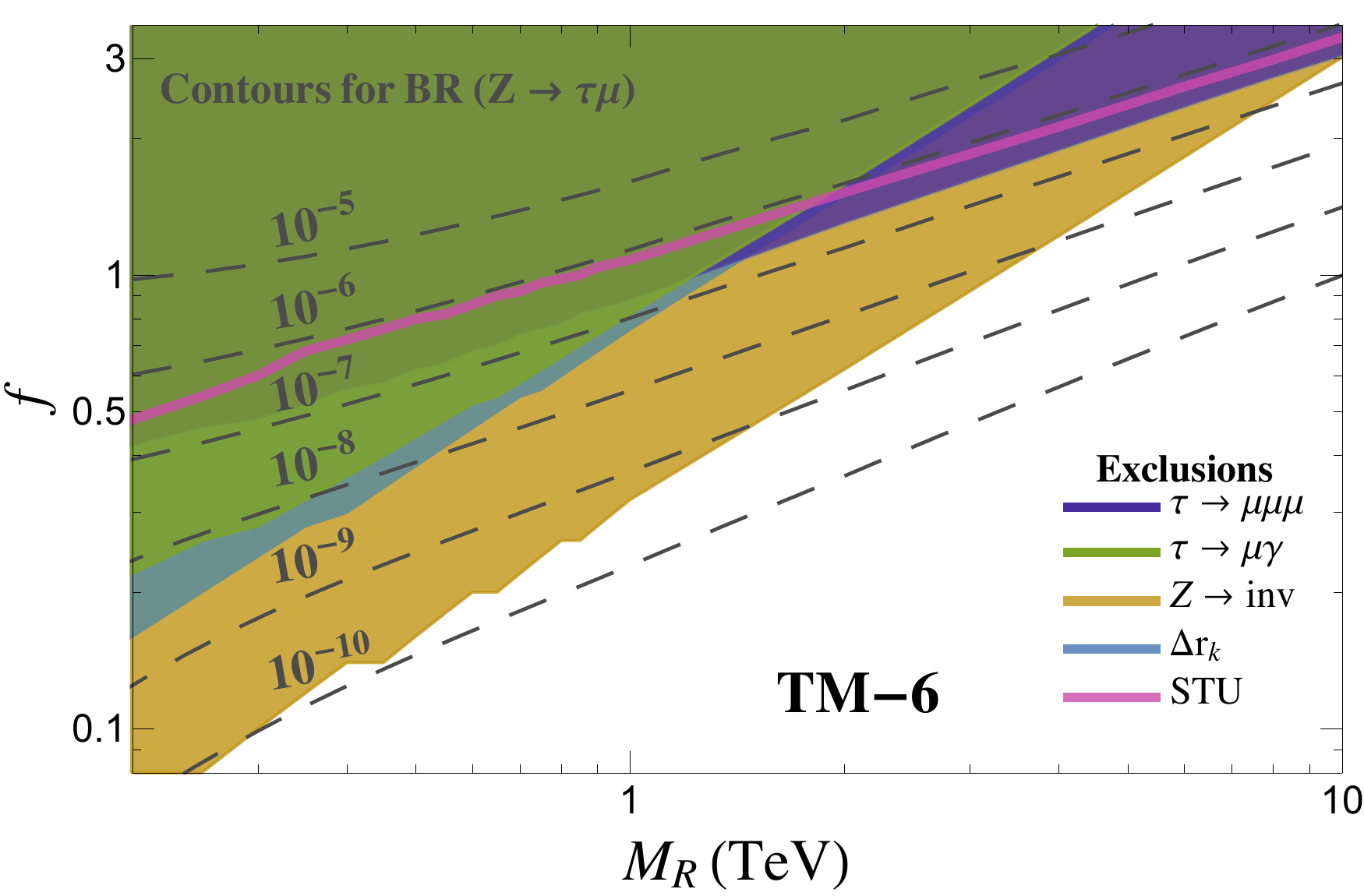}
\includegraphics[height=.22\textheight]{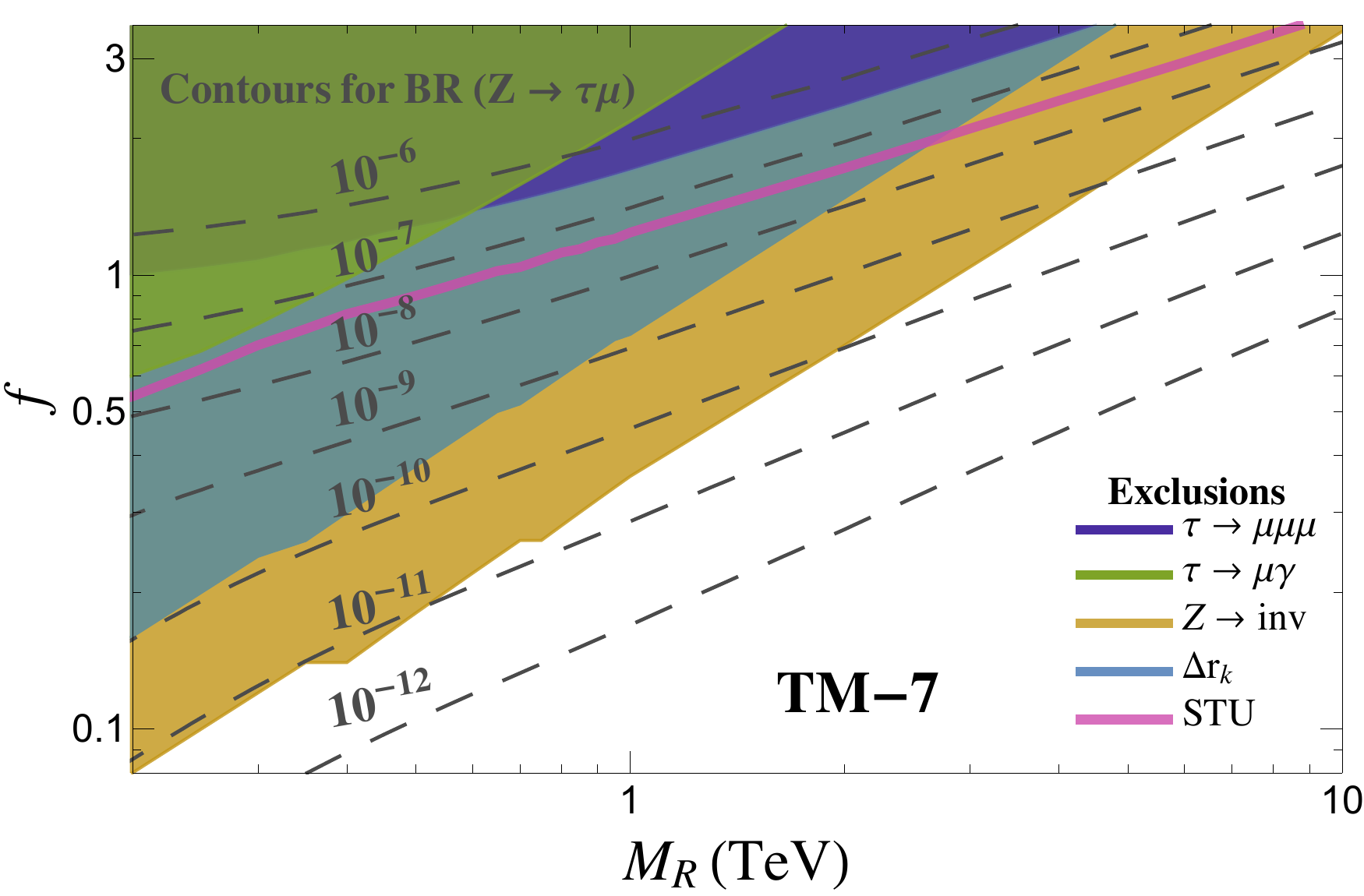}\qquad
\includegraphics[height=.22\textheight]{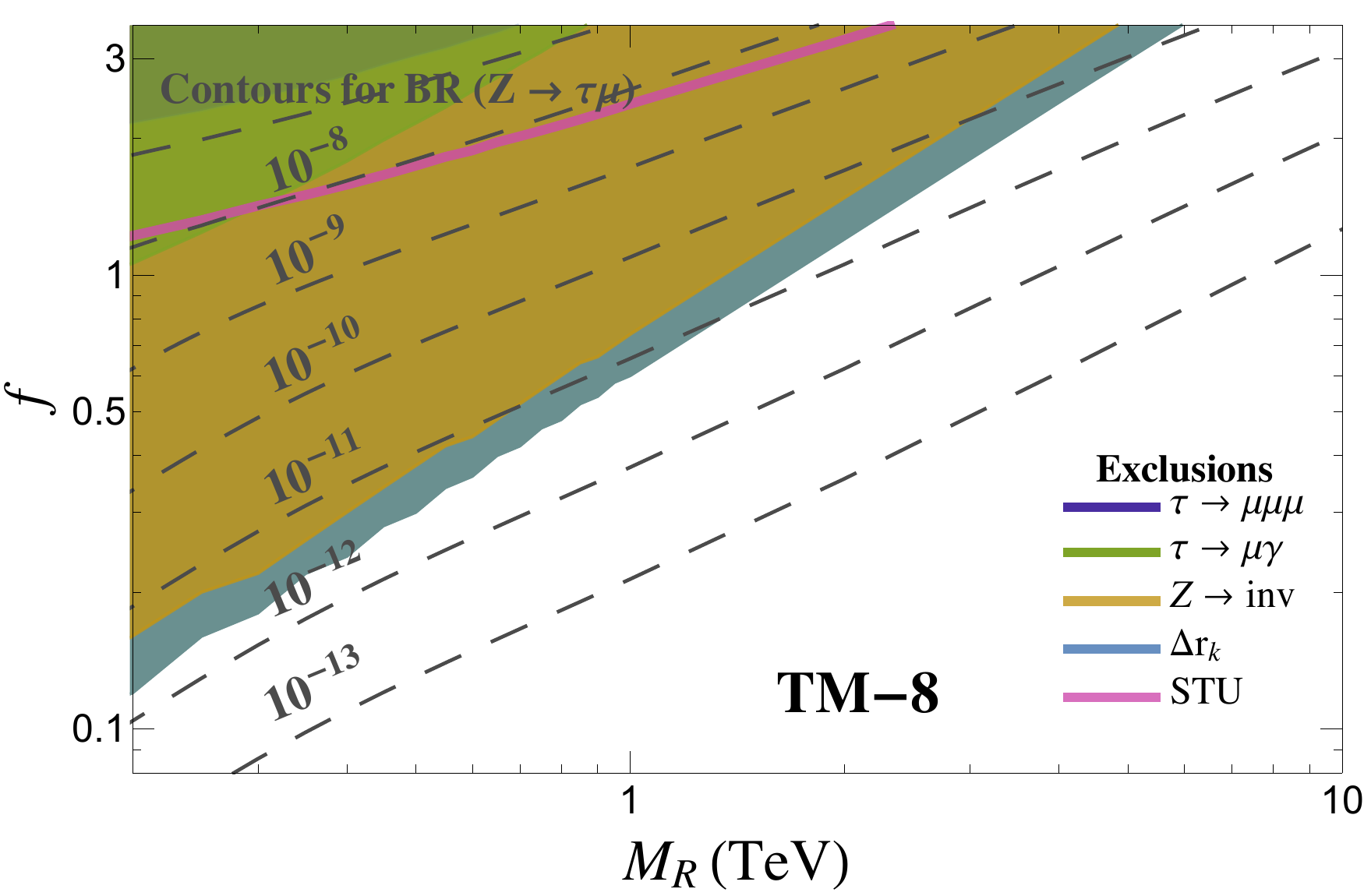}
\caption{Contour lines for BR($Z\to\tau\mu$) (dashed lines) in the ($M_R,f$) plane of the ISS model for the eight TM scenarios in \tabref{TMscenarios}. Shadowed areas are the excluded regions by $\tau\to\mu\mu\mu$ (purple), $\tau\to\mu\gamma$ (green), Z invisible width (yellow) and  $\Delta r_k$ (cyan). The region above the pink solid line is excluded by the $S$, $T$, $U$ parameters. We obtain similar results for BR($Z\to\tau e$) in the TE scenarios by exchanging $\mu$ and $e$ in these plots of the TM scenarios.}\label{ZtaumufMRplane}
\end{center}
\end{figure}

We show in \figref{ZtaumufMRplane} the results for BR($Z \to \tau \mu$) together with the constraints from: $\tau \to \mu\mu\mu$, $\tau\to\mu\gamma$, $Z \to {\rm inv.}$, $\Delta r_K$ and the EWPO ($S$, $T$ and $U$). 
As in the previous Section, we show our results only for the ${\rm LFV}_{\tau\mu}$ sector in the TM scenarios, although the conclusions are very similar for ${\rm LFV}_{\tau e}$ in the TE scenarios.  
We use different colors in the shadowed areas to represent the exclusion regions from each of the constraints listed above.
Specifically, the purple area is excluded by the upper bound on BR($\tau\to\mu\mu\mu$), the green area by BR($\tau\to\mu\gamma$), the yellow area by the $Z$ invisible width, the cyan area by $\Delta r_k$ and the area above the pink solid line is excluded by the $S$, $T$, $U$ parameters. 
Although we are not explicitly showing them here, we have also checked that the total parameter space allowed by all these constraints is also permitted by our requirements on perturbativity and on the validity of the $\mu_X$ parametrization explored in \secref{sec:MUXcheck}. 
Notice that some of the colored areas are hidden below the excluded regions corresponding to the more constraining observables.

On top of all the bounds, we display in \figref{ZtaumufMRplane} the predicted contour lines for BR($Z\to\tau\mu$) as dashed lines.  
As expected from the correlation studied in \figref{Ztaumutau3mu}, we see that these contour lines have approximately the same slope as the border of the exclusion region from BR($\tau\to\mu\mu\mu$), and in particular, the line corresponding to BR$(Z\to\tau\mu)=2 \times 10^{-7}$ is very close to the upper bound line of the three body decay in all the TM scenarios ( i.e., the border of the purple line). 
Furthermore, in the large $M_R$ and large $f$ region of these plots we see that for several TM scenarios, concretely TM-2, TM-3, TM-4 and TM-5, the BR($\tau\to\mu\mu\mu$) is indeed the most constraining observable. 

In contrast, in the low $M_R$ and low $f$ region, the most constraining cLFV observable is the radiative decay $\tau\to\mu\gamma$.
On the other hand, regarding the flavor preserving observables, it is clear that the EWPO do not play a relevant role here, but both $\Delta r_K$ and the invisible $Z$ width put relevant constraints in some scenarios. In particular, $\Delta r_K$ is the most constraining observable in the case of TM-8, and the $Z$ invisible width is so in the scenarios TM-1, TM-6 and TM-7.   
We also learn that, typically, the $Z$ invisible width is  the most constraining observable in the region of low $M_R$ values, whereas BR($\tau\to\mu\mu\mu$) is  the most constraining observable in the region of high $M_R$ values. 
Thus, generically, it is the crossing of these two excluded areas in the ($M_R, f$) plane what gives the focus area of the maximum allowed LFV $Z$ decay rates with a value of BR$(Z\to\tau\mu)\sim 2 \times 10^{-7}$, as we already inferred from \figref{Ztaumutau3mu}.
This crossing occurs at different values of $M_R$ and $f$ in each scenario.  
For example, in the TM-4 and TM-5 scenarios it happens at $M_R\sim 2-4$~TeV and for $f\sim {\cal O}(1)$, while in the TM-6 $M_R$ is around 10~TeV and $f\sim {\cal O}(2)$. 
On the other hand, if we focus our attention on the mass range of interest for present direct neutrino production searches at LHC, say masses around 1~TeV and below, we observe that the allowed BR$(Z\to\tau\mu)$ rates are smaller than this maximum value $2\times10^{-7}$; nevertheless they are still in the reach of future linear colliders ($10^{-9}$) for some scenarios, like TM-4 or TM-5.

Summarizing, in this Chapter we have studied in full detail the LFV $Z$ decays in scenarios with suppressed $\mu$-$e$ transitions that are designed to find large rates for processes including a $\tau$ lepton, and we have  investigated those that are allowed by all the present constraints. 
We have therefore fully explored in parallel also the most relevant constraints within these scenarios of the ISS model.
Important constraints come from experimental upper bounds on the LFV three body lepton decays, since they are strongly correlated to the LFVZD in these scenarios. 
Taking into account all the relevant bounds, we found that 
\noindent
 heavy ISS neutrinos with masses in the few TeV range can induce maximal rates of BR$(Z\to\tau\mu)\sim 2 \times 10^{-7}$ and BR$(Z\to\tau e)\sim 2 \times 10^{-7}$ in the TM and TE scenarios, respectively, larger than what was found in previous studies. 
These rates are potentially measurable at future linear colliders and FCC-ee. 
Therefore, we have shown that searches for LFVZD at future colliders may be a powerful tool to probe cLFV in low scale seesaw models, in complementarity with low-energy (high-intensity) facilities searching for cLFV processes.
Another appealing feature of our results is that the here presented improved sensitivity to LFVZD rates could come together with the possibility that the heavy neutrinos could be directly produced at LHC, as we will explore in the next Chapter.


\chapter{Exotic LFV signals from low scale seesaw neutrinos at the LHC}
\fancyhead[RO] {\scshape Exotic LFV signals from low scale seesaw neutrinos at the lhc}
\label{LHC}

One of the most interesting phenomenological implications of the existence of low scale seesaw neutrinos with masses in the energy range from the hundreds of GeV up to few TeV, is that they can be directly searched for at the CERN LHC and that, if their couplings to the SM particle are large, the probability of producing them can be sizable. 
The most frequently studied signatures of  heavy neutrinos are those related to their Majorana nature~\cite{Pilaftsis:1991ug,Datta:1993nm,delAguila:2007qnc,Atre:2009rg}
and, in particular, the most characteristic signal is the same-sign dilepton plus two jets events which is being searched for at the LHC. 

In the low scale seesaw models we are interested in, as long as they assume an approximated lepton number conservation to fit the observed light neutrino masses, the heavy neutrinos form pseudo-Dirac pairs, with a small Majorana character proportional precisely to the small LN breaking scale. 
Therefore, the rates of the smoking-gun signal of Majorana neutrinos are suppressed in these models.
Alternative searches for this pseudo-Dirac character of the heavy neutrinos have also been explored in the literature in connection with the appearance of other interesting multilepton signals~\cite{delAguila:2008cj} at the LHC, like the trilepton final state~\cite{Chen:2011hc,BhupalDev:2012zg,Das:2012ze,Das:2014jxa,Bambhaniya:2014kga,Bambhaniya:2014hla,Das:2015toa,Das:2016hof}. 
 
In this Chapter, we propose a new exotic signal of the right-handed neutrinos at the LHC that is based on another interesting feature of the low scale seesaw models, the fact that they incorporate large lepton flavor violation for specific choices of the model parameters, as we have extensively studied in the previous Chapters of this Thesis. 
We focus again on  the inverse seesaw model as a specific realization of these low scale seesaw models and study the LFV effects coming from the neutrino Yukawa couplings $Y_\nu$. 
Our specific proposal here is to look at rare LHC events of the type of $\ell^\pm_k\ell^\mp_m jj$, and more specifically with one muon, one tau lepton and two jets in the final state that are produced in these ISS scenarios with large LFV, and that presumably will have a very small SM background. 
This Chapter then summarizes our computation of the rates for these exotic $\mu \tau j j$ events due to the production and decays of the heavy quasi-Dirac neutrinos at the LHC within the ISS.  
The results presented in this Chapter have been published in Ref.~\cite{Arganda:2015ija}.

\section{The flavor of the heavy neutrinos}
\label{sec:HeavyNeutrinoMixins}

\begin{figure}[t!]
\begin{center}
\includegraphics[width=\textwidth]{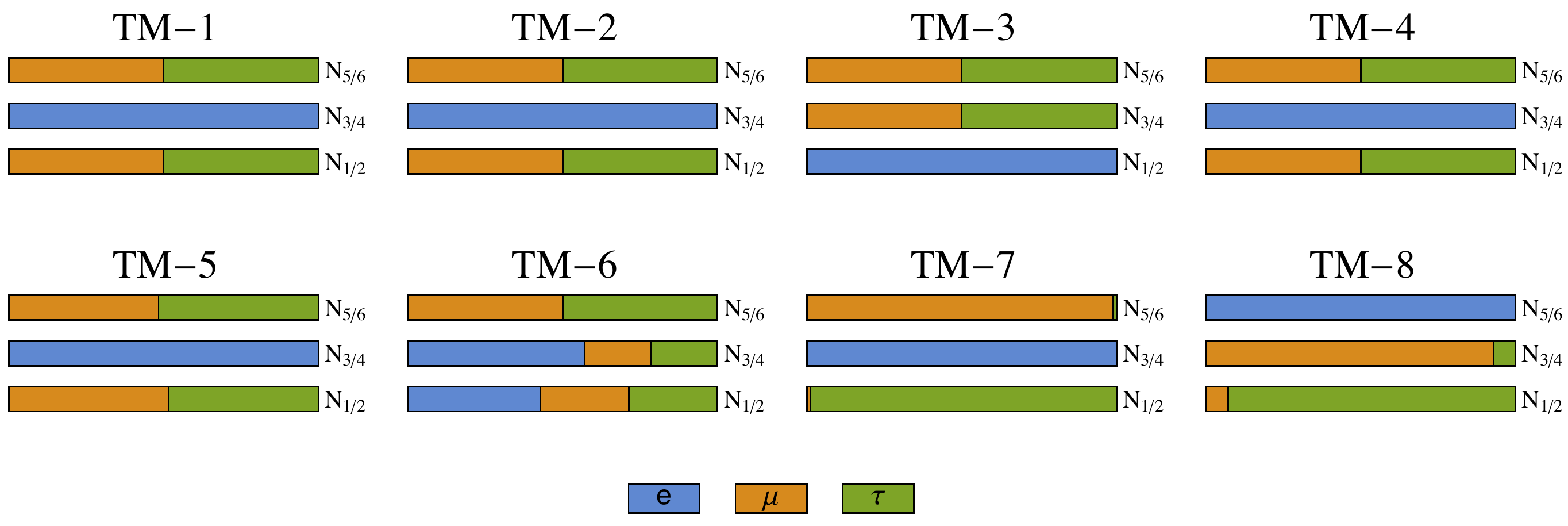}
\end{center}
\caption{Heavy neutrino flavor mixings,  as defined in  \eqref{mixeq}, within some of the ISS scenarios of \tabref{TMscenarios}.
Blue, orange and green colors represent the relative mixing with the electron, muon and tau flavor, respectively. }\label{mixingbars}
\end{figure}

As we have seen in previous Chapters, radiatively induced LFV processes are sensitive to a particular combination of the Yukawa coupling matrix, i.e., to $| Y_\nu^{} Y_\nu^\dagger |$. 
Therefore, in terms of the heavy neutrino flavor mixing, $B_{\ell N}$ as introduced in \eqref{eq:BCmatrices}, they constrain the combination {$|B_{\ell N_i}^{} B_{\ell' N_i}^*|$}, but not the {$B_{\ell N}^{}$} itself, which is the relevant element that controls the flavor pattern of each of the heavy neutrinos.

In  \figref{mixingbars} we show the flavor content of each heavy neutrino for some of the TM scenarios in \tabref{TMscenarios} in the same language of the flavor structure in \figref{Hierarchies} for the light neutrinos. 
Concretely we chose as examples the scenarios from TM-1 to TM-8, and define the length of the colored bars as 
\begin{equation}
S_{\ell N_i}= \dfrac{\displaystyle{|B_{\ell N_i}|^2}}{\displaystyle{\sum_{\ell=e,\mu,\tau} |B_{\ell N_i}|^2}}\,,
\label{mixeq}
\end{equation}
and, therefore, it represents the relative mixing of the heavy neutrino $N_i$ with a given flavor $\ell$.
It should be further noticed, that the values of these ${B_{\ell N}}$ mixing parameters are determined within the ISS model in terms of the input $m_D$ and $M_R$ mass matrices and, therefore, in the range where $m_D \ll M_R$ they are suppressed as ${B_{\ell N}} \sim \mathcal O (m_DM_R^{-1})$. 
This fact implies that, for our assumptions of degenerate entries of the diagonal $M_R$ matrix, the relative mixings defined as in \eqref{mixeq} are independent of $M_R$ in this situation. 

We learn from \figref{mixingbars} that, although these TM scenarios share the property of suppressing the LFV $\mu$-$e$ and $\tau$-$e$ rates while maximizing the $\tau$-$\mu$ ones,  the heavy neutrino flavor mixing pattern is not always the same in each scenario.
We also see that some heavy neutrinos carry an interesting amount of both $\mu$ and $\tau$ flavors, specially in the first six scenarios, pointing towards signals with both $\mu$ and $\tau$ leptons simultaneously. 
Therefore, in the following we will explore the possibility of producing this kind of events at the LHC, concretelly $\tau^\pm\mu^\mp jj$ events, which can be considered as naively background free in the SM, and are therefore interesting exotic events to search for. 
Moreover, we notice that similar results can be obtained for $\tau ejj$ events if the TE scenarios are considered.

\section[Predictions of exotic $\tau\mu jj$ event rates from heavy neutrinos]{Predictions of exotic $\boldsymbol{\tau\mu jj}$ event rates from heavy neutrinos}
\label{sec:LHCresults}

\begin{figure}[t!]
\begin{center}
\includegraphics[width=0.49\textwidth]{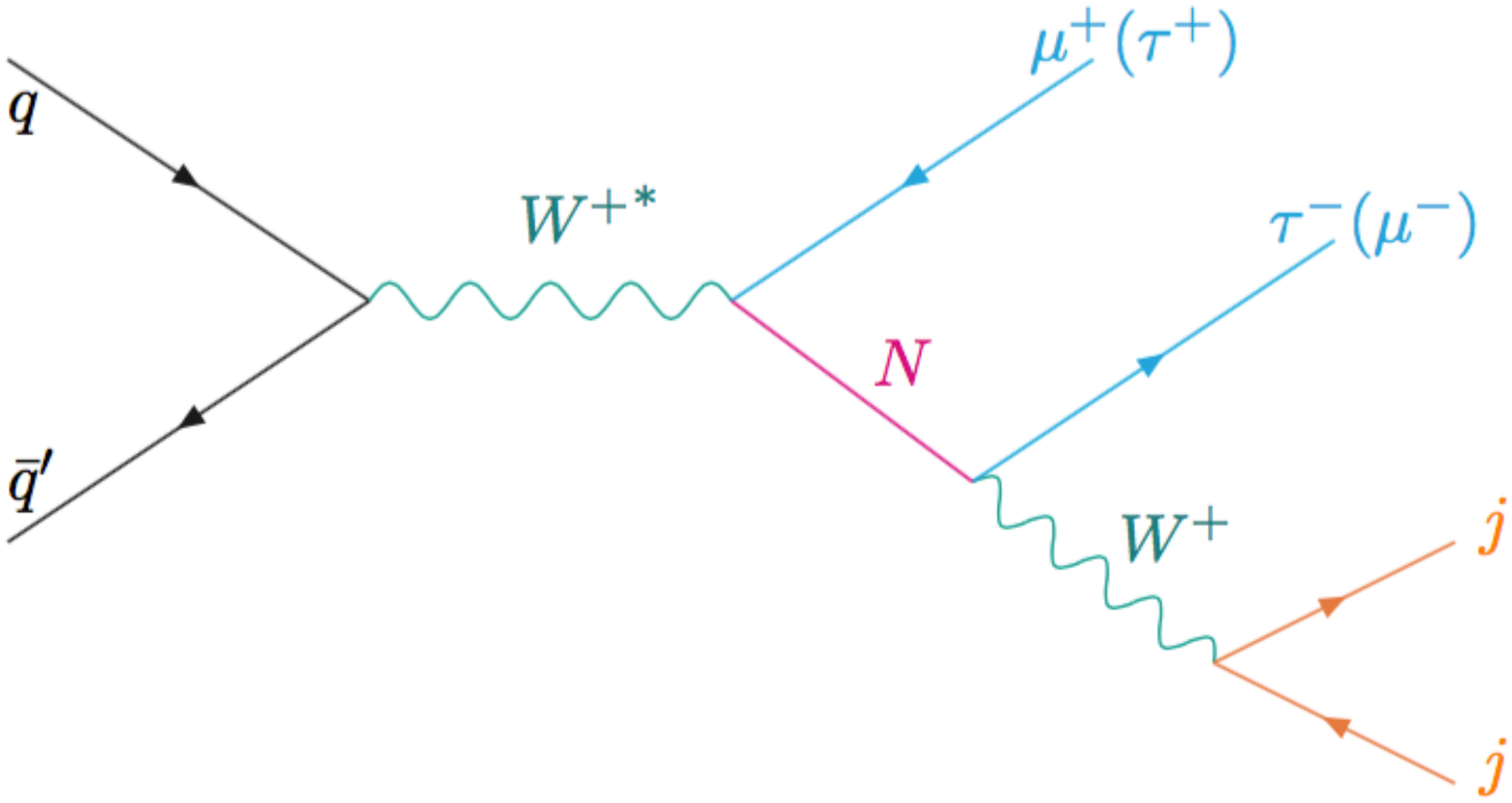} 
\includegraphics[width=0.49\textwidth]{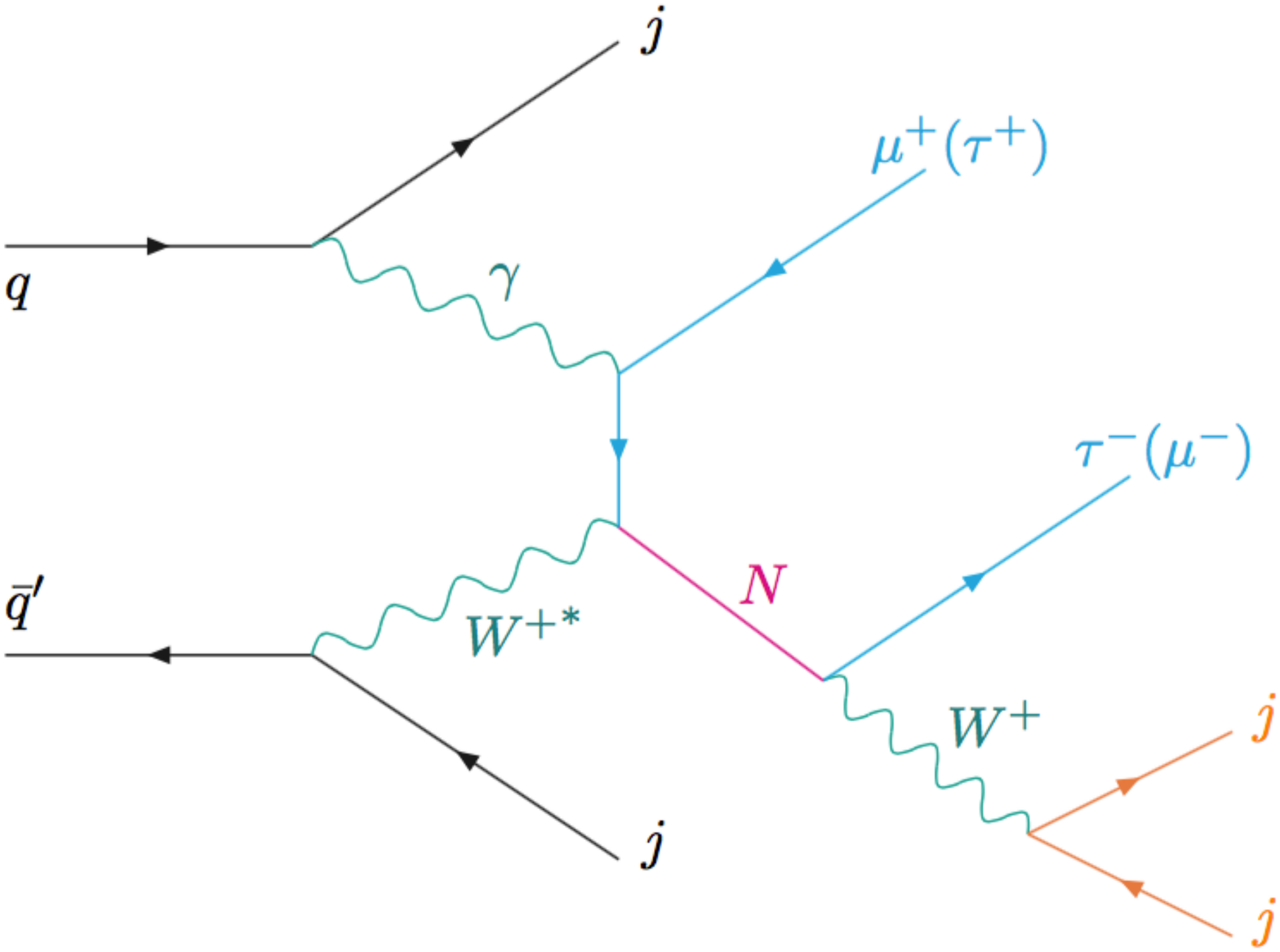}
\caption{The two main  processes, Drell-Yan and $\gamma W$ fusion, producing exotic $\tau^\pm \mu^\mp j j$ events via heavy neutrino production and decay at the LHC.}\label{LHCproductiondiagrams}
\end{center}
\end{figure}

Heavy neutrinos with masses of the TeV order and below can be produced at present and future colliders, in particular in the new runs of the LHC. 
The dominant production mechanism in this case is the Drell-Yan (DY)  process, \figref{LHCproductiondiagrams} left, where the heavy neutrino is produced in association with a charged lepton.
The $\gamma W$ fusion, \figref{LHCproductiondiagrams} right, also produces the same signal with two extra jets and, in fact, can  be also relevant especially for large neutrino masses  in the $\mathcal O(1 ~\rm{TeV})$ energy range~\cite{Dev:2013wba,Alva:2014gxa}.

In order to estimate the heavy neutrino production at the LHC, we have used {\it Feynrules}~\cite{Alloul:2013bka} to implemente the model in {\it MadGraph5}~\cite{Alwall:2014hca}. 
Following Ref.~\cite{Alva:2014gxa}, we have used a $K$-factor of $1.2$ for the DY-process and split the $\gamma W$ process in three regimes characterized by the virtuality of the photon\footnote{We warmly thank Richard Ruiz, Tao Han and Daniel Alva for their generous help  and clarifications in the implementation of the $\gamma W$ process.}: elastic, inelastic and deep inelastic scattering (DIS) regimes. In particular, we have set the boundaries between these three regimes to $\Lambda_\gamma^{\rm Elas}=1.22$~GeV and $\Lambda_\gamma^{\rm DIS}=15$~GeV.
In order to detect them and to regularize possible collinear singularities, we have also imposed the following cuts to the transverse momentum and pseudorapidity of the outgoing leptons:
\begin{equation}
p_T^\ell > 10~ {\rm GeV}, \quad |\eta^\ell|<2.4~.
\end{equation} 

\begin{figure}[H]
\begin{center}
\includegraphics[width=0.48\textwidth]{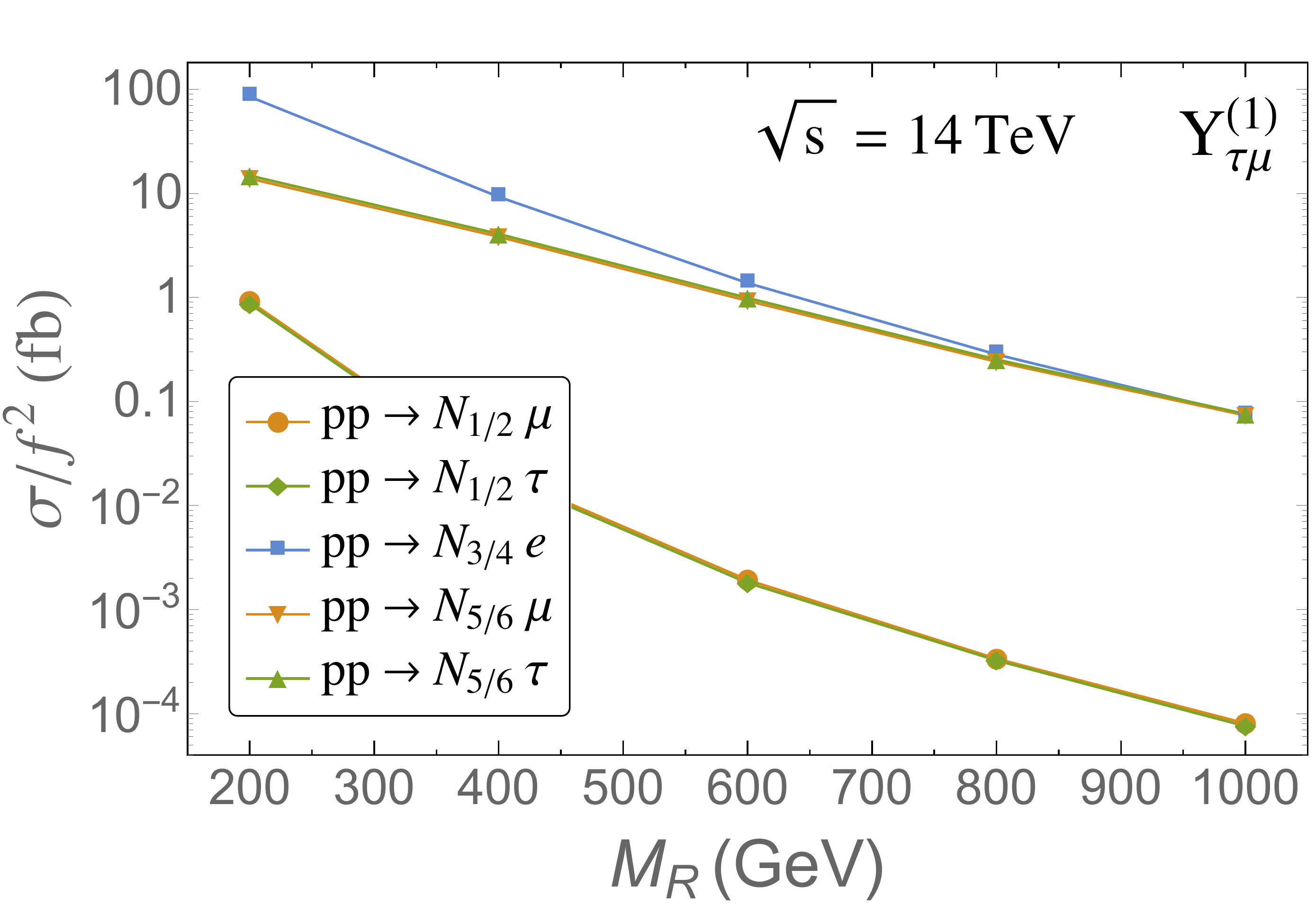} 
\includegraphics[width=0.47\textwidth]{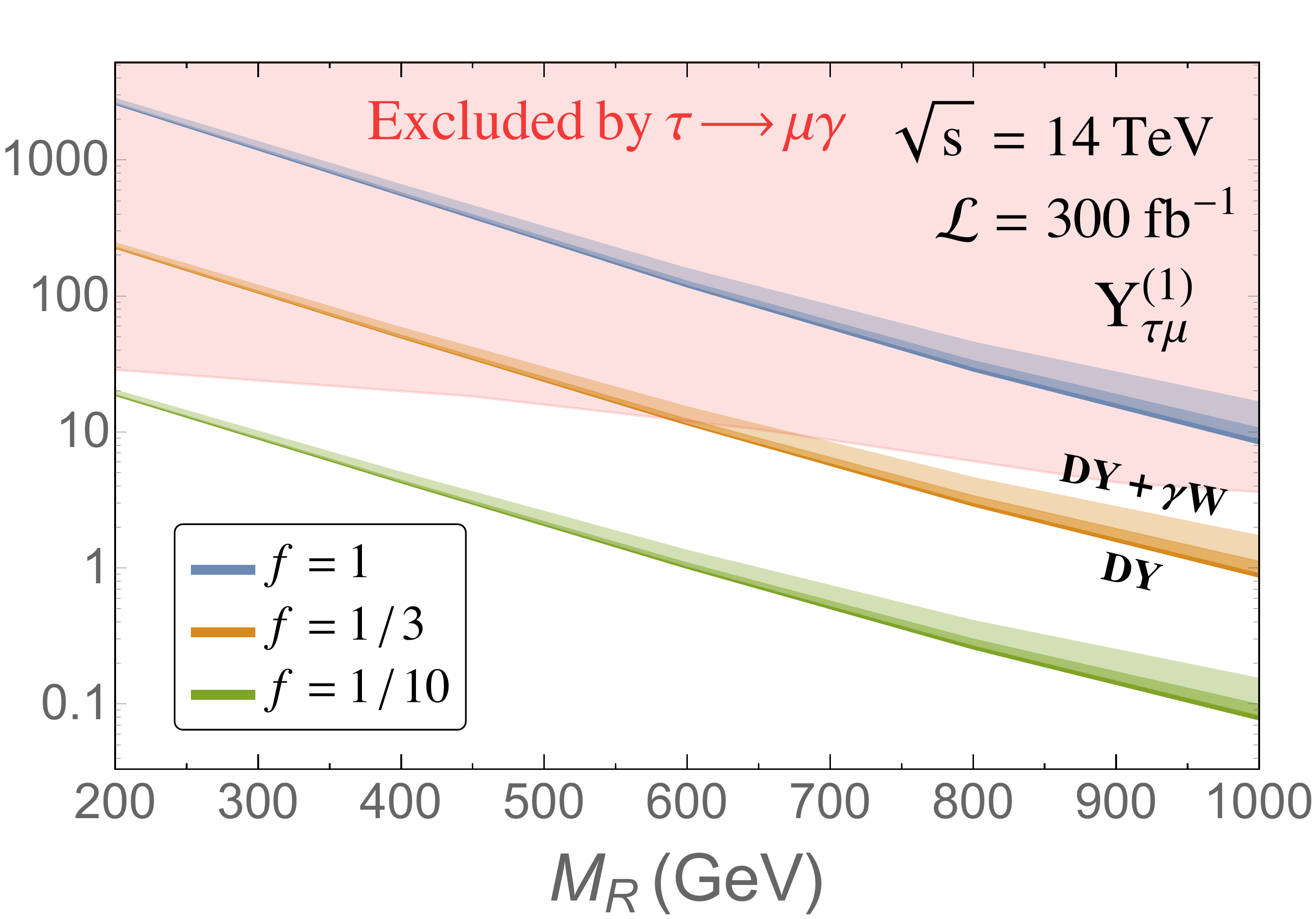} 
\includegraphics[width=0.48\textwidth]{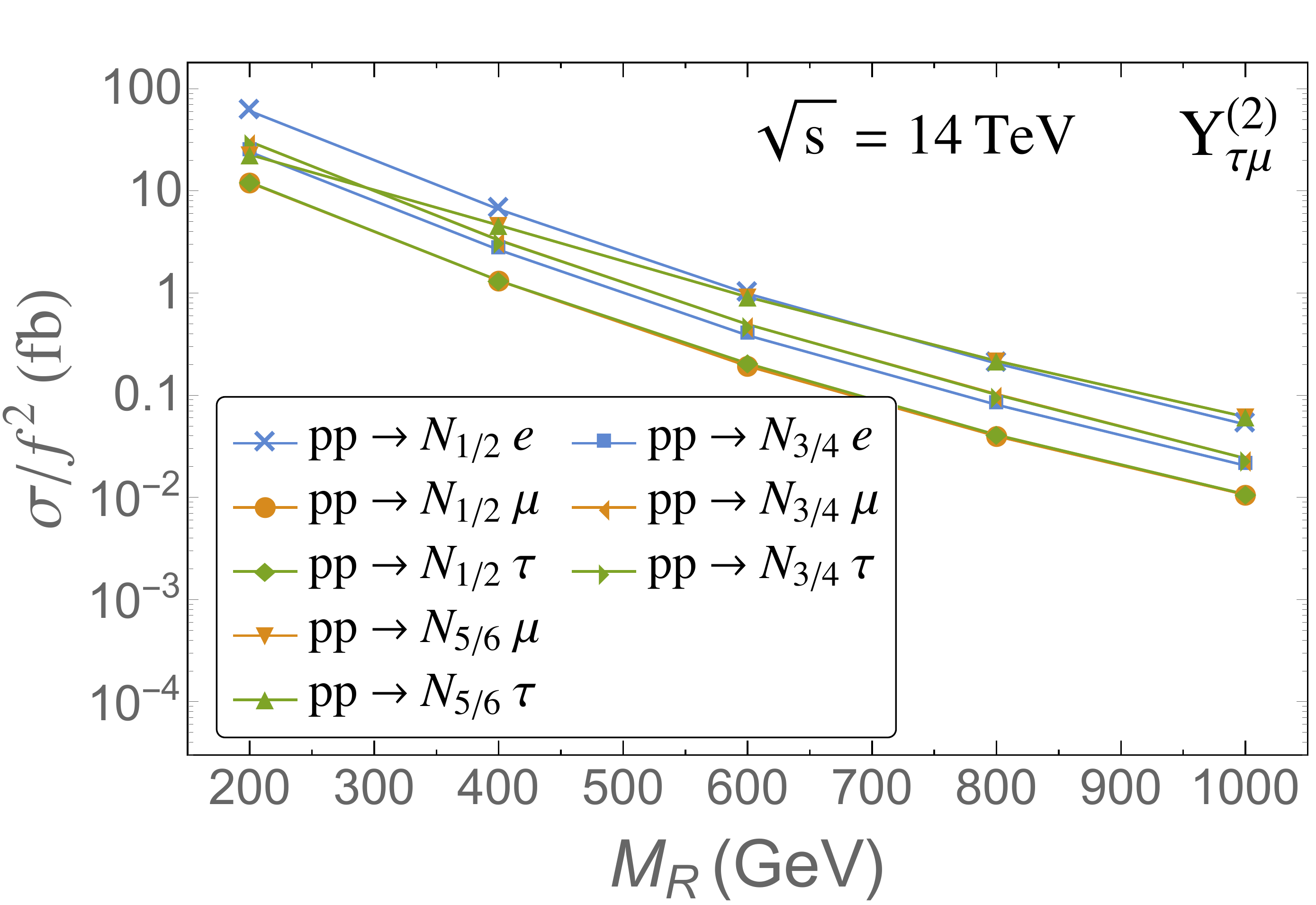} 
\includegraphics[width=0.47\textwidth]{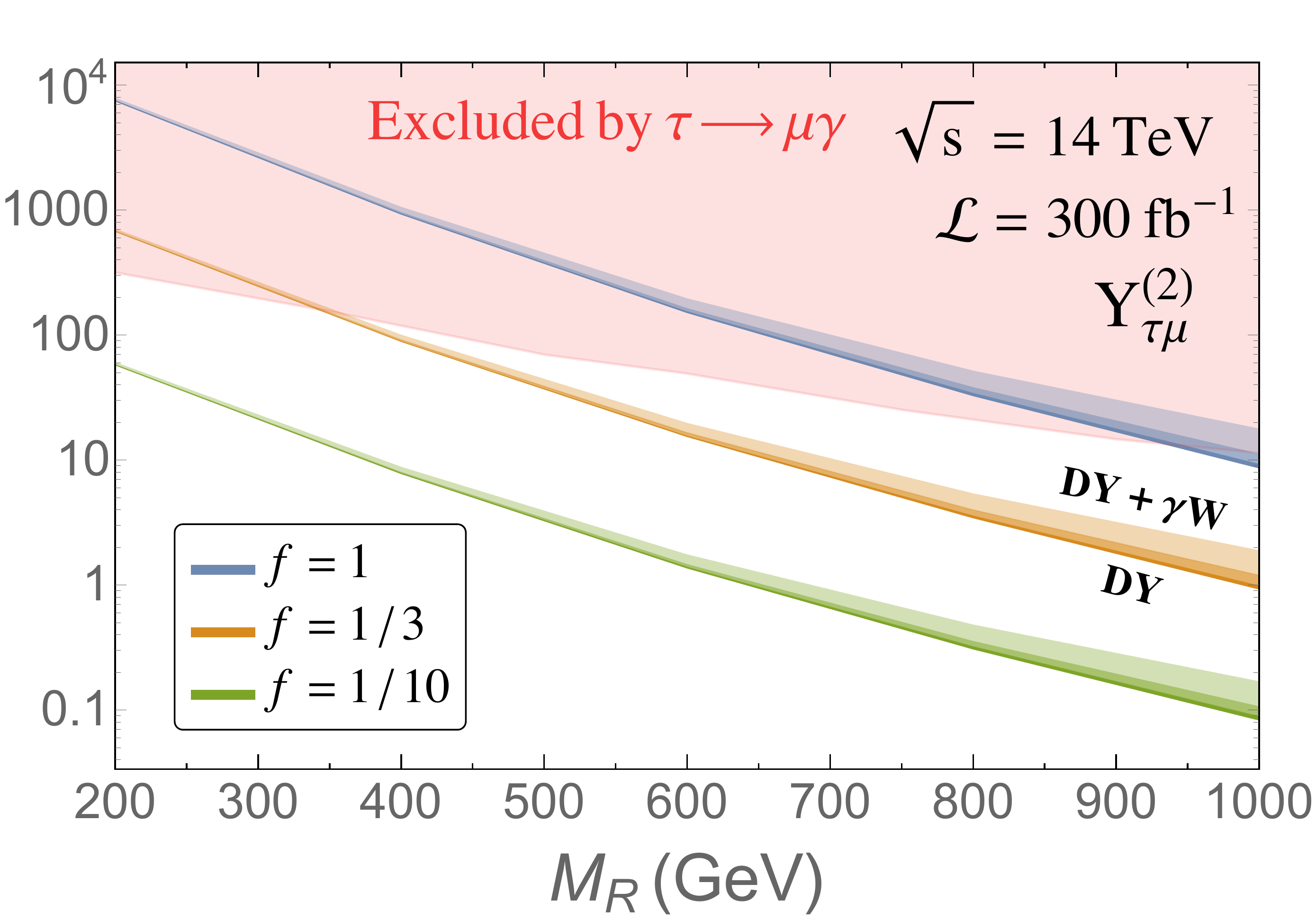} 
\includegraphics[width=0.48\textwidth]{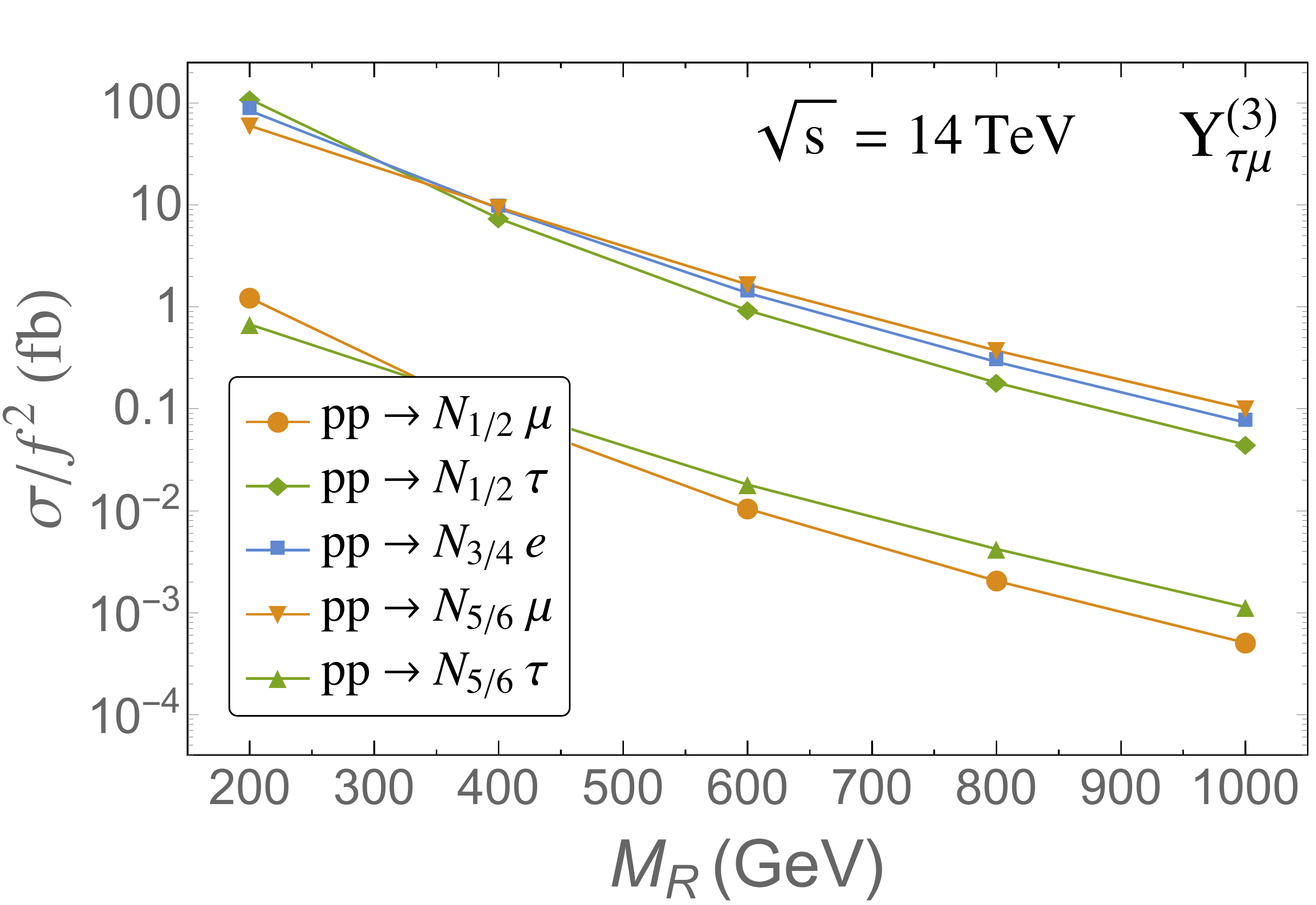}
\includegraphics[width=0.47\textwidth]{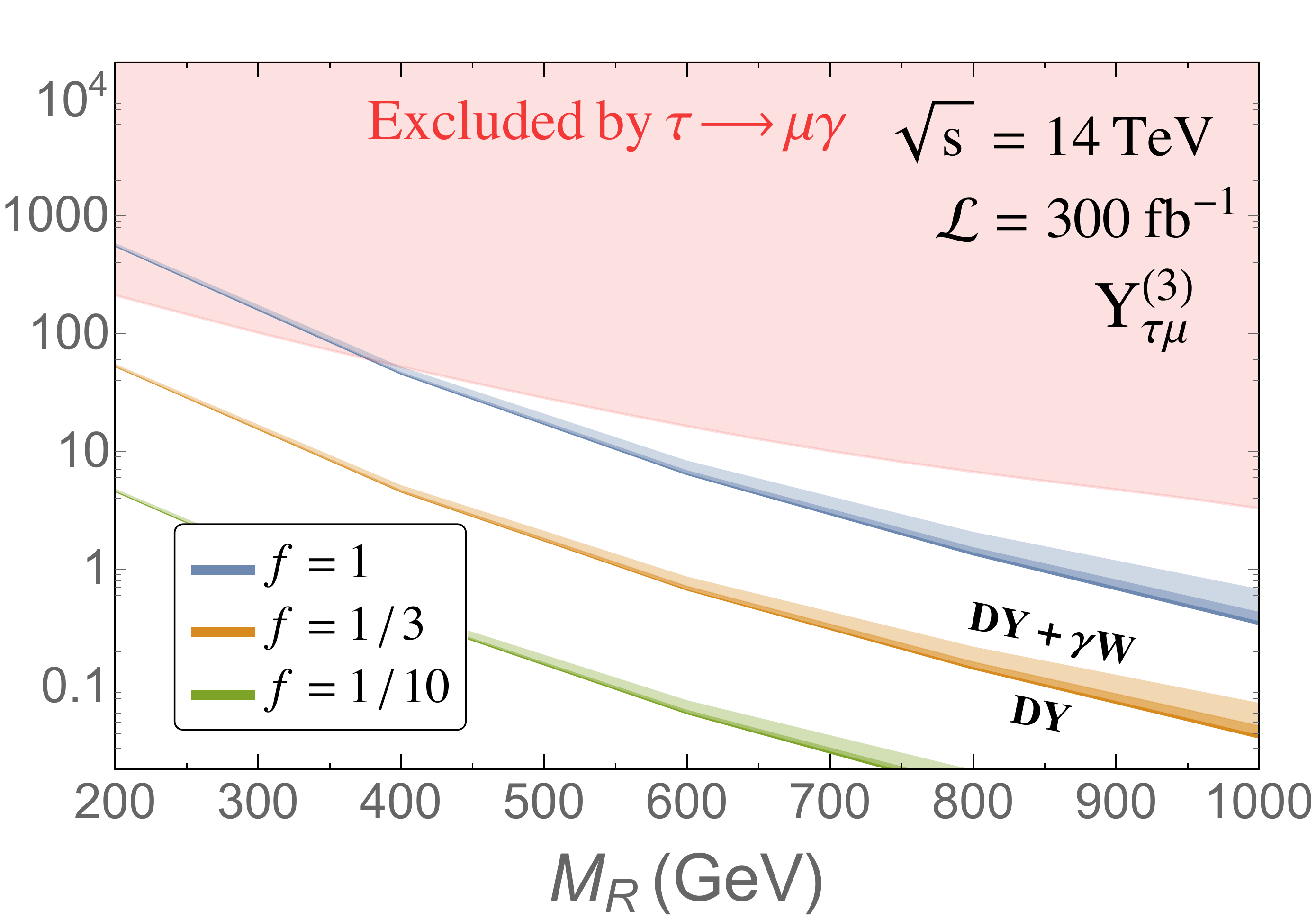} 
\caption{Left panels: Heavy neutrino DY-production, normalized by $f^2$, at the LHC for scenarios TM-5 ($Y_{\tau\mu}^{(1)}$), TM-6 ($Y_{\tau\mu}^{(2)}$), TM-7 ($Y_{\tau\mu}^{(3)}$) from \tabref{TMscenarios}.   Processes not shown are negligible.
Right panels: Number of exotic $\mu\tau j j$ events at the LHC for the same scenarios and for three values of $f$. 
For each $f$, the bottom solid line is the prediction of $\mu\tau j j$ events from DY  and the upper lines on top of each of the three shadowed regions are the predictions after adding the $\mu\tau jj$ events from $\gamma W$, and imposing $p_T^{\rm max}=10,20,40$ GeV, from bottom to top, to the two extra jets. 
The upper red shadowed areas are excluded by  $\tau\to\mu\gamma$.
}\label{LHCproduction}
\end{center}
\end{figure}

The results for the scenarios TM-5, TM-6 and TM-7 from \tabref{TMscenarios} are shown in \figref{LHCproduction}, where the dominant DY production cross sections normalized by $f^2$ are plotted as a function of the heavy mass parameter $M_R$.
We see that the production cross sections can be of the fb order, reachable then at the LHC, for masses $M_R\lesssim600$~GeV.
Notice that the results are always equal for the pseudo-Dirac pairs, since their Majorana character plays a subleading role in their production.

We can also learn that the flavor of the associated charged lepton is different depending on the heavy neutrino produced and the scenario considered, and that this pattern can be understood looking at the mixing in \figref{mixingbars}.
For example, $N_{3/4}$ are mainly electronic neutrinos in the  TM-1,2,4,5,7 scenarios and, therefore, they are basically produced exclusively with electrons. 
$N_{1/2}$ are equally produced with muons and taus in the TM-1,2,4,5 scenarios, dominantly produced with electrons in the TM-3 scenario, and mainly produced with taus in the TM-7 and 8 scenarios. 
On the other hand, $N_{5/6}$ are equally produced with muons and taus in TM-1,2,3,4,5,6 scenarios,  mainly produced with muons in the TM-7 scenario and mostly produced with electrons in the TM-8 scenario.

Once the heavy neutrinos are produced, they will decay inside the detector. 
As mentioned in \secref{sec:HeavyWidhts}, in the limit $M_R\gg m_D$ the heavy neutrino masses are close to $M_R$, with small differences of $\mathcal O(m_D^2M_R^{-1})$ between the different pseudo-Dirac pairs and, therefore, assuming that they are practically degenerate,  their decay into each other should be suppressed, with the dominant channels, then,  being $N_j\to Z\nu_i, H\nu_i, W^\pm \ell_i^\mp$. 
The expressions for these relevant decay channels are given in \eqrefs{Nwidth} and (\ref{NwidthFF}).

It is interesting to study the rich flavor structure of the decay products, which depends on the decaying heavy neutrino  and the scenario we are considering.
Like in the production, the flavor preference of the decays to $W^\pm \ell^\mp$,  which are the ones relevant to this study (see \figref{LHCproductiondiagrams}), also follows the same pattern as in \figref{mixingbars}.
Therefore, we can expect the production and decay of the heavy neutrinos to lead to exotic $\mu\tau jj$ events with no missing energy and $M_{jj}\sim M_W$, with $M_{jj}$ the invariant mass of the two jets.

Using the narrow width approximation, the total cross section of the exotic events we are interested in is given by:
\begin{align}
\sigma(pp\to \mu\tau j j) &=\sum_{i=1}^{6}  \Big\{ 
\sigma(pp \to N_i \mu^\pm) {\rm BR}(N_i\to W^\pm \tau^\mp) +\sigma(pp \to N_i \tau^\pm) {\rm BR}(N_i\to W^\pm \mu^\mp)  \Big\} \nonumber\\
&\times {\rm BR}(W^\pm\to jj)\,,
\end{align}
with $\mu\tau jj=\mu^+\tau^-jj + \mu^-\tau^+jj$.

\figref{LHCproduction} shows the expected number of exotic events $\mu\tau j j$ at the LHC for an integrated luminosity of $\mathcal L~=~300~ {\rm fb}^{-1}$ at $\sqrt s=14$ TeV.
The lower solid lines for each choice of $f$ are the number of events considering only the DY-production. 

Moreover, $\gamma W$ fusion processes can also contribute to this kind of exotic events if the $p_T$ of the extra jets are below a maximum value $p_T^{\rm max}$ and, therefore, they can be considered as soft or collinear jets which can escape detection. 
In this case the predicted total number of exotic events is the sum of the events produced by DY and  $\gamma W$ channels.
These total contributions for different values of $p_T^{\rm max}=10,20$ and $40$ GeV are shown as  the border lines on top of the shadowed areas with gradual decreasing intensity above each solid line.
 In addition we have included in the plots red shadowed areas that represent the regions excluded by the experimental upper bound on BR($\tau\to\mu\gamma$), which, as we saw in the previous Chapters, is the most constraining LFV observable at this range of masses.
We can see that, after considering all the LFV constraints, the three scenarios lead to an  interesting number of $\mathcal O(10-100)$ total $\mu\tau jj$ exotic events for the range of $M_R$ studied here of [200 GeV, 1 TeV]. 
Applying constraints from other observables beyond the LFV ones, like the ones studied in \secref{sec:otherconstraints}, could change this final conclusion. 
For example, looking at \figref{ZtaumufMRplane}, we see that in the TM-5 scenario for $M_R$ values in the (200, 1000) GeV interval, the $Z$ invisible width sets a maximum value for $f$ of about 1.5 times stronger, meaning that the maximum allowed number of events would be about 2-3 times smaller. 
Similarly, considering the global fit constraints as in \secref{sec:LFVHDmax} would increase the excluded red area, reducing the maximum number of events predicted in the allowed parameter space.
A more complete analysis deserves further study and will be done in a forthcoming work, where we will also explore the larger luminosity options for future LHC phases. 

The SM backgrounds for events with two leptons of different flavor have been studied in Ref.~\cite{Aad:2015pfa}.
However, a high efficiency in the $\tau$-tagging and a good reconstruction of the $W$ boson invariant mass from the two leading jets would help in reducing the background.
In that case, the main background would come from processes with photons or jets misidentified as leptons, mainly from $W/Z+\gamma^*$, $W/Z+$ jets and multijet events with at least four jets with one of them misidentified as a muon and another as a tau; and from $Z/\gamma^*+ {\rm jets}\to\mu^+\mu^-$ + jets if one of the muons is misidentified as a $\tau$ candidate. 
Nevertheless, a dedicated background study for these particular $\mu\tau jj$ exotic events is needed and will be done in a future work.

Summarizing, in this Chapter we have proposed a new interesting way to study the production and decay of the heavy neutrinos of the ISS in connection with LFV. 
We have presented the computation of the predicted number of exotic $\mu\tau jj$ events which can be produced in these ISS scenarios with large LFV by the production of heavy pseudo-Dirac neutrinos together with a lepton of flavor $\ell$, both via DY and $\gamma W$ fusion processes, and their subsequent decay into $W$ plus a lepton of different flavor.
We have concluded that, for the three TM-like scenarios studied here, a number of $\mathcal O(10-100)$  total $\mu\tau jj$ exotic events without missing energy can be produced at the next  run of the LHC when $300\, {\rm fb}^{-1}$ are collected, for values of $M_R$ from 200 GeV to 1 TeV. 
Similarly, rare  $\tau e jj$ processes could be produced within the equivalent TE ISS scenarios.
Although in other scenarios with large LFV $\mu e jj$ events could also be produced, which would be interesting since they could provide in addition observable CP asymmetries~\cite{Bray:2007ru}, the number of events would be strongly limited by the $\mu \to e \gamma$ upper bound.
This idea of looking for $\tau\mu jj$ exotic events has been recently explored in the context of a 100 TeV pp collider in Ref.~\cite{Golling:2016gvc}, finding promising results for a luminosity of $L=10 {\rm ab}^{-1}$.
Of course, a more realistic study of these exotic events, including detector simulation, together with a full background study should be done in order to reach a definitive conclusion, but this will be addressed in a future work.

\fancyhead[LE] {}

\chapter*{Conclusions}
\label{Conclusions}
\addcontentsline{toc}{chapter}{Conclusions}
\fancyhead[RO] {\scshape Conclusions }

Flavor violating processes have been, are and will be crucial for the construction and development of Particle Physics theory. 
In the last years, the observation of lepton flavor violation in the neutral sector via neutrino oscillations has established that neutrinos do have masses, which is at present the most clear experimental evidence telling us that the SM must be extended. 
In the same manner, any evidence of LFV transitions in the charged sector would automatically imply the presence of new physics, even beyond the SM with neutrino masses minimally added.
This fact makes charged LFV processes an optimal place to look for new physics. 
Unfortunately, no such cLFV processes have been observed yet, although a strong experimental effort is being made in this direction, and future experiments are planning to improve the sensitivities up to really impressive levels.

In general, any modification of the neutral lepton sector in order to account for neutrino masses
will affect directly or indirectly, mainly via quantum corrections, to the charged lepton sector, leaving a trail of phenomenological implications that experiments could potentially observe. 
Among the many different extensions for addressing neutrino mass generation, we have focused on low scale seesaw models, in particular in the ISS and SUSY-ISS models, which share the appealing feature of adding new right-handed neutrinos with masses at the TeV range, i.e., at the energy scale that present colliders as the LHC are exploring. 
Therefore, in this Thesis we have explored the connection between the presence of right-handed neutrinos at the TeV mass scale and the potential existence of processes with charged LFV.

As we discussed when introducing the inverse seesaw model in \chref{Models}, one of its most important features is that it introduces three different mass scales with three different purposes: a small lepton number violating scale, $\mu_X$, responsible of explaining the smallness of the light neutrino masses; a large $M_R$ scale that governs the masses of the heavy pseudo-Dirac neutrino pairs; and a Dirac mass at the electroweak scale, $m_D=vY_\nu$, which controls the interaction between the (mainly left-handed) light and (mainly right-handed)  heavy neutrinos with the Higgs boson. 
Along this Thesis, we have clearly seen that the most relevant parameters for the cLFV processes that we are interested in are $M_{R}$ and $Y_\nu$.
Consequently, we have introduced a new parametrization for accommodating neutrino oscillation data,  the $\mu_X$ parametrization, alternative to the often used Casas-Ibarra parametrization, that allows to choose precisely $M_R$ and $Y_\nu$ as independent input parameters of the model. 

In order to gain intuition on the general properties of cLFV processes in the ISS model, we have first revisited in \chref{PhenoLFV} the LFV lepton decays, meaning the radiative decays $\ell_m\to\ell_k\gamma$ and the three body decays $\ell_m\to\ell_k\ell_k\ell_k$ with $k\neq m$.
This study has allowed us to establish the basic ideas of our analysis, as well as to understand the main differences of using the Casas-Ibarra parametrization or the $\mu_X$ parametrization.
We have seen that, although physics must not depend on the parametrization one chooses, the efficiency of an analysis in reaching some particular but interesting directions in the parameter space may radically change.
As a particular example of this idea, we have studied the LFV radiative decays when using the $\mu_X$ parametrization, where the Yukawa coupling matrix is one of the independent input parameters.
Using this freedom, and the geometrical interpretation of the Yukawa matrix discussed in \secref{sec:scenarios}, we were able to define directions in the parameter space where the cLFV transitions are favored between two particular flavors, while keeping $\mu$-$e$ transitions always highly suppressed. 
This is particularly interesting in the light of present experimental constraints on cLFV processes, since there are very strong bounds in the $\mu$-$e$ sector, while they are weaker in the $\tau$-$e$ and $\tau$-$\mu$ sectors and, therefore, there is  more room for larger allowed LFV predictions in these two latter sectors.

In \chref{LFVHD} we have studied in full detail the LFV Higgs decays  $H \to \ell_k\bar \ell_m$ induced at the one-loop level from the ISS right-handed neutrinos.
We have presented a full one-loop computation of the BR($H \to \ell_k\bar \ell_m$) rates for the three possible channels, $\ell_k\bar \ell_m=\mu \bar \tau, e \bar \tau, e \bar \mu$, and have also analyzed in full detail the predictions as functions of the various relevant ISS parameters. 
We found, as in the LFV lepton decays, that the most relevant parameters are $M_R$ and $Y_\nu$.
Nevertheless, we have seen that, interestingly, the dependence of the LFVHD rates on these parameters is not the same as that of the LFV radiative decays. 

In order to better understand these differences, we have performed a new and independent computation using a very different approach which turns out to provide simpler and more useful analytical results.
Instead of applying the usual diagrammatic method of  the full one-loop computation, we have used the mass insertion approximation, which works with the chiral EW neutrino basis, including the left- and right-handed states $\nu_L$ and $\nu_R$ and the extra singlets $X$  of the ISS,  instead of dealing with the nine physical neutrino states, $n_{1\dots9}$, of the mass basis.   

To further simplify this MIA computation, we have first prepared the chiral basis in a convenient way, such that all the effects of the singlet $X$ states are collected into a redefinition of the $\nu_R$ propagator,  which we have called here {\it fat propagator},  and then we have derived the set of Feynman rules for these proper chiral states that summarizes the relevant interactions involved in the one-loop computation of the LFVHD rates. 
The peculiarity of using this particular chiral basis is that it leads to a quite generic set of Feynman rules for the subset of interactions involving the neutrino sector, mainly $\nu_L$ and $\nu_R$,  which are the relevant ones for the LFV observables of our interest here, and therefore our results could be valid for other low scale seesaw models sharing these same Feynman rules.  
With the MIA  we have then organized the one-loop computation of the LFVHD rates in terms of a perturbative expansion in powers of the neutrino Yukawa coupling matrix $Y_\nu$. 
It is worth recalling that in the ISS model, the $Y_\nu$ matrix is the unique relevant origin of LFV and, thus, it is the proper expansion parameter in the MIA computation.

 We have presented the analytical results using the MIA for the form factors that define the one-loop LFVHD amplitude, and we have done this computation first to  leading order, ${\cal O}((Y_\nu^{} Y_\nu^\dagger)_{km})$, and later to the next to leading order, i.e., including terms up to ${\cal O}((Y_\nu^{} Y_\nu^\dagger Y_\nu^{} Y_\nu^\dagger)_{km})$. 
 Moreover, we have demonstrated that our analytical results are gauge invariant, obtaining the same result in the Feynman-'t Hooft gauge and in the unitary gauge.  
This is certainly a good check of our analytical results.
 Numerically, we have found that in order to get a good numerical convergence of the MIA with the full results, it is absolutely necessary to include both $\mathcal O(Y_\nu^2)$ and $\mathcal O(Y_\nu^4)$ terms. 
Indeed, the presence of the $\mathcal O(Y_\nu^4)$ terms is what explains the different funcional behavior with the parameters that we observed for the LFVHD rates with respect to the radiative decays, which are well described with only the $\mathcal O(Y_\nu^2)$ terms.
We have then checked numerically that the MIA works pretty well in a big range of the relevant model parameters $Y_\nu$ and $M_R$. 
For a small Yukawa coupling,  given in our notation  by a small global factor, say $f<0.5$, we have obtained an extremely good convergence of the MIA and the full results even for moderate $M_R$ of a few hundred GeV and above. 
For larger Yukawa couplings, say with $0.5< f< 2$ we have also found a good convergence, but for heavier $M_R$ of above ${\cal O}(1 {\rm TeV})$.
 
 In addition to the form factors, we have also derived in \secref{sec:VeffMIA} an analytical expression of the LFV effective vertex describing the $H\ell_k\ell_m$ coupling that is radiatively generated to one-loop from the  heavy right-handed neutrinos. 
 For that computation we have presented our systematic expansion of the form factors in inverse powers of $M_R$, which is valid in the mass range of our interest, $m_\ell \ll m_D,m_W,m_H\ll M_R$, and we have found the most relevant terms of ${\cal O}(v^2/M_R^2)$ in this series. 
 In doing this expansion, we have taken care of the contributions from the external Higgs boson momentum which are relevant since in this observable the Higgs particle is on-shell,   and we  have also followed the track of all the EW masses involved,  like $m_W$ and $m_H$, which are both of order  $v$ and therefore contribute to the wanted ${\cal O}(v^2/M_R^2)$ terms.  
 The lepton masses (except for the global factor from the heaviest lepton $m_{\ell_k}\gg m_{\ell_m}$) do not provide relevant corrections and have been neglected in this computation of the effective vertex. 
 We have shown with several examples that this simple MIA formula works extremely well for the interesting window in the $(Y_\nu, M_R)$ parameter space which is allowed by the present experimental constraints.  
 Therefore, we believe that our final analytical formula for the LFV effective $H\ell_k\ell_m$ vertex given in \eqref{VeffMIA}  is very simple and can be useful for other authors who wish to perform a fast estimate of the LFVHD rates in terms of their own preferred input parameter values for $Y_\nu$  and $M_R$.

For the numerical estimates of the full one-loop results of the LFVHD rates, we have explored the ISS parameter space considering again the two discussed parametrizations for accommodating light neutrino masses and mixings.
First, we have considered the Casas-Ibarra parametrization and explored the LFVHD rates  from the simplest case of diagonal $\mu_X$ and $M_R$ matrices with degenerate entries for $M_{R_i}$, to a more general case with hierarchical heavy neutrinos. 
In these cases, we concluded that the largest maximum LFV Higgs decay rates within the ISS that are allowed
by the constraints on the LFV radiative decays are for BR($H \to e \bar \tau$) and BR($H \to \mu \bar \tau$) and reach at most $10^{-10}$ for the degenerate heavy neutrino case and $10^{-9}$ for the hierarchical case. 
Second, we have considered the $\mu_X$ parametrization and explored the phenomenologically well motivated scenarios that are more promising for LFVHD searches in the $\tau$-$e$ and $\tau$-$\mu$ sectors. 
 We have demonstrated that in this kind of ISS scenarios there are solutions with much larger allowed LFVHD rates than in the previous cases, leading to maximal rates allowed by the bounds on the radiative decays of around $10^{-5}$ for either BR($H \to \mu \bar \tau$) or BR($H \to e \bar \tau$). 

Finally, we have considered  the effects of other kind of constraints to the ISS parameter space by making use of the global fit analysis to present data and the perturbativity requirements on the Yukawa couplings. 
These constraints result in allowed BR($H\to e \bar \tau$) and BR($H\to\mu\bar\tau$) ratios being at most of about $10^{-7}$, which are unfortunately far below the present experimental sensitivities and, therefore, future experiments would be needed for testing these predictions.

In \secref{sec:LFVHDSUSY}, we have also addressed the question of whether the SUSY realization of the ISS model can lead to enhanced predictions for the LFV Higgs decay rates. 
We have considered the MSSM model with the lightest $CP$-even Higgs boson $h$ identified as the SM-like Higgs boson, and extended  with three pairs of ISS neutrinos and their corresponding SUSY partners, the sneutrinos.
We have then presented the results of an updated and full one-loop calculation of the SUSY contributions to lepton flavor violating Higgs decays in the SUSY-ISS model. 
These contributions come from chargino-sneutrino loops with sneutrino couplings off-diagonal in flavor, and from neutralino-slepton loops,  due to the misalignment in flavor between the slepton and lepton sectors caused by running effects. 
We found much larger contributions than in the type-I seesaw model  coming from the lower values of $M_R \sim {\cal O}(1$ TeV), an increased RGE-induced slepton mixing, and the presence of new right-handed sneutrinos at the TeV scale.
Then, the couplings of both sleptons and sneutrinos can transmit sizable LFV due to the large $Y_\nu^2/(4 \pi) \sim  {\cal O} (1) $ we considered. 
We showed that the branching ratio of $h \to \tau \bar \mu$ exhibits different behaviors as a function of the seesaw and SUSY scale if it is dominated by chargino or neutralino loops. 
Moreover, a non-zero trilinear coupling $A_\nu$ leads to increased LFVHD rates. 
Choosing different benchmark points, we found that  BR($h \to \tau \bar\mu$) of the order of $10^{-2}$ can be reached while agreeing with the experimental limits on radiative decays, which can be tested at the present runs of the LHC.
This calls up for a complete study including non-supersymmetric contributions in the SUSY-ISS model, like those from the extended Higgs sector, and a detailed analysis of experimental constraints beyond radiative LFV decays, which will be addressed in a future work.

In \chref{LFVZD} we have revisited the LFV $Z$ decays in presence of right-handed neutrinos with TeV range masses, which are very interesting observables that are currently being searched for at the LHC and will be further explored by the next generation of experiments. 
A first study of these observables within the ISS context with three pairs of fermionic singlets was done in  Ref.~\cite{Abada:2014cca}, finding maximum allowed ratios of about $10^{-9}$.
Here, we have alternatively studied in full detail the LFVZD rates in our selected TM and TE scenarios, which as we said are designed to find large rates for processes including a $\tau$ lepton, and we have  investigated those that are allowed by all the present constraints. 
In addition to the radiative decays, important constraints come from experimental upper bounds on the LFV three body lepton decays, since they are strongly correlated to the LFVZD in these scenarios. 
Taking into account all the relevant bounds, we found that heavy ISS neutrinos with masses in the few TeV range can induce maximal rates of BR$(Z\to\tau\mu)\sim 2 \times 10^{-7}$ and BR$(Z\to\tau e)\sim 2 \times 10^{-7}$ in the TM and TE scenarios, respectively. 
These rates are considerably larger than what was found in previous studies and potentially measurable at future linear colliders and FCC-ee. 
Therefore, we have seen that searches for LFVZD at future colliders may be a powerful tool to probe cLFV in low scale seesaw models, in complementarity with low-energy (high-intensity) facilities searching for cLFV processes.

Another appealing feature of our results is that the predictions for the cLFV processes come together with the possibility that the heavy neutrinos could be directly produced at the LHC.
Being the ISS neutrinos pseudo-Dirac fermions, the standard same-sign dilepton searches for heavy Majorana neutrinos are not effective, implying that new search strategies need to be explored.
In \chref{LHC}  we have proposed a new interesting way of studying  the production and decay of the heavy neutrinos of the ISS in connection with LFV. 
We have presented the computation of the predicted number of exotic $\mu\tau jj$ events, which can be produced in the TM scenarios with large LFV, where the heavy pseudo-Dirac neutrinos are produced together with a lepton of a given flavor, both via Drell-Yan and $\gamma W$ fusion processes, and then decay into a $W$ and a lepton of different flavor.
We have concluded that, for the studied benchmark scenarios, a number of $\mathcal O(10-100)$  total $\mu\tau jj$ exotic events without missing energy can be produced at the next run of the LHC when $300~{\rm fb}^{-1}$ of integrated luminosity are reached, and for values of $M_R$ from 200 GeV to 1 TeV respecting the constraints from LFV violating observables. 
Similarly, other rare processes like $\tau e jj$ or $\mu e jj$ could be produced within other ISS scenarios with large LFV, although for the latter ones 
the number of events would be strongly limited by the $\mu \to e \gamma$ upper bound.
These promising results deserve a more realistic study of these exotic events, including detector simulation, together with a full background study, which should be done in order to reach a definitive conclusion and it will be addressed in a future work.

As an overall conclusion of this Thesis, we can state that searching for charged lepton flavor violating processes is a very powerful strategy for testing the presence of low scale seesaw neutrinos with masses of a few TeV or below, which on the other hand are common in many models for explaining the observed neutrino masses. 
As we have seen along this Thesis, the addition to the SM of these new states, not much heavier that the EW scale and with a potentially complex flavor structure, has an important impact in the phenomenology of the charged leptons, which could be seen at lepton flavor violating processes.
Flavor physics has been crucial in the history of the SM and it will play a major role in the discovery of new physics.


\chapter*{Conclusiones}
\label{Conclusiones}
\addcontentsline{toc}{chapter}{Conclusiones}
\fancyhead[RO] {\scshape Conclusiones }

Los procesos con violaci\'on de sabor han sido, son y ser\'an cruciales para la construcci\'on y desarrollo te\'orico de la F\'isica de Part\'iculas. 
En los \'ultimos a\~nos, la observaci\'on de violaci\'on de sabor lept\'onico en el sector neutro a trav\'es de  las oscilaciones de neutrinos ha demostrado que los neutrinos tienen masas, hecho que supone la evidencia experimental actual m\'as clara de que el SM debe ser extendido. 
De la misma manera, cualquier evidencia de transiciones con LFV en el sector cargado supondr\'ia autom\'aticamente la existencia de nueva f\'isica, m\'as all\'a incluso del SM con las masas de los neutrinos a\~nadidas de manera m\'inima. 
Este hecho hace de los procesos con LFV en el sector cargado un lugar \'optimo para buscar nueva f\'isica.
Desafortunadamente, no se ha observado ning\'un proceso con cLFV todav\'ia, aunque se est\'a realizando un gran esfuerzo experimental en esta l\'inea y los futuros experimentos prev\'en mejorar las sensibilidades a este tipo de procesos hasta niveles realmente impresionantes. 

En general, cualquier modificaci\'on en el sector de leptones neutros tratando de explicar la masa de los neutrinos afectar\'a directa o indirectamente, a trav\'es de correcciones cu\'anticas, al sector de leptones cargados, dejando as\'i una traza de implicaciones fenomenol\'ogicas que los experimentos podr\'ian llegar a observar.
Entre las muchas y diferentes extensiones posibles para explicar la generaci\'on de masas de los neutrinos,  nos hemos centrado en los modelos {\it seesaw} de baja escala, en particular en los modelos ISS y SUSY-ISS, que comparten la interesante cualidad de a\~nadir nuevos neutrinos dextr\'ogiros con masas en el rango del TeV, i.e., a la escala de energ\'ia que los colisionadores actuales como el LHC est\'an explorando. 
Por tanto, en esta Tesis hemos explorado la conexi\'on entre la presencia de neutrinos dextr\'ogiros en la escala de masas del TeV y la posible existencia de procesos con LFV en el sector cargado. 

Como hemos discutido al introducir el modelo de {\it seesaw} inverso en el Cap\'itulo~\ref{Models}, una de sus cualidades m\'as interesantes es que introduce tres escalas diferentes con tres prop\'ositos muy diferentes: una escala ligera que viola el n\'umero lept\'onico, $\mu_X$, la responsable de explicar la ligereza de la masa de los neutrinos que observamos, una escala pesada $M_R$ que gobierna las masas de los pares de neutrinos pesados pseudo-Dirac, y una masa de Dirac en la escala electrod\'ebil, $m_D=vY_\nu$, que controla la interacci\'on entre los neutrinos ligeros (mayormente lev\'ogiros) y los pesados (mayormente dextr\'ogiros) con el bos\'on de Higgs. 
A lo largo de esta Tesis, hemos visto claramente que los par\'ametros m\'as relevantes para los procesos con cLFV en los que estamos interesados son $M_R$ y $Y_\nu$.
Por tanto, hemos introducido una nueva parametrizaci\'on para ajustar los datos de las oscilaciones de neutrinos, la parametrizaci\'on $\mu_X$, alternativa a la  frecuentemente utilizada parametrizaci\'on de Casas-Ibarra, y que permite elegir como par\'ametros libres del modelo precisamente $M_R$ y $Y_\nu$.

Con el objetivo de obtener intuici\'on sobre las propiedades generales de los procesos con cLFV en el modelo ISS, en el Cap\'itulo~\ref{PhenoLFV} hemos empezado por repasar las desintegraciones con LFV de los leptones, centr\'andonos en las radiativas $\ell_m\to\ell_k\gamma$ y las de tres cuerpos $\ell_m\to\ell_k\ell_k\ell_k$ con $k\neq m$.
Este estudio nos ha permitido establecer las ideas b\'asicas de nuestro an\'alisis, as\'i como entender las principales diferencias a la hora de utilizar la parametrizaci\'on de Casas-Ibarra o la de $\mu_X$.
Hemos visto que, aunque la f\'isica no puede depender de la parametrizaci\'on que uno utilice, esta decisi\'on s\'i que puede afectar a la eficiencia de un an\'alisis a la hora de alcanzar ciertas direcciones particulares pero interesantes del espacio de par\'ametros.
Como ejemplo ilustrativo de esta idea, hemos estudiado las desintegraciones radiativas con LFV utilizando la parametrizaci\'on $\mu_X$, lo que nos ha permitido trabajar con la matriz de acoplamiento de Yukawa como par\'ametro libre. 
Utilizando esta libertad, junto con  la interpretaci\'on geom\'etrica de la matriz de Yukawa discutida en la Secci\'on~\ref{sec:scenarios}, hemos podido definir direcciones en el espacio de par\'ametros donde las transiciones con cLFV est\'an favorecidas entre dos sabores particulares, a la vez que est\'an muy suprimidas en el sector $\mu$-$e$.
Esta situaci\'on es especialmente interesante a la luz de las cotas actuales sobre procesos cLFV, dado que las cotas son muy fuertes en el sector $\mu$-$e$, pero m\'as d\'ebiles en los sectores $\tau$-$e$ y $\tau$-$\mu$ y, por tanto, existe m\'as espacio para procesos con LFV con tasas de desintegraci\'on m\'as altas en estos dos \'ultimos sectores. 

En el Cap\'itulo~\ref{LFVHD} hemos estudiado detalladamente las desintegraciones con LFV del bos\'on de Higgs, $H\to\ell_k\bar\ell_m$, inducidas por los neutrinos dextr\'ogiros del modelo ISS a nivel de un {\it loop}. 
Hemos presentado el c\'alculo completo de BR($H\to\ell_k\bar\ell_m$) a un {\it loop} para los tres posibles canales, $\ell_k\bar \ell_m=\mu \bar \tau, e \bar \tau, e \bar \mu$, y hemos analizado en detalle la predici\'on de este observable en funci\'on de los diferentes par\'ametros del modelo ISS.
Hemos encontrado que, al igual que en el caso de las desintegraciones  radiativas con LFV, los par\'ametros m\'as relevantes son $M_R$ y $Y_\nu$.
Sin embargo, hemos visto que, interesantemente, la dependencia de los LFVHD con estos par\'ametros no es la misma que la de las desintegraciones radiativas con LFV. 

Con el objetivo de entender mejor estas diferencias, hemos realizado un c\'alculo nuevo e independiente, abordando el problema de una manera diferente y que proporciona resultados anal\'iticos m\'as simples y \'utiles.
En vez de realizar el c\'alculo completo a nivel de un {\it loop} siguiendo el proceso diagram\'atico habitual, hemos usado la t\'ecnica de la aproximaci\'on de inserci\'on de masa, que trabaja con la base quiral electrod\'ebil de los neutrinos, incluyendo los estados lev\'ogiros $\nu_L$, dextr\'ogiros $\nu_R$ y los singletes extra $X$ del ISS, en vez de lidiar con los nueve neutrinos f\'isicos, $n_{1\dots9}$, de la base de masas.  

De cara a simplificar a\'un m\'as nuestro c\'alculo en la MIA, primero hemos preparado la base quiral en una forma conveniente, de modo que todos los efectos de los singletes $X$ est\'en contenidos en la redefinici\'on de los propagadores de $\nu_R$, a los que hemos llamado {\it propagadores gordos}, y despu\'es hemos derivado el conjunto de reglas de Feynman de las interacciones necesarias para realizar el c\'alculo a un {\it loop} de los LFVHD en esta base quiral. 
La peculiaridad de usar esta base quiral es que lleva a una serie de reglas de Feynman bastante gen\'ericas para este conjunto de interacciones que involucran el sector de los neutrinos, $\nu_L$ y $\nu_R$ mayormente, que son los relevantes para los observables con LFV en los que estamos interesados, y por tanto los resultados obtenidos podr\'ian ser v\'alidos para cualquier otro modelo de {\it seesaw} de baja escala que comparta las mismas reglas de Feynman. 
Usando la t\'ecnica de la MIA hemos organizado el c\'alculo de los LFVHD a un {\it loop} como una expansi\'on perturbativa en potencias de la matriz de acoplamiento Yukawa $Y_\nu$ de los neutrinos.
Merece la pena remarcar que en el modelo ISS, la matriz $Y_\nu$ es el \'unico origen relevante de  LFV y, por tanto, es el par\'ametro adecuado en el que realizar la expansi\'on MIA. 

Hemos presentado los resultados anal\'iticos de nuestro c\'alculo con la MIA de los factores de forma que definen la amplitud del LFVHD, c\'alculo que hemos realizado al orden dominante ${\cal O}((Y_\nu^{} Y_\nu^\dagger)_{km})$ primero, y al siguiente orden despu\'es, i.e., incluyendo los t\'erminos de hasta ${\cal O}((Y_\nu^{} Y_\nu^\dagger Y_\nu^{} Y_\nu^\dagger)_{km})$. 
Asimismo, hemos demostrado la invariancia gauge de nuestros resultados obteni\'endolos tanto en el gauge de Feynman-'t Hooft como en el gauge unitario.
Esta es sin duda una buena comprobaci\'on de nuestros resultados.
Num\'ericamente, hemos visto que para obtener una buena convergencia entre los resultados de la MIA y los completos es necesario incluir tanto los t\'erminos $\mathcal O(Y_\nu^2)$ como los $\mathcal O(Y_\nu^4)$. 
De hecho, son justo estos t\'erminos $\mathcal O(Y_\nu^4)$ los que explican las diferencias en el comportamiento de las tasas de LFVHD con los par\'ametros del modelo con respecto al de las desintegraciones radiativas, el cual es bien descrito s\'olo con los t\'erminos $\mathcal O(Y_\nu^2)$.
Hemos comprobado que estas f\'ormulas dan un resultado muy parecido al del c\'alculo completo en un gran rango de los par\'ametros relevantes $Y_\nu$ y $M_R$.
En el caso de acoplamientos Yukawa peque\~nos, que en nuestra notaci\'on viene definido por un valor peque\~no de $f$, digamos $f<0.5$, hemos obtenido una convergencia muy buena entre los resultados de la MIA y los completos, incluso para valores de $M_R$ moderados, de unos pocos cientos de GeV, y mayores. 
Para acoplamientos Yukawa m\'as grandes, digamos $0.5<f<2$, tambi\'en hemos visto que los resultados convergen, pero esto ocurre a valores m\'as altos de $M_R$ por encima de $\mathcal O(1 {\rm TeV})$.

Adem\'as de los factores de forma, en la Secci\'on~\ref{sec:VeffMIA} hemos derivado una expresi\'on anal\'itica para el v\'ertice efectivo con LFV que describe el acoplamiento $H\ell_k\ell_m$ que se genera radiativamente a nivel de un {\it loop} con neutrinos dextr\'ogiros pesados.
Para este c\'alculo hemos realizado una expansi\'on sistem\'atica de los factores de forma en potencias inversas de $M_R$, v\'alida en el rango de masas en el que estamos interesados $m_\ell \ll m_D,m_W,m_H\ll M_R$, y hemos obtenido los t\'erminos m\'as relevantes a ${\cal O}(v^2/M_R^2)$. 
En esta expansi\'on hemos tenido en cuenta las contribuciones del momento de la pata externa del bos\'on de Higgs, que son necesarias al tratarse de un observable con el Higgs en su capa de masas, y tambi\'en hemos seguido la pista a todas las masas EW involucradas, $m_W$ y $m_H$, ambas de orden $v$ y por tanto importantes para los t\'erminos $\mathcal O(v^2/M_R^2)$.
Las masas de los leptones, sin embargo, no contribuyen de manera relevante (salvo en el factor global con una masa del lept\'on m\'as pesado, $m_{\ell_k}\gg m_{\ell_m}$ en nuestro caso) y las hemos por tanto despreciado al calcular el v\'ertice efectivo. 
Hemos demostrado con varios ejemplos que esta simple f\'ormula de la MIA funciona muy bien en la ventana de inter\'es del espacio de par\'ametros $(Y_\nu, M_R)$ permitido por las cotas experimentales actuales.
Por tanto, consideramos que nuestra f\'ormula anal\'itica final para el v\'ertice efectivo LFV $H\ell_k\ell_m$, dado en la Ec.~(\ref{VeffMIA}), es simple y puede ser \'util para otros autores que deseen estimar de forma r\'apida las tasas de LFVHD para sus valores preferidos de $Y_\nu$ y $M_R$.

A la hora de realizar las estimaciones num\'ericas de los resultados a un {\it loop} completos de las tasas LFVHD, hemos explorado el espacio de par\'ametros del modelo ISS considerando de nuevo las dos parametrizaciones mencionadas anteriormente, que resultan  \'utiles para ajustar los datos de las oscilaciones de neutrinos. 
En primer lugar, hemos considerado la parametrizaci\'on de Casas-Ibarra y explorado las tasas de LFVHD desde el caso m\'as simple en el que las matrices $\mu_X$ y $M_R$ son diagonales y con entradas degeneradas para $M_{R_i}$, hasta un caso m\'as gen\'erico donde los neutrinos pesados son jer\'arquicos. 
En este caso, hemos conclu\'ido que las mayores tasas de desintegraci\'on con LFV del Higgs permitidas por las cotas sobre las desintegraciones radiativas con LFV son para los canales BR($H\to e\bar\tau$) y BR($H\to\mu\bar\tau$) y alcanzan como mucho $10^{-10}$ en el caso con neutrinos pesados degenerados y $10^{-9}$ en el caso jer\'arquico. 
En segundo lugar, hemos considerado la parametrizaci\'on $\mu_X$ y explorado los escenarios reviamente  introducidos y motivados fenomenol\'ogicamente, que son m\'as prometedores de cara a las tasas de LFVHD en los sectores $\tau$-$e$ y $\tau$-$\mu$.
Hemos demostrado que en este tipo de escenarios del ISS existen soluciones, permitidas por las desintegraciones radiativas, con tasas de LFVHD mayores que en el caso anterior, de hasta $10^{-5}$ para BR($H\to\mu\bar\tau$) o BR($H\to e\bar\tau$).

Por \'ultimo, hemos estudiado los efectos de otro tipo de restricciones al espacio de par\'ametros del ISS  considerando los resultados de los an\'alisis globales a los datos actuales, y exigiendo tambi\'en que los acoplamientos Yukawa sean perturbativos. 
Estas cotas resultan en un m\'aximo para BR($H\to e\bar \tau$) y BR($H\to\mu\bar\tau$) del orden de $10^{-7}$, aproximadamente, las cuales, por desgracia, se encuentran lejos de las sensibilidades experimentales actuales y, por ende, se necesitar\'an experimentos futuros para testar estas predicciones.

En la Secci\'on~\ref{sec:LFVHDSUSY}, nos hemos planteado la pregunta de si la realizaci\'on SUSY del ISS podr\'ia llevar a predicciones de las tasas de desintegraci\'on LFV del Higgs mayores. 
Hemos considerado para ello el modelo MSSM con el bos\'on de Higgs con $CP$ par m\'as ligero actuando como el bos\'on de Higgs del SM, y  extendido con tres pares de neutrinos del ISS y sus correspondientes compa\~neros SUSY, los sneutrinos. 
Hemos presentado los resultados del c\'alculo completo y actualizado de las contribuciones SUSY a un {\it loop} a las desintegraciones con violaci\'on de sabor lept\'onico en el modelo SUSY-ISS.
Estas contribuciones vienen de los {\it loops} de charginos-sneutrinos, siendo los acoplamientos de los sneutrinos no diagonales en sabor, y de los {\it loops} de neutralinos-sleptones, debido a la no alineaci\'on entre los sectores de sleptones y los leptones inducida por los efectos de las ecuaciones del grupo de renormalizaci\'on.
Hemos encontrado que estas contribuciones son mucho mayores que en el modelo de {\it seesaw} de tipo-I debido al valor m\'as bajo de $M_R\sim\mathcal O(1$~TeV), al crecimiento de la mezcla entre sleptones inducida via RGE, y a la presencia de sneutrinos dextr\'ogiros a la escala del TeV.
En esta situaci\'on los acoplamientos de tanto los sleptones como los sneutrinos pueden transmitir una LFV grande, dado que estamos considerando valores altos para $Y_\nu^2/(4\pi)\sim\mathcal O(1)$.
Hemos demostrado que las anchuras de desintegraci\'on de $h\to\tau\bar\mu$ se comportan diferente al variar la escala {\it seesaw} o la SUSY, dependiendo de si est\'an dominadas por los {\it loops} de charginos o neutralinos. 
Adem\'as, la presencia de un acoplamiento trilineal $A_\nu$ tiende a aumentar a\'un m\'as las tasas LFVHD.
Eligiendo diferentes escenarios, hemos encontrado que pueden alcanzarse  BR($h \to \tau \bar\mu$) del orden de $10^{-2}$ en puntos permitidos por las cotas a las desintegraciones radiativas, valores que podr\'ian ser observados actualmente por el LHC. 
Este hecho motiva un estudio m\'as completo, incluyendo las contribuciones no supersim\'etricas del modelo SUSY-ISS, como las del sector extendido de Higgs, y un an\'alisis m\'as detallado de las cotas experimentales m\'as all\'a de las desintegraciones radiativas con LFV.
Todo ello ser\'a abordado en un futuro trabajo. 

En el Cap\'itulo~\ref{LFVZD} hemos revisado las desintegraciones con LFV del $Z$ en presencia de neutrinos dextr\'ogiros en el rango de masas del TeV, unos observables muy interesantes ya que est\'an siendo actualmente buscados por el LHC, y que la siguiente generaci\'on de experimentos planea explorar  con gran precisi\'on.
Un primer estudio de estos observables en el contexto del ISS con tres pares de singletes fermi\'onicos fue llevado a cabo en la Ref.~\cite{Abada:2014cca}, donde encontraron tasas de desintegraci\'on m\'aximas de $10^{-9}$.
En este Cap\'itulo, en cambio, hemos estudiado en todo detalle las LFVZD en nuestros escenarios TM y TE, que como dec\'iamos han sido dise\~nados para encontrar tasas altas en los procesos que involucran un lept\'on $\tau$, y hemos investigado el espacio de par\'ametros que est\'a permitido por todas las cotas experimentales actuales. 
Junto con las desintegraciones radiativas, hemos encontrado cotas importantes provenientes de las desintegraciones  a tres cuerpos con LFV, dado que est\'an fuertemente correlacionadas con las LFVZD en estos escenarios. 
Teniendo en cuenta todas las cotas relevantes, hemos encontrado que neutrinos del ISS con masas de unos pocos TeV pueden inducir tasas de BR$(Z\to\tau\mu)\sim 2 \times 10^{-7}$ y BR$(Z\to\tau e)\sim 2 \times 10^{-7}$ en los escenarios TM y TE, respectivamente.
Estos valores son considerablemente m\'as altos que los obtenidos en estudios previos y alcanzables por los colisionadores lineales futuros y los FCC-ee.
Por lo tanto, hemos visto que las b\'usquedas de LFVZD en los colisionadores futuros pueden ser una herramienta \'util a la hora de probar la cLFV en los modelos de {\it seesaw} de bajas energ\'ias,   complementariamente a otros experimentos en busca de cLFV de baja energ\'ia.
 
Otra de las cualidades interesantes de nuestros resultados es que las predicciones para los procesos cLFV vienen acompa\~nados de la posibilidad de que los neutrinos pesados sean producidos directamente en el LHC.
Siendo estos neutrinos fermiones pseudo-Dirac, las b\'usquedas est\'andares de neutrinos de Majorana, basadas en leptones con la misma carga, no son efectivas, por lo que nuevas estrategias son necesarias.
En el Cap\'itulo~\ref{LHC} hemos propuesto una nueva forma interesante de estudiar la producci\'on y desintegraci\'on de los neutrinos pesados del ISS en conexi\'on con la LFV. 
Hemos presentado una predicci\'on del n\'umero de eventos ex\'oticos $\mu\tau jj$, que podr\'ian ocurrir en los escenarios TM con gran LFV, donde los neutrinos pesados pseudo-Dirac se producen junto con un lept\'on de un sabor, mediante Drell-Yan o fusi\'on $\gamma W$, y se desintegran dando un $W$ y otro lept\'on de un sabor distinto. 
Hemos concluido que, para los escenarios estudiados, se podr\'ian producir del orden de $\mathcal O(10-100)$  eventos ex\'oticos del tipo $\mu\tau jj$ sin energ\'ia transversa perdida en la pr\'oxima etapa del LHC cuando se alcance la luminosidad integrada de $300~{\rm fb}^{-1}$, para valores de $M_R$ entre los 200~GeV y 1~TeV y respetando las cotas actuales sobre procesos con LFV.  
De manera parecida, otros procesos ex\'oticos del tipo $\tau e jj$ or $\mu e jj$ podr\'ian producirse en escenarios equivalentes del ISS, aunque en estos \'ultimos el n\'umero total de eventos estar\'ia muy suprimido debido a las fuertes cotas sobre  $\mu \to e \gamma$.
Los resultados obtenidos son prometedores y merecen un estudio m\'as realista de estos eventos ex\'oticos, incluyendo la simulaci\'on del detector, junto con un estudio completo del ruido, el cual es necesario para poder concluir de manera definitiva. Este estudio ser\'a llevado a cabo en un trabajo futuro.

Como conclusi\'on final de esta Tesis, podemos afirmar que las b\'usquedas de procesos con violaci\'on de sabor lept\'onico en el sector cargado son una estrategia muy potente a la hora de testar la presencia de neutrinos de modelos {\it seesaw} a baja energ\'ia, con masas en el entorno del TeV o por debajo, que por otro lado son comunes en muchos modelos que explican las masas observadas de los neutrinos. 
Tal y como hemos visto a lo largo de esta Tesis, el hecho de a\~nadir nuevos estados al SM, no mucho m\'as pesados que la escala EW y con una estructura de sabor compleja, tiene un impacto importante en la fenomenolog\'ia de los leptones cargados, algo que podr\'ia ser observado en procesos con violaci\'on de sabor lept\'onico. 
La f\'isica del sabor ha sido crucial en la historia del SM y jugar\'a un papel esencial en el descubrimiento de nueva f\'isica.

\fancyhead[LE] { \scshape Appendix\ \thechapter}
\begin{appendices}
\chapter{Formulas for LFV lepton decays} 
\fancyhead[RO] {\scshape Formulas for LFV lepton decays  }
\label{App:LFVdecays}
In this Appendix we collect, for completeness, the needed formulas for the full one-loop computation of the LFV lepton decays in the particle mass basis, both the three body $\ell_m\to\ell_k\ell_k\ell_k$ and the radiative $\ell_m\to\ell_k\gamma$ decays, with $k\neq m$.
We have taken these expressions from Refs.~\cite{Ilakovac:1994kj,Alonso:2012ji} and implemented them in our code.

In the case of the three body decays, the branching ratio  BR($\ell_m \to \ell_k \ell_k\ell_k$) can be expressed as~\cite{Ilakovac:1994kj,Alonso:2012ji}:
\begin{align}
\text{BR}
(\ell_m \to \ell_k \ell_k \ell_k)=& \frac{\alpha^4_W
}{24576\pi^3}\frac{m^4_{\ell_m}}{m^4_W}\frac{m_{\ell_m}}{\Gamma_{\ell_m}} 
\times \Bigg\{ 2 \left|\frac{1}{2}F^{\ell_m \ell_k \ell_k \ell_k}_{\rm Box}
+F^{\ell_m \ell_k}_Z-2s^2_W\Big(F^{\ell_m \ell_k}_Z-F^{\ell_m \ell_k}_\gamma\Big)\right|^2\nonumber \\
+& 16 s^2_W \text{Re}\left[ \Big(F^{\ell_m \ell_k}_Z 
+\frac{1}{2}F^{\ell_m \ell_k \ell_k \ell_k}_{\rm Box}\Big) G^{\ell_m \ell_k^{\,\scalebox{.75}{*}}}_\gamma \right]
- 48 s^4_W \text{Re}\left[\Big(F^{\ell_m \ell_k}_Z-F^{\ell_m \ell_k}_\gamma\Big)G^{\ell_m \ell_k^{\,\scalebox{.75}{*}}}_\gamma \right] \nonumber \\ 
+&4 s^4_W \left|F^{\ell_m \ell_k}_Z-F^{\ell_m \ell_k}_\gamma\right|^2 
+32 s^4_W \big|G^{\ell_m \ell_k}_\gamma\big|^2\left[\ln \frac{m^2_{\ell_m}}{m^2_{\ell_k}} -\frac{11}{4}	\right]
\Bigg\}\,.  \label{eq:mueee} 
\end{align}
The BR($\ell_m \to \ell_k \ell_k \ell_k$) contains several form factors, corresponding to the dipole, 
penguin (photon and $Z$) and box diagrams.
The expressions for these form factors are given by:  
\begin{align}
G^{\ell_m \ell_k}_\gamma &= \sum_{i=1}^{9} B_{\ell_k n_i}B^*_{\ell_m
  n_i} G_\gamma(x_i)\,,  \nonumber  \\ 
F^{\ell_m \ell_k}_\gamma &= \sum_{i=1}^{9} B_{\ell_k n_i}B^*_{\ell_m
  n_i} F_\gamma(x_i)\,, \nonumber \\ 
F^{\ell_m \ell_k}_Z &= \sum_{i,j=1}^{9} B_{\ell_k n_i}B^*_{\ell_m n_j}
\left(\delta_{ij} F_Z(x_i) + C_{n_i n_j} G_Z(x_i,x_j) + C^*_{n_i n_j} H_Z(x_i,x_j)   \right)\,, \nonumber \\ 
F^{\ell_m \ell_k \ell_k \ell_k}_{\rm Box}&=  \sum_{i,j=1}^{9}B_{\ell_k n_i} B^*_{\ell_m n_j}\left(B_{\ell_k n_i}B^*_{\ell_k n_j}G_{\rm Box}(x_i,x_j)+2\,B^*_{\ell_k n_i}B_{\ell_k n_j}F_{\rm
  Box}(x_i,x_j)\right)\,, \label{eq:formfact}
\end{align}
where $x_i$ stands for the  dimensionless ratio of masses ($x_i = m^2_{n_i}/m_W^2$).
Moreover, the following loop functions enter in the previous form factors~\cite{Ilakovac:1994kj,Alonso:2012ji}: 
\begin{align}
F_Z(x)&= -\frac{5x}{2(1-x)}-\frac{5x^2}{2(1-x)^2}\ln x \, , \nonumber
\\  
G_Z(x,y)&= -\frac{1}{2(x-y)}\left[	\frac{x^2(1-y)}{1-x}\ln x -
  \frac{y^2(1-x)}{1-y}\ln y	\right]\, ,  \nonumber \\ 
H_Z(x,y)&=  \frac{\sqrt{xy}}{4(x-y)}\left[	\frac{x^2-4x}{1-x}\ln
  x - \frac{y^2-4y}{1-y}\ln y	\right] \, , \nonumber \\ 
F_\gamma(x)&= 	\frac{x(7x^2-x-12)}{12(1-x)^3} -
\frac{x^2(x^2-10x+12)}{6(1-x)^4} \ln x	\, , \nonumber \\ 
G_\gamma(x)&=    -\frac{x(2x^2+5x-1)}{4(1-x)^3} -
\frac{3x^3}{2(1-x)^4} \ln x \,,   \nonumber \\	
F_{\rm Box}(x, y) &= \frac{1}{x - y} \bigg\{
\left(1+\frac{x
y}{4}\right)\left[\frac{1}{1-x}+\frac{x^2}{(1-x)^2}\ln
x\right] - 2x
y\left[\frac{1}{1-x}+\frac{x}{(1-x)^2}\ln
x\right] -(x\to y)\bigg\} \, ,  \nonumber \\	
G_{\rm Box}(x, y) &= -\frac{\sqrt{xy}}{x - y}\bigg\{
(4+xy)\left(\frac{1}{1-x}+x \frac{\ln x}{(1-x)^2} \right) 
-2\left(\frac{1}{1-x}+x^2  \frac{\ln x}{(1-x)^2}\right)-(x\to y)\bigg\}\,. 
\label{eq:formfact2}		
\end{align}
In the limit of degenerate neutrino masses ($x=y$), we get the following expressions: 
\begin{align}
G_Z(x,x ) &=  x (-1 + x - 2 \ln x)/(2 (1-x)) \, ,  \nonumber\\
H_Z(x,x ) & =  -  x (4 - 5x + x^2 + (4 - 2x + x^2)\ln
  x)/(4(1 - x)^2)  \, , \nonumber\\ 
F_{\rm Box}(x,x )& =  (4-19 x^2+16 x^3-x^4-2 x(-4+4 x+3 x^2) \ln x)/(4(1-x)^3)\,   , \nonumber\\ 
G_{\rm Box}(x,x)	 &= x \left(6 - 8 x+4 x^2-2 x^3+(4+x^2+x^3) \ln 
 x \right)/(-1 + x)^3  \, .\label{limitval2} 
\end{align}
For the LFV radiative decay rates, we use the analytical formulas appearing in~\cite{Ilakovac:1994kj,Deppisch:2004fa,Alonso:2012ji} that have also been implemented in our code:
\begin{equation}
\label{BRradiative}
{\rm BR}(\ell_m\to \ell_k\gamma)=\frac{\alpha^3_W s_W^2}{256\pi^2}\left(\frac{m_{\ell_m}}{m_W}\right)^4\frac{m_{\ell_m}}{\Gamma_{\ell_m}}\big|G_{mk}\big|^2,
\end{equation}
where $\Gamma_{\ell_m}$ is the total decay width of the lepton $\ell_m$, and
\begin{align}\label{Gmk}
G_{mk}&=\sum_{i=1}^{9}B_{\ell_kn_i} B^*_{\ell_mn_i}\, G_\gamma\left(x_i\right)\,,
\end{align} 
with $G_\gamma(x)$ defined in \eqref{eq:formfact2} and, again, $x_i\equiv m_{n_i}^2/m_W^2$.

\chapter{Formulas for low energy flavor conserving observables} 
\fancyhead[RO] {\scshape Formulas for low energy flavor conserving observables}
\label{App:constraints}
In this Appendix we collect the expressions needed for computing the low energy observables described in \secref{sec:otherconstraints}. 
These formulas are taken from the literature and summarized here for completeness.

\section*{Lepton Universality: $\boldsymbol{\Delta r_k}$}
We collect here the formulas to calculate the quantity $\Delta r_K$ (see \eqref{deltarkandRK}), which parametrizes the deviation with respect to the SM prediction arising from the sterile neutrinos contribution, as a test of lepton flavor universality.
The expression for  $\Delta r_K$ in a generic SM extension with sterile neutrinos has been given in~\cite{Abada:2012mc}:
\begin{equation}
\label{eq:deltarK}
\Delta r_K \,= \,\frac{m_\mu^2 (m_K^2 - m_\mu^2)^2}{m_e^2 (m_K^2 - m_e^2)^2}\,
\frac{\operatornamewithlimits{\sum}_{i=1}^{N_\text{max}^{(e)}} 
 |B_{e n_i}|^2\, \left[m_K^2 (m_{n_i}^2+m_{e}^2) - 
 (m_{n_i}^2-m_{e}^2)^2 \right] \lambda^{1/2}(m_{K} ,m_{n_i}, m_{e})}
{\operatornamewithlimits{\sum}_{j=1}^{N_\text{max}^{(\mu)}} 
 |B_{\mu 
 n_j}|^2\,  \left[m_K^2 (m_{n_j}^2+m_{\mu}^2) - 
 (m_{n_j}^2-m_{\mu}^2)^2 \right] \lambda^{1/2}(m_{K} ,m_{n_j}, m_{\mu})}-1 \,,
\end{equation}
where $N_\text{max}^{e, \mu}$ is the heaviest neutrino mass
eigenstate kinematically allowed in association with $e$ or $\mu$ respectively, and the kinematical function $\lambda(m_{K} ,m_{n_i}, m_{\ell})$ reads~\cite{Abada:2012mc}:
\begin{equation}\label{lambdabc}
\lambda(a,b,c) \, = \, (a^2 - b^2 -c^2)^2 -
4\,b^2\,c^2\,.
\end{equation}

\vspace{.5cm}
\section*{The $\boldsymbol{Z}$ invisible decay width}

The $Z$ invisible decay width in presence of massive Majorana neutrinos, like it is the case of the present ISS model,  reads~\cite{Abada:2013aba}:
\begin{align}
\Gamma(Z&\to{\rm inv.})_{\rm ISS}=
 \sum_{n} \Gamma(Z\to n n)_{\rm ISS} 
 =\, \operatornamewithlimits{\sum}_{i\leq j=1}^{N_\text{max}} (1 - \frac{1}{2} \delta_{ij})
\frac{\sqrt{2} G_F }{48\, \pi\, m_Z}  \times \lambda^{1/2}(m_{Z} ,m_{n_i}, m_{n_j}) \nonumber \\
&  \times \left[2  |C_{n_i n_j}|^2 \left(2 m_Z^2 - m_{n_i}^2 - m_{n_j}^2 - \frac{(m_{n_i}^2 - m_{n_j}^2)^2}{m_Z^2}\right) - 12 m_{n_i} 
   m_{n_j} \text{Re}\Big[\big(C_{n_i n_j}\big)^2\Big]  \right].
\label{eq:Znunu:sum}
\end{align}
where $N_\text{max}$ is  the heaviest neutrino mass which is kinematically allowed and $\lambda$ is given in \eqref{lambdabc}.

\section*{Oblique parameters: $\boldsymbol{S, T, U}$}

The Majorana neutrino contributions to the $S, T, U$ parameters have been computed in Ref.~\cite{Akhmedov:2013hec}. 
We apply those formulas to compute the sterile neutrinos contributions to the oblique parameters in the ISS model.\\
The equation for the $T$ parameter reads:
\begin{align}
\label{eq:Tpar}
T_\text{tot}=&~ T_{\rm ISS}+T_{\text{SM}}=
\frac{-1}{8\pi s^2_W m_W^2}   \Biggl\{
\sum_{\alpha=1}^{3}  m^2_{\ell_\alpha} B_0 (0,m^2_{\ell_\alpha},m^2_{\ell_\alpha})
- 2\,\sum_{i=1}^{9}\sum_{\alpha=1}^{3} \big|B_{\ell_\alpha n_i}\big|^2 \,Q(0,m^2_{n_i},m^2_{\ell_\alpha}) \nonumber\\
&+ \sum_{i,j=1}^{9}\Big(C_{n_i n_j} C_{n_j n_i} Q(0,m^2_{n_i},m^2_{n_j})
+(C_{n_i n_j})^2 m_{n_i} m_{n_j} B_0(0,m^2_{n_i},m^2_{n_j})\Big) \Bigg\} \,,
\end{align}
with the index $\alpha$ refering to the charged leptons and 
\begin{align}
Q(q^2,&m^2_1,m^2_2)\equiv (D-2)B_{00}(q^2,m_1^2,m_2^2)+q^2\bigl[ 
B_1(q^2,m_1^2,m_2^2)+B_{11}(q^2,m_1^2,m_2^2)
\bigr]\,,
\end{align}
where $D\equiv 4-2\epsilon$  ($\epsilon\rightarrow 0$) and  $B_0$, $B_1$, $B_{11}$ and $B_{00}$ are the Passarino-Veltman functions~\cite{Passarino:1978jh} in the {\it LoopTools}~\cite{Hahn:1998yk} notation. 

The SM contribution can be cast as: 
\begin{equation}
\label{eq:TparSM}
 T_{\text{SM}} = -\frac{1}{8\pi s^2_W m_W^2}  \bigg\{3 Q(0,0,0) - 2 \sum_{\alpha=1}^{3} Q(0,0,m^2_{\ell_\alpha}) +\sum_{\alpha=1}^{3}  m^2_{\ell_\alpha} B_0 (0,m^2_{\ell_\alpha},m^2_{\ell_\alpha})\bigg\}\,,
\end{equation}
where it has been used that the active neutrino masses are zero and the leptonic mixing matrix $U$ is unitary in the SM.

\noindent
The equation for the $S$ parameter is:
\begin{align}
\label{eq:parS}
S_\text{tot}=&~S_{\rm ISS}+S_{\text{SM}}
=-\frac{1}{2\pi m_Z^2} \Bigg\{
 \sum_{i,j=1}^{9} C_{n_i n_j} C_{n_j n_i} \Delta Q(m_Z^2,m^2_{n_i},m^2_{n_j}) \nonumber\\
+& \sum_{i,j=1}^{9}(C_{n_i n_j})^2 m_{n_i} m_{n_j} \Big(B_0(0,m^2_{n_i},m^2_{n_j})-B_0(m_Z^2,m^2_{n_i},m^2_{n_j})\Big)\nonumber\\
+&\sum_{\alpha=1}^{3} m^2_{\ell_\alpha} \big(B_0 (0,m^2_{\ell_\alpha},m^2_{\ell_\alpha})-2B_0 (m_Z^2,m^2_{\ell_\alpha},m^2_{\ell_\alpha})\big) +  Q(m_Z^2,m^2_{\ell_\alpha},m^2_{\ell_\alpha})\Bigg\}\,,
\end{align}
where $\Delta Q(q^2,m^2_1,m^2_2)\equiv Q(0,m^2_1,m^2_2)-Q(q^2,m^2_1,m^2_2)$ and
\begin{align}
\label{eq:SparSM}
 S_{\text{SM}} =& -\frac{1}{2\pi m_Z^2}  \bigg\{
 3 \Delta Q(m_Z^2,0,0) \nonumber\\
& + \sum_{\alpha=1}^{3} m^2_{\ell_\alpha} \big(B_0 (0,m^2_{\ell_\alpha},m^2_{\ell_\alpha})-2B_0 (m_Z^2,m^2_{\ell_\alpha},m^2_{\ell_\alpha})\big) +  Q(m_Z^2,m^2_{\ell_\alpha},m^2_{\ell_\alpha})\bigg\}\,.
\end{align}
Finally, the $U$ parameter is given by:
\begin{align}
\label{eq:parU}
U_\text{tot}=&~U_{\rm ISS}+U_{\text{SM}}
=\frac{1}{2\pi m_Z^2}\Bigg\{
 \sum_{i,j=1}^{9} C_{n_i n_j} C_{n_j n_i} \Delta Q(m_Z^2,m^2_{n_i},m^2_{n_j}) \nonumber\\
+& \sum_{i,j=1}^{9}(C_{n_i n_j})^2 m_{n_i} m_{n_j} \Big(B_0(0,m^2_{n_i},m^2_{n_j})-B_0(m_Z^2,m^2_{n_i},m^2_{n_j})\Big)\nonumber\\
-&\sum_{i=1}^9 \sum_{\alpha=1}^3 2\,\frac{m_Z^2}{m_W^2} \big|B_{\ell_\alpha n_i}\big|^2 \Delta Q(m_W^2,m^2_{n_i},m^2_{\ell_\alpha})\nonumber\\
+&\sum_{\alpha=1}^{3} m^2_{\ell_\alpha} \big(B_0 (0,m^2_{\ell_\alpha},m^2_{\ell_\alpha})-2B_0 (m_Z^2,m^2_{\ell_\alpha},m^2_{\ell_\alpha})\big) -  Q(m_Z^2,m^2_{\ell_\alpha},m^2_{\ell_\alpha})\Bigg\}\,,
\end{align}
and its SM contribution reads:
\begin{align}
\label{eq:UparSM}
 U_{\text{SM}} = \frac{1}{2\pi m_Z^2} &\bigg\{3 \Delta Q(m_Z^2,0,0) 
 + \sum_{\alpha=1}^{3}\Big( m^2_{\ell_\alpha} \big(B_0 (0,m^2_{\ell_\alpha},m^2_{\ell_\alpha})
 -2B_0 (m_Z^2,m^2_{\ell_\alpha},m^2_{\ell_\alpha})\big) \nonumber\\
 &-  Q(m_Z^2,m^2_{\ell_\alpha},m^2_{\ell_\alpha})
-2\,\frac{m_Z^2}{m_W^2} \Delta Q(m_W^2,0,m^2_{\ell_\alpha}) \Big)
\bigg\}\,.
\end{align}

\chapter{Form factors for LFVHD in the ISS model}
\fancyhead[RO] {\scshape Form factors for LFVHD in the ISS model}
\label{App:LFVHD}

In this Appendix we collect the analytical results for the form factors contributing to the LFV Higgs decay $H\to\ell_k\bar\ell_m$, as defined in \eqref{LFVHDamp}. 
They are computed in the Feynman-'t Hooft gauge and expressed in the physical basis. The numbers (1)-(10) correspond to the diagrams in \figref{diagsLFVHDphysbasis}.
These formulas are taken from Ref.~\cite{Arganda:2004bz} and adapted for the case of the ISS model with three pairs of fermionic singlets. Notice that we have corrected the global signs of $F_L^{(1)}$, $F_{L,R}^{(4)}$ and $F_{L,R}^{(5)}$, which were typos in the original expressions given in~\cite{Arganda:2004bz}.

In all these formulas, summation over neutrino indices is understood, which run as
 $i,j=1, ..., 9$. 
The loop functions are the Passarino-Veltman functions~\cite{Passarino:1978jh} in the {\it LoopTools}~\cite{Hahn:1998yk} notation, and they are defined in \eqrefs{loopfunctionB} and (\ref{loopfunctionC}).
\begin{align}
F_L^{(1)}& = \frac{g^2}{4 m_W^3} \frac{1}{16 \pi^2} B_{\ell_k n_i} B_{\ell_m n_j}^* \bigg\{m_{\ell_k} m_{n_j} \Big[(m_{n_i} + m_{n_j}) \mbox{Re}\left(C_{n_i n_j}\right) + i(m_{n_j} - m_{n_i}) \mbox{Im}\left(C_{n_i n_j}\right) \Big] \tilde{C}_0 \non \\
&+ (C_{12} - C_{11}) \bigg[(m_{n_i} + m_{n_j}) \mbox{Re}\left(C_{n_i n_j}\right) \left( -m_{\ell_k}^3 m_{n_j} - m_{n_i} m_{\ell_k} m_{\ell_m}^2 + m_{n_i} m_{n_j}^2 m_{\ell_k} + m_{n_i}^2 m_{n_j} m_{\ell_k} \right) \non \\
&+   i(m_{n_j} - m_{n_i}) \mbox{Im}\left(C_{n_i n_j}\right) \left( -m_{\ell_k}^3 m_{n_j} + m_{n_i} m_{\ell_k} m_{\ell_m}^2 - m_{n_i} m_{n_j}^2 m_{\ell_k} + m_{n_i}^2 m_{n_j} m_{\ell_k} \right) \bigg] \bigg\}\,,\non\\
F_R^{(1)}& = \frac{g^2}{4 m_W^3} \frac{1}{16 \pi^2} B_{\ell_k n_i} B_{\ell_m n_j}^* \bigg\{ m_{n_i} m_{\ell_m} \Big[(m_{n_i} + m_{n_j}) \mbox{Re}\left(C_{n_i n_j}\right) - i(m_{n_j} - m_{n_i}) \mbox{Im}\left(C_{n_i n_j}\right) \Big] \tilde{C}_0  \non\\
&+ C_{12} \bigg[(m_{n_i} + m_{n_j}) \mbox{Re}\left(C_{n_i n_j}\right) \left( m_{\ell_m}^3 m_{n_i} - m_{n_i} m_{n_j}^2 m_{\ell_m} - m_{n_i}^2 m_{n_j} m_{\ell_m} + m_{n_j} m_{\ell_k}^2 m_{\ell_m} \right) \non \\
&+   i(m_{n_j} - m_{n_i}) \mbox{Im}\left(C_{n_i n_j}\right) \left( -m_{\ell_m}^3 m_{n_i} + m_{n_i} m_{n_j}^2 m_{\ell_m} - m_{n_i}^2 m_{n_j} m_{\ell_m} + m_{n_j} m_{\ell_k}^2 m_{\ell_m} \right) \bigg] \bigg\}\,,\non
\end{align}
where $C_{11, 12} = C_{11, 12}(m_{\ell_k}^2, m_H^2, m_W^2, m_{n_i}^2, m_{n_j}^2)$ and $\tilde{C}_0 = \tilde{C}_0(m_{\ell_k}^2, m_H^2, m_{W }^2, m_{n_i}^2, m_{n_j}^2)$.

%
%
\begin{align*}
F_L^{(2)}& = \frac{g^2}{2 m_W} \frac{1}{16 \pi^2} B_{\ell_k n_i} B_{\ell_m n_j}^* m_{\ell_k} \Big\{-m_{n_j} \Big[(m_{n_i} + m_{n_j}) \mbox{Re}\left(C_{n_i n_j}\right) + i(m_{n_j} - m_{n_i}) \mbox{Im}\left(C_{n_i n_j}\right) \Big] C_0  \\
&+  (C_{12} - C_{11}) \Big[(m_{n_i} + m_{n_j})^2 \, \mbox{Re}\left(C_{n_i n_j}\right) + i(m_{n_j} - m_{n_i})^2 \, \mbox{Im}\left(C_{n_i n_j}\right)\Big] \Big\}\,,\\
F_R^{(2)}& = -\frac{g^2}{2 m_W} \frac{1}{16 \pi^2} B_{\ell_k n_i} B_{\ell_m n_j}^* m_{\ell_m} \Big\{m_{n_i} \Big[(m_{n_i} + m_{n_j}) \mbox{Re}\left(C_{n_i n_j}\right) - i(m_{n_j} - m_{n_i}) \mbox{Im}\left(C_{n_i n_j}\right) \Big] C_0  \\
&+  C_{12} \Big[(m_{n_i} + m_{n_j})^2 \, \mbox{Re}\left(C_{n_i n_j}\right)+ i(m_{n_j} - m_{n_i})^2 \, \mbox{Im}\left(C_{n_i n_j}\right) \Big] \Big\}\,,
\end{align*}
where $C_{0, 11, 12} = C_{0, 11, 12}(m_{\ell_k}^2, m_H^2, m_W^2, m_{n_i}^2, m_{n_j}^2)$.
%
%
\begin{align*}
F_L^{(3)}& = \frac{g^2}{16 \pi^2} B_{\ell_k n_i} B_{\ell_m n_i}^* m_{\ell_k} m_W \left(C_{11} - C_{12}\right)\,,\\
F_R^{(3)}& =  \frac{g^2}{16 \pi^2} B_{\ell_k n_i} B_{\ell_m n_i}^* m_{\ell_m} m_W C_{12}\,,
\end{align*}
where $C_{11, 12} = C_{11, 12}(m_{\ell_k}^2, m_H^2, m_{n_i}^2, m_W^2, m_W^2)$.
%
%
\begin{align*}
F_L^{(4)}& = \frac{g^2}{4 m_W} \frac{1}{16 \pi^2} B_{\ell_k n_i} B_{\ell_m n_i}^* m_{\ell_k} \Big\{ m_{\ell_m}^2 (C_{12} - 2 C_{11}) + m_{n_i}^2 (C_{11} - C_{12}) - m_{n_i}^2 C_0 \Big\}\,, \\
F_R^{(4)}& = \frac{g^2}{4 m_W} \frac{1}{16 \pi^2} B_{\ell_k n_i} B_{\ell_m n_i}^* m_{\ell_m} \Big\{ \tilde{C}_0 + 2 m_{\ell_m}^2 C_{11} + m_{n_i}^2 C_{12} + (m_{\ell_k}^2 - 2 m_H^2) (C_{11} - C_{12}) + 2 m_{n_i}^2 C_0 \Big\}\,,
\end{align*}
where $C_{0, 11, 12} = C_{0, 11, 12}(m_{\ell_k}^2, m_H^2, m_{n_i}^2, m_W^2, m_W^2)$ and $\tilde{C}_0 = \tilde{C}_0(m_{\ell_k}^2, m_H^2, m_{n_i}^2, m_W^2, m_W^2)$.
%
%
\begin{align*}
F_L^{(5)}& = \frac{g^2}{4 m_W} \frac{1}{16 \pi^2} B_{\ell_k n_i} B_{\ell_m n_i}^* m_{\ell_k} \Big\{\tilde{C}_0 + 2 m_{n_i}^2 C_0 + (m_{n_i}^2 + 2 m_{\ell_k}^2) C_{11} + (m_{\ell_m}^2 - m_{n_i}^2 - 2 m_H^2) C_{12} \Big\} \,,\\
F_R^{(5)}& = -\frac{g^2}{4 m_W} \frac{1}{16 \pi^2} B_{\ell_k n_i} B_{\ell_m n_i}^* m_{\ell_m} \Big\{m_{n_i}^2 C_0 + m_{\ell_k}^2 C_{11} + (m_{\ell_k}^2 - m_{n_i}^2) C_{12} \Big\}\,,
\end{align*}
where $C_{0, 11, 12} = C_{0, 11, 12}(m_{\ell_k}^2, m_H^2, m_{n_i}^2, m_W^2, m_W^2)$ and $\tilde{C}_0 = \tilde{C}_0(m_{\ell_k}^2, m_H^2, m_{n_i}^2, m_W^2, m_W^2)$.
%
%
\begin{align*}
F_L^{(6)}& = \frac{g^2}{4 m_W^3} \frac{1}{16 \pi^2} B_{\ell_k n_i} B_{\ell_m n_i}^* m_{\ell_k} m_H^2 \Big\{m_{n_i}^2 (C_0 + C_{11}) + (m_{\ell_m}^2 - m_{n_i}^2) C_{12} \Big\}\,, \\
F_R^{(6)}& = \frac{g^2}{4 m_W^3} \frac{1}{16 \pi^2} B_{\ell_k n_i} B_{\ell_m n_i}^* m_{\ell_m} m_H^2 \Big\{m_{n_i}^2 (C_0 + C_{12}) + m_{\ell_k}^2 (C_{11} - C_{12})  \Big\}\,,
\end{align*}
where $C_{0, 11, 12} = C_{0, 11, 12}(m_{\ell_k}^2, m_H^2, m_{n_i}^2, m_W^2, m_W^2)$.
%
%
\begin{align*}
F_L^{(7)}& = \frac{g^2}{2 m_W} \frac{1}{16 \pi^2} B_{\ell_k n_i} B_{\ell_m n_i}^* \frac{m_{\ell_m}^2 m_{\ell_k}}{m_{\ell_k}^2 - m_{\ell_m}^2} B_1 \,,\\
F_R^{(7)}& = \frac{g^2}{2 m_W} \frac{1}{16 \pi^2} B_{\ell_k n_i} B_{\ell_m n_i}^* \frac{m_{\ell_k}^2 m_{\ell_m}}{m_{\ell_k}^2 - m_{\ell_m}^2} B_1\,, \\
F_L^{(8)}& = \frac{g^2}{4 m_W^3} \frac{1}{16 \pi^2} B_{\ell_k n_i} B_{\ell_m n_i}^* \frac{m_{\ell_k}}{m_{\ell_k}^2 - m_{\ell_m}^2} \Big\{ m_{\ell_m}^2 (m_{\ell_k}^2 + m_{n_i}^2) B_1 + 2 m_{n_i}^2 m_{\ell_m}^2 B_0 \Big\} \,,\\
F_R^{(8)}& = \frac{g^2}{4 m_W^3} \frac{1}{16 \pi^2} B_{\ell_k n_i} B_{\ell_m n_i}^* \frac{m_{\ell_m}}{m_{\ell_k}^2 - m_{\ell_m}^2} \Big\{ m_{\ell_k}^2 (m_{\ell_m}^2 + m_{n_i}^2) B_1 + m_{n_i}^2 (m_{\ell_k}^2 + m_{\ell_m}^2) B_0 \Big\}\,,
\end{align*}
where $B_{0, 1} = B_{0, 1}(m_{\ell_k}^2, m_{n_i}^2, m_W^2)$.
%
%
\begin{align*}
F_L^{(9)}& = \frac{g^2}{2 m_W} \frac{1}{16 \pi^2} B_{\ell_k n_i} B_{\ell_m n_i}^* \frac{m_{\ell_m}^2 m_{\ell_k}}{m_{\ell_m}^2 - m_{\ell_k}^2} B_1\,, \\
F_R^{(9)}& = \frac{g^2}{2 m_W} \frac{1}{16 \pi^2} B_{\ell_k n_i} B_{\ell_m n_i}^* \frac{m_{\ell_k}^2 m_{\ell_m}}{m_{\ell_m}^2 - m_{\ell_k}^2} B_1\,, \\
F_L^{(10)}& = \frac{g^2}{4 m_W^3} \frac{1}{16 \pi^2} B_{\ell_k n_i} B_{\ell_m n_i}^* \frac{m_{\ell_k}}{m_{\ell_m}^2 - m_{\ell_k}^2} \Big\{ m_{\ell_m}^2 (m_{\ell_k}^2 + m_{n_i}^2) B_1 + m_{n_i}^2 (m_{\ell_k}^2 + m_{\ell_m}^2) B_0 \Big\}\,, \\
F_R^{(10)}& = \frac{g^2}{4 m_W^3} \frac{1}{16 \pi^2} B_{\ell_k n_i} B_{\ell_m n_i}^* \frac{m_{\ell_m}}{m_{\ell_m}^2 - m_{\ell_k}^2} \Big\{ m_{\ell_k}^2 (m_{\ell_m}^2 + m_{n_i}^2) B_1 + 2 m_{n_i}^2 m_{\ell_k}^2 B_0 \Big\}\,,
\end{align*}
where $B_{0, 1} = B_{0, 1}(m_{\ell_m}^2, m_{n_i}^2, m_W^2)$.

\chapter{Form factors for  LFVHD in the SUSY-ISS model}
\fancyhead[RO] {\scshape Form factors for  LFVHD in the SUSY-ISS model}
\label{App:LFVHD_SUSY}

In this Appendix we present the form factors that correspond to the diagrams of \figref{SUSYdiagrams}, together with the relevant couplings needed for performing the computation. 
The original calculation in the SUSY type-I seesaw was done in Ref.~\cite{Arganda:2004bz} in the mass basis and in the Feynman-'t Hooft gauge, which we have adapted  to the SUSY-ISS model.
In order to do that, we have derived the new relevant couplings with respect to the SUSY type-I seesaw model, which we give in the following. 

When compared with the SUSY type-I seesaw, only the coupling factors $A_{R\alpha j}^{(\ell)}$ and $g_{H_x \tilde{\nu}_{\alpha} \tilde{\nu}_{\beta}}$ are modified. 
In the SUSY inverse seesaw, they are defined in the mass basis with diagonal charged leptons by
\begin{align}\label{SUSYISScouplings}
 A_{R\alpha j}^{(e,\mu,\tau)}=& \tilde U_{(1,2,3)\alpha} V_{j1} - \frac{{m_D}_{(1,2,3)k}}{\sqrt{2}m_W\sin \beta} \tilde U_{k+9,\alpha} V_{j2} \,, \nonumber \\
 g_{H_x \tilde{\nu}_{\alpha} \tilde{\nu}_{\beta}} = &
     -i g \bigg[ (g_{LL, \nu}^{(x)})_{ik} \tilde U_{i\alpha}^{*} \tilde U_{k\beta} 
    + (g_{RR, \nu}^{(x)})_{ik} \tilde U_{i+9,\alpha}^{*} \tilde U_{k+9,\beta}  
    +  (g_{LR, \nu}^{(x)})_{ik} \tilde U_{i,\alpha}^{*} \tilde U_{k+9,\beta} \nonumber\\
    &+ (g_{LR, \nu}^{(x)})^*_{ik} \tilde U^*_{k+9,\alpha} \tilde U_{i,\beta}
    +  (g_{LX, \nu}^{(x)})_{ik} \tilde U_{i,\alpha}^{*} \tilde U_{k+12,\beta} + (g_{LX, \nu}^{(x)})^*_{ik} \tilde U^*_{k+12,\alpha} \tilde U_{i,\beta} \bigg]\,, \nonumber\\
 (g_{LL, \nu}^{(x)})_{ik} =&- \frac{m_Z}{2\cos{\theta_W}} \sigma_3^{(x)} \delta_{ik} +\frac{(m_D m_D^\dagger)_{ik}}{m_W \sin \beta} \sigma_6^{(x)}\,, \nonumber \\
 (g_{RR, \nu}^{(x)})_{ik} =& \frac{(m_D^\dagger m_D)_{ik}}{m_W \sin\beta} \sigma_6^{(x)}\,, \nonumber \\
 (g_{LR, \nu}^{(x)})_{ik} =&   \frac{(m_D A_\nu^\dagger)_{ik}}{2m_W\sin\beta} \sigma_2^{(x)} + \frac{\mu}{ 2m_W\sin \beta} (m_D)_{ik} \sigma_7^{(x)} \,, \nonumber \\
 (g_{LX, \nu}^{(x)})_{ik} =&  \frac{(m_D M_R^*)_{ik}}{2 m_W \sin\beta} \sigma_2^{(x)}\,,
\end{align}
which are summed over the internal indices, with $i\,,k=1\,,...\,,3$ and $x$ referring to $H_x=(h, H,A)$. 
We reproduce below, for completeness, the unmodified coupling factors from Ref.~\cite{Arganda:2004bz} (correcting a typo in $W_{Rij}^{(x)}$)  in the mass basis with diagonal charged leptons
\begin{align}
A_{L\alpha j}^{(e,\mu,\tau)}=& -\frac{m_{e,\mu,\tau}}{\sqrt{2}m_W cos\beta}U_{j2}^* \tilde U_{(1,2,3)\alpha}\,, \nonumber \\
B_{L\alpha a}^{(e,\mu,\tau)}=& \sqrt{2}\left[\frac{m_{e,\mu,\tau}}{2m_W cos\beta}N_{a3}^*R_{(1,3,5)\alpha}^{(\ell)} +\left[\sin\theta_W 
  N_{a1}^{'*}-\frac{\sin^2\theta_W}{cos\theta_W}N_{a2}^{'*}\right]R_{(2,4,6)\alpha}^{(\ell)}\right]\,, \nonumber \\
B_{R\alpha a}^{(e,\mu,\tau)}=& \sqrt{2}\left[\left(-\sin\theta_WN_{a1}^{'}-\frac{1}{\cos\theta_W}(\frac{1}{2}-\sin^2\theta_W)N_{a2}^{'}\right)
  R_{(1,3,5)\alpha}^{(\ell)}+ \frac{m_{e,\mu,\tau}}{2m_W \cos\beta}N_{a3}R_{(2,4,6)\alpha}^{(\ell)}\right]\,, \nonumber \\
W_{Lij}^{(x)}=&\frac{1}{\sqrt{2}}\left(-\sigma_1^{(x)}U_{j2}^*V_{i1}^*+ \sigma_2^{(x)}U_{j1}^*V_{i2}^*\right)\,, \nonumber \\
W_{Rij}^{(x)}=&\frac{1}{\sqrt{2}}\left(-\sigma_1^{(x)*}U_{i2}V_{j1}+\sigma_2^{(x)*}U_{i1}V_{j2}\right)\,, \nonumber \\
D_{Lab}^{(x)}=&\frac{1}{2\cos\theta_W}\left[(\sin\theta_W N_{b1}^*-\cos\theta_W N_{b2}^*)(\sigma_1^{(x)}N_{a3}^*+\sigma_2^{(x)}N_{a4}^*) \right. \nonumber \\
  &+(\sin\theta_W N_{a1}^*-\cos\theta_W   N_{a2}^*)(\sigma_1^{(x)}N_{b3}^*+\sigma_2^{(x)}N_{b4}^*)\left.\right] \,, \nonumber \\
D_{Rab}^{(x)}=&D_{Lab}^{(x)*}\,, \nonumber \\
S_{L,\ell}^{(x)} =& - \frac{m_{\ell}}{2 m_W \cos\beta}{\sigma_1^{(x)*}}\,, \nonumber \\
S_{R,\ell}^{(x)} =&S_{L, \ell}^{(x)*} \,, \nonumber \\
g_{H_x \tilde{\ell}_{\alpha} \tilde{\ell}_{\beta}} =& -i g \left[ g_{LL, e}^{(x)} R_{1\alpha}^{*(\ell)} R_{1\beta}^{(\ell)} + 
   g_{RR, e}^{(x)} R_{2\alpha}^{*(\ell)} R_{2\beta}^{(\ell)} + g_{LR, e}^{(x)} R_{1\alpha}^{*(\ell)} R_{2\beta}^{(\ell)} + g_{RL, e}^{(x)} R_{2\alpha}^{*(\ell)} R_{1\beta}^{(\ell)}\right. \nonumber \\
  &+ g_{LL, \mu}^{(x)} R_{3\alpha}^{*(\ell)} R_{3\beta}^{(\ell)} + g_{RR, \mu}^{(x)} R_{4\alpha}^{*(\ell)} R_{4\beta}^{(\ell)} + g_{LR, \mu}^{(x)} R_{3\alpha}^{*(\ell)} R_{4\beta}^{(\ell)} 
   + g_{RL, \mu}^{(x)} R_{4\alpha}^{*(\ell)} R_{3\beta}^{(\ell)} \nonumber \\
  &+ \left. g_{LL, \tau}^{(x)} R_{5\alpha}^{*(\ell)} R_{5\beta}^{(\ell)} + g_{RR, \tau}^{(x)} R_{6\alpha}^{*(\ell)} R_{6\beta}^{(\ell)} + 
   g_{LR, \tau}^{(x)} R_{5\alpha}^{*(\ell)} R_{6\beta}^{(\ell)} + g_{RL, \tau}^{(x)} R_{6\alpha}^{*(\ell)} R_{5\beta}^{(\ell)} \right] \,, \nonumber \\
g_{LL, \ell}^{(x)} =&  \frac{m_Z}{\cos{\theta_W}} \sigma_3^{(x)} \left( \frac{1}{2}- \sin^2{\theta_W} \right) + \frac{m_{\ell}^2}{m_W \cos{\beta}} \sigma_4^{(x)}\,, \nonumber\\
g_{RR, \ell}^{(x)} =&  \frac{m_Z}{\cos{\theta_W}} \sigma_3^{(x)} \left(  \sin^2{\theta_W} \right) + \frac{m_{\ell}^2}{m_W \cos{\beta}}  \sigma_4^{(x)}\,,\nonumber \\
g_{LR, \ell}^{(x)} =& \left(-\sigma_1^{(x)}A_\ell-\sigma_5^{(x)}\mu\right) \frac{m_{\ell}}{2 m_W \cos{\beta}}\,, \nonumber \\
g_{RL, \ell}^{(x)} =& g_{LR, \ell}^{(x)*}\,.
\end{align}
Here, $\tilde U$ is the sneutrino rotation matrix defined in \eqref{snuRot}, $R^{(\ell)}$ the rotation of charged sleptons according to \eqref{sleptonRot}, $U$ and $V$ are the rotation matrices for the charginos, $N$ the ones that rotates the neutralinos, with $N^{(')}_{a1,a2}$ defined in \eqref{Na1a2}, and,
\begin{align}
&\sigma_1^{(x)} = \begin{pmatrix}
  \sin  \alpha   \\
  -\cos \alpha   \\
  i \sin \beta 
\end{pmatrix}, \quad
\sigma_2^{(x)} = \begin{pmatrix}
 \cos \alpha  \\
 \sin \alpha  \\
 -i \cos \beta
\end{pmatrix}, \quad
\sigma_3^{(x)} = \begin{pmatrix}
 \sin  ( \alpha  +  \beta) \\
 -\cos ( \alpha  +  \beta) \\
 0
\end{pmatrix}, \quad
\sigma_4^{(x)} = \begin{pmatrix}
 -\sin \alpha  \\
 \cos  \alpha  \\
 0
\end{pmatrix}, \quad
\nonumber \\
&\sigma_5^{(x)} = \begin{pmatrix}
 \cos \alpha  \\
 \sin \alpha  \\
 i \cos\beta
\end{pmatrix}, \quad
\sigma_6^{(x)} = \begin{pmatrix}
 \cos \alpha  \\
 \sin \alpha  \\
 0
\end{pmatrix}, \quad
\sigma_7^{(x)} = \begin{pmatrix}
 \sin    \alpha \\
 -\cos   \alpha \\
 -i \sin  \beta 
\end{pmatrix}, \quad
\textrm{ for } 
 H_x=\begin{pmatrix}
 h^0\\
 H^0\\
 A^0
\end{pmatrix}. 
\end{align}

Besides using these new couplings, the only changes required to adapt the original form factors to the SUSY-ISS model are the sum over sneutrinos that has to be extended to the 18 mass eigenstates.
In the following formulas, summation over all indices corresponding to internal propagators is understood, meaning  $\alpha\,, \beta=1\,,...\,,18$ for the sneutrinos, $i\,, j=1\,,2$ for the charginos, $\alpha\,, \beta=1\,,...\,,6$ for the charged sleptons and $a\,,b=1\,,...\,,4$ for the neutralinos.
The contributions from the sneutrino-chargino loops are given by, 
\begin{align}
%
%
F_{L,x}^{(1)} =& - \frac{g^2}{16 \pi^2}
\bigg[\Big(B_0 + m_{\tilde {\nu}_{\alpha}}^2 C_0+m_{\ell_m}^2 C_{12}+m_{\ell_k}^2 (C_{11} - C_{12})\Big)\,
\kappa_{L 1}^{x,\,\tilde \chi^-} \nonumber \\
&+m_{\ell_k} m_{\ell_m} \left(C_{11}+C_0\right)\kappa_{L 2}^{x, \tilde \chi^-}+
m_{\ell_k} m_{\tilde \chi _j^-} \left(C_{11}-C_{12}+C_0\right)\kappa_{L 3}^{x, \tilde \chi^-}\,+ 
m_{\ell_m} m_{\tilde \chi _j^-} C_{12}\,\kappa_{L 4}^{x,\tilde \chi^-}\nonumber \\
& +m_{\ell_k} m_{\tilde \chi _i^-} \left(C_{11}-C_{12}\right)\kappa_{L 5}^{x,\tilde \chi^-} + 
m_{\ell_m} m_{\tilde \chi _i^-} \left(C_{12}+C_0\right)\kappa_{L 6}^{x,\tilde \chi^-}+ 
m_{\tilde \chi _i^-} m_{\tilde \chi _j^-}C_0\,\kappa_{L 7}^{x,\tilde \chi^-} \bigg]\,, \nonumber \\
%
%
F_{L,x}^{(2)} =&
- \frac{igg_{H_x\tilde {\nu}_\alpha \tilde {\nu}_\beta}}{16\pi^2}
\left[-m_{\ell_k}(C_{11}-C_{12})\,\iota_{L 1}^{x,\tilde \chi^-}-
m_{\ell_m} C_{12}\, \iota_{L 2}^{x,\tilde \chi^-}+
m_{\tilde \chi^-_i}C_0\,\iota_{L 3}^{x,\tilde \chi^-}\right]\,,\nonumber \\
%
%
F_{L,x}^{(3)} =&
 \frac{-S_{L,\ell_{m}}^{(x)}}{m_{\ell_k}^2-m_{\ell_m}^2}\left[m_{\ell_k}^2 \Sigma_R^{\tilde
    \chi^-} (m_{\ell_k}^2)+m_{\ell_k}^2 \Sigma_{Rs}^{\tilde \chi^-}(m_{\ell_k}^2)
+ m_{\ell_m}\left(m_{\ell_k} \Sigma_L^{\tilde \chi^-} (m_{\ell_k}^2)+
m_{\ell_k}\Sigma_{Ls}^{\tilde \chi^-} (m_{\ell_k}^2)\right)\right]\,,\nonumber \\
%
%
F_{L,x}^{(4)} =&
\frac{-S_{L,\ell_{k}}^{(x)}}{m_{\ell_m}^2-m_{\ell_k}^2}
\left[m_{\ell_m}^2 \Sigma_L^{\tilde \chi^-} (m_{\ell_m}^2)+
m_{\ell_m} m_{\ell_k} \Sigma_{Rs}^{\tilde \chi^-}(m_{\ell_m}^2)\right.
+ \left. m_{\ell_k}\left(m_{\ell_m} \Sigma_R^{\tilde \chi^-} (m_{\ell_m}^2)
+m_{\ell_k} \Sigma_{Ls}^{\tilde \chi^-} (m_{\ell_m}^2)\right)\right]\,,\nonumber \\
\end{align}
where the contributions from slepton-neutralino loops read as, 
\begin{align}
%
%
F_{L,x}^{(5)} =& - \frac{g^2}{16 \pi^2}
\bigg[\Big(B_0 + m_{\tilde \ell_{\alpha}}^2 C_0+m_{\ell_m}^2 C_{12}+m_{\ell_k}^2 (C_{11} - C_{12})\Big)\,
\kappa_{L 1}^{x,\,\tilde \chi^0} \nonumber \\
&+m_{\ell_k} m_{\ell_m} \left(C_{11}+C_0\right)\kappa_{L 2}^{x, \tilde \chi^0}+
m_{\ell_k} m_{\tilde \chi _b^0} \left(C_{11}-C_{12}+C_0\right)\kappa_{L 3}^{x, \tilde \chi^0}\,+ 
m_{\ell_m} m_{\tilde \chi _b^0} C_{12}\,\kappa_{L 4}^{x,\tilde \chi^0}\nonumber \\
& +m_{\ell_k} m_{\tilde \chi _a^0} \left(C_{11}-C_{12}\right)\kappa_{L 5}^{x,\tilde \chi^0} + 
m_{\ell_m} m_{\tilde \chi _a^0} \left(C_{12}+C_0\right)\kappa_{L 6}^{x,\tilde \chi^0}+ 
m_{\tilde \chi _a^0} m_{\tilde \chi _b^0}C_0\,\kappa_{L 7}^{x,\tilde \chi^0} \bigg]\,, \nonumber \\
%
%
F_{L,x}^{(6)} =&
- \frac{igg_{H_x\tilde \ell_\alpha \tilde \ell_\beta}}{16\pi^2}
\left[-m_{\ell_k}(C_{11}-C_{12})\,\iota_{L 1}^{x,\tilde \chi^0}-
m_{\ell_m} C_{12}\, \iota_{L 2}^{x,\tilde \chi^0}+
m_{\tilde \chi_a^0}C_0\,\iota_{L 3}^{x,\tilde \chi^0}\right]\,,\nonumber \\
%
%
F_{L,x}^{(7)} =&
 \frac{-S_{L,\ell_m}^{(x)}}{m_{\ell_k}^2-m_{\ell_m}^2}\left[m_{\ell_k}^2 \Sigma_R^{\tilde
    \chi^0} (m_{\ell_k}^2)+m_{\ell_k}^2 \Sigma_{Rs}^{\tilde \chi^0}(m_{\ell_k}^2)
+ m_{\ell_m}\left(m_{\ell_k} \Sigma_L^{\tilde \chi^0} (m_{\ell_k}^2)+
m_{\ell_k}\Sigma_{Ls}^{\tilde \chi^0} (m_{\ell_k}^2)\right)\right]\,,\nonumber \\
%
%
F_{L,x}^{(8)} =&
 \frac{-S_{L,\ell_k}^{(x)}}{m_{\ell_m}^2-m_{\ell_k}^2}
\left[m_{\ell_m}^2 \Sigma_L^{\tilde \chi^0} (m_{\ell_m}^2)+
m_{\ell_m} m_{\ell_k} \Sigma_{Rs}^{\tilde \chi^0}(m_{\ell_m}^2)\right.
+ \left. m_{\ell_k}\left(m_{\ell_m} \Sigma_R^{\tilde \chi^0} (m_{\ell_m}^2)
+m_{\ell_k} \Sigma_{Ls}^{\tilde \chi^0} (m_{\ell_m}^2)\right)\right]\,, 
\label{formfactorLbs}
\end{align}

\noindent
where,
 \[B_0=
 \left\{   \begin{array}{l}
            B_0(m_{H_x}^2,m_{\tilde \chi^-_i}^2,m_{\tilde \chi^-_j}^2) \textrm{ in } F_{L,x}^{(1)}\,,\\
            B_0(m_{H_x}^2,m_{\tilde \chi^0_a}^2,m_{\tilde \chi^0_b}^2) \textrm{ in } F_{L,x}^{(5)}\,,
  \end{array}
\right. 
\]
and
 \[C_{0,11,12}=
 \left\{   \begin{array}{l}
C_{0,11,12} (m_{\ell_k}^2,m_{H_x}^2, m_{\tilde {\nu}_ {\alpha}}^2,m_{\tilde \chi^-_i}^2,m_{\tilde \chi^-_j}^2) \textrm{ in } F_{L,x}^{(1)}\,,\\
C_{0,11,12} (m_{\ell_k}^2,m_{H_x}^2, m_{\tilde \chi^-_i}^2,m_{\tilde {\nu}_ {\alpha}}^2,m_{\tilde {\nu}_ {\beta}}^2) \textrm{ in } F_{L,x}^{(2)}\,, \\
C_{0,11,12} (m_{\ell_k}^2,m_{H_x}^2, m_{\tilde {l}_ {\alpha}}^2,m_{\tilde \chi^0_a}^2,m_{\tilde \chi^0_b}^2) \textrm{ in } F_{L,x}^{(5)}\,,  \\
C_{0,11,12} (m_{\ell_k}^2,m_{H_x}^2, m_{\tilde \chi^0_a}^2,m_{\tilde {l}_ {\alpha}}^2,m_{\tilde {l}_ {\beta}}^2) \textrm{ in } F_{L,x}^{(6)}\,.
\end{array}
\right. 
\]

The couplings and self-energies from the neutralino contributions to the form factors were defined as
\begin{eqnarray*}
\kappa_{L 1}^{x,\,\tilde \chi^0} = B_{L \alpha a}^{(\ell_k)}D_{Rab}^{(x)}B_{R \alpha b}^{(\ell_m)*}\,, 
     && \iota_{L 1}^{x, \tilde\chi^0} = B_{R \alpha a}^{(\ell_k)}B_{R \beta a}^{(\ell_m)*}\,,\\
\kappa_{L 2}^{x,\,\tilde \chi^0} = B_{R \alpha a}^{(\ell_k)}D_{Lab}^{(x)}B_{L \alpha b}^{(\ell_m)*} \,,
     && \iota_{L 2}^{x, \tilde\chi^0} = B_{L \alpha a}^{(\ell_k)}B_{L \beta a}^{(\ell_m)*}\,,\\
\kappa_{L 3}^{x,\,\tilde \chi^0} = B_{R \alpha a}^{(\ell_k)}D_{Lab}^{(x)}B_{R \alpha b}^{(\ell_m)*}\,,
     && \iota_{L 3}^{x, \tilde\chi^0} = B_{L \alpha a}^{(\ell_k)}B_{R \beta a}^{(\ell_m)*}\,,\\
\kappa_{L 4}^{x,\,\tilde \chi^0} = B_{L \alpha a}^{(\ell_k)}D_{Rab}^{(x)}B_{L \alpha b}^{(\ell_m)*}\,, && \\
\kappa_{L 5}^{x,\,\tilde \chi^0} = B_{R \alpha a}^{(\ell_k)}D_{R ab}^{(x)}B_{R \alpha b}^{(\ell_m)*}\,, &&\\
\kappa_{L 6}^{x,\,\tilde \chi^0} = B_{L \alpha a}^{(\ell_k)}D_{L ab}^{(x)}B_{L \alpha b}^{(\ell_m)*}\,, &&\\
\kappa_{L 7}^{x,\,\tilde \chi^0} = B_{L \alpha a}^{(\ell_k)}D_{Lab}^{(x)}B_{R \alpha b}^{(\ell_m)*}\,, &&
\end{eqnarray*}\vspace*{-0.7cm}
\begin{eqnarray}
\Sigma_L^{\tilde\chi^0} (k^2) &=& -\frac{g^2}{16\pi^2}B_1(k^2,m_{\tilde\chi_a^0}^2, m_{\tilde \ell_\alpha}^2) B_{R\,\alpha a}^{(\ell_k)}B_{R\,\alpha a}^{(\ell_m)*}\,, \nonumber \\
m_{\ell_k} \Sigma_{Ls}^{\tilde \chi^0} (k^2) &=&  \frac{g^2m_{\tilde\chi_a^0}}{16\pi^2}B_0(k^2,m_{\tilde\chi_a^0}^2, m_{\tilde \ell_\alpha}^2) B_{L\,\alpha a}^{(\ell_k)}B_{R\,\alpha a}^{(\ell_m)*}\,.
\end{eqnarray}

The couplings and self energies from the chargino contributions to the form factors,  $\kappa^{x, \, \tilde\chi^-}$, $\iota^{x, \, \tilde\chi^-}$, and $\Sigma^{\tilde\chi^-}$  can
be obtained from the previous expressions $\kappa^{x, \, \tilde\chi^0}$,  $\iota^{x, \, \tilde\chi^0}$ and $\Sigma^{ \tilde\chi^0}$ by using the following replacement rules
$m_{\tilde\chi_a^0}\rightarrow m_{\tilde\chi_i^-}$, $m_{\tilde \ell_\alpha}\rightarrow m_{\tilde \nu_\alpha}$, $B^{(l)}\rightarrow A^{(l)}$, 
$D^{(x)}\rightarrow W^{(x)}$, $a \rightarrow i$, and $b \rightarrow j$.
 
The form factors $F_{R,x}^{(i)},i=1,...,8$ can be obtained from $F_{L,x}^{(i)},i=1,...,8$ through the exchange $L\leftrightarrow R$ in all places.

\chapter{Formulas for the MIA computation}
\fancyhead[RO] {\scshape Formulas for the MIA computation}
\label{App:MIA}

In this Appendix we give the technical details of our MIA computation of the LFV Higgs decay rates in \secref{sec:LFVHDMIA}.
More concretely, we explain the derivation of the {\it fat propagators}, used in our MIA computation in the Feynman-'t Hooft and unitary gauges. 
We give our results for the form factors, up to $\mathcal O(Y_\nu^4)$ order in the MIA expansion, showing explicitly that we obtain the same in both gauges. 
Furthermore, we explore the large $M_R$ limit and derive useful approximate expressions for the loop integrals needed for computing the $H\ell_k\ell_m$ effective vertex of \eqref{VeffMIA}.

\section{Modified neutrino propagators}
\label{App:modifiedpropagators}

Here we derive the right-handed neutrino  {\it fat propagators} used for the computations in \secref{sec:LFVHDMIA}. The idea is to resum all possible large flavor diagonal $M_R$ mass insertions, which we denote with a dot in order to distinguish them from the flavor off-diagonal ones, in a way such that the large mass appears effectively in the denominator of the propagators of the new states. 

\begin{figure}[t!]
\begin{center}
\includegraphics[width=\textwidth]{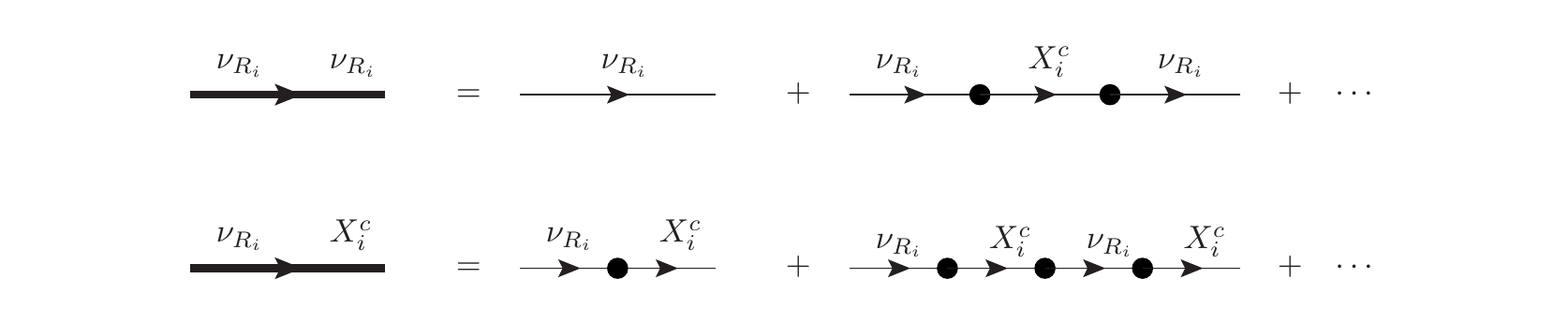}
\caption{Propagators and mass insertions in the electroweak basis. Big black dots denote flavor diagonal mass insertions.}\label{PropsandInsertions}
\end{center}
\end{figure}

In order to make a MIA computation in the electroweak basis $(\nu_L\,,\;\nu_R^c\,,\;X^c)$, we need to take into account all the propagators and mass insertions given by the neutrino mass matrix. In the ISS model we are considering, this mass matrix is given by \eqref{ISSmatrix}, which we repeat here for completeness:
\begin{equation}
\label{ISSmatrixApp}
 M_{\mathrm{ISS}}=\left(\begin{array}{c c c} 0 & m_D & 0 \\ m_D^T & 0 & M_R \\ 0 & M_R^T & \mu_X \end{array}\right)\,.
\end{equation}
From this mass matrix, we obtain the propagators and mass insertions summarized in \figref{PropsandInsertions}.
It is important to notice the presence of the $P_L$ and $P_R$ projectors for the chiral fields, which have been properly added according to: 
\begin{equation}
\begin{array}{lcl}
\nu_L^c , \nu_R , X  &\longrightarrow& {\rm \bf RH~ fields} , \\
\nu_L , \nu_R^c , X^c  &\longrightarrow& {\rm \bf LH~ fields} .
\end{array}
\end{equation}
As previously mentioned, there are three types of mass insertions and they are controlled by the matrices $m_D$, $M_R$ and $\mu_X$. 
The mass insertions $M_R$ that relate $\nu_R$ and $X$ fields are taken to be flavor diagonal  and are denoted by a big dot in \figref{PropsandInsertions}.
On the other hand, crosses indicate flavor non-diagonal insertions coming from $m_D$ (big cross) and $\mu_X$ (small cross), which connect the fields $\nu_L$-$\nu_R$ and two $X$'s respectively.
Nevertheless, given that we work under the assumption that $\mu_X$ is a tiny scale, we neglect $\mu_X$ mass insertions for our LFVHD computations and, therefore, we consider $m_D$ as the only relevant LFV insertion.

\begin{figure}[t!]
\begin{center}
\includegraphics[width=\textwidth]{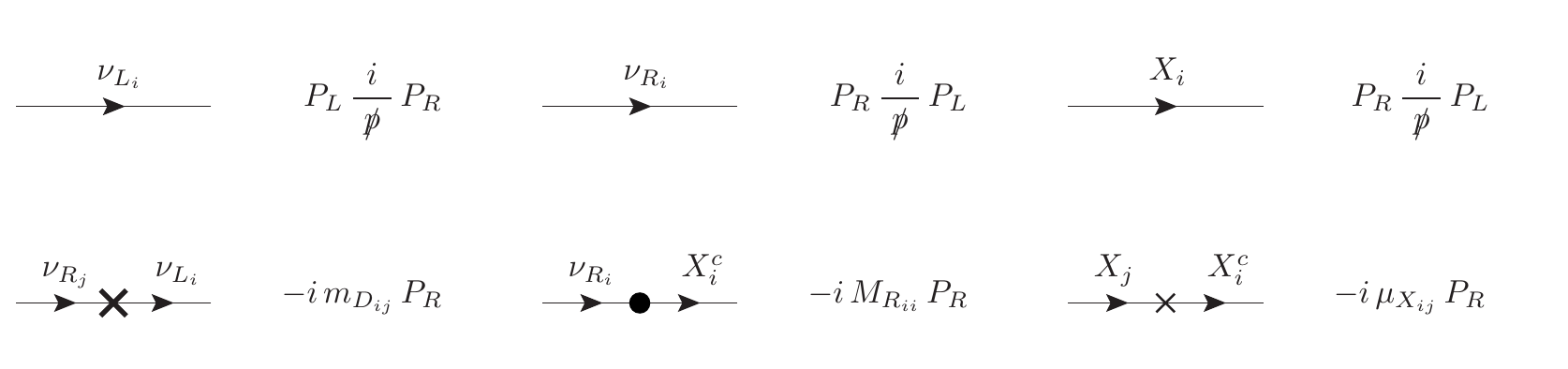}
\caption{Modified neutrino propagators after resuming an infinite number of $M_R$ mass insertions, denoted here by big black dots. We use fat arrow lines  with same (different) initial and final states to denote that all possible even (odd) number of $M_R$ insertions have been considered.  The fat lines with same initial and final $\nu_R$ states are referred to in this work as {\it fat propagators}.}\label{FatProps}
\end{center}
\end{figure}

Since our motivation in this work is to make a MIA computation for LFV H decays by perturbatively inserting LFV mass insertions, we find convenient to take into account first the effects of all possible flavor diagonal $M_R$ insertions. 
Moreover, this procedure allows us to consider $M_R$ also as a heavy scale so we can define an effective vertex for the $H$-$\ell_i$-$\ell_j$ interaction.
This can be done by defining two types of modified propagators, one for same initial and final state consisting of all possible even number of $M_R$ insertions (which we call {\it fat propagator}), and one for different initial and final states with an odd number of $M_R$ insertions, as it is schematically shown in \figref{FatProps}.
We can then define two modified propagators starting with $\nu_R$ by adding the corresponding series:
\begin{align}
{\rm Prop}_{\, \nu_{R_i}\to\nu_{R_i}}
&= P_R \dfrac i{\slashed p}  P_L + P_R \dfrac i{\slashed p}  P_L ~ \Big(-i M_{R_i}^* P_L \Big)~ P_L \dfrac i{\slashed p}  P_R ~ \Big(-i M_{R_i} P_R \Big)~ P_R \dfrac i{\slashed p}  P_L + \cdots\nonumber\\
& = P_R~ \dfrac i{\slashed p}\sum_{n\geq0}\bigg(\dfrac{|M_{R_i}|^2}{p^2}\bigg)^n ~ P_L= P_R~ \dfrac{ i\slashed p}{p^2-|M_{R_i}|^2}~ P_L\,, \label{FatpropnnuR}
\\\nonumber\\
{\rm Prop}_{\, \nu_{R_i}\to X_{i}^c}
&= P_L \dfrac i{\slashed p}  P_R ~  \Big(-i M_{R_i} P_R \Big)~ P_R \dfrac i{\slashed p}  P_L
\nonumber\\
&\hspace{-1cm}+ ~ P_L \dfrac i{\slashed p}  P_R ~  \Big(-i M_{R_i} P_R \Big)~ P_R \dfrac i{\slashed p}  P_L~   \Big(-i M_{R_i}^* P_L \Big)~ P_L \dfrac i{\slashed p}  P_R ~  \Big(-i M_{R_i} P_R \Big)~ P_R \dfrac i{\slashed p}  P_L + \cdots \nonumber\\
&= P_L~ \dfrac {i M_{R_i}}{p^2}\sum_{n\geq0}\bigg(\dfrac{|M_{R_i}|^2}{p^2}\bigg)^n ~ P_L= P_L~ \dfrac{ i M_{R_i}}{p^2-|M_{R_i}|^2}~ P_L\,.
\end{align}
And we can similarly define other modified propagators considering also the $\nu_R^c$ and $X$ states. In the present study of LFVHD, it happens that the $X$ fields do not interact with any of the external legs involved in the LFV process we want to compute. Consequently, to take into account the effects from $X$ in the LFVHD, it is enough to consider the {\it fat propagator} in \eqref{FatpropnnuR} when computing the one-loop contributions to $H\to\ell_k\bar\ell_m$.

\section{MIA Form Factors in the Feynman-'t Hooft gauge} 
\label{App:FormFactorsMIA}

Here we present the analytical results for the form factors $F_{L,R}$ involved in the computation of the LFVHD decay rates when computed with the MIA to one-loop order.
 We consider the leading order corrections, ${\cal O} (Y_\nu^2)$, and the next to leading corrections, ${\cal O} (Y_\nu^4)$, as explained in the text. This means the computation of all the one-loop diagrams in \figrefs{LFVHDMIAdiagsvertexY2}, \ref{LFVHDMIAdiagslegY2}, \ref{LFVHDMIAdiagsvertexY4} and \ref{LFVHDMIAdiagslegY4}. 
They are written in terms of the usual one-loop functions for the two-point $B's$,  three-point $C's$, and four-point $D's$ functions. 
We follow these definitions and conventions:
\begin{equation}\label{loopfunctionB}
    \mu^{4-D}~ \int \frac{d^D k}{(2\pi)^D} \frac{\{1; k^{\mu}\}}
    {[k^2 - m_1^2][(k+p_1)^2 - m_2^2]}
    = \frac{i}{16\pi^2} \left\{ B_0; p_1^{\mu} B_1 \right\}
    (p_1,m_1,m_2)\,,
\end{equation}
\begin{align}
    \mu^{4-D}& \int \frac{d^D k}{(2\pi)^D}
    \frac{\{1; k^2; k^{\mu}\}}
    {[k^2 - m_1^2][(k + p_1)^2 - m_2^2][(k + p_1 + p_2)^2 - m_3^2]}
    \nonumber \\
     & = \frac{i}{16\pi^2}
    \left\{ C_0; {\tilde C}_0; p_1^{\mu}C_{11} + p_2^{\mu}C_{12} \right\}
    (p_1, p_2, m_1, m_2, m_3)\,, \label{loopfunctionC}
\end{align}
\begin{align}
   \mu^{4-D}& \int \frac{d^D k}{(2\pi)^D}
    \frac{\{1; k^2; k^{\mu}\}}
    {[k^2 - m_1^2][(k + p_1)^2 - m_2^2][(k + p_1 + p_2)^2 - m_3^2][(k + p_1 + p_2+ p_3)^2 - m_4^2]}
    \nonumber \\
     & = \frac{i}{16\pi^2}
    \left\{ D_0; {\tilde D}_0; p_1^{\mu}D_{11} + p_2^{\mu}D_{12} +p_3^{\mu}D_{13}\right\}
    (p_1, p_2, p_3, m_1, m_2, m_3,m_4)\,. \label{loopfunctionD}
\end{align}

We start with the left-handed form factors and present the contributions diagram by diagram, following the notation explained in the text and shortening $m_{k,m}\equiv m_{\ell_{k,m}}$.
We will restrict ourselves to the dominant contributions, meaning those that will provide $\mathcal O(v^2/M_R^2)$ terms when doing the large $M_R$ expansion, as explained in the next Appendix. 
For instance, contributions from loop functions of type $D_i$ where $M_R$ appears in two of the mass arguments go as $1/M_R^4$ and provide subdominant corrections that are not considered here. 

The results of the ${\cal O} (Y_\nu^2)$ contributions are:
\begin{align}
 F_{L}^{{\rm MIA (1)\,\, (Y^2)}}=& \frac{1}{32 \pi^{2}} \frac{m_{k}}{m_{W}}   \left(Y_{\nu}^{} Y_{\nu}^{\dagger}  \right)^{km} \Big( (\tilde{C}_{0}+m_{k}^{2}(C_{11}-C_{12})+m_{m}^{2}C_{12})_{(1a)}   \nonumber \\
& +m_{m}^{2}(C_{0}+C_{11})_{(1b)} -m_{m}^{2}(C_{12})_{(1c)} -m_{m}^{2}(C_{0}+C_{12})_{(1d)} \Big)     \,,\nonumber \\
F_{L}^{{\rm MIA (2)\,\, (Y^2)}}=& \frac{-1}{16 \pi^{2}} m_{k}m_{W} \left(Y_{\nu}^{} Y_{\nu}^{\dagger} \right)^{km} \Big( (C_{0}+C_{11}-C_{12})_{(2a)} +(C_{11}-C_{12})_{(2b)} \Big) \,,\nonumber \\
 F_{L}^{{\rm MIA (3)\,\, (Y^2)}}=& \frac{1}{8 \pi^{2}} m_{k}m_{W}^{3} \left(Y_{\nu}^{} Y_{\nu}^{\dagger} \right)^{km} ( D_{12}-D_{13} )_{(3a)} \,,\nonumber \\
 F_{L}^{{\rm MIA (4)\,\, (Y^2)}}=& \frac{-1}{32 \pi^{2}} m_{k}m_{W} \left(Y_{\nu}^{} Y_{\nu}^{\dagger} \right)^{km} \Big( (C_{0}-C_{11}+C_{12})_{(4a)} +m_{m}^{2}(2D_{12}-D_{13})_{(4b)} \Big)\,, \nonumber \\
 F_{L}^{{\rm MIA (5)\,\, (Y^2)}}=& \frac{1}{32 \pi^{2}} m_{k}m_{W}  \left(Y_{\nu}^{} Y_{\nu}^{\dagger}  \right)^{km} \Big( (2C_{0}+C_{11}-C_{12})_{(5a)} \nonumber\\
 &+(C_{0}+2m_{k}^{2}D_{12}-(2m_{H}^{2}-m_{m}^{2})D_{13})_{(5b)} \Big) \,,\nonumber \\
 F_{L}^{{\rm MIA (6)\,\, (Y^2)}}=& \frac{1}{32 \pi^{2}} \frac{m_{k}}{m_{W}} m_{H}^{2} \left( Y_{\nu}^{} Y_{\nu}^{\dagger} \right)^{km} \Big( (C_{11}-C_{12})_{(6a)} +(C_{0})_{(6c)} +m_{m}^{2}(D_{13})_{(6d)} \Big) \,,\nonumber \\
 F_{L}^{{\rm MIA (7)\,\, (Y^2)}}=& \frac{1}{16 \pi^{2}} m_{k}m_{W} \frac{m_{m}^{2}}{m_{k}^{2}-m_{m}^{2}}  \left(Y_{\nu}^{} Y_{\nu}^{\dagger} \right)^{km} ( C_{12} )_{(7a)} \,, \nonumber \\
  F_{L}^{{\rm MIA (8)\,\, (Y^2)}}=& \frac{1}{32 \pi^{2}} \frac{m_{k}}{m_{W}} \frac{m_{m}^{2}}{m_{k}^{2}-m_{m}^{2}}  \left( Y_{\nu}^{} Y_{\nu}^{\dagger} \right)^{km} \Big( (B_{1})_{(8a)} +(B_{0})_{(8b)} +(B_{0})_{(8c)} +m_{k}^{2}(C_{12})_{(8d)} \Big)\,, \nonumber \\
 F_{L}^{{\rm MIA (9)\,\, (Y^2)}}=& \frac{-1}{16 \pi^{2}} m_{k}m_{W} \frac{m_{m}^{2}}{m_{k}^{2}-m_{m}^{2}}  \left(Y_{\nu}^{} Y_{\nu}^{\dagger} \right)^{km} ( C_{12} )_{(9a)}\,, \nonumber \\
 F_{L}^{{\rm MIA (10)\,\, (Y^2)}}=& \frac{-1}{32 \pi^{2}} \frac{m_{k}}{m_{W}} \frac{m_{m}^{2}}{m_{k}^{2}-m_{m}^{2}}  \left( Y_{\nu}^{} Y_{\nu}^{\dagger} \right)^{km} \Big( (B_{1})_{(10a)} +(B_{0})_{(10b)}+\frac{m_{k}^{2}}{m_{m}^{2}}(B_{0})_{(10c)}  \nonumber\\
 &+m_{k}^{2}(C_{12})_{(10d)} \Big) \,   .
\label{FLtot_op2_Y2}
\end{align}
The results of the dominant ${\cal O} (Y_\nu^4)$ contributions are:
\begin{align}
F_{L}^{{\rm MIA (1)\,\, (Y^4)}}=&  \frac{1}{32 \pi^{2}} \frac{m_{k}}{m_{W}}
\left(Y_{\nu}^{} Y_{\nu}^{\dagger} Y_{\nu}^{} Y_{\nu}^{\dagger} \right)^{km} v^{2} \Big( -(C_{11}-C_{12})_{(1e)} -(C_{11}-C_{12}+C_{0})_{(1f)}    \nonumber \\
&+(\tilde{D}_{0})_{(1g)} +(\tilde{D}_{0})_{(1h)} +(C_{0})_{(1j)} \Big)\,,
\nonumber \\
 F_{L}^{{\rm MIA (8)\,\, (Y^4)}}=& \frac{1}{32 \pi^{2}} \frac{m_{k}}{m_{W}} \frac{m_{m}^{2}}{m_{k}^{2}-m_{m}^{2}}
 \left( Y_{\nu}^{} Y_{\nu}^{\dagger} Y_{\nu}^{} Y_{\nu}^{\dagger} \right)^{km} v^{2}\Big( (C_{12})_{(8e)}+(C_{0})_{(8f)}+(C_{0})_{(8g)} \Big) \,,  \nonumber \\
F_{L}^{{\rm MIA (10)\,\, (Y^4)}}=& \frac{-1}{32 \pi^{2}} \frac{m_{k}}{m_{W}} \frac{m_{m}^{2}}{m_{k}^{2}-m_{m}^{2}}
\left( Y_{\nu}^{} Y_{\nu}^{\dagger} Y_{\nu}^{} Y_{\nu}^{\dagger} \right)^{km} v^{2}\Big( (C_{12})_{(10e)}+(C_{0})_{(10f)}+\frac{m_{k}^{2}}{m_{m}^{2}}(C_{0})_{(10g)} \Big) \, .  
\label{FLtot_op2_Y4}
\end{align}
Next, we present the right-handed form factors. The results of the  ${\cal O} (Y_\nu^2)$ contributions  are: 
\begin{align}
 F_{R}^{{\rm MIA (1)\,\, (Y^2)}}=& \frac{1}{32 \pi^{2}} \frac{m_{m}}{m_{W}}   \left(Y_{\nu}^{} Y_{\nu}^{\dagger}  \right)^{km} \Big( m_{k}^{2}(C_{0}+C_{11})_{(1a)} +(\tilde{C}_{0}+m_{k}^{2}(C_{11}-C_{12})+m_{m}^{2}C_{12})_{(1b)}  \nonumber \\
& -m_{k}^{2}(C_{0}+C_{11}-C_{12})_{(1c)} -m_{k}^{2}(C_{11}-C_{12})_{(1d)} \Big)  \,,   \nonumber \\
F_{R}^{{\rm MIA (2)\,\, (Y^2)}}=& \frac{-1}{16 \pi^{2}} m_{m}m_{W} \left(Y_{\nu}^{} Y_{\nu}^{\dagger} \right)^{km} \Big( (C_{12})_{(2a)} +(C_{0}+C_{12})_{(2b)} \Big)\,, \nonumber \\
 F_{R}^{{\rm MIA (3)\,\, (Y^2)}}=& \frac{1}{8 \pi^{2}} m_{m}m_{W}^{3} \left(Y_{\nu}^{} Y_{\nu}^{\dagger} \right)^{km} (D_{13})_{(3a)} \,,\nonumber \\
F_{R}^{{\rm MIA (4)\,\, (Y^2)}}=& \frac{1}{32 \pi^{2}} m_{m}m_{W}  \left(Y_{\nu}^{} Y_{\nu}^{\dagger}  \right)^{km} \Big( (2C_{0}+C_{12})_{(4a)} \nonumber\\
&+\big(C_{0}+2m_{m}^{2}D_{12}-(2m_{H}^{2}-m_{k}^{2})(D_{12}-D_{13})\big)_{(4b)} \Big) \,,\nonumber \\
 F_{R}^{{\rm MIA (5)\,\, (Y^2)}}=& \frac{-1}{32 \pi^{2}} m_{m}m_{W} \left(Y_{\nu}^{} Y_{\nu}^{\dagger} \right)^{km} \Big( (C_{0}-C_{12})_{(5a)} +m_{k}^{2}(D_{12}+D_{13})_{(5b)} \Big)\,, \nonumber \\
 F_{R}^{{\rm MIA (6)\,\, (Y^2)}}=& \frac{1}{32 \pi^{2}} \frac{m_{m}}{m_{W}} m_{H}^{2} \left( Y_{\nu}^{} Y_{\nu}^{\dagger} \right)^{km} \Big( (C_{12})_{(6a)} +(C_{0})_{(6b)} +m_{k}^{2}(D_{12}-D_{13})_{(6d)} \Big) \,,\nonumber \\
 F_{R}^{{\rm MIA (7)\,\, (Y^2)}}=& \frac{1}{16 \pi^{2}} m_{m}m_{W} \frac{m_{k}^{2}}{m_{k}^{2}-m_{m}^{2}}  \left(Y_{\nu}^{} Y_{\nu}^{\dagger} \right)^{km} ( C_{12} )_{(7a)}  \,,\nonumber \\
  F_{R}^{{\rm MIA (8)\,\, (Y^2)}}=& \frac{1}{32 \pi^{2}} \frac{m_{m}}{m_{W}} \frac{m_{k}^{2}}{m_{k}^{2}-m_{m}^{2}}  \left( Y_{\nu}^{} Y_{\nu}^{\dagger} \right)^{km} \Big( (B_{1})_{(8a)} +\frac{m_{m}^{2}}{m_{k}^{2}}(B_{0})_{(8b)} +(B_{0})_{(8c)} +m_{m}^{2}(C_{12})_{(8d)} \Big) \,,\nonumber \\
 F_{R}^{{\rm MIA (9)\,\, (Y^2)}}=& \frac{-1}{16 \pi^{2}} m_{m}m_{W} \frac{m_{k}^{2}}{m_{k}^{2}-m_{m}^{2}}  \left(Y_{\nu}^{} Y_{\nu}^{\dagger} \right)^{km} ( C_{12} )_{(9a)}\,, \nonumber \\
 F_{R}^{{\rm MIA (10)\,\, (Y^2)}}=& \frac{-1}{32 \pi^{2}} \frac{m_{m}}{m_{W}} \frac{m_{k}^{2}}{m_{k}^{2}-m_{m}^{2}}  \left( Y_{\nu}^{} Y_{\nu}^{\dagger} \right)^{km} \Big( (B_{1})_{(10a)} +(B_{0})_{(10b)} +(B_{0})_{(10c)}   \nonumber \\ 
& +m_{m}^{2}(C_{12})_{(10d)} \Big) \, .\label{FRtot_op2_Y2}
\end{align}
The results of the dominant ${\cal O} (Y_\nu^4)$ contributions are:
\begin{align}
F_{R}^{{\rm MIA (1)\,\, (Y^4)}}=&  \frac{1}{32 \pi^{2}} \frac{m_{m}}{m_{W}}
\left(Y_{\nu}^{} Y_{\nu}^{\dagger} Y_{\nu}^{} Y_{\nu}^{\dagger} \right)^{km} v^{2} \Big( -(C_{0}+C_{12})_{(1e)} -(C_{12})_{(1f)}    \nonumber \\
&+(C_{0})_{(1i)} +(\tilde{D}_{0})_{(1k)} +(\tilde{D}_{0})_{(1l)}  \Big)\,,
\nonumber \\
 F_{R}^{{\rm MIA (8)\,\, (Y^4)}}=& \frac{1}{32 \pi^{2}} \frac{m_{m}}{m_{W}} \frac{m_{k}^{2}}{m_{k}^{2}-m_{m}^{2}}
 \left( Y_{\nu}^{} Y_{\nu}^{\dagger} Y_{\nu}^{} Y_{\nu}^{\dagger} \right)^{km} v^{2}\Big( (C_{12})_{(8e)}+\frac{m_{m}^{2}}{m_{k}^{2}}(C_{0})_{(8f)}+(C_{0})_{(8g)} \Big) \,,  \nonumber \\
F_{R}^{{\rm MIA (10)\,\, (Y^4)}}=& \frac{-1}{32 \pi^{2}} \frac{m_{m}}{m_{W}} \frac{m_{k}^{2}}{m_{k}^{2}-m_{m}^{2}}
\left( Y_{\nu}^{} Y_{\nu}^{\dagger} Y_{\nu}^{} Y_{\nu}^{\dagger} \right)^{km} v^{2}\Big( (C_{12})_{(10e)}+(C_{0})_{(10f)}+(C_{0})_{(10g)} \Big) \, .
\label{FRtot_op2_Y4}
\end{align}
The arguments of the above one-loop integrals are the following: 
\be\begin{array}{rll}
 \tilde{C}_{0}, C_{i} =& \tilde{C}_{0}, C_{i} (p_{2},p_{1},m_{W},0,M_{R}) & \text{in } (1a), (1c), (2a) \nonumber \\
 \tilde{C}_{0}, C_{i} =& \tilde{C}_{0}, C_{i} (p_{2},p_{1},m_{W},M_{R},0) & \text{in } (1b), (1d), (2b)  \nonumber \\
 C_{i} =& C_{i} (p_{2},p_{1},m_{W},M_{R},M_{R}) & \text{in } (1e), (1f),(1i), (1j)  \nonumber \\
 \tilde{D}_{0} =& \tilde{D}_{0} (p_{2},0,p_{1},m_{W},0,M_{R},M_{R}) & \text{in } (1g)  \nonumber \\
 \tilde{D}_{0} =& \tilde{D}_{0} (p_{2},p_{1},0,m_{W},0,M_{R},M_{R}) & \text{in } (1h)  \nonumber \\
 \tilde{D}_{0} =& \tilde{D}_{0} (p_{2},p_{1},0,m_{W},M_{R},M_{R},0) & \text{in } (1k)  \nonumber \\
 \tilde{D}_{0} =& \tilde{D}_{0} (p_{2},0,p_{1},m_{W},M_{R},M_{R},0) & \text{in } (1l)  \nonumber \\
 D_{i} =& D_{i} (0,p_{2},p_{1},0,M_{R},m_{W},m_{W}) & \text{in } (3a), (4b), (5b), (6d)  \nonumber\\
 C_{i} =&  C_{i} (p_{2},p_{1},M_{R},m_{W},m_{W}) & \text{in } (4a), (4b), (5a), (5b), (6a), (6b), (6c)  \nonumber \\
 C_{12} =& C_{12} (0,p_{2},0,M_{R},m_{W}) & \text{in } (7a), (8d)  \nonumber \\
 B_{i} =& B_{i} (p_{2},M_{R},m_{W}) & \text{in } (8a), (8b), (8c)  \nonumber \\
 C_{i} =& C_{i} (0,p_{2},M_{R},M_{R},m_{W}) & \text{in } (8e), (8f),(8g)   \nonumber \\
 C_{12} =& C_{12} (0,p_{3},0,M_{R},m_{W}) & \text{in } (9a), (10d)  \nonumber \\
 B_{i} =& B_{i} (p_{3},M_{R},m_{W}) & \text{in } (10a), (10b), (10c)  \nonumber \\
 C_{i} =& C_{i} (0,p_{3},M_{R},M_{R},m_{W}) & \text{in } (10e), (10f),(10g) \,.
\label{argfloops_op2}
\end{array}\ee
We want to remark that the above formulas are valid for the degenerate $M_{R_i}=M_R$ case. Nevertheless, they can be easily generalized to the non-degenerate case by properly including the summation indices. For example, it would be enough to change 
\begin{align}
(Y_\nu Y_\nu^\dagger)^{km} C_\alpha(p_2,p_1,m_W,0,M_R) 	&\rightarrow (Y_\nu^{ka} Y_\nu^{\dagger am}) C_\alpha(p_2,p_1,m_W,0,M_{R_a})\,, \nonumber\\
(Y_\nu Y_\nu^\dagger Y_\nu Y_\nu^\dagger)^{km} C_\alpha(p_2,p_1,m_W,M_R,M_R) &\rightarrow	 (Y_\nu^{ka} Y_\nu^{\dagger ai} Y_\nu^{ib} Y_\nu^{\dagger bm}) C_\alpha(p_2,p_1,m_W,M_{R_a},M_{R_b})\,,
\end{align}
and similarly for all the terms. 

\section[The large $M_R$ expansion]{The large $\boldsymbol{M_R}$ expansion} 
\label{App:LoopIntegrals}

Here we present our analytical results for the loop-functions and form factors involved in our computation of LFVHD rates in the large $M_R$ limit. To reach this limit we perform a systematic expansion of the amplitude in powers of $(v^2/M_R^2)$.  Generically,  the first order in this expansion is ${\cal O} (v^2/M_R^2)$ the next order is ${\cal O} (v^4/M_R^4)$, etc. The logarithmic dependence with $M_R$ is left unexpanded. In the final expansion we will keep just the dominant terms in the form factors of ${\cal O} (v^2/M_R^2)$ which have been shown to be sufficient to describe successfully the final amplitude for LFVHD in the heavy right-handed neutrino mass region of our interest, i.e., $M_R \gg v$.

We first calculate the large $M_{R}$ expansions of all the one-loop functions and second we plug these expansions in the form factors formulas. To do this, we perform first the integration of the Feynman's parameters and next  expand  them for large $M_{R}\gg v$. 
Since the mass of the Higgs boson enters here,  we cannot take the most used approximation of neglecting external momentum particles. 
In fact our expansions presented in this Appendix will apply to the present case of on-shell Higgs boson, i.e., with $p_1^2=m_H^2$ and $m_H$ being the realistic Higgs boson mass. Furthermore, it should be noticed that in principle
there are three very different scales of masses involved in the computation: the lepton sector masses ($m_{\ell_m}$ and $m_{\ell_k}$), the electroweak sector masses ($m_W$ and $m_H$) and the new physics scale $M_R$.
As we said, in a good approximation we can neglect the lepton masses in the one-loop functions at the beginning. However, both electroweak masses $m_W$ and $m_H$ must be retained in order to calculate the ${\cal O} (M_{R}^{-2})$ terms of the one-loop functions. Actually, in practice we consider the vacuum expectation value $v$, which is the common scale entering in both electroweak masses within the SM, and as we said above, we perform a well defined expansion in powers of an unique dimensionless parameter that is given by the ratio $v^{2}/M_{R}^{2}$.

At the numerical level,  we have checked that all the expansions presented in the following are in very good accordance with the numerical results from {\it LoopTools}. 
The analytical expansions that we get for the dominant terms of the loop functions, i.e., up to $\mathcal O(M_R^{-2})$, are summarized as,
\begin{align}
&B_{0} \left(p,M_R,m_W\right) = \Delta +1 -\log \Big(\frac{M_{R}^{2}}{\mu^2}\Big) +\frac{m_{W}^{2} \log \left(\frac{m_{W}^{2}}{M_{R}^{2}}\right)}{M_{R}^{2}} +\frac{p^2}{2M_R^2}  \,,\nonumber\\
&C_{0} \left(p_2,p_1,m_W,0,M_R\right) = C_{0} \left(p_2,p_1,m_W,M_R,0\right) = \frac{\log \left(\frac{m_{W}^{2}}{M_{R}^{2}}\right)}{M_{R}^{2}}  \,, \nonumber\\
&C_{0} \left(p_2,p_1,M_{R,}m_W,m_W\right) = \frac{2 \sqrt{4 \lambda-1} \arctan \left(\sqrt{\frac{1}{4 \lambda-1}}\right)-1+\log \left(\frac{m_{W}^{2}}{M_{R}^{2}}\right)}{M_{R}^{2}} \,,  \nonumber\\
&C_{0} \left(p_2,p_1,m_W,M_R,M_R\right)= -\frac{1}{M_{R}^{2}} \,, \nonumber\\
&C_{0} \left(0,p_{lep},M_{R},M_{R},m_W\right) = -\frac{1}{M_{R}^{2}}  \,,\nonumber\\
&\tilde{C}_{0} \left(p_2,p_1,m_W,M_R,0\right) = \tilde{C}_{0} \left(p_2,p_1,m_W,0,M_R\right) \nonumber\\
&= \Delta 
+1 -\log \Big(\frac{M_{R}^{2}}{\mu^2}\Big)  + \frac{m_{W}^{2} \log \left(\frac{m_{W}^{2}}{M_{R}^{2}}\right)}{M_{R}^{2}}+\frac{m_{H}^{2}}{2 M_{R}^{2}} \,,  \nonumber\\
&\tilde{D}_{0} \left(p_2,0,p_1,m_W,0,M_{R},M_{R}\right) = \tilde{D}_{0} \left(p_2,p_1,0,m_W,0,M_{R},M_{R}\right) = -\frac{1}{M_{R}^{2}}  \,,\nonumber\\
&\tilde{D}_{0} \left(p_2,0,p_1,m_W,M_{R},M_{R},0\right) = \tilde{D}_{0} \left(p_2,p_1,0,m_W,M_{R},M_{R},0\right) = -\frac{1}{M_{R}^{2}} \,, \nonumber\\
&B_{1} \left(p,M_R,m_W\right) = -\frac{\Delta }{2}-\frac{3}{4}+\frac12\log\Big(\frac{M_{R}^{2}}{\mu^2}\Big)-\frac{m_{W}^{2} \left(2 \log \left(\frac{m_{W}^{2}}{M_{R}^{2}}\right)+1\right)}{2 M_{R}^{2}}  -\frac{p^2}{3M_R^2}\,, \nonumber\\
&C_{11} \left(p_2,p_1,m_W,0,M_R\right) = \frac{1-\log \left(\frac{m_{W}^{2}}{M_{R}^{2}}\right)}{2 M_{R}^{2}}  \,, \nonumber\\
&C_{12} \left(p_2,p_1,m_W,0,M_R\right) = \frac{1}{2 M_{R}^{2}}  \,, \nonumber\\
&C_{11} \left(p_2,p_1,m_W,M_R,0\right) = \frac{1-\log \left(\frac{m_{W}^{2}}{M_{R}^{2}}\right)}{2 M_{R}^{2}}   \,,\nonumber\\
&C_{12} \left(p_2,p_1,m_W,M_R,0\right) = -\frac{\log \left(\frac{m_{W}^{2}}{M_{R}^{2}}\right)}{2 M_{R}^{2}}   \,,\nonumber\\
&C_{11} \left(p_2,p_1,M_R,m_W,m_W\right) = 2 C_{12} \left(p_2,p_1,M_R,m_W,m_W\right)  \nonumber\\
&\hspace{4.6cm}= -\frac{4 \sqrt{4 \lambda-1} \arctan \left(\sqrt{\frac{1}{4 \lambda-1}}\right)+2 \log \left(\frac{m_{W}^{2}}{M_{R}^{2}}\right)-1}{2 M_{R}^{2}} \,,  \nonumber\\
&C_{11} \left(p_2,p_1,m_W,M_R,M_R\right) = 2 C_{12} \left(p_2,p_1,m_W,M_R,M_R\right) = \frac{1}{2M_{R}^{2}}\,,  \nonumber\\
&C_{12} \left(0,p_{lep},0,M_R,m_W\right)= \frac{-\log \left(\frac{m_{W}^{2}}{M_{R}^{2}}\right)-1}{2 M_{R}^{2}}  \,,\nonumber\\
&C_{12} \left(0,p_{lep},M_R,M_R,m_W\right)= \frac{1}{2 M_{R}^{2}}  \,,\nonumber\\
&D_{12} \left(0,p_2,p_1,0,M_R,m_W,m_W\right) = 2 D_{13} \left(0,p_2,p_1,0,M_R,m_W,m_W\right)   \nonumber\\
&\hspace{2cm}= \frac{2\left( -4 \lambda \arctan^2 \left(\sqrt{\frac{1}{4 \lambda-1}}\right)+2 \sqrt{4 \lambda-1} \arctan \left(\sqrt{\frac{1}{4 \lambda-1}}\right)-1 \right)}{ M_{R}^{2} m_{H}^{2}} \, ,
\label{allfloops}
\end{align}
where we have used the usual definitions in dimensional regularization, $\Delta= 2/\epsilon-\gamma_E +{\rm Log}(4\pi)$ with $D=4-\epsilon$ and $\mu$ the usual scale,  and we have denoted the mass ratio $\lambda=m_{W}^{2}/m_{H}^{2}$ to shorten the result.

Taking into account the formulas  in \eqref{allfloops},  plugging them into the results of the form factors in the  \appref{App:FormFactorsMIA}, neglecting the tiny terms with lepton masses, and pairing diagrams conveniently, we finally get the results for the dominant terms of the various type diagrams (i), see \figref{diagsLFVHDphysbasis}, of the MIA form factors valid in the large $M_R\gg v$ regime:
\begin{eqnarray}
F_{L}^{(1)} &=& \frac{1}{32 \pi^{2}} \frac{m_{k}}{m_{W}} \Bigg[ \left(Y_{\nu}^{} Y_{\nu}^{\dagger}\right)^{km} \left( \Delta +1 -\log (\frac{M_{R}^{2}}{\mu^2}) + \frac{m_{W}^{2} \log \left(\frac{m_{W}^{2}}{M_{R}^{2}}\right)}{M_{R}^{2}}+\frac{m_{H}^{2}}{2 M_{R}^{2}} \right)   \nonumber\\
&& -\frac{5}{2}\frac{v^{2}}{M_{R}^{2}}\left(Y_{\nu}^{} Y_{\nu}^{\dagger} Y_{\nu}^{} Y_{\nu}^{\dagger} \right)^{km}  \Bigg] \,,\nonumber\\
F_{L}^{(2)} &=& -\frac{1}{32 \pi^{2}} \frac{m_{k}}{m_{W}}  \left(Y_{\nu}^{} Y_{\nu}^{\dagger}\right)^{km} \frac{m_{W}^{2}}{M_{R}^{2}} \left( 1+\log \left(\frac{m_{W}^{2}}{M_{R}^{2}}\right) \right)\,,   \nonumber\\
F_{L}^{(3)} &=& \frac{1}{8 \pi^{2}} \frac{m_{k}}{m_{W}} \left(Y_{\nu}^{} Y_{\nu}^{\dagger} \right)^{km} \frac{\lambda m_{W}^{2}}{M_{R}^{2}} \left( -4\lambda \arctan^2 \left(\frac{1}{\sqrt{4\lambda-1}}\right) \right. \nonumber\\
&& \left. +2\sqrt{4\lambda-1}\arctan \left(\frac{1}{\sqrt{4\lambda-1}}\right) -1  \right) \,,\nonumber\\
F_{L}^{(4+5)} &=& \frac{1}{32 \pi^{2}} \frac{m_{k}}{m_{W}} \left(Y_{\nu}^{} Y_{\nu}^{\dagger} \right)^{km} \frac{m_{W}^{2}}{M_{R}^{2}} \left( 8\lambda \arctan^2 \left(\frac{1}{\sqrt{4\lambda-1}}\right) \right. \nonumber\\
&& \left. -2\sqrt{4\lambda-1}\arctan \left(\frac{1}{\sqrt{4\lambda-1}}\right) +\frac{1}{2} +\log \left(\frac{m_{W}^{2}}{M_{R}^{2}}\right)  \right)\,, \nonumber\\
F_{L}^{(6)} &=& \frac{1}{32 \pi^{2}} \frac{m_{k}}{m_{W}} \left(Y_{\nu}^{} Y_{\nu}^{\dagger} \right)^{km} \frac{m_{H}^{2}}{M_{R}^{2}} \left( \sqrt{4\lambda-1}\arctan \left(\frac{1}{\sqrt{4\lambda-1}}\right) -\frac{3}{4} +\frac{\log \left(\frac{m_{W}^{2}}{M_{R}^{2}}\right)}{2}  \right) \,, \nonumber\\
F_{L}^{(7+9)} &=& 0 \,,\nonumber\\
F_{L}^{(8+10)} &=& -\frac{1}{32 \pi^{2}} \frac{m_{k}}{m_{W}} \Bigg[ \left(Y_{\nu}^{} Y_{\nu}^{\dagger}\right)^{km} \left( \Delta +1 -
\log (\frac{M_{R}^{2}}{\mu^2}) + \frac{m_{W}^{2} \log \left(\frac{m_{W}^{2}}{M_{R}^{2}}\right)}{M_{R}^{2}} \right)   \nonumber\\
&& -\frac{v^{2}}{M_{R}^{2}}\left(Y_{\nu}^{} Y_{\nu}^{\dagger} Y_{\nu}^{} Y_{\nu}^{\dagger} \right)^{km}  \Bigg] \, .  
\label{FLsimple_totdom}
\end{eqnarray}
And similar formulas can be obtained for the $F_R$ form factors.
Notice that in the results above we have included all the relevant contributions, i.e., up  to ${\cal O}(Y_\nu^2+Y_\nu^4)$  and it turns out, as announced in  \secref{sec:LFVHDMIA}, that they are just the diagrams (1)+(8)+(10) that provide contributions of  ${\cal O}(Y_\nu^4)$ with a $v^2/M_R^2$ dependence. 
The other diagrams will also give  ${\cal O}(Y_\nu^4)$ contributions but they will be suppressed since they go with a $v^4/M_R^4$ dependence, and we do not keep  these small contributions in our expansions.
\section{MIA form factors in the unitary gauge} 
\label{App:othergauge}
In order to check the gauge invariance of our results for the LFVHD form factors (and therefore the partial width) that we have computed in the MIA  by using the Feynman-'t Hooft gauge, we present here the computation of these same form factors but using a different gauge choice, in particular the unitary gauge (UG).   
We will demonstrate that when computing the MIA form factor $F_L$ to ${\cal O}(Y_{\nu}^{2}+Y_{\nu}^{4})$ we get the same result as in \eqref{FLdominant}. A similar demonstration can be done for $F_R$ but we do not include it here for shortness. For this exercise, we ignore the tiny terms suppressed by factors of the lepton masses as we did in \eqref{FLdominant}.

\begin{figure}[t!]
\begin{center}
\includegraphics[scale=0.8]{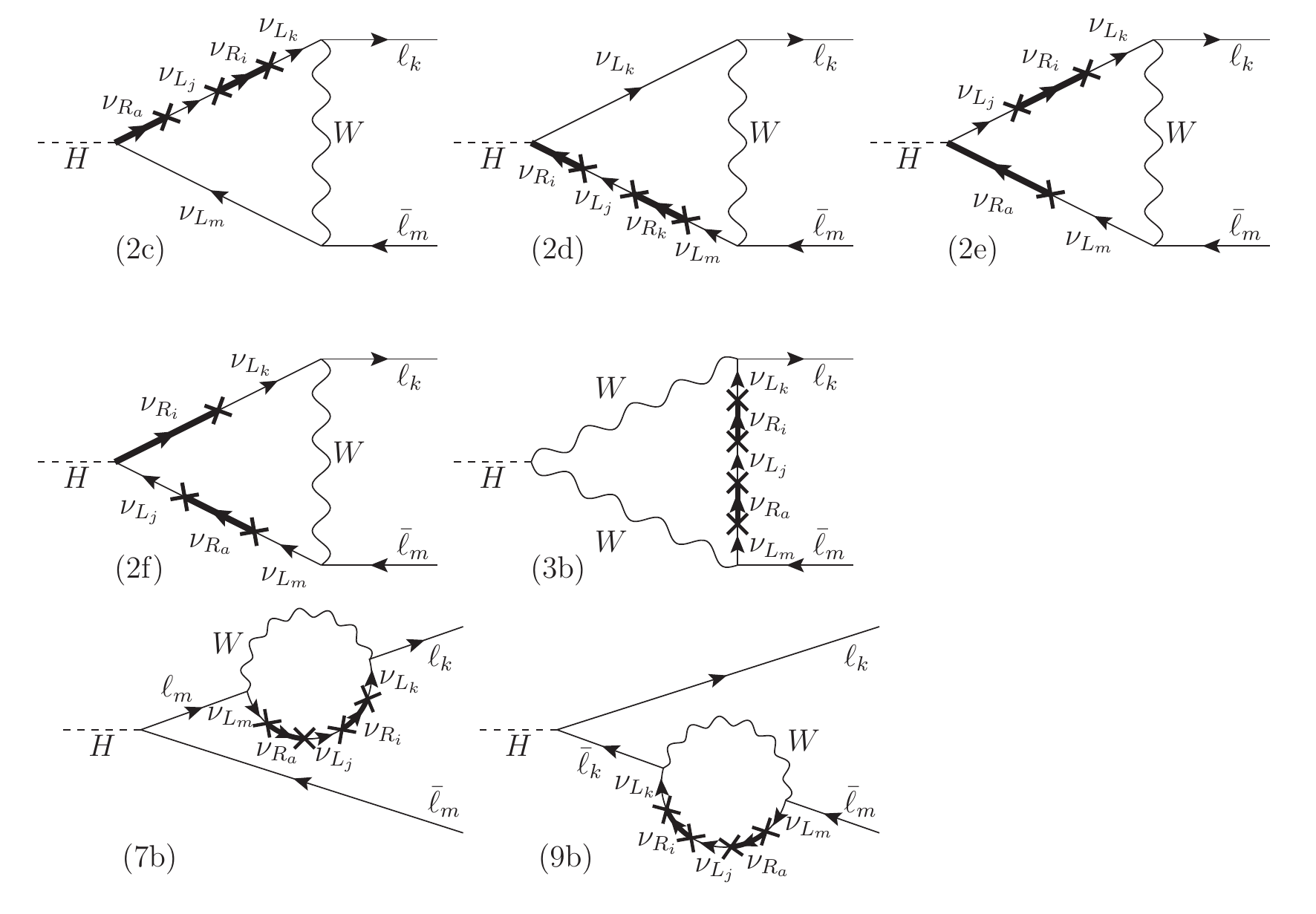}
\caption{Relevant diagrams for the form factors to ${\cal O}(Y_\nu^4)$ in the unitary gauge}
\label{diagsUG}
\end{center}
\end{figure}

First, we list the relevant one-loop diagrams contributing to the form factor $F_L$ in the UG. Since, in this gauge there are not Goldstone bosons, there will be just diagrams of type: (2), (3), (7) and (9).  
Generically, each of these diagrams will get contributions of  ${\cal O}(Y_{\nu}^{2})$ and  ${\cal O}(Y_{\nu}^{4})$. Second, we write the propagator of the $W$ gauge boson in the UG, $P_W^{\rm UG}$,  by splitting it into two parts, $P_W^{a}$ and $P_W^{b}$:
\be
P_W^{\rm UG}=P_W^{a}+P_W^{b}=-\frac{i g_{\mu \nu}}{p^{2}-m_{W}^{2}} +\frac{i p_{\mu}p_{\nu}}{m_{W}^{2}(p^{2}-m_{W}^{2})} \, ,
\ee
such that, $P_W^{a}$ coincides with the $W$ propagator in the Feynman-'t Hooft gauge. Then, each diagram of type (i), i=2,3,7,9,  will receive  three kind of contributions: 1) from the part $P_W^{a}$ one gets the same contributions to  ${\cal O}(Y_{\nu}^{2})$ as those we got  in the Feynman-'t Hooft gauge from the five diagrams (2a), (2b), (3a), (7a) and (9a)  in \figrefs{LFVHDMIAdiagsvertexY2} and \ref{LFVHDMIAdiagslegY2};  2) new contributions to ${\cal O}(Y_{\nu}^{2})$ that come from considering the new propagator term $P_W^b$ in these same diagrams (2a), (2b), (3a), (7a) and (9a); 3) contributions to ${\cal O}(Y_{\nu}^{4})$ that come from new diagrams which were not relevant in the Feynman-'t Hooft gauge, but they are relevant in the UG. By relevant we mean leading to dominant ${\cal O}(M_R^{-2})$ contributions in the large $M_R$ expansion. These new diagrams 
contributing to order ${\cal O}(Y_{\nu}^{4})$ in the UG are the seven diagrams shown in figure~\ref{diagsUG}. Thus, we get in total twelve one-loop diagrams contributing in the UG: (2a), (2b), (2c), (2d), (2e), (2f), (3a), (3b), (7a), (7b), (9a) and (9b). 

Next we present the results in the UG for each type of diagram (i), specifying the various contributions explained above, which for clarity we present correspondingly ordered in three lines,  the first line is for kind 1), the second line is for kind 2) and the third line is for kind 3). 
The UG $F_L$ form factors to  ${\cal O}(Y_{\nu}^{2}+Y_{\nu}^{4})$ that we get are, as follows:

 \begin{align}
F_{L}^{\rm UG (2)} &= -\frac{1}{16 \pi^{2}} m_{k}m_{W} \left(Y_{\nu}^{} Y_{\nu}^{\dagger} \right)^{km} \left( (C_{0}+C_{11}-C_{12})_{(2a)} +(C_{11}-C_{12})_{(2b)} \right) \nonumber \\
& +\frac{1}{32 \pi^{2}} \frac{m_{k}}{m_{W}} \left[ \left(Y_{\nu}^{} Y_{\nu}^{\dagger} \right)^{km} \left( (\tilde{C}_{0})_{(2a)}-(B_{1})_{(2b)} \right) \right. \nonumber\\
& \left. +\left(Y_{\nu}^{} Y_{\nu}^{\dagger} Y_{\nu}^{} Y_{\nu}^{\dagger} \right)^{km} v^{2} \left( -(C_{11})_{(2c)} +(\tilde{D}_{0})_{(2d)} +(\tilde{D}_{0}-(C_{11}-C_{12}))_{(2e)} -(C_{11}-C_{12})_{(2f)} \right) \right] \,, \nonumber \\  
F_{L}^{\rm UG(3)} &= \frac{1}{8 \pi^{2}} m_{k}m_{W}^{3} \left(Y_{\nu}^{} Y_{\nu}^{\dagger} \right)^{km} ( D_{12}-D_{13} )_{(3a)} \nonumber \\
  &  -\frac{1}{32 \pi^{2}} \left(Y_{\nu}^{} Y_{\nu}^{\dagger} \right)^{km} \frac{m_{k}}{m_{W}} \left[ 2B_{0}+B_{1} -(2m_{W}^{2}+m_{H}^{2})(C_{0}+C_{11}-C_{12}) +2m_{W}^{2}m_{H}^{2}D_{13} \right]_{(3a)} \nonumber\\
& -\frac{1}{32 \pi^{2}} \left(Y_{\nu}^{} Y_{\nu}^{\dagger} Y_{\nu}^{} Y_{\nu}^{\dagger} \right)^{km} \frac{m_{k}}{m_{W}} v^{2} \left[ 2C_{0}+C_{12} \right]_{(3b)} \,, \nonumber \\
F_{L}^{\rm UG(7)} &= \frac{1}{16 \pi^{2}} m_{k}m_{W} \frac{m_{m}^{2}}{m_{k}^{2}-m_{m}^{2}}  \left(Y_{\nu}^{} Y_{\nu}^{\dagger} \right)^{km} ( C_{12} )_{(7a)}  \nonumber \\
  & \frac{1}{32 \pi^{2}} \frac{m_{k}m_{m}^{2}}{m_{W}(m_{k}^{2}-m_{m}^{2})} \left[ \left(Y_{\nu}^{} Y_{\nu}^{\dagger} \right)^{km}( 2B_{0}+B_{1} )_{(7a)} \right. \nonumber\\
& \left. +\left(Y_{\nu}^{} Y_{\nu}^{\dagger} Y_{\nu}^{} Y_{\nu}^{\dagger} \right)^{km} v^{2}(2C_{0}+C_{12})_{(7b)} \right] \,, \nonumber \\
F_{L}^{\rm UG(9)} &= -\frac{1}{16 \pi^{2}} m_{k}m_{W} \frac{m_{m}^{2}}{m_{k}^{2}-m_{m}^{2}}  \left(Y_{\nu}^{} Y_{\nu}^{\dagger} \right)^{km} ( C_{12} )_{(9a)} \,,\nonumber \\ 
  & -\frac{1}{32 \pi^{2}} \frac{m_{k}m_{m}^{2}}{m_{W}(m_{k}^{2}-m_{m}^{2})} \left[ \left(Y_{\nu}^{} Y_{\nu}^{\dagger} \right)^{km}( 2B_{0}+B_{1} )_{(9a)} \right. \nonumber\\
& \left. +\left(Y_{\nu}^{} Y_{\nu}^{\dagger} Y_{\nu}^{} Y_{\nu}^{\dagger} \right)^{km} v^{2}(2C_{0}+C_{12})_{(9b)} \right] \, ,
\label{FL_Uxi1}
\end{align}
where the arguments of the one-loop functions are:
\be\begin{array}{rll}
 \tilde{C}_{0}, C_{i} =& \tilde{C}_{0}, C_{i}(p_{2},p_{1},m_{W},0,M_{R})  & \text{in } (2a) \nonumber \\
 B_{i} = & B_{i}(p_{lep},m_{W},M_{R})  & \text{in } (2b)  \nonumber \\
 C_{i} = & C_{i}(p_{2},p_{1},m_{W},M_{R},0)  & \text{in } (2b) \nonumber \\
 C_{i} =  & C_{i}(p_{lep},0,m_{W},M_{R},M_{R})  & \text{in } (2c)  \nonumber \\
 \tilde{D}_{0} = & \tilde{D}_{0}(p_{2},p_{1},0,m_{W},0,M_{R},M_{R})  & \text{in } (2d)  \nonumber \\
 C_{i} = &  C_{i}(p_{2},p_{1},m_{W},M_R,M_{R}) & \text{in } (2e),(2f)  \nonumber \\
 \tilde{D}_{0} =&  \tilde{D}_{0}(p_{2},0,p_{1},m_{W},0,M_{R},M_{R})  & \text{in } (2e)  \nonumber \\
 B_{i} = & B_{i}(p_{lep},M_{R},m_{W})  & \text{in } (3a), (7a), (9a)  \nonumber \\
 C_{i} = & C_{i}(p_{2},p_{1},M_{R},m_{W},m_{W})  & \text{in } (3a) \nonumber \\
 D_{i} = & D_{i}(0,p_{2},p_{1},0,M_{R},m_{W},m_{W})  & \text{in } (3a)  \nonumber \\
 C_{i} = & C_{i}(0,p_{lep},M_{R},M_R,m_{W})  & \text{in } (3b), (7b), (9b) \nonumber \\
 C_{i} = & C_{i}(0,p_{lep},0,M_R,m_{W})  & \text{in } (7a), (9a) \, . \nonumber 
\label{argfloops_U}
\end{array}\ee
The comparison of the previous results  with that in  \eqref{FLdominant} then goes as follows.  First, it is clear from the above results, that once again the contributions from diagrams (7) and (9) cancel out fully, as it happened in the Feynman-'t Hooft gauge.  
Therefore, $F_{L}^{\rm UG} =F_{L}^{\rm UG(2)}+ F_{L}^{\rm UG(3)}$. Then, the first line in $F_{L}^{\rm UG(2)}$ and the first line in $F_{L}^{\rm UG(3)}$ match correspondingly with the contributions from (2) and (3) in the Feynman-'t Hooft gauge.  Next, by using the relation,  
\be
 B_{0}(p_{lep},M_{R},m_{W})+B_{1}(p_{lep},M_{R},m_{W}) +B_1(p_{lep},m_{W},M_{R})= 0  \,,
\ee
we get that the sum of the second line in $F_{L}^{\rm UG(2)}$ and the second line in $F_{L}^{\rm UG(3)}$ gives exactly the contributions to 
${\cal O}(Y_\nu^2)$ from (1)+(8)+(10)+(4)+(5)+(6) in the Feynman-'t Hooft gauge. Finally, by using the relation 
\be
 C_{11}(p_{lep},0,m_{W},M_{R},M_{R})+(C_{0}+C_{12})(0,p_{lep},M_{R},M_{R},m_{W}) = 0  \, ,
\label{iden_compU}
\ee
we get that the sum of the third line in $F_{L}^{\rm UG(2)}$ and the third line in $F_{L}^{\rm UG(3)}$ gives exactly the contributions to  ${\cal O}(Y_\nu^4)$ from (1)+(8)+(10). Therefore, in summary, we get the identity of the total result for $F_L$ computed in both gauges, leading to the gauge invariant result of \eqref{FLdominant}.

\chapter{Form factors for LFVZD in the ISS model} 
\fancyhead[RO] {Form factors for LFVZD in the ISS model}
\label{App:LFVZD}

In this Appendix we give the analytical expressions for the form factors contributing to $Z\to\ell_k\bar\ell_m$, as they are defined in \eqrefs{LFVZDwidth} and (\ref{LFVZDFF}). 
In the Feynman-t'Hooft gauge, they are obtained by computing the ten diagrams shown in \figref{LFVZDDiagrams}. 
We take the results from~\cite{Illana:1999ww,Illana:2000ic,Abada:2014cca} and adapt them to our notation and to the convection of {\it LoopTools}~\cite{Hahn:1998yk} for the loop functions. 

The form factors of the different diagrams are
\begin{equation}
\mathcal F_Z^{(1)} = \frac12\, B_{\ell_k n_i}^{} B^*_{\ell_m n_j}\left\{ -C_{n_i n_j} \,x_i x_j \, m_W^2 C_0 + C_{n_i n_j}^* \sqrt{x_ix_j} \Big[m_Z^2\, C_{12} - 2 C_{00}+\frac12\Big]\right\},
\end{equation}
where $C_{0,12,00}\equiv C_{0,12,00}(0,m_Z^2,0,m_W^2,m_{n_i}^2, m_{n_j}^2)$;
\begin{equation}
\mathcal F_Z^{(2)}  = B_{\ell_k n_i}^{}B^*_{\ell_m n_j}\left\{ -C_{n_i n_j} \Big[m_Z^2 \Big( C_0 + C_1 +C_2 + C_{12} \Big) - 2 C_{00} +1  \Big]+ C_{n_i n_j}^* \sqrt{x_ix_j} \, m_W^2 C_0\right\},
\end{equation}
where $C_{0,1,2,12,00}\equiv C_{0,1,2,12,00}(0,m_Z^2,0,m_W^2,m_{n_i}^2, m_{n_j}^2)$;
\begin{equation}
\mathcal F_Z^{(3)} = 2c_W^2 B_{\ell_k n_i}^{} B^*_{\ell_m n_i}\left\{  m_Z^2 \Big(C_1+C_2+C_{12}\Big)-6 C_{00}+1\right\} ,
\end{equation}
where $C_{1,2,12,00}\equiv C_{1,2,12,00}(0,m_Z^2,0,m_{n_i}^2,m_W^2,m_W^2)$;
\begin{equation}
\mathcal F_Z^{(4)}+\mathcal F_Z^{(5)}  = -2s_W^2\, B_{\ell_k n_i}^{} B^*_{\ell_m n_i}  \, x_i\, m_W^2  C_0 ,
\end{equation}
where $C_{0}\equiv C_{0}(0,m_Z^2,0,m_{n_i}^2,m_W^2,m_W^2)$;
\begin{equation}
\mathcal F_Z^{(6)} = -(1-2s_W^2)\,  B_{\ell_k n_i}^{} B^*_{\ell_m n_i}\,  x_i\, C_{00},
\end{equation}
where  $C_{00}\equiv C_{00}(0,m_Z^2,0,m_{n_i}^2,m_W^2,m_W^2)$;
\begin{equation}
\mathcal F_Z^{(7)}+\mathcal F_Z^{(8)}+\mathcal F_Z^{(9)}+\mathcal F_Z^{(10)}  = \frac12(1-2c_W^2)\, B_{\ell_k n_i}^{} B^*_{\ell_m n_i}\left\{(2+x_i)B_1+1\right\},
\end{equation}
where $B_1\equiv B_1(0,m_{n_i}^2,m_W^2)$.

In all these formulas,  sum over neutrino indices, $i,j=1,... ,9$ has to be understood, $x_i\equiv~m_{n_i}^2/m_W^2$ and  the charged lepton masses have been neglected.  

\end{appendices}

\fancyhead{}

\chapter*{Agradecimientos}
 
 Se acaba aqu\'i un largo viaje. 
 Uno se siente como Frodo y Sam en el Monte del Destino, agotado tras un duro camino, pero satisfecho por haberlo completado. 
 Durante este tiempo me he pegado con libros, c\'alculos y programas, he discutido con gente hasta la saciedad y he invertido  horas como para haber ido y vuelto a Mordor un par de veces. 
 A\'un as\'i, mirando atr\'as me doy cuenta de todo lo que he aprendido y, sobre todo, de todo lo que me he divertido. 
 Porque de eso deber\'ia tratarse todo esto, de divertirse. 
Y yo lo he podido hacer gracias a mucha gente a la que, llegado este momento, no tengo m\'as que palabras de agradecimiento. 
 
La primera responsable de todo esto ha sido la jefa, Mar\'ia Jos\'e.  
Puede que hayamos tenido momentos malos, como todos, pero tambi\'en los hemos tenido buenos. 
Desde el m\'aster, han sido cinco a\~nos de discusiones por hacerse con el trono de la cabezoner\'ia. 
Reconozco que a d\'ia de hoy no s\'e qui\'en ha ganado esa batalla. 
Eso s\'i, me ha servido para aprender f\'isica y mucho m\'as.
Me has ense\~nado a pelear por lo que creo, por lo que pienso, a seguir lo que dice mi intuici\'on aunque parezca un camino sin salida, porque as\'i es como al final se encuentra la soluci\'on a cualquier problema en la f\'isica y en la vida.
Contigo he dado los primeros pasos como investigador, y eso es algo que llevar\'e siempre conmigo.
 
 Echando la vista atr\'as, me veo en la carrera, enfrent\'andome a mi primeros problemas de f\'isica junto a mis compa\~neros  fisikis de clase, y con esto me refiero obviamente a lanzar tizas sobre una diana con quien-no-ha-de-ser-nombrado en la pizarra. Eso s\'i que se nos daba bien. 
 Disfrut\'e mucho estudiando con Alize, Ander y Mar\'ia. Form\'abamos un gran grupo, el mejor en el que he estado, capaces de resolver hasta los problemas que no ten\'ian soluci\'on.
 Fueron muchas horas juntos, sobre todo con Reto. 
Juntos demostramos que se puede sacar una carrera de Bikain con especializaci\'on en ping-pong bajo climatolog\'ia extrema. 
 
Despu\'es aterric\'e en el IFT, un entorno inmejorable para hacer estas cosas raras que hacemos y que tanto me gustan.
Sobre todo por la gente que vive en \'el. 
Aqu\'i he conocido a colegas y amigos para toda la vida, porque pasar por todo esto juntos une, mucho. 
Y si con alguien he pasado tiempo ha sido con Josu. 
M\'aster, escuelas, congresos... nos hemos sentido m\'as tontos que nadie muchas veces, pero haberlo hecho juntos ha ayudado a salir adelante y a solucionar los problemas de una patada. 

Javi, t\'u llegaste el \'ultimo al despacho, pero pronto te hiciste merecedor de un sitio en \'el. 
Has contribuido a las  cartas, Simpsons y f\'utbol, incluso como asesor de comunio, poco m\'as se puede pedir. 
Me quedo tranquilo sabiendo que dejamos el despacho y la corona de King in the Poch en buenas manos. Cu\'idalos y, sobre todo, no los pierdas en una apuesta loca. 
 
V\'ictor, t\'u has sido un ejemplo a seguir. 
En ti he visto siempre c\'omo sobrevivir en este  phenomundo, a veces tan duro y casi injusto para los estudiantes, que curran m\'as que nadie y parece que no les salen las cosas nunca. 
Pero t\'u eras capaz de afrontarlo con una sonrisa, vi\'endole el lado positivo, y demostrando que si curras al final salen las cosas.
Me ha gustado tenerte por aqu\'i, me he divertido mucho contigo hablando de f\'isica y preparando las clases, y espero que por fin nos pongamos a currar juntos.
 
Irene, Miguel... erais bichos raros, siempre jugando con cuerdas, pero os quer\'iamos igual y nos re\'iamos de, o sea con, vosotros. 
Eso s\'i, todav\'ia sigo esperando el v\'ideo...
Junto con Ana, Leyre, Aitor y Santi me he divertido mucho, tanto dentro como fuera del IFT, jugando a la Pocha o en la Pocha. 
Tambi\'en con los miembros del IFuTsal, y espero que ahora que no me tendr\'eis de lastre pod\'ais llevar una copa a la quinta planta.
 
M\'as all\'a de los estudiantes, el IFT me ha dado la oportunidad de discutir con expertos de este mundillo y de fre\'irles a preguntas. 
Gracias a Enrique por su paciencia, por haber contestado a todas mis dudas, por haberme generado unas nuevas y por volver a contestarlas.  
A Carlos, por saber de todo, y por haberme dado la oportunidad de asomarme al mundillo de la divulgavi\'on. 
A Sven, por su superilusi\'on por la fenomenolog\'ia y por la cerveza, algo que espero consiga plasmar en un {\it Pheno-Beer} alg\'un d\'ia.
A Jes\'us, por compartir conmigo su infinita bibliograf\'ia. 
A Luca, Choco {\it et al}, por demostrarme que se puede ser un gran f\'isico y a la vez tener una gran parte humana, aunque todav\'ia no te he perdonado el l\'io del {\it Pheno-Coffee}...
Y, sobre todo, a Juanjo.
Eres un torbellino de sabidur\'ia, de pasi\'on por la f\'isica.
De ti he aprendido mucha f\'isica, pero tambi\'en a disfrutar de ella. 
Reconozco que la cuarta planta del IFT no ha sido lo mismo sin ti. 
  
Gracias a los miembros del tribunal, por haber aceptado mi invitaci\'on y dedicado tiempo a valorar esta Tesis. 
  
Quiero agradec\'erselo tambi\'en a mis hermanos mayores. 
Miguel, t\'u nos acogiste en tu despacho y nos ense\~naste las cosas b\'asicas para sobrevivir en este mundo, como la de hacerse con un sof\'a y un proyector para el despacho.
Ana, nos diste el primer empujoncito para echar a volar.
Y Ernesto, que te voy a decir a ti, has estado ah\'i siempre, ayud\'andome con todo, en el trabajo y fuera de \'el.
Me lo he pasado muy bien currando contigo y espero que sigamos haci\'endolo, que si no te quedas sin entradas para Ipurua.

Durante estos a\~nos he podido trabajar con gente asombrosa. 
Adem\'as de a los ya mencionados, debo agradec\'erselo  a C\'edric, pues en sus dos a\~nos en Madrid tuvo la paciencia de compartir sus conocimientos sobre lo que al final han sido los cimientos de esta Tesis.
A mi madrina Valentina, por todos sus consejos. 
A Ale y Roberto, por haberos calculado hasta los loops de los cordones de los zapatos. 
A Richard, aunque no hayamos trabajado juntos todav\'ia, por ense\~narme que la {\it collider  physics} puede ser divertida. 
A la Tropa Quiral y sus reuniones que bien podr\'ian inspirar una historia de 13 Rue del Percebe. 
Gracias a todos por ayudarme a sacar esto adelante. 

Fuera de la f\'isica he contado con el apoyo de mucha gente. 
En Madrid he tenido a mis amigos medio daneses, en especial a los Sergios, Miguel y Blabla, y este \'ultimo hay que incluirlo por todo lo que habl\'ais.
Gracias tambi\'en a Carmen y Luis, por haberme acogido y haberme hecho sentir como en casa desde el primer momento. 

En Eibar se qued\'o mi familia, la que nunca se ha olvidado de m\'i pese a tantos a\~nos fuera. 
Quiero agradecerle a mi Ama, por haber peleado siempre y tanto por todos nosotros; a mi Aita, por ense\~narme tantas cosas, como el f\'utbol, la cocina, los puzzles... y, sobre todo, el Athletic; a mi hermana, por hacernos re\'ir con sus extravagantes historias; a mis t\'ias y t\'ios, por haberme mostrado siempre todo su apoyo, hiciese lo que hiciese; a mi Amama, por hacer los mejores fritos del mundo; y sobre todo a mi Aitxitxa, por haberme metido los n\'umeros en la cabeza con todas  esas tardes ense\~n\'andome a multiplicar en el sal\'on de su casa,  por hacerme ver lo divertido que es preguntarse el por qu\'e de las cosas. Eskerrik asko.

Luego est\'a la otra familia, la que se elige. 
La mosca, Ibargain, Argatxa... el nombre es lo de menos, la gente no cambia. 
Gracias por estar ah\'i siempre, por ofrecerme un sitio donde ir a olvidarme de todo, a hacer el gandul sin pensar en nada m\'as, algo en lo que somos de primera. 
Gracias por no dejarme perder la Ilusi\'on de volver a Tierra Santa y por pintarme la cara color esperanza. 
MGHP.

 Ahora bien, hay una persona a la que tengo que agradec\'erselo m\'as que a nadie y, por desgracia, las palabras se quedan cortas para ello.
 Me has ayudado dentro y fuera de la f\'isica, hasta me has llevado al Zoo a ver mapaches...
 Claudia, hablo de ti, y de toda tu ayuda para hacer realidad esta tesis, de tu paciencia por haber aguantado mis momentos malos y ayudado a que fuesen buenos, de lo bien que lo pasamos juntos.
 He descubierto que la f\'isica me gusta m\'as si es contigo, que todo me gusta m\'as si es contigo. 
 O simplemente, Fringe.
 Intensa es la fuerza en nosotros, aprovech\'emoslo.


\bibliography{bibliography}



\end{document}